\documentclass[10pt,letterpaper]{article}
\pdfoutput=1
\usepackage[margin=1in]{geometry}
\usepackage[utf8]{inputenc}
\usepackage[T1]{fontenc}
\usepackage[nopatch=footnote]{microtype}
\usepackage{amsfonts}
\usepackage{amssymb}
\usepackage{amsmath}
\usepackage{amsthm}
\usepackage[inline]{enumitem}
\usepackage{setspace}
\usepackage[hidelinks,bookmarks,bookmarksopen,bookmarksnumbered]{hyperref}
\usepackage[capitalize]{cleveref}
\usepackage{mathtools}
\usepackage{tikz}
\usepackage[font=small,labelfont=bf]{caption}
\usepackage{titlesec}
\usepackage{thmtools}
\makeatletter
\IfFormatAtLeastTF{2026-06-01}{%
  \renewcommand\thmt@autorefsetup{%
    \@xa\def\csname\thmt@envname autorefname\@xa\endcsname\@xa{\thmt@thmname}%
  }%
}{}
\makeatother
\usepackage{wrapfig}
\usepackage{framed}
\usepackage{soul}

\usetikzlibrary{trees}
\usetikzlibrary{arrows}
\usetikzlibrary{arrows.meta,arrows}
\usetikzlibrary{decorations.pathreplacing}
\usetikzlibrary{positioning}
\usetikzlibrary{calc}

\newtheorem{theorem}{Theorem}[section]

\newtheorem{lemma}[theorem]{Lemma}

\newtheorem{observation}[theorem]{Observation}
\crefname{observation}{Observation}{Observations}
\newtheorem{proposition}[theorem]{Proposition}
\theoremstyle{definition}
\newtheorem{definition}[theorem]{Definition}

\crefname{construction}{Construction}{Constructions}
\crefname{claim}{Claim}{Claims}
\theoremstyle{remark}
\newtheorem{remark}[theorem]{Remark}
\newtheorem{example}[theorem]{Example}

\makeatletter
\renewcommand\thmt@autorefsetup{\@xa\def\csname\thmt@envname autorefname\@xa\endcsname\@xa{\thmt@thmname}}
\makeatother

\makeatletter
\@ifundefined{if@skiphyperref}
  {\PackageError{paper}{Hyperref target suppression is unavailable}{}}
  {}
\newcommand{\paper@refstepcounterwithouttarget}[2]{%
  \let\paper@saved@skiphyperref\if@skiphyperref
  \@skiphyperreftrue
  #2%
  \let\if@skiphyperref\paper@saved@skiphyperref
  \hyper@makecurrent{#1}%
}
\newcommand{\thm@deferanchor}[2]{%
  \paper@refstepcounterwithouttarget{#1}{#2}%
  \edef\thm@anchor{%
    \noexpand\Hy@raisedlink{%
      \noexpand\hyper@anchorstart{\@currentHref}\noexpand\hyper@anchorend}}%
}
\let\thm@anchor\@empty
\patchcmd{\cref@thmnoarg}
  {\refstepcounter{#2}}
  {\thm@deferanchor{#2}{\refstepcounter{#2}}}
  {}{\PackageError{paper}{Cannot patch theorem destinations}{}}
\patchcmd{\cref@thmoptarg}
  {\refstepcounter[#1]{#3}}
  {\thm@deferanchor{#3}{\refstepcounter[#1]{#3}}}
  {}{\PackageError{paper}{Cannot patch shared-counter theorem destinations}{}}
\let\thm@original@deferredhead\deferred@thm@head
\def\deferred@thm@head#1{%
  \thm@original@deferredhead{\thm@anchor#1}%
  \let\thm@anchor\@empty
}
\AtBeginDocument{%
\@ifundefined{ttl@Hy@refstepcounter}
  {\PackageError{paper}{Cannot defer section destinations}{}}
  {\renewcommand{\ttl@Hy@refstepcounter}[1]{%
    \paper@refstepcounterwithouttarget{#1}{\refstepcounter{#1}}%
    \edef\ttl@Hy@saveanchor{%
      \noexpand\Hy@raisedlink{%
        \noexpand\hyper@anchorstart{\@currentHref}\noexpand\hyper@anchorend}}%
  }}
}
\makeatother

\crefname{subsection}{Section}{Sections}

\definecolor{mypink}{RGB}{199, 21, 133}
\hypersetup{
  colorlinks,
  urlcolor = blue,
  linkcolor = mypink, %
  citecolor = blue %
}

\setlist[enumerate]{nosep, topsep=1ex}
\setlist[itemize]{nosep, topsep=1ex}
\setlist[description]{nosep,topsep=1ex,before=\setlength{\parskip}{0pt}}

\newenvironment{problem}[1][]{%
  \setlength{\FrameSep}{1.5ex}%
  \begin{framed}%
  \crefalias{theorem}{problem}%
  \refstepcounter{theorem}%
  \noindent\textbf{Problem \thetheorem:} \textsc{#1}%
}{%
  \vspace{-1.3ex}%
  \end{framed}%
}
\crefname{problem}{Problem}{Problems}
\Crefname{problem}{Problem}{Problems}

\newcommand{\DSComponents}{%
  \par\medskip%
  \noindent%
  \textbf{\textit{Data structure}}\quad%
}
\newcommand{\DSQueries}{%
  \par\medskip%
  \noindent%
  \textbf{\textit{Queries}}\quad%
}
\newcommand{\DSConstruction}{%
  \par\medskip%
  \noindent%
  \textbf{\textit{Construction}}\quad%
}

\allowdisplaybreaks

\setul{2pt}{0.6pt}

\let\oldabstract\abstract
\let\oldendabstract\endabstract
\makeatletter
\renewenvironment{abstract}
{%
  {\list{}{\addtolength{\leftmargin}{0.2em}%
    \listparindent 1.5em%
     \itemindent    \listparindent%
     \rightmargin   \leftmargin%
     \parsep        \z@ \@plus\p@}%
     \item\relax}%
  {\endlist}%
\oldabstract}
{\oldendabstract}
\makeatother

\newcommand{\bigO}{\mathcal{O}}

\newcommand{\floor}[1]{\left\lfloor #1 \right\rfloor}
\newcommand{\dd}{\mathinner{.\,.}}
\newcommand{\probname}[1]{\text{\sc #1}}

\newcommand{\Pat}{P}
\newcommand{\Text}{T}
\newcommand{\Textinf}{\Text^{\infty}}
\newcommand{\Textlen}{n}
\newcommand{\AlphabetSize}{\sigma}
\newcommand{\IntegerAlphabet}{[0 \dd \AlphabetSize)}
\newcommand{\BinaryAlphabet}{\{{\tt 0}, {\tt 1}\}}
\newcommand{\emptystring}{\varepsilon}
\newcommand{\deltatext}{\delta_{\rm text}}
\newcommand{\SSS}{\mathsf{S}}

\newcommand{\PackedRepresentation}[3]{\mathrm{packed}_{#1,#2}(#3)}
\newcommand{\PackedSeqRepresentation}[3]{\PackedRepresentation{#1}{#2}{#3}}

\newcommand{\Z}{\mathbb{Z}}
\newcommand{\Zz}{\Z_{\geq 0}}

\newcommand{\Zp}{\Z_{\geq 1}}

\newcommand{\SA}[1]{\mathrm{SA}_{#1}}
\newcommand{\ISA}[1]{\mathrm{ISA}_{#1}}
\newcommand{\LexSorted}[2]{\mathrm{LexSorted}(#1,#2)}

\newcommand{\LPF}[1]{\mathrm{LPF}_{#1}}
\newcommand{\LPnF}[1]{\mathrm{LPnF}_{#1}}

\newcommand{\Rank}[3]{\mathsf{rank}_{#1}(#2,#3)}
\newcommand{\LCE}[3]{\mathrm{LCE}_{#1}(#2,#3)}
\newcommand{\lcp}[2]{\mathrm{lcp}(#1,#2)}
\newcommand{\per}[1]{\mathrm{per}(#1)}
\newcommand{\revstr}[1]{\overline{#1}}
\newcommand{\Successor}[2]{\mathrm{succ}_{#1}(#2)}
\newcommand{\RL}[1]{\mathrm{RL}(#1)}
\newcommand{\BWT}[1]{\mathrm{BWT}_{#1}}
\newcommand{\LCF}[2]{\mathrm{LCF}(#1, #2)}
\newcommand{\Val}[2]{\mathrm{val}_{#1}(#2)}
\newcommand{\InversionCount}[1]{\mathsf{inv\mbox{-}count}(#1)}
\newcommand{\ColoredInversionCount}[2]{\mathsf{colored\mbox{-}inv\mbox{-}count}(#1,#2)}
\newcommand{\bin}[2]{{\rm bin}_{#1}(#2)}
\newcommand{\pad}[1]{{\rm pad}(#1)}

\newcommand{\InsertSubseq}[2]{\mathsf{insert}(#1,#2)}
\newcommand{\DeleteSubseq}[2]{\mathsf{delete}(#1,#2)}

\newcommand{\LZSizeSym}{z}
\newcommand{\LZNonOvSizeSym}{z_{\rm no}}
\newcommand{\RLBWTSizeSym}{r}
\newcommand{\LZSize}[1]{\LZSizeSym(#1)}
\newcommand{\LZNonOvSize}[1]{\LZNonOvSizeSym(#1)}
\newcommand{\RLBWTSize}[1]{\RLBWTSizeSym(#1)}

\newcommand{\OccTwo}[2]{\mathrm{Occ}(#1, #2)}
\newcommand{\RangeBegTwo}[2]{\mathrm{RangeBeg}(#1, #2)}
\newcommand{\RangeEndTwo}[2]{\mathrm{RangeEnd}(#1, #2)}

\newcommand{\DistPrefixPos}[4]{\mathrm{DistPrefix}(#1,#2,#3,#4)}
\newcommand{\DistPrefixes}[3]{\mathcal{D}(#1, #2, #3)}

\newcommand{\RootPos}[3]{\mathrm{root}(#1,#2,#3)}
\newcommand{\RootPat}[2]{\mathrm{root}(#1,#2)}
\newcommand{\HeadPos}[3]{\mathrm{head}(#1,#2,#3)}
\newcommand{\HeadPat}[2]{\mathrm{head}(#1,#2)}
\newcommand{\TailPos}[3]{\mathrm{tail}(#1,#2,#3)}
\newcommand{\TailPat}[2]{\mathrm{tail}(#1,#2)}
\newcommand{\ExpPos}[3]{\mathrm{exp}(#1,#2,#3)}
\newcommand{\ExpPat}[2]{\mathrm{exp}(#1,#2)}
\newcommand{\TypePos}[3]{\mathrm{type}(#1,#2,#3)}
\newcommand{\TypePat}[2]{\mathrm{type}(#1,#2)}
\newcommand{\RunEndFullPos}[3]{e^{\rm full}(#1,#2,#3)}
\newcommand{\RunEndFullPat}[2]{e^{\rm full}(#1,#2)}
\newcommand{\RunEndPos}[3]{e(#1,#2,#3)}
\newcommand{\RunEndPat}[2]{e(#1,#2)}

\newcommand{\RName}{\mathsf{R}}
\newcommand{\RTwo}[2]{\RName(#1, #2)}
\newcommand{\RThree}[3]{\RName_{#1}(#2, #3)}
\newcommand{\RFour}[4]{\RName_{#1,#2}(#3, #4)}
\newcommand{\RFive}[5]{\RName_{#1,#2,#3}(#4, #5)}

\newcommand{\RPrimName}{\RName'}
\newcommand{\RPrimTwo}[2]{\RPrimName(#1, #2)}

\newcommand{\dol}{\text{\normalfont $\texttt{\$}$}}
\newcommand{\hash}{\text{\normalfont $\texttt{\#}$}}
\newcommand{\zero}{{\tt 0}}
\newcommand{\one}{{\tt 1}}
\newcommand{\two}{{\tt 2}}
\newcommand{\three}{{\tt 3}}
\newcommand{\four}{{\tt 4}}
\newcommand{\five}{{\tt 5}}
\newcommand{\six}{{\tt 6}}
\newcommand{\seven}{{\tt 7}}
\newcommand{\eight}{{\tt 8}}

\hypersetup{
  pdftitle = {On the Hardness Hierarchy for the O(n sqrt(log n)) Complexity
    in the Word RAM},
  pdfauthor = {Dominik Kempa, Tomasz Kociumaka}
}

\begin{document}
\pagenumbering{roman}

\title{On the Hardness Hierarchy for the $\mathcal{O}(n \sqrt{\log n})$ Complexity\\ in the Word RAM\thanks{
  The conference version of this paper appeared in Proceedings of the
  57th Annual ACM Symposium on Theory of Computing (STOC) 2025,
  pp. 290--300; see \url{https://doi.org/10.1145/3717823.3718291}.
}}

\author{
  \large Dominik Kempa\thanks{Partially funded by the
  NSF CAREER Award 2337891 and the
  Simons Foundation Junior Faculty Fellowship.}\\[-0.3ex]
  \normalsize Department of Computer Science,\\[-0.3ex]
  \normalsize Stony Brook University,\\[-0.3ex]
  \normalsize Stony Brook, NY, USA\\[-0.3ex]
  \normalsize \texttt{kempa@cs.stonybrook.edu}
  \and
  \large Tomasz Kociumaka\thanks{A part of this work was prepared when
  T.\ Kociumaka was funded by the Ministry of Education and Science of
  Bulgaria's support for INSAIT, Sofia University ``St. Kliment Ohridski'',
  as part of the Bulgarian National Roadmap for Research Infrastructure.}\\[-0.3ex]
  \normalsize Max Planck Institute for Informatics,\\[-0.3ex]
  \normalsize Saarland Informatics Campus,\\[-0.3ex]
  \normalsize Saarbrücken, Germany\\[-0.3ex]
  \normalsize \texttt{tomasz.kociumaka@mpi-inf.mpg.de}
}

\date{\vspace{-0.5cm}}
\maketitle

\begin{abstract}
  In this work, we study the relative hardness of fundamental problems
  with state-of-the-art word RAM algorithms taking $\bigO(n\sqrt{\log
    n})$ time for instances described in $\Theta(n)$ machine words
  (i.e., $\Theta(n\log n)$ bits).  The word RAM model nowadays serves
  as the default model of computation for sequential algorithms, and
  understanding its limitations lies at the core of theoretical
  computer science.  The class of problems solvable in $\bigO(n
  \sqrt{\log n})$ time is one of the six \emph{levels of hardness}
  listed in the seminal paper of Chan and P\u{a}tra\c{s}cu [SODA
    2010]. According to the current state of knowledge, this class
  characterizes problems from several domains, including counting
  inversions, string processing problems (BWT Construction, LZ77 Factorization,
  Longest Common Substring, Batched Longest Previous Factor Queries,
  Batched Inverse Suffix Array Queries), and computational geometry problems
  (Orthogonal Range Counting, Orthogonal Segment Intersection Counting).  Our
  contribution is twofold:

  \begin{itemize}
  \item We present several new connections between the aforementioned
    string problems and the classical \emph{Dictionary Matching} problem,
    which asks whether a given text contains an exact occurrence of
    at least one of the given patterns.  A solution based on the
    Aho--Corasick automaton dates back to 1975.  In this work, we restrict Dictionary
    Matching to instances with $\Theta(n)$ binary patterns of common length
    $m=\Theta(\log n)$, short enough to be stored using $\bigO(1)$
    machine words each, and we prove that, unless this problem can be
    solved faster than the current bound of $\bigO(n\sqrt{\log n})$,
    most fundamental string problems cannot be solved faster either.
 
  \item With further reductions, we extend this hierarchy beyond
    string problems, proving that computational tasks like counting
    inversions---a fundamental component in geometric
    algorithms---inherit this hardness. This, in turn, establishes the
    hardness of Orthogonal Range Counting and Orthogonal Segment
    Intersection Counting.  The key to extending our results to other domains
    is a surprising equivalent characterization of Dictionary Matching
    in terms of a new problem we call \emph{String Nesting}, which,
    through a chain of three more reductions, can be solved by
    counting inversions.
  \end{itemize}

  Put together, our results unveil a single hard problem, with two
  different but equivalent formulations, that underlies the hardness
  of nearly all known major problems, coming from different domains,
  currently occupying the $\bigO(n \sqrt{\log n})$ level of hardness.
  These results drastically funnel further efforts to improve the
  complexity of near-linear problems.
  
  Many of our reductions hold even for simpler versions of basic
  problems, such as determining the \emph{parity of the number of
    phrases} in the LZ77 factorization or the \emph{number of runs} in
  the BWT.  This yields stronger results that can be used to design
  future reductions more easily.  As an auxiliary outcome of our
  framework, we also prove that several central string problems in the
  word RAM model do not get easier when limited to strings over the binary
  alphabet.  Our reductions to the binary case simplify the currently
  fastest algorithms for many classical problems, including LZ77
  Factorization and Longest Common Substring.
\end{abstract}

\thispagestyle{empty}
\clearpage

\thispagestyle{empty}
\newcommand{\ReductionFigureScale}{0.85}

\usetikzlibrary{decorations.markings}

\newcommand{\ReductionLinkRectangle}[3]{%
  \pdfannot width #2 height #3 depth 0pt {%
    /Subtype /Link /Border [0 0 0]
    /A << /S /GoTo /D (\pdfescapestring{\getrefbykeydefault{#1}{anchor}{}}) >>
  }%
  \phantom{\rule{#2}{#3}}%
}
\newcommand{\ReductionNodeLink}[2]{%
  \begin{scope}[reset cm]
  \path let \p1=(#1.south west), \p2=(#1.north east) in
    node[overlay,anchor=south west,draw=none,fill=none,
      inner sep=0pt,outer sep=0pt] at (\p1)
    {\ReductionLinkRectangle{#2}{\dimexpr\x2-\x1\relax}{\dimexpr\y2-\y1\relax}};
  \end{scope}
}

\tikzset{reduction link/.style={overlay,decorate,decoration={markings,
  mark=between positions 2pt and -2pt step 2pt with {
    \node[anchor=center,draw=none,fill=none,inner sep=0pt,outer sep=0pt]
      {\ReductionLinkRectangle{#1}{4pt}{4pt}};
  }}}}

\definecolor{myblue}{HTML}{ffffff}
\definecolor{mygreen}{HTML}{ffffff}
\definecolor{myred}{HTML}{ffffff}

\begin{figure}[p]
  \centering
  \resizebox{\ReductionFigureScale\linewidth}{!}{%
  \begin{tikzpicture}[
      edge from parent/.style={draw, thin, -{Stealth[scale=1.2]}},
      level 1/.style={sibling distance=25mm, level distance=22mm},
      level 2/.style={sibling distance=29mm, level distance=21mm},
      level 3/.style={sibling distance=27mm, level distance=20mm},
      every node/.style={rectangle, rounded corners, draw, thin, align=center},
      scale = 0.8
    ]
    \normalsize
    \setstretch{0.9}

    \node[fill=mygreen] (dm) {\textsc{Dictionary} \textsc{Matching}}
      child {node[fill=myblue] (lcf) {\textsc{Longest} \\  \textsc{Common}\\ \textsc{Factor}}
      }
      child { node[fill=myblue] (rlbwt) {\textsc{RLBWT}\\ \textsc{Size}}
        child {
          node[fill=myred!60] (bwt) {\textsc{BWT}\\ \textsc{Construction}}
        }
      }
      child {
        node[fill=myblue] (isa) {\textsc{Batched} \\ \textsc{ISA}\\ \textsc{Queries}}
      }
      child { node[fill=mygreen] (lzparity) {\textsc{LZ77}\\ \textsc{Parity}}
        child {node[fill=myblue] (lzsize) {\textsc{LZ77}\\ \textsc{Size}}
          child { node[fill=myred!60] (lzfactor) {\textsc{LZ77}\\ \textsc{Factorization}}}
        }
      }
      child {
        node[fill=myblue] (lpf) {\textsc{Batched} \\ \textsc{LPF}\\ \textsc{Queries}}
      };
    \node[left=4.8cm of dm,fill=mygreen] (bn) {\textsc{String} \textsc{Nesting}}
      child { node[fill=mygreen] (rps) {\textsc{Range} \\ \textsc{Prefix} \\ \textsc{Search}}
        child {
          node[fill=myblue] (cic) {\textsc{Counting}\\ \textsc{Colored}\\ \textsc{Inversions}}
        }
      };

    \node[right=1.2cm of cic,fill=myblue] (ic) {\textsc{Counting} \\ \textsc{Inversions}};
    \node[below left=0.6cm and -0.8cm of ic,fill=myblue] (orc) {\textsc{Orthogonal} \\ \textsc{Range} \\ \textsc{Counting}};
    \node[below right=0.6cm and -0.8cm of ic,fill=myblue] (osi) {\textsc{Orthogonal} \\ \textsc{Segment} \\ \textsc{Intersection} \\ \textsc{Counting}};

    \draw[->,thin, -{Stealth[scale=1.2]}] ($(bn.east)+(0.03,0.2)$) to[bend left=25] ($(dm.west)+(0.03,0.2)$);
    \draw[->,thin, -{Stealth[scale=1.2]}] ($(dm.west)+(0.03,-0.2)$) to[bend left=25] ($(bn.east)+(0.03,-0.2)$);
    \draw[->,thin, -{Stealth[scale=1.2]}] ($(cic.east)+(0.03,0.2)$) to[bend left=25] ($(ic.west)+(0.03,0.2)$);
    \draw[->,thin, -{Stealth[scale=1.2]}] ($(ic.west)+(0.03,-0.2)$) to[bend left=25] ($(cic.east)+(0.03,-0.2)$);

    \draw[->,dashed,thin, -{Stealth[scale=1.2]}] (ic) -- (orc);
    \draw[->,dashed,thin, -{Stealth[scale=1.2]}] (ic) -- (osi);
    \draw[->,dashed,thin, -{Stealth[scale=1.2]}] (ic) -- (bwt);

    \path[reduction link=sec:lcf] (dm) -- (lcf);
    \path[reduction link=sec:bwt] (dm) -- (rlbwt);
    \path[reduction link=rm:rlbwt-size-to-bwt-construction] (rlbwt) -- (bwt);
    \path[reduction link=sec:isa] (dm) -- (isa);
    \path[reduction link=sec:lz] (dm) -- (lzparity);
    \path[reduction link=rm:lz77-parity-size-factorization] (lzparity) -- (lzsize);
    \path[reduction link=rm:lz77-parity-size-factorization] (lzsize) -- (lzfactor);
    \path[reduction link=sec:lpf] (dm) -- (lpf);
    \path[reduction link=sec:sn-to-rpm] (bn) -- (rps);
    \path[reduction link=sec:rpm-to-cci] (rps) -- (cic);
    \path[reduction link=sec:sn-to-dm-problem-reduction]
      ($(bn.east)+(0.03,0.2)$) to[bend left=25] ($(dm.west)+(0.03,0.2)$);
    \path[reduction link=sec:dm-to-sn-problem-reduction]
      ($(dm.west)+(0.03,-0.2)$) to[bend left=25] ($(bn.east)+(0.03,-0.2)$);
    \path[reduction link=sec:cci-to-ci]
      ($(cic.east)+(0.03,0.2)$) to[bend left=25] ($(ic.west)+(0.03,0.2)$);
    \path[reduction link=sec:ci-to-cci]
      ($(ic.west)+(0.03,-0.2)$) to[bend left=25] ($(cic.east)+(0.03,-0.2)$);

    \ReductionNodeLink{dm}{prob:dm}
    \ReductionNodeLink{bn}{prob:sn}
    \ReductionNodeLink{lcf}{prob:lcf}
    \ReductionNodeLink{rlbwt}{prob:rlbwt-size}
    \ReductionNodeLink{bwt}{prob:bwt-construction}
    \ReductionNodeLink{isa}{prob:batched-isa}
    \ReductionNodeLink{lzparity}{prob:lz77-parity}
    \ReductionNodeLink{lzsize}{prob:lz77-size}
    \ReductionNodeLink{lzfactor}{prob:lz77-factorization}
    \ReductionNodeLink{lpf}{prob:batched-lpf}
    \ReductionNodeLink{rps}{prob:rps}
    \ReductionNodeLink{cic}{prob:cci}
    \ReductionNodeLink{ic}{prob:ci}

  \end{tikzpicture}%
  }
  \caption{\small
    Our hierarchy of hardness among problems with a state-of-the-art
    time complexity $\bigO(n \sqrt{\log n})$ for an input of $\Theta(n)$
    machine words (i.e., $\Theta(n \log n)$ bits).
    For string problems,
    this implies that the length of the input string(s) is $\Theta(n \log n)$.
    An arrow $P_1 \rightarrow P_2$ indicates that problem $P_1$ can be
    reduced to problem $P_2$, which implies that $P_2$ is \emph{at
    least as hard} as $P_1$.  Solid arrows correspond to reductions
    proved in this paper, while dashed arrows illustrate the reductions
    in~\cite{ChanP10,sss}.
    The reductions for \probname{LZ77 Parity}, \probname{LZ77
    Size}, \probname{LZ77 Factorization}, and \probname{Batched LPF
    Queries} hold for both the overlapping and non-overlapping
    variants of LZ77 and LPF.
    Nodes representing problems formally defined in this paper link to
    their definitions, and solid arrows link to the corresponding sections
    or remarks.
  }\label{fig:reductions}
\end{figure}

\clearpage
\pagenumbering{arabic}
\setcounter{page}{1}

\section{Introduction}\label{sec:intro}

Word RAM~\cite{Hagerup98} is a classical model in computational
complexity, in which we assume that memory consists of $w$-bit
cells called \emph{(machine) words} (with $w \geq \log n$, where $n$
is the input size) arranged in an array. Random access to any
machine word, as well as standard arithmetic, logical, shift, and
bitwise operations on a constant number of machine words, take $\bigO(1)$
time.

Word RAM is widely used to analyze algorithms, and understanding its
capabilities and limitations lies at the core of theoretical computer
science.  In this paper, we address the problem of understanding one
of the six main ``levels'' of hardness in the word RAM model
characterized by Chan and P\u{a}tra\c{s}cu~\cite{ChanP10}, namely,
algorithms with time complexity $\bigO(n \sqrt{\log n})$.  As
listed in~\cite{ChanP10} and numerous follow-up works, the above
complexity appears in several domains, most commonly in string
processing\footnote{For simplicity, we focus on strings over binary
  alphabet $\Sigma = \BinaryAlphabet$. As proved later in this
  paper, for several of these problems a larger alphabet can be
  efficiently reduced to the binary case; see, e.g.,
  Theorems~\ref{th:lz-alphabet-reduction-poly-to-binary},
  \ref{th:lcf-alphabet-reduction-poly-to-binary},
  \ref{th:isa-alphabet-reduction-poly-to-binary}, and
  \ref{th:lpf-alphabet-reduction-poly-to-binary}.}, data compression,
computational geometry, and combinatorial algorithms:
\begin{itemize}
\item $\bigO(n \sqrt{\log n})$ time is currently the best known complexity
  for constructing the Burrows--Wheeler Transform (BWT)~\cite{sss},\footnote{To
    streamline the presentation of our results, throughout \cref{sec:intro},
    unless stated otherwise, we use $n$ to denote the
    \emph{total input size in \textbf{machine words}, each storing
    a number of bits logarithmic in the input size}.  In string
    processing, $\Textlen$ typically denotes the length of the
    (binary) string. The complexity considered in this paper is then
    expressed as $\bigO(\Textlen / \sqrt{\log \Textlen})$~\cite{sss,breaking,sublinearlz}.}
  which is used in many applications, including text
  indexing~\cite{FerraginaM05,cst,Gagie2020}, sequence
  alignment~\cite{bowtie2,bwa}, DNA and natural text
  compression~\cite{cox2012large,adjeroh2002dna,moffat2005word,Manzini01},
  image compression~\cite{rawat2014evaluation,vo2009image}, and many others;
  see~\cite{bwtbook}.

\item
  $\bigO(n \sqrt{\log n})$ is the complexity of the fastest known
  algorithm for Lempel--Ziv (LZ77) factorization~\cite{sublinearlz}.
  This factorization is used in algorithms for approximating the smallest
  grammar~\cite{Rytter03}, constructing dictionary-compressed
  indexes~\cite{breaking,collapsing}, computing
  maximal periodic substrings~\cite{main1989detecting,KolpakovK99,ChenPS07,CrochemoreI08},
  repeats with a fixed gap~\cite{KolpakovK00}, approximate
  repetitions~\cite{KolpakovK03}, tandem repeats~\cite{GusfieldS04},
  sequence alignments~\cite{CrochemoreLZ02}, local
  periods~\cite{DuvalKKLL04}, and seeds~\cite{KociumakaKRRW20}.

\item $\bigO(n \sqrt{\log n})$ time is the runtime of the fastest
  algorithm for the problem of \emph{longest common substring
    (LCS)}~\cite{Charalampopoulos21} (also called \emph{longest common
    factor (LCF)}) of two strings.

\item $\bigO(n \sqrt{\log n})$ is the fastest known complexity to
  solve several \emph{offline} problems, where we aim to answer a
  batch of $n$ queries of some type on an input of $\Theta(n)$ words.
  This includes longest previous factor (LPF)
  queries~\cite{sublinearlz}, inverse suffix array (ISA)
  queries~\cite{breaking}, pattern matching
  queries~\cite{MunroNN20,breaking}, bi-colored range
  counting~\cite{ChanP10}, dynamic ranking
  queries~\cite{ChanP10}, and various suffix tree
  queries~\cite{breaking}.  These methods, in turn, yield the fastest
  algorithms for counting inversions and 2D orthogonal segment
  intersection~\cite{ChanP10}.

\item $\bigO(n \sqrt{\log n})$ is the fastest construction time of
  optimal-space data structures efficiently answering (in contrast to
  the above, in an \emph{online} manner) many different types of queries,
  including rank and select
  queries~\cite{WaveletSuffixTree,MunroNV16}, range
  predecessor~\cite{BelazzouguiP16}, range reporting~\cite{Gao0N20},
  range median~\cite{ChanP10}, orthogonal range
  counting~\cite{ChanP10}, prefix rank/select~\cite{breaking}, and
  prefix RMQ~\cite{sublinearlz}.
\end{itemize}

At this point, it is natural to ask whether there exists some
fundamental barrier in all of the above problems or the
correlation is coincidental and problem-specific approaches should be
applied in attempts to obtain $o(n\sqrt{\log n})$-time solutions.
This is important for at least two reasons:
\begin{itemize}
\item For several of the above problems, such progress would have
  enormous impact.  For example, the aforementioned LZ77
  factorization, in addition to algorithmic applications listed above,
  is also the \emph{most commonly} used compression algorithm, with
  59 of the 234 lossless compressors listed in the Large Text Compression
  Benchmark~\cite{ltcb} using LZ77 as their main algorithm.
  Thus, improving the $\bigO(n\sqrt{\log n})$ time of the fastest LZ77
  algorithm~\cite{sublinearlz} would be a breakthrough in the field of
  data compression. Such a big payoff, however, also implies that any
  notion of \emph{hardness} would save a substantial amount of research
  effort.

\item There exist several deceptively similar pairs of problems, where
  one problem has an optimal $\bigO(n)$-time algorithm while the other
  has a best-known complexity of $\bigO(n \sqrt{\log n})$.
  For example, computing the longest palindromic substring of a binary
  string occupying $n$ words\footnote{Note that the length of the string
  is then $\Theta(n \log n)$.} takes $\bigO(n)$ time~\cite{CPR22},
  whereas computing the longest common
  substring~\cite{Charalampopoulos21} of two such binary strings takes $\bigO(n \sqrt{\log n})$
  time. Similarly, preprocessing a binary string for $\bigO(1)$-time
  longest-common-extension (LCE) queries takes $\bigO(n)$
  time~\cite{sss}, but the fastest preprocessing for efficient
  longest-previous-factor (LPF) queries takes $\bigO(n \sqrt{\log n})$
  time \cite{sublinearlz}. Currently, there are no insights
  into \emph{why} these problems fall into one class or the other, and
  numerous problems are independently pursued in hopes of
  discovering an $\bigO(n)$-time
  algorithm~\cite{sss,sublinearlz,ChanP10,Charalampopoulos21,breaking,MunroNV16,MunroNN20}.
\end{itemize}

Despite the tremendous progress in establishing lower bounds for data
structures in the word RAM model~\cite{larsen2013models,Patrascu08},
proving unconditional lower bounds for complex \emph{algorithms} in
this model still seems outside of reach.
The best-known approach to making progress in such situations is to study
hierarchies of problems ordered by their computational hardness.
This allows researchers to focus their efforts on problems at the very top.
Two classical examples of this approach are the theory of
NP-hardness (for problems with no known polynomial-time algorithms) and fine-grained complexity
(for problems with runtime $\bigO(n^{1+\epsilon})$, $\epsilon > 0$).
However, no such hierarchy exists for near-linear problems. We thus ask:
\begingroup
\setlength{\topsep}{6pt}
\begin{center}
  \begin{tikzpicture}
    \node[
      draw=black!48,
      fill=black!2,
      line width=.35pt,
      rounded corners=1pt,
      inner xsep=1.25em,
      inner ysep=.85em,
      text width=.85\linewidth,
      align=center
    ] {
      \textbf{Question.} \emph{Do the above problems share a\\
        common feature that explains their difficulty?}
    };
  \end{tikzpicture}
\end{center}
\endgroup

\subsection{Our Results}\label{sec:intro-results}

We answer the above question affirmatively by developing a
comprehensive hierarchy of hardness for near-linear problems with time
complexity $\bigO(n \sqrt{\log n})$ in the word RAM model
(Figure~\ref{fig:reductions}).  In more detail, our contribution is
twofold:

First, in \cref{sec:dm-to-string-problems}, we prove that at the core
of all central \emph{string problems} with complexity $\bigO(n
\sqrt{\log n})$ lies the following \probname{Dictionary Matching} problem
(\cref{prob:dm}):
given the packed representation\footnote{See \cref{def:packed-representation} for the formal definition.}
of a binary string $\Text$ with $|\Text|\geq2$ and a dictionary
$\mathcal D\subseteq\BinaryAlphabet^m$ containing
$|\mathcal D|=\lfloor|\Text|/\log|\Text|\rfloor$ patterns of common
length $m=2\lfloor\log|\Text|\rfloor$,\footnote{Choosing the factor $2$
already captures the difficulty of binary dictionary matching with arbitrary
dictionary sizes and arbitrary common pattern lengths
(\cref{th:dm-length-reduction}; see also \cref{rm:dm-sn-factor-two}).}
decide whether any pattern
$\Pat\in\mathcal D$ occurs as a substring of $\Text$.
For an input of $n$ machine words, the dictionary contains $\Theta(n)$
patterns, each of length $\Theta(\log n)$.
Dictionary Matching is a classical problem that can be solved using the \mbox{Aho--Corasick
  (AC)} automaton \cite{AhoC75}. However, five decades after the
discovery of the AC automaton, the fastest known solution to \probname{Dictionary
Matching} still runs in $\bigO(n \sqrt{\log n})$ time~\cite{breaking}, making
it a strong ``central'' problem.

The string problems in our reductions take different forms of input:
a single string, two strings, or a text together with a collection
of patterns.
To compare their complexities, we introduce a parameter $N$, defined
as an upper bound on the total number of binary symbols in the input
strings.
We can then express the running time and space usage of each problem
as functions of $N$.
With this notation, our results for string problems can be summarized
in a single theorem.

\begin{theorem}[Dictionary Matching $\rightarrow$ String Problems]
  Let $N\in\Z_{\geq2}$. Consider the word RAM model with word size
  $w$ satisfying $1+\log N\leq w=\bigO(\log N)$.\footnote{As usual
  in the word RAM model, we assume $w=\Theta(\log N)$. The lower bound
  $w\geq1+\log N$ ensures that integers in $[0\dd2N)$ fit in one
  machine word.}
  Suppose that there exists an algorithm for one of the problems
  \probname{RLBWT}
  \probname{Size}, \probname{LZ77} \probname{Parity},
  \probname{Longest} \probname{Common} \probname{Factor},
  \probname{Batched} \probname{ISA} \probname{Queries}, or
  \probname{Batched} \probname{LPF} \probname{Queries}
  (\cref{prob:rlbwt-size,prob:lz77-parity,prob:lcf,prob:batched-isa,prob:batched-lpf})
  that, given $N$ and any valid input of
  at most $N$ symbols over alphabet $\BinaryAlphabet$,\footnote{See
  Theorems~\ref{th:equivalence-dm-sn}, \ref{th:lz}, \ref{th:bwt},
  \ref{th:isa}, \ref{th:lpf}, and~\ref{th:lcf}
  for the formal descriptions of valid inputs.} achieves
  the following complexities:
  \begin{itemize}
  \item running time $T(N)$,
  \item space usage $S(N)$ bits.
  \end{itemize}
  Then, there exists an
  algorithm for \probname{Dictionary} \probname{Matching}
  (\cref{prob:dm}) that, given $N$ and any valid input of
  at most $N$ symbols over alphabet $\BinaryAlphabet$, achieves
  the following complexities:
  \begin{itemize}
  \item running time $\bigO(T(N))$,
  \item space usage $\bigO(S(N))$ bits.
  \end{itemize}
\end{theorem}

In terms of the input size in words, this means that, unless
\probname{Dictionary} \probname{Matching} can be solved in
$o(n\sqrt{\log n})$ time, none of the above problems can either.

Note that our reductions include problems that are easier than those
discussed above (leading to a \emph{stronger} result).  For example,
instead of reducing the \probname{Dictionary} \probname{Matching} problem directly to
\probname{LZ77 Factorization}, we show that determining the
\emph{parity of the number of phrases} in LZ77 parsing
(see \cref{prob:lz77-parity})
is already at least as hard as \probname{Dictionary Matching}
(\cref{th:lz}).
The hardness of full LZ77 factorization then follows
as a simple corollary. Similarly, computing the number of runs in the
Burrows--Wheeler Transform (BWT),
rather than constructing the BWT itself,
is at least as hard as \probname{Dictionary Matching}
(\cref{th:bwt}).

\subsection{Beyond String Problems}\label{sec:intro-beyond-string-problems}

Our second contribution is extending the hierarchy of hardness to
problems outside the string-processing domain.  Our central goal is to
prove the hardness of \probname{Counting} \probname{Inversions} (given an array
$A$ of size $n$, compute the number of pairs $(i,j)$ with $i<j$ and
$A[i]>A[j]$),\footnote{A reduction \emph{in the opposite direction}
(reducing counting inversions to a string problem) was presented
in~\cite{sss}, which proved that
\probname{Counting} \probname{Inversions} $\rightarrow$ \probname{BWT
  Construction}. This construction does not, however, directly yield
a reduction in the reverse direction.}  as it was established in~\cite{ChanP10} to be
reducible to many fundamental geometric problems, including \probname{Orthogonal}
\probname{Range} \probname{Counting} and \probname{Orthogonal}
\probname{Segment} \probname{Intersection} \probname{Counting}, which are themselves
building blocks of many more algorithms~\cite{Chazelle88}.  \probname{Counting Inversions}
is a textbook problem solvable in $\bigO(n \log n)$ time
with a folklore \textsc{MergeSort} modification.  A completely new and
nontrivial approach presented in~\cite{ChanP10} improved this to
$\bigO(n \sqrt{\log n})$.  Our transfer of hardness from string
problems to counting inversions is obtained by putting together four
steps corresponding to
\cref{sec:equiv-dm-and-sn,sec:sn-to-rpm,sec:rpm-to-cci,sec:equiv-ci-and-cci}.

First, in \cref{sec:equiv-dm-and-sn},
we prove a surprising
equivalence between the \probname{Dictionary} \probname{Matching}
problem and a problem we call \probname{String} \probname{Nesting}:
given collections $\mathcal{P} \subseteq
\BinaryAlphabet^{m} \times \BinaryAlphabet^{m}$ and $\mathcal{Q}
\subseteq \BinaryAlphabet^{\leq m} \times \BinaryAlphabet^{\leq m}$
of string pairs
represented in the packed form, such that every $(A,B) \in
\mathcal{Q}$ satisfies $|A| + |B| = m$, output a
$\texttt{YES}/\texttt{NO}$ answer indicating whether we can ``nest''
some $(A,B) \in \mathcal{Q}$ in some $(X,Y) \in \mathcal{P}$, which
occurs when $A$ is a suffix of $X$ and $B$ is a prefix of $Y$.
We assume that $|\mathcal P|=|\mathcal Q|=k\geq2$ and
$m=2\lfloor\log k\rfloor$ (\cref{prob:sn}). We prove the following theorem
(see \cref{th:equivalence-dm-sn}):

\begin{theorem}[Dictionary Matching $\xleftrightarrow{\ \ }$ String Nesting]\label{th:step-1}
  Let $N\in\Z_{\geq2}$. Consider the word RAM model with word size
  $w$ satisfying $1+\log N\leq w=\bigO(\log N)$.
  The problems of \probname{Dictionary} \probname{Matching} and
  \probname{String} \probname{Nesting}
  (\cref{prob:dm,prob:sn}) are asymptotically equivalent:
  if there exists an algorithm for one of these two problems that,
  given $N$ and any valid input of at most $N$ symbols over alphabet
  $\BinaryAlphabet$,\footnote{See \cref{th:equivalence-dm-sn} for the
  complete description of a valid input to each problem.}
  achieves the following complexities:
  \begin{itemize}
  \item running time $T(N)$,
  \item space usage $S(N)$ bits,
  \end{itemize}
  then there exists an algorithm
  for the other problem that, given $N$ and any valid input of at most $N$
  symbols over alphabet $\BinaryAlphabet$, achieves the following
  complexities:
  \begin{itemize}
  \item running time $\bigO(T(N))$,
  \item space usage $\bigO(S(N))$ bits.
  \end{itemize}
\end{theorem}

Observe that \probname{String} \probname{Nesting} is a seemingly
easier problem than \probname{Dictionary} \probname{Matching}:
given an instance $\mathcal{P} =
\{(X_1,Y_1), \ldots, \allowbreak (X_k,Y_k)\}$ and $\mathcal{Q} = \{(A_1,B_1),
\ldots, \allowbreak (A_k,B_k)\}$ of \probname{String Nesting}, we first form
$T = X_1 \dol Y_1 X_2 \dol Y_2 \cdots X_k \dol Y_k$
and $\mathcal{D} = \{A_i \dol B_i : i \in [1 \dd k]\}$,
where $\dol$ is a fresh symbol.
A pattern in $\mathcal D$ occurs in $T$ if and only if some
$(A,B)\in\mathcal Q$ can be nested in some $(X,Y)\in\mathcal P$.
The alphabet and length reductions of
\cref{th:dm-alphabet-reduction,pr:dm-length-reduction-general} then produce
valid binary \probname{Dictionary Matching} instances.

The reduction in the opposite direction requires
a more involved construction. For this overview, denote $\Textlen=|\Text|$.
We compute a set $C\subseteq[1\dd\Textlen]$ of
$\bigO(\Textlen/\log\Textlen)$ text positions and split every dictionary
pattern $\Pat_i$ into $A_iB_i$. These splits have the following property:
$\Pat_i$ occurs in $\Text$ if and only if some position $c\in C$ has
$A_i$ immediately before it and $B_i$ beginning at it.
For each selected position, we collect the text fragments on its two sides.
The resulting pairs of text
fragments and pattern splits give the suffix--prefix comparisons required
by \probname{String Nesting}.
To construct $C$ and the pattern splits, we combine the locally consistent
sampling provided by \emph{string synchronizing sets}~\cite{sss} with an
efficient algorithm for matching periodic patterns.
See \cref{th:dm-to-sn} for the complete reduction.

Note that \probname{String Nesting} also has a geometric
interpretation. For binary strings $B$ and $Y$ with $|B|\leq |Y|=m$,
the condition of $B$ being a prefix of $Y$ is equivalent to
$b \leq y < b'$, where $b$ and $y$ are the integers represented in
binary by $B\cdot\texttt{0}^{m-|B|}$ and $Y$, respectively, and
$b'=b+2^{m-|B|}$.
Applying the same observation to $\revstr{A}$ and $\revstr{X}$ accounts
for the condition of $A$ being a suffix of $X$. We can therefore
view $\mathcal{P}$ as a set of points on a plane and $\mathcal{Q}$ as
a set of axis-aligned rectangles (whose projections on both axes are
dyadic intervals). The problem then reduces to checking whether
any point in $\mathcal{P}$ lies within any of the rectangles in $\mathcal{Q}$.

Reducing a problem on a single string to a problem on a set of strings
brings our problem closer to counting inversions.  However, this is
still an existential problem (checking if \emph{some} $(A,B) \in
\mathcal{Q}$ is nested in \emph{some} $(X,Y) \in \mathcal{P}$) and so
far has little to do with counting inversions.  To establish these
further connections while preserving reducibility to counting
inversions, we proceed in three steps:
\begin{enumerate}
\item We begin by showing, in \cref{sec:sn-to-rpm},
  that \probname{String Nesting} can be reduced to a slightly more general
  problem of \probname{Range} \probname{Prefix} \probname{Search}, in
  which we are given an array $S[1 \dd m]$ of $m$ equal-length strings
  of length $\ell = 2\lfloor\log m\rfloor$ and a sequence $(b_1,e_1,Q_1),
  \ldots, (b_m,e_m,Q_m)$ of $m$ triples, where $0 \le b_i \le e_i \le m$ and $Q_i \in
  \BinaryAlphabet^{\leq \ell}$. Our goal is to output a
  $\texttt{YES}/\texttt{NO}$ answer indicating whether there exists $i
  \in [1 \dd m]$ such that $Q_i$ is a prefix of at least one string in
  $S(b_i \dd e_i]$.

\item In the next step
  (\cref{sec:rpm-to-cci}),
  we prove that
  \probname{Range} \probname{Prefix} \probname{Search} reduces to a
  problem we call \probname{Counting} \probname{Colored}
  \probname{Inversions}, in which, given arrays $A[1 \dd m]$ and $C[1
  \dd m]$, our goal is to compute $|\{(i,j) \in [1 \dd m]^2 :
  i<j,\,A[i]>A[j],\text{ and }C[i] \neq C[j]\}|$, i.e., the number of
  ``colored'' inversions.  To design the reduction, we show that,
  given $S[1 \dd m]$ and a sequence $(b_i,e_i,Q_i)_{i \in [1 \dd m]}$
  (i.e., an instance of \probname{Range} \probname{Prefix}
  \probname{Search}), we can insert modified copies of $Q_i$ into $S$
  and color the resulting arrays of strings such that performing this
  reduction twice, with a different modification of $Q_i$ each time,
  ensures that the colored inversions in two instances will always cancel
  out unless $Q_i$ is a prefix of some string in $S(b_i \dd
  e_i]$; see \cref{lm:inv}.

\item To finish our reduction chain, in the last step
  (\cref{sec:equiv-ci-and-cci}),
  we show that computationally, the
  problem of \probname{Counting} \probname{Colored}
  \probname{Inversions} and the problem of \probname{Counting}
  \probname{Inversions} are equivalent
  (\cref{th:inv-problems-equivalence}).
\end{enumerate}

Combining
\cref{th:equivalence-dm-sn,th:rpm,th:rpm-to-cci,th:inv-problems-equivalence}
gives the following theorem:

\begin{theorem}[Dictionary Matching $\rightarrow$ Counting Inversions]
  Let $N\in\Z_{\geq2}$. Consider the word RAM model with word size
  $w$ satisfying $1+\log N\leq w=\bigO(\log N)$.
  Suppose that there exists an algorithm for \probname{Counting Inversions}
  (\cref{prob:ci}) that, given $N$ and any valid input of at most
  $N$ bits,\footnote{See \cref{th:inv-problems-equivalence} for the
  complete input description, including its size in bits.}
  achieves the following complexities:
  \begin{itemize}
  \item running time $T(N)$,
  \item space usage $S(N)$ bits.
  \end{itemize}
  Then, there exists an algorithm
  for \probname{Dictionary} \probname{Matching}
  (\cref{prob:dm}) that, given $N$ and any valid input of
  at most $N$ symbols over alphabet $\BinaryAlphabet$, achieves
  the following complexities:
  \begin{itemize}
  \item running time $\bigO(T(N))$,
  \item space usage $\bigO(S(N))$ bits.
  \end{itemize}
\end{theorem}

\subsection{Alphabet Reductions}\label{sec:intro-alphabet-reductions}

As part of the above reductions, we obtain a useful auxiliary
result: We establish that several of the string problems considered
above, naturally generalized to alphabets $\IntegerAlphabet$, reach full hardness
already for $\AlphabetSize = 2$. To illustrate this, consider the
problem of computing the longest common factor $\LCF{S_1}{S_2}$, where
$S_1, S_2 \in \IntegerAlphabet^{*}$, and denote their total length by
$\Textlen=|S_1|+|S_2|$. Assume that
$2\leq\AlphabetSize\leq\min\{|S_1|,|S_2|\}$.
The fully general algorithm for
LCF presented in~\cite{Charalampopoulos21} achieves the runtime of
$\bigO((\Textlen \log \AlphabetSize) / \sqrt{\log \Textlen})$. For
$\AlphabetSize = 2$, we thus indeed obtain the complexity
$\bigO(\Textlen / \sqrt{\log \Textlen})$, which matches the above bound
when the input size is measured in words. Taking into account
the variable alphabet size, however, introduces a considerable
complication in the design and analysis of the algorithm
in~\cite{Charalampopoulos21} (just like in the state-of-the-art
algorithms for other problems~\cite{breaking,sss,sublinearlz}).  In
this paper, we show black-box alphabet reductions for nearly all of
the studied problems (\cref{sec:alphabet-reductions}). For example, in the case of the LCF problem, given the
packed representations of any two strings $S_1 \in
\IntegerAlphabet^{\Textlen_1}$ and $S_2 \in
\IntegerAlphabet^{\Textlen_2}$ (where $\Textlen_1,\Textlen_2\geq1$,
$\Textlen=\Textlen_1+\Textlen_2$, and
$3\leq\AlphabetSize\leq\min\{\Textlen_1,\Textlen_2\}$), we can, in $\bigO(\Textlen /
\log_{\AlphabetSize} \Textlen)$ time,\footnote{Here we apply
  \cref{th:lcf-alphabet-reduction-poly-to-binary} with $N$ equal to the total
  length of the encoded strings. Since $N=\Theta(\Textlen\log\AlphabetSize)$
  and $\log N=\Theta(\log\Textlen)$, its $\bigO(N/\log N)$ bound yields
  the stated running time.}
construct the packed representations of strings $S^{\rm
  bin}_1, S^{\rm bin}_2 \in \BinaryAlphabet^*$ with $|S^{\rm bin}_1| =
\bigO(\Textlen_1 \log \AlphabetSize)$ and $|S^{\rm bin}_2| =
\bigO(\Textlen_2 \log \AlphabetSize)$, as well as integers $\alpha$
and $\beta$ such that
\[
  \LCF{S_1}{S_2} = \left\lfloor
    \frac{\LCF{S^{\rm bin}_1}{S^{\rm bin}_2} - \alpha}{\beta} \right\rfloor.
\]
This implies that, under the stated alphabet-size assumption, it
suffices to design an algorithm for LCF working for the binary
alphabet, and then the desired complexity of $\bigO((\Textlen \log
\AlphabetSize) / \sqrt{\log \Textlen})$ can be obtained in an entirely
black-box fashion.  Our reductions thus also prove that the binary
alphabet is the hardest case for numerous fundamental problems,
simplify many
algorithms~\cite{Charalampopoulos21,breaking,sublinearlz} for larger
alphabets, and provide a blueprint for alphabet
reductions for other problems.

\subsection{Related Work}\label{sec:related-work}

Many of the problems studied in this paper (including \probname{BWT} \probname{Construction} and
\probname{LZ77} \probname{Factorization}) have been extensively examined in various
settings, such as construction on highly compressible input
strings~\cite{Kempa19,Ellert23,DiazDominguezN23}, construction in
compressed time (i.e., proportional to the size of the input text in
compressed form)~\cite{collapsing,resolution}, approximate
variants~\cite{FischerGGK15,Ellert23,KosolobovVNP20},
online computation~\cite{Starikovskaya12,PolicritiP15,Kosolobov15,KempaK17b,OhnoSTIS18},
dynamic settings~\cite{BannaiCR24,BonehGK26}, and the general alphabet setting (where the set
of allowed symbol manipulations is
restricted)~\cite{EllertGG23,Kosolobov15b}.  In many of these
settings, the problems are understood much more clearly.  For example,
the optimal bound for computing LZ77 on a length-$\Textlen$ string
when the input is over an \emph{ordered alphabet} (where the alphabet is
some abstract ordered set, and we can only check if one symbol is
smaller than another) was proved to be $\Theta(\Textlen \log
\AlphabetSize)$ in~\cite{Kosolobov15b}. The optimal bound for
an \emph{unordered alphabet} (where we can only test symbols for
equality, and the number $\AlphabetSize$ of distinct symbols is not
known in advance), on the other hand, is known to be $\Theta(\Textlen
\AlphabetSize)$~\cite{EllertGG23}.

Similarly, many of the above problems have been studied in parallel,
external-memory, GPU, or quantum models, including
\probname{LZ77} \probname{Factorization}~\cite{KleinW05,ShunZ13,HanLN22,OzsoyS11,OzsoySC14,ZuH14,KarkkainenKP14,KosolobovVNP20,quantumlz}
and \probname{BWT Construction}~\cite{SepulvedaNN20,EgidiLMT19,FerraginaGM12,quantumlz}.

\subsection{Organization of the Paper}\label{sec:organization}

First, in \cref{sec:prelim}, we
formally introduce the notation and basic definitions used in the paper.
In \cref{sec:overview-lz,sec:overview-bwt}, we give an overview
of two reductions from \probname{Dictionary Matching}.
In \cref{sec:tools}, we collect auxiliary results about packed strings,
periodicity testing, and sparse suffix sorting.
In \cref{sec:sparse-anchors}, we develop the construction of sparse anchor
positions for pattern matching.
We present reductions to the binary alphabet in \cref{sec:alphabet-reductions}.
In \cref{sec:equiv-dm-and-sn}, we prove equivalence of
\probname{Dictionary Matching} and \probname{String Nesting}.
In \cref{sec:dm-to-string-problems}, we describe reductions from
\probname{Dictionary Matching} to string problems, including
\probname{LZ77 Parity} (\cref{sec:lz}), \probname{RLBWT Size} (\cref{sec:bwt}),
\probname{Batched ISA Queries} (\cref{sec:isa}), \probname{Batched LPF Queries}
(\cref{sec:lpf}), and \probname{Longest Common Factor} (\cref{sec:lcf}).  In the
final three sections, we develop the branch of the reduction tree
corresponding to non-string problems. In
\cref{sec:sn-to-rpm}, we reduce \probname{String Nesting} to \probname{Range Prefix
Search}. In \cref{sec:rpm-to-cci}, we reduce \probname{Range} \probname{Prefix}
\probname{Search} to \probname{Counting} \probname{Colored} \probname{Inversions}.
Finally, in \cref{sec:equiv-ci-and-cci},
we show that \probname{Counting Colored Inversions} is computationally equivalent
to \probname{Counting Inversions}.

\section{Preliminaries}\label{sec:prelim}

\subsection{Model of Computation}\label{sec:model}

We use the standard word RAM model of computation~\cite{Hagerup98}
with $w$-bit \emph{machine words}, where $w = \Omega(\log N)$ and $N$ is
the input size. All standard bitwise and arithmetic operations take $\bigO(1)$ time.
We assume that the word size is sufficiently large to address all memory used
by the algorithm, including its input and output.
Unless explicitly stated otherwise, we measure space complexity in machine words.
When expressed in bits, a space bound counts $w$ bits per occupied word.
Space denotes the maximum total storage used at any time, including the input,
the output, and all auxiliary data.

\subsection{Basic Definitions}\label{sec:basic-definitions}

A \emph{string} is a finite sequence of characters from a given
\emph{alphabet} $\Sigma$.  The length of a string $S$ is denoted $|S|$. For $i
\in [1\dd |S|]$, the $i$th character of $S$ is denoted $S[i]$.
A~\emph{substring} or a \emph{factor} of $S$ is a string $S[i \dd j]$,
where $i\in[1\dd |S|+1]$ and $j\in[0\dd |S|]$.
If $i\leq j$, then $S[i\dd j]=S[i]S[i+1]\cdots S[j]$;
otherwise, $S[i\dd j]=\emptystring$, the \emph{empty string}.
We also write $S[i\dd j)=S[i\dd j-1]$ and
$S(i\dd j]=S[i+1\dd j]$.
Substrings of the form $S[1\dd j]$ and $S[i\dd |S|]$ are called
\emph{prefixes} and \emph{suffixes}, respectively.
We use $\revstr{S}$ to denote the \emph{reverse} of $S$, i.e.,
$S[|S|]\cdots S[2]S[1]$.
We denote the \emph{concatenation} of two strings $U$ and $V$, that
is, the string $U[1]\cdots U[|U|]V[1]\cdots V[|V|]$, by $UV$ or $U\cdot V$.
Furthermore, $S^k = \bigodot_{i=1}^k S$ is the concatenation of $k \in
\Zz$ copies of $S$; note that $S^0 = \emptystring$.
An integer $p\in
[1\dd |S|]$ is a \emph{period} of $S$ if $S[i] = S[i + p]$ holds for
every $i \in [1 \dd |S|-p]$. We denote the shortest period of $S$ as
$\per{S}$.  For every $S \in \Sigma^{+}$, we define the
infinite power $S^{\infty}$ so that $S^{\infty}[i] = S[1 + (i-1) \bmod |S|]$
for $i \in \Z$.  In particular, $S = S^{\infty}[1 \dd |S|]$.
A \emph{run} in a string is a maximal nonempty substring consisting of
equal symbols. For $S=c_1^{\ell_1}\cdots c_h^{\ell_h}$, where
$c_i\in\Sigma$, $\ell_i\geq1$, and adjacent symbols $c_i,c_{i+1}$ differ,
the \emph{run-length encoding} of $S$ is the sequence
$\RL{S}=(c_1,\ell_1),\ldots,(c_h,\ell_h)$.
For example, $\RL{\texttt{aaabbc}}=(\texttt{a},3),(\texttt{b},2),(\texttt{c},1)$.
The run-length encoding of $\emptystring$ is the empty sequence.
For $U,V\in\Sigma^*$, we denote by $\lcp{U}{V}$ the length of their
longest common prefix.
For any $U,V\in\Sigma^*$, we denote by $\LCF{U}{V}$ the length
of a longest common factor of $U$ and $V$.
We use $\preceq$ to denote the order on $\Sigma$, extended to the
\emph{lexicographic} order on $\Sigma^*$ so that $U,V\in \Sigma^*$
satisfy $U \preceq V$ if and only if either
\begin{enumerate*}[label=(\alph*)]
  \item $U$ is a prefix of~$V$, or
  \item $U[1 \dd i) = V[1 \dd i)$ and
    $U[i]\prec V[i]$ holds for some $i\in [1\dd \min(|U|,|V|)]$.
\end{enumerate*}
We write $U\prec V$ if $U\preceq V$ and $U\neq V$.
All logarithms are base two unless a different base is specified.

\begin{definition}[Pattern occurrences and suffix-range boundaries]\label{def:occ}
  For any $\Pat \in \Sigma^{*}$ and $\Text \in \Sigma^*$, we define
  \begin{align*}
    \OccTwo{\Pat}{\Text}
      &= \{j \in [1 \dd |\Text|] : j + |\Pat| \leq |\Text| + 1\text{ and }\Text[j \dd j + |\Pat|) = \Pat\},\\
    \RangeBegTwo{\Pat}{\Text}
      &= |\{j \in [1 \dd |\Text|] : \Text[j \dd |\Text|] \prec \Pat\}|,\\
    \RangeEndTwo{\Pat}{\Text}
      &= \RangeBegTwo{\Pat}{\Text} + |\OccTwo{\Pat}{\Text}|.
  \end{align*}
\end{definition}

\subsection{Packed Representation of Strings and String Sequences}\label{sec:packed-representation}

\begin{definition}[Packed representation of a string]\label{def:packed-representation}
  Let $\AlphabetSize \in \Z_{\geq 2}$ satisfy $\log \AlphabetSize \leq w$,
  and denote $b = \lceil \log \AlphabetSize \rceil$ and $s = \lfloor w/b \rfloor$.
  The \emph{packed representation} of a string $S \in \IntegerAlphabet^{*}$,
  denoted $\PackedRepresentation{w}{\AlphabetSize}{S}$, is the sequence
  $(x_1, \ldots, x_k)$ of $w$-bit words, where $k = \lceil |S|/s \rceil$ and,
  for every $i \in [1 \dd k]$,
  \[
    x_i = \sum_{\delta=1}^{\min(s, |S|-(i-1)s)}
      S[(i-1)s+\delta] \cdot 2^{(\delta-1)b}.
  \]
  The representation uses $\bigO(1 + |S|\log\AlphabetSize/w)$ words.
  In other words, each symbol is encoded using $\lceil \log \AlphabetSize \rceil$
  bits, and consecutive symbols are packed into each word from least to most
  significant bits, with $s$ symbols per word except possibly the last.
  All unused bits are set to zero.
\end{definition}

\begin{definition}[Packed representation of a sequence of strings]\label{def:packed-sequence-representation}
  Let $\AlphabetSize\in\Z_{\geq2}$ satisfy $\log\AlphabetSize\leq w$,
  and let $m,\ell\in\Z_{\geq0}$. The \emph{packed representation}
  of a sequence $W[1\dd m]$ of strings in $\IntegerAlphabet^\ell$,
  denoted $\PackedSeqRepresentation{w}{\AlphabetSize}{W}$, is the
  packed representation of their concatenation:
  \[
    \PackedSeqRepresentation{w}{\AlphabetSize}{W}
      :=\PackedRepresentation{w}{\AlphabetSize}{\textstyle\bigodot_{i=1}^m W[i]}.
  \]
  The common length $\ell$ determines the boundaries between strings,
  which need not coincide with machine-word boundaries.
  By \cref{def:packed-representation}, this representation uses
  $\bigO(1+m\ell\log\AlphabetSize/w)$ words.
\end{definition}

\begin{example}\label{ex:packed-representation}
  Let $w=8$, $\AlphabetSize=5$, and $S=\texttt{413}$.
  Each symbol occupies $b=3$ bits. A word therefore holds $s=2$ symbols.
  The packed representation $\PackedRepresentation{8}{5}{S}$ of $S$
  consists of the words $x_1=4+1\cdot2^3=12$ and $x_2=3$.
  Written from most to least significant bit, they are
  $\texttt{00\,001\,100}$ and $\texttt{00\,000\,011}$.
  Thus, $S[1]=4$ occupies the lowest three bits of the first word,
  and $S[2]=1$ occupies the next three bits. The highest two bits of
  each word and the unused symbol slot in the second word are zero.
  For the sequence $W=(\texttt{413},\texttt{204})$, we have
  $\PackedSeqRepresentation{8}{5}{W}
    =\PackedRepresentation{8}{5}{\texttt{413204}}=(12,19,32)$.
  The second word contains the last symbol of $W[1]$ and the first
  symbol of $W[2]$. Thus, the boundary between the two strings lies
  inside a machine word.
\end{example}

\begin{remark}\label{rm:packed-representation-interface}
  To avoid repeating bit manipulations in our algorithms, we use
  the basic operations of \cref{pr:packed-representation}, all of
  which can be implemented using standard word RAM operations.
  We use packed representations for string inputs and, where required,
  outputs.
  Internally, our algorithms and data structures often represent a
  short string $X$ by its length and integer value
  $\Val{\AlphabetSize}{X}$ (\cref{def:val}), independently of how
  its symbols are packed into machine words.
  The structures from \cref{th:val-encoding,th:val-decoding} convert
  between these representations.

  The remaining constructions in \cref{sec:tools,sec:sparse-anchors}
  and the later algorithmic reductions manipulate packed strings only through these basic
  operations and conversions, or through tools built from them.

  To use a different implementation of packed strings, for example
  one that allows symbols to cross word boundaries or stores symbol
  fields in a different order within words, it suffices to adapt only
  the basic operations and conversions, leaving the remaining
  algorithms unchanged.
  The remaining constructions retain their correctness and time
  and space bounds, provided that the operations and conversions
  satisfy the same guarantees and the representation satisfies
  the same asymptotic size bound.
\end{remark}

\begin{proposition}\label{pr:packed-representation}
  Let $\AlphabetSize,N\in\Z_{\geq2}$, and assume that $\AlphabetSize$
  is known and that $b = \lceil \log \AlphabetSize \rceil$ has been computed.
  In the word RAM model with word size $w$ satisfying $w\geq b$ and
  $w>\log N$, the following operations on packed representations
  (\cref{def:packed-representation}) are supported:
  \begin{enumerate}[itemsep=1ex]

  \item\label{pr:packed-representation-initialize}
    \ul{Initialize:}
    Given any $m \in [0 \dd N]$ and any symbol
    $a \in \IntegerAlphabet$, the packed representation
    $\PackedRepresentation{w}{\AlphabetSize}{a^m}$ of $a^m$ can be
    constructed in $\bigO(1 + m\log\AlphabetSize/w)$ time.

  \item\label{pr:packed-representation-access}
    \ul{Access:}
    Given the packed representation $\PackedRepresentation{w}{\AlphabetSize}{S}$
    of $S$ and its length $|S|$, where
    $S \in \IntegerAlphabet^{\leq N}$, and any
    $i \in [1 \dd |S|]$, $S[i]$ can be computed in $\bigO(1)$ time.

  \item\label{pr:packed-representation-update}
    \ul{Update:}
    Given the packed representation $\PackedRepresentation{w}{\AlphabetSize}{S}$
    of $S$ and its length $|S|$, where
    $S \in \IntegerAlphabet^{\leq N}$, any $i \in [1 \dd |S|]$,
    and any $a \in \IntegerAlphabet$,
    the packed representation $\PackedRepresentation{w}{\AlphabetSize}{S}$
    can be updated in $\bigO(1)$ time to reflect the assignment $S[i] := a$.

  \item\label{pr:packed-representation-substring}
    \ul{Substring:}
    Given the packed representation $\PackedRepresentation{w}{\AlphabetSize}{S}$
    of $S$ and its length $|S|$, where
    $S \in \IntegerAlphabet^{\leq N}$, any $\ell \in [0 \dd |S|]$,
    and any $p \in [1 \dd |S|-\ell+1]$,
    the packed representation $\PackedRepresentation{w}{\AlphabetSize}{S[p \dd p+\ell)}$
    can be computed in
    $\bigO(1 + \ell\log\AlphabetSize/w)$ time.

  \item\label{pr:packed-representation-concat}
    \ul{Concatenate:}
    Given $m \in [1 \dd N]$ and, for every $i \in [1 \dd m]$,
    the packed representation $\PackedRepresentation{w}{\AlphabetSize}{S_i}$
    of $S_i$ and its length $|S_i|$, where
    $S_i \in \IntegerAlphabet^{\leq N}$ and
    $\sum_{i=1}^{m}|S_i| \leq N$,
    the packed representation $\PackedRepresentation{w}{\AlphabetSize}{S}$
    of $S = S_1 \cdots S_m$ can be computed in
    $\bigO(m + |S|\log\AlphabetSize/w)$ time.

  \item\label{pr:packed-representation-split}
    \ul{Split:}
    Given the packed representation $\PackedRepresentation{w}{\AlphabetSize}{S}$
    of $S$ and its length $|S|$, where
    $S \in \IntegerAlphabet^{\leq N}$, an integer
    $m \in [1 \dd N]$, and nonnegative integers
    $\ell_1, \ldots, \ell_m$ summing to $|S|$, the packed representations
    $\PackedRepresentation{w}{\AlphabetSize}{S_i}$ of the strings
    $S_i \in \IntegerAlphabet^{\ell_i}$, $i \in [1 \dd m]$, satisfying
    $S = S_1 \cdots S_m$ can be computed in
    $\bigO(m + |S|\log\AlphabetSize/w)$ time.

  \item\label{pr:packed-representation-increase}
    \ul{Increase:}
    Given the packed representation $\PackedRepresentation{w}{\AlphabetSize}{S}$
    of $S$, its length $|S|$, and an integer $\delta\in[0\dd\AlphabetSize)$,
    where $S\in\IntegerAlphabet^{\leq N}$ and
    $S[i]+\delta<\AlphabetSize$ for every $i\in[1\dd|S|]$,
    the packed representation $\PackedRepresentation{w}{\AlphabetSize}{S'}$
    of the string $S'$ satisfying $|S'|=|S|$ and
    $S'[i]=S[i]+\delta$ for every $i\in[1\dd|S|]$ can be computed in
    $\bigO(1+|S|\log\AlphabetSize/w)$ time.

  \item\label{pr:packed-representation-compare}
    \ul{Compare:}
    Given the lengths $|X|,|Y|$ and the packed representations
    $\PackedRepresentation{w}{\AlphabetSize}{X}$ and
    $\PackedRepresentation{w}{\AlphabetSize}{Y}$ of strings
    $X,Y\in\IntegerAlphabet^{\leq N}$, we can determine
    whether $X\prec Y$, $X=Y$, or $Y\prec X$ in
    $\bigO(1+\min\{|X|,|Y|\}\log\AlphabetSize/w)$ time.

  \end{enumerate}
\end{proposition}

\subsection{Suffix Arrays}\label{sec:suffix-array}

\begin{definition}[Suffix array]\label{def:sa}
  Let $\Textlen\in\Z_{\geq1}$ and $\Text\in\Sigma^{\Textlen}$.
  The \emph{suffix array} of $\Text$ is an array
  $\SA{\Text}[1\dd\Textlen]$ containing a permutation of
  $[1\dd\Textlen]$ such that
  \[
    \Text[\SA{\Text}[1]\dd\Textlen]
      \prec \Text[\SA{\Text}[2]\dd\Textlen]
      \prec \cdots \prec \Text[\SA{\Text}[\Textlen]\dd\Textlen],
  \]
  i.e., $\SA{\Text}[i]$ is the starting position of the
  $i$th lexicographically smallest nonempty suffix of $\Text$ for every
  $i\in[1\dd\Textlen]$.
\end{definition}

\begin{definition}[Sequence of positions in lex-order]\label{def:lex-sorted}
  Let $\Text \in \Sigma^{\Textlen}$, $Q \subseteq [1 \dd \Textlen+1]$, and $q = |Q|$. By
  $\LexSorted{Q}{\Text}$, we denote a sequence $(a_i)_{i \in [1 \dd q]}$
  containing all positions from $Q$ such that for every
  $i, j \in [1 \dd q]$, $i < j$ implies
  $\Text[a_i \dd \Textlen] \prec \Text[a_j \dd \Textlen]$.
  Position $\Textlen+1$ represents the empty suffix.
\end{definition}

\begin{figure}[t!]
  \centering
  \begin{tikzpicture}[yscale=0.2975]
    \small
    \foreach \x [count=\i] in {a, aababa, aababababaababa,
      aba, abaababa, abaababababaababa, ababa, ababaababa,
      abababaababa, ababababaababa, ba, baababa,
      baababababaababa, baba, babaababa, babaababababaababa,
      bababaababa, babababaababa, bbabaababababaababa}
      \draw (2.3,-\i) node[right] {$\texttt{\x}$};
    \draw (2.3,0) node[right] {\scriptsize $\Text[\SA{\Text}[i]\dd\Textlen]$};
    \foreach \x [count=\i] in {b,b,b,b,b,b,a,b,b,a,a,a,a,a,a,b,a,a,a}{
      \draw (-0.3,-\i) node {\footnotesize $\i$};
      \draw (0.5,-\i) node {$\texttt{\x}$};
    }
    \draw (-0.3,0) node {\scriptsize $i$};
    \draw (0.5,0) node {\scriptsize $\BWT{\Text}[i]$};
    \foreach \x [count=\i] in {19,14,5,17,12,3,15,10,8,6,18,13,4,16,11,2,9,7,1}
      \draw (1.7,-\i) node {$\x$};
    \draw (1.7,0) node {\scriptsize $\SA{\Text}[i]$};
  \end{tikzpicture}
  \caption{The sorted suffixes of $\Text=\texttt{\upshape bbabaababababaababa}$
    together with its suffix array and BWT.}\label{fig:example}
\end{figure}

\begin{example}\label{ex:suffix-arrays}
  Let $\Text=\texttt{bbabaababababaababa}$, as in \cref{fig:example}.
  The seventh suffix in lexicographic order is
  $\Text[15\dd19]=\texttt{ababa}$. Thus, $\SA{\Text}[7]=15$.
  For $Q=\{15,17,19\}$, we have
  $\LexSorted{Q}{\Text}=(19,17,15)$.
\end{example}

\begin{definition}[Inverse suffix array]\label{def:isa}
  Let $\Textlen\in\Z_{\geq1}$ and $\Text\in\Sigma^{\Textlen}$.
  The \emph{inverse suffix array} of $\Text$ is the array
  $\ISA{\Text}[1\dd\Textlen]$ containing the inverse permutation
  of $\SA{\Text}$ (\cref{def:sa}), i.e., the unique array
  satisfying $\ISA{\Text}[\SA{\Text}[i]]=i$ for all
  $i\in[1\dd\Textlen]$.
  Equivalently, for every $j\in[1\dd\Textlen]$, the entry $\ISA{\Text}[j]$
  is the lexicographic rank of $\Text[j\dd\Textlen]$ among the
  nonempty suffixes of $\Text$.
\end{definition}

\begin{example}\label{ex:isa}
  For $\Text$ from \cref{fig:example}, the inverse suffix array is as follows:
  \[
    \begin{array}{c|*{19}{r}}
      j & 1&2&3&4&5&6&7&8&9&10&11&12&13&14&15&16&17&18&19\\\hline
      \Text[j] & \texttt{b}&\texttt{b}&\texttt{a}&\texttt{b}&\texttt{a}&\texttt{a}&\texttt{b}&\texttt{a}&\texttt{b}&\texttt{a}&\texttt{b}&\texttt{a}&\texttt{b}&\texttt{a}&\texttt{a}&\texttt{b}&\texttt{a}&\texttt{b}&\texttt{a}\\
      \ISA{\Text}[j] & 19&16&6&13&3&10&18&9&17&8&15&5&12&2&7&14&4&11&1
    \end{array}
  \]
  In particular, $\ISA{\Text}[15]=7$ because $\SA{\Text}[7]=15$.
\end{example}

\begin{remark}\label{rm:occ}
  For every $\Text\in\Sigma^+$ and $\Pat\in\Sigma^*$, the suffixes
  beginning with $\Pat$ occupy consecutive positions in $\SA{\Text}$.
  With the notation of \cref{def:occ}, we have
  \[
    \OccTwo{\Pat}{\Text}
      =\{\SA{\Text}[i]:i\in(\RangeBegTwo{\Pat}{\Text}\dd
        \RangeEndTwo{\Pat}{\Text}]\}.
  \]
  This also holds for $\Pat=\emptystring$, for which
  $\RangeBegTwo{\Pat}{\Text}=0$ and
  $\RangeEndTwo{\Pat}{\Text}=|\Text|$.
\end{remark}

\begin{example}\label{ex:occ}
  For $\Text$ from \cref{fig:example} and $\Pat=\texttt{ababa}$, we have
  $\OccTwo{\Pat}{\Text}=\{6,8,10,15\}$.
  These suffixes occupy positions $7$ through $10$ in the suffix array.
  Consequently, $\RangeBegTwo{\Pat}{\Text}=6$ and $\RangeEndTwo{\Pat}{\Text}=10$.
\end{example}

\subsection{Burrows--Wheeler Transform}\label{sec:burrows-wheeler-transform}

\begin{definition}[Burrows--Wheeler transform~\cite{bwt}]\label{def:bwt}
  Let $\Textlen\in\Z_{\geq1}$ and $\Text\in\Sigma^{\Textlen}$.
  The \emph{Burrows--Wheeler transform} (BWT) of $\Text$ is the
  string $\BWT{\Text}[1\dd\Textlen]\in\Sigma^{\Textlen}$ defined,
  for every $i\in[1\dd\Textlen]$, by
  \[
    \BWT{\Text}[i]=
    \begin{cases}
      \Text[\SA{\Text}[i]-1] & \text{if }\SA{\Text}[i]>1,\\
      \Text[\Textlen] & \text{if }\SA{\Text}[i]=1.
    \end{cases}
  \]
  Here, $\SA{\Text}$ is the suffix array of $\Text$ (\cref{def:sa}).
  Equivalently, $\BWT{\Text}[i]=\Textinf[\SA{\Text}[i]-1]$
  for every $i\in[1\dd\Textlen]$.
  We denote the number of runs in the BWT by
  $\RLBWTSize{\Text}:=|\RL{\BWT{\Text}[1\dd\Textlen]}|$.
\end{definition}

\begin{example}\label{ex:bwt}
  For $\Text$ from \cref{fig:example}, we have
  $\BWT{\Text}[1\dd19]=\texttt{b}^6\texttt{a}\texttt{b}^2
    \texttt{a}^6\texttt{b}\texttt{a}^3$, and hence $\RLBWTSize{\Text}=6$.
  In particular, $\SA{\Text}[7]=15$ gives
  $\BWT{\Text}[7]=\Text[14]=\texttt{a}$, while
  $\SA{\Text}[19]=1$ gives $\BWT{\Text}[19]=\Text[19]=\texttt{a}$.
\end{example}

\subsection{Lempel--Ziv Factorization}\label{sec:lempel-ziv-factorization}

\begin{definition}[Longest previous factor array]\label{def:lpf}
  Let $\Textlen\in\Z_{\geq1}$ and $\Text\in\Sigma^{\Textlen}$.
  The \emph{longest previous factor array}
  $\LPF{\Text}[1\dd\Textlen]$ of $\Text$ is defined by
  $\LPF{\Text}[1]=0$ and, for every $j\in[2\dd\Textlen]$,
  \[
    \LPF{\Text}[j]=\max\{\ell\in[0\dd\Textlen-j+1]:
      \min\OccTwo{\Text[j\dd j+\ell)}{\Text}<j\}
  \]
  (see \cref{def:occ}).
\end{definition}

\begin{definition}[Longest previous non-overlapping factor array]\label{def:lpnf}
  Let $\Textlen\in\Z_{\geq1}$ and $\Text\in\Sigma^{\Textlen}$.
  The \emph{longest previous non-overlapping factor array}
  $\LPnF{\Text}[1\dd\Textlen]$ of $\Text$ is defined by
  $\LPnF{\Text}[1]=0$ and, for every $j\in[2\dd\Textlen]$,
  \[
    \LPnF{\Text}[j]=\max\{\ell\in[0\dd\Textlen-j+1]:
      \min\OccTwo{\Text[j\dd j+\ell)}{\Text}+\ell-1<j\}
  \]
  (see \cref{def:occ}).
\end{definition}

\begin{example}\label{ex:lpf}
  For $\Text=\texttt{abababab}$, the two arrays are as follows:
  \[
    \begin{array}{c|rrrrrrrr}
      j & 1&2&3&4&5&6&7&8\\\hline
      \Text[j] & \texttt{a}&\texttt{b}&\texttt{a}&\texttt{b}&\texttt{a}&\texttt{b}&\texttt{a}&\texttt{b}\\
      \LPF{\Text}[j] & 0&0&6&5&4&3&2&1\\
      \LPnF{\Text}[j] & 0&0&2&2&4&3&2&1
    \end{array}
  \]
  The suffix $\Text[3\dd8]=\texttt{ababab}$ also occurs at position $1$,
  giving $\LPF{\Text}[3]=6$. Since only its first two symbols fit before
  position $3$, we have $\LPnF{\Text}[3]=2$.
\end{example}

\begin{observation}\label{ob:LPF}
  Let $\Textlen\in\Z_{\geq1}$ and $\Text \in \Sigma^{\Textlen}$.
  For every $i \in [2 \dd \Textlen]$, it holds
  $\LPF{\Text}[i] \geq \LPF{\Text}[i-1] - 1$ and
  $\LPnF{\Text}[i] \geq \LPnF{\Text}[i-1] - 1$.
\end{observation}

\begin{lemma}\label{lm:LPF}
  Let $\Textlen\in\Z_{\geq1}$ and $\Text \in \Sigma^{\Textlen}$.
  For every $i, j \in [1 \dd \Textlen]$ satisfying
  $i \leq j$, it holds
  \[
    i + \LPF{\Text}[i]
      \leq j + \LPF{\Text}[j],
    \qquad
    i + \LPnF{\Text}[i]
      \leq j + \LPnF{\Text}[j].
  \]
\end{lemma}
\begin{proof}
  By repeatedly applying \cref{ob:LPF}, we obtain
  $i + \LPF{\Text}[i] \leq i + 1 + \LPF{\Text}[i + 1] \leq
  i + 2 + \LPF{\Text}[i + 2] \leq \dots \leq i + (j-i) + \LPF{\Text}[i + (j-i)]
  = j + \LPF{\Text}[j]$. The other inequality follows by an analogous argument.
\end{proof}

\begin{definition}[LZ77 factorization~\cite{LZ77}]\label{def:lz77}
  Let $\Textlen\in\Z_{\geq1}$ and $\Text\in\Sigma^{\Textlen}$.
  The \emph{LZ77 factorization} of $\Text$ is the greedy left-to-right
  partition $\Text=f_1\cdots f_z$ into nonempty substrings, called
  \emph{phrases}. For every $i\in[1\dd z]$, the length of $f_i$
  is defined by
  \[
    |f_i|=\max(1,\LPF{\Text}[j_i]),
  \]
  where $\LPF{\Text}$ is the longest previous factor array
  (\cref{def:lpf}) and
  $j_i=1+\sum_{k=1}^{i-1}|f_k|$.
  A phrase $f_i$ is a \emph{literal phrase} if $\LPF{\Text}[j_i]=0$,
  and a \emph{repeat phrase} otherwise.
  We denote the number of phrases by $\LZSize{\Text}=z$.
\end{definition}

\begin{definition}[LZ77 representation]\label{def:lz77-representation}
  Let $\Text\in\Sigma^+$, and let $\Text=f_1\cdots f_z$ be its LZ77
  factorization (\cref{def:lz77}). An \emph{LZ77 representation}
  of $\Text$ is a sequence of $z$ pairs encoding the phrases
  $f_1,\ldots,f_z$ in this order. For every $i\in[1\dd z]$, let
  $j=1+\sum_{k=1}^{i-1}|f_k|$ be the starting position of $f_i$.
  A literal phrase $f_i$ is encoded as $(\Text[j],0)$.
  A repeat phrase $f_i$ is encoded as $(p,|f_i|)$, where
  $p\in[1\dd j)$ is any position satisfying
  $\Text[p\dd p+|f_i|)=f_i$.
  When symbols, positions, and lengths each fit in a machine word,
  the representation occupies $\bigO(\LZSize{\Text})$ words.
\end{definition}

\begin{remark}\label{rm:lz77-size}
  Although this is not immediate from the definition, for every string
  $\Text\in\IntegerAlphabet^{\Textlen}$ with $\Textlen,\AlphabetSize\geq2$,
  the number of phrases $\LZSize{\Text}$ in its LZ77 factorization satisfies
  $\LZSize{\Text}=\bigO(\Textlen/\log_{\AlphabetSize}\Textlen)$~\cite[Theorem~2]{LZ76}.
  This bound on the output size allows for sublinear-time algorithms
  over sufficiently small alphabets, such as the algorithm
  from~\cite{sublinearlz}.
\end{remark}

\begin{example}\label{ex:lz77}
  The string $\Text=\texttt{abababab}$ from \cref{ex:lpf} has the LZ77
  factorization $\Text=\texttt{a}\cdot\texttt{b}\cdot\texttt{ababab}$,
  and hence $\LZSize{\Text}=3$. One LZ77 representation is
  $(\texttt{a},0),(\texttt{b},0),(1,6)$.
  The last pair refers to $\Text[1\dd6]$, which overlaps the phrase
  $\Text[3\dd8]$.
\end{example}

\begin{definition}[Non-overlapping LZ77 factorization]\label{def:nonoverlapping-lz77}
  Let $\Textlen\in\Z_{\geq1}$ and $\Text\in\Sigma^{\Textlen}$.
  The \emph{non-overlapping LZ77 factorization} of $\Text$ is the
  greedy left-to-right partition $\Text=f_1\cdots f_z$ into nonempty
  phrases defined as in \cref{def:lz77}, with $\LPF{\Text}$ replaced
  by $\LPnF{\Text}$ (\cref{def:lpnf}). Thus, for every
  $i\in[1\dd z]$, we have
  \[
    |f_i|=\max(1,\LPnF{\Text}[j_i]),
  \]
  where $j_i=1+\sum_{k=1}^{i-1}|f_k|$.
  A phrase $f_i$ is literal if $\LPnF{\Text}[j_i]=0$,
  and a repeat phrase otherwise. We denote the number of phrases by
  $\LZNonOvSize{\Text}=z$.
\end{definition}

\begin{definition}[Non-overlapping LZ77 representation]\label{def:nonoverlapping-lz77-representation}
  Let $\Text\in\Sigma^+$, and let $\Text=f_1\cdots f_z$ be its
  non-overlapping LZ77 factorization (\cref{def:nonoverlapping-lz77}).
  A \emph{non-overlapping LZ77 representation} of $\Text$ is a sequence
  of $z$ pairs encoding these phrases in order as in
  \cref{def:lz77-representation}.
  For every repeat phrase $f_i$ starting at position $j$, the source
  position $p$ in its pair $(p,|f_i|)$ additionally satisfies
  $p+|f_i|\leq j$.
  When symbols, positions, and lengths each fit in a machine word,
  the representation occupies $\bigO(\LZNonOvSize{\Text})$ words.
\end{definition}

\begin{example}\label{ex:nonoverlapping-lz77}
  The non-overlapping LZ77 factorization of $\Text=\texttt{abababab}$
  is $\Text=\texttt{a}\cdot\texttt{b}\cdot\texttt{ab}\cdot\texttt{abab}$.
  Thus, $\LZNonOvSize{\Text}=4$.
  One representation is $(\texttt{a},0),(\texttt{b},0),(1,2),(1,4)$.
  The repeat phrases start at positions $3$ and $5$, and their source
  occurrences $\Text[1\dd2]$ and $\Text[1\dd4]$ end before those positions.
\end{example}

\subsection{Longest Common Extensions}\label{sec:longest-common-extensions}

\begin{definition}[Longest common extension]\label{def:lce}
  Let $S\in\Sigma^*$ and $i,j\in[1\dd|S|+1]$.
  The \emph{longest common extension} of $S$ at positions $i$ and $j$,
  denoted $\LCE{S}{i}{j}$, is the length of the longest common prefix
  of the suffixes $S[i\dd|S|]$ and $S[j\dd|S|]$.
\end{definition}

\begin{example}
  For $S=\texttt{cabdcabac}$, we have $\LCE{S}{1}{5}=3$ (the common prefix
  is $\texttt{cab}$), while $\LCE{S}{1}{10}=0$ because $S[10\dd9]=\emptystring$.
\end{example}

\begin{theorem}[{\cite[Theorem~5.4]{sss}}]\label{th:lce}
  Let $N,\AlphabetSize\in\Z_{\geq2}$ satisfy $\AlphabetSize\leq N$.
  Consider the word RAM model with word size $w\geq1+\log N$.
  Given $N$, $\AlphabetSize$, and the packed representation
  $\PackedRepresentation{w}{\AlphabetSize}{\Text}$
  (\cref{def:packed-representation}) of a text
  $\Text\in\IntegerAlphabet^N$, we can deterministically construct,
  in $\bigO(N/\log_{\AlphabetSize}N)$ time and using
  $\bigO(N/\log_{\AlphabetSize}N)$ words of space, a data structure
  of size $\bigO(N/\log_{\AlphabetSize}N)$ words that, given
  $i,j\in[1\dd N]$, computes $\LCE{\Text}{i}{j}$
  (\cref{def:lce}) in $\bigO(1)$ time.
\end{theorem}

\subsection{String Synchronizing Sets}\label{sec:string-synchronizing-sets}

\begin{definition}[$\tau$-periodic positions~\cite{sss}]
  \label{def:periodic-positions}
  Let $\Textlen\in\Z_{\geq0}$, $\Text\in\Sigma^{\Textlen}$, and
  $\tau\in\Z_{\geq1}$. We define the set of \emph{$\tau$-periodic
  positions} of $\Text$ as
  \[
    \RTwo{\tau}{\Text}
      := \{i \in [1 \dd \Textlen - 3\tau + 2] :
        \per{\Text[i \dd i + 3\tau - 1)} \leq \tfrac{1}{3}\tau\}.
  \]
\end{definition}

\begin{definition}[$\tau$-periodic and $\tau$-nonperiodic patterns]\label{def:periodic-pattern}
  Let $m\in\Z_{\geq0}$, $\Pat \in \Sigma^{m}$, and $\tau\in\Z_{\geq1}$.
  We say that $\Pat$ is
  \emph{$\tau$-periodic} if it holds $m \geq 3\tau - 1$ and
  $\per{\Pat[1 \dd 3\tau - 1]} \leq \tfrac{1}{3}\tau$. Otherwise, it
  is called \emph{$\tau$-nonperiodic}.
\end{definition}

\begin{definition}[$\tau$-synchronizing set~\cite{sss}]\label{def:sss}
  Let $\Textlen\in\Z_{\geq2}$, $\Text\in \Sigma^{\Textlen}$, and $\tau \in
  [1 \dd \lfloor\frac{\Textlen}{2}\rfloor]$. A set $\SSS
  \subseteq [1 \dd \Textlen - 2\tau + 1]$ is called a
  \emph{$\tau$-synchronizing set} of $\Text$ if it satisfies the
  following \emph{consistency} and \emph{density} conditions:
  \begin{enumerate}
  \item\label{def:sss-consistency}
    If $\Text[i \dd i + 2\tau) = \Text[j\dd j + 2\tau)$, then $i \in \SSS$
    holds if and only if $j \in \SSS$
    (for $i, j \in [1 \dd \Textlen - 2\tau + 1]$),
  \item\label{def:sss-density}
    $\SSS\cap[i \dd i + \tau) = \emptyset$ if and only if
    $i \in \RTwo{\tau}{\Text}$ (for
    $i \in [1 \dd \Textlen - 3\tau + 2]$; see
    \cref{def:periodic-positions}).
  \end{enumerate}
\end{definition}

\begin{example}\label{ex:synchronizing-set}
  Let $\tau=3$ and $\Text=\texttt{a}^{10}\texttt{b}\texttt{a}^{10}$.
  A substring of length $3\tau-1=8$ has period at most $\tau/3=1$
  exactly when it consists entirely of $\texttt{a}$'s. Hence,
  $\RTwo{\tau}{\Text}=[1\dd3]\cup[12\dd14]$.
  For example, $\Text[3\dd10]=\texttt{a}^8$ is a $\tau$-periodic pattern,
  while $\Text[4\dd11]=\texttt{a}^7\texttt{b}$ is a $\tau$-nonperiodic pattern.
  The set $\SSS=\{6,9,11\}$ is a $\tau$-synchronizing set of $\Text$.
  For consistency, every occurrence of $\texttt{a}^6$ starts outside
  $\SSS$, while each length-$6$ substring containing $\texttt{b}$
  occurs only once.
  For density, the intervals $[i\dd i+3)$ with $i\in[1\dd14]$
  that contain an element of $\SSS$ are precisely those with
  $i\in[4\dd11]$, the complement of $\RTwo{\tau}{\Text}$ in $[1\dd14]$.
\end{example}

\begin{theorem}[{\cite{sss}}]\label{th:sss-existence-and-construction}
  Let $\Textlen\in\Z_{\geq2}$. For every $\Text\in\Sigma^{\Textlen}$ and $\tau \in
  [1 \dd \lfloor\frac{\Textlen}{2}\rfloor]$, there exists a $\tau$-synchronizing
  set $\SSS$ (\cref{def:sss}) of size at most $30\Textlen/\tau$.
  Moreover, in the word RAM model with word size $w\geq1+\log\Textlen$,
  if $\Text \in \IntegerAlphabet^{\Textlen}$, where
  $\AlphabetSize\in\Z_{\geq1}$ satisfies $\AlphabetSize = \Textlen^{\bigO(1)}$,
  then, given $\Textlen$, $\AlphabetSize$, $\tau$, and $\Text$, such $\SSS$
  can be deterministically constructed in $\bigO(\Textlen)$ time.
\end{theorem}

\begin{theorem}[{\cite{sss}}]\label{th:sss-packed-construction}
  Let $\Textlen,\AlphabetSize\in\Z_{\geq2}$ satisfy
  $\AlphabetSize\leq\Textlen$, let $\mu\in(0,\tfrac15)$ be a
  constant, and let $\tau\in\Z_{\geq1}$ satisfy
  $\tau\leq\mu\log_{\AlphabetSize}\Textlen$.
  Consider the word RAM model with word size $w\geq1+\log\Textlen$.
  Given $\Textlen$, $\AlphabetSize$, $\tau$, and the packed
  representation $\PackedRepresentation{w}{\AlphabetSize}{\Text}$
  (\cref{def:packed-representation}) of a text
  $\Text\in\IntegerAlphabet^{\Textlen}$, we can deterministically
  construct an array containing a $\tau$-synchronizing set of
  $\Text$ (\cref{def:sss}) of size at most $30\Textlen/\tau$ in
  $\bigO(\Textlen/\tau)$ time using
  $\bigO(\Textlen/\tau)$ words of space.
\end{theorem}

\begin{theorem}[{\cite{sss}}]\label{th:sss-lex-sort}
  Let $\AlphabetSize,\Textlen\in\Z_{\geq2}$ satisfy
  $\AlphabetSize\leq\Textlen$, and let
  $\tau\in[1\dd\lfloor\tfrac{\Textlen}{2}\rfloor]$ satisfy
  $\tau=\bigO(\log_{\AlphabetSize}\Textlen)$.
  Consider the word RAM model with word size $w\geq1+\log\Textlen$.
  Let $\SSS$ be a $\tau$-synchronizing set (\cref{def:sss}) of a text
  $\Text\in\IntegerAlphabet^{\Textlen}$, with
  $|\SSS|=\bigO(\Textlen/\tau)$. Given $\Textlen$, $\AlphabetSize$, $\tau$,
  the packed representation
  $\PackedRepresentation{w}{\AlphabetSize}{\Text}$
  (\cref{def:packed-representation}) and an array listing the elements
  of $\SSS$, we can deterministically compute the sequence
  $\LexSorted{\SSS}{\Text}$ (\cref{def:lex-sorted}) in
  $\bigO(\Textlen/\tau)$ time.
\end{theorem}

\subsection{Rank Queries on Bitvectors}\label{sec:bitvector-rank}

\begin{definition}[Rank queries on bitvectors]\label{def:bitvector-rank}
  Let $m\in\Z_{\geq0}$ and $B\in\BinaryAlphabet^m$.
  For every $a\in\BinaryAlphabet$ and $j\in[0\dd m]$, we define
  $\Rank{B}{j}{a}:=|\{i\in[1\dd j]:B[i]=a\}|$.
\end{definition}

\begin{example}
  For $B=\texttt{010110010}$, we have $\Rank{B}{5}{\one}=3$,
  since the ones in $B[1\dd5]$ are at positions $2$, $4$, and $5$,
  while $\Rank{B}{0}{\one}=0$.
\end{example}

\begin{theorem}[{\cite[Lemma~2.3]{WaveletSuffixTree};
    see also~\cite{Clark96,Jac89,MunroNV16}}]
  \label{th:bitvector-rank}
  Let $n\in\Z_{\geq2}$, and consider the word RAM model with
  word size $w>\log n$.
  Given $n$ and the packed representation
  $\PackedRepresentation{w}{2}{B}$ (\cref{def:packed-representation})
  of a bitvector $B\in\BinaryAlphabet^n$, we can deterministically
  construct, in $\bigO(n/\log n)$ time, a data structure of size
  $\bigO(n/\log n)$ words answering rank queries
  (\cref{def:bitvector-rank}) in $\bigO(1)$ time.
\end{theorem}

\subsection{Range Maximum Queries}\label{sec:range-maximum-queries}

\begin{theorem}[{\cite{FischerH11}}]\label{th:range-maximum-queries}
  Let $m\in\Z_{\geq1}$, and consider the word RAM model with word size
  $w>\log m$. Given $m$ and an array $A[1\dd m]$ of integers that fit
  in a machine word, we can deterministically construct, in $\bigO(m)$
  time, a data structure of size $\bigO(m)$ words that, given
  $b,e\in[0\dd m]$ with $b<e$, returns the smallest index
  $j\in(b\dd e]$ satisfying
  $A[j]=\max\{A[i]:i\in(b\dd e]\}$ in $\bigO(1)$ time.
\end{theorem}

\subsection{Inversion Counts}\label{sec:inversion-counts}

\begin{definition}[Inversion count]\label{def:inv-count}
  Let $m\in\Z_{\geq0}$, and let $\mathcal{X}$ be a set equipped with a total order $\prec$.
  For every array $A[1 \dd m]$ of elements from $\mathcal{X}$, we denote
  \[
    \InversionCount{A} := |\{(i,j) \in [1 \dd m]^2 : i < j\text{ and }A[i] \succ A[j]\}|.
  \]
\end{definition}

\begin{definition}[Colored inversion count]\label{def:colored-inv-count}
  Let $m\in\Z_{\geq0}$, and let $\mathcal{X}$ be a set equipped with a total order $\prec$.
  For every array $A[1 \dd m]$ of elements from $\mathcal{X}$, and every
  array $C[1 \dd m]$ of integers, we denote
  \[
    \ColoredInversionCount{C}{A} := |\{(i,j) \in [1 \dd m]^2 : i < j,\,C[i] \neq C[j],\text{ and }A[i] \succ A[j]\}|.
  \]
\end{definition}

\begin{example}
  For arrays $A[1\dd3]$ and $C[1\dd3]$ containing the sequences
  $(2,1,0)$ and $(0,0,1)$, respectively, we have
  $\InversionCount{A}=3$ and $\ColoredInversionCount{C}{A}=2$.
\end{example}

\section{Technical Overview}\label{sec:overview}

\subsection{Reducing Dictionary Matching to LZ77 Factorization}\label{sec:overview-lz}

In this section, we sketch the key ideas in our reduction of the
\probname{Dictionary} \probname{Matching} problem to \probname{LZ77}
\probname{Factorization} (or, more precisely, \probname{LZ77}
\probname{Parity}).

\paragraph{Towards the Basic Idea}

Assume that we are given an instance of the \probname{Dictionary}
\probname{Matching} problem, i.e., a packed representation of a nonempty text
$\Text \in \BinaryAlphabet^{\Textlen}$ and the packed representation
of a nonempty collection $\mathcal{D} = \{\Pat_1, \ldots, \Pat_k\}$ of
$k=\lfloor\Textlen/\log\Textlen\rfloor$ distinct patterns of common
length $m=2\lfloor\log\Textlen\rfloor$. We assume $\Textlen\geq8$
(for smaller $\Textlen$, direct comparison solves the instance in
$\bigO(1)$ time).

Let us first explore naive attempts to reduce \probname{Dictionary}
\probname{Matching} to \probname{LZ77} \probname{Factorization}.
Arguably, the string to
be factorized should consist of the text $\Text$ followed by the
concatenation of patterns $\Pat_i$.  The underlying intuition is that
$\Pat_i$ should constitute a single phrase if and only if $\Pat_i$
occurs in $\Text$.  If we could afford a large alphabet with a fresh
sentinel symbol ${\dol_i}$ for each pattern $\Pat_i$, the constructed string
could be of the form $\Text \cdot \dol_0 \cdot \big(\bigodot_{i=1}^k
\Pat_i\cdot {\dol_i}\big)$.  In this string, every sentinel symbol
occurs exactly once and thus constitutes a single-character phrase.  If
$\Pat_i$ occurs in $\Text$, then the phrase starting immediately after
$\dol_{i-1}$ consists of the entire $\Pat_i$ (it cannot be any longer
due to $\dol_{i}$).  Otherwise, due to the assumption that the
patterns are distinct strings of the same length, $\Pat_i$ does not
have an earlier occurrence, and its factorization consists of at least
two phrases.

Unfortunately, we cannot afford an alphabet of size $\Textlen^{\Theta(1)}$:
its reduction back to the binary case would then incur a $\Theta(\log
\Textlen)$-factor overhead.  The natural idea would be to replace each
$\dol_i$ with a string $\Pat'_i$ consisting of $\Theta(\log \Textlen)$
characters from an auxiliary constant-sized alphabet such as
$\{\two,\three\}$.  To this end, we let $\Pat'_i$ be a ``copy'' of
$\Pat_i$ with all occurrences of $\zero$ and $\one$ replaced by $\two$
and $\three$, respectively.  This serves as an easy way of ensuring
that the strings $\Pat'_i$ are distinct and of the same length.

A remaining challenge is ensuring that each block $\Pat_i \cdot
\Pat'_i$ consists of full phrases.  For this, we introduce a new
gadget located between $\Text$ and $\bigodot_{i=1}^k \Pat_i\cdot
\Pat'_i$, whose purpose is to introduce more substrings as possible
phrases.  Specifically, we will ensure that the set of possible
phrases includes $\Pat_i[1\dd m)$, $\Pat_i[m]\cdot \Pat'_i$, and~$\Pat'_i$.
This way, if $\Pat_i$ occurs in $\Text$, then the block
$\Pat_i\cdot \Pat'_i$ will be decomposed into $\Pat_i$ and $\Pat'_i$;
otherwise, the decomposition will consist of $\Pat_i[1\dd m)$ and
$\Pat_i[m]\cdot \Pat'_i$.  Formally, the string constructed as the
result of our discussion is as follows:
\[
  T \cdot \five \cdot \Big(\bigodot_{i=1}^k \Pat_i[1\dd m) \cdot \four
    \cdot \Pat_i[m]\cdot \Pat'_i \cdot \four\Big)\cdot \six \cdot
    \Big(\bigodot_{i=1}^k \Pat_i \cdot \Pat'_i\Big).
\]
Here, the occurrences of $\four$ guarantee that, among all strings in
$\{\zero,\one,\two,\three\}^*$, only the desired strings and their substrings are introduced as
possible phrases.  The unique occurrence of $\six$ constitutes a
synchronization point so that we need to reason only about the
factorization of the suffix $\bigodot_{i=1}^k \Pat_i \cdot \Pat'_i$.

\paragraph{The Complete Reduction}

In the reduction above, the factorization of the constructed string
lets us determine which patterns $\Pat_i$ occur in $\Text$.
Nevertheless, we cannot tell anything solely based on the number of
LZ77 phrases.  To alleviate this issue, we further extend the middle
gadget so that, once we encounter a pattern $\Pat_i$ that occurs in
$\Text$, the factorization is ``shifted by one character'' and
independent of whether subsequent patterns occur in $\Text$.
Specifically, for $j>i$, the phrases starting within $\Pat_j\cdot
\Pat'_j$ will be $\Pat_j(1\dd m]$ and $\Pat'_j\cdot \Pat_{j+1}[1]$.
Now, the answer to the \probname{Dictionary Matching} instance can be
recovered based on whether the extra character $\Pat_{k+1}[1]\coloneqq
\zero$, which we append at the end of the constructed string, needs an
extra phrase to be included in the LZ77 factorization.

\begin{figure}[t!]
  \centering

  \begin{tikzpicture}[scale=0.34]
    \scriptsize

    \def\blockwidth{16.0}
    \def\blockheight{1.0}

    \foreach \i/\j in {1/2, 2/3, 3/4} {

      \pgfmathtruncatemacro{\blocknum}{\i - 1}
      \pgfmathtruncatemacro{\xpos}{\blocknum * \blockwidth}

      \draw[fill=white] (\xpos, 0) rectangle ++(2, 1);
      \node at (\xpos + 1, 0.5*\blockheight) {$m_{\i}e_{\i}$};
      \draw[fill=gray!80] (\xpos+2.0, 0) rectangle ++(1, 1);
      \node[white] at (\xpos + 2.5, 0.5*\blockheight) {${\tt 4}$};
      \draw[fill=white] (\xpos+3.0, 0) rectangle ++(2, 1);
      \node at (\xpos + 4.0, 0.5*\blockheight) {$b_{\i}m_{\i}$};
      \draw[fill=gray!80] (\xpos+5.0, 0) rectangle ++(1, 1);
      \node[white] at (\xpos + 5.5, 0.5*\blockheight) {${\tt 4}$};
      \fill[white] (\xpos+6.0, 0) rectangle ++(1, 1);
      \node at (\xpos + 6.5, 0.5*\blockheight) {$e_{\i}$};
      \fill[gray!25] (\xpos+7.0, 0) rectangle ++(3, 1);
      \node at (\xpos + 8.5, 0.5*\blockheight) {$b'_{\i}m'_{\i}e'_{\i}$};
      \draw (\xpos+6.0, 0) rectangle ++(4, 1);
      \draw[fill=gray!80] (\xpos+10.0, 0) rectangle ++(1, 1);
      \node[white] at (\xpos + 10.5, 0.5*\blockheight) {${\tt 4}$};
      \fill[gray!25] (\xpos+11.0, 0) rectangle ++(3, 1);
      \fill[white] (\xpos+14.0, 0) rectangle ++(1, 1);
      \node at (\xpos + 12.5, 0.5*\blockheight) {$b'_{\i}m'_{\i}e'_{\i}$};
      \node at (\xpos + 14.5, 0.5*\blockheight) {$b_{\j}$};
      \draw (\xpos+11.0, 0) rectangle ++(4, 1);
      \draw[fill=gray!80] (\xpos+15.0, 0) rectangle ++(1, 1);
      \node[white] at (\xpos + 15.5, 0.5*\blockheight) {${\tt 4}$};
    }

    \def\xoff{0.1}
    \draw [decorate,decoration={brace,amplitude=4pt,mirror,raise=0.3ex}] (3+\xoff,0) -- ++(2-2*\xoff,0) node[midway,yshift=-1.3em]{$s_1$};
    \draw [decorate,decoration={brace,amplitude=4pt,mirror,raise=0.3ex}] (6+\xoff,0) -- ++(4-2*\xoff,0) node[midway,yshift=-1.3em]{$s_2$};
    \draw [decorate,decoration={brace,amplitude=4pt,mirror,raise=0.3ex}] (19+\xoff,0) -- ++(2-2*\xoff,0) node[midway,yshift=-1.3em]{$s_3$};
    \draw [decorate,decoration={brace,amplitude=4pt,mirror,raise=0.3ex}] (22+\xoff,0) -- ++(4-2*\xoff,0) node[midway,yshift=-1.3em]{$s_4$};
    \draw [decorate,decoration={brace,amplitude=4pt,mirror,raise=0.3ex}] (35+\xoff,0) -- ++(2-2*\xoff,0) node[midway,yshift=-1.3em]{$s_5$};
    \draw [decorate,decoration={brace,amplitude=4pt,mirror,raise=0.3ex}] (38+\xoff,0) -- ++(4-2*\xoff,0) node[midway,yshift=-1.3em]{$s_6$};

    \draw [decorate,decoration={brace,amplitude=4pt,raise=0.3ex}] (3+\xoff,1) -- ++(2-2*\xoff,0) node[midway,yshift=1.3em]{$s_1$};
    \draw [decorate,decoration={brace,amplitude=4pt,raise=0.3ex}] (6+\xoff,1) -- ++(4-2*\xoff,0) node[midway,yshift=1.3em]{$s_2$};
    \draw [decorate,decoration={brace,amplitude=4pt,raise=0.3ex}] (27+\xoff,1) -- ++(4-2*\xoff,0) node[midway,yshift=1.3em]{$s_4$};
    \draw [decorate,decoration={brace,amplitude=4pt,raise=0.3ex}] (32+\xoff,1) -- ++(2-2*\xoff,0) node[midway,yshift=1.3em]{$s_5$};
    \draw [decorate,decoration={brace,amplitude=4pt,raise=0.3ex}] (43+\xoff,1) -- ++(4-2*\xoff,0) node[midway,yshift=1.3em]{$s_6$};
  \end{tikzpicture}

  \begin{tikzpicture}[scale=0.34]
    \scriptsize

    \def\blockwidth{3.0}
    \def\blockheight{1.0}

    \foreach \i in {1, 2, ..., 6} {

      \pgfmathtruncatemacro{\blocknum}{\i - 1}
      \pgfmathtruncatemacro{\xpos}{\blocknum * \blockwidth}
    
      \pgfmathtruncatemacro{\labelnum}{(\i + 1) / 2}
      \ifodd\i
        \fill[white] (\xpos, 0) rectangle ++(\blockwidth, \blockheight);
        \def\labeltext{$b_{\labelnum}m_{\labelnum}e_{\labelnum}$}
      \else
        \fill[gray!25] (\xpos, 0) rectangle ++(\blockwidth, \blockheight);
        \def\labeltext{$b'_{\labelnum}m'_{\labelnum}e'_{\labelnum}$}
      \fi
    
      \node at (\xpos + 0.5*\blockwidth, 0.5*\blockheight) {\labeltext};
    }

    \pgfmathsetmacro{\xpos}{6 * \blockwidth}
    \fill[white] (\xpos, 0) rectangle ++(0.3*\blockwidth, \blockheight);
    \node at (\xpos + 0.15*\blockwidth, 0.5*\blockheight) {$b_4$};
    \draw (0, 0) rectangle ++(19, \blockheight);

    \def\xoff{0.1}
    \draw [decorate,decoration={brace,amplitude=4pt,mirror,raise=0.3ex}] (0+\xoff,0) -- ++(2-2*\xoff,0) node[midway,yshift=-1.3em]{$f_1$};
    \draw [decorate,decoration={brace,amplitude=4pt,mirror,raise=0.3ex}] (2+\xoff,0) -- ++(4-2*\xoff,0) node[midway,yshift=-1.3em]{$f_2$};
    \draw [decorate,decoration={brace,amplitude=4pt,mirror,raise=0.3ex}] (6+\xoff,0) -- ++(2-2*\xoff,0) node[midway,yshift=-1.3em]{$f_3$};
    \draw [decorate,decoration={brace,amplitude=4pt,mirror,raise=0.3ex}] (8+\xoff,0) -- ++(4-2*\xoff,0) node[midway,yshift=-1.3em]{$f_4$};
    \draw [decorate,decoration={brace,amplitude=4pt,mirror,raise=0.3ex}] (12+\xoff,0) -- ++(2-2*\xoff,0) node[midway,yshift=-1.3em]{$f_5$};
    \draw [decorate,decoration={brace,amplitude=4pt,mirror,raise=0.3ex}] (14+\xoff,0) -- ++(4-2*\xoff,0) node[midway,yshift=-1.3em]{$f_6$};
    \draw [decorate,decoration={brace,amplitude=4pt,mirror,raise=0.3ex}] (18+\xoff,0) -- ++(1-2*\xoff,0) node[midway,yshift=-1.3em]{$f_7$};

    \draw [decorate,decoration={brace,amplitude=4pt,raise=0.3ex}] (0+\xoff,1) -- ++(2-2*\xoff,0) node[midway,yshift=1.3em]{$f_1$};
    \draw [decorate,decoration={brace,amplitude=4pt,raise=0.3ex}] (2+\xoff,1) -- ++(4-2*\xoff,0) node[midway,yshift=1.3em]{$f_2$};
    \draw [decorate,decoration={brace,amplitude=4pt,raise=0.3ex}] (6+\xoff,1) -- ++(3-2*\xoff,0) node[midway,yshift=1.3em]{$f_3$};
    \draw [decorate,decoration={brace,amplitude=4pt,raise=0.3ex}] (9+\xoff,1) -- ++(4-2*\xoff,0) node[midway,yshift=1.3em]{$f_4$};
    \draw [decorate,decoration={brace,amplitude=4pt,raise=0.3ex}] (13+\xoff,1) -- ++(2-2*\xoff,0) node[midway,yshift=1.3em]{$f_5$};
    \draw [decorate,decoration={brace,amplitude=4pt,raise=0.3ex}] (15+\xoff,1) -- ++(4-2*\xoff,0) node[midway,yshift=1.3em]{$f_6$};
  \end{tikzpicture}
  
  \vspace{-1ex}
  \caption{The factorization $f_1f_2 \cdots f_{z'}$ of $S$ (with $k=3$)
    after discarding the phrases corresponding to the prefix $S_1S_2$ in
    each of the two cases discussed in \cref{sec:overview-lz}.
    The top string is $S_2$ with its last symbol $\six$ removed,
    and the bottom string is $S_3$. The phrases $f_1, \ldots, f_7$ shown below $S_3$ correspond to the case where no pattern from $\mathcal{D}$
    occurs in $\Text$.  The phrases shown above $S_3$ correspond to the case
    where some pattern $\Pat_i$ occurs in $\Text$; here, $i=2$.
    Below (resp.\ above) $S_2$, $s_j$ marks the source of the phrase $f_j$
    in the first (resp.\ second) case; in the second case, the source
    of $f_3=\Pat_2$ lies in $\Text$.}\label{fig:lz-basic}
\end{figure}

Formally, we construct the final string to be $S=S_1\cdot S_2 \cdot
S_3$, where:
\begin{itemize}
\item $S_1 := \Text \cdot {\tt 5}$,
\item $S_2 := (\bigodot_{i=1}^{k}
  \Pat_i(1 \dd m] \cdot {\tt 4} \cdot
  \Pat_i[1 \dd m) \cdot {\tt 4} \cdot
  \Pat_i[m] \cdot \Pat'_i \cdot {\tt 4} \cdot
  \Pat'_i \cdot \Pat_{i+1}[1] \cdot {\tt 4})
  \cdot {\tt 6}$,
\item $S_3 := (\bigodot_{i=1}^{k} \Pat_i \cdot \Pat'_i ) \cdot \Pat_{k+1}[1]$.
\end{itemize}
In \cref{sec:lz}, we prove that the number of LZ77 phrases in
the suffix $S_3$ of $S$, which can be retrieved as $z(S)-z(S_1S_2)$,
equals $2k$ if some pattern $\Pat_i$ occurs in $\Text$, and equals
$2k+1$ otherwise.  Note that these two cases can be distinguished
based on the parity of $z(S)$ and $z(S_1S_2)$.

The two cases are visualized in \cref{fig:lz-basic}, where $f_1 f_2
\cdots f_{z'}$ denotes the LZ77 factorization of $S$ after
discarding phrases corresponding to the prefix $S_1S_2$. Moreover, for
every $i \in [1 \dd k]$, $b_i = \Pat_i[1]$ and $e_i = \Pat_i[m]$
denote the first and last symbols of $\Pat_i$, respectively.
We also let $b_{k+1}=\zero$ denote the terminal symbol.
Let $m_i$ be such that $\Pat_i = b_i m_i e_i$. Note that $m_i \neq \emptystring$
since $m \geq 3$. Analogously, we define $b'_i$,
$m'_i$, and $e'_i$ for $\Pat'_i$.

To implement the reduction, we use the packed-string tools from
\cref{sec:tools} to construct the packed representation of $S$ in
$\bigO(\Textlen / \log \Textlen)$ time.

\paragraph{Alphabet Reduction}

The above reduction shows that we can solve \probname{Dictionary}
\probname{Matching} using \probname{LZ77 Parity}, but it increases the
alphabet size from two to seven.  To bring the alphabet for the LZ77
instance back to two, we design an efficient black-box procedure that
reduces the LZ77 factorization for a string over an integer alphabet
to the binary case. Binary inputs require no alphabet reduction.
For a text
$\Text \in \IntegerAlphabet^{\Textlen}$, where
$3\leq\AlphabetSize\leq\Textlen$, the reduction algorithm computes in
$\bigO(\Textlen / \log_{\AlphabetSize}\Textlen)$ time an integer $\delta
\geq 1$ and a binary string $S$ of length $|S| = \bigO(\Textlen \log
\AlphabetSize)$ such that $\LZSize{\Text} = \LZSize{S} - \LZSize{S[1
    \dd \delta]}$.

Arguably, it is natural to associate a binary string $C_a$ of fixed
length $m=\bigO(\log \AlphabetSize)$ with each symbol $a\in \IntegerAlphabet$ and
then set $S(\delta\dd |S|]=\bigodot_{i=1}^{\Textlen} C_{\Text[i]}$.
Ideally, we would like to ensure that, for each phrase $\Text[i\dd j]$
in the factorization of $\Text$, we have a matching phrase
$C_{\Text[i]}\cdots C_{\Text[j]}$ in the factorization of $S$.  Let us
optimistically assume that the previous phrase indeed ends immediately
before $C_{\Text[i]}$.  Suppose that $\Text[i]=a$ forms a
single-character phrase with $i<\Textlen$ and an earlier occurrence $T[i']$, and denote
$b\coloneqq T[i+1]\ne T[i'+1]\eqqcolon c$.  The next phrase in the
decomposition of $S$ consists of $C_a$ followed by (at least) the
longest common prefix of $C_b$ and $C_c$.  With $\AlphabetSize>2$, we
cannot guarantee that this common prefix is empty. We therefore need to
weaken our invariant: the endpoints of the phrase $C_{\Text[i]}\cdots
C_{\Text[j]}$ might be moved forward by up to $\ell-1$ characters,
where $\ell$ is the minimum length such that $C_x[1\dd \ell]$ uniquely
identifies $x\in\IntegerAlphabet$.  In particular, we can only assume
that the current phrase of $S$ starts somewhere within the first
$\ell$ characters of $C_a$.  Thus, to keep the invariant, we need to
guarantee that $C_a[\ell \dd m]C_b[1\dd \ell]$ has an earlier
occurrence in $S$ if and only if $ab$ has one in $T$.  For this, we
should also guarantee that $C_x[\ell\dd m]$ uniquely identifies $x\in
\IntegerAlphabet$ and that $C_x[\ell \dd m]C_y[1\dd \ell]$ does not
have any ``misaligned'' occurrence within $C_{x'}\cdot C_{y'}$ for any
$x,y,x',y'\in \IntegerAlphabet$.

With the three desiderata provided above, we can prove that
if $\Text[i\dd j]$ is a phrase in the factorization of $\Text$ and $j<\Textlen$, then a
phrase in the factorization of $S$ starting within the first $\ell$
characters of $C_{\Text[i]}$ does not reach the $\ell$-th character of
$C_{\Text[j+1]}$: since there are no misaligned occurrences, any
previous occurrence of $C_{\Text[i]}[\ell\dd m]\cdot
C_{\Text[i+1]}\cdots C_{\Text[j]}C_{\Text[j+1]}[1\dd \ell]$ has
the form $C_{\Text[i']}[\ell\dd m]\cdot C_{\Text[i'+1]}\cdots
C_{\Text[j']}C_{\Text[j'+1]}[1\dd \ell]$.  Since $C_x[\ell\dd m]$ and
$C_x[1\dd \ell]$ already distinguish $x\in \IntegerAlphabet$, we must
have $T[i'\dd j'+1]=T[i\dd j+1]$, contradicting the assumption that
$\Text[i\dd j]$ was a phrase.

It is also easy to see that if $\Text[i\dd j]$ has a previous
occurrence, then so does $C_{\Text[i]}\cdots C_{\Text[j]}$, and the
current phrase of $S$ reaches at least the end of $C_{\Text[j]}$.  A
problematic case is when $\Text[i]$ is the leftmost occurrence of a
character; in this case, we must ensure that $C_{\Text[i]}$ occurs
somewhere in $S[1\dd \delta]$, but this occurrence is not followed by
$C_x[1\dd \ell]$ for any $x\in \IntegerAlphabet$.
We meet this requirement by placing one copy of each $C_a$, followed
by ${\tt 110}$, in $S[1\dd\delta]$. In the encoding below, every
$C_x$ starts with ${\tt 111}$.

\begin{figure}[t!]
  \centering

  \begin{tikzpicture}[yscale=0.34,xscale=0.31]
    \centering
    \small
    \def\blockwidth{6.68}
    \pgfmathsetmacro{\separatorwidth}{\blockwidth/4}
    \pgfmathsetmacro{\pairwidth}{\blockwidth+\separatorwidth}

    \draw (0*\pairwidth,0) rectangle ++(\blockwidth,1);
    \draw (0*\pairwidth+\blockwidth,0) rectangle ++(\separatorwidth,1);
    \draw (1*\pairwidth,0) rectangle ++(\blockwidth,1);
    \draw (1*\pairwidth+\blockwidth,0) rectangle ++(\separatorwidth,1);
    \draw (2*\pairwidth,0) rectangle ++(\blockwidth,1);
    \draw (2*\pairwidth+\blockwidth,0) rectangle ++(\separatorwidth,1);
    \draw (3*\pairwidth,0) rectangle ++(0.8,1);

    \node at (-0.7, 0.5) {$L$:};
    \node at (0*\pairwidth+0.5*\blockwidth, 0.5) {${\tt 1110 0000 0000}$};
    \node at (0*\pairwidth+\blockwidth+0.5*\separatorwidth, 0.5) {${\tt 110}$};
    \node at (1*\pairwidth+0.5*\blockwidth, 0.5) {${\tt 1110 0010 0010}$};
    \node at (1*\pairwidth+\blockwidth+0.5*\separatorwidth, 0.5) {${\tt 110}$};
    \node at (2*\pairwidth+0.5*\blockwidth, 0.5) {${\tt 1110 1000 1000}$};
    \node at (2*\pairwidth+\blockwidth+0.5*\separatorwidth, 0.5) {${\tt 110}$};
    \node at (3*\pairwidth+0.4, 0.5) {${\tt 1}$};

    {
      \small
      \node at (0*\pairwidth+0.5*\blockwidth, 1.6) {$C_0$};
      \node at (1*\pairwidth+0.5*\blockwidth, 1.6) {$C_1$};
      \node at (2*\pairwidth+0.5*\blockwidth, 1.6) {$C_2$};
    }
  \end{tikzpicture}

  \vspace{1ex}

  \begin{tikzpicture}[yscale=0.34,xscale=0.31]
    \centering
    \small
    \def\blockwidth{6.68}

    \fill[gray!35] (0*\blockwidth,0) rectangle ++(\blockwidth*0.662,1);
    \fill[gray!35] (1*\blockwidth,0) rectangle ++(\blockwidth*0.662,1);
    \fill[gray!35] (2*\blockwidth,0) rectangle ++(\blockwidth*0.662,1);
    \fill[gray!35] (3*\blockwidth,0) rectangle ++(\blockwidth*0.662,1);
    \fill[gray!35] (4*\blockwidth,0) rectangle ++(\blockwidth*0.662,1);
    \fill[gray!35] (5*\blockwidth,0) rectangle ++(\blockwidth*0.662,1);
    \fill[gray!35] (6*\blockwidth,0) rectangle ++(\blockwidth*0.662,1);

    \draw (0*\blockwidth,0) rectangle ++(\blockwidth,1);
    \draw (1*\blockwidth,0) rectangle ++(\blockwidth,1);
    \draw (2*\blockwidth,0) rectangle ++(\blockwidth,1);
    \draw (3*\blockwidth,0) rectangle ++(\blockwidth,1);
    \draw (4*\blockwidth,0) rectangle ++(\blockwidth,1);
    \draw (5*\blockwidth,0) rectangle ++(\blockwidth,1);
    \draw (6*\blockwidth,0) rectangle ++(\blockwidth,1);
 
    \node at (-0.7, 0.5) {$R$:};
    \node at (0*\blockwidth+0.5*\blockwidth, 0.5) {${\tt 1110 0000 0000}$};
    \node at (1*\blockwidth+0.5*\blockwidth, 0.5) {${\tt 1110 0010 0010}$};
    \node at (2*\blockwidth+0.5*\blockwidth, 0.5) {${\tt 1110 1000 1000}$};
    \node at (3*\blockwidth+0.5*\blockwidth, 0.5) {${\tt 1110 0000 0000}$};
    \node at (4*\blockwidth+0.5*\blockwidth, 0.5) {${\tt 1110 0010 0010}$};
    \node at (5*\blockwidth+0.5*\blockwidth, 0.5) {${\tt 1110 1000 1000}$};
    \node at (6*\blockwidth+0.5*\blockwidth, 0.5) {${\tt 1110 1000 1000}$};

    {
      \small
      \node at (0*\blockwidth+0.5*\blockwidth, 1.6) {$C_0$};
      \node at (1*\blockwidth+0.5*\blockwidth, 1.6) {$C_1$};
      \node at (2*\blockwidth+0.5*\blockwidth, 1.6) {$C_2$};
      \node at (3*\blockwidth+0.5*\blockwidth, 1.6) {$C_0$};
      \node at (4*\blockwidth+0.5*\blockwidth, 1.6) {$C_1$};
      \node at (5*\blockwidth+0.5*\blockwidth, 1.6) {$C_2$};
      \node at (6*\blockwidth+0.5*\blockwidth, 1.6) {$C_2$};
    }

    \def\xoff{0.06}
    \pgfmathsetmacro{\bitwidth}{\blockwidth/12}
    \draw [decorate,decoration={brace,amplitude=7pt,mirror,raise=0.1ex}] (2*\bitwidth+\xoff,0) -- ++(12*\bitwidth-2*\xoff,0) node[midway,yshift=-1.5em]{$f_1$};
    \draw [decorate,decoration={brace,amplitude=7pt,mirror,raise=0.1ex}] (14*\bitwidth+\xoff,0) -- ++(12*\bitwidth-2*\xoff,0) node[midway,yshift=-1.5em]{$f_2$};
    \draw [decorate,decoration={brace,amplitude=7pt,mirror,raise=0.1ex}] (26*\bitwidth+\xoff,0) -- ++(12*\bitwidth-2*\xoff,0) node[midway,yshift=-1.5em]{$f_3$};
    \draw [decorate,decoration={brace,amplitude=7pt,mirror,raise=0.1ex}] (38*\bitwidth+\xoff,0) -- ++(38*\bitwidth-2*\xoff,0) node[midway,yshift=-1.5em]{$f_4$};
    \draw [decorate,decoration={brace,amplitude=7pt,mirror,raise=0.1ex}] (76*\bitwidth+\xoff,0) -- ++(8*\bitwidth-2*\xoff,0) node[midway,yshift=-1.5em]{$f_5$};
  \end{tikzpicture}

  \vspace{-0.5ex}
  \caption{The strings $L$ and $R$ from the LZ77 alphabet reduction in
    \cref{sec:overview-lz} for the text $\Text = {\tt 0120122}$ with
    parameters $\AlphabetSize = 3$ and $k = 2$. The five phrases $f_1,
    \ldots, f_5$ from the LZ77 factorization of the string $S = LR$,
    after discarding the phrases that start in the prefix $L$, are
    underlined.  Note that the LZ77 factorization of $\Text$ is $\Text
    = {\tt 0} \cdot {\tt 1} \cdot {\tt 2} \cdot {\tt 012} \cdot {\tt
    2}$, and the phrases $f_1, \ldots, f_5$ in the figure above
    correspond to this factorization. In particular, $\LZSize{\Text} =
    5 = \LZSize{LR} - \LZSize{L}$.  The key property used to prove
    this is that every phrase in the LZ77 parsing of $S$ that is
    entirely contained in $R$ starts with one of the shaded
    symbols.}\label{fig:lz-alphabet-reduction}
\end{figure}

Based on the above requirements, we construct the string $S$ as
follows. Let $k = \lceil \log \AlphabetSize \rceil$.
We use the binary representation $\bin{k}{\cdot}$ and padding
operation $\pad{\cdot}$ from \cref{def:bin,def:pad}.
The padding operation inserts ${\tt 0}$ after each bit;
e.g., $\pad{\bin{2}{3}} = {\tt 1010}$. For every $a \in [0
\dd \AlphabetSize)$, let $C_a := {\tt 1}^{2k-1}\cdot {\tt 0} \cdot
\pad{\bin{k}{a}} \cdot \pad{\bin{k}{a}}$. Then, let $S[1\dd
\delta] =L$ and $S(\delta\dd |S|]=R$, where:
\[
  L := \Big(\bigodot_{a=0}^{\AlphabetSize-1} C_{a} \cdot
    {\tt 110}\Big) \cdot {\tt 1} \quad\text{ and }\quad R :=
    \bigodot_{i=1}^{\Textlen} C_{\Text[i]}.
\]
An example is shown in \cref{fig:lz-alphabet-reduction}.
We note that the extra $\one$ at the end of $L$ ensures that the last
phrase starting within $L$ does not extend beyond the first $2k-1$
positions of $R$ (since $\one^{2k}$ has a unique occurrence in $S$).
With this as the base case and the argument above for the inductive
step, we prove in \cref{sec:lz-alphabet-reduction} that $\LZSize{S} -
\LZSize{L} = \LZSize{\Text}$.  The packed-string tools from \cref{sec:tools}
let us construct the packed representations of $L$ and $R$ in
$\bigO(\Textlen / \log_{\AlphabetSize} \Textlen)$ time.
By putting everything together, we obtain
our final reduction from \probname{Dictionary Matching} to
\probname{LZ77 Parity} for binary strings (\cref{th:lz}).

\subsection{Reducing Dictionary Matching to Burrows--Wheeler Transform}\label{sec:overview-bwt}

We now present an overview of the reduction from \probname{Dictionary} \probname{Matching} to
\probname{BWT} \probname{Construction} (or, more precisely, to \probname{RLBWT}
\probname{Size}).

\paragraph{Preliminary Reduction}

\begin{wrapfigure}{R}{0.265\textwidth}
  \centering
  \vspace{-0.5cm}
  \begin{tikzpicture}[scale=0.33]
    \centering
    \scriptsize
    \def\symbolwidth{1}
    \def\blockwidth{3}
    \def\height{1.2}
    \def\col{gray!35}

    \draw (0,1*\height) rectangle ++(\blockwidth,\height);
    \draw (0,2*\height) rectangle ++(\blockwidth,\height);
    \draw (0,3*\height) rectangle ++(\blockwidth,\height);
    \draw[fill=\col] (\blockwidth,1*\height) rectangle ++(\symbolwidth,\height);
    \draw[fill=\col] (\blockwidth,2*\height) rectangle ++(\symbolwidth,\height);
    \draw[fill=\col] (\blockwidth,3*\height) rectangle ++(\symbolwidth,\height);
    \draw[fill=\col] (-2,1*\height) rectangle ++(\symbolwidth,\height);
    \draw[fill=\col] (-2,2*\height) rectangle ++(\symbolwidth,\height);
    \draw[fill=\col] (-2,3*\height) rectangle ++(\symbolwidth,\height);
    \node at (0.5*\blockwidth,0.5*\height+1*\height) {$\widetilde{P}_i$};
    \node at (0.5*\blockwidth,0.5*\height+2*\height) {$\widetilde{P}_i$};
    \node at (0.5*\blockwidth,0.5*\height+3*\height) {$\widetilde{P}_i$};
    \node at (\blockwidth+0.5*\symbolwidth,0.5*\height+1*\height) {${\tt 6}$};
    \node at (\blockwidth+0.5*\symbolwidth,0.5*\height+2*\height) {${\tt 5}$};
    \node at (\blockwidth+0.5*\symbolwidth,0.5*\height+3*\height) {${\tt 4}$};
    \node at (-2+0.5*\symbolwidth,0.5*\height+1*\height) {${\tt 4}$};
    \node at (-2+0.5*\symbolwidth,0.5*\height+2*\height) {${\tt 8}$};
    \node at (-2+0.5*\symbolwidth,0.5*\height+3*\height) {${\tt 4}$};
    \node at (\blockwidth+\symbolwidth+1,0.5*\height+1*\height) {$\cdots$};
    \node at (\blockwidth+\symbolwidth+1,0.5*\height+2*\height) {$\cdots$};
    \node at (\blockwidth+\symbolwidth+1,0.5*\height+3*\height) {$\cdots$};
    \node at (\blockwidth+\symbolwidth+1,0.5*\height+5*\height) {$\cdots$};
    \node at (\blockwidth+\symbolwidth+1,0.5*\height+6*\height) {$\cdots$};
    \node at (\blockwidth+\symbolwidth+1,0.5*\height+7*\height) {$\cdots$};
    \foreach \i in {1,2,3,4,5,6,7,8} {
      \draw (\blockwidth+\symbolwidth,\i*\height) -- (\blockwidth+\symbolwidth+0.5,\i*\height);
    }

    \def\yoff{5*\height}
    \draw (0,\yoff+0*\height) rectangle ++(\blockwidth,\height);
    \draw (0,\yoff+1*\height) rectangle ++(\blockwidth,\height);
    \draw (0,\yoff+2*\height) rectangle ++(\blockwidth,\height);
    \draw[fill=\col] (\blockwidth,\yoff+0*\height) rectangle ++(\symbolwidth,\height);
    \draw[fill=\col] (\blockwidth,\yoff+1*\height) rectangle ++(\symbolwidth,\height);
    \draw[fill=\col] (\blockwidth,\yoff+2*\height) rectangle ++(\symbolwidth,\height);
    \draw[fill=\col] (-2,\yoff+0*\height) rectangle ++(\symbolwidth,\height);
    \draw[fill=\col] (-2,\yoff+1*\height) rectangle ++(\symbolwidth,\height);
    \draw[fill=\col] (-2,\yoff+2*\height) rectangle ++(\symbolwidth,\height);
    \node at (0.5*\blockwidth,\yoff+0.5*\height) {$\widetilde{P}_i$};
    \node at (0.5*\blockwidth,\yoff+0.5*\height+1*\height) {$\widetilde{P}_i$};
    \node at (0.5*\blockwidth,\yoff+0.5*\height+2*\height) {$\widetilde{P}_i$};
    \node at (\blockwidth+0.5*\symbolwidth,\yoff+0.5*\height) {${\tt 4}$};
    \node at (\blockwidth+0.5*\symbolwidth,\yoff+0.5*\height+1*\height) {${\tt 3}$};
    \node at (\blockwidth+0.5*\symbolwidth,\yoff+0.5*\height+2*\height) {${\tt 2}$};
    \node at (-2+0.5*\symbolwidth,\yoff+0.5*\height) {${\tt 4}$};
    \node at (-2+0.5*\symbolwidth,\yoff+0.5*\height+1*\height) {${\tt 7}$};
    \node at (-2+0.5*\symbolwidth,\yoff+0.5*\height+2*\height) {${\tt 4}$};

    \node at (0.1 + 0.5*\symbolwidth,0.5*\height) {$\cdots$};
    \node at (-1.9+0.5*\symbolwidth,0.5*\height) {$\cdots$};
    \node at (0.1 +0.5*\symbolwidth, 4.5*\height) {$\cdots$};
    \node at (-1.9+0.5*\symbolwidth, 4.5*\height) {$\cdots$};
    \node at (0.1 +0.5*\symbolwidth, 8.5*\height) {$\cdots$};
    \node at (-1.9+0.5*\symbolwidth, 8.5*\height) {$\cdots$};

    \node at (-2+0.5*\symbolwidth, 9.5*\height) {$B$};
    \node[right] at (-0.3, 9.5*\height) {$S[\SA{S}[j] \dd |S|]$};

    \draw [thin, <->] (-3,3*\height) -- (-3,6*\height) node[midway,left] {$n_i$};
    \draw[dashed] (-3,3*\height) -- (-2,3*\height);
    \draw[dashed] (-3,6*\height) -- (-2,6*\height);
  \end{tikzpicture}
  \caption{The range in $\SA{S}$ corresponding to suffixes starting with $\widetilde{\Pat_i}$. We also
    show $B = \BWT{S}$.}\label{fig:bwt-simple-range}
  \vspace{0.0cm}
\end{wrapfigure}

Assume that we are given an instance of the \probname{Dictionary} \probname{Matching}
problem, i.e., a text $\Text \in \BinaryAlphabet^{\Textlen}$ and a set
$\mathcal{D} = \{\Pat_1, \ldots, \Pat_k\}$ of
$k=\lfloor\Textlen/\log\Textlen\rfloor$ distinct patterns of length
$m=2\lfloor\log\Textlen\rfloor$.

The main idea in our reduction is to construct a text $S$ containing
both $\Text$ and the patterns $\Pat_1, \ldots, \Pat_k$ embedded in
a way that: (a) brings both the explicit occurrence of $\Pat_i$ in $S$
and a potential occurrence of $\Pat_i$ in $\Text$ close in the suffix
array, and (b) allows us to detect such an occurrence by inspecting the
BWT alone.  Simply defining $S = \Text \cdot \bigodot_{i=1}^{k}
\Pat_i$ will not work because it does not yield a discernible way to
detect occurrences of $\Pat_i$ in $\Text$.

The first idea is to spread symbols of the original text to allow for
creating clusters of positions in the suffix array.  For any string $A
= a_1a_2 \dots a_q\in \BinaryAlphabet^q$, we denote $\widetilde{A} :=
a_1{\tt 4}a_2{\tt 4} \cdots {\tt 4}a_q$.  Let us then consider the
following text:
\[
  S = \four \widetilde{\Text} \four  \cdot \Big(\bigodot_{i=1}^{k}
    \four  \widetilde{\Pat_i} \two   \cdot
    \seven \widetilde{\Pat_i} \three \cdot
    \eight \widetilde{\Pat_i} \five  \cdot
    \four \widetilde{\Pat_i}  \six\Big).
\]

Consider now the range in the suffix array of $S$ corresponding to
suffixes starting with $\widetilde{\Pat_i}$. The BWT in this range is
always of the form $\four\seven\four^{n_i}\eight\four$, where $n_i$ is
the number of occurrences of $\widetilde{\Pat_i}$ in
$\widetilde{\Text}$ (which is the same as the number of occurrences of
$\Pat_i$ in $\Text$); see \cref{fig:bwt-simple-range}. In particular,
the BWT of $S$ contains a substring of the form
$\seven\four^{t}\eight$ with $t > 0$ if and only if, for some $i \in [1
  \dd k]$, the pattern $\Pat_i$ occurs in $\Text$. This check can be
implemented using lookup tables in $\bigO(\Textlen / \log \Textlen)$
time.

This construction can be used to prove that \probname{BWT
  Construction} is at least as hard as \probname{Dictionary}
\probname{Matching}.  However, recall that our goal is to solve
dictionary matching using only $\RLBWTSize{S}$.  The runs in $\BWT{S}$
do not have any predictable structure as they largely depend on the
structure of $\BWT{\Text}$.

\paragraph{The General Reduction}

To address the above challenge, instead of constructing a single text
$S$, we construct two strings $S_1$ and $S_2$, both of length
$\Theta(\Textlen)$ and over $\{\zero,\one,\ldots,{\tt 9}\}$, in which the
gadgets responsible for creating clusters in the suffix array are
organized slightly differently. More precisely:
\begin{itemize}
\item $S_1 := {\tt 4}\widetilde{T}{\tt 4} \cdot
  \bigodot_{i=1}^{k} ({\tt 4}\widetilde{\Pat_i}{\tt 2}\cdot
                      {\tt 8}\widetilde{\Pat_i}{\tt 5}\cdot
                      {\tt 4}\widetilde{\Pat_i}{\tt 6}\cdot
                      {\tt 4}\widetilde{\Pat_i}{\tt 7}\cdot
                      {\tt 9}\widetilde{\Pat_i}{\tt 3})$,
\item $S_2 := {\tt 4}\widetilde{T}{\tt 4} \cdot
  \bigodot_{i=1}^{k} ({\tt 4}\widetilde{\Pat_i}{\tt 2}\cdot
                      {\tt 8}\widetilde{\Pat_i}{\tt 6}\cdot
                      {\tt 4}\widetilde{\Pat_i}{\tt 5}\cdot
                      {\tt 4}\widetilde{\Pat_i}{\tt 7}\cdot
                      {\tt 9}\widetilde{\Pat_i}{\tt 3})$.
\end{itemize}
We argue in \cref{lm:dm-to-bwt} that $\RLBWTSize{S_2} -
\RLBWTSize{S_1}$ is the number of patterns $\Pat_i$ that \emph{do not}
occur in $\Text$.  Thus, we can solve the input
\probname{Dictionary Matching} instance by testing whether
$\RLBWTSize{S_2}-\RLBWTSize{S_1} = |\mathcal{D}|$.

The challenge in proving the above formula for $\RLBWTSize{S_2} -
\RLBWTSize{S_1}$ is that suffix arrays (and hence BWTs) are very
sensitive to string edits: A single change of a symbol can permute the
entire suffix array in an unpredictable and complex
manner~\cite{dynsa}, and in our reduction, $S_1$ and $S_2$ differ on
$\Theta(k)$ positions. Thus, to understand the value $\RLBWTSize{S_2}
- \RLBWTSize{S_1}$, we need to prove that certain complex
fragments of the suffix array are preserved in both strings, while some
others change in a known and predictable way.  The proof consists
of six steps, each of which we outline below.

\begin{figure}
  \centering
  \scalebox{1.15}{%
  \begin{tikzpicture}[
      scale=0.27,
      myarrow/.style={->, thin, -{Stealth[scale=1.1]}}
    ]

    \centering
    \scriptsize
    \def\symbolwidth{1}
    \def\blockwidth{2}
    \def\textwidth{4}
    \def\height{1.2}
    \def\col{gray!35}
    \def\yoff{-4}
    \pgfmathtruncatemacro{\groupwidth}{2*\symbolwidth+\blockwidth}

    \fill[gray!20] (0,\yoff+\height) rectangle (11,0);
    \fill[gray!20] (14,\yoff+\height) rectangle (15,0);
    \fill[gray!20] (18,\yoff+\height) rectangle (31,0);
    \fill[gray!20] (34,\yoff+\height) rectangle (35,0);
    \fill[gray!20] (38,\yoff+\height) rectangle (46,0);
    \fill[gray!20] (11,0) -- (14,0) -- (18,\yoff+\height) -- (15,\yoff+\height) -- cycle;
    \fill[gray!20] (15,0) -- (18,0) -- (14,\yoff+\height) -- (11,\yoff+\height) -- cycle;
    \fill[gray!20] (31,0) -- (34,0) -- (38,\yoff+\height) -- (35,\yoff+\height) -- cycle;
    \fill[gray!20] (35,0) -- (38,0) -- (34,\yoff+\height) -- (31,\yoff+\height) -- cycle;

    \draw[dashed] (0,\yoff+\height) -- (0,0);
    \draw[dashed] (11,\yoff+\height) -- (11,0);

    \draw[dashed] (14,\yoff+\height) -- (14,0);
    \draw[dashed] (15,\yoff+\height) -- (15,0);

    \draw[dashed] (18,\yoff+\height) -- (18,0);
    \draw[dashed] (31,\yoff+\height) -- (31,0);

    \draw[dashed] (34,\yoff+\height) -- (34,0);
    \draw[dashed] (35,\yoff+\height) -- (35,0);

    \draw[dashed] (38,\yoff+\height) -- (38,0);
    \draw[dashed] (46,\yoff+\height) -- (46,0);

    \draw[dashed] (15,\yoff+\height) -- (11,0);
    \draw[dashed] (18,\yoff+\height) -- (14,0);
    \draw[dashed] (11,\yoff+\height) -- (15,0);
    \draw[dashed] (14,\yoff+\height) -- (18,0);

    \draw[dashed] (35,\yoff+\height) -- (31,0);
    \draw[dashed] (38,\yoff+\height) -- (34,0);
    \draw[dashed] (31,\yoff+\height) -- (35,0);
    \draw[dashed] (34,\yoff+\height) -- (38,0);

    {
      \node at (-0.8, 0.5*\height) {\small $S_1$:};
      \draw[fill=\col] (0*\symbolwidth+0*\textwidth+0* \blockwidth,0) rectangle ++(\symbolwidth,\height);
      \draw (1*\symbolwidth+0*\textwidth+0* \blockwidth,0) rectangle ++(\textwidth,\height);
      \draw[fill=\col] (1*\symbolwidth+1*\textwidth+0* \blockwidth,0) rectangle ++(\symbolwidth,\height);
      \node at (\symbolwidth+0.5*\textwidth,0.5*\height) {$\widetilde{\Text}$};
      \node at (0.5*\symbolwidth,0.5*\height) {${\tt 4}$};
      \node at (\symbolwidth+\textwidth+0.5*\symbolwidth,0.5*\height) {${\tt 4}$};
      \foreach \i/\j/\syma/\symb in {0/1/4/2,
                                     1/1/8/5,
                                     2/1/4/6,
                                     3/1/4/7,
                                     4/1/9/3,
                                     5/2/4/2,
                                     6/2/8/5,
                                     7/2/4/6,
                                     8/2/4/7,
                                     9/2/9/3} {
        \pgfmathtruncatemacro{\offset}{2 * \symbolwidth + \textwidth + \i * \groupwidth}
        \draw[fill=\col] (\offset,0) rectangle ++(\symbolwidth,\height);
        \draw (\offset+\symbolwidth,0) rectangle ++(\blockwidth,\height);
        \draw[fill=\col] (\offset+\symbolwidth+\blockwidth,0) rectangle ++(\symbolwidth,\height);
        \node at (\offset+\symbolwidth+0.5*\blockwidth,0.5*\height) {$\widetilde{P}_{\j}$};
        \node at (\offset+0.5*\symbolwidth,0.5*\height) {${\tt \syma}$};
        \node at (\offset+\symbolwidth+\blockwidth+0.5*\symbolwidth,0.5*\height) {${\tt \symb}$};
      }
    }

    {
      \node at (-0.8, 0.5*\height+\yoff) {\small $S_2$:};
      \draw[fill=\col] (0*\symbolwidth+0*\textwidth+0* \blockwidth,0+\yoff) rectangle ++(\symbolwidth,\height);
      \draw (1*\symbolwidth+0*\textwidth+0* \blockwidth,0+\yoff) rectangle ++(\textwidth,\height);
      \draw[fill=\col] (1*\symbolwidth+1*\textwidth+0* \blockwidth,0+\yoff) rectangle ++(\symbolwidth,\height);
      \node at (\symbolwidth+0.5*\textwidth,0.5*\height+\yoff) {$\widetilde{\Text}$};
      \node at (0.5*\symbolwidth,0.5*\height+\yoff) {${\tt 4}$};
      \node at (\symbolwidth+\textwidth+0.5*\symbolwidth,0.5*\height+\yoff) {${\tt 4}$};
      \foreach \i/\j/\syma/\symb in {0/1/4/2,
                                     1/1/8/6,
                                     2/1/4/5,
                                     3/1/4/7,
                                     4/1/9/3,
                                     5/2/4/2,
                                     6/2/8/6,
                                     7/2/4/5,
                                     8/2/4/7,
                                     9/2/9/3} {
        \pgfmathtruncatemacro{\offset}{2 * \symbolwidth + \textwidth + \i * \groupwidth}
        \draw[fill=\col] (\offset,0+\yoff) rectangle ++(\symbolwidth,\height);
        \draw (\offset+\symbolwidth,0+\yoff) rectangle ++(\blockwidth,\height);
        \draw[fill=\col] (\offset+\symbolwidth+\blockwidth,0+\yoff) rectangle ++(\symbolwidth,\height);
        \node at (\offset+\symbolwidth+0.5*\blockwidth,0.5*\height+\yoff) {$\widetilde{P}_{\j}$};
        \node at (\offset+0.5*\symbolwidth,0.5*\height+\yoff) {${\tt \syma}$};
        \node at (\offset+\symbolwidth+\blockwidth+0.5*\symbolwidth,0.5*\height+\yoff) {${\tt \symb}$};
      }
    }

  \end{tikzpicture}%
  }
  \caption{The strings $S_1$ and $S_2$ from the BWT reduction in
    \cref{sec:overview-bwt} for $k = 2$. The function $f$, used in the
    analysis, is shown between the strings. It maps positions in $S_1$
    to positions in $S_2$ such that, for every $i, j \in [1 \dd
    \Textlen']$ (where $\Textlen' = |S_1| = |S_2|$), the relation
    $S_1[i \dd \Textlen'] \prec S_1[j \dd \Textlen']$ holds if and
    only if $S_2[f(i) \dd \Textlen'] \prec S_2[f(j) \dd
    \Textlen']$.}\label{fig:bwt-mapping}
\end{figure}

\textbf{1.} Denote $\Textlen' = |S_1| = |S_2|$ and $\delta = 2m + 1$.
Let $\mathcal{L}$ (resp.\ $\mathcal{R}$) be the set of all positions $i
\in [1 \dd \Textlen' - 2m + 1]$ such that $S_1[i \dd i + 2m)$ contains the
symbol ${\tt 5}$ (resp.\ ${\tt 6}$).  We define the mapping $f : [1
\dd \Textlen'] \rightarrow [1 \dd \Textlen']$ such that
$f(x)=x+\delta$ for $x\in \mathcal{L}$, $f(x)=x-\delta$ for $x\in
\mathcal{R}$, and $f(x)=x$ for $x\in [1\dd \Textlen']\setminus
(\mathcal{L} \cup \mathcal{R})$; see \cref{fig:bwt-mapping} for an
illustration.  The first step of the proof is to observe that, for
every $i, j \in [1 \dd \Textlen']$, $S_1[i \dd \Textlen'] \prec S_1[j
\dd \Textlen']$ holds if and only if $S_2[f(i) \dd \Textlen'] \prec
S_2[f(j) \dd \Textlen']$ does. For $i\neq j$, first suppose that one
suffix is a proper prefix of the other. Since the final block
${\tt 9}\widetilde{\Pat}_k{\tt 3}$ occurs only once, the shorter suffix
lies within this block. Its matching prefix lies within another block
of the form ${\tt 9}\widetilde{\Pat}_t{\tt 3}$. These blocks are unchanged
between $S_1$ and $S_2$, and $f$ fixes both starting positions. These properties
preserve the proper-prefix relation. Otherwise, the proof distinguishes
two cases:
\begin{itemize}
\item If $i, j \in \mathcal{L} \cup \mathcal{R}$, then the compared
  suffixes of $S_1$ and $S_2$ must have a very short common prefix,
  resulting in consistent lexicographical comparisons.  This is
  because, for every $t \in [1 \dd k]$, the strings ${\tt 5}{\tt
  4}\widetilde{P}_{t}$ and ${\tt 6}{\tt 4}\widetilde{P}_{t}$ each occur
  exactly once in $S_1$ and exactly once in $S_2$; see
  Case~\ref{lm:dm-to-bwt-case-1b} in the proof of
  \cref{lm:dm-to-bwt} for details.
\item If at least one of the positions $i,j$ is not in $\mathcal{L}
  \cup \mathcal{R}$, then, due to the location of all occurrences of
  ${\tt 5}$ and ${\tt 6}$ in $S_1$ and $S_2$, it follows that the
  longest common prefix of $S_1[i \dd \Textlen']$ and $S_1[j \dd
    \Textlen']$ contains neither ${\tt 5}$ nor ${\tt 6}$.  Moreover,
  ${\tt 5}$ or ${\tt 6}$ can occur as the mismatching character for at
  most one of the positions. This proves that the same will hold
  after mapping the positions $i$ and $j$ to $S_2$, except the
  mismatching ${\tt 5}$ (resp.\ ${\tt 6}$) may become a ${\tt 6}$
  (resp.\ ${\tt 5}$); see Case~\ref{lm:dm-to-bwt-case-1c} in the proof of
  \cref{lm:dm-to-bwt} for details.
\end{itemize}

\textbf{2.} The suffix-order equivalence from Step 1 and the bijectivity
of $f$ imply that $\ISA{S_1}[i] = \ISA{S_2}[f(i)]$ holds for every $i
\in [1 \dd \Textlen']$. In other words, the number of suffixes of
$S_1$ that are smaller than $S_1[i \dd \Textlen']$ is the same as the
number of suffixes of $S_2$ that are smaller than $S_2[f(i) \dd
\Textlen']$.

\begin{figure}[t]
  \centering

  \begin{minipage}[b]{0.35\textwidth}
  \centering

  \begin{tikzpicture}[scale=0.33]
    \centering
    \scriptsize
    \def\symbolwidth{1}
    \def\blockwidth{3}
    \def\height{1.2}
    \def\col{gray!35}

    \draw (0,0*\height) rectangle ++(\blockwidth,\height);
    \draw (0,1*\height) rectangle ++(\blockwidth,\height);
    \draw (0,2*\height) rectangle ++(\blockwidth,\height);
    \draw (0,3*\height) rectangle ++(\blockwidth,\height);
    \draw[fill=\col] (\blockwidth,0*\height) rectangle ++(\symbolwidth,\height);
    \draw[fill=\col] (\blockwidth,1*\height) rectangle ++(\symbolwidth,\height);
    \draw[fill=\col] (\blockwidth,2*\height) rectangle ++(\symbolwidth,\height);
    \draw[fill=\col] (\blockwidth,3*\height) rectangle ++(\symbolwidth,\height);
    \draw[fill=\col] (-2,0*\height) rectangle ++(\symbolwidth,\height);
    \draw[fill=\col] (-2,1*\height) rectangle ++(\symbolwidth,\height);
    \draw[fill=\col] (-2,2*\height) rectangle ++(\symbolwidth,\height);
    \draw[fill=\col] (-2,3*\height) rectangle ++(\symbolwidth,\height);
    \node at (0.5*\blockwidth,0.5*\height) {$\widetilde{P}_i$};
    \node at (0.5*\blockwidth,0.5*\height+1*\height) {$\widetilde{P}_i$};
    \node at (0.5*\blockwidth,0.5*\height+2*\height) {$\widetilde{P}_i$};
    \node at (0.5*\blockwidth,0.5*\height+3*\height) {$\widetilde{P}_i$};
    \node at (\blockwidth+0.5*\symbolwidth,0.5*\height) {${\tt 7}$};
    \node at (\blockwidth+0.5*\symbolwidth,0.5*\height+1*\height) {${\tt 6}$};
    \node at (\blockwidth+0.5*\symbolwidth,0.5*\height+2*\height) {${\tt 5}$};
    \node at (\blockwidth+0.5*\symbolwidth,0.5*\height+3*\height) {${\tt 4}$};
    \node at (-2+0.5*\symbolwidth,0.5*\height) {${\tt 4}$};
    \node at (-2+0.5*\symbolwidth,0.5*\height+1*\height) {${\tt 4}$};
    \node at (-2+0.5*\symbolwidth,0.5*\height+2*\height) {${\tt 8}$};
    \node at (-2+0.5*\symbolwidth,0.5*\height+3*\height) {${\tt 4}$};
    \node at (\blockwidth+\symbolwidth+1,0.5*\height+0*\height) {$\cdots$};
    \node at (\blockwidth+\symbolwidth+1,0.5*\height+1*\height) {$\cdots$};
    \node at (\blockwidth+\symbolwidth+1,0.5*\height+2*\height) {$\cdots$};
    \node at (\blockwidth+\symbolwidth+1,0.5*\height+3*\height) {$\cdots$};
    \node at (\blockwidth+\symbolwidth+1,0.5*\height+5*\height) {$\cdots$};
    \node at (\blockwidth+\symbolwidth+1,0.5*\height+6*\height) {$\cdots$};
    \node at (\blockwidth+\symbolwidth+1,0.5*\height+7*\height) {$\cdots$};
    \foreach \i in {0,1,2,3,4,5,6,7,8} {
      \draw (\blockwidth+\symbolwidth,\i*\height) -- (\blockwidth+\symbolwidth+0.5,\i*\height);
    }

    \def\yoff{5*\height}
    \draw (0,\yoff+0*\height) rectangle ++(\blockwidth,\height);
    \draw (0,\yoff+1*\height) rectangle ++(\blockwidth,\height);
    \draw (0,\yoff+2*\height) rectangle ++(\blockwidth,\height);
    \draw[fill=\col] (\blockwidth,\yoff+0*\height) rectangle ++(\symbolwidth,\height);
    \draw[fill=\col] (\blockwidth,\yoff+1*\height) rectangle ++(\symbolwidth,\height);
    \draw[fill=\col] (\blockwidth,\yoff+2*\height) rectangle ++(\symbolwidth,\height);
    \draw[fill=\col] (-2,\yoff+0*\height) rectangle ++(\symbolwidth,\height);
    \draw[fill=\col] (-2,\yoff+1*\height) rectangle ++(\symbolwidth,\height);
    \draw[fill=\col] (-2,\yoff+2*\height) rectangle ++(\symbolwidth,\height);
    \node at (0.5*\blockwidth,\yoff+0.5*\height) {$\widetilde{P}_i$};
    \node at (0.5*\blockwidth,\yoff+0.5*\height+1*\height) {$\widetilde{P}_i$};
    \node at (0.5*\blockwidth,\yoff+0.5*\height+2*\height) {$\widetilde{P}_i$};
    \node at (\blockwidth+0.5*\symbolwidth,\yoff+0.5*\height) {${\tt 4}$};
    \node at (\blockwidth+0.5*\symbolwidth,\yoff+0.5*\height+1*\height) {${\tt 3}$};
    \node at (\blockwidth+0.5*\symbolwidth,\yoff+0.5*\height+2*\height) {${\tt 2}$};
    \node at (-2+0.5*\symbolwidth,\yoff+0.5*\height) {${\tt 4}$};
    \node at (-2+0.5*\symbolwidth,\yoff+0.5*\height+1*\height) {${\tt 9}$};
    \node at (-2+0.5*\symbolwidth,\yoff+0.5*\height+2*\height) {${\tt 4}$};

    \node at (0.1 + 0.5*\symbolwidth,-0.5*\height) {$\cdots$};
    \node at (-1.9+0.5*\symbolwidth,-0.5*\height) {$\cdots$};
    \node at (0.1 +0.5*\symbolwidth, 4.5*\height) {$\cdots$};
    \node at (-1.9+0.5*\symbolwidth, 4.5*\height) {$\cdots$};
    \node at (0.1 +0.5*\symbolwidth, 8.5*\height) {$\cdots$};
    \node at (-1.9+0.5*\symbolwidth, 8.5*\height) {$\cdots$};

    \node at (-2+0.5*\symbolwidth, 9.5*\height) {$B_1$};
    \node[right] at (-0.3, 9.5*\height) {$S_1[\SA{S_1}[j] \dd \Textlen']$};

    \draw [thin, <->] (-3,3*\height) -- (-3,6*\height) node[midway,left] {$n_i$};
    \draw[dashed] (-3,3*\height) -- (-2,3*\height);
    \draw[dashed] (-3,6*\height) -- (-2,6*\height);
  \end{tikzpicture}

  \end{minipage}\hspace{0.04\textwidth}%
  \begin{minipage}[b]{0.35\textwidth}
  \centering

  \begin{tikzpicture}[scale=0.33]
    \centering
    \scriptsize
    \def\symbolwidth{1}
    \def\blockwidth{3}
    \def\height{1.2}
    \def\col{gray!35}

    \draw (0,0*\height) rectangle ++(\blockwidth,\height);
    \draw (0,1*\height) rectangle ++(\blockwidth,\height);
    \draw (0,2*\height) rectangle ++(\blockwidth,\height);
    \draw (0,3*\height) rectangle ++(\blockwidth,\height);
    \draw[fill=\col] (\blockwidth,0*\height) rectangle ++(\symbolwidth,\height);
    \draw[fill=\col] (\blockwidth,1*\height) rectangle ++(\symbolwidth,\height);
    \draw[fill=\col] (\blockwidth,2*\height) rectangle ++(\symbolwidth,\height);
    \draw[fill=\col] (\blockwidth,3*\height) rectangle ++(\symbolwidth,\height);
    \draw[fill=\col] (-2,0*\height) rectangle ++(\symbolwidth,\height);
    \draw[fill=\col] (-2,1*\height) rectangle ++(\symbolwidth,\height);
    \draw[fill=\col] (-2,2*\height) rectangle ++(\symbolwidth,\height);
    \draw[fill=\col] (-2,3*\height) rectangle ++(\symbolwidth,\height);
    \node at (0.5*\blockwidth,0.5*\height) {$\widetilde{P}_i$};
    \node at (0.5*\blockwidth,0.5*\height+1*\height) {$\widetilde{P}_i$};
    \node at (0.5*\blockwidth,0.5*\height+2*\height) {$\widetilde{P}_i$};
    \node at (0.5*\blockwidth,0.5*\height+3*\height) {$\widetilde{P}_i$};
    \node at (\blockwidth+0.5*\symbolwidth,0.5*\height) {${\tt 7}$};
    \node at (\blockwidth+0.5*\symbolwidth,0.5*\height+1*\height) {${\tt 6}$};
    \node at (\blockwidth+0.5*\symbolwidth,0.5*\height+2*\height) {${\tt 5}$};
    \node at (\blockwidth+0.5*\symbolwidth,0.5*\height+3*\height) {${\tt 4}$};
    \node at (-2+0.5*\symbolwidth,0.5*\height) {${\tt 4}$};
    \node at (-2+0.5*\symbolwidth,0.5*\height+1*\height) {${\tt 8}$};
    \node at (-2+0.5*\symbolwidth,0.5*\height+2*\height) {${\tt 4}$};
    \node at (-2+0.5*\symbolwidth,0.5*\height+3*\height) {${\tt 4}$};
    \node at (\blockwidth+\symbolwidth+1,0.5*\height+0*\height) {$\cdots$};
    \node at (\blockwidth+\symbolwidth+1,0.5*\height+1*\height) {$\cdots$};
    \node at (\blockwidth+\symbolwidth+1,0.5*\height+2*\height) {$\cdots$};
    \node at (\blockwidth+\symbolwidth+1,0.5*\height+3*\height) {$\cdots$};
    \node at (\blockwidth+\symbolwidth+1,0.5*\height+5*\height) {$\cdots$};
    \node at (\blockwidth+\symbolwidth+1,0.5*\height+6*\height) {$\cdots$};
    \node at (\blockwidth+\symbolwidth+1,0.5*\height+7*\height) {$\cdots$};
    \foreach \i in {0,1,2,3,4,5,6,7,8} {
      \draw (\blockwidth+\symbolwidth,\i*\height) -- (\blockwidth+\symbolwidth+0.5,\i*\height);
    }

    \def\yoff{5*\height}
    \draw (0,\yoff+0*\height) rectangle ++(\blockwidth,\height);
    \draw (0,\yoff+1*\height) rectangle ++(\blockwidth,\height);
    \draw (0,\yoff+2*\height) rectangle ++(\blockwidth,\height);
    \draw[fill=\col] (\blockwidth,\yoff+0*\height) rectangle ++(\symbolwidth,\height);
    \draw[fill=\col] (\blockwidth,\yoff+1*\height) rectangle ++(\symbolwidth,\height);
    \draw[fill=\col] (\blockwidth,\yoff+2*\height) rectangle ++(\symbolwidth,\height);
    \draw[fill=\col] (-2,\yoff+0*\height) rectangle ++(\symbolwidth,\height);
    \draw[fill=\col] (-2,\yoff+1*\height) rectangle ++(\symbolwidth,\height);
    \draw[fill=\col] (-2,\yoff+2*\height) rectangle ++(\symbolwidth,\height);
    \node at (0.5*\blockwidth,\yoff+0.5*\height) {$\widetilde{P}_i$};
    \node at (0.5*\blockwidth,\yoff+0.5*\height+1*\height) {$\widetilde{P}_i$};
    \node at (0.5*\blockwidth,\yoff+0.5*\height+2*\height) {$\widetilde{P}_i$};
    \node at (\blockwidth+0.5*\symbolwidth,\yoff+0.5*\height) {${\tt 4}$};
    \node at (\blockwidth+0.5*\symbolwidth,\yoff+0.5*\height+1*\height) {${\tt 3}$};
    \node at (\blockwidth+0.5*\symbolwidth,\yoff+0.5*\height+2*\height) {${\tt 2}$};
    \node at (-2+0.5*\symbolwidth,\yoff+0.5*\height) {${\tt 4}$};
    \node at (-2+0.5*\symbolwidth,\yoff+0.5*\height+1*\height) {${\tt 9}$};
    \node at (-2+0.5*\symbolwidth,\yoff+0.5*\height+2*\height) {${\tt 4}$};

    \node at (0.1 + 0.5*\symbolwidth,-0.5*\height) {$\cdots$};
    \node at (-1.9+0.5*\symbolwidth,-0.5*\height) {$\cdots$};
    \node at (0.1 +0.5*\symbolwidth, 4.5*\height) {$\cdots$};
    \node at (-1.9+0.5*\symbolwidth, 4.5*\height) {$\cdots$};
    \node at (0.1 +0.5*\symbolwidth, 8.5*\height) {$\cdots$};
    \node at (-1.9+0.5*\symbolwidth, 8.5*\height) {$\cdots$};

    \node at (-2+0.5*\symbolwidth, 9.5*\height) {$B_2$};
    \node[right] at (-0.3, 9.5*\height) {$S_2[\SA{S_2}[j] \dd \Textlen']$};

    \draw [thin, <->] (-3,2*\height) -- (-3,6*\height) node[midway,left] {$n_i+1$};
    \draw[dashed] (-3,2*\height) -- (-2,2*\height);
    \draw[dashed] (-3,6*\height) -- (-2,6*\height);
  \end{tikzpicture}

  \end{minipage}

  \caption{Ranges of the suffix arrays of $S_1$ and $S_2$ consisting of the
    starting positions of all suffixes prefixed with $\widetilde{\Pat}_i$, and the associated substrings of
    $B_1$ and $B_2$.}\label{fig:bwt-ranges}
\end{figure}

\textbf{3.} Denote $B_i = \BWT{S_i}$ for $i \in \{1,2\}$. For every $c
\in \{{\tt 0}, \ldots, {\tt 9}\}$, let $b_c$ and $e_c$ be such that
$\SA{S_1}(b_c \dd e_c]$ consists of the starting positions of all suffixes
of $S_1$ starting with $c$. The same property then holds for $\SA{S_2}$ since $S_1$ and $S_2$
are permutations of each other.

In this step, we prove that $|\RL{B_1(b_2 \dd e_9]}| = |\RL{B_2(b_2
\dd e_9]}|$.  To see this, first observe that, for every $i \in [1 \dd
k]$, the strings ${\tt 4}\widetilde{P}_{i}{\tt 6}$ and ${\tt
4}\widetilde{P}_{i}{\tt 7}$ each occur exactly once in $S_1$.
Since ${\tt 6}$ and ${\tt 7}$ are consecutive alphabet symbols, the
suffixes beginning at these two occurrences are adjacent in lexicographic
order. Thus, the symbols preceding the two
substrings (which are ${\tt 5}$ and ${\tt 6}$, respectively) occur
consecutively in $B_1$. This exhausts all occurrences of ${\tt 5}$ and
${\tt 6}$ in $S_1$, and hence, for some strings $X_0, \ldots, X_k$
containing neither symbol ${\tt 5}$ nor ${\tt 6}$,
\[
  B_1(b_2 \dd e_9] = X_0{\tt 5}{\tt 6}X_1{\tt 5}{\tt 6}X_2 \cdots X_{k-1}{\tt 5}{\tt 6}X_k.
\]
A similar analysis applies to $S_2$, except the symbols preceding
${\tt 4}\widetilde{P}_i{\tt 5}$ and ${\tt 4}\widetilde{P}_i{\tt 7}$
are ${\tt 6}$ and ${\tt 5}$, respectively. Moreover, using the
relation of $\ISA{S_1}$ and $\ISA{S_2}$, we can prove that $X_0,
\ldots, X_k$ remain the same in the formula $B_2(b_2 \dd
e_9]=X_0{\tt 6}{\tt 5}X_1{\tt 6}{\tt 5}X_2 \cdots X_{k-1}{\tt 6}{\tt
5}X_k$.  Since, for every $i\in [0\dd k]$, the string $X_i$
contains neither ${\tt 5}$ nor ${\tt 6}$, it follows that
\[
  |\RL{B_1(b_2 \dd e_9]}| =
    2k + \textstyle\sum_{i=0}^{k}|\RL{X_i}| =
    |\RL{B_2(b_2 \dd e_9]}|.
\]

\textbf{4.} Next, we prove that $|\RL{B_1(b_0 \dd e_1]}| = 3k + 1 +
|\{i \in [1 \dd k] : \Pat_i\text{ occurs in }\Text\}|$.  To this end,
consider $i \in [1 \dd k]$ and the range $\SA{S_1}(b \dd e]$
consisting of the starting positions of all suffixes of $S_1$ having
$\widetilde{\Pat}_i$ as a prefix.
Observe that, by the definition of $S_1$ and since every occurrence of
$\Pat_i$ in $\Text$ induces an occurrence of ${\tt
4}\widetilde{\Pat}_i{\tt 4}$ in ${\tt 4}\widetilde{\Text}{\tt 4}$,
it holds $B_1(b \dd e] = {\tt 4}{\tt 9}{\tt 4}^{n_i}{\tt 8}{\tt 4}{\tt
4}$, where $n_i$ is the number of occurrences of $\Pat_i$ in
$\Text$ (see \cref{fig:bwt-ranges}).  This implies that
\[
  B_1(b_0 \dd e_1] =
      {\tt 4}^{r_0}{\tt 9}{\tt 4}^{n_{a_1}}{\tt 8}{\tt
      4}^{r_1}{\tt 9}{\tt 4}^{n_{a_2}}{\tt 8}{\tt 4}^{r_2} \cdots {\tt
      4}^{r_{k-1}}{\tt 9}{\tt 4}^{n_{a_k}}{\tt 8}{\tt 4}^{r_{k}},
\]
where $(r_0, r_1, \ldots, r_k)$ is some sequence of positive integers and
$(a_1, \ldots, a_k)$ is a permutation of $\{1, \ldots, k\}$ satisfying
$\widetilde{\Pat}_{a_1} \prec \dots \prec
\widetilde{\Pat}_{a_k}$. Therefore, $|\RL{B_1(b_0 \dd e_1]}| = 3k + 1
+ |\{i \in [1 \dd k] : n_i > 0\}|$, and the claim follows.

\textbf{5.} Using a similar argument as above, next we prove that
$|\RL{B_2(b_0 \dd e_1]}| = 4k + 1$. To see this, let $i \in [1 \dd k]$
and $b, e$ be such that $\SA{S_2}(b \dd e]$ consists of the starting
positions of all suffixes of $S_2$ prefixed with $\widetilde{\Pat}_i$.
Then, $B_2(b \dd e] = {\tt 4}{\tt
9}{\tt 4}^{n_i+1}{\tt 8}{\tt 4}$, where $n_i$ is the number of
occurrences of $\Pat_i$ in $\Text$ (see \cref{fig:bwt-ranges}). Hence,
\[
  B_2(b_0 \dd e_1] = {\tt 4}^{r'_0}{\tt 9}{\tt 4}^{n_{a_1}+1}{\tt
    8}{\tt 4}^{r'_1}{\tt 9}{\tt 4}^{n_{a_2}+1}{\tt 8}{\tt 4}^{r'_2}
    \cdots {\tt 4}^{r'_{k-1}}{\tt 9}{\tt 4}^{n_{a_k}+1}{\tt 8}{\tt
    4}^{r'_{k}},
\]
where $(r'_0, r'_1, \ldots, r'_k)$ is a sequence of positive integers
and $(a_1, \ldots, a_k)$ is as above. Thus, $|\RL{B_2(b_0 \dd e_1]}| = 4k + 1$.

\textbf{6.} In the last step, we put everything together. Since
$B_1(b_0 \dd e_1], B_2(b_0 \dd e_1] \in \{{\tt 4}, {\tt 8},
{\tt 9}\}^{+}$ and $B_1(b_2 \dd e_2], \allowbreak B_2(b_2 \dd e_2] \in
\BinaryAlphabet^{+}$, there is a run ending at
position $e_1$ in both $B_1$ and $B_2$. This implies that
\allowbreak$|\RL{B_1(b_0 \dd e_9]}| = |\RL{B_1(b_0 \dd e_1]}| + |\RL{B_1(b_2 \dd
e_9]}|$, and analogously for~$B_2$. By the above formulas, we obtain
the claim, i.e., $|\RL{B_2(b_0 \dd e_9]}| - |\RL{B_1(b_0 \dd e_9]}| =
k-|\{i \in [1 \dd k] : \Pat_i\text{ occurs in }\Text\}|=|\{i \in [1
\dd k] : \Pat_i\text{ does not occur in }\Text\}|$.

To complete the reduction, it remains to observe that the strings
$S_1$ and $S_2$ can be efficiently constructed given the packed
representation of $\Text$ and $\mathcal{D}$.  In \cref{lm:dm-to-bwt-binary}, we
further show how to modify this reduction so that both strings $S_1$
and $S_2$ are over the binary alphabet.

\section{Tools}
\label{sec:tools}

\subsection{Integer Values of Packed Strings}\label{sec:tools-integer-values}

\begin{definition}[Integer value of a string]\label{def:val}
  Let $\AlphabetSize\in\Z_{\geq2}$. For every
  $X\in\IntegerAlphabet^+$, we define
  $\Val{\AlphabetSize}{X}:=\sum_{i=1}^{|X|}X[i]\cdot
    \AlphabetSize^{|X|-i}\in[0\dd\AlphabetSize^{|X|})$.
  Thus, $X$ is the base-$\AlphabetSize$ representation of
  $\Val{\AlphabetSize}{X}$, padded on the left with zeros to
  length $|X|$. We also define $\Val{\AlphabetSize}{\emptystring}=0$.
\end{definition}

\begin{example}\label{ex:val}
  The string $S=\texttt{413}$ from \cref{ex:packed-representation} has
  integer value $\Val{5}{S}=4\cdot5^2+1\cdot5+3=108$,
  while its packed representation $\PackedRepresentation{8}{5}{S}$
  consists of the two words $12$ and $3$.
\end{example}

\begin{observation}\label{ob:val}
  Let $\AlphabetSize\in\Z_{\geq2}$ and $\ell\in\Z_{\geq0}$.
  For any $X,Y\in\IntegerAlphabet^\ell$, we have
  $X\prec Y$ if and only if
  $\Val{\AlphabetSize}{X}<\Val{\AlphabetSize}{Y}$.
  Moreover, $X=Y$ if and only if
  $\Val{\AlphabetSize}{X}=\Val{\AlphabetSize}{Y}$.
\end{observation}

\begin{theorem}\label{th:val-encoding}
  Let $N,\AlphabetSize\in\Z_{\geq2}$ satisfy $\AlphabetSize\leq N$,
  and denote $\ell=\lfloor\log_{\AlphabetSize}N\rfloor$.
  In the word RAM model with word size $w\geq1+\log N$, given $N$ and
  $\AlphabetSize$, we can in $\bigO(\sqrt N)$ time construct a data structure of size
  $\bigO(\sqrt N)$ words that, given $|X|$ and the packed representation
  $\PackedRepresentation{w}{\AlphabetSize}{X}$ (\cref{def:packed-representation})
  of any string $X\in\IntegerAlphabet^{\leq\ell}$, returns
  $\Val{\AlphabetSize}{X}$ (\cref{def:val}) in $\bigO(1)$ time.
\end{theorem}
\begin{proof}
  Compute $b=\lceil\log\AlphabetSize\rceil$ and $\ell$ in
  $\bigO(\log N)$ time.
  If $\ell\leq4$, it suffices to store $b$ and $\AlphabetSize$.
  A query initializes $v=0$ and, for $j=1,\ldots,|X|$, retrieves
  $X[j]$ using \cref{pr:packed-representation}\eqref{pr:packed-representation-access}
  and sets $v:=\AlphabetSize v+X[j]$.
  The final value is $\Val{\AlphabetSize}{X}$, and this takes
  $\bigO(1)$ time since $|X|\leq4$.
  Assume henceforth that $\ell\geq5$, and let
  $\tau=\lfloor\frac14\log_{\AlphabetSize}N\rfloor=\lfloor\ell/4\rfloor$.
  Thus, $\tau\geq1$ and $\ell/\tau=\bigO(1)$.
  Moreover,
  $b\tau\leq\frac12\log N<w$.
  Thus, the packed representation of any nonempty string of length
  at most $\tau$ fits in one word.

  Let $A_{\rm pow}[0\dd\ell]$ be an array satisfying
  $A_{\rm pow}[j]=\AlphabetSize^j$ for every $j\in[0\dd\ell]$.
  For every $i\in[1\dd\tau]$, let $L_i[0\dd2^{bi})$ be an array
  defined as follows. For each $Y\in\IntegerAlphabet^i$, if $y$ is
  the only word in $\PackedRepresentation{w}{\AlphabetSize}{Y}$, then
  $L_i[y]=\Val{\AlphabetSize}{Y}$. All other entries are zero.

  \DSComponents
  The data structure consists of the integers $b$, $\AlphabetSize$,
  $\ell$, and $\tau$, together with the following arrays:
  \begin{enumerate}
  \item The array $A_{\rm pow}$. It occupies $\bigO(\log N)$ words.
  \item We also store the arrays $L_i$ for every $i\in[1\dd\tau]$.
    Their total size is
    $\bigO(\sum_{i=1}^{\tau}2^{bi})=\bigO(2^{b\tau})=\bigO(\sqrt N)$
    words.
  \end{enumerate}
  The total size is $\bigO(\sqrt N)$ words.

  \DSQueries
  For $|X|=0$, return $0$. Otherwise, proceed as follows.
  \begin{enumerate}

  \item First, we obtain the packed representations of consecutive
    blocks $X_1,\ldots,X_q$ whose concatenation is $X$.
    Each block has length $\tau$, except possibly the last,
    which has length in $[1\dd\tau]$.
    Apply \cref{pr:packed-representation}\eqref{pr:packed-representation-split}
    to obtain these representations, and let $x_j$ be the only word
    in the representation of $X_j$.
    Since $q=\lceil|X|/\tau\rceil=\bigO(1)$, this takes
    $\bigO(q+|X|\log\AlphabetSize/w)=\bigO(1)$ time.

  \item Next, we combine the integer values of the blocks.
    Initialize $v=0$ and, for $j=1,\ldots,q$, set
    \[
      v:=v\cdot A_{\rm pow}[|X_j|]+L_{|X_j|}[x_j].
    \]
    After iteration $j$, we have $v=\Val{\AlphabetSize}{X_1\cdots X_j}$.
    Thus, we return $v=\Val{\AlphabetSize}{X}$ after $\bigO(q)=\bigO(1)$ time.
  \end{enumerate}
  All powers and intermediate values fit in one word because
  $\AlphabetSize^\ell\leq N<2^w$.
  The packed operations use length bound $N$, as permitted by
  $\AlphabetSize\leq N$ and $w>\log N$.

  \DSConstruction
  We construct the arrays as follows.
  \begin{enumerate}

  \item First, we compute $A_{\rm pow}$ in
    $\bigO(\ell)=\bigO(\log N)$ time.

  \item We now fill the tables $L_i$ for $i\in[1\dd\tau]$.
    For each $i$, initialize $L_i$ to zeros.
    For every $v\in[0\dd\AlphabetSize^i)$, compute the length-$i$
    base-$\AlphabetSize$ representation $Y$ of $v$, including leading
    zeros, by repeated division. Then compute its packed word
    $y=\sum_{j=1}^{i}Y[j]2^{(j-1)b}$ and set $L_i[y]=v$.
    This takes $\bigO(i)$ time per value of $v$, yielding a total time of
    \[
      \bigO\Big(\sum_{i=1}^{\tau}(2^{bi}+i\AlphabetSize^i)\Big)
        =\bigO(2^{b\tau}+\tau\AlphabetSize^\tau)
        =\bigO(\sqrt N+N^{1/4}\log N)
        =\bigO(\sqrt N).
    \]
  \end{enumerate}
  Including the computation of the parameters, construction takes
  $\bigO(\sqrt N)$ time.
\end{proof}

\begin{theorem}\label{th:val-decoding}
  Let $N,\AlphabetSize\in\Z_{\geq2}$ satisfy $\AlphabetSize\leq N$,
  and denote $\ell=\lfloor\log_{\AlphabetSize}N\rfloor$.
  In the word RAM model with word size $w\geq1+\log N$, given $N$ and
  $\AlphabetSize$, we can in $\bigO(\sqrt N)$ time construct a data structure of size
  $\bigO(\sqrt N)$ words that, given $|X|$ and the integer
  $\Val{\AlphabetSize}{X}$ (\cref{def:val}) for any string
  $X\in\IntegerAlphabet^{\leq\ell}$, returns the packed representation
  $\PackedRepresentation{w}{\AlphabetSize}{X}$
  (\cref{def:packed-representation}) in $\bigO(1)$ time.
\end{theorem}
\begin{proof}
  Compute $b=\lceil\log\AlphabetSize\rceil$ and $\ell$ in
  $\bigO(\log N)$ time.
  If $\ell\leq4$, the structure stores $N$, $\AlphabetSize$, and $b$.
  Given $m=|X|$ and $v=\Val{\AlphabetSize}{X}$, initialize the
  packed representation of $\zero^m$.
  For $j=m,m-1,\ldots,1$, set its $j$th symbol to
  $v\bmod\AlphabetSize$ and replace $v$ by
  $\lfloor v/\AlphabetSize\rfloor$, using the initialization and
  update operations of \cref{pr:packed-representation}.
  This constructs the required representation in $\bigO(1)$ time,
  including when $m=0$.
  Assume henceforth that $\ell\geq5$, and let
  $\tau=\lfloor\frac14\log_{\AlphabetSize}N\rfloor=\lfloor\ell/4\rfloor$.
  Thus, $\tau\geq1$ and $\ell/\tau=\bigO(1)$.
  Since $\tau\log\AlphabetSize\leq\frac14\log N$, the packed
  representation of any string of length at most $\tau$ occupies
  $\bigO(1)$ words.

  Let $A_{\rm pow}[0\dd\ell]$ be an array satisfying
  $A_{\rm pow}[j]=\AlphabetSize^j$ for every $j\in[0\dd\ell]$.
  For every $i\in[0\dd\tau]$, let $D_i[0\dd\AlphabetSize^i)$ be
  an array such that, for every $Y\in\IntegerAlphabet^i$, the entry
  $D_i[\Val{\AlphabetSize}{Y}]$ stores the packed representation
  $\PackedRepresentation{w}{\AlphabetSize}{Y}$.
  In particular, $D_0[0]$ stores the empty representation.

  \DSComponents
  We store $N$, $\AlphabetSize$, $b$, $\ell$, and $\tau$, together
  with the following arrays:
  \begin{enumerate}
  \item The array $A_{\rm pow}$. It occupies $\bigO(\log N)$ words.
  \item We also store the arrays $D_0,\ldots,D_\tau$.
    Each entry occupies $\bigO(1)$ words. Hence, these arrays together
    use $\bigO(\sum_{i=0}^{\tau}\AlphabetSize^i)
      =\bigO(\AlphabetSize^\tau)=\bigO(N^{1/4})$ words.
  \end{enumerate}
  The total size is $\bigO(\sqrt N)$ words.

  \DSQueries
  Given $m=|X|$ and $v=\Val{\AlphabetSize}{X}$, return the empty
  representation if $m=0$. Otherwise, proceed in three steps.
  \begin{enumerate}

  \item First, we compute the integer values of consecutive blocks of $X$.
    Let $X_1,\ldots,X_q$ be the blocks whose concatenation is $X$,
    each of length $\tau$ except possibly the last, where
    $q=\lceil m/\tau\rceil$.
    Denote their lengths by $\ell_1,\ldots,\ell_q$.
    For every $i\in[1\dd q]$, let
    $r_i=\sum_{j=i+1}^q\ell_j$ and compute
    \[
      v_i=\left\lfloor\frac{v}{A_{\rm pow}[r_i]}\right\rfloor
        \bmod A_{\rm pow}[\ell_i].
    \]
    This extracts the base-$\AlphabetSize$ digits of $X_i$, establishing
    $v_i=\Val{\AlphabetSize}{X_i}$.
    Since $q=\bigO(1)$, the step takes $\bigO(1)$ time.

  \item Next, we obtain the packed representation of each block.
    For every $i\in[1\dd q]$, retrieve the representation stored
    in $D_{\ell_i}[v_i]$. By definition, it represents $X_i$,
    including any leading zeros.
    This takes $\bigO(q)=\bigO(1)$ time.

  \item Finally, we concatenate the block representations in their
    original order, using
    \cref{pr:packed-representation}\eqref{pr:packed-representation-concat}
    with lengths $\ell_1,\ldots,\ell_q$.
    Since $X=X_1\cdots X_q$, this gives the required packed
    representation of $X$.
    The operation takes
    $\bigO(q+m\log\AlphabetSize/w)=\bigO(1)$ time.
  \end{enumerate}
  The total query time is $\bigO(1)$.
  All powers and integer values fit in one word because
  $\AlphabetSize^\ell\leq N<2^w$.
  All packed operations use alphabet size $\AlphabetSize$ and
  length bound $N$, as permitted by the assumptions.

  \DSConstruction
  We store the parameters computed above and construct the arrays
  as follows.
  \begin{enumerate}

  \item First, we compute $A_{\rm pow}$ in
    $\bigO(\ell)=\bigO(\log N)$ time.

  \item We now construct the arrays $D_0,\ldots,D_\tau$ by appending
    one symbol at a time.
    Store the empty packed representation in $D_0[0]$.
    For $i=0,\ldots,\tau-1$, every $v\in[0\dd\AlphabetSize^i)$,
    and every $c\in[0\dd\AlphabetSize)$, construct the packed
    representation of the one-symbol string $c$ using the
    initialization and update operations of \cref{pr:packed-representation}.
    Append it to the representation stored in $D_i[v]$ using
    \cref{pr:packed-representation}\eqref{pr:packed-representation-concat},
    and store the result in $D_{i+1}[\AlphabetSize v+c]$.
    If $v=\Val{\AlphabetSize}{Y}$ for $Y\in\IntegerAlphabet^i$,
    this constructs the representation of $Y\cdot c$, whose
    integer value is $\AlphabetSize v+c$.
    Each entry takes $\bigO(1)$ time to construct because its
    string has length at most $\tau$.
    Thus, filling the arrays takes
    $\bigO(\AlphabetSize^\tau)=\bigO(N^{1/4})$ time.
  \end{enumerate}
  Including the computation of the parameters, construction takes
  $\bigO(\log N+N^{1/4})=\bigO(\sqrt N)$ time and uses
  $\bigO(\sqrt N)$ words of space.
\end{proof}

\subsection{Packed String Reversal}\label{sec:tools-string-reversal}

\begin{theorem}\label{th:string-reversal}
  Let $N,\AlphabetSize\in\Z_{\geq2}$ satisfy $\AlphabetSize\leq N$.
  In the word RAM model with word size $w\geq1+\log N$, given $N$ and
  $\AlphabetSize$, we can in $\bigO(\sqrt N)$ time construct a data structure of size
  $\bigO(\sqrt N)$ words that, given $|S|$ and the packed representation
  $\PackedRepresentation{w}{\AlphabetSize}{S}$
  (\cref{def:packed-representation}) of any string
  $S\in\IntegerAlphabet^{\leq N}$, returns the packed representation
  $\PackedRepresentation{w}{\AlphabetSize}{\revstr{S}}$ of $\revstr{S}$
  (\cref{sec:basic-definitions}) in $\bigO(1+|S|/\log_{\AlphabetSize}N)$ time.
\end{theorem}
\begin{proof}
  Compute $b=\lceil\log\AlphabetSize\rceil$ and
  $\ell=\lfloor\log_{\AlphabetSize}N\rfloor$ in
  $\bigO(\log N)$ time.
  If $\ell\leq4$, the structure stores only $N$, $\AlphabetSize$, and $b$.
  To answer a query, initialize the packed representation of
  $\zero^{|S|}$ and, for $j=1,\ldots,|S|$, retrieve $S[j]$ and
  set the symbol at position $|S|-j+1$ in the new representation
  to this value, using \cref{pr:packed-representation}.
  This takes $\bigO(1+|S|)=\bigO(1+|S|/\log_{\AlphabetSize}N)$
  time because $\log_{\AlphabetSize}N<5$.
  Assume henceforth that $\ell\geq5$, and let
  $h=\lfloor\frac14\log_{\AlphabetSize}N\rfloor=\lfloor\ell/4\rfloor\geq1$.
  Since $h\log\AlphabetSize\leq\frac14\log N<w$, the packed
  representation of any string of length at most $h$ occupies
  $\bigO(1)$ words.

  For every $r\in[0\dd h]$, let $L_r[0\dd\AlphabetSize^r)$ be an
  array such that, for each $X\in\IntegerAlphabet^r$, the entry
  $L_r[\Val{\AlphabetSize}{X}]$ stores the packed representation
  $\PackedRepresentation{w}{\AlphabetSize}{\revstr{X}}$.
  In particular, $L_0[0]$ stores the empty representation.

  \DSComponents
  We store $N$, $\AlphabetSize$, $b$, and $h$, together with the following
  components:
  \begin{enumerate}
  \item The data structure from \cref{th:val-encoding} with size
    parameter $N$ and alphabet size $\AlphabetSize$.
    It occupies $\bigO(\sqrt N)$ words.
  \item We also store the arrays $L_0,\ldots,L_h$.
    Since each entry occupies $\bigO(1)$ words, the arrays together use
    $\bigO(\sum_{r=0}^h\AlphabetSize^r)=\bigO(\AlphabetSize^h)
      =\bigO(N^{1/4})$ words.
  \end{enumerate}
  The total size is $\bigO(\sqrt N)$ words.

  \DSQueries
  For $|S|=0$, return the empty packed representation.
  Otherwise, proceed as follows.
  \begin{enumerate}

  \item First, we obtain the packed representations of consecutive
    blocks $S_1,\ldots,S_q$ whose concatenation is $S$.
    Each block has length $h$, except possibly the last,
    which has length in $[1\dd h]$.
    Apply \cref{pr:packed-representation}\eqref{pr:packed-representation-split}
    with these block lengths.
    This takes $\bigO(q+|S|\log\AlphabetSize/w)$ time,
    where $q=\lceil|S|/h\rceil$.

  \item Next, we retrieve the packed representation of each reversed block.
    For every $i\in[1\dd q]$, use the integer-value data structure
    to compute $v_i=\Val{\AlphabetSize}{S_i}$ and retrieve
    $L_{|S_i|}[v_i]$, which stores the packed representation of
    $\revstr{S_i}$.
    Since $|S_i|\leq h\leq\ell$, each query takes $\bigO(1)$ time.
    This step therefore takes $\bigO(q)$ time.

  \item Finally, we concatenate these representations in reverse order.
    Since
    \[
      \revstr{S}=\revstr{S_q}\,\revstr{S_{q-1}}\cdots\revstr{S_1},
    \]
    applying \cref{pr:packed-representation}\eqref{pr:packed-representation-concat}
    to the retrieved representations in this order, with lengths
    $|S_q|,\ldots,|S_1|$, gives the required packed representation
    of $\revstr{S}$.
    This takes $\bigO(q+|S|\log\AlphabetSize/w)$ time.
  \end{enumerate}
  Since $h=\Theta(\log_{\AlphabetSize}N)$ and $w>\log N$,
  the total query time is $\bigO(1+|S|/\log_{\AlphabetSize}N)$.
  All packed operations use alphabet size $\AlphabetSize$ and
  length bound $N$, as permitted by the assumptions.

  \DSConstruction
  We construct the stored components as follows.
  \begin{enumerate}

  \item First, we construct the integer-value data structure from
    \cref{th:val-encoding} in $\bigO(\sqrt N)$ time.
    We also allocate the arrays $L_0,\ldots,L_h$ and initialize
    $L_0[0]$ to the empty representation.
    The arrays have $\bigO(\AlphabetSize^h)=\bigO(N^{1/4})$ entries
    in total.

  \item We now fill the tables in increasing order of $r$.
    For $r=0,\ldots,h-1$, every $v\in[0\dd\AlphabetSize^r)$,
    and every $c\in[0\dd\AlphabetSize)$, construct the packed
    representation of the one-symbol string $c$ using the
    initialization and update operations of \cref{pr:packed-representation}.
    Prepend it to the representation stored in $L_r[v]$ using
    \cref{pr:packed-representation}\eqref{pr:packed-representation-concat},
    and store the result in $L_{r+1}[\AlphabetSize v+c]$.
    If $v=\Val{\AlphabetSize}{X}$ for $X\in\IntegerAlphabet^r$, then
    \[
      \Val{\AlphabetSize}{X\cdot c}=\AlphabetSize v+c,
      \qquad
      \revstr{X\cdot c}=c\cdot\revstr{X}.
    \]
    These identities show that the stored representation is the required one.
    Each entry takes $\bigO(1+(r+1)\log\AlphabetSize/w)=\bigO(1)$
    time to construct, and every entry is constructed exactly once.
    Thus, filling the tables takes $\bigO(\AlphabetSize^h)=\bigO(N^{1/4})$
    time.
  \end{enumerate}
  Including the computation of the parameters, construction takes
  $\bigO(\sqrt N)$ time and uses $\bigO(\sqrt N)$ words of space.
\end{proof}

\subsection{Packed Pointwise Concatenation and Partition}
\label{sec:tools-packed-pointwise-concatenation-partition}

\begin{theorem}\label{th:packed-pointwise-concatenation-partition}
  Let $N,\AlphabetSize\in\Z_{\geq2}$ satisfy $\AlphabetSize\leq N$.
  Consider the word RAM model with word size $w\geq1+\log N$.
  Given $N$ and $\AlphabetSize$, we can construct in $\bigO(\sqrt N)$ time
  a data structure of size
  $\bigO(\sqrt N)$ words supporting pointwise concatenation and partition
  as follows. Each query specifies $m\in[0\dd N]$ and
  $\ell_1,\ell_2\in\Z_{\geq0}$ satisfying
  $1\leq\ell_1+\ell_2\leq N$ and $m(\ell_1+\ell_2)\leq N$. For sequences
  $U\in(\IntegerAlphabet^{\ell_1})^m$ and
  $V\in(\IntegerAlphabet^{\ell_2})^m$, let
  $W\in(\IntegerAlphabet^{\ell_1+\ell_2})^m$ be the sequence satisfying
  \[
    W[i]=U[i]V[i]\qquad\text{for every }i\in[1\dd m].
  \]
  The structure supports the following two operations on packed representations of string sequences
  (\cref{def:packed-sequence-representation}):
  \begin{enumerate}
  \item Given $\PackedSeqRepresentation{w}{\AlphabetSize}{U}$ and
    $\PackedSeqRepresentation{w}{\AlphabetSize}{V}$, construct
    $\PackedSeqRepresentation{w}{\AlphabetSize}{W}$.
  \item Given $\PackedSeqRepresentation{w}{\AlphabetSize}{W}$, construct
    $\PackedSeqRepresentation{w}{\AlphabetSize}{U}$ and
    $\PackedSeqRepresentation{w}{\AlphabetSize}{V}$.
  \end{enumerate}
  Each operation takes
  $\bigO(1+m(\ell_1+\ell_2)/\log_{\AlphabetSize}N)$ time.
\end{theorem}
\begin{proof}
  Denote $b=\lceil\log\AlphabetSize\rceil$ and
  $h=\lfloor\frac{1}{16}\log_{\AlphabetSize}N\rfloor$.

  For every triple of positive integers $\ell_1,\ell_2,g$ satisfying
  $g(\ell_1+\ell_2)\leq h$, we define three lookup tables as follows.
  Denote $\ell=\ell_1+\ell_2$.
  The first table is the word array
  \[
    A_{\rm con}^{\ell_1,\ell_2,g}
      [0\dd\AlphabetSize^{g\ell_1})[0\dd\AlphabetSize^{g\ell_2}).
  \]
  Its indices are the integer values of the concatenations of two
  sequences of $g$ strings. More precisely, for
  $x\in[0\dd\AlphabetSize^{g\ell_1})$ and
  $y\in[0\dd\AlphabetSize^{g\ell_2})$, let
  $X\in(\IntegerAlphabet^{\ell_1})^g$ and
  $Y\in(\IntegerAlphabet^{\ell_2})^g$ be the unique sequences satisfying
  \[
    x=\Val{\AlphabetSize}{X[1]\cdots X[g]},\qquad
    y=\Val{\AlphabetSize}{Y[1]\cdots Y[g]}.
  \]
  Denote
  \[
    z=\Val{\AlphabetSize}{X[1]Y[1]\cdots X[g]Y[g]}.
  \]
  We define $A_{\rm con}^{\ell_1,\ell_2,g}[x,y]=z$.
  Splitting $X[1]Y[1]\cdots X[g]Y[g]$ into $g$ strings of length
  $\ell$, and each of those after its first $\ell_1$ symbols,
  recovers $X$ and $Y$.
  Thus, the table gives a bijection from its domain to
  $[0\dd\AlphabetSize^{g\ell})$.
  The other two tables are word arrays
  $A_1^{\ell_1,\ell_2,g}[0\dd\AlphabetSize^{g\ell})$ and
  $A_2^{\ell_1,\ell_2,g}[0\dd\AlphabetSize^{g\ell})$ satisfying
  \[
    A_1^{\ell_1,\ell_2,g}[z]=x,
    \qquad
    A_2^{\ell_1,\ell_2,g}[z]=y.
  \]
  They recover the integer values of $X[1]\cdots X[g]$ and
  $Y[1]\cdots Y[g]$.
  To locate the tables, let $A_{\rm tab}$ be an array indexed by
  triples in $[0\dd h]^3$. Its entry at $(\ell_1,\ell_2,g)$ stores pointers
  to these three tables whenever the triple satisfies the conditions
  above. All other entries are zero.

  \DSComponents
  We store $N$, $\AlphabetSize$, $b$, and $h$, together with the
  following components:
  \begin{enumerate}
  \item The integer-value encoding and decoding structures from
    \cref{th:val-encoding,th:val-decoding}, both with size parameter
    $N$ and alphabet size $\AlphabetSize$. Together they occupy
    $\bigO(\sqrt N)$ words.
  \item We also store the three lookup tables for every valid triple
    $(\ell_1,\ell_2,g)$.
    For each triple, they have $3\AlphabetSize^{g\ell}\leq
    3\AlphabetSize^h\leq3N^{1/16}$ entries altogether.
    There are at most $h^3$ triples. The tables therefore occupy
    $\bigO(N^{1/16}\log^3 N)$ words.
  \item The array $A_{\rm tab}$. It occupies $\bigO((h+1)^3)$ words.
  \end{enumerate}
  The total size is $\bigO(\sqrt N)$ words.

  \DSQueries
  If $m=0$, return the empty representations.
  If $\ell_1=0$, concatenation returns the supplied representation of
  $V$, while partition returns the empty representation of $U$ and
  the supplied representation of $W$ as that of $V$.
  The case $\ell_2=0$ is symmetric.
  Assume henceforth that $m,\ell_1,\ell_2\geq1$, and denote
  $\ell=\ell_1+\ell_2$. We distinguish two cases.
  \begin{enumerate}

  \item Suppose first that $\ell\leq h$. We process the input in
    groups of $r=\lfloor h/\ell\rfloor$ consecutive strings, with a
    possibly smaller last group. Let $q=\lceil m/r\rceil$ and
    $r_i=\min\{r,m-(i-1)r\}$ for $i\in[1\dd q]$.
    Both operations proceed in three steps.
    \begin{enumerate}
    \item First, we compute the integer values of the input groups.
      For concatenation, split the representation of $U$ into pieces
      of lengths $(r_1\ell_1,\ldots,r_q\ell_1)$, and that of $V$
      into pieces of lengths $(r_1\ell_2,\ldots,r_q\ell_2)$, using
      \cref{pr:packed-representation}\eqref{pr:packed-representation-split}.
      Query the integer-value encoding structure on each piece, and
      denote the values for group $i$ by $x_i$ and $y_i$, respectively.
      For partition, split the representation of $W$ into pieces
      of lengths $(r_1\ell,\ldots,r_q\ell)$, and use the same
      structure to obtain their integer values $z_i$.
    \item Next, we obtain the packed representations of the output groups.
      For every $i\in[1\dd q]$, retrieve the table pointers from
      $A_{\rm tab}[\ell_1,\ell_2,r_i]$.
      For concatenation, retrieve
      $z_i=A_{\rm con}^{\ell_1,\ell_2,r_i}[x_i,y_i]$ and query
      the decoding structure with length $r_i\ell$ and value $z_i$.
      For partition, retrieve
      $x_i=A_1^{\ell_1,\ell_2,r_i}[z_i]$ and
      $y_i=A_2^{\ell_1,\ell_2,r_i}[z_i]$, and decode them with
      lengths $r_i\ell_1$ and $r_i\ell_2$, respectively.
      The tables use the actual group size $r_i$, including in the
      shorter final group.
    \item Finally, we concatenate the output groups in their original order.
      For concatenation, their lengths are
      $(r_1\ell,\ldots,r_q\ell)$, and the result is
      $\PackedSeqRepresentation{w}{\AlphabetSize}{W}$.
      For partition, concatenate the first returned representations with lengths
      $(r_1\ell_1,\ldots,r_q\ell_1)$ and the second returned representations with
      lengths $(r_1\ell_2,\ldots,r_q\ell_2)$ to obtain
      $\PackedSeqRepresentation{w}{\AlphabetSize}{U}$ and
      $\PackedSeqRepresentation{w}{\AlphabetSize}{V}$, respectively.
      These operations use
      \cref{pr:packed-representation}\eqref{pr:packed-representation-concat}.
    \end{enumerate}
    Since every group has total length at most $h\leq
    \lfloor\log_{\AlphabetSize}N\rfloor$, each integer-value
    conversion and table lookup takes $\bigO(1)$ time.
    The running time of either operation is
    $\bigO(q+m\ell\log\AlphabetSize/w)$.
    Since $r=\lfloor\frac{1}{16\ell}\log_{\AlphabetSize}N\rfloor\geq1$,
    we have $q=\bigO(1+m\ell/\log_{\AlphabetSize}N)$.
    Together with $w\geq1+\log N$, this gives
    $\bigO(1+m\ell/\log_{\AlphabetSize}N)$ time.

  \item Suppose now that $\ell>h$. We construct the result using
    the individual strings, as follows.
    \begin{enumerate}
    \item We first obtain the packed representations of $U[i]$ and
      $V[i]$ for every $i\in[1\dd m]$.
      For concatenation, split the representations of $U$ and $V$
      into $m$ pieces of lengths $\ell_1$ and $\ell_2$, respectively.
      For partition, split the representation of $W$ using the
      alternating lengths $(\ell_1,\ell_2,\ldots,\ell_1,\ell_2)$.
      These operations use
      \cref{pr:packed-representation}\eqref{pr:packed-representation-split}.
    \item We now assemble the required packed sequence representations.
      For concatenation, concatenate the pieces in the order
      $U[1],V[1],\ldots,U[m],V[m]$, using the alternating lengths
      $\ell_1,\ell_2$. For partition, concatenate the pieces
      $U[1],\ldots,U[m]$ and $V[1],\ldots,V[m]$ into two representations,
      using their respective lengths $\ell_1$ and $\ell_2$.
      By \cref{pr:packed-representation}\eqref{pr:packed-representation-concat},
      the results are precisely the required representations of
      $W$, or of $U$ and $V$, respectively.
    \end{enumerate}
    The running time is $\bigO(m+m\ell\log\AlphabetSize/w)$.
    Since $\ell>h$ implies $\ell>\frac{1}{16}\log_{\AlphabetSize}N$, this is
    $\bigO(1+m\ell/\log_{\AlphabetSize}N)$.
  \end{enumerate}
  All basic packed operations in these queries use alphabet size
  $\AlphabetSize$ and length bound $N$.
  The total length is $m\ell\leq N$. After the zero-length cases,
  $\ell\geq2$, and hence the number of pieces satisfies $2m\leq m\ell\leq N$.
  Thus, their length and word-size assumptions hold.

  \DSConstruction
  We construct the stored components as follows.
  \begin{enumerate}

  \item First, we compute $b$ and $h$ and construct the integer-value
    encoding and decoding structures.
    Initialize $A_{\rm tab}$ with zeros and compute
    $\AlphabetSize^0,\ldots,\AlphabetSize^h$ by repeated multiplication.
    This step takes $\bigO(\sqrt N+\log N+(h+1)^3)=\bigO(\sqrt N)$ time.

  \item We now construct the three tables for every
    $\ell_1\in[1\dd h)$, $\ell_2\in[1\dd h-\ell_1]$, and
    $g\in[1\dd\lfloor h/(\ell_1+\ell_2)\rfloor]$.
    For each triple, let $\ell=\ell_1+\ell_2$ and allocate
    $A_{\rm con}^{\ell_1,\ell_2,g}$, $A_1^{\ell_1,\ell_2,g}$, and
    $A_2^{\ell_1,\ell_2,g}$.
    For every $x\in[0\dd\AlphabetSize^{g\ell_1})$ and
    $y\in[0\dd\AlphabetSize^{g\ell_2})$, we fill the corresponding
    entries in three steps.
    \begin{enumerate}
    \item First, we obtain the symbols of the sequences $X$ and $Y$
      defined above. Compute the base-$\AlphabetSize$ representations
      of $x$ and $y$, with lengths $g\ell_1$ and $g\ell_2$ and
      including leading zeros, by repeated division.
      Group the digits into $g$ strings of lengths $\ell_1$ and
      $\ell_2$, respectively.
    \item Next, we compute the integer value
      \[
        z=\Val{\AlphabetSize}{X[1]Y[1]\cdots X[g]Y[g]}.
      \]
      Initialize $z=0$ and process the symbols in this order,
      replacing $z$ by $\AlphabetSize z+c$ for every symbol $c$.
    \item We then fill one entry in each of the three tables:
      \[
        A_{\rm con}^{\ell_1,\ell_2,g}[x,y]:=z,\qquad
        A_1^{\ell_1,\ell_2,g}[z]:=x,\qquad
        A_2^{\ell_1,\ell_2,g}[z]:=y.
      \]
    \end{enumerate}
    The bijection established above ensures that, as $(x,y)$ ranges
    over all index pairs, every entry of each table is assigned exactly
    once. Store the three table pointers in
    $A_{\rm tab}[\ell_1,\ell_2,g]$.
  \end{enumerate}
  Each pair $(x,y)$ takes $\bigO(g\ell)=\bigO(h)$ time to process.
  There are at most $h^3$ parameter triples, each with at most
  $\AlphabetSize^h\leq N^{1/16}$ pairs $(x,y)$.
  Thus, the total construction time is
  \[
    \bigO(\sqrt N+N^{1/16}\log^4 N)=\bigO(\sqrt N).
  \]
  The construction uses $\bigO(\sqrt N)$ words of space.
\end{proof}

\begin{theorem}\label{th:packed-pointwise-concatenation-partition-offline}
  Let $N,\AlphabetSize\in\Z_{\geq2}$ satisfy $\AlphabetSize\leq N$.
  Consider the word RAM model with word size $w\geq1+\log N$.
  Let $m\in[0\dd N]$ and $\ell_1,\ell_2\in\Z_{\geq0}$ satisfy
  $1\leq\ell_1+\ell_2\leq N$ and $m(\ell_1+\ell_2)\leq N$. For sequences
  $U\in(\IntegerAlphabet^{\ell_1})^m$ and
  $V\in(\IntegerAlphabet^{\ell_2})^m$, let
  $W\in(\IntegerAlphabet^{\ell_1+\ell_2})^m$ be the sequence satisfying
  \[
    W[i]=U[i]V[i]\qquad\text{for every }i\in[1\dd m].
  \]
  Given $N$, $\AlphabetSize$, and $m,\ell_1,\ell_2$, we can perform the following
  two tasks on packed representations of string sequences
  (\cref{def:packed-sequence-representation}):
  \begin{enumerate}
  \item Given $\PackedSeqRepresentation{w}{\AlphabetSize}{U}$ and
    $\PackedSeqRepresentation{w}{\AlphabetSize}{V}$, construct
    $\PackedSeqRepresentation{w}{\AlphabetSize}{W}$.
  \item Given $\PackedSeqRepresentation{w}{\AlphabetSize}{W}$, construct
    $\PackedSeqRepresentation{w}{\AlphabetSize}{U}$ and
    $\PackedSeqRepresentation{w}{\AlphabetSize}{V}$.
  \end{enumerate}
  Each task takes
  $\bigO(\sqrt N+m(\ell_1+\ell_2)/\log_{\AlphabetSize}N)$ time
  and uses
  $\bigO(\sqrt N+m(\ell_1+\ell_2)/\log_{\AlphabetSize}N)$ words of space,
  including the input and output representations.
\end{theorem}
\begin{proof}
  Construct the data structure from
  \cref{th:packed-pointwise-concatenation-partition} with size
  parameter $N$ and alphabet size $\AlphabetSize$, and perform the
  requested operation. Construction takes $\bigO(\sqrt N)$ time,
  and the operation takes
  $\bigO(1+m(\ell_1+\ell_2)/\log_{\AlphabetSize}N)$ time.
  This gives the claimed total time and bounds the working space
  in words. The input and output occupy
  $\bigO(1+m(\ell_1+\ell_2)\log\AlphabetSize/w)$ words,
  which also fit the claimed bound because $w\geq1+\log N$.
\end{proof}

\subsection{Packed Alphabet Lifting}
\label{sec:tools-packed-alphabet-lifting}

\begin{theorem}\label{th:alphabet-lifting}
  Let $N,\AlphabetSize_1,\AlphabetSize_2\in\Z_{\geq2}$ satisfy
  $\AlphabetSize_1\leq\AlphabetSize_2\leq N$.
  In the word RAM model with word size $w\geq1+\log N$, given $N$,
  $\AlphabetSize_1$, and $\AlphabetSize_2$, we can in
  $\bigO(\sqrt N)$ time construct a data structure of size
  $\bigO(\sqrt N)$ words that, given $|S|$ and the packed representation
  $\PackedRepresentation{w}{\AlphabetSize_1}{S}$
  (\cref{def:packed-representation}) of any string
  $S\in[0\dd\AlphabetSize_1)^{\leq N}$, returns the packed representation
  $\PackedRepresentation{w}{\AlphabetSize_2}{S}$ of $S$ interpreted
  over alphabet $[0\dd\AlphabetSize_2)$ in
  $\bigO(1+|S|/\log_{\AlphabetSize_2}N)$ time.
  More generally, given also an integer
  $\delta\in[0\dd\AlphabetSize_2-\AlphabetSize_1]$, the data structure
  returns $\PackedRepresentation{w}{\AlphabetSize_2}{R}$ within the same
  time bound, where $R\in[0\dd\AlphabetSize_2)^{|S|}$ satisfies
  $R[j]=S[j]+\delta$ for every $j\in[1\dd|S|]$.
\end{theorem}
\begin{proof}
  We first describe how to obtain
  $\PackedRepresentation{w}{\AlphabetSize_2}{S}$.
  Denote $b_1=\lceil\log\AlphabetSize_1\rceil$,
  $b_2=\lceil\log\AlphabetSize_2\rceil$, and
  $\ell=\lfloor\log_{\AlphabetSize_2}N\rfloor$.
  These parameters can be computed in $\bigO(\log N)$ time.
  If $\ell\leq4$, the structure stores these parameters together
  with $N$, $\AlphabetSize_1$, and $\AlphabetSize_2$.
  To answer a query, initialize the packed representation of
  $\zero^{|S|}$ over $[0\dd\AlphabetSize_2)$, and, for
  $j=1,\ldots,|S|$, retrieve $S[j]$ from the input and set the
  $j$th symbol of the new representation to this value, using
  \cref{pr:packed-representation}.
  This takes $\bigO(1+|S|)=\bigO(1+|S|/\log_{\AlphabetSize_2}N)$
  time because $\log_{\AlphabetSize_2}N<5$.
  Assume henceforth that $\ell\geq5$, and let
  $h=\lfloor\frac14\log_{\AlphabetSize_2}N\rfloor=\lfloor\ell/4\rfloor$.
  We have $h\geq1$ and $h\log\AlphabetSize_2\leq\frac14\log N$.
  Thus, a string of length at most $h$ occupies $\bigO(1)$ words
  in either packed representation.

  For every $r\in[0\dd h]$, let $L_r[0\dd\AlphabetSize_1^r)$
  be an array such that, for every $X\in[0\dd\AlphabetSize_1)^r$,
  the entry $L_r[\Val{\AlphabetSize_1}{X}]$ stores the packed
  representation $\PackedRepresentation{w}{\AlphabetSize_2}{X}$.
  In particular, $L_0[0]$ stores the empty representation.
  For a fixed length $r$, distinct strings have distinct integer
  values by \cref{ob:val}, which makes the entries well-defined.

  \DSComponents
  We store $N$, $\AlphabetSize_1$, $\AlphabetSize_2$, $b_1$, $b_2$,
  and $h$, together with the following components:
  \begin{enumerate}
  \item The data structure from \cref{th:val-encoding} with size
    parameter $N$ and alphabet size $\AlphabetSize_1$.
    It uses $\bigO(\sqrt N)$ words.
  \item We store the arrays $L_0,\ldots,L_h$.
    Since each entry occupies $\bigO(1)$ words, these arrays together use
    $\bigO(\sum_{r=0}^h\AlphabetSize_1^r)=\bigO(\AlphabetSize_1^h)
      =\bigO(N^{1/4})$ words.
  \end{enumerate}
  The total size is $\bigO(\sqrt N)$ words.

  \DSQueries
  For $|S|=0$, return the empty packed representation.
  Otherwise, proceed as follows.
  \begin{enumerate}

  \item First, we obtain the packed representations of consecutive
    blocks $S_1,\ldots,S_q$ whose concatenation is $S$.
    Each block has length $h$, except possibly the last,
    which has length in $[1\dd h]$.
    Apply \cref{pr:packed-representation}\eqref{pr:packed-representation-split}
    with these block lengths to obtain their packed representations
    over $[0\dd\AlphabetSize_1)$.
    This takes $\bigO(q+|S|\log\AlphabetSize_1/w)$ time,
    where $q=\lceil|S|/h\rceil$.

  \item Next, we convert the representation of each block to
    alphabet $[0\dd\AlphabetSize_2)$.
    For every $i\in[1\dd q]$, use the integer-value data structure
    to compute $v_i=\Val{\AlphabetSize_1}{S_i}$ and retrieve
    the packed representation $\PackedRepresentation{w}{\AlphabetSize_2}{S_i}$
    stored in $L_{|S_i|}[v_i]$.
    Since $|S_i|\leq h\leq\lfloor\log_{\AlphabetSize_1}N\rfloor$,
    each conversion takes $\bigO(1)$ time, for $\bigO(q)$ time in total.

  \item Finally, we concatenate the converted representations.
    Apply \cref{pr:packed-representation}\eqref{pr:packed-representation-concat}
    to the representations of $S_1,\ldots,S_q$ over
    $[0\dd\AlphabetSize_2)$, in this order, obtaining
    $\PackedRepresentation{w}{\AlphabetSize_2}{S}$.
    This takes $\bigO(q+|S|\log\AlphabetSize_2/w)$ time.
  \end{enumerate}
  Since $h=\Theta(\log_{\AlphabetSize_2}N)$ and $w>\log N$,
  the total query time is $\bigO(1+|S|/\log_{\AlphabetSize_2}N)$.
  All packed operations use length bound $N$ and alphabet size
  $\AlphabetSize_1$ or $\AlphabetSize_2$, as permitted by the assumptions.

  \DSConstruction
  We construct the data structure as follows.
  \begin{enumerate}

  \item First, we construct the integer-value data structure
    from \cref{th:val-encoding} in $\bigO(\sqrt N)$ time.

  \item We now construct the arrays $L_0,\ldots,L_h$.
    Store the empty packed representation in $L_0[0]$.
    For $r=0,\ldots,h-1$, allocate $L_{r+1}$ and process every
    $v\in[0\dd\AlphabetSize_1^r)$ and $c\in[0\dd\AlphabetSize_1)$.
    Construct the packed representation of the one-symbol string $c$
    over $[0\dd\AlphabetSize_2)$ using the initialization and update
    operations of \cref{pr:packed-representation}.
    Append it to the representation stored in $L_r[v]$ using
    \cref{pr:packed-representation}\eqref{pr:packed-representation-concat},
    and store the result in $L_{r+1}[\AlphabetSize_1v+c]$.
    If $v=\Val{\AlphabetSize_1}{X}$ for a length-$r$ string $X$,
    this constructs the representation of $X\cdot c$, whose integer
    value over the source alphabet is $\AlphabetSize_1v+c$.
    Each entry takes $\bigO(1)$ time to construct because
    $(r+1)\log\AlphabetSize_2=\bigO(\log N)$.
    Thus, all tables are constructed in
    $\bigO(\AlphabetSize_1^h)=\bigO(N^{1/4})$ time.
  \end{enumerate}
  Including the computation of the parameters, construction takes
  $\bigO(\sqrt N)$ time.

  To support the offset $\delta$, first obtain
  $\PackedRepresentation{w}{\AlphabetSize_2}{S}$ as above and then apply
  \cref{pr:packed-representation}\eqref{pr:packed-representation-increase}
  to add $\delta$ to every symbol.
  The operation is applicable because
  $S[j]+\delta\leq\AlphabetSize_1-1+\delta\leq\AlphabetSize_2-1$.
  It takes $\bigO(1+|S|\log\AlphabetSize_2/w)$ time, preserving the total
  query bound of $\bigO(1+|S|/\log_{\AlphabetSize_2}N)$.
\end{proof}

\subsection{Packed Binary Mapping}
\label{sec:tools-packed-binary-encoding}

\begin{definition}[Binary representation of an integer]\label{def:bin}
  For any $k \in \Zp$ and $x \in [0 \dd 2^k)$, by $\bin{k}{x} \in \BinaryAlphabet^{k}$
  we denote the length-$k$ string containing the binary representation of $x$ (with leading zeros).
\end{definition}

\begin{theorem}\label{th:packed-binary-encoding}
  Let $N,\AlphabetSize\in\Z_{\geq2}$ satisfy $\AlphabetSize\leq N$,
  and denote $k=\lceil\log\AlphabetSize\rceil$.
  Consider the word RAM model with word size $w\geq1+\log N$.
  Given $N$ and $\AlphabetSize$, in $\bigO(\sqrt N)$ time and using
  $\bigO(\sqrt N)$ words of space, we can construct a data structure
  of size $\bigO(\sqrt N)$ words supporting the following query.
  Given the length $|S|$ and packed representation
  $\PackedRepresentation{w}{\AlphabetSize}{S}$ of any string
  $S\in\IntegerAlphabet^*$ satisfying $k|S|\leq N$
  (\cref{def:packed-representation}),
  it computes in $\bigO(1+|S|/\log_{\AlphabetSize}N)$ time the
  packed representation $\PackedRepresentation{w}{2}{S_{\rm bin}}$
  of the binary string
  \[
    S_{\rm bin}=\bin{k}{S[1]}\cdots\bin{k}{S[|S|]}
      \in\BinaryAlphabet^{|S|\cdot k}
    \qquad\text{(\cref{def:bin}).}
  \]
\end{theorem}
\begin{proof}
  We use the bound $M=2^{\lceil\log N\rceil}$ for the auxiliary
  structures. We have $N\leq M<2N$. Since $w$ is an integer,
  the word-size assumption implies
  $w\geq1+\lceil\log N\rceil=1+\log M$.
  Denote $\alpha=2^k\leq M$ and
  $h=\lfloor\log_{\alpha}M\rfloor\geq1$.

  \DSComponents
  We store $N$, $M$, $\AlphabetSize$, $k$, $\alpha$, and $h$, together
  with the following three data structures, each with size parameter $M$:
  \begin{enumerate}
  \item We store the alphabet-lifting structure from
    \cref{th:alphabet-lifting} with source alphabet size
    $\AlphabetSize$ and target alphabet size $\alpha$.
  \item The integer-value encoding structure from
    \cref{th:val-encoding}, with alphabet size $\alpha$.
  \item We also store the decoding structure from
    \cref{th:val-decoding} with alphabet size $2$.
  \end{enumerate}
  Together, these components occupy $\bigO(\sqrt M)=\bigO(\sqrt N)$ words.

  \DSQueries
  Let $m=|S|$. If $m=0$, return the empty representation.
  Otherwise, we proceed in three steps.
  \begin{enumerate}

  \item First, we obtain the packed representations of consecutive
    blocks $S_1,\ldots,S_q$ whose concatenation is $S$.
    Each block has length $h$ except possibly the last.
    Denote $q=\lceil m/h\rceil$ and $r_i=|S_i|$.
    Apply \cref{pr:packed-representation}\eqref{pr:packed-representation-split}
    with these block lengths.

  \item Next, we construct the packed representation of
    \[
      B_i=\bin{k}{S_i[1]}\cdots\bin{k}{S_i[r_i]}
    \]
    for every $i\in[1\dd q]$.
    To this end, use the lifting structure to obtain
    $\PackedRepresentation{w}{\alpha}{S_i}$, and then compute
    $v_i=\Val{\alpha}{S_i}$ using the integer-value structure.
    Since $\alpha=2^k$, we have $\Val{2}{B_i}=\Val{\alpha}{S_i}=v_i$.
    Querying the binary-decoding structure with length $kr_i$
    and value $v_i$ therefore returns
    $\PackedRepresentation{w}{2}{B_i}$, including any leading zeros.
    Each conversion takes $\bigO(1)$ time because
    $r_i\leq h$ and $kr_i\leq kh\leq\log M$.

  \item Finally, we concatenate the packed representations of
    $B_1,\ldots,B_q$, in this order and with lengths
    $kr_1,\ldots,kr_q$, using
    \cref{pr:packed-representation}\eqref{pr:packed-representation-concat}.
    Since $B_1\cdots B_q=S_{\rm bin}$, this gives the required output.
  \end{enumerate}
  The total query time is $\bigO(q+mk/w)$.
  Since $h=\Theta(\log_{\AlphabetSize}N)$ and $k=\Theta(\log\AlphabetSize)$,
  this is $\bigO(1+m/\log_{\AlphabetSize}N)$.
  The basic packed operations use length bound $N$, since the input and
  output lengths are at most $mk\leq N$.

  \DSConstruction
  We construct the data structure as follows.
  \begin{enumerate}

  \item First, we compute $M$, $k$, $\alpha$, and $h$, and store them
    together with $N$ and $\AlphabetSize$.
    This takes $\bigO(\log N)$ time.

  \item We now construct the three auxiliary data structures.
    Their assumptions hold because $\AlphabetSize\leq\alpha\leq M$
    and $w\geq1+\log M$.
    Thus, construction takes $\bigO(\sqrt M)=\bigO(\sqrt N)$ time in total.
  \end{enumerate}
  Including parameter computation, construction takes
  $\bigO(\sqrt N)$ time and uses $\bigO(\sqrt N)$ words of space.
\end{proof}

\subsection{Packed Extended Binary Mapping}
\label{sec:tools-packed-extended-binary-mapping}

\begin{proposition}\label{pr:basic-alphabet-mapping}
  Let $N,\AlphabetSize\in\Z_{\geq2}$ satisfy $\AlphabetSize\leq N$,
  and denote $b:=\lceil\log\AlphabetSize\rceil$.
  For every $a\in\IntegerAlphabet$, let
  $C(a):={\tt 1}^{b+1} \cdot {\tt 0} \cdot \bin{b}{a} \cdot {\tt 0}$
  (\cref{def:bin}). Consider the word RAM model with word size
  $w\geq1+\log N$. Given $N$ and $\AlphabetSize$, we can in
  $\bigO(\sqrt N)$ time and using
  $\bigO(\sqrt N)$ words of space construct a data structure
  of size $\bigO(\sqrt N)$ words supporting the following query.
  Given $|S|$ and the packed representation
  $\PackedRepresentation{w}{\AlphabetSize}{S}$ of any string
  $S\in\IntegerAlphabet^*$ satisfying $(2b+3)|S|\leq N$
  (\cref{def:packed-representation}), it constructs the packed representation
  $\PackedRepresentation{w}{2}{S_{\rm bin}}$ of the binary string
  $S_{\rm bin}=C(S[1])\cdots C(S[|S|])$, in
  $\bigO(1+|S|/\log_{\AlphabetSize}N)$ time.
\end{proposition}
\begin{proof}
  Denote $\delta:=2b+3$.

  \DSComponents
  We store $N$, $\AlphabetSize$, $b$, and $\delta$, together with
  the following two data structures:
  \begin{enumerate}
  \item The data structure from \cref{th:packed-binary-encoding}
    with size parameter $N$ and alphabet size $\AlphabetSize$.
  \item The data structure from
    \cref{th:packed-pointwise-concatenation-partition} with size
    parameter $N$ and alphabet size $2$.
  \end{enumerate}
  Together, these components occupy $\bigO(\sqrt N)$ words.

  \DSQueries
  Let $m=|S|$. If $m=0$, return the empty representation.
  Otherwise, we proceed in three steps.
  \begin{enumerate}

  \item First, we construct the packed representation of the binary string
    \[
      X=\bin{b}{S[1]}\cdots\bin{b}{S[m]}.
    \]
    To this end, query the binary-encoding structure on the supplied
    packed representation of $S$ and its length.
    The query takes $\bigO(1+m/\log_{\AlphabetSize}N)$ time.
    Its output also represents the sequence
    $\bin{b}{S[1]},\ldots,\bin{b}{S[m]}$ of strings of common length $b$.

  \item Next, we construct the packed representation of the sequence
    of $m$ copies of $Q=\one^{b+1}\zero$.
    We first construct the packed representations of $\one^{m(b+1)}$
    and $\zero^m$ using
    \cref{pr:packed-representation}\eqref{pr:packed-representation-initialize}.
    View these representations as those of sequences of $m$ copies
    of $\one^{b+1}$ and $\zero$, respectively.
    Using the structure from
    \cref{th:packed-pointwise-concatenation-partition}, pointwise
    concatenate these sequences to obtain the required packed representation.
    This step takes $\bigO(1+mb/\log N)$ time.

  \item Finally, we assemble the packed representation of $S_{\rm bin}$.
    Using the same structure, pointwise concatenate the sequence of
    copies of $Q$ with the sequence $\bin{b}{S[1]},\ldots,\bin{b}{S[m]}$.
    Then append a zero to each resulting string by another pointwise
    concatenation, using the packed representation of $\zero^m$
    from the preceding step.
    The result represents the sequence
    $C(S[1]),\ldots,C(S[m])$. By the definition of packed sequence
    representations, it is also the packed representation of
    $S_{\rm bin}$. We return this representation.
    The two pointwise concatenations take $\bigO(1+mb/\log N)$ time.
  \end{enumerate}
  All intermediate representations have total length at most
  $m\delta\leq N$. Since $m\geq1$, all pointwise concatenations
  also use at most $N$ strings of common length at most $\delta\leq N$.
  Thus, the binary-encoding and pointwise queries satisfy their length
  bounds. The basic packed operations use alphabet size $2$ and length
  bound $N$.
  Since $b=\Theta(\log\AlphabetSize)$, the total query time is
  \[
    \bigO(1+mb/\log N)=\bigO(1+m/\log_{\AlphabetSize}N).
  \]

  \DSConstruction
  Store $N$ and $\AlphabetSize$, compute and store $b$ and $\delta$
  in $\bigO(\log N)$ time, and construct the two data structures
  with the parameters specified above.
  Both apply under the assumed word size $w\geq1+\log N$.
  The total construction takes $\bigO(\sqrt N)$ time using
  $\bigO(\sqrt N)$ words of space.
\end{proof}

\subsection{Packed Morphisms}
\label{sec:tools-packed-morphisms}

\begin{theorem}\label{th:packed-morphism}
  Let $d\in\Z_{\geq1}$ be a fixed constant, and let
  $N,\AlphabetSize\in\Z_{\geq2}$ satisfy $\AlphabetSize\leq N$.
  Consider the word RAM model with word size $w\geq1+\log N$, and let
  $\mu:\IntegerAlphabet\to\IntegerAlphabet^d$.
  Given $N$, $\AlphabetSize$, and an array containing the packed representations of
  $\mu(0),\ldots,\mu(\AlphabetSize-1)$
  (\cref{def:packed-representation}), we can in
  $\bigO(\sqrt N+\AlphabetSize)$ time construct a data structure of
  size $\bigO(\sqrt N+\AlphabetSize)$ words supporting the following query.
  Given $|S|$ and the packed representation
  $\PackedRepresentation{w}{\AlphabetSize}{S}$ of
  $S\in\IntegerAlphabet^*$ satisfying $d|S|\leq N$, it computes in
  $\bigO(1+|S|/\log_{\AlphabetSize}N)$ time the packed representation of
  \[
    \mu(S)=\mu(S[1])\cdots\mu(S[|S|])
      \in\IntegerAlphabet^{d|S|}.
  \]
\end{theorem}
\begin{proof}
  If $N<d$, the only admissible query string is empty, and a structure
  returning the empty representation suffices. Assume henceforth that $N\geq d$.
  Let $A[0\dd\AlphabetSize)$ denote the supplied array, where
  $A[a]$ contains the packed representation of $\mu(a)$.
  Compute $b=\lceil\log\AlphabetSize\rceil$ and
  $h=\lfloor\frac{1}{4d}\log_{\AlphabetSize}N\rfloor$ in $\bigO(\log N)$ time.
  If $h=0$, the structure stores $N$, $\AlphabetSize$, $d$, $b$, and $A$.
  An empty query returns the empty representation.
  Otherwise, retrieve $S[1],\ldots,S[|S|]$ using
  \cref{pr:packed-representation}\eqref{pr:packed-representation-access},
  and concatenate the representations in
  $A[S[1]],\ldots,A[S[|S|]]$, in this order, using
  \cref{pr:packed-representation}\eqref{pr:packed-representation-concat}.
  This takes $\bigO(1+|S|)$ time, which satisfies the claimed bound
  because $\log_{\AlphabetSize}N<4d$ and $d$ is constant.
  Assume henceforth that $h\geq1$.

  For every $r\in[0\dd h]$, let $L_r[0\dd\AlphabetSize^r)$ be an
  array such that, for every $X\in\IntegerAlphabet^r$, the entry
  $L_r[\Val{\AlphabetSize}{X}]$ contains the packed representation
  $\PackedRepresentation{w}{\AlphabetSize}{\mu(X)}$.
  In particular, $L_0[0]$ contains the empty representation.
  By \cref{ob:val}, distinct strings of the same length have distinct integer
  values, which makes the entries well-defined.
  Each entry occupies
  $\bigO(1+dh\log\AlphabetSize/w)=\bigO(1)$ words because $d$ is
  constant and $dh\log\AlphabetSize\leq\frac14\log N$.

  \DSComponents
  We store $N$, $\AlphabetSize$, $d$, $b$, and $h$, together with
  the following components:
  \begin{enumerate}
  \item The array $A[0\dd\AlphabetSize)$.
    It occupies $\bigO(\AlphabetSize)$ words.
  \item The integer-value encoding structure from \cref{th:val-encoding},
    with size parameter $N$ and alphabet size $\AlphabetSize$.
    It occupies $\bigO(\sqrt N)$ words.
  \item We store the arrays $L_0,\ldots,L_h$.
    Their total size is
    $\bigO(\sum_{r=0}^h\AlphabetSize^r)
      =\bigO(\AlphabetSize^h)=\bigO(N^{1/4})$ words.
  \end{enumerate}
  Together, these components occupy $\bigO(\sqrt N+\AlphabetSize)$ words.

  \DSQueries
  Denote $n=|S|$. If $n=0$, return the empty representation.
  Otherwise, proceed as follows.
  \begin{enumerate}

  \item First, we obtain the packed representations of consecutive
    blocks $S_1,\ldots,S_q$ whose concatenation is $S$.
    Each block has length $h$ except possibly the last, and
    $q=\lceil n/h\rceil$.
    To this end, split the supplied packed representation at these
    boundaries using
    \cref{pr:packed-representation}\eqref{pr:packed-representation-split}.

  \item Next, we obtain the packed representation of $\mu(S_i)$
    for every $i\in[1\dd q]$.
    Compute $v_i=\Val{\AlphabetSize}{S_i}$ using the structure from
    \cref{th:val-encoding}, and then retrieve the representation stored
    in $L_{|S_i|}[v_i]$.
    Each conversion and lookup takes $\bigO(1)$ time because
    $|S_i|\leq h\leq\lfloor\log_{\AlphabetSize}N\rfloor$.

  \item Finally, we construct the packed representation of $\mu(S)$.
    Using \cref{pr:packed-representation}\eqref{pr:packed-representation-concat},
    concatenate the representations of $\mu(S_1),\ldots,\mu(S_q)$,
    with lengths $d|S_1|,\ldots,d|S_q|$, in this order.
    Their concatenation equals $\mu(S)$ by its definition.
  \end{enumerate}
  Splitting and concatenation take $\bigO(q+dn\log\AlphabetSize/w)$
  time, and all conversions and lookups take $\bigO(q)$ time.
  Since $d$ is constant, $h=\Theta(\log_{\AlphabetSize}N)$.
  Together with $w\geq1+\log N$, this gives
  a total query time of $\bigO(1+n/\log_{\AlphabetSize}N)$.

  \DSConstruction
  We proceed in two steps.
  \begin{enumerate}
  \item First, copy the supplied array $A$ in $\bigO(\AlphabetSize)$
    time and construct the integer-value encoding structure from
    \cref{th:val-encoding} in $\bigO(\sqrt N)$ time.
  \item We now construct $L_0,\ldots,L_h$ as follows.
    Store the empty representation in $L_0[0]$.
    For $r=0,\ldots,h-1$, and for every
    $v\in[0\dd\AlphabetSize^r)$ and $a\in[0\dd\AlphabetSize)$,
    concatenate the representations in $L_r[v]$ and $A[a]$ using
    \cref{pr:packed-representation}\eqref{pr:packed-representation-concat},
    and store the result in $L_{r+1}[\AlphabetSize v+a]$.
    If $v=\Val{\AlphabetSize}{X}$ for a string $X$ of length $r$,
    the result represents $\mu(X)\mu(a)=\mu(Xa)$, and
    $\Val{\AlphabetSize}{Xa}=\AlphabetSize v+a$.
    Each entry takes $\bigO(1)$ time to construct because
    $(r+1)d\log\AlphabetSize=\bigO(\log N)$.
    Thus, all tables are constructed in $\bigO(\AlphabetSize^h)
    =\bigO(N^{1/4})$ time.
  \end{enumerate}
  Including parameter computation, the total construction time is
  $\bigO(\sqrt N+\AlphabetSize)$.

  The basic packed operations use alphabet size $\AlphabetSize$ and
  length bound $N$. The query output has length $dn\leq N$, and every
  string stored in a table has length at most
  $dh\leq\frac14\log_{\AlphabetSize}N\leq N$.
  The remaining strings have length at most $N$ as well.
  Thus, the basic packed operations and the integer-value structure
  apply under the assumed word size $w\geq1+\log N$.
\end{proof}

\subsection{Packed Longest Common Prefix Queries}
\label{sec:tools-packed-lcp}

\begin{theorem}\label{th:packed-lcp}
  Let $N,\AlphabetSize\in\Z_{\geq2}$ satisfy $\AlphabetSize\leq N$.
  In the word RAM model with word size $w\geq1+\log N$, given $N$ and
  $\AlphabetSize$, we can in $\bigO(\sqrt N)$ time construct a data structure of size
  $\bigO(\sqrt N)$ words that, given the lengths $|X|,|Y|$ and the
  packed representations $\PackedRepresentation{w}{\AlphabetSize}{X}$
  and $\PackedRepresentation{w}{\AlphabetSize}{Y}$
  (\cref{def:packed-representation}) of any strings
  $X,Y\in\IntegerAlphabet^{\leq N}$, computes $\lcp{X}{Y}$ in
  \[
    \bigO\left(1+\frac{\min\{|X|,|Y|\}}{\log_{\AlphabetSize}N}\right)
  \]
  time.
\end{theorem}
\begin{proof}
  Compute $b=\lceil\log\AlphabetSize\rceil$ and
  $h=\lfloor\frac18\log_{\AlphabetSize}N\rfloor$ in
  $\bigO(\log N)$ time.
  If $h=0$, the structure stores only $N$, $\AlphabetSize$, and $b$.
  A query compares corresponding symbols using
  \cref{pr:packed-representation}\eqref{pr:packed-representation-access}
  until finding a mismatch or reaching the end of a string.
  Since $\log_{\AlphabetSize}N<8$, this takes
  $\bigO(1+\min\{|X|,|Y|\}/\log_{\AlphabetSize}N)$ time.
  Assume henceforth that $h\geq1$.

  For every $r\in[1\dd h]$, let $L_r$ be an array indexed by
  pairs in $[0\dd\AlphabetSize^r)^2$, satisfying
  \[
    L_r[\Val{\AlphabetSize}{U},\Val{\AlphabetSize}{V}]
      =\lcp{U}{V}
    \qquad\text{for every }U,V\in\IntegerAlphabet^r.
  \]
  The entries are well-defined by \cref{ob:val}.

  \DSComponents
  We store $N$, $\AlphabetSize$, $b$, and $h$, together with the
  following components:
  \begin{enumerate}
  \item The integer-value data structure from \cref{th:val-encoding},
    with size parameter $N$ and alphabet size $\AlphabetSize$.
    It occupies $\bigO(\sqrt N)$ words.
  \item We also store the arrays $L_1,\ldots,L_h$.
    Together they occupy
    $\bigO(\sum_{r=1}^h\AlphabetSize^{2r})
      =\bigO(\AlphabetSize^{2h})=\bigO(N^{1/4})$ words.
  \end{enumerate}
  The total size is $\bigO(\sqrt N)$ words.

  \DSQueries
  Denote $m=\min\{|X|,|Y|\}$ and initialize $a=0$.
  While $a<m$, perform the following steps.
  \begin{enumerate}

  \item First, we obtain the packed representations of the next two
    blocks. Let $r=\min\{h,m-a\}$, and denote
    $U=X[a+1\dd a+r]$ and $V=Y[a+1\dd a+r]$.
    Extract their packed representations using
    \cref{pr:packed-representation}\eqref{pr:packed-representation-substring}.

  \item Next, we compute the common-prefix length of $U$ and $V$.
    Use the integer-value structure to obtain
    $u=\Val{\AlphabetSize}{U}$ and $v=\Val{\AlphabetSize}{V}$,
    and retrieve $\lambda=L_r[u,v]$.
    If $\lambda<r$, return $a+\lambda$.
    Otherwise, set $a:=a+r$ and continue.
  \end{enumerate}
  If the loop ends, return $m$.
  At the start of each iteration, $X[1\dd a]=Y[1\dd a]$.
  Thus, the first block pair with $\lambda<r$ determines the
  first mismatch. If no such pair exists, one string is a prefix
  of the other. This proves correctness, including when $m=0$.
  Each iteration takes $\bigO(1)$ time: extracting a block costs
  $\bigO(1+h\log\AlphabetSize/w)=\bigO(1)$, and
  $r\leq h\leq\lfloor\log_{\AlphabetSize}N\rfloor$ permits the
  integer-value queries.
  Since $h=\Theta(\log_{\AlphabetSize}N)$, the total query time is
  $\bigO(1+m/h)=\bigO(1+m/\log_{\AlphabetSize}N)$.

  \DSConstruction
  We construct the stored components in two steps.
  \begin{enumerate}

  \item First, we construct the integer-value structure.
    By \cref{th:val-encoding}, this takes $\bigO(\sqrt N)$ time.

  \item We now fill the arrays $L_1,\ldots,L_h$.
    Compute the powers $\AlphabetSize^0,\ldots,\AlphabetSize^h$
    by repeated multiplication.
    For every $r\in[1\dd h]$ and $u,v\in[0\dd\AlphabetSize^r)$,
    obtain the length-$r$ base-$\AlphabetSize$ representations
    $U$ and $V$ of $u$ and $v$, including leading zeros, by
    repeated division. Compare their digits from left to right
    and set $L_r[u,v]:=\lcp{U}{V}$.
    This takes $\bigO(r)$ time per entry, giving a total construction
    time of $\bigO(h\AlphabetSize^{2h})$ for all arrays.
  \end{enumerate}
  Including the computation of the parameters, construction takes
  \[
    \bigO(\sqrt N+h\AlphabetSize^{2h})
      =\bigO(\sqrt N+N^{1/4}\log N)
      =\bigO(\sqrt N)
  \]
  time and uses $\bigO(\sqrt N)$ words of space.
\end{proof}

\begin{remark}
  We give a data-structure formulation to avoid assuming that the
  position of the least or most significant set bit of a machine
  word can be found in constant time without preprocessing.
  The lookup tables above provide the common-prefix lengths of
  short strings using only the operations of the stated word RAM model.
\end{remark}

\subsection{Periods of Short Packed Strings}
\label{sec:tools-periodicity-testing}

\begin{proposition}\label{pr:periods-of-short-strings}
  Let $N,\AlphabetSize\in\Z_{\geq2}$ satisfy $\AlphabetSize\leq N$, and
  let $d\in[1\dd\lfloor\log_{\AlphabetSize}N\rfloor]$. Consider the word
  RAM model with word size $w\geq1+\log N$. Given $N$, $\AlphabetSize$,
  and $d$, in $\bigO(\AlphabetSize^d d)$ time and using
  $\bigO(\AlphabetSize^d)$ words of space, we can construct a data
  structure of size $\bigO(\AlphabetSize^d)$ words that, given $|X|$
  and the packed representation
  $\PackedRepresentation{w}{\AlphabetSize}{X}$
  (\cref{def:packed-representation}) of any nonempty string
  $X\in\IntegerAlphabet^+$ satisfying $|X|\leq d$, returns
  $\per{X}$ in $\bigO(1)$ time.
\end{proposition}
\begin{proof}
  Denote $M=\AlphabetSize^d\leq N$. For every $m\in[1\dd d]$, let
  $L_m[0\dd\AlphabetSize^m)$ be an array satisfying
  \[
    L_m[\Val{\AlphabetSize}{X}]=\per{X}
      \qquad\text{for every }X\in\IntegerAlphabet^m
  \]
  (see \cref{def:val}). These arrays are well defined because the length
  $m$ and the integer value uniquely determine $X$.

  \DSComponents
  We store the arrays $L_1,\ldots,L_d$ and the integer-value structure
  from \cref{th:val-encoding} with size parameter $M$ and alphabet
  size $\AlphabetSize$. Every table entry belongs to $[1\dd d]$ and
  fits in one word. Since
  $\sum_{m=1}^d\AlphabetSize^m=\bigO(M)$, the components together
  occupy $\bigO(M)=\bigO(\AlphabetSize^d)$ words.

  \DSQueries
  Given $m=|X|$ and the packed representation of $X$, use the
  integer-value structure to compute $z=\Val{\AlphabetSize}{X}$,
  and return $L_m[z]$. The query takes $\bigO(1)$ time because
  $m\leq d=\log_{\AlphabetSize}M$.

  \DSConstruction
  We construct the components in two steps.
  \begin{enumerate}

  \item\label{step:periodicity-testing-prepare} First, we prepare the powers and the integer-value structure.
    Compute an auxiliary array $A_{\rm pow}[0\dd d]$ satisfying
    $A_{\rm pow}[m]=\AlphabetSize^m$ for every $m\in[0\dd d]$.
    This gives $M=A_{\rm pow}[d]$. Construct the integer-value
    structure with the parameters specified above and allocate the
    arrays $L_1,\ldots,L_d$. This takes $\bigO(M)$ time and words
    of space.

  \item Next, we fill the period tables. For $m=1,\ldots,d$ and
    every $z\in[0\dd A_{\rm pow}[m])$, test $p=1,\ldots,m$
    in increasing order until
    \[
      \left\lfloor\frac{z}{A_{\rm pow}[p]}\right\rfloor
        =z\bmod A_{\rm pow}[m-p].
    \]
    Set $L_m[z]:=p$ for the first value that satisfies this equality.
    Such a value always exists, since both sides are zero for $p=m$.
  \end{enumerate}

  To prove correctness, consider any $X\in\IntegerAlphabet^m$ with
  $m\in[1\dd d]$, and denote $z=\Val{\AlphabetSize}{X}$.
  For every $p\in[1\dd m)$, the base-$\AlphabetSize$ representations
  of $\lfloor z/\AlphabetSize^p\rfloor$ and
  $z\bmod\AlphabetSize^{m-p}$, padded to length $m-p$, are
  respectively $X[1\dd m-p]$ and $X[p+1\dd m]$.
  Thus, the tested equality holds exactly when $p$ is a period of
  $X$. Since $m$ is also a period, the first successful candidate
  is $\per{X}$, as required.

  All powers, array indices, and arithmetic operands are at most
  $M\leq N$ and fit in one word. The integer-value structure applies
  because $\AlphabetSize\leq M$ and
  $w\geq1+\log N\geq1+\log M$.
  Filling the tables takes
  $\bigO(\sum_{m=1}^d\AlphabetSize^m m)
    =\bigO(\AlphabetSize^d d)$ time.
  Together with Step~\ref{step:periodicity-testing-prepare}, this gives the claimed construction
  time. The auxiliary array uses $\bigO(d)$ words, giving a total
  space bound of $\bigO(\AlphabetSize^d)$ words.
\end{proof}

\subsection{Packed Run Counting}
\label{sec:tools-packed-run-counting}

\begin{theorem}\label{th:packed-run-counting-structure}
  Let $N,\AlphabetSize\in\Z_{\geq2}$ satisfy $\AlphabetSize\leq N$.
  In the word RAM model with word size $w\geq1+\log N$, given $N$ and
  $\AlphabetSize$, we can in $\bigO(\sqrt N)$ time construct a data structure of size
  $\bigO(\sqrt N)$ words that, given $|S|$ and the packed representation
  $\PackedRepresentation{w}{\AlphabetSize}{S}$
  (\cref{def:packed-representation}) of any string
  $S\in\IntegerAlphabet^{\leq N}$, computes $|\RL{S}|$ in
  $\bigO(1+|S|/\log_{\AlphabetSize}N)$ time.
\end{theorem}
\begin{proof}
  Compute $b=\lceil\log\AlphabetSize\rceil$ and
  $h=\lfloor\frac14\log_{\AlphabetSize}N\rfloor$ in $\bigO(\log N)$ time.
  If $h=0$, the structure stores only $N$, $\AlphabetSize$, and $b$.
  A query on the empty string returns zero. For a nonempty string,
  scan its symbols using
  \cref{pr:packed-representation}\eqref{pr:packed-representation-access}
  and count one run for the first symbol and one for every change
  between consecutive symbols. Since $h=0$ implies
  $\log_{\AlphabetSize}N<4$, the query takes
  $\bigO(1+|S|/\log_{\AlphabetSize}N)$ time.
  Assume henceforth that $h\geq1$.
  For every $t\in[1\dd h]$, let $L_t[0\dd\AlphabetSize^t)$ be an
  array satisfying
  \[
    L_t[\Val{\AlphabetSize}{X}]=|\RL{X}|
    \qquad\text{for every }X\in\IntegerAlphabet^t.
  \]
  The entries are well-defined by \cref{ob:val}.

  \DSComponents
  We store $N$, $\AlphabetSize$, $b$, and $h$, together with the
  following components:
  \begin{enumerate}
  \item We use the integer-value data structure from
    \cref{th:val-encoding} with size parameter $N$
    and alphabet size $\AlphabetSize$. It supports strings of
    length at most $h$ and occupies $\bigO(\sqrt N)$ words.
  \item The arrays $L_1,\ldots,L_h$. Together they occupy
    $\bigO(\sum_{t=1}^h\AlphabetSize^t)=\bigO(\AlphabetSize^h)$ words.
  \end{enumerate}
  Since $\AlphabetSize^h\leq N^{1/4}$, the total size
  is $\bigO(\sqrt N)$ words.

  \DSQueries
  For $|S|=0$, return zero. Otherwise, proceed as follows.
  \begin{enumerate}
  \item First, we obtain the packed representations of consecutive
    blocks $S_1,\ldots,S_q$ whose concatenation is $S$.
    Each block has length $h$, except possibly the last,
    which has length in $[1\dd h]$.
    Apply \cref{pr:packed-representation}\eqref{pr:packed-representation-split}
    with these block lengths.
    Here, $q=\lceil|S|/h\rceil$.
  \item Next, we sum the run counts of the blocks.
    For every $j\in[1\dd q]$, query the integer-value structure
    to obtain $v_j=\Val{\AlphabetSize}{S_j}$ in $\bigO(1)$ time.
    Then compute
    \[
      r=\sum_{j=1}^q L_{|S_j|}[v_j].
    \]
  \item Finally, we account for runs that cross block boundaries.
    For every $j\in[1\dd q)$, retrieve $S[jh]$ and $S[jh+1]$
    using \cref{pr:packed-representation}\eqref{pr:packed-representation-access}.
    If these symbols are equal, subtract one from $r$.
    Joining consecutive blocks merges exactly one pair of runs
    when their adjacent symbols agree. Thus, the final value of $r$
    is $|\RL{S}|$. We return $r$.
  \end{enumerate}
  Since $h=\Theta(\log_{\AlphabetSize}N)$, splitting and processing
  the blocks take
  $\bigO(q+|S|b/w)=\bigO(1+|S|/\log_{\AlphabetSize}N)$ time.
  All packed operations use alphabet size $\AlphabetSize$ and
  length bound $N$, as permitted by the assumptions.
  Since the sum of the block counts is at most $|S|\leq N<2^w$,
  the computation uses single-word arithmetic.

  \DSConstruction
  We construct the stored components in two steps.
  \begin{enumerate}
  \item First, we construct the integer-value structure.
    By \cref{th:val-encoding}, this takes $\bigO(\sqrt N)$ time.
  \item We now fill the arrays $L_1,\ldots,L_h$.
    For every $t\in[1\dd h]$ and $v\in[0\dd\AlphabetSize^t)$,
    compute the length-$t$ base-$\AlphabetSize$ representation
    $X$ of $v$, including leading zeros, by repeated division.
    Count its runs by comparing consecutive digits, and set
    $L_t[v]:=|\RL{X}|$.
    Each entry takes $\bigO(t)$ time, yielding a total construction time of
    $\bigO(\sum_{t=1}^h t\AlphabetSize^t)=\bigO(h\AlphabetSize^h)$.
  \end{enumerate}
  Including the computation and storage of the parameters, the total
  construction time is
  \[
    \bigO(\sqrt N+h\AlphabetSize^h)
      =\bigO(\sqrt N+N^{1/4}\log N)
      =\bigO(\sqrt N).
  \]
  The construction uses $\bigO(\sqrt N)$ words of space.
\end{proof}

\begin{theorem}\label{th:packed-run-counting}
  Let $\AlphabetSize,\Textlen\in\Z_{\geq2}$ satisfy
  $\AlphabetSize\leq\Textlen$, and consider the word RAM model
  with word size $w\geq1+\log\Textlen$. Given $\Textlen$, $\AlphabetSize$,
  and the packed representation
  $\PackedRepresentation{w}{\AlphabetSize}{\Text}$ of a string
  $\Text\in\IntegerAlphabet^{\Textlen}$
  (\cref{def:packed-representation}), we can compute $|\RL{\Text}|$
  in $\bigO(\Textlen/\log_{\AlphabetSize}\Textlen)$ time using
  $\bigO(\Textlen/\log_{\AlphabetSize}\Textlen)$ words of space.
\end{theorem}
\begin{proof}
  Construct the data structure from \cref{th:packed-run-counting-structure}
  with size parameter $N=\Textlen$ and alphabet size $\AlphabetSize$,
  and query it on $\Text$.
  Since $\AlphabetSize\geq2$, the total time is
  \[
    \bigO\left(\sqrt{\Textlen}
      +\Textlen/\log_{\AlphabetSize}\Textlen\right)
      =\bigO(\Textlen/\log_{\AlphabetSize}\Textlen).
  \]
  The packed input occupies
  $\bigO(1+\Textlen\log\AlphabetSize/w)$ words, which is within
  the same bound. Together with the construction and query time
  bounds, this also gives the claimed space bound.
\end{proof}

\subsection{Packed String Repetition}
\label{sec:tools-packed-string-repetition}

\begin{proposition}\label{pr:packed-string-repetition}
  Let $N,\AlphabetSize\in\Z_{\geq2}$ satisfy $\AlphabetSize\leq N$,
  and assume that $\AlphabetSize$ is known and that
  $b=\lceil\log\AlphabetSize\rceil$ has been computed.
  Consider the word RAM model with word size $w>\log N$.
  Given the length $|S|$ and packed representation
  $\PackedRepresentation{w}{\AlphabetSize}{S}$ of a nonempty string
  $S\in\IntegerAlphabet^{\leq N}$ (\cref{def:packed-representation}),
  and an integer $m\in[0\dd N]$, we can compute the packed
  representation of the length-$m$ prefix
  \[
    R=S^\infty[1\dd m]\in\IntegerAlphabet^m,
  \]
  where $S^\infty$ denotes the infinite repetition of $S$
  (\cref{sec:basic-definitions}).
  The algorithm takes
  \[
    \bigO\left(1+\log\left(1+\frac{m}{|S|}\right)
      +\frac{m\log\AlphabetSize}{w}\right)
  \]
  time and uses
  $\bigO(1+(|S|+m)\log\AlphabetSize/w)$ words of space,
  including the input and output.
\end{proposition}
\begin{proof}
  If $m\leq|S|$, extract the length-$m$ prefix of $S$ using
  \cref{pr:packed-representation}\eqref{pr:packed-representation-substring},
  within the claimed bounds. Assume henceforth that $m>|S|$.

  Starting with $X=S$ and its supplied packed representation,
  we extend $X$ while $|X|<m$. In each iteration, compute
  $h=\min\{|X|,m-|X|\}$ and extract the packed representation of
  $H=X[1\dd h]$ using
  \cref{pr:packed-representation}\eqref{pr:packed-representation-substring}.
  Then construct the packed representation of $XH$ using
  \cref{pr:packed-representation}\eqref{pr:packed-representation-concat}
  and replace $X$ by $XH$, retaining the new representation.
  Return the resulting representation when $|X|=m$.
  Every iteration except possibly the last doubles $|X|$.
  Thus, at the start of each iteration, $X$ consists of whole copies
  of $S$, and appending $H$ continues this repetition. The returned
  string therefore equals $R$.

  Geometric growth gives $\bigO(1+\log(1+m/|S|))$ iterations
  and $\bigO(m)$ total length of constructed strings, proving
  the time bound. At any point, we keep only a constant number of
  intermediate strings of length at most $m$. Together with the
  input and the working space of the packed operations, these occupy
  $\bigO(1+(|S|+m)\log\AlphabetSize/w)$ words.
\end{proof}

\subsection{Packed Substring Frequencies}
\label{sec:tools-packed-substring-frequencies}

\begin{theorem}[{\cite{sss}}]\label{th:packed-substring-frequencies}
  Let $N,\AlphabetSize\in\Z_{\geq2}$ satisfy $\AlphabetSize\leq N$,
  and let $\ell\in[1\dd\lfloor\frac12\log_{\AlphabetSize}N\rfloor]$.
  Consider the word RAM model with word size $w\geq1+\log N$.
  Given $N$, $\AlphabetSize$, $\ell$, $|\Text|$, and the packed representation
  $\PackedRepresentation{w}{\AlphabetSize}{\Text}$
  (\cref{def:packed-representation}) of any nonempty string
  $\Text\in\IntegerAlphabet^{\leq N}$, we can
  deterministically compute arrays $F_p[0\dd\AlphabetSize^p)$,
  for $p\in[1\dd\ell]$, satisfying
  \[
    F_p[\Val{\AlphabetSize}{X}]=|\OccTwo{X}{\Text}|
    \qquad\text{for every }X\in\IntegerAlphabet^p
  \]
  (\cref{def:val,def:occ}). The algorithm takes
  $\bigO(N/\log_{\AlphabetSize}N)$ time and uses
  $\bigO(N/\log_{\AlphabetSize}N)$ words of space.
\end{theorem}
\begin{proof}
  Our construction uses the method for counting short-substring
  frequencies from~\cite{sss}. We give a detailed construction in the
  notation of this paper, including the implementation details needed for
  the stated bounds. We also adjust the block lengths to support every
  $\ell\leq\lfloor\frac12\log_{\AlphabetSize}N\rfloor$.

  Denote $\Textlen=|\Text|$,
  $t=\lceil\frac14\log_{\AlphabetSize}N\rceil$, and
  $h=\ell+t-1$. We first count occurrences of length-$\ell$
  substrings. To process $t$ consecutive starting positions together,
  we use blocks of length $h$, each containing exactly $t$ substrings
  of length $\ell$. The frequencies for shorter lengths are then
  obtained from those for longer lengths.
  The algorithm proceeds in three steps.
  \begin{enumerate}
  \item First, we prepare the arrays and the integer-value structure.
    Compute $b=\lceil\log\AlphabetSize\rceil$, $t$, and $h$ in
    $\bigO(\log N)$ time. Construct the data structure from
    \cref{th:val-encoding} with size parameter $N$ and alphabet size
    $\AlphabetSize$ in $\bigO(\sqrt N)$ time and words of space.
    Since $h<\frac34\log_{\AlphabetSize}N$, this structure supports
    all strings of length at most $h$.
    Compute the array $A_{\rm pow}[0\dd h]$ with
    $A_{\rm pow}[r]=\AlphabetSize^r$ for every $r\in[0\dd h]$
    by successive multiplication, and initialize
    $F_1,\ldots,F_\ell$ to zeros.
    Whenever we need the integer value of a substring of $\Text$
    below, we extract its packed representation using
    \cref{pr:packed-representation}\eqref{pr:packed-representation-substring}
    and apply the structure from \cref{th:val-encoding}.
    Each such substring has length at most $h$, and both operations
    take $\bigO(1)$ time.

  \item\label{step:substring-frequencies-longest} We now compute $F_\ell$ by grouping consecutive starting
    positions into groups of size $t$.
    If $\Textlen<\ell$, all entries of $F_\ell$ remain zero,
    and we proceed to the next step.
    Otherwise, denote $q=\lfloor(\Textlen-\ell+1)/t\rfloor$ and
    proceed as follows.
    \begin{enumerate}
    \item First, we count the length-$h$ blocks starting at positions
      $it+1$ for $i\in[0\dd q)$.
      Let $H[0\dd\AlphabetSize^h)$ denote the array in which
      $H[x]$ counts the indices $i\in[0\dd q)$ with
      $\Val{\AlphabetSize}{\Text[it+1\dd it+h]}=x$.
      To compute $H$, initialize it to zeros and, for every
      $i\in[0\dd q)$, compute
      $x=\Val{\AlphabetSize}{\Text[it+1\dd it+h]}$ and increase
      $H[x]$ by one. For every $i\in[0\dd q)$, the block ends at
      $it+h\leq qt+\ell-1\leq\Textlen$.

    \item Next, we count the length-$\ell$ substrings within these
      blocks, using the counts in $H$.
      For every $x\in[0\dd\AlphabetSize^h)$ with $H[x]>0$ and
      every $j\in[0\dd t)$, compute
      \[
        y=\left\lfloor\frac{x}{A_{\rm pow}[t-1-j]}\right\rfloor
          \bmod A_{\rm pow}[\ell]
      \]
      and increase $F_\ell[y]$ by $H[x]$.
      To see what this counts, let $X\in\IntegerAlphabet^h$ be
      the unique string satisfying $\Val{\AlphabetSize}{X}=x$.
      We have $y=\Val{\AlphabetSize}{X[j+1\dd j+\ell]}$.
      Thus, each block $\Text[it+1\dd it+h]$ contributes one
      occurrence for each starting position in $[it+1\dd(i+1)t]$.
      The intervals partition $[1\dd qt]$.

    \item Finally, we count the substrings starting at the remaining
      positions. For every $j\in[qt+1\dd\Textlen-\ell+1]$,
      compute $y=\Val{\AlphabetSize}{\Text[j\dd j+\ell)}$ and
      increase $F_\ell[y]$ by one.
      There are fewer than $t$ such positions.
    \end{enumerate}
    Together, these computations count every occurrence of every
    length-$\ell$ string exactly once.

  \item\label{step:substring-frequencies-shorter} Finally, we compute the frequencies for shorter strings.
    For $p=\ell-1,\ldots,1$, first set
    \[
      F_p[v]:=\sum_{a=0}^{\AlphabetSize-1}
        F_{p+1}[\AlphabetSize v+a]
      \qquad\text{for every }v\in[0\dd\AlphabetSize^p).
    \]
    If $p\leq\Textlen$, compute
    $v=\Val{\AlphabetSize}{\Text[\Textlen-p+1\dd\Textlen]}$
    and increase $F_p[v]$ by one.
    Every occurrence of a length-$p$ string extends to a
    length-$(p+1)$ substring, except for the length-$p$ suffix
    of $\Text$. The sum counts all occurrences that extend,
    and the subsequent increment counts this suffix.
    For $p>\Textlen$, all counts remain zero.
    This also proves correctness when $\Textlen<\ell$.
  \end{enumerate}
  Since $\ell\geq1$, we have $\log_{\AlphabetSize}N\geq2$ and
  $t=\Theta(\log_{\AlphabetSize}N)$.
  Step~\ref{step:substring-frequencies-longest} takes $\bigO(\AlphabetSize^h+\Textlen/t+t\AlphabetSize^h+t)$
  time, while Step~\ref{step:substring-frequencies-shorter} takes
  $\bigO(\sum_{p=1}^{\ell-1}\AlphabetSize^{p+1}+\ell)
    =\bigO(\AlphabetSize^\ell+\ell)$ time.
  Including preprocessing, the total time is
  \[
    \bigO(\sqrt N+N/t+t\AlphabetSize^h)
      =\bigO(N/\log_{\AlphabetSize}N).
  \]
  Here, $\AlphabetSize^h\leq N^{3/4}$ and
  $t=\bigO(\log N)$, and the last bound follows from
  $(\log N)^2=\bigO(N^{1/4})$.
  The arrays $H$, $F_1,\ldots,F_\ell$, and $A_{\rm pow}$,
  together with the integer-value structure, occupy
  $\bigO(\AlphabetSize^h+\AlphabetSize^\ell+h+\sqrt N)
    =\bigO(N/\log_{\AlphabetSize}N)$ words.
  The packed input fits within the same bound.
  All powers, indices, and frequency counts fit in one word.
  The packed operations use alphabet size $\AlphabetSize$ and
  length bound $N$, as permitted by the assumptions.
\end{proof}

\subsection{Packed String Sorting}\label{sec:tools-packed-string-sorting}

\begin{proposition}\label{pr:packed-string-sorting}
  Let $N,k,\AlphabetSize\in\Z_{\geq2}$ and $\ell\in[0\dd N]$
  satisfy $k,\AlphabetSize\leq N$.
  Consider the word RAM model with word size $w\geq1+\log N$.
  Given $N$, $\AlphabetSize$, and a sequence $X_1,\ldots,X_k$ of strings in
  $\IntegerAlphabet^\ell$, supplied by their common length $\ell$ and
  packed representations $\PackedRepresentation{w}{\AlphabetSize}{X_i}$
  (\cref{def:packed-representation}), we can deterministically
  compute arrays $\pi[1\dd k]$ and $R[1\dd k]$ satisfying:
  \begin{enumerate}
  \item\label{pr:packed-string-sorting-order}
    The sequence $(\pi[1],\ldots,\pi[k])$ is a permutation of
    $[1\dd k]$, and
    $X_{\pi[1]}\preceq X_{\pi[2]}\preceq\cdots\preceq X_{\pi[k]}$.
  \item\label{pr:packed-string-sorting-stability}
    For every $i,j\in[1\dd k]$ with $i<j$, if
    $X_{\pi[i]}=X_{\pi[j]}$, then $\pi[i]<\pi[j]$.
  \item\label{pr:packed-string-sorting-ranks}
    For every $i\in[1\dd k]$, it holds
    $R[i]=|\{X_j:j\in[1\dd k],\ X_j\prec X_i\}|$, i.e.,
    $R[i]$ is the number of distinct input strings
    lexicographically smaller than $X_i$.
  \end{enumerate}
  The algorithm takes
  $\bigO(k(1+\ell\log\AlphabetSize/\log k))$ time and uses
  $\bigO(k(1+\ell\log\AlphabetSize/\log k))$ words of space,
  including all preprocessing.
  In particular, if $\ell=\bigO(\log_{\AlphabetSize}k)$, the
  algorithm takes $\bigO(k)$ time and uses $\bigO(k)$ words of space.
\end{proposition}
\begin{proof}
  If $\ell=0$, return $\pi[i]=i$ and $R[i]=0$ for every
  $i\in[1\dd k]$ in $\bigO(k)$ time. Assume henceforth that $\ell\geq1$.
  Compute $b=\lceil\log\AlphabetSize\rceil$ in
  $\bigO(\log\AlphabetSize)$ time.
  All packed operations below use \cref{pr:packed-representation}
  with alphabet size $\AlphabetSize$ and length bound $N$.
  The algorithm proceeds as follows.
  \begin{enumerate}

  \item First, we compute the sorting permutation $\pi$ by stable radix sort.
    Initialize $\pi[i]=i$ for every $i\in[1\dd k]$.
    We distinguish two cases.
    \begin{enumerate}
    \item If $\AlphabetSize\leq k$, we sort blocks of
      $h=\lfloor\log_{\AlphabetSize}k\rfloor\geq1$ symbols at a time.
      Construct the data structure from \cref{th:val-encoding}
      with size parameter $k$ and alphabet size $\AlphabetSize$,
      in $\bigO(\sqrt k)$ time.
      Let $t=\lceil\ell/h\rceil$ and, for each $i\in[1\dd k]$,
      let $X_{i,1},\ldots,X_{i,t}$ be the consecutive blocks of $X_i$,
      each of length $h$ except possibly the last.
      For $j=t,t-1,\ldots,1$, extract the packed representation of
      $X_{i,j}$ for every $i\in[1\dd k]$ and compute
      $a_i=\Val{\AlphabetSize}{X_{i,j}}$ using the data structure.
      Then apply stable counting sort to the indices in $\pi$,
      using $a_i$ for index $i$.
      For a fixed $j$, all blocks have the same length. Hence,
      their integer values preserve their order by \cref{ob:val}.
      Moreover, $a_i<\AlphabetSize^h\leq k$.
      Extracting and converting a block takes $\bigO(1)$ time,
      and each counting-sort pass takes $\bigO(k)$ time.
      Since $h=\Theta(\log_{\AlphabetSize}k)$, the total time is
      $\bigO(k(1+\ell\log\AlphabetSize/\log k))$.
    \item If $\AlphabetSize>k$, we sort one symbol at a time,
      representing it in digits of $d=\lfloor\log k\rfloor$ bits.
      For $j=\ell,\ell-1,\ldots,1$, retrieve the symbols
      $X_1[j],\ldots,X_k[j]$ and stably sort the indices in $\pi$
      by these symbols using radix sort in base $2^d$.
      Within each symbol, process the digits from least to most
      significant using stable counting sort.
      Each pass takes $\bigO(k)$ time because $2^d\leq k$.
      There are $\lceil b/d\rceil=\bigO(\log\AlphabetSize/\log k)$
      passes per symbol, giving a total time of
      $\bigO(k\ell\log\AlphabetSize/\log k)$.
    \end{enumerate}
    Processing the blocks or symbols from right to left produces
    lexicographic order. Since every pass is stable, equal strings
    retain their original order.

  \item Finally, we assign the ranks in $R$.
    Initialize $r=0$ and $R[\pi[1]]=0$.
    For $i=2,\ldots,k$, compare $X_{\pi[i-1]}$ and $X_{\pi[i]}$
    using \cref{pr:packed-representation}\eqref{pr:packed-representation-compare}.
    If the strings differ, increment $r$ by one.
    Set $R[\pi[i]]=r$.
    Since equal strings are consecutive in $\pi$, the counter
    increases exactly once for each new distinct string,
    giving the required ranks.
    This step takes $\bigO(k(1+\ell\log\AlphabetSize/w))$ time.
  \end{enumerate}
  The parameters $h$ or $d$ can be computed in
  $\bigO(\log k)$ time. Together with
  $w\geq1+\log N\geq1+\log k$, this shows that all steps,
  including preprocessing, take
  $\bigO(k(1+\ell\log\AlphabetSize/\log k))$ total time.
  The input representations use
  $\bigO(k(1+\ell\log\AlphabetSize/w))$ words, and the
  data structure and all auxiliary and output arrays use
  $\bigO(k)$ additional words, proving the claimed space bound.
\end{proof}

\subsection{Sparse Suffix Sorting}\label{sec:tools-sparse-suffix-sorting}

\subsubsection{Nonperiodic Positions}

\begin{proposition}\label{pr:sparse-suffix-sorting-nonperiodic}
  Let $N,\AlphabetSize\in\Z_{\geq2}$ satisfy $\AlphabetSize\leq N$,
  let $\mu\in(0,\tfrac15)$ be a constant, and let
  $\tau\in\Z_{\geq1}$ satisfy $\tau\leq\mu\log_{\AlphabetSize}N$.
  Consider the word RAM model with word size $w\geq1+\log N$.
  Given $N$, $\AlphabetSize$, $\tau$, an integer $q\in[0\dd N]$, the packed
  representation $\PackedRepresentation{w}{\AlphabetSize}{\Text}$
  (\cref{def:packed-representation}) of a text
  $\Text\in\IntegerAlphabet^N$, and an array listing the elements
  of a set $Q\subseteq[1\dd N]\setminus\RTwo{\tau}{\Text}$
  (\cref{def:periodic-positions}) of size $q$, we can
  deterministically compute the sequence $\LexSorted{Q}{\Text}$
  (\cref{def:lex-sorted}).
  The algorithm takes $\bigO(N/\tau+q)$ time and uses
  $\bigO(N/\tau+q)$ words of space.
\end{proposition}
\begin{proof}
  Denote the input array by $A[1\dd q]$.
  If $q\leq1$, return the sequence listed in $A$.
  Denote $h=3\tau-1$, and compute
  $b=\lceil\log\AlphabetSize\rceil$ in
  $\bigO(\log\AlphabetSize)$ time.
  We first order the suffixes by their prefixes of length $h$,
  using the entire suffix when it is shorter.
  For suffixes with equal length-$h$ prefixes, ranks at synchronizing
  positions determine their order.
  The algorithm proceeds in three steps.
  \begin{enumerate}
  \item First, we construct the synchronizing set, its suffix ranks,
    and the supporting data structures.
    We proceed in three steps.
    \begin{enumerate}
    \item\label{step:sparse-sorting-nonperiodic-ranks}
      We begin by computing arrays of synchronizing positions
      and their suffix ranks.
      Using \cref{th:sss-packed-construction} with the same constant
      $\mu$, compute an array $U[1\dd t]$ listing a $\tau$-synchronizing
      set $\SSS$ of $\Text$, where $t=|\SSS|=\bigO(N/\tau)$.
      We now arrange $U$ in ascending numerical order and construct
      an array $R[1\dd t]$ in which $R[i]$ is the rank of
      $\Text[U[i]\dd N]$ among the suffixes starting in $\SSS$,
      with ranks starting at one.
      To this end, compute an array $D[1\dd t]$ containing
      $\LexSorted{\SSS}{\Text}$ using \cref{th:sss-lex-sort}.
      Construct an array $E[1\dd t]$ with $E[i]=(D[i],i)$ for
      every $i\in[1\dd t]$, and sort the pairs in $E$ in ascending
      order of their first coordinates.
      For every $i\in[1\dd t]$, set $U[i]$ and $R[i]$ to the
      first and second coordinates of $E[i]$, respectively.
    \item Next, we construct a bitvector $B[1\dd N]$ indicating
      which positions belong to $\SSS$, together with its rank
      structure.
      To this end, initialize the packed representation of $\zero^N$ and set
      $B[U[i]]=\one$ for every $i\in[1\dd t]$, using the
      initialization and update operations of
      \cref{pr:packed-representation}.
      Construct the rank structure from \cref{th:bitvector-rank}
      for $B$.
    \item Finally, construct the integer-value structure from
      \cref{th:val-encoding}.
    \end{enumerate}
  \item Next, we compute the prefix and suffix-rank information
    used to sort the entries of $A$.
    We do this in two steps.
    \begin{enumerate}
    \item We begin with the short prefixes.
      For every $j\in[1\dd q]$, denote
      \[
        a_j=\min\{h,N-A[j]+1\},\qquad
        X_j=\Text[A[j]\dd A[j]+a_j),\qquad
        v_j=\Val{\AlphabetSize}{X_j\zero^{h-a_j}}.
      \]
      Extract the packed representation of $X_j$ and append the
      packed representation of $\zero^{h-a_j}$ using
      \cref{pr:packed-representation}.
      Compute $v_j$ using the integer-value structure.
    \item\label{step:sparse-sorting-nonperiodic-sync}
      Next, we compute, for every $j\in[1\dd q]$, a rank $r_j$ used
      to order suffixes with equal prefixes of length $h$.
      For each $j$ with $a_j<h$, set $r_j=0$.
      For every remaining $j$, we compute the first position
      $s_j\in\SSS$ at or after $A[j]$, together with the rank $r_j$
      of $\Text[s_j\dd N]$ among the suffixes starting in $\SSS$,
      as follows. Query the rank structure to compute
      \[
        c_j=\Rank{B}{A[j]-1}{\one}+1,
      \]
      and set $s_j=U[c_j]$ and $r_j=R[c_j]$.
    \end{enumerate}
  \item\label{step:sparse-sorting-nonperiodic-sort}
    Finally, we sort the positions in $A$ by their suffixes.
    Initialize an array $\pi[1\dd q]$ by setting $\pi[j]=j$
    for every $j\in[1\dd q]$.
    Stably sort the indices in $\pi$ first by $r_j$, then by
    $a_j$, and finally by $v_j$, all in ascending order.
    Return the sequence $(A[\pi[1]],\ldots,A[\pi[q]])$.
  \end{enumerate}

  To prove correctness, we first verify the values computed in
  Step~\ref{step:sparse-sorting-nonperiodic-sync}. Since $U$ lists $\SSS$ in numerical order
  after Step~\ref{step:sparse-sorting-nonperiodic-ranks}, we have
  \[
    \Rank{B}{U[i]}{\one}=i
      \qquad\text{for every }i\in[1\dd t].
  \]
  Consider $j$ with $a_j=h$. The rank query counts the synchronizing
  positions smaller than $A[j]$. Thus, $s_j$ is the first
  synchronizing position at or after $A[j]$.
  Since $A[j]\notin\RTwo{\tau}{\Text}$, the density condition
  in \cref{def:sss} guarantees $s_j\in[A[j]\dd A[j]+\tau)$.
  Moreover, since the sorts in Step~\ref{step:sparse-sorting-nonperiodic-sort} are stable,
  $v_j$ determines the resulting order first, with ties resolved
  by $a_j$ and then by $r_j$.

  Next, observe that comparing $(v_j,a_j)$
  lexicographically gives the same order as comparing $X_j$.
  A mismatch within both prefixes determines the order of their
  padded integer values. If one prefix is shorter and is a proper
  prefix of the other, its padded value is smaller, or the values
  are equal and its length is smaller.
  Distinct prefixes therefore determine the order of the full
  suffixes, since every prefix shorter than $h$ is already an
  entire suffix. Equal prefixes shorter than $h$ are equal suffixes.

  It remains to compare indices $j,j'$ with $a_j=a_{j'}=h$ and
  $X_j=X_{j'}$. Denote $i=A[j]$ and $i'=A[j']$.
  For every $d\in[0\dd\tau)$, equality of these prefixes gives
  \[
    \Text[i+d\dd i+d+2\tau)=\Text[i'+d\dd i'+d+2\tau).
  \]
  By consistency of $\SSS$, we have $i+d\in\SSS$ if and only if
  $i'+d\in\SSS$. Density guarantees a position of $\SSS$ in each
  of the intervals $[i\dd i+\tau)$ and $[i'\dd i'+\tau)$.
  Consequently, their first synchronizing positions satisfy
  $s_j-i=s_{j'}-i'$.
  The prefixes $\Text[i\dd s_j)$ and $\Text[i'\dd s_{j'})$ agree,
  and the order of the full suffixes is determined by the ranks
  $R[c_j]=r_j$ and $R[c_{j'}]=r_{j'}$ of the suffixes starting
  at $s_j$ and $s_{j'}$, respectively.
  This proves that the returned sequence is $\LexSorted{Q}{\Text}$.

  All basic packed operations use length bound $N$, with alphabet
  size two for $B$ and $\AlphabetSize$ for the text prefixes.
  Every other result applied in this proof uses size parameter $N$,
  and all of them except \cref{th:bitvector-rank} use alphabet size
  $\AlphabetSize$.
  Since $h=3\tau-1<3\mu\log_{\AlphabetSize}N<\log_{\AlphabetSize}N$,
  we have $h\leq\lfloor\log_{\AlphabetSize}N\rfloor$. Each prefix
  extraction, concatenation, and integer-value query takes
  $\bigO(1)$ time.
  Constructing $\SSS$ and sorting its suffixes take
  $\bigO(N/\tau)$ time, while the integer-value structure takes
  $\bigO(\sqrt N)$ time and words of space.
  Initializing $B$ and setting its $t$ ones take
  $\bigO(t+N/\log N)$ time. Constructing its rank structure takes
  $\bigO(N/\log N)$ time and words, and all rank queries together
  take $\bigO(q)$ time.
  All integers used in sorting are at most $N$.
  Using digits of $\lfloor\frac12\log N\rfloor$ bits, every
  integer sort takes a constant number of counting-sort passes.
  Sorting $E$ and $\pi$ therefore takes
  $\bigO(t+q+\sqrt N)$ total time.
  The parameter computations take $\bigO(\log N)$ time.
  Together with $\tau\leq\mu\log_{\AlphabetSize}N$, these
  bounds give $\bigO(N/\tau+q)$ total time.
  The input, structures, and auxiliary arrays occupy
  $\bigO(N/\tau+q)$ words.
\end{proof}

\subsubsection{Periodic Positions}

\begin{proposition}\label{pr:sparse-suffix-sorting-periodic}
  Let $N,\AlphabetSize\in\Z_{\geq2}$ satisfy $\AlphabetSize\leq N$,
  let $\mu\in(0,\tfrac15)$ be a constant, and let
  $\tau\in\Z_{\geq1}$ satisfy $\tau\leq\mu\log_{\AlphabetSize}N$.
  Consider the word RAM model with word size $w\geq1+\log N$.
  Given $N$, $\AlphabetSize$, $\tau$, an integer $q\in[0\dd N]$, the packed
  representation $\PackedRepresentation{w}{\AlphabetSize}{\Text}$
  (\cref{def:packed-representation}) of a text
  $\Text\in\IntegerAlphabet^N$, and an array listing the elements
  of a set $Q\subseteq\RTwo{\tau}{\Text}$
  (\cref{def:periodic-positions}) of size $q$, we can
  deterministically compute the sequence $\LexSorted{Q}{\Text}$
  (\cref{def:lex-sorted}).
  The algorithm takes $\bigO(N/\tau+q)$ time and uses
  $\bigO(N/\tau+q)$ words of space.
\end{proposition}
\begin{proof}
  Denote the input array by $A[1\dd q]$.
  If $q\leq1$, return the sequence listed in $A$.
  Denote $h=3\tau-1$, and compute
  $b=\lceil\log\AlphabetSize\rceil$ in
  $\bigO(\log\AlphabetSize)$ time.
  Suffixes with equal length-$h$ prefixes have the same infinite
  periodic continuation. We compare these suffixes by where they
  end or first differ from this continuation, using suffix ranks
  to resolve the remaining ties.
  The algorithm proceeds in four steps.
  \begin{enumerate}
  \item First, we construct the synchronizing set and the structures
    used to compare the suffixes.
    We do this in three steps.
    \begin{enumerate}
    \item\label{step:sparse-sorting-periodic-ranks}
      We first compute arrays of synchronizing positions and
      their suffix ranks.
      Using \cref{th:sss-packed-construction} with the same constant
      $\mu$, compute an array $U[1\dd t]$ listing a $\tau$-synchronizing
      set $\SSS$ of $\Text$, where $t=|\SSS|=\bigO(N/\tau)$.
      We now arrange $U$ in ascending numerical order and construct
      an array $R[1\dd t]$ in which $R[i]$ is the rank of
      $\Text[U[i]\dd N]$ among the suffixes starting in $\SSS$,
      with ranks starting at one.
      To this end, compute an array $D[1\dd t]$ containing
      $\LexSorted{\SSS}{\Text}$ using \cref{th:sss-lex-sort}.
      Construct an array $E[1\dd t]$ with $E[i]=(D[i],i)$ for
      every $i\in[1\dd t]$, and sort the pairs in $E$ in ascending
      order of their first coordinates.
      For every $i\in[1\dd t]$, set $U[i]$ and $R[i]$ to the
      first and second coordinates of $E[i]$, respectively.
    \item Next, we construct a bitvector $B[1\dd N]$ indicating
      membership in $\SSS$ and a structure answering rank queries
      on $B$, as follows.
      Initialize the packed representation of $\zero^N$ and set
      $B[U[i]]=\one$ for every $i\in[1\dd t]$, using the
      initialization and update operations of
      \cref{pr:packed-representation}.
      Construct the rank structure from \cref{th:bitvector-rank}
      for $B$.
    \item We then construct the LCE structure from \cref{th:lce} for $\Text$,
      the integer-value structure from \cref{th:val-encoding},
      and the shortest-period structure from
      \cref{pr:periods-of-short-strings} with length bound $h$.
    \end{enumerate}
  \item\label{step:sparse-sorting-periodic-runs}
    Next, we determine where the initial periodic portion of
    each suffix ends.
    For every $j\in[1\dd q]$, denote
    $X_j=\Text[A[j]\dd A[j]+h)$.
    Extract its packed representation and compute
    $v_j=\Val{\AlphabetSize}{X_j}$ and $p_j=\per{X_j}$ using
    the integer-value and shortest-period structures, respectively.
    Compute also
    \[
      e_j=A[j]+p_j+\LCE{\Text}{A[j]}{A[j]+p_j},
      \qquad \lambda_j=e_j-A[j].
    \]
  \item\label{step:sparse-sorting-periodic-sync}
    We now compute the suffix ranks needed when the periodic
    prefixes have equal lengths.
    For every $j\in[1\dd q]$, proceed as follows.
    \begin{enumerate}
    \item If $e_j=N+1$, set $r_j=0$.
    \item Otherwise, query the rank structure to compute
      \[
        c_j=\Rank{B}{e_j-2\tau+1}{\one},
      \]
      and set $r_j=R[c_j]$.
    \end{enumerate}
  \item\label{step:sparse-sorting-periodic-sort}
    Finally, we sort the positions in $A$ by their suffixes.
    We use stable integer sorting in the following order.
    \begin{enumerate}
    \item First, stably sort the sequence $1,\ldots,q$ in
      ascending order of $r_j$.
      Scan the resulting sequence and append each index $j$
      to one of two initially empty arrays $I_-$ and $I_+$:
      append it to $I_-$ if $e_j=N+1$ or
      $\Text[e_j]\prec\Text[e_j-p_j]$, and to $I_+$ otherwise.
      Denote their lengths by $q_-$ and $q_+$, respectively.
    \item\label{step:sparse-sorting-periodic-lengths}
      Next, stably sort $I_-[1\dd q_-]$ in ascending order
      of $\lambda_j$ and $I_+[1\dd q_+]$ in descending order
      of $\lambda_j$.
    \item Concatenate $I_-$ followed by $I_+$, and stably sort
      the resulting sequence in ascending order of $v_j$.
      Store this sequence in $\pi[1\dd q]$.
      Return the sequence $(A[\pi[1]],\ldots,A[\pi[q]])$.
    \end{enumerate}
  \end{enumerate}

  To prove correctness, we first verify the values computed in
  Steps~\ref{step:sparse-sorting-periodic-runs} and~\ref{step:sparse-sorting-periodic-sync}.
  By the definition of $e_j$, the fragment $\Text[A[j]\dd e_j)$ is
  the longest prefix of the suffix with period $p_j$, and its length
  is $\lambda_j$. Since $U$ lists $\SSS$ in numerical order after
  Step~\ref{step:sparse-sorting-periodic-ranks}, the query $\Rank{B}{U[i]}{\one}$ returns $i$,
  the index of $U[i]$ in the array $U$, for every $i\in[1\dd t]$.
  Consider $j$ with $e_j\leq N$. By \cite[Fact~3.2]{sss}, the
  longest prefix of $\Text[A[j]\dd N]$ with period $p_j$ ends at
  position $s+2\tau-2$, where $s$ is the smallest element of
  $\SSS\cup\{N-2\tau+2\}$ that is at least $A[j]$. Thus,
  $s=e_j-2\tau+1\leq N-2\tau+1$, and hence $s$ is not the sentinel
  $N-2\tau+2$ and belongs to $\SSS$. Consequently,
  $U[c_j]=e_j-2\tau+1$, and $r_j$ is the rank of the suffix starting
  at this position. Moreover, for $e_j\leq N$, the symbol
  $\Text[e_j-p_j]$ is the next symbol in the periodic continuation of
  $X_j$ and differs from $\Text[e_j]$ by the definition of $e_j$.
  Since the sorts in Step~\ref{step:sparse-sorting-periodic-sort} are stable, equal lengths
  in $I_-$ and $I_+$ remain ordered by $r_j$. Within each group of
  equal prefixes $X_j$, the indices from $I_-$ precede those from
  $I_+$, with their orders from Step~\ref{step:sparse-sorting-periodic-lengths} preserved.

  Next, we compare two indices $j,j'$.
  If $X_j\neq X_{j'}$, then $v_j$ and $v_{j'}$ determine the
  order of the full suffixes.
  Otherwise, $p_j=p_{j'}$, and the suffixes agree with the same
  infinite periodic continuation for $\lambda_j$ and
  $\lambda_{j'}$ symbols, respectively.
  Each suffix then ends or differs from the continuation at the
  next symbol. We distinguish three cases.
  \begin{itemize}
  \item If one index belongs to $I_-$ and the other to $I_+$,
    the suffix whose index belongs to $I_-$ is smaller.
    Indeed, at the first position where either suffix ceases
    to agree with the continuation, the former suffix ends or
    has a smaller symbol, or the latter has a larger symbol.
  \item If both indices belong to $I_-$ and
    $\lambda_j\neq\lambda_{j'}$, the suffix with the smaller
    value of $\lambda$ ends or has a smaller symbol while the
    other still agrees with the continuation.
    If both belong to $I_+$, the suffix with the smaller value
    of $\lambda$ has a larger symbol there.
    Thus, ascending and descending order of $\lambda$ give
    the required orders within $I_-$ and $I_+$, respectively.
  \item Suppose that the indices belong to the same array and
    $\lambda_j=\lambda_{j'}$.
    If neither suffix ends after its periodic prefix, the positions
    $e_j-2\tau+1$ and $e_{j'}-2\tau+1$ have equal offsets
    $\lambda_j-2\tau+1$ from their respective starting positions.
    The prefixes before these positions agree, and the ranks
    $r_j$ and $r_{j'}$ determine the order of the full suffixes.
    If exactly one suffix ends after its periodic prefix, it is
    a proper prefix of the other. Its rank is zero, while the
    other rank is positive.
    If both suffixes end there, equality of their lengths gives
    $A[j]=A[j']$.
  \end{itemize}
  These cases justify all comparisons in Step~\ref{step:sparse-sorting-periodic-sort} and prove
  that the returned sequence is $\LexSorted{Q}{\Text}$.

  All basic packed operations use length bound $N$, with alphabet
  size two for $B$ and $\AlphabetSize$ for the text prefixes.
  Every other result applied in this proof uses size parameter $N$,
  and all of them except \cref{th:bitvector-rank} use alphabet size
  $\AlphabetSize$.
  Since $h=3\tau-1<3\mu\log_{\AlphabetSize}N<\log_{\AlphabetSize}N$,
  we have $h\leq\lfloor\log_{\AlphabetSize}N\rfloor$. All prefix
  extractions and queries take $\bigO(1)$ time.
  Constructing $\SSS$ and sorting its suffixes take
  $\bigO(N/\tau)$ time. The LCE and integer-value structures
  take $\bigO(N/\log_{\AlphabetSize}N)$ and $\bigO(\sqrt N)$
  time and words, respectively.
  Constructing the packed bitvector $B$ takes $\bigO(t+N/\log N)$
  time, and its rank structure takes $\bigO(N/\log N)$ time
  and words. Computing all $r_j$ then takes $\bigO(q)$ time.
  The shortest-period structure takes
  \[
    \bigO(\AlphabetSize^h h)
      \subseteq\bigO(N^{3\mu}\log N)
      \subseteq\bigO(N/\tau)
  \]
  time and at most this many words, since $3\mu<1$ and
  $\tau=\bigO(\log N)$.
  All integers used in sorting are at most $N$ and fit in one word.
  As in \cref{pr:sparse-suffix-sorting-nonperiodic}, radix sorting
  with digits of $\lfloor\frac12\log N\rfloor$ bits takes
  $\bigO(t+q+\sqrt N)$ total time for all the arrays.
  The scans and parameter computations fit within the same bound.
  This proves the claimed $\bigO(N/\tau+q)$ time and space bounds,
  including the input.
\end{proof}

\subsubsection{Arbitrary Positions}

\begin{theorem}\label{th:sparse-suffix-sorting}
  Let $N,\AlphabetSize\in\Z_{\geq2}$ satisfy $\AlphabetSize\leq N$.
  Consider the word RAM model with word size $w\geq1+\log N$.
  Given $N$, $\AlphabetSize$, an integer $q\in[0\dd N+1]$, the packed representation
  $\PackedRepresentation{w}{\AlphabetSize}{\Text}$
  (\cref{def:packed-representation}) of a text
  $\Text\in\IntegerAlphabet^N$, and an array listing the elements
  of a set $Q\subseteq[1\dd N+1]$ of size $q$, we can
  deterministically compute the sequence $\LexSorted{Q}{\Text}$
  (\cref{def:lex-sorted}).
  The algorithm takes $\bigO(N/\log_{\AlphabetSize}N+q)$ time and
  uses $\bigO(N/\log_{\AlphabetSize}N+q)$ words of space.
\end{theorem}
\begin{proof}
  First suppose that $Q\subseteq[1\dd N]$, and denote the input
  array by $A[1\dd q]$.
  If $q\leq1$, return the sequence listed in $A$.
  Compute $b=\lceil\log\AlphabetSize\rceil$ in
  $\bigO(\log\AlphabetSize)$ time.
  First consider $\log_{\AlphabetSize}N<7$.
  The set $\SSS=[1\dd N-1]$ is a $1$-synchronizing set of $\Text$.
  Construct an array listing $\SSS$ in $\bigO(N)$ time, and compute
  $\LexSorted{\SSS}{\Text}$ from it in $\bigO(N)$ time using
  \cref{th:sss-lex-sort} with $\tau=1$.
  Insert position $N$ before the first position $s$ in this sequence
  with $\Text[s]\succeq\Text[N]$, or append it if there is no such
  position. This is correct because $\Text[N\dd N]$ is a proper prefix
  of every other suffix starting with $\Text[N]$, and every suffix
  starting at a position $s$ with $\Text[s]\neq\Text[N]$ is ordered
  relative to $\Text[N\dd N]$ by its first symbol.
  This gives the order of all nonempty suffixes in $\bigO(N)$ time.
  Store their ranks in an array indexed by starting position.
  Sort the positions in $A$ by these ranks using counting sort,
  and return the resulting sequence $\LexSorted{Q}{\Text}$.
  The time and space are
  $\bigO(N+q)=\bigO(N/\log_{\AlphabetSize}N+q)$.

  Assume henceforth that $\log_{\AlphabetSize}N\geq7$, and denote
  \[
    \tau=\left\lfloor\frac17\log_{\AlphabetSize}N\right\rfloor,
    \qquad h=3\tau-1.
  \]
  We sort the two groups of suffixes and merge their orders as follows.
  \begin{enumerate}
  \item First, we separate the positions in $Q$ according to
    whether they belong to $\RTwo{\tau}{\Text}$.
    Construct the shortest-period structure from
    \cref{pr:periods-of-short-strings} with length bound $h$.
    Initialize two empty arrays $A_n$ and $A_p$.
    For $j=1,\ldots,q$, append $A[j]$ to $A_n$ if
    $N-A[j]+1<h$.
    Otherwise, extract the packed representation of
    $\Text[A[j]\dd A[j]+h)$ and query its shortest period.
    Append $A[j]$ to $A_p$ if this period is at most $\tau/3$,
    and to $A_n$ otherwise.
    Denote the lengths of these arrays by $q_n$ and $q_p$.
    The arrays list the sets
    $Q_n=Q\setminus\RTwo{\tau}{\Text}$ and
    $Q_p=Q\cap\RTwo{\tau}{\Text}$, respectively.
  \item Next, we sort the suffixes in each group.
    Compute arrays $D_n[1\dd q_n]$ and $D_p[1\dd q_p]$ containing
    $\LexSorted{Q_n}{\Text}$ and $\LexSorted{Q_p}{\Text}$, respectively,
    using \cref{pr:sparse-suffix-sorting-nonperiodic} and
    \cref{pr:sparse-suffix-sorting-periodic} with $\mu=1/7$.
  \item Finally, we merge $D_n$ and $D_p$ to obtain
    $\LexSorted{Q}{\Text}$.
    Construct the LCE structure from \cref{th:lce} for $\Text$.
    To compare the suffixes starting at positions $j$ and $j'$
    from the two arrays, compute $\ell=\LCE{\Text}{j}{j'}$.
    If one suffix has length $\ell$, it is the smaller suffix.
    Otherwise, compare $\Text[j+\ell]$ and
    $\Text[j'+\ell]$ using \cref{pr:packed-representation}.
    Merge the two sorted arrays using these constant-time
    comparisons and return the resulting sequence.
  \end{enumerate}

  The first step classifies the positions by the definition of
  $\RTwo{\tau}{\Text}$. The two propositions sort the suffixes
  within their groups, and the LCE comparisons give their order
  across the groups. Since $Q_n$ and $Q_p$ partition $Q$, the
  returned sequence is $\LexSorted{Q}{\Text}$.

  All basic packed operations use alphabet size $\AlphabetSize$
  and length bound $N$.
  Every other result applied in this proof uses size parameter $N$
  and alphabet size $\AlphabetSize$.
  The shortest-period structure takes
  $\bigO(N^{3/7}\log N)=\bigO(N/\log_{\AlphabetSize}N)$
  time and at most this many words, and classifying the positions
  takes $\bigO(q)$ additional time.
  Since $\tau=\Theta(\log_{\AlphabetSize}N)$ and $q_n+q_p=q$,
  the two sorting propositions together take
  $\bigO(N/\log_{\AlphabetSize}N+q)$ time and words.
  Constructing the LCE structure takes
  $\bigO(N/\log_{\AlphabetSize}N)$ time and words, and the merge
  takes $\bigO(q)$ time. These bounds prove the claim in this case.

  If $N+1\in Q$, scan the input array to construct an array listing
  $Q\setminus\{N+1\}$. Apply the algorithm above to this array,
  and prepend $N+1$ to the resulting sequence. This gives
  $\LexSorted{Q}{\Text}$, since the empty suffix is lexicographically
  smaller than every nonempty suffix. The additional work takes
  $\bigO(q)$ time and words, and $N+1<2^w$ under the word-size
  assumption.
\end{proof}

\begin{theorem}\label{th:stable-sparse-suffix-sorting}
  Let $N,\AlphabetSize\in\Z_{\geq2}$ satisfy $\AlphabetSize\leq N$.
  Consider the word RAM model with word size $w\geq1+\log N$.
  Given $N$, $\AlphabetSize$, an integer $q\in\Z_{\geq0}$, the packed representation
  $\PackedRepresentation{w}{\AlphabetSize}{\Text}$
  (\cref{def:packed-representation}) of a text
  $\Text\in\IntegerAlphabet^N$, and an array $A[1\dd q]$ with
  entries in $[1\dd N+1]$, we can deterministically compute arrays
  $\pi[1\dd q]$ and $R[1\dd q]$ satisfying:
  \begin{enumerate}
  \item\label{th:stable-sparse-suffix-sorting-order}
    The sequence $(\pi[1],\ldots,\pi[q])$ is a permutation of
    $[1\dd q]$, and
    \[
      \Text[A[\pi[i]]\dd N]\preceq\Text[A[\pi[i+1]]\dd N]
        \qquad\text{for every }i\in[1\dd q).
    \]
  \item\label{th:stable-sparse-suffix-sorting-stability}
    For every $i,j\in[1\dd q]$ with $i<j$, if
    $A[\pi[i]]=A[\pi[j]]$, then $\pi[i]<\pi[j]$.
  \item\label{th:stable-sparse-suffix-sorting-ranks}
    For every $i\in[1\dd q]$, $R[i]$ is the number of distinct
    suffixes in $\{\Text[A[j]\dd N]:j\in[1\dd q]\}$ that are
    lexicographically smaller than $\Text[A[i]\dd N]$.
  \end{enumerate}
  The entries of $A$ need not be distinct.
  The algorithm takes $\bigO(N/\log_{\AlphabetSize}N+q)$ time and
  uses $\bigO(N/\log_{\AlphabetSize}N+q)$ words of space,
  including all preprocessing.
\end{theorem}
\begin{proof}
  If $q=0$, return two empty arrays. If $q=1$, return
  $\pi[1]=1$ and $R[1]=0$.
  Assume henceforth that $q\geq2$.
  We first sort the suffixes at the distinct positions supplied in $A$.
  Their ranks then allow us to sort the original indices, including
  those corresponding to repeated positions.
  The algorithm proceeds in three steps.
  \begin{enumerate}
  \item First, we list the distinct positions in numerical order and
    determine where each position from $A$ occurs in this list.
    Initialize an array $I[1\dd q]$ with $I[j]=j$ for every
    $j\in[1\dd q]$, and sort the indices $j$ stored in $I$
    in ascending order of $A[j]$ using radix sort.
    Initialize an empty array $U$ and a counter $r=0$.
    For $t=1,\ldots,q$, if $t=1$ or $A[I[t]]\neq A[I[t-1]]$,
    append $A[I[t]]$ to $U$ and increment $r$ by one.
    In either case, set $G[I[t]]=r$ in an array $G[1\dd q]$.
    At the end of the scan, $U[1\dd r]$ lists the set
    $Q=\{A[j]:j\in[1\dd q]\}$ in ascending numerical order,
    and $A[j]=U[G[j]]$ for every $j\in[1\dd q]$.

  \item Next, we compute the suffix ranks in $R$.
    Using \cref{th:sparse-suffix-sorting}, compute an array
    $D[1\dd r]$ containing $\LexSorted{Q}{\Text}$.
    Construct an array $E[1\dd r]$ with $E[t]=(D[t],t-1)$
    for every $t\in[1\dd r]$.
    Sort the pairs in $E$ in ascending order of their first
    coordinates using radix sort.
    The first coordinate of $E[t]$ is now $U[t]$, and its second
    coordinate is the number of suffixes starting in $Q$ that are
    lexicographically smaller than $\Text[U[t]\dd N]$.
    For every $j\in[1\dd q]$, set $R[j]$ to the second
    coordinate of $E[G[j]]$.

  \item Finally, compute the stable sorting permutation.
    Initialize $\pi[j]=j$ for every $j\in[1\dd q]$.
    Sort the indices $j$ in $\pi$ in ascending order of $R[j]$
    using stable counting sort, and return $\pi$ and $R$.
  \end{enumerate}
  Suffixes starting at distinct positions have different lengths and
  are therefore distinct. Consequently, the ranks assigned by the
  second step count exactly the distinct supplied suffixes that are
  lexicographically smaller. Sorting by these ranks gives the required
  suffix order. Repeated positions receive equal ranks, and the stable
  sort preserves the original order of their indices, proving stability.

  The call to \cref{th:sparse-suffix-sorting} uses size parameter $N$ and
  alphabet size $\AlphabetSize$ and takes
  $\bigO(N/\log_{\AlphabetSize}N+r)$ time and words.
  Both radix sorts use keys in $[1\dd N+1]$.
  With digits of $\lceil\frac12\log N\rceil$ bits, each sort
  uses a constant number of counting-sort passes and takes
  $\bigO(q+\sqrt N)$ time and words.
  All ranks belong to $[0\dd r)$, and hence the final counting sort
  takes $\bigO(q+r)$ time and words.
  The scans and array initialization take $\bigO(q+r)$ time.
  Computing the digit length takes $\bigO(\log N)$ time.
  Since $r\leq q$ and
  $\sqrt N+\log N=\bigO(N/\log_{\AlphabetSize}N)$,
  the total time and space satisfy the claimed bounds.
  Positions, ranks, and the counter $r$ are at most $N+1$ and fit in
  one word because $w\geq1+\log N$. Indices in $[1\dd q]$ fit in one
  word because the input array $A[1\dd q]$ is stored in memory, which
  the word RAM model assumes to be addressable (\cref{sec:model}).
  The space bound includes the input and output.
\end{proof}

\section{Sparse Anchors for Pattern Matching}
\label{sec:sparse-anchors}

\subsection{Preliminaries}
\label{sec:sparse-anchor-preliminaries}

\begin{lemma}\label{lm:pat-occ}
  Let $\Textlen,m\in\Z_{\geq1}$ and $\Text \in \Sigma^{\Textlen}$.
  Consider a pattern $\Pat \in \Sigma^{m}$ such that no nonempty suffix
  of $\Text$ is a proper prefix of $\Pat$. Let $j \in \Z$ be such that
  $\Pat = \Textinf[j \dd j + m)$, $j \leq \Textlen$, and $j + m > 1$.
  Then, $j \in \OccTwo{\Pat}{\Text}$.
\end{lemma}
\begin{proof}
  First, we prove that $j \geq 1$. Suppose that $j < 1$.
  Let $\ell = 1 + ((-j) \bmod \Textlen)$. Then
  $1 \leq \ell \leq \Textlen$, $\ell \leq 1-j < m$, and
  $j+\ell \equiv 1 \pmod{\Textlen}$. Consequently,
  $\Pat[1 \dd \ell] = \Textinf[j \dd j+\ell)
  = \Text[\Textlen-\ell+1 \dd \Textlen]$.
  Thus, a nonempty suffix of $\Text$ is a proper prefix of $\Pat$,
  contrary to the assumption.

  Next, we prove that $j+m\leq\Textlen+1$. Suppose otherwise,
  and denote $\ell=\Textlen+1-j$. Since $1\leq j\leq\Textlen$
  and $j+m>\Textlen+1$, we have $0<\ell<m$. Moreover,
  $\Pat[1\dd\ell]=\Text[j\dd\Textlen]$.
  Thus, a nonempty suffix of $\Text$ is a proper prefix of $\Pat$,
  again contradicting the assumption.

  By the above, $\Textinf[j \dd j + m) = \Pat$, $j \geq 1$, and $j + m \leq \Textlen + 1$. Thus,
  $j \in \OccTwo{\Pat}{\Text}$.
\end{proof}

\subsection{Nonperiodic Patterns}
\label{sec:sparse-anchors-nonperiodic}

\subsubsection{Preliminaries}
\label{sec:sparse-anchors-nonperiodic-definitions}

\begin{definition}[Successor]\label{def:succ}
  Consider any totally ordered set $\mathcal{U}$, and let $A \subseteq \mathcal{U}$ be a finite nonempty subset of $\mathcal{U}$.
  For every $x \in \mathcal{U}$ satisfying $x \leq \max A$,
  we denote $\Successor{A}{x} = \min\{x' \in A : x \leq x'\}$.
\end{definition}

\begin{definition}[Distinguishing prefix of a suffix~\cite{sss,breaking}]\label{def:dist-prefixes}
  Let $\Textlen\in\Z_{\geq2}$ and $\Text \in \Sigma^{\Textlen}$.
  For $\tau \in [1 \dd \lfloor \tfrac{\Textlen}{2} \rfloor]$, let
  $\SSS$ be a $\tau$-synchronizing set of $\Text$ (\cref{def:sss}).
  For every $j \in [1 \dd \Textlen - 3\tau + 2] \setminus \RTwo{\tau}{\Text}$,
  we denote (see \cref{def:periodic-positions,def:succ})
  \[
    \DistPrefixPos{j}{\tau}{\Text}{\SSS} := \Text[j \dd \Successor{\SSS}{j} + 2\tau).
  \]
  We then let
  \[
    \DistPrefixes{\tau}{\Text}{\SSS}
      := \{\DistPrefixPos{j}{\tau}{\Text}{\SSS} : j \in [1 \dd \Textlen -
          3\tau + 2] \setminus \RTwo{\tau}{\Text}\}.
  \]
\end{definition}

\begin{remark}\label{rm:dist-prefixes}
  Note that $\Successor{\SSS}{j}$ in \cref{def:dist-prefixes}
  is well-defined for every $j \in [1 \dd \Textlen - 3\tau + 2]
  \setminus \RTwo{\tau}{\Text}$, because by
  \cref{def:sss}\eqref{def:sss-density}, for such $j$
  we have $[j \dd j + \tau) \cap \SSS \neq \emptyset$.
\end{remark}

\begin{example}\label{ex:distinguishing-prefixes}
  Let $\Text=\texttt{a}^{10}\texttt{b}\texttt{a}^{10}$, $\tau=3$, and
  $\SSS=\{6,9,11\}$, as in \cref{ex:synchronizing-set}.
  We have $\Successor{\SSS}{5}=6$, while
  $\Successor{\SSS}{6}=6$ because $6\in\SSS$.
  The corresponding distinguishing prefixes are
  \[
    \DistPrefixPos{5}{\tau}{\Text}{\SSS}
      =\Text[5\dd12)=\texttt{a}^6\texttt{b},
    \qquad
    \DistPrefixPos{6}{\tau}{\Text}{\SSS}
      =\Text[6\dd12)=\texttt{a}^5\texttt{b}.
  \]
  Both belong to $\DistPrefixes{\tau}{\Text}{\SSS}$.
\end{example}

\begin{lemma}[{\cite{sss,breaking}; see also~\cite{sublinearlz}}]\label{lm:dist-prefixes}
  Let $\Textlen\in\Z_{\geq2}$ and $\Text \in \Sigma^{\Textlen}$.
  For $\tau \in [1 \dd \lfloor \tfrac{\Textlen}{2} \rfloor]$,
  let $\SSS$ be a $\tau$-synchronizing set of $\Text$ (\cref{def:sss}). Then:
  \begin{enumerate}
  \item\label{lm:dist-prefixes-it-1}
    It holds $\DistPrefixes{\tau}{\Text}{\SSS} \subseteq \Sigma^{\leq 3\tau - 1}$
    (\cref{def:dist-prefixes}).
  \item\label{lm:dist-prefixes-it-2}
    $\DistPrefixes{\tau}{\Text}{\SSS}$ is prefix-free, i.e.,
    for $D, D' \in \DistPrefixes{\tau}{\Text}{\SSS}$,
    $D \neq D'$ implies that $D$ is not a prefix~of~$D'$.
  \end{enumerate}
\end{lemma}

\begin{lemma}[{\cite{breaking}; see also~\cite{sublinearlz}}]\label{lm:dist-prefix-existence}
  Let $\Textlen\in\Z_{\geq2}$ and $\Text \in \Sigma^{\Textlen}$.
  For $\tau \in [1 \dd \lfloor \tfrac{\Textlen}{2} \rfloor]$, let
  $\SSS$ be a $\tau$-synchronizing set of $\Text$ (\cref{def:sss}).
  Let $m\in\Z_{\geq1}$ and $\Pat \in \Sigma^{m}$ be a $\tau$-nonperiodic pattern (\cref{def:periodic-pattern}) such
  that $m \geq 3\tau - 1$ and $\OccTwo{\Pat}{\Text} \neq \emptyset$.
  Then, there exists a unique
  $D \in \DistPrefixes{\tau}{\Text}{\SSS}$ (\cref{def:dist-prefixes})
  that is a prefix of $\Pat$.
\end{lemma}

\begin{remark}\label{rm:dist-prefixes-arbitrary-text}
  Although~\cite{sublinearlz} assumes that the final symbol of $\Text$
  occurs only once, its proofs of the claims in
  \cref{lm:dist-prefixes,lm:dist-prefix-existence} do not use this assumption.
  These lemmas therefore apply to arbitrary texts.
\end{remark}

\subsubsection{Basic Navigation}
\label{sec:sparse-anchors-nonperiodic-navigation}

\begin{proposition}[{Based on~\cite{breaking}}]\label{pr:nonperiodic-dist-pref}
  Let $N,\AlphabetSize\in\Z_{\geq2}$ satisfy $\AlphabetSize\leq N$.
  Consider the word RAM model with word size $w\geq1+\log N$.
  Given $N$, $\AlphabetSize$, an integer $\tau\geq1$ satisfying
  $\tau\leq\tfrac16\log_{\AlphabetSize}N$, the packed representation
  $\PackedRepresentation{w}{\AlphabetSize}{\Text}$
  (\cref{def:packed-representation}) of a text
  $\Text\in\IntegerAlphabet^N$, and a
  $\tau$-synchronizing set $\SSS$ of $\Text$ (\cref{def:sss}) specified by its
  size and an array listing its elements,
  we can in $\bigO(N/\log_{\AlphabetSize}N+|\SSS|)$ time
  and using $\bigO(N/\log_{\AlphabetSize}N+|\SSS|)$ words
  of space construct a data structure of size $\bigO(\sqrt N)$
  words that supports the following query in $\bigO(1)$ time.
  Given $|X|$ and the packed representation
  $\PackedRepresentation{w}{\AlphabetSize}{X}$ of any string
  $X\in\IntegerAlphabet^{\leq3\tau-1}$, it returns $|D|$ and
  $\PackedRepresentation{w}{\AlphabetSize}{D}$, where $D$ is the
  unique string in $\DistPrefixes{\tau}{\Text}{\SSS}$
  (\cref{def:dist-prefixes}) that is a prefix of $X$, if such a
  string exists. Otherwise, it reports that no such prefix exists.
\end{proposition}
\begin{proof}
  The data structure is based on the distinguishing-prefix lookup table
  of~\cite{breaking}.
  We give a construction for arbitrary texts using the substring
  frequencies from \cref{th:packed-substring-frequencies}.

  Denote $K=\lceil\sqrt{N}\rceil$,
  $b=\lceil\log\AlphabetSize\rceil$, and $h=3\tau-1$.
  For every $p\in[0\dd h]$, let $L_p[0\dd\AlphabetSize^p)$ be
  an array such that, for every $X\in\IntegerAlphabet^p$,
  $L_p[\Val{\AlphabetSize}{X}]$ equals $|D|$ if some
  $D\in\DistPrefixes{\tau}{\Text}{\SSS}$ is a prefix of $X$,
  and zero otherwise.
  Such $D$ is nonempty and unique by
  \cref{def:dist-prefixes,lm:dist-prefixes}. Thus, the arrays are
  well-defined, and zero denotes the absence of such a prefix.

  \DSComponents
  We store $N$, $\AlphabetSize$, $b$, and $h$, together with
  the following components:
  \begin{enumerate}
  \item The integer-value structure from \cref{th:val-encoding}
    with size parameter $K$ and alphabet size $\AlphabetSize$.
    It occupies $\bigO(\sqrt K)=\bigO(N^{1/4})$ words.
    Since $\AlphabetSize^h<\sqrt N\leq K$, it supports
    every string of length at most $h$.
  \item The arrays $L_0,\ldots,L_h$. Together they occupy
    $\bigO(\sum_{p=0}^h\AlphabetSize^p)=\bigO(\AlphabetSize^h)$ words.
  \end{enumerate}
  The total size is
  $\bigO(\AlphabetSize^h+N^{1/4})=\bigO(\sqrt N)$ words.

  \DSQueries
  Given $m=|X|$ and $\PackedRepresentation{w}{\AlphabetSize}{X}$,
  we answer a query as follows.
  \begin{enumerate}
  \item First, use the structure from \cref{th:val-encoding} to
    compute $x=\Val{\AlphabetSize}{X}$, and retrieve $d=L_m[x]$.
  \item If $d=0$, report that no distinguishing prefix exists.
    Otherwise, use
    \cref{pr:packed-representation}\eqref{pr:packed-representation-substring}
    to extract $\PackedRepresentation{w}{\AlphabetSize}{X[1\dd d]}$,
    and return it together with $d$.
  \end{enumerate}
  Correctness follows from the definition of $L_m$.
  The conversion takes $\bigO(1)$ time, and extracting the prefix
  takes $\bigO(1+d\log\AlphabetSize/w)=\bigO(1)$ time because
  $d\leq h=\bigO(\log_{\AlphabetSize}N)$.

  \DSConstruction
  We construct the integer-value structure and identify all
  distinguishing prefixes in auxiliary Boolean arrays. We then
  use these arrays to fill $L_0,\ldots,L_h$.
  The construction proceeds in three steps.
  \begin{enumerate}

  \item\label{step:nonperiodic-dist-pref-support}
    First, we prepare the integer-value structure and the powers
    of $\AlphabetSize$ used below.
    Compute $K$, $b$, and $h$ in $\bigO(\log N)$ time.
    Construct the structure from \cref{th:val-encoding}.
    By successive multiplication, compute the array
    $A_{\rm pow}[0\dd h]$ satisfying
    $A_{\rm pow}[p]=\AlphabetSize^p$ for every $p\in[0\dd h]$.

  \item\label{step:nonperiodic-dist-pref-marks}
    Next, we identify the strings in
    $\DistPrefixes{\tau}{\Text}{\SSS}$.
    For every $p\in[0\dd h]$, let $M_p[0\dd\AlphabetSize^p)$
    denote the Boolean array satisfying
    \[
      M_p[\Val{\AlphabetSize}{D}]=1
        \quad\Longleftrightarrow\quad
      D\in\DistPrefixes{\tau}{\Text}{\SSS}
      \qquad\text{for every }D\in\IntegerAlphabet^p.
    \]
    We compute these arrays in three substeps.
    \begin{enumerate}
    \item\label{step:nonperiodic-dist-pref-frequencies}
      First, we compute the frequencies of length-$h$ substrings
      of $\Text$. Apply \cref{th:packed-substring-frequencies}
      to $\Text$ with $\ell=h$.
      This gives, in particular, the array $F_h[0\dd\AlphabetSize^h)$
      with $F_h[\Val{\AlphabetSize}{X}]=|\OccTwo{X}{\Text}|$
      for every $X\in\IntegerAlphabet^h$.
    \item\label{step:nonperiodic-dist-pref-sync}
      We now construct a table identifying the length-$2\tau$
      substrings that start at positions in $\SSS$.
      Let $C[0\dd\AlphabetSize^{2\tau})$ be the Boolean array such that,
      for every $Y\in\IntegerAlphabet^{2\tau}$,
      $C[\Val{\AlphabetSize}{Y}]=1$ if and only if
      $Y=\Text[s\dd s+2\tau)$ for some $s\in\SSS$.
      Initialize $C$ to zero. For every position $s$ in the supplied
      array, extract
      $\PackedRepresentation{w}{\AlphabetSize}{\Text[s\dd s+2\tau)}$
      using \cref{pr:packed-representation}\eqref{pr:packed-representation-substring}.
      Use the structure from \cref{th:val-encoding} to compute
      $y=\Val{\AlphabetSize}{\Text[s\dd s+2\tau)}$, and set $C[y]:=1$.
    \item\label{step:nonperiodic-dist-pref-scan}
      Finally, we fill the arrays $M_p$ by examining the
      length-$h$ strings that occur in $\Text$.
      Initialize $M_0,\ldots,M_h$ to zero.
      For every $x\in[0\dd A_{\rm pow}[h])$ with $F_h[x]>0$,
      denote by $X\in\IntegerAlphabet^h$ the unique string with
      $\Val{\AlphabetSize}{X}=x$.
      We search for the smallest offset $d\in[0\dd\tau)$ for which
      $C[\Val{\AlphabetSize}{X[d+1\dd d+2\tau]}]=1$ as follows.
      For $d=0,\ldots,\tau-1$, compute
      $z=\lfloor x/A_{\rm pow}[\tau-1-d]\rfloor$.
      If $C[z\bmod A_{\rm pow}[2\tau]]=1$, set
      $M_{2\tau+d}[z]:=1$ and proceed to the next $x$.
      If all tests fail, leave the arrays unchanged for this $x$.
    \end{enumerate}

  \item\label{step:nonperiodic-dist-pref-tables}
    Finally, we compute the lookup tables $L_0,\ldots,L_h$.
    Set $L_0[0]:=0$. For $p=1,\ldots,h$ and every
    $x\in[0\dd A_{\rm pow}[p])$, set
    \[
      L_p[x]:=
      \begin{cases}
        p,&\text{if }M_p[x]=1,\\
        L_{p-1}[\lfloor x/\AlphabetSize\rfloor],&\text{otherwise}.
      \end{cases}
    \]
  \end{enumerate}

  We now show that the construction computes the arrays
  $L_0,\ldots,L_h$ defined above. First, consider an integer $x$ and
  an offset $d$ examined in Step~\ref{step:nonperiodic-dist-pref-scan}.
  By \cref{def:val}, the computed integer satisfies
  $z=\Val{\AlphabetSize}{X[1\dd2\tau+d]}$, while
  $z\bmod\AlphabetSize^{2\tau}
    =\Val{\AlphabetSize}{X[d+1\dd d+2\tau]}$.
  Thus, at the first successful offset $d$, the step marks the prefix
  $X[1\dd2\tau+d]$.
  We verify that the strings marked in the arrays $M_p$ are
  exactly those in $\DistPrefixes{\tau}{\Text}{\SSS}$.
  \begin{itemize}
  \item Suppose first that a string is marked while processing
    a length-$h$ string $X$ in
    Step~\ref{step:nonperiodic-dist-pref-scan}.
    Let $j\in\OccTwo{X}{\Text}$. Since $|X|=h$, we have
    $j\leq N-h+1$, as required in \cref{def:dist-prefixes}.
    For every $d\in[0\dd\tau)$, the position $j+d$ lies in
    $[1\dd N-2\tau+1]$. The consistency condition in
    \cref{def:sss} therefore implies
    \[
      C[\Val{\AlphabetSize}{X[d+1\dd d+2\tau]}]=1
        \quad\Longleftrightarrow\quad j+d\in\SSS.
    \]
    If $d$ is the first successful offset, then
    $\Successor{\SSS}{j}=j+d<j+\tau$.
    By the density condition, $j\notin\RTwo{\tau}{\Text}$.
    Consequently, the marked string
    $X[1\dd2\tau+d]=\Text[j\dd j+d+2\tau)$ is
    $\DistPrefixPos{j}{\tau}{\Text}{\SSS}$.
  \item Conversely, every $D\in\DistPrefixes{\tau}{\Text}{\SSS}$
    is defined at some
    $j\in[1\dd N-h+1]\setminus\RTwo{\tau}{\Text}$.
    The length-$h$ string $X=\Text[j\dd j+h)$ therefore occurs
    in $\Text$ and is examined in
    Step~\ref{step:nonperiodic-dist-pref-scan}.
    By the same equivalence, its first successful offset is
    $d=\Successor{\SSS}{j}-j$, and the algorithm marks
    $X[1\dd2\tau+d]=D$.
  \end{itemize}
  It remains to justify the assignments in
  Step~\ref{step:nonperiodic-dist-pref-tables}. Consider the
  length-$p$ string $X$ with $\Val{\AlphabetSize}{X}=x$.
  If $M_p[x]=1$, then $X$ itself is a distinguishing prefix,
  and prefix-freeness rules out a shorter one.
  Otherwise, the distinguishing prefixes of $X$ are exactly
  those of $X[1\dd p)$, whose integer value is
  $\lfloor x/\AlphabetSize\rfloor$.
  Induction on $p$, starting with the empty string, proves that
  the tables have their defining values.

  We now analyze the construction.
  Step~\ref{step:nonperiodic-dist-pref-support} takes
  $\bigO(N^{1/4}+\log N)$ time.
  \Cref{th:packed-substring-frequencies} applies in
  Step~\ref{step:nonperiodic-dist-pref-frequencies} with size
  parameter $N$ and alphabet size $\AlphabetSize$ because
  $1\leq h<\frac12\log_{\AlphabetSize}N$, and this substep takes
  $\bigO(N/\log_{\AlphabetSize}N)$ time and words of space.
  Since $2\tau\leq h$, each extraction and conversion in
  Step~\ref{step:nonperiodic-dist-pref-sync} takes $\bigO(1)$ time,
  and this substep takes $\bigO(\AlphabetSize^{2\tau}+|\SSS|)$ time.
  Step~\ref{step:nonperiodic-dist-pref-scan} takes
  $\bigO(\AlphabetSize^h\tau)$ time, and
  Step~\ref{step:nonperiodic-dist-pref-tables} takes
  $\bigO(\AlphabetSize^h)$ time.
  All basic packed operations use alphabet size $\AlphabetSize$ and
  length bound $N$. The integer-value structure applies because
  $\AlphabetSize\leq K\leq N$ and
  $h\leq\lfloor\log_{\AlphabetSize}K\rfloor$.
  Since $\tau=\bigO(\log_{\AlphabetSize}N)$ and
  $\AlphabetSize^h<\sqrt N$, construction takes
  \[
    \bigO(N/\log_{\AlphabetSize}N
      +|\SSS|+\AlphabetSize^h\tau)
      =\bigO(N/\log_{\AlphabetSize}N+|\SSS|).
  \]
  The last bound follows from $(\log N)^2=\bigO(\sqrt N)$.
  Including the packed input, the array listing $\SSS$, and the
  temporary arrays, construction uses
  $\bigO(N/\log_{\AlphabetSize}N+|\SSS|)$ words of space.
  Only the components listed above are stored in the resulting structure.
\end{proof}

\subsubsection{Occurrence Characterizations}
\label{sec:sparse-anchors-nonperiodic-occurrences}

\begin{lemma}\label{lm:nonperiodic-occ}
  Let $\Textlen\in\Z_{\geq2}$, $m\in\Z_{\geq1}$ and $\Text \in \Sigma^{\Textlen}$.
  For $\tau \in [1 \dd \lfloor \tfrac{\Textlen}{2} \rfloor]$, let
  $\SSS$ be a $\tau$-synchronizing set of $\Text$ (\cref{def:sss}).
  Let $\Pat\in\Sigma^m$ be a pattern with a prefix
  $D\in\DistPrefixes{\tau}{\Text}{\SSS}$ (\cref{def:dist-prefixes}),
  and denote $\deltatext=|D|-2\tau$.
  Then, $\deltatext\in[0\dd\tau)$ and
  \[
    \OccTwo{\Pat}{\Text}
      =\left\{s-\deltatext:
        \begin{array}{l}
          s\in\SSS,\quad 1\leq s-\deltatext\leq\Textlen-m+1,\\
          \Pat=\Text[s-\deltatext\dd s-\deltatext+m)
        \end{array}
      \right\}.
  \]
\end{lemma}
\begin{proof}
  By \cref{def:dist-prefixes}, there exists
  $i\in[1\dd\Textlen-3\tau+2]\setminus\RTwo{\tau}{\Text}$ such that
  $D=\Text[i\dd s+2\tau)$, where $s=\Successor{\SSS}{i}$.
  The density condition in \cref{def:sss} gives $i\leq s<i+\tau$.
  Hence, $\deltatext=s-i\in[0\dd\tau)$.
  We prove the equality of the two sets by showing both inclusions.
  \begin{itemize}
  \item First, consider $j\in\OccTwo{\Pat}{\Text}$. We show that
    $j+\deltatext\in\SSS$.
    Since $D$ is a prefix of $\Pat$, it occurs at position $j$.
    Using $\deltatext+2\tau=|D|$, we obtain
    \[
      \Text[j+\deltatext\dd j+\deltatext+2\tau)
        =D[\deltatext+1\dd|D|]
        =\Text[s\dd s+2\tau).
    \]
    Both substrings lie within $\Text$, and $s\in\SSS$.
    The consistency condition in \cref{def:sss} therefore gives
    $j+\deltatext\in\SSS$.
    Together with the occurrence at $j$, this proves that $j$
    belongs to the set on the right-hand side.
  \item Conversely, every element $j=s-\deltatext$ of the set on
    the right-hand side satisfies $1\leq j\leq\Textlen-m+1$ and
    $\Pat=\Text[j\dd j+m)$. Thus,
    $j\in\OccTwo{\Pat}{\Text}$.\qedhere
  \end{itemize}
\end{proof}

\begin{lemma}\label{lm:nonperiodic-occ-cyclic}
  Let $\Textlen\in\Z_{\geq2}$, $m\in\Z_{\geq1}$, $\Text \in \Sigma^{\Textlen}$,
  $\tau \in [1 \dd \lfloor \tfrac{\Textlen}{2} \rfloor]$, and let
  $\SSS$ be a $\tau$-synchronizing set of $\Text$.
  Let $D \in \DistPrefixes{\tau}{\Text}{\SSS}$ (\cref{def:dist-prefixes}),
  and let $\Pat \in \Sigma^{m}$ be a $\tau$-nonperiodic pattern
  (\cref{def:periodic-pattern}) such that no nonempty suffix of $\Text$ is a proper prefix of $\Pat$,
  and $D$ is a prefix of $\Pat$. Denote $\deltatext = |D| - 2\tau$,
  and let $A$ and $B$ be such that $\Pat = A B$ and $|A| = \deltatext$.
  Then, it holds
  \begin{align*}
    \OccTwo{\Pat}{\Text}
      &= \{s - \deltatext : s \in \SSS,\ A\text{ is a suffix of }\Textinf[s - m \dd s),\text{ and }\\
      &\hspace{3.55cm}B\text{ is a prefix of }\Textinf[s \dd s + m)\}.
  \end{align*}
\end{lemma}
\begin{proof}
  By \cref{lm:nonperiodic-occ}, we have $\deltatext\in[0\dd\tau)$,
  and every $j\in\OccTwo{\Pat}{\Text}$ satisfies
  $j+\deltatext\in\SSS$. Since $\Pat=AB$ and $|A|=\deltatext$,
  the suffix and prefix conditions in the claim hold for
  $s=j+\deltatext$. Thus, every occurrence belongs to the set on
  the right-hand side.

  Conversely, consider $s\in\SSS$ satisfying those conditions,
  and denote $j=s-\deltatext$. Since $m\geq|D|=\deltatext+2\tau$,
  we have
  \[
    \Pat=\Textinf[j\dd j+m),\qquad
    j\leq\Textlen,\qquad j+m\geq s+2\tau>1.
  \]
  By \cref{lm:pat-occ}, we obtain $j\in\OccTwo{\Pat}{\Text}$.
\end{proof}

\subsubsection{Main Algorithms}
\label{sec:sparse-anchors-nonperiodic-construction}

\begin{proposition}\label{pr:sparse-anchors-nonperiodic-short-offsets}
  Let $N,\AlphabetSize\in\Z_{\geq2}$ satisfy $\AlphabetSize\leq N$.
  Consider the word RAM model with word size $w\geq1+\log N$.
  Given $N$, $\AlphabetSize$, an integer $\tau\geq1$ satisfying
  $\tau\leq\tfrac16\log_{\AlphabetSize}N$, integers $n\in[1\dd N]$ and
  $k\in\Z_{\geq0}$, the packed representation
  $\PackedRepresentation{w}{\AlphabetSize}{\Text}$
  (\cref{def:packed-representation}) of a text
  $\Text\in\IntegerAlphabet^n$, and arrays containing the lengths
  $m_1,\ldots,m_k$ and packed representations of $k$ $\tau$-nonperiodic
  patterns $\Pat_1,\ldots,\Pat_k\in\IntegerAlphabet^+$
  (\cref{def:periodic-pattern}), in this order, where
  \[
    n+\sum_{i=1}^k m_i\leq N,
    \qquad m_i=|\Pat_i|\geq3\tau-1
      \quad\text{for every }i\in[1\dd k],
  \]
  we can deterministically compute a set $C\subseteq[1\dd n]$
  of anchor positions with $|C|\leq30N/\tau$ and an array
  $\Delta[1\dd k]$ with entries in $[0\dd\tau)$ such that,
  for every $i\in[1\dd k]$, the pattern $\Pat_i$ occurs in
  $\Text$ if and only if there exists $c\in C$ satisfying
  \[
    \begin{aligned}
      1&\leq c-\Delta[i]\leq n-m_i+1,\\
      \Pat_i&=\Text[c-\Delta[i]\dd c-\Delta[i]+m_i).
    \end{aligned}
  \]
  The set $C$ is returned as an array listing its elements.
  The algorithm takes $\bigO(N/\tau)$ time
  and uses $\bigO(N/\tau)$ words of space.
  The patterns need not be distinct.
\end{proposition}
\begin{proof}
  If $k=0$, return an empty set and an empty array.
  Compute $\lceil\log\AlphabetSize\rceil$ in
  $\bigO(\log N)$ time, and initialize every entry of $\Delta$ to zero.
  If $\tau\leq2$, return $C=[1\dd n]$ and $\Delta$.
  The occurrence equivalence holds with $c$ equal to an occurrence
  start, and $|C|=n\leq N\leq30N/\tau$.
  In this case, the construction
  takes $\bigO(n+k+\log N)=\bigO(N/\tau)$ time
  and fits the claimed space bound, including the input.
  Henceforth, assume that $\tau\geq3$. We proceed in three steps.
  \begin{enumerate}
  \item First, we construct the anchor set.
    To this end, denote
    \[
      \Text_{\rm aux}=\Text\zero^{N-n},
    \]
    and construct its packed representation using initialization and
    concatenation from \cref{pr:packed-representation}.
    Using \cref{th:sss-packed-construction} with $\mu=1/6$, compute an
    array listing a $\tau$-synchronizing set $\SSS$ of $\Text_{\rm aux}$
    with $|\SSS|\leq30N/\tau$.
    Denote $C=\SSS\cap[1\dd n]$.
    Construct the array listing $C$ by scanning the returned array
    and copying its entries that are at most $n$.
  \item Next, we construct the distinguishing-prefix data structure
    from \cref{pr:nonperiodic-dist-pref} for $\Text_{\rm aux}$ and
    $\SSS$.
  \item We now compute the split positions.
    For every $i\in[1\dd k]$, extract the packed representation of
    $\Pat_i[1\dd3\tau-1]$ using
    \cref{pr:packed-representation}\eqref{pr:packed-representation-substring}.
    Query the structure with $3\tau-1$ and this representation.
    If it returns a distinguishing-prefix length $\ell_i$, set
    $\Delta[i]:=\ell_i-2\tau$. Otherwise, leave $\Delta[i]=0$.
    Return the array listing $C$ together with $\Delta$.
  \end{enumerate}

  By construction, $C\subseteq[1\dd n]$ and $|C|\leq30N/\tau$.
  By \cref{lm:dist-prefixes}\eqref{lm:dist-prefixes-it-1}, every
  string in $\DistPrefixes{\tau}{\Text_{\rm aux}}{\SSS}$ has length
  at most $3\tau-1$. Thus, $\Pat_i$ and $\Pat_i[1\dd3\tau-1]$
  have exactly the same prefixes in this set for every $i\in[1\dd k]$.
  By \cref{lm:nonperiodic-occ}, the offset computed from any returned
  length belongs to $[0\dd\tau)$. An unsuccessful query leaves
  $\Delta[i]=0$.
  Consider a pattern $\Pat_i$ that occurs at position $j$ in $\Text$.
  It also occurs at $j$ in $\Text_{\rm aux}$, and
  \cref{lm:dist-prefix-existence} ensures that the query returns
  its distinguishing-prefix length.
  Applying \cref{lm:nonperiodic-occ} to $\Text_{\rm aux}$ gives
  $c=j+\Delta[i]\in\SSS$.
  Since $0\leq\Delta[i]<\tau\leq m_i$, we have
  $1\leq c\leq j+m_i-1\leq n$.
  Thus, $c\in C$, and the occurrence at $j=c-\Delta[i]$
  satisfies both conditions in the claim.
  Conversely, those conditions explicitly specify an occurrence
  of $\Pat_i$ in $\Text$ starting at $c-\Delta[i]$.

  We finish with the time and space bounds.
  The minimum pattern length gives
  $k\leq N/(3\tau-1)=\bigO(N/\tau)$, and
  $\tau\leq\tfrac16\log_{\AlphabetSize}N$ implies
  $N/\log_{\AlphabetSize}N=\bigO(N/\tau)$.
  All basic packed operations use alphabet size $\AlphabetSize$
  and length bound $N$.
  The synchronizing-set construction and the structure from
  \cref{pr:nonperiodic-dist-pref} use size parameter $N$ and
  alphabet size $\AlphabetSize$.
  Constructing the packed representation of $\Text_{\rm aux}$ takes
  $\bigO(1+N\log\AlphabetSize/w)=\bigO(N/\log_{\AlphabetSize}N)$ time.
  The synchronizing-set construction takes $\bigO(N/\tau)$ time
  and words of space.
  Since $\tau\leq\tfrac16\log_{\AlphabetSize}N$ and
  $w\geq1+\log N$, the distinguishing-prefix
  structure applies with text length $N$.
  Its construction takes
  $\bigO(N/\log_{\AlphabetSize}N+|\SSS|)
    =\bigO(N/\tau)$ time
  and words of space. Extracting the pattern prefixes and answering
  the $k$ queries takes $\bigO(k)$ time.
  Copying the elements of $C$ takes $\bigO(|\SSS|)$ time.
  The input occupies $\bigO(k+N\log\AlphabetSize/w)$ words.
  The auxiliary text, arrays, and construction workspace together
  occupy $\bigO(N/\tau)$ words.
  Finally, the $\bigO(\log N)$ time to compute $\lceil\log\AlphabetSize\rceil$
  is within the claimed time bound.
\end{proof}

\subsection{Periodic Patterns}
\label{sec:sparse-anchors-periodic}

\subsubsection{Preliminaries}
\label{sec:sparse-anchors-periodic-definitions}

\begin{definition}[Root, type, and end of the periodic prefix]\label{def:pat-root}
  Let $\tau\in\Z_{\geq1}$ and let $\Pat \in \Sigma^{+}$ be a $\tau$-periodic
  pattern (\cref{def:periodic-pattern}). Denote $p = \per{\Pat[1 \dd 3\tau - 1]}$.
  We then define
  \begin{itemize}
  \item $\RootPat{\Pat}{\tau}
    := \min\{\Pat[1 + t \dd 1 + t + p) : t \in [0 \dd p)\}$,
  \item $\RunEndPat{\Pat}{\tau} := 1 + p +
    \lcp{\Pat[1 \dd |\Pat|]}{\Pat[1 + p \dd |\Pat|]}$.
  \end{itemize}
  Furthermore, we then let
  \[
    \TypePat{\Pat}{\tau} =
      \begin{cases}
       +1 & \text{if }\RunEndPat{\Pat}{\tau} \leq |\Pat|\text{ and }
            \Pat[\RunEndPat{\Pat}{\tau}] \succ \Pat[\RunEndPat{\Pat}{\tau} - p],\\
       -1 & \text{otherwise}.
      \end{cases}
  \]
\end{definition}

\begin{remark}\label{rm:pat-root}
  Observe that in \cref{def:pat-root}, we can
  write $\Pat[1 \dd \RunEndPat{\Pat}{\tau}) = H' H^{k} H''$, where $H =
  \RootPat{\Pat}{\tau}$, $k\in\Z_{\geq0}$, and $H'$ (resp.\ $H''$) is a proper
  suffix (resp.\ proper prefix) of $H$. This factorization is unique, since the
  opposite would contradict the synchronization property of primitive
  strings~\cite[Lemma~1.11]{AlgorithmsOnStrings}.
\end{remark}

\begin{definition}[Head, exponent, and tail of a periodic pattern]\label{def:pat-head}
  Let $\tau\in\Z_{\geq1}$ and let $\Pat \in \Sigma^{+}$ be a $\tau$-periodic pattern.
  Let $\Pat[1 \dd \RunEndPat{\Pat}{\tau}) = H' H^{k} H''$ be such that
  $H = \RootPat{\Pat}{\tau}$, $k\in\Z_{\geq0}$, and $H'$ (resp.\ $H''$) is a proper
  suffix (resp.\ proper prefix) of $H$ (see \cref{rm:pat-root}).
  We then define:
  \begin{itemize}
  \item $\HeadPat{\Pat}{\tau} := |H'|$,
  \item $\ExpPat{\Pat}{\tau} := k$,
  \item $\TailPat{\Pat}{\tau} := |H''|$,
  \item $\RunEndFullPat{\Pat}{\tau} :=
    1 + |H'| + k \cdot |H| =
    \RunEndPat{\Pat}{\tau} - \TailPat{\Pat}{\tau}$.
  \end{itemize}
\end{definition}

\begin{example}\label{ex:periodic-pattern-parameters}
  Let $\tau=6$ and $\Pat=(\texttt{ba})^{10}\texttt{c}$, with
  $\texttt{a}\prec\texttt{b}\prec\texttt{c}$.
  The prefix $\Pat[1\dd17]$ has shortest period $2=\tau/3$,
  and hence $\Pat$ is $\tau$-periodic.
  Its root is $\RootPat{\Pat}{\tau}=\texttt{ab}$, and its
  longest prefix of period $2$ factors as
  \[
    \Pat[1\dd20]=\texttt{b}\cdot(\texttt{ab})^9\cdot\texttt{a}.
  \]
  Thus, $\HeadPat{\Pat}{\tau}=1$, $\ExpPat{\Pat}{\tau}=9$, and
  $\TailPat{\Pat}{\tau}=1$.
  Moreover, $\RunEndPat{\Pat}{\tau}=21$,
  $\RunEndFullPat{\Pat}{\tau}=20$, and
  $\TypePat{\Pat}{\tau}=+1$, since
  $\Pat[21]=\texttt{c}\succ\Pat[19]=\texttt{b}$.
  The string $\Pat'=(\texttt{ba})^{10}$ has the same root, head,
  exponent, tail, and endpoints, but $\TypePat{\Pat'}{\tau}=-1$
  because $\RunEndPat{\Pat'}{\tau}=21=|\Pat'|+1$.
\end{example}

\begin{definition}[Parameters of a periodic position]\label{def:pos-root}
  Let $\Textlen\in\Z_{\geq2}$, $\Text \in \Sigma^{\Textlen}$,
  $\tau \in [1 \dd \floor{\frac{\Textlen}{2}}]$,
  and $j \in \RTwo{\tau}{\Text}$.
  Letting $\Pat = \Text[j \dd \Textlen]$, we
  define:
  \begin{itemize}
  \item $\RootPos{j}{\tau}{\Text} := \RootPat{\Pat}{\tau}$,
  \item $\HeadPos{j}{\tau}{\Text} := \HeadPat{\Pat}{\tau}$,
  \item $\ExpPos{j}{\tau}{\Text} := \ExpPat{\Pat}{\tau}$,
  \item $\TailPos{j}{\tau}{\Text} := \TailPat{\Pat}{\tau}$,
  \item $\RunEndPos{j}{\tau}{\Text} := j + \RunEndPat{\Pat}{\tau} - 1$,
  \item $\RunEndFullPos{j}{\tau}{\Text} := j + \RunEndFullPat{\Pat}{\tau} - 1$,
  \item $\TypePos{j}{\tau}{\Text} := \TypePat{\Pat}{\tau}$.
  \end{itemize}
\end{definition}

\begin{remark}\label{rm:pos-root}
  Note that applying the notation for $\tau$-periodic patterns (\cref{def:periodic-pattern}) to $\Pat$ in
  \cref{def:pos-root} is well-defined since it is easy to see that for
  every $j \in \RTwo{\tau}{\Text}$ (\cref{def:periodic-positions}), the string $\Text[j \dd \Textlen]$
  is $\tau$-periodic. Note also that, letting
  $s = \HeadPos{j}{\tau}{\Text}$,
  $H = \RootPos{j}{\tau}{\Text}$,
  $p = |H|$, and
  $k = \ExpPos{j}{\tau}{\Text}$, it holds:
  \begin{itemize}
  \item $\RunEndPos{j}{\tau}{\Text} = j + p + \LCE{\Text}{j}{j+p}$,
  \item $\RunEndFullPos{j}{\tau}{\Text} = j + s + kp = \RunEndPos{j}{\tau}{\Text} - \TailPos{j}{\tau}{\Text}$.
  \end{itemize}
\end{remark}

\begin{definition}[Subsets of periodic positions]\label{def:R-subsets}
  Let $\Textlen\in\Z_{\geq2}$, $\Text \in \Sigma^{\Textlen}$, and
  $\tau \in [1 \dd \floor{\frac{\Textlen}{2}}]$.
  For every $H \in \Sigma^{+}$, $s \in \Zz$, and $k \in \Zz$,
  we define the following subsets of $\RTwo{\tau}{\Text}$:
  \begin{itemize}
  \item $\RThree{H}{\tau}{\Text} := \{j \in \RTwo{\tau}{\Text} : \RootPos{j}{\tau}{\Text} = H\}$,
  \item $\RFour{s}{H}{\tau}{\Text} := \{j \in \RThree{H}{\tau}{\Text} : \HeadPos{j}{\tau}{\Text} = s\}$,
  \item $\RFive{s}{k}{H}{\tau}{\Text} := \{j \in \RFour{s}{H}{\tau}{\Text} : \ExpPos{j}{\tau}{\Text} = k\}$.
  \end{itemize}
\end{definition}

\begin{definition}[Starts of maximal blocks of periodic positions]\label{def:R-prim}
  Let $\Textlen\in\Z_{\geq2}$, $\Text \in \Sigma^{\Textlen}$, and
  $\tau \in [1 \dd \floor{\frac{\Textlen}{2}}]$.
  The set of starting positions of maximal blocks in $\RTwo{\tau}{\Text}$
  is denoted
  \[
    \RPrimTwo{\tau}{\Text} := \{j \in \RTwo{\tau}{\Text} : j-1 \not\in \RTwo{\tau}{\Text}\}.
  \]
\end{definition}

\begin{example}\label{ex:periodic-position-parameters}
  Let $\tau=6$ and
  $\Text=\texttt{c}(\texttt{ba})^{10}\texttt{c}(\texttt{ba})^{10}$,
  with $\texttt{a}\prec\texttt{b}\prec\texttt{c}$.
  This text has length $42$, and its periodic positions form two blocks:
  \[
    \RTwo{\tau}{\Text}=[2\dd5]\cup[23\dd26],
    \qquad
    \RPrimTwo{\tau}{\Text}=\{2,23\}.
  \]
  At position $j=2$, the factorization
  $\Text[2\dd22)=\texttt{b}(\texttt{ab})^9\texttt{a}$ gives
  $\RootPos{2}{\tau}{\Text}=\texttt{ab}$,
  $\HeadPos{2}{\tau}{\Text}=\TailPos{2}{\tau}{\Text}=1$, and
  $\ExpPos{2}{\tau}{\Text}=9$.
  The endpoints are $\RunEndPos{2}{\tau}{\Text}=22$ and
  $\RunEndFullPos{2}{\tau}{\Text}=21$, and
  $\TypePos{2}{\tau}{\Text}=+1$ because
  $\Text[22]=\texttt{c}\succ\Text[20]=\texttt{b}$.
  Position $j=23$ has the same root, head, exponent, and tail,
  but its endpoints are $\RunEndPos{23}{\tau}{\Text}=43$ and
  $\RunEndFullPos{23}{\tau}{\Text}=42$, while
  $\TypePos{23}{\tau}{\Text}=-1$ because the periodic part reaches
  the end of $\Text$.
  Every periodic position has root $H=\texttt{ab}$, and hence
  $\RThree{H}{\tau}{\Text}=\RTwo{\tau}{\Text}$.
  Restricting to head $1$ gives
  $\RFour{1}{H}{\tau}{\Text}=\{2,4,23,25\}$, and further
  restricting to exponent $9$ gives
  $\RFive{1}{9}{H}{\tau}{\Text}=\{2,23\}$.
\end{example}

\begin{lemma}[{\cite{collapsing}}]\label{lm:R-text-block}
  Let $\Textlen\in\Z_{\geq2}$, $\Text \in \Sigma^{\Textlen}$, and
  $\tau \in [1 \dd \floor{\frac{\Textlen}{2}}]$. For every $j \in
  \RTwo{\tau}{\Text}$ such that $j-1 \in \RTwo{\tau}{\Text}$, it holds (see \cref{def:periodic-positions,def:pos-root})
  \begin{itemize}
  \item $\RootPos{j-1}{\tau}{\Text} = \RootPos{j}{\tau}{\Text}$,
  \item $\RunEndPos{j-1}{\tau}{\Text} = \RunEndPos{j}{\tau}{\Text}$,
  \item $\TailPos{j-1}{\tau}{\Text} = \TailPos{j}{\tau}{\Text}$,
  \item $\RunEndFullPos{j-1}{\tau}{\Text} = \RunEndFullPos{j}{\tau}{\Text}$,
  \item $\TypePos{j-1}{\tau}{\Text} = \TypePos{j}{\tau}{\Text}$.
  \end{itemize}
\end{lemma}

\begin{lemma}[{\cite{collapsing}}]\label{lm:periodic-pos-lce}
  Let $\Textlen\in\Z_{\geq2}$, $\Text \in \Sigma^{\Textlen}$,
  $\tau \in [1 \dd \lfloor \tfrac{\Textlen}{2} \rfloor]$, and
  $j \in [1 \dd \Textlen]$. We use the notation from
  \cref{def:periodic-positions,def:pat-root,def:pat-head,def:pos-root}.
  \begin{enumerate}
  \item\label{lm:periodic-pos-lce-it-1}
    Let $\Pat \in \Sigma^{+}$ be a $\tau$-periodic pattern (\cref{def:periodic-pattern}). Then,
    the following conditions are equivalent:
    \begin{itemize}
    \item $\lcp{\Pat}{\Text[j \dd \Textlen]} \geq 3\tau - 1$,
    \item $j \in \RTwo{\tau}{\Text}$, $\RootPos{j}{\tau}{\Text} =
      \RootPat{\Pat}{\tau}$, and $\HeadPos{j}{\tau}{\Text} =
      \HeadPat{\Pat}{\tau}$.
    \end{itemize}
    Moreover, if $\lcp{\Pat}{\Text[j \dd \Textlen]} > \RunEndPat{\Pat}{\tau} - 1$, then:
    \begin{itemize}
    \item $\RunEndPat{\Pat}{\tau} - 1 = \RunEndPos{j}{\tau}{\Text} - j$,
    \item $\TailPat{\Pat}{\tau} = \TailPos{j}{\tau}{\Text}$,
    \item $\RunEndFullPat{\Pat}{\tau} - 1 =
      \RunEndFullPos{j}{\tau}{\Text} - j$,
    \item $\ExpPat{\Pat}{\tau} = \ExpPos{j}{\tau}{\Text}$,
    \item $\TypePat{\Pat}{\tau} = \TypePos{j}{\tau}{\Text}$.
    \end{itemize}
  \item\label{lm:periodic-pos-lce-it-2}
    Let $j' \in \RTwo{\tau}{\Text}$. Then, the following conditions
    are equivalent:
    \begin{itemize}
    \item $\LCE{\Text}{j'}{j} \geq 3\tau - 1$,
    \item $j \in \RTwo{\tau}{\Text}$, with
      $\RootPos{j}{\tau}{\Text} = \RootPos{j'}{\tau}{\Text}$, and
      $\HeadPos{j'}{\tau}{\Text} = \HeadPos{j}{\tau}{\Text}$.
    \end{itemize}
    Moreover, if $\LCE{\Text}{j}{j'} > \RunEndPos{j'}{\tau}{\Text} - j'$, then:
    \begin{itemize}
    \item $\RunEndPos{j'}{\tau}{\Text} - j' = \RunEndPos{j}{\tau}{\Text} - j$,
    \item $\TailPos{j'}{\tau}{\Text} = \TailPos{j}{\tau}{\Text}$,
    \item $\RunEndFullPos{j'}{\tau}{\Text} - j' =
      \RunEndFullPos{j}{\tau}{\Text} - j$,
    \item $\ExpPos{j'}{\tau}{\Text} = \ExpPos{j}{\tau}{\Text}$,
    \item $\TypePos{j'}{\tau}{\Text} = \TypePos{j}{\tau}{\Text}$.
    \end{itemize}
  \end{enumerate}
\end{lemma}

\begin{remark}\label{rm:collapsing-arbitrary-text}
  In~\cite{collapsing}, \cref{lm:R-text-block,lm:periodic-pos-lce} are
  stated for roots chosen by an arbitrary function that assigns the
  same rotation to all rotations of a primitive string, and for texts
  whose last symbol does not occur elsewhere. Our root
  (\cref{def:pat-root,def:pos-root}) is the lexicographically minimal
  rotation, which is such a choice, and the proofs in~\cite{collapsing}
  do not use the assumption on the last symbol. The two lemmas therefore
  apply to arbitrary texts.
\end{remark}

\begin{lemma}[{\cite{breaking}}]\label{lm:runs}
  Let $\Textlen\in\Z_{\geq2}$. For every $\Text \in \Sigma^{\Textlen}$ and
  $\tau \in [1 \dd \lfloor \tfrac{\Textlen}{2} \rfloor]$,
  it holds (see \cref{def:R-prim,def:pos-root})
  $|\RPrimTwo{\tau}{\Text}| \leq \tfrac{\Textlen}{2\tau}$
  and $\sum_{i \in \RPrimTwo{\tau}{\Text}} (\RunEndPos{i}{\tau}{\Text} - i) \leq 2\Textlen$.
\end{lemma}
\begin{proof}
  The bound on the sum and the inequality $a'-a\geq2\tau$ for all
  $a,a'\in\RPrimTwo{\tau}{\Text}$ with $a<a'$ are shown
  in~\cite{breaking}. Since every element of
  $\RPrimTwo{\tau}{\Text}$ belongs to $[1\dd\Textlen-3\tau+2]$
  (\cref{def:periodic-positions}), the latter inequality gives
  $|\RPrimTwo{\tau}{\Text}|\leq\lfloor(\Textlen-3\tau+1)/(2\tau)\rfloor+1
    \leq\Textlen/(2\tau)$.
\end{proof}

\subsubsection{Basic Navigation}
\label{sec:sparse-anchors-periodic-navigation}

\begin{proposition}[{Based on~\cite{breaking}}]\label{pr:efull}
  Let $N,\AlphabetSize\in\Z_{\geq2}$ satisfy
  $\AlphabetSize\leq N$.
  Consider the word RAM model with word size $w\geq1+\log N$.
  Given $N$ and $\AlphabetSize$, we can in $\bigO(\sqrt N)$ time and using
  $\bigO(\sqrt N)$ words of space construct a data structure of
  size $\bigO(\sqrt N)$ words that, given an integer
  $\tau\geq1$ satisfying $\tau\leq\tfrac12\log_{\AlphabetSize}N$,
  $m\in[1\dd N]$, and
  the packed representation $\PackedRepresentation{w}{\AlphabetSize}{\Pat}$
  of a $\tau$-periodic pattern $\Pat\in\IntegerAlphabet^m$
  (\cref{def:packed-representation,def:periodic-pattern}), supports
  the following queries (see \cref{def:pat-root,def:pat-head}):
  \begin{enumerate}
  \item\label{pr:efull-root-head} Compute $\HeadPat{\Pat}{\tau}$,
    $|\RootPat{\Pat}{\tau}|$, and
    $\PackedRepresentation{w}{\AlphabetSize}{\RootPat{\Pat}{\tau}}$
    in $\bigO(1)$ time.
  \item\label{pr:efull-parameters} Compute $\RunEndPat{\Pat}{\tau}$,
    $\RunEndFullPat{\Pat}{\tau}$, $\ExpPat{\Pat}{\tau}$,
    $\TailPat{\Pat}{\tau}$, and $\TypePat{\Pat}{\tau}$ in
    $\bigO(1+m/\log_{\AlphabetSize}N)$ time.
  \end{enumerate}
\end{proposition}
\begin{proof}
  The navigation structure in~\cite{breaking} supports these queries
  for $\tau=\lfloor\tfrac17\log_{\AlphabetSize}N\rfloor$ after
  $\bigO(N/\log_{\AlphabetSize}N)$-time preprocessing.
  We give a construction supporting every value of $\tau$ in the claim
  with the smaller $\bigO(\sqrt N)$ preprocessing bound.

  Denote $b=\lceil\log\AlphabetSize\rceil$ and
  $h=\lfloor\tfrac16\log_{\AlphabetSize}N\rfloor$.
  For every $r\in[1\dd h]$ and $u\in[0\dd\AlphabetSize^{2r})$, let
  $X_{r,u}\in\IntegerAlphabet^{2r}$ be the unique string satisfying
  $\Val{\AlphabetSize}{X_{r,u}}=u$.
  For every $r\in[1\dd h]$, let $A_r[0\dd\AlphabetSize^{2r})$
  be an array defined as follows.
  If $p=\per{X_{r,u}}\leq r$, its entry $A_r[u]$ stores
  $p$, the starting offset $s\in[0\dd p)$ of the lexicographically
  smallest cyclic rotation of $X_{r,u}[1\dd p]$, and the packed
  representation of $X_{r,u}[1+s\dd1+s+p)$.
  Otherwise, the entry stores two zeros and the empty packed representation.

  \DSComponents
  We store $N$, $\AlphabetSize$, $b$, and $h$, together with
  the following components:
  \begin{enumerate}
  \item The integer-value structure from \cref{th:val-encoding},
    with size parameter $N$ and alphabet size $\AlphabetSize$.
    It occupies $\bigO(\sqrt N)$ words.
  \item We also store the packed LCP structure from \cref{th:packed-lcp}
    with the same parameters. It occupies $\bigO(\sqrt N)$ words.
  \item The arrays $A_r$ for every $r\in[1\dd h]$.
    Each stored string has length at most $r$, and its packed
    representation occupies $\bigO(1+r\log\AlphabetSize/w)=\bigO(1)$ words.
    Thus, the arrays use
    $\bigO(\sum_{r=1}^h\AlphabetSize^{2r})=\bigO(N^{1/3})$ words.
  \end{enumerate}

  The total size is $\bigO(\sqrt N)$ words.

  \DSQueries
  Given $\tau$, $m$, and the packed representation
  $\PackedRepresentation{w}{\AlphabetSize}{\Pat}$ of a
  $\tau$-periodic pattern $\Pat\in\IntegerAlphabet^m$,
  we answer the queries as follows:
  \begin{enumerate}
  \item To answer a query of type~\eqref{pr:efull-root-head},
    compute $r=\lfloor\tau/3\rfloor$ and extract the packed
    representation of $\Pat[1\dd2r]$ using
    \cref{pr:packed-representation}\eqref{pr:packed-representation-substring}.
    Use the integer-value structure to compute
    $u=\Val{\AlphabetSize}{\Pat[1\dd2r]}$, and return the head,
    root length, and packed root stored in $A_r[u]$.
    To prove correctness, denote $p=\per{\Pat[1\dd3\tau-1]}$.
    Since $\Pat$ is $\tau$-periodic, we have $1\leq p\leq r\leq h$.
    The prefix $\Pat[1\dd2r]$ has shortest period $p$ as well.
    Indeed, a smaller period $q<p$ would give
    $\Pat[i]=\Pat[i+q]$ for every $i\in[1\dd p]$, because
    $p+q\leq2r$.
    By the $p$-periodicity of $\Pat[1\dd3\tau-1]$, reducing positions
    modulo $p$ extends these equalities throughout this longer prefix,
    contradicting its shortest period $p$.
    The first $p$ symbols determine the root and head, which are
    therefore returned correctly.
    Since $2r\leq\lfloor\log_{\AlphabetSize}N\rfloor$ and
    $2r\log\AlphabetSize=\bigO(\log N)$, extraction, conversion,
    and returning the stored values take $\bigO(1)$ time.
  \item To answer a query of type~\eqref{pr:efull-parameters},
    we proceed in three steps:
    \begin{enumerate}
    \item First, we obtain $p=|\RootPat{\Pat}{\tau}|$ and
      $s=\HeadPat{\Pat}{\tau}$ using the query of
      type~\eqref{pr:efull-root-head} in $\bigO(1)$ time.
    \item Next, we compute
      \[
        \lambda=\lcp{\Pat}{\Pat[p+1\dd m]}.
      \]
      To this end, extract the packed representation of
      $\Pat[p+1\dd m]$ using
      \cref{pr:packed-representation}\eqref{pr:packed-representation-substring}.
      Query the packed LCP structure on $\Pat$ and this suffix.
      The extraction takes $\bigO(1+m\log\AlphabetSize/w)$ time,
      and the LCP query takes $\bigO(1+m/\log_{\AlphabetSize}N)$ time.
    \item Finally, we derive the requested parameters from $p$, $s$,
      and $\lambda$. Compute
      \[
        e=1+p+\lambda,\qquad
        k=\left\lfloor\frac{e-1-s}{p}\right\rfloor,\qquad
        t=(e-1-s)\bmod p.
      \]
      By \cref{def:pat-root,def:pat-head}, these give
      $\RunEndPat{\Pat}{\tau}=e$,
      $\RunEndFullPat{\Pat}{\tau}=e-t$,
      $\ExpPat{\Pat}{\tau}=k$, and $\TailPat{\Pat}{\tau}=t$.
      If $e\leq m$, use
      \cref{pr:packed-representation}\eqref{pr:packed-representation-access}
      to compare $\Pat[e]$ and $\Pat[e-p]$.
      The type is $+1$ if $e\leq m$ and $\Pat[e]\succ\Pat[e-p]$,
      and $-1$ otherwise. In particular, no symbol is read at $e=m+1$.
      All endpoints are at most $N+1<2^w$.
      This step takes $\bigO(1)$ time.
    \end{enumerate}
    Since $w>\log N$, the total query time is
    $\bigO(1+m/\log_{\AlphabetSize}N)$.
  \end{enumerate}

  \DSConstruction
  We construct the stored components in two steps.
  \begin{enumerate}
  \item First, we compute $b$ and $h$ in
    $\bigO(\log N)$ time. We then construct the integer-value and
    packed LCP structures in $\bigO(\sqrt N)$ time.
  \item Next, we construct the arrays $A_r$.
    For $r=1,\ldots,h$, proceed as follows.
    \begin{enumerate}
    \item Compute $\AlphabetSize^{2r}$ by successive multiplication,
      and initialize each entry of $A_r$ to two zeros and the empty
      packed representation.
    \item For every $u\in[0\dd\AlphabetSize^{2r})$, compute the symbols
      of $X_{r,u}$ as the length-$2r$ base-$\AlphabetSize$
      representation of $u$, including leading zeros, by repeated division.
      Test candidate periods $q\in[1\dd r]$ in ascending order,
      checking whether $X_{r,u}[i]=X_{r,u}[i-q]$ for every
      $i\in[q+1\dd2r]$.
      If none passes, leave $A_r[u]$ unchanged.
      Otherwise, denote by $p$ the first candidate that passes.
      Compare the $p$ cyclic rotations of $X_{r,u}[1\dd p]$ symbol by
      symbol to find the starting offset $s$ of the smallest one.
      Construct the packed representation of $X_{r,u}[1+s\dd1+s+p)$
      using the initialization and update operations of
      \cref{pr:packed-representation}, and store it together with
      $p$ and $s$ in $A_r[u]$.
      This takes $\bigO(r^2)$ time per entry, using
      $\bigO(r)$ auxiliary words for the symbols of $X_{r,u}$.
    \end{enumerate}
  \end{enumerate}
  Since $\AlphabetSize^{2h}\leq N^{1/3}$ and $h=\bigO(\log N)$,
  the total construction time is
  \[
    \bigO\Big(\sqrt N+\sum_{r=1}^h\AlphabetSize^{2r}r^2\Big)
      =\bigO(\sqrt N+N^{1/3}\log^2 N)
      =\bigO(\sqrt N).
  \]
  All packed operations use alphabet size $\AlphabetSize$ and length
  bound $N$, as permitted by the word-size assumption.
  The input, components, and auxiliary data use $\bigO(\sqrt N)$ words.
\end{proof}

\begin{proposition}[{Based on~\cite{breaking}}]\label{pr:periodic-pos}
  Let $N,\AlphabetSize\in\Z_{\geq2}$ satisfy
  $\AlphabetSize\leq N$.
  Consider the word RAM model with word size $w\geq1+\log N$.
  Given $N$, $\AlphabetSize$, and the packed representation
  $\PackedRepresentation{w}{\AlphabetSize}{\Text}$
  (\cref{def:packed-representation}) of a text
  $\Text\in\IntegerAlphabet^{N}$, we can in
  $\bigO(N/\log_{\AlphabetSize}N)$ time
  and using $\bigO(N/\log_{\AlphabetSize}N)$ words
  of space construct a data structure of size
  $\bigO(N/\log_{\AlphabetSize}N)$ words that,
  given an integer $\tau\geq1$ satisfying
  $\tau\leq\tfrac12\log_{\AlphabetSize}N$ and a position
  $j\in\RTwo{\tau}{\Text}$ (\cref{def:periodic-positions}),
  computes all the following in $\bigO(1)$ time (see \cref{def:pos-root}):
  \begin{itemize}
  \item The length and packed representation
    $\PackedRepresentation{w}{\AlphabetSize}{H}$ of
    $H=\RootPos{j}{\tau}{\Text}$.
  \item The values $\HeadPos{j}{\tau}{\Text}$, $\RunEndPos{j}{\tau}{\Text}$,
    $\RunEndFullPos{j}{\tau}{\Text}$, $\ExpPos{j}{\tau}{\Text}$,
    $\TailPos{j}{\tau}{\Text}$, and $\TypePos{j}{\tau}{\Text}$.
  \end{itemize}
\end{proposition}
\begin{proof}
  The cited result assumes that the final text symbol is unique.
  We give a construction for arbitrary texts and all values of $\tau$
  permitted by the claim.

  \DSComponents
  We store $N$, $\AlphabetSize$, $b=\lceil\log\AlphabetSize\rceil$,
  and the packed representation of $\Text$, together with the structure
  from \cref{pr:efull} and the LCE structure
  from \cref{th:lce} for $\Text$.
  The latter supports LCE queries in $\bigO(1)$ time.

  \DSQueries
  Given $\tau$ and $j\in\RTwo{\tau}{\Text}$, we proceed in two steps.
  \begin{enumerate}
  \item First, we obtain the root and head of position $j$.
    Extract the packed representation of
    $\Pat=\Text[j\dd j+3\tau-1)$ using
    \cref{pr:packed-representation}\eqref{pr:packed-representation-substring},
    and query the structure from \cref{pr:efull}\eqref{pr:efull-root-head}
    with $\tau$, $3\tau-1$, and this representation.
    The pattern $\Pat$ is $\tau$-periodic and has the same root and head
    as $\Text[j\dd N]$ by \cref{def:periodic-positions,def:pat-root,def:pat-head}.
    Thus, the query returns $s=\HeadPos{j}{\tau}{\Text}$,
    $p=|H|$, and the packed representation of $H=\RootPos{j}{\tau}{\Text}$.
  \item Next, we compute the remaining parameters. Using the LCE structure,
    compute
    \[
      e=j+p+\LCE{\Text}{j}{j+p},\qquad
      k=\left\lfloor\frac{e-j-s}{p}\right\rfloor,
      \qquad t=(e-j-s)\bmod p.
    \]
    By \cref{rm:pos-root,def:pos-root}, these give
    $\RunEndPos{j}{\tau}{\Text}=e$,
    $\RunEndFullPos{j}{\tau}{\Text}=e-t$,
    $\ExpPos{j}{\tau}{\Text}=k$, and $\TailPos{j}{\tau}{\Text}=t$.
    If $e=N+1$, return type $-1$.
    Otherwise, use
    \cref{pr:packed-representation}\eqref{pr:packed-representation-access}
    to compare $\Text[e]$ and $\Text[e-p]$, returning type $+1$
    if $\Text[e]\succ\Text[e-p]$, and $-1$ otherwise.
  \end{enumerate}
  The condition $j\in\RTwo{\tau}{\Text}$ ensures that the extracted
  prefix lies in $\Text$ and that $j+p\leq N$, as required by the LCE query.
  Since $(3\tau-1)\log\AlphabetSize=\bigO(\log N)$, extracting the
  packed prefix takes $\bigO(1)$ time. All other query operations take
  $\bigO(1)$ time. The computed positions are at most $N+1$ and fit
  in a machine word under the assumed word-size bound.

  \DSConstruction
  Compute $b$ in $\bigO(\log N)$ time, and construct the
  two structures with size parameter $N$ and alphabet size $\AlphabetSize$.
  The LCE structure takes $\bigO(N/\log_{\AlphabetSize}N)$ time and
  words of space, and the structure from \cref{pr:efull} takes
  $\bigO(\sqrt N)$ time and words of space.
  Since $\sqrt N=\bigO(N/\log_{\AlphabetSize}N)$, the total
  construction time and space, including the packed text, satisfy
  the claimed bounds.
\end{proof}

\begin{theorem}\label{th:periodic-position-intervals}
  Let $N,\AlphabetSize\in\Z_{\geq2}$ satisfy
  $\AlphabetSize\leq N$.
  Consider the word RAM model with word size $w\geq1+\log N$.
  Given $N$, $\AlphabetSize$, an integer $\tau\geq1$ satisfying
  $\tau\leq\tfrac12\log_{\AlphabetSize}N$, and the packed representation
  $\PackedRepresentation{w}{\AlphabetSize}{\Text}$
  (\cref{def:packed-representation}) of any text
  $\Text\in\IntegerAlphabet^N$, we can deterministically compute
  an integer $k\in\Z_{\geq0}$ satisfying $k=\bigO(N/\tau)$ and
  arrays $A_{\rm beg}[1\dd k]$ and $A_{\rm end}[1\dd k]$ such that
  the intervals $[A_{\rm beg}[i]\dd A_{\rm end}[i])$, for
  $i\in[1\dd k]$, are precisely the maximal intervals of
  $\RTwo{\tau}{\Text}$ (\cref{def:periodic-positions}), listed in
  increasing order. The construction takes $\bigO(N/\tau)$ time
  and uses $\bigO(N/\tau)$ words of space.
\end{theorem}
\begin{proof}
  For $\tau=\lfloor\tfrac17\log_{\AlphabetSize}N\rfloor\geq1$
  and texts ending in a unique symbol, this result can also be obtained
  from~\cite{breaking} by explicitly constructing a string synchronizing set.
  The proof below uses regularly spaced blocks and longest common extension
  queries, and does not require a unique final symbol.

  Compute $b=\lceil\log\AlphabetSize\rceil$ in
  $\bigO(\log N)$ time. If $\tau\leq2$, then
  $\RTwo{\tau}{\Text}=\emptyset$, because every nonempty string
  has period at least one. In this case, return $k=0$ and two empty
  arrays. Henceforth, assume that $\tau\geq3$, and denote
  $h=3\tau-1$ and $m=\lfloor N/\tau\rfloor$.

  We construct the output in two steps.
  \begin{enumerate}

  \item First, we prepare the structures used to extend short periodic
    blocks. Construct the shortest-period structure from
    \cref{pr:periods-of-short-strings} with length bound $d=\tau$.
    This takes $\bigO(\AlphabetSize^\tau\tau)$
    time and uses $\bigO(\AlphabetSize^\tau)$ words of space.
    Next, use \cref{th:string-reversal} to obtain the packed
    representation of $\revstr{\Text}$. Finally, construct the LCE
    data structures from \cref{th:lce} for $\Text$ and $\revstr{\Text}$.
    Each structure answers LCE queries in $\bigO(1)$ time and can be
    constructed in $\bigO(N/\log_{\AlphabetSize}N)$ time and words
    of space. All constructions use size parameter $N$ and alphabet
    size $\AlphabetSize$.

  \item We now compute the interval endpoints.
    Initialize two empty arrays $A_{\rm beg}$ and $A_{\rm end}$,
    and set $k:=0$. For $t=0,\ldots,m-1$, denote
    $s=1+t\tau$ and $X=\Text[s\dd s+\tau)$, and proceed as follows:
    \begin{enumerate}
    \item First, we compute $p=\per{X}$. To this end, extract the packed
      representation of $X$ using
      \cref{pr:packed-representation}\eqref{pr:packed-representation-substring}
      and query the shortest-period structure with $|X|=\tau$ and
      this representation.
      If $p>\tau/3$, proceed to the next value of $t$.
    \item Next, we extend $X$ to the left and right while preserving
      period $p$. To this end, use the two LCE structures to compute
      \[
        r=\LCE{\Text}{s}{s+p},\qquad
        \ell=
        \begin{cases}
          0 & \text{if }s=1,\\
          \LCE{\revstr{\Text}}{N-s+2}{N-s-p+2}
            & \text{otherwise}.
        \end{cases}
      \]
      Set $a:=s-\ell$ and $e:=s+p+r$.
    \item Finally, if $e-a<h$, proceed to the next value of $t$.
      Otherwise, if $k=0$ or $A_{\rm beg}[k]\neq a$, append $a$ to
      $A_{\rm beg}$ and $e-h+1$ to $A_{\rm end}$, and increase
      $k$ by one.
    \end{enumerate}
  \end{enumerate}

  We recall two consequences of the periodicity
  lemma~\cite{periodicitylemma}. If two fragments with periods $p$
  and $q$ overlap in at least $p+q$ symbols, their union has period
  $g=\gcd(p,q)$. Indeed, their intersection has period $g$, and
  each fragment extends the same $g$-periodic string, since the
  intersection has length at least $p+g$ and $q+g$.
  Consequently, every substring of length at least $2p$ of a string
  with shortest period $p$ also has shortest period $p$: a smaller
  period $q$ would give period $\gcd(p,q)<p$ for the whole string.

  We prove correctness in three steps.
  \begin{enumerate}
  \item First, we prove that every generated interval is a maximal interval of
    $\RTwo{\tau}{\Text}$. Consider an iteration with $p\leq\tau/3$.
    The reverse-text query computes the longest common suffix of
    $\Text[1\dd s)$ and $\Text[1\dd s+p)$, with length zero when
    $s=1$. Together with $r=\LCE{\Text}{s}{s+p}$, this shows that
    $\Text[a\dd e)$ is the maximal $p$-periodic fragment containing
    $X$. Its shortest period is $p$, because it contains $X$.
    If $e-a\geq h$, every position in $[a\dd e-h+1)$ belongs to
    $\RTwo{\tau}{\Text}$. Suppose that a neighboring position also
    belongs to this set. Its length-$h$ window has a period
    $q\leq\tau/3$ and overlaps $\Text[a\dd e)$ in $h-1\geq p+q$
    symbols. The union therefore has period $\gcd(p,q)$. Since
    $\Text[a\dd e)$ has shortest period $p$, this greatest common
    divisor equals $p$, contradicting the maximality of the
    $p$-periodic fragment. Thus, the generated interval is maximal.

  \item Next, we show that every position in $\RTwo{\tau}{\Text}$ belongs
    to a generated interval. Let $j\in\RTwo{\tau}{\Text}$, and denote
    $Y=\Text[j\dd j+h)$ and $q=\per{Y}\leq\tau/3$.
    The first position $s$ of the form $1+t\tau$ at or after $j$
    satisfies $s\leq j+\tau-1$. Hence, $X=\Text[s\dd s+\tau)$
    lies within $Y$ and is one of the blocks processed by the
    algorithm. Since $|X|=\tau\geq2q$, the shortest period of $X$
    is $q$. The maximal $q$-periodic fragment found in this
    iteration contains $Y$. It follows that the generated interval contains $j$.

  \item Finally, we show that the intervals are generated in increasing
    order and that all repetitions of the same interval are consecutive.
    Distinct maximal periodic fragments
    found by the algorithm overlap in fewer than $\tau$ symbols.
    Otherwise, their periods $p,q\leq\tau/3$ and the overlap
    property imply that their union has period $\gcd(p,q)$.
    Since both periods are shortest, we obtain $p=q$, contradicting
    maximality unless the fragments are equal. Suppose now that
    blocks beginning at $s<s'$ yield distinct fragments $[a\dd e)$
    and $[a'\dd e')$ with $a'\leq a$. The second fragment contains
    the entire block $\Text[s\dd s+\tau)$, because $a'\leq a\leq s$
    and $e'\geq s'+\tau>s+\tau$. The two fragments would
    therefore overlap in at least $\tau$ symbols, a contradiction.
    Thus, their left endpoints increase strictly whenever the
    fragments differ. The algorithm consequently removes exactly
    the repetitions and returns all maximal intervals in order.
  \end{enumerate}

  Since every sampled block lies within $\Text$ and has length
  $\tau\leq\lfloor\log_{\AlphabetSize}N\rfloor$, extraction
  and the shortest-period query take $\bigO(1)$ time.
  Whenever we extend a block, $p\leq\tau/3<\tau$, ensuring that all positions
  supplied to the LCE structures belong to $[1\dd N]$.
  The endpoints satisfy $1\leq a<e\leq N+1$ and fit in one word.
  Thus, each iteration takes $\bigO(1)$ time, and each output array
  has at most $m=\lfloor N/\tau\rfloor$ entries.
  Since $\AlphabetSize^\tau\leq\sqrt N$ and
  $\tau^2=\bigO(\log^2 N)=\bigO(\sqrt N)$, the period-table
  construction takes $\bigO(\AlphabetSize^\tau\tau)=\bigO(N/\tau)$ time.
  Reversing $\Text$ and constructing the LCE structures take
  $\bigO(N/\log_{\AlphabetSize}N)$ time and words of space.
  Together with the $m$ iterations, the total time and space are
  \[
    \bigO(\AlphabetSize^\tau\tau+N/\log_{\AlphabetSize}N+N/\tau)
      =\bigO(N/\tau),
  \]
  with space measured in words. Here we used
  $\tau\leq\tfrac12\log_{\AlphabetSize}N$ to bound the reversal
  and LCE costs.
\end{proof}

\subsubsection{Occurrence Characterizations}
\label{sec:sparse-anchors-periodic-occurrences}

\begin{lemma}\label{lm:fully-periodic-pat-occ}
  Let $\Textlen\in\Z_{\geq2}$, $m\in\Z_{\geq1}$, $\Text \in \Sigma^{\Textlen}$,
  and $\tau \in [1 \dd \lfloor \tfrac{\Textlen}{2} \rfloor]$.
  Let $\Pat \in \Sigma^{m}$ be a $\tau$-periodic pattern (\cref{def:periodic-pattern}) satisfying
  $\RunEndPat{\Pat}{\tau} = m + 1$. Denote $s = \HeadPat{\Pat}{\tau}$,
  $H = \RootPat{\Pat}{\tau}$, $k = \ExpPat{\Pat}{\tau}$, and $t = \TailPat{\Pat}{\tau}$
  (\cref{def:pat-root,def:pat-head}).
  Then, it holds (see \cref{def:R-subsets,def:pos-root})
  \[
    \OccTwo{\Pat}{\Text}
      = \{x \in \RFive{s}{k}{H}{\tau}{\Text} : \TailPos{x}{\tau}{\Text} \geq t\}
      \cup \{x \in \RFour{s}{H}{\tau}{\Text} : \ExpPos{x}{\tau}{\Text} > k\}.
  \]
\end{lemma}
\begin{proof}
  Denote $p=|H|$, and let $H'$ be the suffix of $H$ of length $s$.
  Since $\RunEndPat{\Pat}{\tau}=m+1$, we have
  $\Pat=H'H^kH[1\dd t]$ and $m=s+kp+t$.
  Since $m\geq3\tau-1$,
  \cref{lm:periodic-pos-lce}\eqref{lm:periodic-pos-lce-it-1} gives
  $\OccTwo{\Pat}{\Text}\subseteq\RFour{s}{H}{\tau}{\Text}$.

  Let $j\in\RFour{s}{H}{\tau}{\Text}$, and denote
  $e_j=\RunEndPos{j}{\tau}{\Text}$,
  $k_j=\ExpPos{j}{\tau}{\Text}$, and
  $t_j=\TailPos{j}{\tau}{\Text}$.
  By \cref{def:pat-head,def:pos-root},
  $\Text[j\dd e_j)=H'H^{k_j}H[1\dd t_j]$.
  We first show that
  \[
    j\in\OccTwo{\Pat}{\Text}
      \quad\Longleftrightarrow\quad e_j-j\geq m.
  \]
  \begin{itemize}
  \item If $j\in\OccTwo{\Pat}{\Text}$, then
    $\Text[j\dd j+m)=\Pat$ has period $p$.
    Since $\Text[j\dd e_j)$ is the longest prefix of
    $\Text[j\dd\Textlen]$ with period $p$
    (\cref{def:pat-root,def:pos-root}), we have $e_j-j\geq m$.
  \item Conversely, suppose that $e_j-j\geq m$.
    The factorizations of $\Pat$ and $\Text[j\dd e_j)$ both begin
    with $H'$ and continue with copies of $H$, possibly ending
    with a proper prefix of $H$.
    Their first $m$ symbols are therefore equal, giving
    $\Text[j\dd j+m)=\Pat$ and $j\in\OccTwo{\Pat}{\Text}$.
  \end{itemize}
  By the above factorizations, we have $e_j-j=s+k_jp+t_j$
  and $m=s+kp+t$.
  Since $0\leq t,t_j<p$, the inequality $e_j-j\geq m$ holds
  if and only if $k_j>k$ or both $k_j=k$ and $t_j\geq t$.
  By \cref{def:R-subsets}, these are exactly the membership
  conditions in the two sets in the claim.
\end{proof}

\begin{lemma}\label{lm:partially-periodic-pat-occ}
  Let $\Textlen\in\Z_{\geq2}$, $m\in\Z_{\geq1}$, $\Text \in \Sigma^{\Textlen}$,
  and $\tau \in [1 \dd \lfloor \tfrac{\Textlen}{2} \rfloor]$. Let
  $\Pat \in \Sigma^{m}$ be a $\tau$-periodic pattern (\cref{def:periodic-pattern})
  such that $\RunEndPat{\Pat}{\tau} \leq m$. Denote
  $H=\RootPat{\Pat}{\tau}$ and $\deltatext = \RunEndFullPat{\Pat}{\tau} - 1$
  (\cref{def:pat-root,def:pat-head}), and let $\Pat=AB$ with $|A|=\deltatext$.
  For every $r\in\RPrimTwo{\tau}{\Text}$, let
  $e_r^{\rm full}=\RunEndFullPos{r}{\tau}{\Text}$ (\cref{def:pos-root,def:R-prim}). Then,
  \begin{align*}
    \OccTwo{\Pat}{\Text}=\{e_r^{\rm full}-\deltatext : {}&
      r\in\RPrimTwo{\tau}{\Text},\ \RootPos{r}{\tau}{\Text}=H,\\*
      &\deltatext\leq e_r^{\rm full}-r,\ B\text{ is a prefix of }\Text[e_r^{\rm full}\dd\Textlen]\}.
  \end{align*}
\end{lemma}
\begin{proof}
  Let $Q$ denote the set on the right-hand side.

  First, we prove that $\OccTwo{\Pat}{\Text}\subseteq Q$.
  Let $j\in\OccTwo{\Pat}{\Text}$. Since $\Pat$ is $\tau$-periodic,
  we have $m\geq3\tau-1$. By
  \cref{lm:periodic-pos-lce}\eqref{lm:periodic-pos-lce-it-1},
  $j\in\RTwo{\tau}{\Text}$ and $\RootPos{j}{\tau}{\Text}=H$.
  Moreover, $m>\RunEndPat{\Pat}{\tau}-1$ allows us to apply
  the same lemma to obtain $\RunEndFullPos{j}{\tau}{\Text}=j+\deltatext$.
  Let $r$ be the start of the maximal block of $\RTwo{\tau}{\Text}$
  containing $j$. Then, $r\in\RPrimTwo{\tau}{\Text}$, and applying
  \cref{lm:R-text-block} to consecutive positions in $[r\dd j]$ gives
  $\RootPos{r}{\tau}{\Text}=H$ and $e_r^{\rm full}=j+\deltatext$.
  Since $r\leq j$, we have $\deltatext\leq e_r^{\rm full}-r$.
  Furthermore, the occurrence of $\Pat=AB$ at $j$ implies that
  $B$ is a prefix of $\Text[e_r^{\rm full}\dd\Textlen]$.
  Thus, $j=e_r^{\rm full}-\deltatext\in Q$.

  Next, we prove that $Q\subseteq\OccTwo{\Pat}{\Text}$.
  Let $j\in Q$, and choose $r\in\RPrimTwo{\tau}{\Text}$ satisfying
  the conditions in the definition of $Q$, with $j=e_r^{\rm full}-\deltatext$.
  By \cref{def:pat-head,def:pos-root}, both $A$ and
  $\Text[r\dd e_r^{\rm full})$ consist of a proper suffix of $H$ followed by
  full copies of $H$. Since $|A|=\deltatext\leq e_r^{\rm full}-r$,
  $A$ is the length-$\deltatext$ suffix of $\Text[r\dd e_r^{\rm full})$.
  Together with $B$ being a prefix of $\Text[e_r^{\rm full}\dd\Textlen]$,
  this gives $\Pat=AB=\Text[j\dd j+m)$ and
  $j\in\OccTwo{\Pat}{\Text}$.
\end{proof}

\subsubsection{Main Algorithms}
\label{sec:sparse-anchors-periodic-matching}

\begin{proposition}[Batched matching for fully periodic patterns]\label{pr:fully-periodic-pattern-occurrences}
  Let $N,\AlphabetSize\in\Z_{\geq2}$ satisfy $\AlphabetSize\leq N$.
  Consider the word RAM model with word size $w\geq1+\log N$.
  Given $N$, $\AlphabetSize$, an integer $\tau\geq1$ satisfying
  $\tau\leq\tfrac12\log_{\AlphabetSize}N$, integers $n\in[1\dd N]$ and
  $k\in\Z_{\geq0}$, the packed representation
  $\PackedRepresentation{w}{\AlphabetSize}{\Text}$
  (\cref{def:packed-representation}) of a text
  $\Text\in\IntegerAlphabet^n$, and arrays containing the lengths
  and packed representations of $\tau$-periodic patterns
  $\Pat_1,\ldots,\Pat_k\in\IntegerAlphabet^+$
  (\cref{def:periodic-pattern}), in this order, where
  $n+\sum_{i=1}^{k}|\Pat_i|\leq N$ and
  $\RunEndPat{\Pat_i}{\tau}=|\Pat_i|+1$ for every
  $i\in[1\dd k]$ (see \cref{def:pat-root}),
  we can deterministically compute an array $W[1\dd k]$ such that,
  for every $i\in[1\dd k]$,
  \begin{itemize}
  \item if $\OccTwo{\Pat_i}{\Text}=\emptyset$, then $W[i]=0$;
  \item otherwise, $W[i]\in\OccTwo{\Pat_i}{\Text}$.
  \end{itemize}
  The algorithm takes $\bigO(N/\tau)$ time and
  uses $\bigO(N/\tau)$ words of space.
  The patterns need not be distinct.
\end{proposition}
\begin{proof}
  Denote $m_i=|\Pat_i|$ for every $i\in[1\dd k]$.
  Compute $\lceil\log\AlphabetSize\rceil$ in
  $\bigO(\log N)$ time.
  Initialize every entry of $W$ to zero. If $k=0$ or $n<3\tau-1$,
  return $W$, since every pattern has length at least $3\tau-1$.
  Henceforth, assume that $k\geq1$ and $n\geq3\tau-1$.
  Since a $\tau$-periodic pattern has positive period at most $\tau/3$
  in its first $3\tau-1$ symbols, we also have $\tau\geq3$.
  The algorithm proceeds in two steps.
  \begin{enumerate}
  \item First, we compute the periodic parameters of the patterns
    and of the positions in $\RPrimTwo{\tau}{\Text}$.
    \begin{enumerate}
    \item We begin by constructing the structures needed to compute
      the periodic parameters. Denote $\Text'=\Text\zero^{N-n}$.
      Construct the packed representation of $\Text'$ using initialization
      and concatenation from \cref{pr:packed-representation}, and
      construct the periodic-position structure from
      \cref{pr:periodic-pos} for $\Text'$.
      Construct also the periodic-pattern structure from
      \cref{pr:efull} and the integer-value structure from
      \cref{th:val-encoding}.
    \item\label{step:fully-periodic-text}
      Next, we compute the parameters associated with the text.
      Apply \cref{th:periodic-position-intervals} to $\Text'$ with parameter $\tau$.
      Retain the returned interval starts that lie in
      $[1\dd n-3\tau+2]$.
      Denote the resulting increasing sequence by $r_1,\ldots,r_t$.
      For every $j\in[1\dd t]$, denote also
      \[
        H_j=\RootPos{r_j}{\tau}{\Text},\qquad
        h_j=\HeadPos{r_j}{\tau}{\Text},\qquad
        e_j=\RunEndPos{r_j}{\tau}{\Text}.
      \]
      Query the periodic-position structure with $\tau$ and $r_j$ to obtain
      $h_j$, the length and packed representation of $H_j$, and
      $e'_j=\RunEndPos{r_j}{\tau}{\Text'}$, and set
      $e_j:=\min\{e'_j,n+1\}$.
    \item We now compute the roots and heads of the patterns.
      For every $i\in[1\dd k]$, denote
      $H'_i=\RootPat{\Pat_i}{\tau}$ and
      $h'_i=\HeadPat{\Pat_i}{\tau}$.
      Query the structure from \cref{pr:efull} with $\tau$, $m_i$, and
      the packed representation of $\Pat_i$ to obtain
      $h'_i$ and the length and packed representation of $H'_i$.
      For any root $X$, the pair $(|X|,\Val{\AlphabetSize}{X})$
      is denoted by $\kappa(X)$.
      Compute $\kappa(H_j)$ for every $j\in[1\dd t]$ and
      $\kappa(H'_i)$ for every $i\in[1\dd k]$ using the integer-value
      structure. We use these pairs to group text positions
      and patterns by root in the next step.
    \end{enumerate}
  \item We now compute $W$ using the roots, heads, and ends of the periodic prefixes.
    For every $j\in[1\dd t]$, compute $\ell_j=e_j-r_j$.
    To find suitable text indices for all patterns, we proceed
    as follows.
    \begin{enumerate}
    \item First, we sort the pattern indices $i\in[1\dd k]$ and
      the indices $j\in[1\dd t]$ of the text positions $r_j$
      by the corresponding roots and heads.
      To this end, construct an array $U[1\dd k]$ containing the
      indices $1,\ldots,k$, and sort its entries $i$
      lexicographically by $(\kappa(H'_i),h'_i)$.
      Similarly, construct an array $D[1\dd t]$ containing the
      indices $1,\ldots,t$, and sort its entries $j$ lexicographically by
      $(\kappa(H_j),h_j)$. Equal root pairs form consecutive
      groups in each array.
    \item We now process the corresponding groups in $U$ and $D$.
      Traverse the groups in both arrays in ascending order of
      their root pairs. When the current root pairs differ, advance
      past the group with the smaller pair. When they agree, process
      the two groups as described below and then advance past both.
      For such a pair of groups, denote the entries of the group
      in $U$ by $i_1,\ldots,i_v$, in their array order.
      The entries of the group in $D$ are denoted by
      $j_1,\ldots,j_g$, also in their array order.
      Denote $p=|H_{j_1}|$ and proceed in three steps.
      \begin{enumerate}
      \item First, we construct a range-maximum structure for the
        values $\ell_j-h_j$ in this text group.
        To this end, construct an array $F[1\dd g]$ with
        $F[y]=\ell_{j_y}-h_{j_y}$ for every $y\in[1\dd g]$,
        and build the range-maximum structure from
        \cref{th:range-maximum-queries} for $F$.
      \item\label{step:fully-periodic-boundaries}
        Next, we compute, for every pattern index $i$ in this group,
        the boundary
        \[
          q_i=|\{y\in[1\dd g]:h_{j_y}<h'_i\}|.
        \]
        Initialize $q=0$. For $y=1,\ldots,v$, denote $i=i_y$.
        Increase $q$ by one while $q<g$ and $h_{j_{q+1}}<h'_i$,
        and then set $q_i:=q$.
      \item\label{step:fully-periodic-queries}
        Finally, we determine the entries of $W$ for this group.
        For $y=1,\ldots,v$, denote $i=i_y$ and proceed as follows.
        \begin{enumerate}
        \item If $q_i>0$, query the range $(0\dd q_i]$.
          Denote the returned index by $y_*$ and the corresponding
          text index by $j_*=j_{y_*}$.
          If $h'_i+F[y_*]-p\geq m_i$, set
          \[
            W[i]:=r_{j_*}+h_{j_*}-h'_i+p.
          \]
        \item If $q_i<g$, query the range $(q_i\dd g]$.
          Denote the returned index by $y_*$ and the corresponding
          text index by $j_*=j_{y_*}$.
          If $h'_i+F[y_*]\geq m_i$, set
          \[
            W[i]:=r_{j_*}+h_{j_*}-h'_i.
          \]
        \end{enumerate}
        Leave $W[i]$ unchanged when a range is empty or its
        inequality fails.
      \end{enumerate}
      Pattern indices in groups with no matching group in $D$
      retain output zero.
    \end{enumerate}
  \end{enumerate}

  We now prove that the computed array $W$ has the required properties.
  We first justify the parameters computed in
  Step~\ref{step:fully-periodic-text}.
  Padding leaves every length-$(3\tau-1)$ window starting in
  $[1\dd n-3\tau+2]$ unchanged. Hence
  $\RTwo{\tau}{\Text}=\RTwo{\tau}{\Text'}\cap[1\dd n-3\tau+2]$.
  Restricting the periodic positions to this initial interval
  preserves the starts of their nonempty blocks. Thus,
  $r_1,\ldots,r_t$ are precisely the elements of
  $\RPrimTwo{\tau}{\Text}$.
  The root and head depend only on the unchanged initial
  length-$(3\tau-1)$ window, and the queries therefore return
  $H_j$ and $h_j$. Padding can extend the periodic prefix only
  beyond the end of $\Text$, which gives $e_j=\min\{e'_j,n+1\}$.
  Moreover, two roots are equal if and only if their pairs
  $\kappa$ are equal.

  Next, consider any $i\in[1\dd k]$.
  The pattern $\Pat_i$ has period $|H'_i|$, and
  \cref{lm:fully-periodic-pat-occ} gives
  \begin{equation}\label{eq:periodic-batch-full}
    \OccTwo{\Pat_i}{\Text}
      =\{x\in\RFour{h'_i}{H'_i}{\tau}{\Text} :
        \RunEndPos{x}{\tau}{\Text}-x\geq m_i\}.
  \end{equation}
  Here the exponent and tail conditions in the lemma express
  that the periodic prefix at $x$ has length at least $m_i$
  (\cref{def:pat-head,def:pos-root}).
  Consider $j\in[1\dd t]$ with $H_j=H'_i$, and denote $p=|H'_i|$.
  By \cref{lm:R-text-block}, every position $x$ in the block
  of consecutive positions from $\RTwo{\tau}{\Text}$ beginning
  at $r_j$ has root $H'_i$ and endpoint $e_j$.
  Consequently, \cref{def:pat-head,def:pos-root} show that the head
  decreases by one modulo $p$ whenever $x$ increases by one within
  this block, and the head of $x$ is $(h_j-(x-r_j))\bmod p$.
  An occurrence of $\Pat_i$ in this block must therefore begin at an offset
  congruent to $h_j-h'_i$ modulo $p$.
  The smallest nonnegative such offset is
  \[
    z_{ij}=(h_j-h'_i)\bmod p.
  \]
  If $\ell_j-z_{ij}\geq m_i$, every length-$(3\tau-1)$ window
  starting in $[r_j\dd r_j+z_{ij}]$ lies inside the period-$p$
  fragment $\Text[r_j\dd e_j)$. Hence $r_j+z_{ij}$ belongs
  to the same block of $\RTwo{\tau}{\Text}$.
  By \eqref{eq:periodic-batch-full}, this block therefore contains
  an occurrence of $\Pat_i$ if and only if $\ell_j-z_{ij}\geq m_i$, in which
  case $r_j+z_{ij}$ is an occurrence. Equivalently, $\Pat_i$ has
  an occurrence in this block if and only if
  \[
    \begin{cases}
      h'_i+\ell_j-h_j\geq m_i &\text{if }h_j\geq h'_i,\\
      h'_i+\ell_j-h_j-p\geq m_i &\text{if }h_j<h'_i.
    \end{cases}
  \]
  Moreover, by \eqref{eq:periodic-batch-full}, every occurrence of
  $\Pat_i$ belongs to a block whose root is $H'_i$.
  If there is no text group with root $H'_i$, the initial value
  $W[i]=0$ is therefore correct.
  Otherwise, when the forward scan in
  Step~\ref{step:fully-periodic-boundaries} assigns $q_i$, the pointer has
  passed exactly the text indices in this group with $h_j<h'_i$.
  Thus, the text indices at positions $(0\dd q_i]$ and
  $(q_i\dd g]$ are precisely those with $h_j<h'_i$ and
  $h_j\geq h'_i$, respectively.
  For each nonempty range, the query in
  Step~\ref{step:fully-periodic-queries} selects a text index $j_*$
  maximizing $\ell_j-h_j$ among its indices.
  This is the only expression depending on $j$ in the relevant
  inequality. Hence, a block satisfying that inequality exists
  if and only if the test for the corresponding maximum succeeds.
  Each assigned value is $r_{j_*}+z_{i j_*}$ and belongs to
  $\OccTwo{\Pat_i}{\Text}$. If neither test succeeds, the pattern
  has no occurrence and $W[i]$ remains zero.

  We finish by checking the bounds and the assumptions of the tools.
  All basic packed operations use length bound $N$ and alphabet
  size $\AlphabetSize$. The periodic-navigation tools and the
  integer-value structure also use these parameters, with $|\Text'|=N$,
  $\tau\leq\tfrac12\log_{\AlphabetSize}N$, and $w\geq1+\log N$.
  Furthermore,
  \[
    k\leq\frac{N}{3\tau-1},\qquad
    t\leq\frac{n}{2\tau},
  \]
  where the bound on $k$ follows from $m_i\geq3\tau-1$ for every
  $i\in[1\dd k]$, and the bound on $t$ follows from \cref{lm:runs}.
  Storing each pattern length in one word therefore uses
  $\bigO(k)=\bigO(N/\tau)$ words of space.
  Since $\tau\leq\tfrac12\log_{\AlphabetSize}N$, both
  $N/\log_{\AlphabetSize}N$ and $\sqrt N$ are $\bigO(N/\tau)$.
  Constructing the packed representation of $\Text'$ and the structures, and
  enumerating the periodic intervals, take
  $\bigO(N/\tau)$ total time and words of space.
  The root and head queries for the patterns and the
  periodic-position queries take $\bigO(1)$ time each.
  Every root has length at most $\tau/3$, allowing its integer
  value to be computed in $\bigO(1)$ time.
  The sorting operations use a constant number of integer fields
  of $\bigO(\log N)$ bits. Radix sorting with digits of
  $\lfloor\frac12\log N\rfloor$ bits takes
  $\bigO(k+t+\sqrt N)=\bigO(N/\tau)$ total time.
  Matching the root groups and computing all boundaries $q_i$
  take $\bigO(k+t)$ total time, since the pointers never decrease.
  Each array $F$ contains integers in $[0\dd n]$, and
  $w>\log n\geq\log g$, as required by
  \cref{th:range-maximum-queries}.
  Constructing these arrays and their range-maximum structures
  takes $\bigO(t)$ total time and words of space.
  There are at most two queries per pattern, taking $\bigO(k)$
  total time. All arrays, including the input and output,
  and all invoked structures occupy
  $\bigO(N/\tau)$ words.
\end{proof}

\begin{proposition}[Batched matching for partially periodic patterns]\label{pr:partially-periodic-pattern-occurrences}
  Let $N,\AlphabetSize\in\Z_{\geq2}$ satisfy $\AlphabetSize\leq N$.
  Consider the word RAM model with word size $w\geq1+\log N$.
  Given $N$, $\AlphabetSize$, an integer $\tau\geq1$ satisfying
  $\tau\leq\tfrac12\log_{\AlphabetSize}N$, integers $n\in[1\dd N]$ and
  $k\in\Z_{\geq0}$, the packed representation
  $\PackedRepresentation{w}{\AlphabetSize}{\Text}$
  (\cref{def:packed-representation}) of a text
  $\Text\in\IntegerAlphabet^n$, and arrays containing the lengths
  and packed representations of $\tau$-periodic patterns
  $\Pat_1,\ldots,\Pat_k\in\IntegerAlphabet^+$
  (\cref{def:periodic-pattern}), in this order, where
  $n+\sum_{i=1}^{k}|\Pat_i|\leq N$ and
  $\RunEndPat{\Pat_i}{\tau}\leq|\Pat_i|$ for every
  $i\in[1\dd k]$ (see \cref{def:pat-root}),
  we can deterministically compute an array $W[1\dd k]$ such that,
  for every $i\in[1\dd k]$,
  \begin{itemize}
  \item if $\OccTwo{\Pat_i}{\Text}=\emptyset$, then $W[i]=0$;
  \item otherwise, $W[i]\in\OccTwo{\Pat_i}{\Text}$.
  \end{itemize}
  The algorithm takes $\bigO(N/\tau)$ time and
  uses $\bigO(N/\tau)$ words of space.
  The patterns need not be distinct.
\end{proposition}
\begin{proof}
  Denote $m_i=|\Pat_i|$ for every $i\in[1\dd k]$ and
  $d=\AlphabetSize+1$. Compute $\lceil\log\AlphabetSize\rceil$ and
  $\lceil\log(d+1)\rceil$ in $\bigO(\log N)$ time.
  Initialize every entry of $W$ to zero. If $k=0$ or $n<3\tau-1$,
  return $W$, since every pattern has length at least $3\tau-1$.
  Henceforth, assume that $k\geq1$ and $n\geq3\tau-1$.
  Since a $\tau$-periodic pattern has positive period at most $\tau/3$
  in its first $3\tau-1$ symbols, we also have $\tau\geq3$.
  The algorithm proceeds in two steps.
  \begin{enumerate}
  \item First, we compute the periodic parameters of the patterns
    and of the positions in $\RPrimTwo{\tau}{\Text}$.
    \begin{enumerate}
    \item We begin by constructing the structures needed to compute
      the periodic parameters. Denote $\Text'=\Text\zero^{N-n}$.
      Construct the packed representation of $\Text'$ using initialization
      and concatenation from \cref{pr:packed-representation}.
      Construct the periodic-position structure from
      \cref{pr:periodic-pos} for $\Text'$, the periodic-pattern
      structure from \cref{pr:efull}, and the integer-value
      structure from \cref{th:val-encoding}.
    \item Next, we compute the parameters associated with the text.
      Apply \cref{th:periodic-position-intervals} to $\Text'$ with parameter $\tau$.
      Retain the returned interval starts that lie in
      $[1\dd n-3\tau+2]$.
      Denote the resulting increasing sequence by $r_1,\ldots,r_t$.
      Padding leaves every length-$(3\tau-1)$ window starting in
      this interval unchanged. Hence
      $\RTwo{\tau}{\Text}=\RTwo{\tau}{\Text'}\cap[1\dd n-3\tau+2]$.
      Restricting the periodic positions to this initial interval
      preserves the starts of their nonempty blocks. Thus,
      $r_1,\ldots,r_t$ are precisely the elements of
      $\RPrimTwo{\tau}{\Text}$.
      For every $j\in[1\dd t]$, denote also
      \[
        H_j=\RootPos{r_j}{\tau}{\Text},\qquad
        \delta_j=\RunEndFullPos{r_j}{\tau}{\Text}-r_j.
      \]
      Query the periodic-position structure with $\tau$ and $r_j$ to obtain
      the length and packed representation of $H_j$, the head
      $h_j=\HeadPos{r_j}{\tau}{\Text'}$, and the end
      $e'_j=\RunEndPos{r_j}{\tau}{\Text'}$ of the periodic prefix.
      The root and head depend only on the unchanged initial
      length-$(3\tau-1)$ window. Padding may extend the periodic
      prefix beyond $\Text$. We first compute its original
      endpoint $e_j:=\min\{e'_j,n+1\}$, and then set
      \[
        \delta_j:=e_j-r_j-\big((e_j-r_j-h_j)\bmod |H_j|\big).
      \]
      By \cref{def:pos-root}, $r_j+\delta_j$ is the full-root endpoint in $\Text$.
    \item We now compute the corresponding parameters for each pattern.
      For every $i\in[1\dd k]$, denote
      \[
        H'_i=\RootPat{\Pat_i}{\tau},\qquad
        \delta'_i=\RunEndFullPat{\Pat_i}{\tau}-1.
      \]
      Query the structure from \cref{pr:efull} with $\tau$, $m_i$, and
      the packed representation of $\Pat_i$ to obtain the length
      and packed representation of $H'_i$ and the full-root endpoint
      $\RunEndFullPat{\Pat_i}{\tau}$. Subtract one from this
      endpoint to obtain $\delta'_i$.
      For any root $X$, the pair $(|X|,\Val{\AlphabetSize}{X})$
      is denoted by $\kappa(X)$.
      Finally, compute $\kappa(H_j)$ for every $j\in[1\dd t]$ and
      $\kappa(H'_i)$ for every $i\in[1\dd k]$ using the integer-value
      structure.
      Two roots are equal if and only if their pairs $\kappa$ are equal.
      We use these pairs to group text positions and patterns by root
      in the next step.
    \end{enumerate}
  \item Next, we find an occurrence of each pattern that occurs
    in the text. For every $i\in[1\dd k]$, denote
    \[
      Y'_i=\Pat_i[\delta'_i+1\dd m_i].
    \]
    For every $j\in[1\dd t]$, denote also
    \[
      Y_j=\Text[r_j+\delta_j\dd n].
    \]
    For every $i\in[1\dd k]$,
    \cref{lm:partially-periodic-pat-occ} gives
    \begin{equation}\label{eq:periodic-batch-partial}
      \OccTwo{\Pat_i}{\Text}
        =\{r_j+\delta_j-\delta'_i : j\in[1\dd t],\ H_j=H'_i,
          \ \delta'_i\leq \delta_j,\ Y'_i\text{ is a prefix of }Y_j\}.
    \end{equation}
    We find the interval of text suffixes having prefix $Y'_i$ in
    the lexicographic order of each root group, and then use a
    range-maximum query to choose an index $j$ maximizing $\delta_j$
    in this interval. For any string $X$ over $\IntegerAlphabet$,
    the string obtained by increasing every symbol by one is
    denoted by $\widehat X$.
    Since $d=\AlphabetSize+1$ is larger than every symbol of the
    lifted strings, $Y'_i$ is a prefix of $Y_j$ if and only if
    $\widehat Y'_i\preceq\widehat Y_j\prec\widehat Y'_i d$.
    We therefore compute the lower and upper endpoints using the
    orders of the strings $\widehat Y'_i$ and $\widehat Y'_i d$,
    respectively. We proceed as follows.
    \begin{enumerate}
    \item We first construct a text supporting suffix sorting and
      LCE queries on the strings $Y'_i$ and $Y_j$.
      For every $i\in[1\dd k]$, compute $|Y'_i|=m_i-\delta'_i$
      and extract the packed representation of $Y'_i$ using
      \cref{pr:packed-representation}\eqref{pr:packed-representation-substring}.
      Use \cref{th:alphabet-lifting} with offset one to
      construct the packed representations of $\widehat\Text$
      and $\widehat Y'_1,\ldots,\widehat Y'_k$.
      Denote
      \[
        S_0=\widehat\Text\zero
          \widehat Y'_1\zero\cdots\widehat Y'_k\zero,
        \qquad S=S_0\zero^{N-|S_0|}.
      \]
      Construct the packed representation of $S$ using initialization
      and concatenation from \cref{pr:packed-representation}.
      The text $\widehat\Text$ starts at position $1$ in $S$,
      and $\widehat Y'_i$ starts at
      $s_i=n+2+\sum_{h=1}^{i-1}(|Y'_h|+1)$.
      Compute these starting positions by scanning the lengths $|Y'_i|$,
      and construct the LCE structure from \cref{th:lce} for $S$.
    \item Next, we sort the pattern indices by $Y'_i$ and the text
      indices by $Y_j$.
      To this end, construct arrays $A'_Y[1\dd k]$ and $A_Y[1\dd t]$
      satisfying
      \[
        \begin{aligned}
          A'_Y[i]&=s_i &&\text{for every }i\in[1\dd k],\\
          A_Y[j]&=r_j+\delta_j &&\text{for every }j\in[1\dd t].
        \end{aligned}
      \]
      The strings $\widehat Y'_i$ and $\widehat Y_j$ occur
      at these positions in $S$, and each is followed by zero.
      Both arrays list their positions in ascending numerical order,
      as verified below.
      Apply \cref{th:sparse-suffix-sorting} to $S$ and $A'_Y$ to
      obtain $\LexSorted{\{A'_Y[i]:i\in[1\dd k]\}}{S}$
      in an array $B'[1\dd k]$.
      Independently, apply the same theorem to $S$ and $A_Y$ to
      obtain $\LexSorted{\{A_Y[j]:j\in[1\dd t]\}}{S}$
      in an array $B[1\dd t]$.
      We now construct arrays $Q[1\dd k]$ and $D[1\dd t]$ containing
      the corresponding pattern and text indices, as follows.
      Construct arrays $E'[1\dd k]$ and $E[1\dd t]$ satisfying
      $E'[h]=(B'[h],h)$ for $h\in[1\dd k]$ and
      $E[h]=(B[h],h)$ for $h\in[1\dd t]$.
      Sort each array in ascending order of the first coordinates
      of its pairs. For every $i\in[1\dd k]$, set $Q[h]:=i$,
      where $h$ is the second coordinate of $E'[i]$.
      For every $j\in[1\dd t]$, set $D[h]:=j$,
      where $h$ is the second coordinate of $E[j]$.
      Since zero precedes every symbol of the lifted strings,
      $Q$ lists the pattern indices in lexicographic order of $Y'_i$,
      while $D$ lists the text indices in lexicographic order of $Y_j$.
      Stably sort the indices $i$ in $Q$ by $\kappa(H'_i)$ and
      the indices $j$ in $D$ by $\kappa(H_j)$.
      Each array is now grouped by root, with the strings in each
      group still in lexicographic order.
    \item We then construct an array $Q'[1\dd k]$ that orders
      the pattern indices by $\widehat Y'_i d$ within each root group.
      Denote by $S'$ the string obtained from $S$ by replacing the
      zero at $s_i+|Y'_i|$ with $d$ for every $i\in[1\dd k]$.
      Make a copy of the packed representation of $S$.
      In this copy, replace the symbols at these $k$ positions using the update
      operation of \cref{pr:packed-representation}, obtaining the
      packed representation of $S'$.
      Apply \cref{th:sparse-suffix-sorting} to $S'$ and $A'_Y$,
      replacing $B'[1\dd k]$ by
      $\LexSorted{\{A'_Y[i]:i\in[1\dd k]\}}{S'}$.
      Set $E'[h]=(B'[h],h)$ for every $h\in[1\dd k]$, and sort
      the pairs in $E'$ in ascending order of their first coordinates.
      For every $i\in[1\dd k]$, set $Q'[h]:=i$,
      where $h$ is the second coordinate of $E'[i]$.
      Stably sort its indices $i$ by $\kappa(H'_i)$.
      Distinct strings $\widehat Y'_i d$ cannot be prefixes of one
      another. Thus, the continuation of each suffix after this
      string does not affect their order.
    \item We now compute the interval endpoints and the answers
      for each root group. Traverse the groups in $Q$ and $Q'$
      together, in ascending order of their root pairs, while
      advancing a pointer through the groups in $D$ to find the
      same root. If there is no such group in $D$, leave the
      corresponding entries of $W$ equal to zero.
      Otherwise, denote the pattern indices in the two groups by
      $i_1,\ldots,i_v$ and $i'_1,\ldots,i'_v$, respectively,
      and the text indices by $j_1,\ldots,j_g$, in their array order.
      For every pattern index $i$ in this group, we will compute
      \[
        \begin{aligned}
          b_i&=|\{y\in[1\dd g]:Y_{j_y}\prec Y'_i\}|,\\
          c_i&=|\{y\in[1\dd g]:\widehat Y_{j_y}\prec\widehat Y'_i d\}|.
        \end{aligned}
      \]
      The text suffixes having prefix $Y'_i$ are precisely those
      with indices $j_{b_i+1},\ldots,j_{c_i}$.
      We compute these endpoints in two scans and then find a
      maximum in each nonempty interval.
      \begin{enumerate}
      \item We first compute the left endpoints $b_i$.
        Initialize $x=1$. For $y=1,\ldots,g$, while $x\leq v$
        and $Y'_{i_x}\preceq Y_{j_y}$, set $b_{i_x}:=y-1$ and
        increase $x$ by one.
        After all $g$ iterations, set $b_{i_h}:=g$ for every
        $h\in[x\dd v]$.
      \item The right endpoints $c_i$ are computed using the order
        in $Q'$. Initialize $x=1$. For $y=1,\ldots,g$, while
        $x\leq v$ and $\widehat Y'_{i'_x}d\prec\widehat Y_{j_y}$,
        set $c_{i'_x}:=y-1$ and increase $x$ by one.
        After all $g$ iterations, set $c_{i'_h}:=g$ for every
        $h\in[x\dd v]$.
      \item Finally, we determine the entries of $W$ for this group.
        Construct an array $F[1\dd g]$ with $F[y]=\delta_{j_y}$ for
        every $y\in[1\dd g]$, and build the range-maximum structure
        from \cref{th:range-maximum-queries} for $F$.
        For each pattern index $i$ in the group with $b_i<c_i$,
        query the range $(b_i\dd c_i]$ and denote the returned
        index by $y_*$ and the corresponding text index by $j_*=j_{y_*}$.
        If $\delta'_i\leq \delta_{j_*}$, set
        $W[i]:=r_{j_*}+\delta_{j_*}-\delta'_i$.
        Leave $W[i]$ unchanged when the interval is empty or the
        inequality fails.
      \end{enumerate}
    \end{enumerate}
  \end{enumerate}

  We first justify that $A'_Y$ and $A_Y$ list distinct positions
  in ascending order. Each $A'_Y[i]=s_i$ is the starting position
  of $\widehat Y'_i$ in $S$, and these strings are stored in the
  order of their indices, proving this property for $A'_Y$.
  Suppose that $j<j'$ but $r_j+\delta_j\geq r_{j'}+\delta_{j'}$.
  The periodic fragments $\Text[r_j\dd e_j)$ and
  $\Text[r_{j'}\dd e_{j'})$ then overlap in at least
  $\delta_{j'}\geq3\tau-1-(\tau/3-1)>2\tau/3$ symbols.
  Since their periods are at most $\tau/3$, the periodicity
  lemma~\cite{periodicitylemma} implies that their union has period
  at most $\tau/3$. Consequently, every position in $[r_j\dd r_{j'}]$
  belongs to $\RTwo{\tau}{\Text}$, contradicting that $r_j$ and
  $r_{j'}$ start different maximal blocks. Thus, $A_Y$ is also
  in ascending order. The pair sorts therefore recover precisely
  the indices corresponding to the sorted suffixes.

  We now prove that the computed array $W$ has the required properties.
  Consider any $i\in[1\dd k]$. If there is no text group
  with root $H'_i$, the set in \eqref{eq:periodic-batch-partial}
  is empty, and the initial value $W[i]=0$ is correct.
  Otherwise, the first scan assigns $b_i=y-1$
  at the smallest $y$ with $Y'_i\preceq Y_{j_y}$, or $b_i=g$
  if no such $y$ exists. This is precisely the number of text
  suffixes in the group that precede $Y'_i$.
  The second scan assigns $c_i=y-1$ at the smallest $y$ with
  $\widehat Y'_i d\prec\widehat Y_{j_y}$, or $c_i=g$ if no such
  $y$ exists. Since $\widehat Y'_i d$ cannot equal any
  $\widehat Y_{j_y}$, this is the number of text suffixes whose
  lifted strings precede $\widehat Y'_i d$.
  Consequently, for every $y\in[1\dd g]$,
  \[
    Y'_i\text{ is a prefix of }Y_{j_y}
    \quad\Longleftrightarrow\quad b_i<y\leq c_i.
  \]
  The text index $j_*$ identified by the range-maximum query
  therefore maximizes $\delta_j$ among those with root $H'_i$ for
  which $Y'_i$ is a prefix of $Y_j$.
  By \eqref{eq:periodic-batch-partial}, the pattern occurs if and
  only if the interval $(b_i\dd c_i]$ is nonempty and
  $\delta_{j_*}\geq\delta'_i$. In that case, $r_{j_*}+\delta_{j_*}-\delta'_i$ is an
  occurrence, as required.

  We finish by checking the bounds and the assumptions of the tools.
  The periodic prefix of each $\Pat_i$ has length at least
  $3\tau-1$, and its final incomplete root copy has length at most
  $|H'_i|-1$. Hence $\delta'_i\geq3\tau-|H'_i|\geq2$ for every
  $i\in[1\dd k]$. Since $k\geq1$, we obtain
  \[
    |S_0|=n+\sum_{i=1}^{k}(m_i-\delta'_i)+k+1
      \leq N-2k+k+1\leq N.
  \]
  Moreover, $\RunEndPat{\Pat_i}{\tau}\leq m_i$ implies
  $\delta'_i<m_i$, and hence every $Y'_i$ is nonempty.
  The bound on $\tau$ gives
  $d+1=\AlphabetSize+2\leq\AlphabetSize^2\leq N$.
  LCE and sparse suffix sorting therefore apply with size parameter
  $N$ and alphabet size $d+1$, and alphabet lifting applies with size
  parameter $N$ and alphabet sizes $\AlphabetSize$ and $d+1$.
  All basic packed operations use length bound $N$ and alphabet
  size $\AlphabetSize$ or $d+1$; their word-size assumptions follow
  from $w\geq1+\log N$.
  The periodic-navigation tools and the integer-value structure use
  size parameter $N$ and alphabet size $\AlphabetSize$, with
  $|\Text'|=N$ and $\tau\leq\tfrac12\log_{\AlphabetSize}N$. Furthermore,
  \[
    k\leq\frac{N}{3\tau-1},\qquad
    t\leq\frac{n}{2\tau},
  \]
  where the bound on $k$ follows from $m_i\geq3\tau-1$ for every
  $i\in[1\dd k]$, and the bound on $t$ follows from \cref{lm:runs}.
  Since $\tau\leq\tfrac12\log_{\AlphabetSize}N$, both
  $N/\log_{\AlphabetSize}N$ and $\sqrt N$ are $\bigO(N/\tau)$.
  Storing each pattern length in one word therefore uses
  $\bigO(k)=\bigO(N/\tau)$ words of space.
  Extracting the packed representations of $Y'_1,\ldots,Y'_k$ takes
  $\bigO(k+N\log\AlphabetSize/w)=\bigO(N/\tau)$ time.
  Constructing the packed representations of $\Text'$, $S$, and $S'$,
  the structures, and the periodic intervals takes
  $\bigO(N/\tau)$ total time and words of space.
  Every root has length at most $\tau/3$, allowing its integer
  value to be computed in $\bigO(1)$ time.
  All periodic-position queries take $\bigO(1)$ time.
  The queries for the patterns take
  $\bigO(k+\sum_{i=1}^{k}m_i/\log_{\AlphabetSize}N)
    =\bigO(N/\tau)$ total time.

  Each comparison in the two endpoint scans also takes
  $\bigO(1)$ time using the LCE structure for $S$ and symbol
  access from \cref{pr:packed-representation}.
  Specifically, for a pattern index $i$ and a text index $j$, compute
  $h=\min\{|Y'_i|,|Y_j|,\LCE{S}{s_i}{r_j+\delta_j}\}$.
  If $h<\min\{|Y'_i|,|Y_j|\}$, both comparisons are determined
  by the symbols at positions $s_i+h$ and $r_j+\delta_j+h$.
  Otherwise, $Y'_i\preceq Y_j$ holds exactly when $|Y'_i|\leq|Y_j|$,
  while $\widehat Y'_i d\prec\widehat Y_j$ is false.

  Since $\log(d+1)=\Theta(\log\AlphabetSize)$, the three calls to
  sparse suffix sorting take
  $\bigO(N/\log_{\AlphabetSize}N+k+t)
    =\bigO(N/\tau)$ time and words of space.
  The other sorting operations use a constant number of integer
  fields of $\bigO(\log N)$ bits. Radix sorting with digits of
  $\lfloor\frac12\log N\rfloor$ bits takes
  $\bigO(k+t+\sqrt N)=\bigO(N/\tau)$ total time.
  The endpoint scans take $\bigO(k+t)$ total time, since each
  pointer advances at most once per pattern and each scan visits
  every text index once. The range-maximum structures take
  $\bigO(t)$ total time to construct, and their queries take
  $\bigO(k)$ total time.
  All arrays, including the input and output, and all invoked
  structures occupy $\bigO(N/\tau)$ words.
  This proves the claimed time and space bounds.
\end{proof}

\begin{theorem}[Batched matching for arbitrary periodic patterns]\label{th:periodic-pattern-occurrences}
  Let $N,\AlphabetSize\in\Z_{\geq2}$ satisfy $\AlphabetSize\leq N$.
  Consider the word RAM model with word size $w\geq1+\log N$.
  Given $N$, $\AlphabetSize$, an integer $\tau\geq1$ satisfying
  $\tau\leq\tfrac12\log_{\AlphabetSize}N$, integers $n\in[1\dd N]$ and
  $k\in\Z_{\geq0}$, the packed representation
  $\PackedRepresentation{w}{\AlphabetSize}{\Text}$
  (\cref{def:packed-representation}) of a text
  $\Text\in\IntegerAlphabet^n$, and arrays containing the lengths
  and packed representations of $\tau$-periodic patterns
  $\Pat_1,\ldots,\Pat_k\in\IntegerAlphabet^+$
  (\cref{def:periodic-pattern}), in this order, where
  $n+\sum_{i=1}^{k}|\Pat_i|\leq N$,
  we can deterministically compute an array $W[1\dd k]$ such that,
  for every $i\in[1\dd k]$,
  \begin{itemize}
  \item if $\OccTwo{\Pat_i}{\Text}=\emptyset$, then $W[i]=0$;
  \item otherwise, $W[i]\in\OccTwo{\Pat_i}{\Text}$.
  \end{itemize}
  The algorithm takes $\bigO(N/\tau)$ time and
  uses $\bigO(N/\tau)$ words of space.
  The patterns need not be distinct.
\end{theorem}
\begin{proof}
  Denote $m_i=|\Pat_i|$ for every $i\in[1\dd k]$.
  If $k=0$, return the empty array.
  Compute $\lceil\log\AlphabetSize\rceil$ in
  $\bigO(\log N)$ time.
  We partition the patterns according to whether their initial
  short period continues throughout the pattern, and then apply
  the two preceding propositions.
  \begin{enumerate}
  \item First, we compute $f_i=\RunEndPat{\Pat_i}{\tau}$ for
    every $i\in[1\dd k]$.
    Construct the structure from \cref{pr:efull}.
    For each $i\in[1\dd k]$, query this structure with $\tau$, $m_i$,
    and the packed representation of $\Pat_i$ to obtain $f_i$.
  \item Next, we find occurrences for both groups of patterns.
    Construct an array $I[1\dd u]$ listing the indices $i$ with
    $f_i=m_i+1$ and an array $J[1\dd v]$ listing those with
    $f_i\leq m_i$, both in ascending order.
    Construct the corresponding arrays of lengths and original
    packed representations, and apply
    \cref{pr:fully-periodic-pattern-occurrences} with parameter $\tau$ to
    $\Pat_{I[1]},\ldots,\Pat_{I[u]}$ and
    \cref{pr:partially-periodic-pattern-occurrences} with parameter $\tau$ to
    $\Pat_{J[1]},\ldots,\Pat_{J[v]}$.
    Use the original text in both calls.
    Denote the returned arrays by $U[1\dd u]$
    and $V[1\dd v]$, respectively.
    Set $W[I[r]]:=U[r]$ for every $r\in[1\dd u]$ and
    $W[J[r]]:=V[r]$ for every $r\in[1\dd v]$.
  \end{enumerate}
  Every pattern belongs to exactly one group, and both calls use
  the same value of $\tau$ as the claim. The two propositions
  therefore give the required answers.

  Every result applied in this proof uses size parameter $N$ and
  alphabet size $\AlphabetSize$.
  The assumptions imply $\tau\leq\tfrac12\log_{\AlphabetSize}N$ and
  $w\geq1+\log N$, as required by \cref{pr:efull}.
  Furthermore,
  $k\leq N/(3\tau-1)=\bigO(N/\tau)$.
  Since $\tau\leq\tfrac12\log_{\AlphabetSize}N$, both
  $N/\log_{\AlphabetSize}N$ and $\sqrt N$ are $\bigO(N/\tau)$.
  Constructing the structure and computing the ends of all periodic prefixes take
  \[
    \bigO\!\left(\sqrt N+k+
      \frac{\sum_{i=1}^{k}m_i}{\log_{\AlphabetSize}N}\right)
      =\bigO(N/\tau)
  \]
  time and words of space. Constructing the two input collections
  and placing their answers in $W$ take
  $\bigO(k+\sum_{i=1}^{k}m_i/\log_{\AlphabetSize}N)$ time.
  Both calls preserve the total-length bound and take
  $\bigO(N/\tau)$ time and words of space,
  proving both bounds.
\end{proof}

\subsection{The General Construction}
\label{sec:sparse-anchor-construction}

\begin{theorem}\label{th:sparse-anchors-short-offsets}
  Let $N,\AlphabetSize\in\Z_{\geq2}$ satisfy $\AlphabetSize\leq N$.
  Consider the word RAM model with word size $w\geq1+\log N$.
  Given $N$, $\AlphabetSize$, an integer $\tau\geq1$ satisfying
  $\tau\leq\tfrac16\log_{\AlphabetSize}N$, integers $n\in[1\dd N]$ and
  $k\in\Z_{\geq0}$, the packed representation
  $\PackedRepresentation{w}{\AlphabetSize}{\Text}$
  (\cref{def:packed-representation}) of a text
  $\Text\in\IntegerAlphabet^n$, and arrays containing the lengths
  $m_1,\ldots,m_k$ and packed representations of patterns
  $\Pat_1,\ldots,\Pat_k\in\IntegerAlphabet^+$, in this order, where
  \[
    n+\sum_{i=1}^k m_i\leq N,
    \qquad m_i=|\Pat_i|\geq3\tau-1
      \quad\text{for every }i\in[1\dd k],
  \]
  we can deterministically compute a set $C\subseteq[1\dd n]$
  of anchor positions with $|C|\leq31N/\tau$ and an array
  $\Delta[1\dd k]$ with entries in $[0\dd\tau)$ such that,
  for every $i\in[1\dd k]$, the pattern $\Pat_i$ occurs in
  $\Text$ if and only if there exists $c\in C$ satisfying
  \[
    \begin{aligned}
      1&\leq c-\Delta[i]\leq n-m_i+1,\\
      \Pat_i&=\Text[c-\Delta[i]\dd c-\Delta[i]+m_i).
    \end{aligned}
  \]
  The set $C$ is returned as an array listing its elements.
  The algorithm takes $\bigO(N/\tau)$ time
  and uses $\bigO(N/\tau)$ words of space.
  The patterns need not be distinct.
\end{theorem}
\begin{proof}
  If $k=0$, return an empty set and an empty array.
  Compute $\lceil\log\AlphabetSize\rceil$ in
  $\bigO(\log N)$ time.
  If $\tau\leq2$, every input pattern is $\tau$-nonperiodic because
  $\per{\Pat_i[1\dd3\tau-1]}\geq1>\tau/3$.
  In this case, the claim follows from
  \cref{pr:sparse-anchors-nonperiodic-short-offsets} with parameter $\tau$.
  Henceforth, assume that $\tau\geq3$. We proceed in three steps.
  \begin{enumerate}
  \item\label{step:sparse-anchors-separate} First, we separate the indices of the nonperiodic and periodic
    patterns into arrays $I_n$ and $I_p$, respectively.
    To this end, denote $d=3\tau-1$ and construct the data structure
    from \cref{pr:periods-of-short-strings} with length bound $d$.
    Initialize both arrays to be empty.
    For $i=1,\ldots,k$, extract the packed representation of
    $\Pat_i[1\dd d]$ using \cref{pr:packed-representation}
    and query the structure to obtain $p_i=\per{\Pat_i[1\dd d]}$.
    Append $i$ to $I_p$ if $p_i\leq\tau/3$, and to
    $I_n$ otherwise.
    Denote $k_n=|I_n|$ and $k_p=|I_p|$.
  \item\label{step:sparse-anchors-anchors} Next, we obtain anchor positions for both groups.
    Use the original text in both calls below, supplying each group's
    pattern lengths and packed representations in the order of
    its index array.
    \begin{enumerate}
    \item Apply \cref{pr:sparse-anchors-nonperiodic-short-offsets}
      with parameter $\tau$ to the patterns indexed by $I_n$.
      Denote the returned set by $C_n$ and the returned
      array by $\Delta_n[1\dd k_n]$.
    \item Apply \cref{th:periodic-pattern-occurrences} with parameter $\tau$
      to the patterns indexed by $I_p$.
      Denote the returned array by $W[1\dd k_p]$.
    \end{enumerate}
  \item We now combine the outputs.
    Construct an array $B$ by copying the array listing $C_n$
    and appending $W[r]$ for every $r\in[1\dd k_p]$
    with $W[r]>0$.
    Sort $B$ in ascending order and remove consecutive equal entries
    to obtain the array listing $C$.
    Initialize every entry of $\Delta[1\dd k]$ to zero, and set
    $\Delta[I_n[r]]:=\Delta_n[r]$ for every
    $r\in[1\dd k_n]$.
    Return $C$ and $\Delta$.
  \end{enumerate}

  Step~\ref{step:sparse-anchors-separate} partitions the patterns according to
  \cref{def:periodic-pattern}. Consider a pattern $\Pat_i$
  that occurs in $\Text$.
  \begin{itemize}
  \item If $i=I_n[r]$, the nonperiodic proposition supplies
    an anchor $c\in C_n$ satisfying both conditions in the
    claim with offset $\Delta_n[r]$.
    The third step retains this anchor and assigns
    $\Delta[i]=\Delta_n[r]$.
  \item If $i=I_p[r]$, the periodic-pattern theorem returns
    $W[r]\in\OccTwo{\Pat_i}{\Text}$.
    The third step places this position in $C$ and sets $\Delta[i]=0$.
    Thus, $c=W[r]$ satisfies both conditions in the claim.
  \end{itemize}
  Conversely, those conditions specify an occurrence of $\Pat_i$
  in $\Text$ starting at $c-\Delta[i]$.
  Every entry of $\Delta$ belongs to $[0\dd\tau)$, and
  \[
    |C|\leq|C_n|+k_p
      \leq\frac{30N}{\tau}+\frac{N}{3\tau-1}
      \leq\frac{31N}{\tau}.
  \]

  We finish with the time and space bounds.
  Every result applied in this proof uses size parameter $N$ and
  alphabet size $\AlphabetSize$.
  Since $d<\frac12\log_{\AlphabetSize}N$, the shortest-period
  structure applies and takes
  $\bigO(\AlphabetSize^d d)=\bigO(\sqrt N\log N)
    =\bigO(N/\tau)$ time to construct.
  The last bound follows from $\tau=\bigO(\log N)$ and
  $(\log N)^2=\bigO(\sqrt N)$.
  Substring extraction uses alphabet size $\AlphabetSize$ and length
  bound $N$, and every extraction and period query takes $\bigO(1)$ time.
  Both calls in Step~\ref{step:sparse-anchors-anchors} preserve the total-length bound and
  satisfy the word-size and $\tau$ assumptions of their respective results.
  Each takes $\bigO(N/\tau)$ time and words of space.
  To sort $B$, use $\lfloor\frac12\log N\rfloor$-bit digits
  and a constant number of counting-sort passes, taking
  $\bigO(|C_n|+k_p+\sqrt N)=\bigO(N/\tau)$ time.
  Since $k\leq N/(3\tau-1)$ and
  $N/\log_{\AlphabetSize}N=\bigO(N/\tau)$,
  constructing the two input collections, the remaining scans and assignments,
  as well as the parameter computation, fit within
  $\bigO(N/\tau)$ time.
  The input occupies $\bigO(k+N\log\AlphabetSize/w)=\bigO(N/\tau)$ words,
  and the total running time bounds all additional space.
  This proves both claimed bounds.
\end{proof}

\section{Alphabet Reductions}\label{sec:alphabet-reductions}

\subsection{Preliminaries}\label{sec:alphabet-reductions-preliminaries}

\begin{definition}[Padding of a binary string]\label{def:pad}
  For any $k \in \Zp$ and $X \in \BinaryAlphabet^{k}$, by $\pad{X}$
  we denote the length-$2k$ string $Y \in \BinaryAlphabet^{2k}$ such
  that for every $i \in [1 \dd k]$, it holds $Y[2i-1] = X[i]$ and $Y[2i] = {\tt 0}$.
\end{definition}

\begin{lemma}\label{lm:sn-to-dm-alphabet-reduction}
  Let $\AlphabetSize\in\Z_{\geq2}$, $\Textlen\in\Z_{\geq0}$, and $m\in\Z_{\geq1}$,
  and let $\Text \in \IntegerAlphabet^{\Textlen}$ and $\Pat \in \IntegerAlphabet^{m}$.
  Denote $b = \lceil \log \AlphabetSize \rceil$. For every $a \in \IntegerAlphabet$,
  denote $C(a) = {\tt 1}^{b+1} \cdot {\tt 0} \cdot \bin{b}{a} \cdot {\tt 0}$ (see \cref{def:bin}). Let
  \begin{align*}
    \Text_{\rm bin} &= C(\Text[1])C(\Text[2])\cdots C(\Text[\Textlen]),\\
    \Pat_{\rm bin} &= C(\Pat[1])C(\Pat[2])\cdots C(\Pat[m]).
  \end{align*}
  Let also $\delta = 2b+3$. Then, it holds
  \[
    \OccTwo{\Pat_{\rm bin}}{\Text_{\rm bin}} = \{1 + \delta \cdot (j-1) : j \in \OccTwo{\Pat}{\Text}\}.
  \]
\end{lemma}
\begin{proof}
  Denote $Q = \{1 + \delta \cdot (j-1) : j \in \OccTwo{\Pat}{\Text}\}$.

  The inclusion $Q \subseteq \OccTwo{\Pat_{\rm bin}}{\Text_{\rm bin}}$ follows immediately
  by definition of $\Text_{\rm bin}$ and $\Pat_{\rm bin}$.

  We now show that $\OccTwo{\Pat_{\rm bin}}{\Text_{\rm bin}} \subseteq Q$. Let
  $j' \in \OccTwo{\Pat_{\rm bin}}{\Text_{\rm bin}}$. Observe that
  $\OccTwo{{\tt 1}^{b+1}}{\Text_{\rm bin}} = \{1 + \delta \cdot (j-1) : j \in [1 \dd \Textlen]\}$.
  Since ${\tt 1}^{b+1}$ is a prefix of $\Pat_{\rm bin}$,
  we thus obtain that there exists $j \in [1 \dd \Textlen]$ such that $j' = 1 + \delta \cdot (j-1)$. By
  definition of $\Text_{\rm bin}$, this implies that $C(\Text[j])C(\Text[j+1]) \cdots C(\Text[j+m-1]) = \Pat_{\rm bin}
  = C(\Pat[1])C(\Pat[2]) \cdots C(\Pat[m])$, which in turn, by definition of $C(\cdot)$, implies that
  $\Text[j \dd j +m) = \Pat$. We therefore obtain $j \in \OccTwo{\Pat}{\Text}$, and hence $j' \in Q$.
\end{proof}

\begin{proposition}\label{pr:lpf-encoding-construction}
  Let $N,\AlphabetSize\in\Z_{\geq3}$ satisfy
  $\AlphabetSize\leq N$.
  Denote
  $k=\lceil\log\AlphabetSize\rceil$ and $B={\tt 1}^{2k-1}{\tt 0}$.
  Consider the word RAM model with word size $w\geq1+\log N$.
  For every $a\in\IntegerAlphabet$, let
  $C(a)=\pad{\bin{k}{a}}$ (\cref{def:bin,def:pad}).
  For a string $\Text\in\IntegerAlphabet^{\Textlen}$ with
  $\Textlen\geq1$ and $4k\Textlen\leq N$, let
  \[
    \Text'=\textstyle\bigodot_{i=1}^{\Textlen}\big(B\cdot C(\Text[i])\big)
      \in\BinaryAlphabet^{4k\Textlen}.
  \]
  Given $N$, $\AlphabetSize$, $\Textlen$, and the packed representation
  $\PackedRepresentation{w}{\AlphabetSize}{\Text}$ of $\Text$
  (\cref{def:packed-representation}), we can in
  $\bigO(N/\log N)$ time and using
  $\bigO(N/\log N)$ words of space
  compute $k$, the length $|\Text'|=4k\Textlen$, and the packed
  representation $\PackedRepresentation{w}{2}{\Text'}$ of $\Text'$.
\end{proposition}
\begin{proof}
  We compute $k$ and the output length $U=4k\Textlen$ in
  $\bigO(\log N)$ time, and proceed as follows.
  \begin{enumerate}

  \item First, we construct the packed representation of the binary string
    \[
      X=\bin{k}{\Text[1]}\cdots\bin{k}{\Text[\Textlen]}.
    \]
    To this end, build the data structure from
    \cref{th:packed-binary-encoding}, and query it on the supplied
    packed representation of $\Text$ and its length.
    Preprocessing takes $\bigO(\sqrt N)$ time, and the query
    takes $\bigO(1+\Textlen/\log_{\AlphabetSize}N)$ time.

  \item Next, we construct the packed representation of
    $Y=C(\Text[1])\cdots C(\Text[\Textlen])$ by inserting a zero
    after every bit of $X$.
    Let $\mu$ be the morphism with $\mu(\zero)=\zero\zero$ and
    $\mu(\one)=\one\zero$, so that $\mu(X)=Y$.
    Using \cref{pr:packed-representation}, construct an array containing
    the packed representations of $\mu(\zero)$ and $\mu(\one)$.
    Build the data structure from \cref{th:packed-morphism}
    with this array in $\bigO(\sqrt N)$ time.
    Query the structure on the packed representation of $X$ and its
    length to obtain the packed representation of $Y=\mu(X)$ in
    $\bigO(1+k\Textlen/\log N)$ time.

  \item Finally, we construct the packed representation of $\Text'$
    by prepending $B$ to every string $C(\Text[i])$.
    To this end, initialize the packed representation of $\one^{2k}$
    and change its last symbol to $\zero$ using
    \cref{pr:packed-representation}, obtaining the representation of $B$.
    Apply \cref{pr:packed-string-repetition} to obtain the packed
    representation of $B^{\Textlen}$ in $\bigO(1+\log\Textlen+k\Textlen/w)$ time.
    View it as the representation of a sequence of $\Textlen$ copies of $B$.
    The packed representation of $Y$ also represents the sequence
    $C(\Text[1]),\ldots,C(\Text[\Textlen])$, whose strings have
    common length $2k$.
    Apply \cref{th:packed-pointwise-concatenation-partition-offline}
    to prepend each copy of $B$ to the corresponding string, giving the required
    packed representation of $\Text'$.
    These operations take
    $\bigO(\sqrt N+\log\Textlen+k\Textlen/\log N)$ time.
    Return $k$, $U$, and the resulting representation.
  \end{enumerate}
  All constructed strings have length at most $U\leq N$, which is also the
  total output length of the pointwise concatenation.
  The binary mapping and morphism produce strings of lengths
  $k\Textlen$ and $2k\Textlen$, respectively, both at most $U$.
  Thus, all three tools apply with size parameter $N$ and the
  stated word size, the binary mapping with alphabet size
  $\AlphabetSize$ and the other two with alphabet size $2$.
  Repetition and the basic packed operations
  also use length bound $N$.
  Since $k\Textlen\leq N/4$, the total time, including preprocessing
  and parameter computation, is
  \[
    \bigO\left(\sqrt N+\log N+\frac{k\Textlen}{\log N}\right)
      =\bigO(N/\log N).
  \]
  The data structures and all working space, including the input and
  output, occupy $\bigO(N/\log N)$ words altogether.
\end{proof}

\subsection{Dictionary Matching}\label{sec:dm-alphabet-reduction}

\begin{theorem}\label{th:dm-alphabet-reduction}
  Let $N,\AlphabetSize\in\Z_{\geq3}$ satisfy
  $\AlphabetSize\leq N$.
  Denote $b:=\lceil\log\AlphabetSize\rceil$ and $\delta:=2b+3$.
  Consider the word RAM model with word size $w\geq1+\log N$.
  Given the integers $N$, $\AlphabetSize$, $\Textlen$, $k$, $m$, the packed
  representation $\PackedRepresentation{w}{\AlphabetSize}{\Text}$
  of a text $\Text\in\IntegerAlphabet^{\Textlen}$, and the
  packed representation $\PackedRepresentation{w}{\AlphabetSize}{\Pat_i}$
  of every pattern in a set
  $\mathcal D=\{\Pat_1,\ldots,\Pat_k\}\subseteq\IntegerAlphabet^m$
  (\cref{def:packed-representation}), where $\Textlen\geq2$,
  $k,m\geq1$, and
  $\delta(\Textlen+km)\leq N$, we can in
  $\bigO(k+N/\log N)$ time and using
  $\bigO(k+N/\log N)$ words of space
  construct the packed representation
  $\PackedRepresentation{w}{2}{\Text_{\rm bin}}$ of a binary text
  $\Text_{\rm bin}$ and the packed representation
  $\PackedRepresentation{w}{2}{\Pat_{i,\rm bin}}$ of every
  pattern in a set
  $\mathcal D_{\rm bin}=\{\Pat_{1,\rm bin},\ldots,\Pat_{k,\rm bin}\}
    \subseteq\BinaryAlphabet^*$ such that:
  \begin{itemize}
  \item $|\mathcal D_{\rm bin}|=k$,
    $|\Text_{\rm bin}|=\delta\Textlen=\Theta(\Textlen\log\AlphabetSize)$,
    and $|\Pat_{i,\rm bin}|=\delta m=\Theta(m\log\AlphabetSize)$
    for every $i\in[1\dd k]$.
  \item For every $i\in[1\dd k]$, the pattern $\Pat_i$ occurs in
    $\Text$ if and only if $\Pat_{i,\rm bin}$ occurs in $\Text_{\rm bin}$.
  \end{itemize}
  The algorithm also returns the lengths of the constructed strings.
\end{theorem}
\begin{proof}
  Let $C$ be defined as in \cref{lm:sn-to-dm-alphabet-reduction}.
  We proceed in two steps.
  \begin{enumerate}

  \item First, we prepare the data structure for encoding strings.
    Compute $M=\Textlen+km$, $b$, $\delta=2b+3$, and $U=\delta M$ in
    $\bigO(\log N)$ time, and then build the data structure from
    \cref{pr:basic-alphabet-mapping} with size parameter $N$ and
    alphabet size $\AlphabetSize$.
    The construction takes $\bigO(\sqrt N)$ time and uses
    $\bigO(\sqrt N)$ words of space.

  \item Next, we construct the binary text and dictionary.
    Query the structure on the supplied packed representations of
    $\Text,\Pat_1,\ldots,\Pat_k$ and their lengths to obtain the
    packed representations of
    \begin{align*}
      \Text_{\rm bin}&=C(\Text[1])\cdots C(\Text[\Textlen]),\\
      \Pat_{i,\rm bin}&=C(\Pat_i[1])\cdots C(\Pat_i[m])
        \qquad\text{for }i\in[1\dd k].
    \end{align*}
    Compute and return the lengths $\delta\Textlen$ and $\delta m$
    together with the resulting representations.
  \end{enumerate}

  The encoded strings have total length $U=\delta M\leq N$.
  Thus, every query satisfies the output-length bound, and the encoding
  tool applies with the stated alphabet and word size.
  The total running time is
  \[
    \bigO(\sqrt N+\log N+k+M/\log_{\AlphabetSize}N)
      =\bigO(k+N/\log N),
  \]
  since $M\log\AlphabetSize\leq U\leq N$.
  All original and constructed representations occupy
  $\bigO(k+M\log\AlphabetSize/w)=\bigO(k+N/\log N)$ words.
  Together with the data structure and the space used during
  one query, this gives the claimed total space bound.
  The injectivity and fixed length $\delta$ of the code $C$ ensure that
  the output patterns are distinct and have the claimed lengths.
  The occurrence equivalence follows from
  \cref{lm:sn-to-dm-alphabet-reduction} applied to $\Text$ and each $\Pat_i$.
\end{proof}

\subsection{String Nesting}\label{sec:sn-alphabet-reduction}

\begin{theorem}\label{th:sn-alphabet-reduction}
  Let $N,\AlphabetSize\in\Z_{\geq3}$ satisfy
  $\AlphabetSize\leq N$.
  Denote $b:=\lceil\log\AlphabetSize\rceil$.
  Consider the word RAM model with word size $w\geq1+\log N$.
  Let $p,k,m\in\Z_{\geq1}$ satisfy $b(2p+k)m\leq N$,
  and let $\mathcal P\subseteq\IntegerAlphabet^m\times\IntegerAlphabet^m$
  and $\mathcal Q\subseteq\IntegerAlphabet^{\leq m}\times
  \IntegerAlphabet^{\leq m}$ satisfy $|\mathcal P|=p$,
  $|\mathcal Q|=k$, and $|A|+|B|=m$ for every
  $(A,B)\in\mathcal Q$. Given $N$, $\AlphabetSize$, $p$, $k$, $m$, and arrays listing
  these sets, with every string $S$ specified by its length and
  packed representation $\PackedRepresentation{w}{\AlphabetSize}{S}$
  (\cref{def:packed-representation}), we can compute $m'=mb$ and sets
  $\mathcal P_{\rm bin}\subseteq\BinaryAlphabet^{m'}\times\BinaryAlphabet^{m'}$
  and $\mathcal Q_{\rm bin}\subseteq\BinaryAlphabet^{\leq m'}\times
  \BinaryAlphabet^{\leq m'}$ satisfying the following properties:
  \begin{itemize}
  \item $|\mathcal P_{\rm bin}|=|\mathcal P|$ and
    $|\mathcal Q_{\rm bin}|=|\mathcal Q|$.
  \item For every $(A',B')\in\mathcal Q_{\rm bin}$,
    $|A'|+|B'|=m'$.
  \item There exist $(X,Y)\in\mathcal P$ and $(A,B)\in\mathcal Q$
    such that $A$ is a suffix of $X$ and $B$ is a prefix of $Y$
    if and only if there exist $(X',Y')\in\mathcal P_{\rm bin}$ and
    $(A',B')\in\mathcal Q_{\rm bin}$ such that $A'$ is a suffix
    of $X'$ and $B'$ is a prefix of $Y'$.
  \end{itemize}
  The output sets are returned in arrays listing their pairs, with each
  string $S'$ supplied by its length and packed representation
  $\PackedRepresentation{w}{2}{S'}$.
  The algorithm takes
  $\bigO(p+k+N/\log N)$ time and uses
  $\bigO(p+k+N/\log N)$ words of space.
\end{theorem}
\begin{proof}
  For every $a\in\IntegerAlphabet$, let $C(a):=\bin{b}{a}$
  (\cref{def:bin}). For every $S\in\IntegerAlphabet^*$, denote
  $C(S):=C(S[1])\cdots C(S[|S|])$.
  This fixed-length code is injective. Hence, $A$ is a suffix of
  $X$ if and only if $C(A)$ is a suffix of $C(X)$, and the
  analogous equivalence holds for prefixes.
  The target collections are
  \[
    \begin{aligned}
      \mathcal P_{\rm bin}&:=\{(C(X),C(Y)):(X,Y)\in\mathcal P\},\\
      \mathcal Q_{\rm bin}&:=\{(C(A),C(B)):(A,B)\in\mathcal Q\}.
    \end{aligned}
  \]
  Their cardinalities, string lengths, and the required
  equivalence follow immediately from these code properties.
  We construct their packed representations in two steps:
  \begin{enumerate}

  \item We first prepare the binary encoding structure.
    To this end, compute $b$ and $m'=mb$.
    Construct the data structure from \cref{th:packed-binary-encoding}
    with size parameter $N$ and alphabet size $\AlphabetSize$.
    This step takes $\bigO(\log N+\sqrt N)=\bigO(\sqrt N)$ time.

  \item Next, we construct $\mathcal P_{\rm bin}$ and
    $\mathcal Q_{\rm bin}$. For each string $S$ in the input pairs,
    query the structure to obtain the packed representation of $C(S)$,
    and store it together with its length $b|S|$ in the corresponding
    output pair. Each query takes
    $\bigO(1+|S|/\log_{\AlphabetSize}N)$ time.
    There are $2p+2k$ queries, and the input strings have total length
    $(2p+k)m\leq N/b$. Thus, this step takes
    $\bigO(p+k+N/\log N)$ time.
  \end{enumerate}
  Every query string $S$ satisfies $b|S|\leq bm\leq N$,
  as required by the encoding tool, and its word-size assumption
  is the same as in the claim.
  Since $\sqrt N=\bigO(N/\log N)$, the total running time
  satisfies the claimed bound.
  The input occupies $\bigO(p+k+N/w)$ words,
  which also satisfies this bound, and the running time bounds the
  remaining space, including the output.
\end{proof}

\subsection{Inverse Suffix Arrays}\label{sec:isa-alphabet-reduction}

\begin{lemma}\label{lm:isa-alphabet-reduction}
  Let $\Textlen\in\Z_{\geq1}$ and $\AlphabetSize\in\Z_{\geq3}$, and let
  $\Text \in \IntegerAlphabet^{\Textlen}$ be a string.
  Denote $k = \lceil \log \AlphabetSize \rceil$. For
  every $a \in \IntegerAlphabet$, let
  \[
     C(a) = {\tt 1}^{k+1} \cdot {\tt 0} \cdot \bin{k}{a} \cdot {\tt 0}
  \]
  (see \cref{def:bin}).
  Let $S = \bigodot_{i=1}^{\Textlen} C(\Text[i])$,
  $\Delta = |S| - |\Text|$, and $\delta = 2k + 3$.
  Then, for every $j \in [1 \dd \Textlen]$, it holds
  \[
    \ISA{\Text}[j] = \ISA{S}[(j-1)\delta + 1] - \Delta.
  \]
\end{lemma}
\begin{proof}
  Let $\mathcal P=\{(j-1)\delta+1:j\in[1\dd\Textlen]\}$ denote
  the set of starting positions of the codewords in $S$.
  We first establish two properties of the suffix order.
  \begin{enumerate}
  \item First, we prove that, for every $i,j\in[1\dd\Textlen]$,
    \[
      \Text[i\dd\Textlen]\prec\Text[j\dd\Textlen]
      \quad\Longleftrightarrow\quad
      S[(i-1)\delta+1\dd|S|]\prec S[(j-1)\delta+1\dd|S|].
    \]
    All codewords have length $\delta$, and
    $C(a)\prec C(a')$ if and only if $a\prec a'$.
    Thus, if two suffixes of $\Text$ differ at some position, their
    first differing symbols determine the order of both suffixes
    and of their corresponding concatenations of codewords.
    If one suffix is a proper prefix of the other, the same is true
    of the two concatenations.
    Since $S[(j-1)\delta+1\dd|S|]=C(\Text[j])\cdots C(\Text[\Textlen])$
    for every $j\in[1\dd\Textlen]$, the equivalence follows.
  \item Next, we show that every suffix starting outside $\mathcal P$
    is smaller than every suffix starting in $\mathcal P$.
    Each codeword begins with ${\tt 1}^{k+1}{\tt 0}$ and ends
    with $\bin{k}{a}{\tt 0}$ for some $a\in\IntegerAlphabet$.
    In particular, each codeword ends in zero.
    A suffix starting outside $\mathcal P$ therefore begins with
    at most $k$ consecutive ones followed by a zero, whereas
    every suffix starting in $\mathcal P$ begins with $k+1$ ones.
    This proves the required comparison.
  \end{enumerate}
  Consider any $j\in[1\dd\Textlen]$.
  Among the suffixes starting in $\mathcal P$, exactly
  $\ISA{\Text}[j]-1$ are smaller than $S[(j-1)\delta+1\dd|S|]$.
  All $|S|-|\mathcal P|=\Delta$ remaining suffixes are also smaller.
  By the definition of the inverse suffix array (\cref{def:isa}),
  we obtain
  \[
    \ISA{S}[(j-1)\delta+1]
      =1+\Delta+(\ISA{\Text}[j]-1)
      =\Delta+\ISA{\Text}[j],
  \]
  which is equivalent to the claim.
\end{proof}

\begin{theorem}
  \label{th:isa-alphabet-reduction-linear-to-binary}
  \label{th:isa-alphabet-reduction-poly-to-binary}
  Let $N,\AlphabetSize\in\Z_{\geq3}$ satisfy
  $\AlphabetSize\leq N$, and denote
  $k=\lceil\log\AlphabetSize\rceil$.
  Consider the word RAM model with word size $w\geq1+\log N$.
  Given $N$, $\AlphabetSize$, $\Textlen$, and the packed representation
  $\PackedRepresentation{w}{\AlphabetSize}{\Text}$ of a string
  $\Text\in\IntegerAlphabet^{\Textlen}$
  (\cref{def:packed-representation}), where $\Textlen\geq1$ and
  $(2k+3)\Textlen\leq N$, we can in
  $\bigO(N/\log N)$ time and using
  $\bigO(N/\log N)$ words of space
  compute an integer $\Delta\geq0$, the length
  $|\Text'|$, and the packed representation
  $\PackedRepresentation{w}{2}{\Text'}$ of a binary
  string $\Text'$ such that $|\Text'|=(2k+3)\Textlen=
  \Theta(\Textlen\log\AlphabetSize)$ and, for every
  $j\in[1\dd\Textlen]$,
  \[
    \ISA{\Text}[j]
      =\ISA{\Text'}[(2k+3)(j-1)+1]-\Delta.
  \]
\end{theorem}
\begin{proof}
  Let $C$, $S$, $\Delta$, and $\delta$ be defined as in
  \cref{lm:isa-alphabet-reduction}.
  We proceed in three steps.
  \begin{enumerate}

  \item First, we prepare the encoding structure.
    Compute $k=\lceil\log\AlphabetSize\rceil$ and $U=(2k+3)\Textlen$
    in $\bigO(\log N)$ time.
    Build the structure from \cref{pr:basic-alphabet-mapping}.
    This takes $\bigO(\sqrt N)$ time and uses
    $\bigO(\sqrt N)$ words of space.

  \item Next, we construct the packed representation of
    \[
      S=C(\Text[1])\cdots C(\Text[\Textlen]).
    \]
    To this end, query the structure on the supplied packed
    representation of $\Text$ and its length.
    Since its codewords are precisely $C(a)$, the query returns the required
    representation in $\bigO(1+\Textlen/\log_{\AlphabetSize}N)$ time.

  \item Finally, compute $\Delta=U-\Textlen$ and return $|S|=U$, $\Delta$, and
    the packed representation of $S$, taking $\Text'=S$.
  \end{enumerate}

  The encoded string has length $U=\delta\Textlen\leq N$.
  Hence, the encoding tool applies with size parameter $N$,
  alphabet size $\AlphabetSize$, and the stated word size.
  The output length and offset are at most $N$ and fit in one word.
  The total running time is
  \[
    \bigO(\sqrt N+\Textlen/\log_{\AlphabetSize}N+\log N)
      =\bigO(N/\log N),
  \]
  since $\Textlen\log\AlphabetSize\leq U\leq N$.
  The input and output representations occupy
  $\bigO(1+\Textlen\log\AlphabetSize/w)$ words.
  Together with the data structure and the space used during
  the query, this gives the claimed total space bound.
  Correctness follows from \cref{lm:isa-alphabet-reduction}.
\end{proof}

\subsection{Longest Previous Factors}\label{sec:lpf-alphabet-reduction}

\begin{lemma}\label{lm:lpf-alphabet-reduction}
  Let $\Textlen\in\Z_{\geq1}$ and $\AlphabetSize\in\Z_{\geq3}$, and let
  $\Text \in \IntegerAlphabet^{\Textlen}$ be a string.
  Denote $k = \lceil \log \AlphabetSize \rceil$. For every $a \in \IntegerAlphabet$,
  let $C(a) = \pad{\bin{k}{a}}$ (see \cref{def:bin,def:pad}).
  Let also $B = {\tt 1}^{2k-1} {\tt 0}$. Consider a string
  \[
    \Text' = B \cdot C(\Text[1]) \cdot B \cdot C(\Text[2]) \cdots B \cdot C(\Text[\Textlen]).
  \]
  Denote $\delta = 4k$. Then, for every $i \in [1 \dd \Textlen]$, it holds
  \[
    \LPF{\Text}[i] = \left\lfloor \frac{\LPF{\Text'}[(i-1)\delta + 1]}{\delta} \right\rfloor\text{ and }
    \LPnF{\Text}[i] = \left\lfloor \frac{\LPnF{\Text'}[(i-1)\delta + 1]}{\delta} \right\rfloor.
  \]
\end{lemma}
\begin{proof}
  Let $i\in[1\dd\Textlen]$, and denote $j=(i-1)\delta+1$.
  For every $r\in\Z_{\geq1}$, we prove that
  \[
    \LPF{\Text'}[j]\geq r\delta
    \quad\Longleftrightarrow\quad
    \LPF{\Text}[i]\geq r.
  \]
  \begin{itemize}
  \item Suppose first that $\LPF{\Text}[i]\geq r$.
    There exists $i'\in[1\dd i)$ such that
    $\Text[i'\dd i'+r)=\Text[i\dd i+r)$.
    Denote $j'=(i'-1)\delta+1$.
    By the construction of $\Text'$, replacing each symbol $a$ of these
    equal factors with $B\cdot C(a)$ gives
    $\Text'[j'\dd j'+r\delta)=\Text'[j\dd j+r\delta)$.
    Since $j'<j$, we obtain $\LPF{\Text'}[j]\geq r\delta$.
  \item Conversely, suppose that $\LPF{\Text'}[j]\geq r\delta$.
    There exists $j'\in[1\dd j)$ such that
    $\Text'[j'\dd j'+r\delta)=\Text'[j\dd j+r\delta)$.
    The latter factor starts with ${\tt 1}^{2k-1}$.
    Every $C(a)$ contains no consecutive ones and ends in zero,
    while $B={\tt 1}^{2k-1}{\tt 0}$ and $2k-1\geq3$.
    Thus, $\OccTwo{{\tt 1}^{2k-1}}{\Text'}
      =\{1+(t-1)\delta:t\in[1\dd\Textlen]\}$.
    In particular, $j'=(i'-1)\delta+1$ for some $i'\in[1\dd i)$.
    The two equal factors consist of $r$ complete blocks of length $\delta$,
    giving $C(\Text[i'+t])=C(\Text[i+t])$ for every $t\in[0\dd r)$.
    Since $C$ is injective by the definitions of $\bin{k}{\cdot}$ and
    $\pad{\cdot}$, we obtain
    $\Text[i'\dd i'+r)=\Text[i\dd i+r)$.
    Hence, $\LPF{\Text}[i]\geq r$.
  \end{itemize}

  The argument also applies to non-overlapping previous occurrences. Indeed, the positions
  $i'$ and $j'=(i'-1)\delta+1$ in both implications satisfy
  $j'+r\delta\leq j$ if and only if $i'+r\leq i$.
  These are precisely the additional conditions in the definitions of
  $\LPnF{\Text'}[j]$ and $\LPnF{\Text}[i]$. Consequently,
  $\LPnF{\Text'}[j]\geq r\delta$ if and only if
  $\LPnF{\Text}[i]\geq r$.

  Both equivalences also hold for $r=0$.
  Denote $r'=\LPF{\Text}[i]$. Applying the first equivalence with
  $r=r'$ and $r=r'+1$, we obtain
  \[
    r'\delta\leq\LPF{\Text'}[j]<(r'+1)\delta.
  \]
  The only integer solution is
  $r'=\lfloor\tfrac{\LPF{\Text'}[j]}{\delta}\rfloor$,
  proving the first identity.
  For $r'=\LPnF{\Text}[i]$, the second equivalence gives
  $r'\delta\leq\LPnF{\Text'}[j]<(r'+1)\delta$,
  whose only integer solution is
  $r'=\lfloor\tfrac{\LPnF{\Text'}[j]}{\delta}\rfloor$.
\end{proof}

\begin{theorem}
  \label{th:lpf-alphabet-reduction-linear-to-binary}
  \label{th:lpf-alphabet-reduction-poly-to-binary}
  Let $N,\AlphabetSize\in\Z_{\geq3}$ satisfy
  $\AlphabetSize\leq N$, and denote
  $k=\lceil\log\AlphabetSize\rceil$.
  Consider the word RAM model with word size $w\geq1+\log N$.
  Given $N$, $\AlphabetSize$, $\Textlen$, and the packed representation
  $\PackedRepresentation{w}{\AlphabetSize}{\Text}$ of a string
  $\Text\in\IntegerAlphabet^{\Textlen}$
  (\cref{def:packed-representation}), where $\Textlen\geq1$ and
  $4k\Textlen\leq N$, we can in
  $\bigO(N/\log N)$ time and using
  $\bigO(N/\log N)$ words of space
  compute the length $|\Text'|$ and the packed representation
  $\PackedRepresentation{w}{2}{\Text'}$ of a binary string
  $\Text'$ such that $|\Text'|=4k\Textlen=
  \Theta(\Textlen\log\AlphabetSize)$ and, for every
  $i\in[1\dd\Textlen]$,
  \[
    \LPF{\Text}[i]
      =\left\lfloor\frac{\LPF{\Text'}[4k(i-1)+1]}{4k}\right\rfloor,
    \qquad
    \LPnF{\Text}[i]
      =\left\lfloor\frac{\LPnF{\Text'}[4k(i-1)+1]}{4k}\right\rfloor.
  \]
\end{theorem}
\begin{proof}
  Apply \cref{pr:lpf-encoding-construction} with size parameter
  $N$ and alphabet size $\AlphabetSize$ to the supplied length and
  packed representation of $\Text$. Its assumptions are the same
  as those of the theorem, and it constructs the packed representation
  of the string $\Text'$ from \cref{lm:lpf-alphabet-reduction}, together
  with its length, within the claimed time and space bounds.
  Both identities follow from that lemma, and
  $|\Text'|=4k\Textlen=\Theta(\Textlen\log\AlphabetSize)$.
\end{proof}

\subsection{Longest Common Factors}\label{sec:lcf-alphabet-reduction}

\begin{lemma}\label{lm:lcf-alphabet-reduction}
  Let $\AlphabetSize\in\Z_{\geq3}$ and $S_1, S_2 \in \IntegerAlphabet^{+}$.
  Denote $\Textlen_1 = |S_1|$
  and $\Textlen_2 = |S_2|$. Let $k = \lceil \log \AlphabetSize \rceil$.
  For every $a \in \IntegerAlphabet$, let
  $C(a) = \pad{\bin{k}{a}}$ (see \cref{def:bin,def:pad}). Let also $B = {\tt 1}^{2k-1}{\tt 0}$.
  Consider strings
  \begin{align*}
    S'_1
      &= B \cdot C(S_1[1]) \cdot B \cdot C(S_1[2]) \cdot B \cdots B \cdot C(S_1[\Textlen_1]) \cdot B,\\
    S'_2
      &= B \cdot C(S_2[1]) \cdot B \cdot C(S_2[2]) \cdot B \cdots B \cdot C(S_2[\Textlen_2]) \cdot B.
  \end{align*}
  Then, it holds
  \[
    \LCF{S_1}{S_2} = \left\lfloor \frac{\LCF{S'_1}{S'_2} - 2k}{4k} \right\rfloor.
  \]
\end{lemma}
\begin{proof}
  Denote $\delta=4k$, $\ell=\LCF{S_1}{S_2}$, and
  $\ell'=\LCF{S'_1}{S'_2}$. We prove that
  \[
    \delta\ell+2k\leq\ell'<\delta(\ell+1)+2k.
  \]
  This implies the claim since the inequalities can be rewritten as
  $\tfrac{\ell'-2k}{4k}-1<\ell\leq\tfrac{\ell'-2k}{4k}$,
  for which $\ell=\lfloor\tfrac{\ell'-2k}{4k}\rfloor$ is the only
  integer solution.
  It thus remains to show the inequalities.

  For the lower bound, let $X$ be a common factor of $S_1$ and
  $S_2$ of length $\ell$. By construction, the string
  $\big(\bigodot_{t=1}^{\ell}(B\cdot C(X[t]))\big)\cdot B$
  occurs in both $S'_1$ and $S'_2$ and has length $\delta\ell+2k$.
  This also covers $\ell=0$.

  To prove the upper bound, suppose that
  $\ell'\geq\delta(\ell+1)+2k$. Let
  $X=S'_1[i_1\dd i_1+\ell')=S'_2[i_2\dd i_2+\ell')$
  be a common factor of length $\ell'$, where
  $i_r\in[1\dd|S'_r|-\ell'+1]$ for $r\in\{1,2\}$.
  We reach a contradiction as follows.
  \begin{enumerate}
  \item\label{step:lcf-alphabet-reduction-congruence} First, we prove that $i_1\equiv i_2\pmod\delta$.
    Every $C(a)$ contains no consecutive ones and ends in zero,
    while $B={\tt 1}^{2k-1}{\tt 0}$ and $2k-1\geq3$.
    Consequently, for each $r\in\{1,2\}$,
    $\OccTwo{{\tt 1}^{2k-1}}{S'_r}
      =\{1+t\delta:t\in[0\dd\Textlen_r]\}$.
    These occurrences start $\delta$ positions apart, and
    $|X|\geq\delta+2k$. Thus, $X$ contains a complete occurrence
    of ${\tt 1}^{2k-1}$.
    Let $a\in[0\dd\ell'-2k+1]$ satisfy
    $X[a+1\dd a+2k)={\tt 1}^{2k-1}$.
    Its starting positions in the two strings are $i_1+a$ and $i_2+a$.
    Both are congruent to $1$ modulo $\delta$, proving the desired
    congruence.
  \item Next, we show that $S_1$ and $S_2$ have a common factor of
    length $\ell+1$.
    Let $d\in[0\dd\delta)$ satisfy
    $i_1+d\equiv2k+1\pmod\delta$.
    The same congruence holds for $i_2+d$ by Step~\ref{step:lcf-alphabet-reduction-congruence}.
    For each $r\in\{1,2\}$, consider the length-$2k$ blocks
    starting at positions $i_r+d+t\delta$ for $t\in[0\dd\ell]$.
    All these blocks lie inside the chosen occurrence of $X$, since
    \[
      d+\ell\delta+2k<\delta(\ell+1)+2k\leq\ell'.
    \]
    By their starting positions modulo $\delta$, they are complete
    blocks $C(S_r[x_r+t])$, where
    $x_r=1+(i_r+d-2k-1)/\delta$.
    In particular, $1\leq x_r\leq x_r+\ell\leq\Textlen_r$.
    Equality of the two occurrences of $X$ gives
    $C(S_1[x_1+t])=X[d+t\delta+1\dd d+t\delta+2k]
      =C(S_2[x_2+t])$ for every $t\in[0\dd\ell]$.
    The mapping $C$ is injective by the definitions of
    $\bin{k}{\cdot}$ and $\pad{\cdot}$.
    Hence, $S_1[x_1\dd x_1+\ell+1)=S_2[x_2\dd x_2+\ell+1)$,
    contradicting the definition of $\ell$.
    \qedhere
  \end{enumerate}
\end{proof}

\begin{theorem}\label{th:lcf-alphabet-reduction-linear-to-binary}
  \label{th:lcf-alphabet-reduction-poly-to-binary}
  Let $N,\AlphabetSize\in\Z_{\geq3}$ satisfy
  $\AlphabetSize\leq N$, and denote
  $k=\lceil\log\AlphabetSize\rceil$.
  Consider the word RAM model with word size $w\geq1+\log N$.
  Given $N$, $\AlphabetSize$, the lengths $\Textlen_1$, $\Textlen_2$,
  and the packed representations
  $\PackedRepresentation{w}{\AlphabetSize}{S_1}$ and
  $\PackedRepresentation{w}{\AlphabetSize}{S_2}$
  of strings $S_i\in\IntegerAlphabet^{\Textlen_i}$ for $i\in\{1,2\}$
  (\cref{def:packed-representation}), where $\Textlen_1,\Textlen_2\geq1$
  and $4k(\Textlen_1+\Textlen_2+1)\leq N$, we can in
  $\bigO(N/\log N)$ time and using
  $\bigO(N/\log N)$ words of space compute
  the lengths $|S'_i|$ and the packed representations
  $\PackedRepresentation{w}{2}{S'_i}$ of binary strings $S'_i$
  for $i\in\{1,2\}$ such that
  $|S'_i|=4k\Textlen_i+2k=\Theta(\Textlen_i\log\AlphabetSize)$
  for every $i\in\{1,2\}$, and
  \[
    \LCF{S_1}{S_2}
      =\left\lfloor\frac{\LCF{S'_1}{S'_2}-2k}{4k}\right\rfloor.
  \]
\end{theorem}
\begin{proof}
  Let $B$, $C$, $S'_1$, and $S'_2$ be defined as in
  \cref{lm:lcf-alphabet-reduction}. Denote $\Textlen=\Textlen_1+\Textlen_2$.
  For every $i\in\{1,2\}$, denote
  $U_i=\bigodot_{j=1}^{\Textlen_i}(B\cdot C(S_i[j]))$.
  The binary packed operations below use \cref{pr:packed-representation}
  with length bound $N$ and alphabet size $2$.
  We proceed in three steps.
  \begin{enumerate}

  \item\label{step:lcf-alphabet-reduction-construct} First, we construct the packed representations of $U_1$ and $U_2$.
    To this end, apply \cref{pr:lpf-encoding-construction} to each
    $S_i$ with size parameter $N$ and alphabet size $\AlphabetSize$.
    The proposition uses the same $B$ and $C$. It therefore constructs
    the required strings in packed form, together with $k$ and
    their lengths $|U_i|=4k\Textlen_i$.

  \item We now construct the packed representation of the string
    $B={\tt 1}^{2k-1}{\tt 0}$.
    Initialize the packed representation of ${\tt 1}^{2k}$ and
    change its last symbol to ${\tt 0}$, using
    \cref{pr:packed-representation}\eqref{pr:packed-representation-initialize}
    and \eqref{pr:packed-representation-update}.
    This takes $\bigO(1+k/w)=\bigO(1)$ time.

  \item Finally, we append the packed representation of $B$ to those of
    $U_1$ and $U_2$.
    For each $i\in\{1,2\}$, use
    \cref{pr:packed-representation}\eqref{pr:packed-representation-concat}
    to append the packed representation of $B$ to that of $U_i$,
    obtaining $S'_i=U_iB$ in packed form.
    Compute the two output lengths $4k\Textlen_i+2k$, and return
    them with the constructed representations.
    The two concatenations take $\bigO(1+\Textlen k/w)$ time.
  \end{enumerate}

  The shared construction applies because
  $\AlphabetSize\leq N$ and $4k\Textlen_i\leq N$ for each $i\in\{1,2\}$.
  The two output strings have total length $4k(\Textlen+1)\leq N$.
  Hence, the length bounds of all packed operations are satisfied,
  and $w\geq1+\log N$ meets all word-size assumptions.
  The two calls in Step~\ref{step:lcf-alphabet-reduction-construct} take $\bigO(N/\log N)$ time
  and words of space. Constructing and appending $B$ takes
  $\bigO(1+\Textlen k/w)=\bigO(N/\log N)$ time, since
  $\Textlen k\leq N/4$ and $w\geq1+\log N$.
  Including the input representations, both encoded strings, and the
  outputs, the total running time and space usage satisfy the claimed bounds.
  The length formulas follow from $|B|=|C(a)|=2k$, and the LCF identity
  follows from \cref{lm:lcf-alphabet-reduction}.
\end{proof}

\subsection{LZ77 Factorization}\label{sec:lz-alphabet-reduction}

\begin{lemma}\label{lm:lz-alphabet-reduction}
  Let $\Textlen\in\Z_{\geq1}$, $\AlphabetSize\in\Z_{\geq3}$, and
  $\Text \in \IntegerAlphabet^{\Textlen}$. Denote
  $k = \lceil \log \AlphabetSize \rceil$.
  For every $a \in [0 \dd \AlphabetSize)$, let also
  \begin{align*}
    C(a) &= {\tt 1}^{2k-1} \cdot {\tt 0} \cdot \pad{\bin{k}{a}} \cdot \pad{\bin{k}{a}}.
  \end{align*}
  Denote $S = L R \in \BinaryAlphabet^{+}$, where (brackets added for clarity):
  \[
    L = \big(\textstyle\bigodot_{a=0}^{\AlphabetSize-1} C(a) \cdot {\tt 110}\big) \cdot {\tt 1},
    \qquad
    R = \textstyle\bigodot_{i=1}^{\Textlen} C(\Text[i]).
  \]
  Then, it holds $\LZSize{S} - \LZSize{L} = \LZSize{\Text}$ and $\LZNonOvSize{S} - \LZNonOvSize{L} = \LZNonOvSize{\Text}$.
\end{lemma}
\begin{proof}
  Denote $\delta=|L|$ and $b=6k\geq12$, the common length of
  the codewords $C(a)$. We prove the two identities in three steps.
  \begin{enumerate}
  \item\label{lm:lz-alphabet-reduction-step-1}
    First, we establish a property of the encoded substrings that
    we will use to bound the longest previous factors.
    Let $i,\ell\in\Z_{\geq1}$ satisfy $i+\ell\leq\Textlen$,
    denote $j=\delta+(i-1)b$, and consider
    \[
      \begin{aligned}
        X&=S[j+4k+1\dd j+\ell b+4k]\\
         &=\pad{\bin{k}{\Text[i]}}\cdot
           \big(\textstyle\bigodot_{t=i+1}^{i+\ell-1}C(\Text[t])\big)
           \cdot{\tt 1}^{2k-1}{\tt 0}\cdot
           \pad{\bin{k}{\Text[i+\ell]}}.
      \end{aligned}
    \]
    We claim that every $x\in\OccTwo{X}{S}$ has the form
    $x=\delta+(y-1)b+4k+1$ for some
    $y\in\OccTwo{\Text[i\dd i+\ell]}{\Text}$.
    To prove this, let $x\in\OccTwo{X}{S}$. The
    expression for $X$ shows that ${\tt 1}^{2k-1}{\tt 0}$ occurs
    at position $x+2k$ in $S$.
    Every padded string $\pad{\bin{k}{a}}$ ends in ${\tt 0}$
    and contains no ${\tt 11}$, while $2k-1\geq3$ and
    ${\tt 110}$ contains only two consecutive ones.
    Consequently, every occurrence of ${\tt 1}^{2k-1}{\tt 0}$
    in $S$ starts at the beginning of a codeword.
    The occurrence at $x+2k$ cannot start at the first position of $L$,
    since $x+2k>1$.
    Any other codeword in $L$ is immediately preceded by
    ${\tt 110}$, whose consecutive ones prevent the preceding
    $2k$ symbols from being $\pad{\bin{k}{\Text[i]}}$.
    Nor can it start at the first position of $R$, because its preceding
    symbol is $S[\delta]={\tt 1}$, whereas $X[2k]={\tt 0}$.
    Thus, $x+2k=\delta+yb+1$ for some $y\in[1\dd\Textlen)$,
    giving $x=\delta+(y-1)b+4k+1$.
    Since $|X|=\ell b$ and this occurrence lies in $S$,
    we have $y+\ell\leq\Textlen$.
    Comparing the padded strings in the two occurrences of $X$
    gives $\pad{\bin{k}{\Text[y+t]}}=
    \pad{\bin{k}{\Text[i+t]}}$ for every $t\in[0\dd\ell]$.
    The mapping $a\mapsto\pad{\bin{k}{a}}$ is injective
    (\cref{def:bin,def:pad}), and hence
    $\Text[y\dd y+\ell]=\Text[i\dd i+\ell]$, as claimed.
  \item\label{lm:lz-alphabet-reduction-step-2}
    Next, we bound the values of $\LPF{S}$ and $\LPnF{S}$
    near the beginning of each codeword.
    Let $i\in[1\dd\Textlen]$, and denote $j=\delta+(i-1)b$,
    $\ell=\max(1,\LPF{\Text}[i])$, and
    $\ell_{\rm no}=\max(1,\LPnF{\Text}[i])$.
    We prove that, for every $u\in[j+1\dd j+4k]$,
    \[
      \begin{aligned}
        j+\ell b+1&\leq u+\LPF{S}[u]\leq j+\ell b+4k,\\
        j+\ell_{\rm no}b+1&\leq u+\LPnF{S}[u]
          \leq j+\ell_{\rm no}b+4k.
      \end{aligned}
    \]
    \begin{enumerate}
    \item We first prove the lower bounds at position $j+1$.
      Consider any $r\in[1\dd\Textlen-i+1]$ and denote
      $Z=S[j+1\dd j+rb]$.
      \begin{itemize}
      \item If $r=1$, then $Z=C(\Text[i])$ occurs entirely
        in $L$. Hence, $\LPF{S}[j+1]\geq b$ and
        $\LPnF{S}[j+1]\geq b$.
      \item If $r>1$ and $\LPF{\Text}[i]\geq r$, then
        there exists $i'<i$ such that
        $\Text[i'\dd i'+r)=\Text[i\dd i+r)$.
        Denote $a=\delta+(i'-1)b+1$.
        By the definition of $R$, the equality of these two
        substrings of $\Text$ implies $S[a\dd a+rb)=Z$.
        Since $a<j+1$, we obtain $\LPF{S}[j+1]\geq rb$.
        If $\LPnF{\Text}[i]\geq r$, we can additionally
        choose $i'$ satisfying $i'+r\leq i$.
        Then $a+rb\leq j+1$, giving $\LPnF{S}[j+1]\geq rb$.
      \end{itemize}
      Taking $r=\ell$ for $\LPF{S}$ and $r=\ell_{\rm no}$
      for $\LPnF{S}$ gives $\LPF{S}[j+1]\geq\ell b$ and
      $\LPnF{S}[j+1]\geq\ell_{\rm no}b$.
    \item We now prove the upper bounds at position $j+4k$.
      Let $r\in[1\dd\Textlen-i]$ and suppose that
      $\LPF{S}[j+4k]\geq rb+1$.
      Then $S[j+4k\dd j+rb+4k]$ has an earlier occurrence
      at some position $a<j+4k$.
      Its suffix $X=S[j+4k+1\dd j+rb+4k]$ occurs at
      $x=a+1$. Applying Step~\ref{lm:lz-alphabet-reduction-step-1}
      with $\ell=r$ gives $a=\delta+(y-1)b+4k$ and
      $\Text[y\dd y+r]=\Text[i\dd i+r]$ for some
      $y\in[1\dd\Textlen-r]$.
      Since $a<j+4k$, we have $y<i$, and hence
      $\LPF{\Text}[i]\geq r+1$.
      If $\LPnF{S}[j+4k]\geq rb+1$, we can additionally
      choose $a$ satisfying $a+rb+1\leq j+4k$.
      Substituting the expressions for $a$ and $j$ gives
      $(y+r-i)b+1\leq0$, and therefore $y+r<i$.
      This occurrence of $\Text[i\dd i+r]$ ends before $i$,
      proving $\LPnF{\Text}[i]\geq r+1$.
      For $r=\Textlen-i+1$, the suffix starting at $j+4k$
      has length $rb-4k+1\leq rb$, which bounds both
      $\LPF{S}[j+4k]$ and $\LPnF{S}[j+4k]$.
      Applying the implication for $\LPF{S}$ with $r=\ell$
      and that for $\LPnF{S}$ with $r=\ell_{\rm no}$,
      and using the suffix-length bound when $r=\Textlen-i+1$,
      we obtain
      $\LPF{S}[j+4k]\leq\ell b$ and
      $\LPnF{S}[j+4k]\leq\ell_{\rm no}b$.
      Otherwise, the corresponding implication would contradict
      the definition of $\ell$ or $\ell_{\rm no}$.
    \end{enumerate}
    By \cref{lm:LPF}, the bounds at $j+1$ and $j+4k$
    imply both claimed inequalities for every $u\in[j+1\dd j+4k]$.
  \item\label{lm:lz-alphabet-reduction-step-3}
    Finally, we use these bounds to count the phrases.
    Consider either the LZ77 factorization (\cref{def:lz77})
    or its non-overlapping variant (\cref{def:nonoverlapping-lz77}),
    with the same choice for $\Text$, $L$, and $S$.
    Denote by $z_L$, $z_S$, and $z_{\Text}$ the numbers of
    phrases in $L$, $S$, and $\Text$, respectively, under the
    chosen convention.
    For $U\in\{S,\Text\}$, let $e^U_r$ denote the last position
    of its $r$th phrase, and set $e^U_0=0$.
    We prove by induction that, for every $p\in[0\dd z_{\Text}]$,
    we have $z_S\geq z_L+p$ and
    \[
      \delta+be^{\Text}_p\leq e^S_{z_L+p}
        <\delta+be^{\Text}_p+4k.
    \]
    \begin{itemize}
    \item For $p=0$, observe first that the factorizations of $L$
      and $S$ agree through the first $z_L-1$ phrases.
      Indeed, if any nonfinal phrase of $L$ could be extended
      in $S$, its one-symbol extension and an earlier occurrence
      of that extension would both lie in $L$, contradicting
      the definition of the factorization of $L$.
      The $z_L$th phrase of $S$ therefore starts at or
      before $\delta$ and ends at or after $\delta$.
      The only occurrence of ${\tt 1}^{2k}$ in $S$ starts at
      position $\delta$. To see this, each codeword begins with
      exactly $2k-1$ consecutive ones, its padded strings end
      in zero and contain no consecutive ones, and the copies
      of ${\tt 110}$ contain only two consecutive ones.
      The final ${\tt 1}$ of $L$ extends the initial run of
      ones in $R$ to length $2k$.
      If the $z_L$th phrase of $S$ ended at or after
      $\delta+4k$, it would contain this unique occurrence and
      have length greater than one. An earlier occurrence of
      the phrase would give another occurrence of ${\tt 1}^{2k}$,
      a contradiction. Hence, $\delta\leq e^S_{z_L}<\delta+4k$,
      proving the base case.
    \item Let $p\in[1\dd z_{\Text}]$ and assume the claim
      for $p-1$. Denote $i=e^{\Text}_{p-1}+1$,
      $\ell=e^{\Text}_p-e^{\Text}_{p-1}$, and
      $j=\delta+(i-1)b$.
      By the definition of the chosen factorization,
      $\ell=\max(1,\LPF{\Text}[i])$ in the overlapping case
      and $\ell=\max(1,\LPnF{\Text}[i])$ in the
      non-overlapping case.
      Since $i\leq\Textlen$ and $4k<b$, the inductive bound
      implies $e^S_{z_L+p-1}<|S|$.
      Thus, the next phrase of $S$ exists, and its starting
      position $u=e^S_{z_L+p-1}+1$ belongs to
      $[j+1\dd j+4k]$.
      The lower bounds in Step~\ref{lm:lz-alphabet-reduction-step-2}
      imply that $\LPF{S}[u]$ and $\LPnF{S}[u]$ are positive.
      Thus, the next phrase ends at $u+\LPF{S}[u]-1$ in
      the overlapping case and at $u+\LPnF{S}[u]-1$ in
      the non-overlapping case.
      Applying the corresponding bounds from that step gives
      $j+\ell b\leq e^S_{z_L+p}<j+\ell b+4k$.
      Since $j+\ell b=\delta+be^{\Text}_p$, this is the
      desired inductive bound.
    \end{itemize}
    Applying the bound with $p=z_{\Text}$ and
    $e^{\Text}_{z_{\Text}}=\Textlen$ gives
    $e^S_{z_L+z_{\Text}}\geq\delta+b\Textlen=|S|$.
    This phrase must therefore be the last one, giving
    $z_S=z_L+z_{\Text}$ and proving the required identity
    for either convention. \qedhere
  \end{enumerate}
\end{proof}

\begin{theorem}\label{th:lz-alphabet-reduction-poly-to-binary}
  Let $N,\AlphabetSize\in\Z_{\geq3}$ satisfy
  $\AlphabetSize\leq N$. Denote
  $k=\lceil\log\AlphabetSize\rceil$.
  Consider the word RAM model with word size $w\geq1+\log N$.
  Given $N$, $\AlphabetSize$, $\Textlen$, and
  the packed representation $\PackedRepresentation{w}{\AlphabetSize}{\Text}$
  (\cref{def:packed-representation}) of a string
  $\Text\in\IntegerAlphabet^{\Textlen}$, where $\Textlen\geq1$ and
  $6k\Textlen+(6k+3)\AlphabetSize+1\leq N$, we can in
  $\bigO(N/\log N)$ time and using
  $\bigO(N/\log N)$ words of space
  compute an integer $\delta\in[1\dd|\Text'|]$, the length
  $|\Text'|$, and the packed representation
  $\PackedRepresentation{w}{2}{\Text'}$ of a string
  $\Text'\in\BinaryAlphabet^+$ such that
  $|\Text'|=6k\Textlen+(6k+3)\AlphabetSize+1
    =\Theta((\Textlen+\AlphabetSize)\log\AlphabetSize)$ and
  \[
    \LZSize{\Text}=\LZSize{\Text'}-\LZSize{\Text'[1\dd\delta]},
    \qquad
    \LZNonOvSize{\Text}=\LZNonOvSize{\Text'}
      -\LZNonOvSize{\Text'[1\dd\delta]}.
  \]
\end{theorem}
\begin{proof}
  Let $C(a)$, $L$, $R$, and $S=LR$ be defined as in
  \cref{lm:lz-alphabet-reduction}.
  Denote $B=\one^{2k-1}\zero$ and
  $D(a)=\pad{\bin{k}{a}}$ for every $a\in[0\dd\AlphabetSize)$,
  so that $C(a)=BD(a)D(a)$.
  Compute $k$ and
  $U=6k\Textlen+(6k+3)\AlphabetSize+1$ in $\bigO(\log N)$ time.
  Build the data structure from
  \cref{th:packed-pointwise-concatenation-partition}.
  Its construction takes
  $\bigO(\sqrt N)$ time and uses $\bigO(\sqrt N)$ words.
  The construction proceeds in three steps.
  \begin{enumerate}

  \item\label{step:lz-alphabet-reduction-text} First, we construct the packed representation of $R$.
    To this end, apply \cref{pr:lpf-encoding-construction} to $\Text$.
    The result is the packed representation
    of $E[1]\cdots E[\Textlen]$, where
    \[
      E[i]=BD(\Text[i])\in\BinaryAlphabet^{4k}
      \qquad\text{for }i\in[1\dd\Textlen].
    \]
    This also represents the sequence $E[1\dd\Textlen]$.
    Using the pointwise-partition operation, split each $E[i]$ after
    its first $2k$ symbols, obtaining the packed representation of
    the sequence $D(\Text[1]),\ldots,D(\Text[\Textlen])$.
    Then use the pointwise-concatenation operation to append each
    $D(\Text[i])$ to $E[i]$. The resulting sequence consists of
    $E[i]D(\Text[i])=C(\Text[i])$ for $i\in[1\dd\Textlen]$,
    and its packed representation is that of $R$.
    The application of \cref{pr:lpf-encoding-construction} and the two
    pointwise operations take $\bigO(N/\log N)$ total time.

  \item\label{step:lz-alphabet-reduction-codewords} Next, we compute the packed representation of $C(a)$ for every
    $a\in[0\dd\AlphabetSize)$.
    First, construct the packed representation of
    $A\in\IntegerAlphabet^{\AlphabetSize}$ satisfying $A[i]=i-1$
    for every $i\in[1\dd\AlphabetSize]$, using the initialization and
    update operations of \cref{pr:packed-representation}.
    Apply the construction from Step~\ref{step:lz-alphabet-reduction-text} to $A$, reusing the pointwise
    data structure, to obtain the packed representation of
    $C(0)\cdots C(\AlphabetSize-1)$.
    Split this representation into $\AlphabetSize$ strings of length
    $6k$ using \cref{pr:packed-representation}\eqref{pr:packed-representation-split}.
    The application of the construction from Step~\ref{step:lz-alphabet-reduction-text} takes
    $\bigO(N/\log N)$ time. Since $k=\bigO(\log N)=\bigO(w)$,
    initializing $A$ and splitting the result take
    $\bigO(\AlphabetSize)$ time. Thus, this step takes
    $\bigO(N/\log N+\AlphabetSize)$ time.

  \item Finally, we assemble the output string.
    Construct the packed representations of ${\tt 110}$ and ${\tt 1}$
    in $\bigO(1)$ time using
    \cref{pr:packed-representation}\eqref{pr:packed-representation-initialize}
    and \cref{pr:packed-representation}\eqref{pr:packed-representation-update}.
    Then concatenate these representations in the order
    \[
      C(0),{\tt 110},C(1),{\tt 110},\ldots,
      C(\AlphabetSize-1),{\tt 110},\one,
    \]
    using \cref{pr:packed-representation}\eqref{pr:packed-representation-concat},
    to obtain the packed representation of $L$.
    Append the packed representation of $R$ to that of $L$ and
    return the result with $\Text'=S$, $|\Text'|=U$, and
    $\delta=|L|=(6k+3)\AlphabetSize+1$.
    Assembling $L$ takes $\bigO(\AlphabetSize)$ time,
    and the final concatenation takes
    $\bigO(1+U/w)$ time.
  \end{enumerate}
  The pointwise structure uses size parameter $N$ and alphabet size $2$.
  Both applications of \cref{pr:lpf-encoding-construction} use
  size parameter $N$ and alphabet size $\AlphabetSize$.
  Their encoded lengths are $4k\Textlen$ and
  $4k\AlphabetSize$, both at most $U$.
  The pointwise operations in Steps~\ref{step:lz-alphabet-reduction-text}
  and~\ref{step:lz-alphabet-reduction-codewords} produce strings of total
  length at most $6k\Textlen$ and $6k\AlphabetSize$, respectively,
  both at most $U$. The basic packed operations also use length bound $N$.
  Since $U\leq N$, all word-size assumptions follow from
  $w\geq1+\log N$.
  Moreover, $k=\Theta(\log\AlphabetSize)$ gives
  $\AlphabetSize=\Theta(k\AlphabetSize/\log(k\AlphabetSize))
    =\bigO(N/\log N)$, since $k\AlphabetSize\leq N$ and
  $x/\log x$ is increasing for $x\geq3$.
  Thus, the total running time is $\bigO(N/\log N)$.
  The data structures, input, output, and intermediate representations
  together occupy $\bigO(N/\log N)$ words.
  The output length is $U=\Theta((\Textlen+\AlphabetSize)\log\AlphabetSize)$,
  and the two identities follow from \cref{lm:lz-alphabet-reduction}.
\end{proof}

\section{Equivalence of Dictionary Matching and String Nesting}\label{sec:equiv-dm-and-sn}

\subsection{Problem Definitions}\label{sec:dm-to-sn-problem-def}

\begin{problem}[Dictionary Matching]\label{prob:dm}
  \begin{description}[style=sameline,itemsep=1ex,font={\normalfont\bf}]
  \item[Input:]
    Let $\Textlen\in\Z_{\geq2}$,
    $k=\lfloor\Textlen/\log\Textlen\rfloor$, and
    $m=2\lfloor\log\Textlen\rfloor$.
    Consider the word RAM model with word size $w>\log\Textlen$.
    The input consists of $\Textlen$, $k$, $m$, the packed representation
    $\PackedRepresentation{w}{2}{\Text}$ (\cref{def:packed-representation}) of a text
    $\Text \in \BinaryAlphabet^{\Textlen}$, and
    an array containing the packed representation $\PackedRepresentation{w}{2}{\Pat}$ of every pattern
    $\Pat$ in a set $\mathcal D\subseteq\BinaryAlphabet^m$
    of size $k$.
  \item[Output:]
  A $\texttt{YES}/\texttt{NO}$ answer indicating whether there exists $\Pat \in \mathcal{D}$ that occurs in $\Text$.
  \end{description}
\end{problem}

\noindent\begin{minipage}{\linewidth}
\begin{problem}[String Nesting]\label{prob:sn}
  \begin{description}[style=sameline,itemsep=1ex,font={\normalfont\bf}]
  \item[Input:]
    Let $k\in\Z_{\geq2}$ and $m=2\lfloor\log k\rfloor$.
    Consider the word RAM model with word size $w>\log k$.
    The input consists of $k$, $m$, and arrays listing the sets of string pairs
    $\mathcal{P} \subseteq \BinaryAlphabet^{m} \times \BinaryAlphabet^{m}$ and
    $\mathcal{Q} \subseteq \BinaryAlphabet^{\leq m} \times \BinaryAlphabet^{\leq m}$,
    where $|\mathcal P|=|\mathcal Q|=k$.
    Each string $S$ in a pair is given by its packed representation
    $\PackedRepresentation{w}{2}{S}$ (\cref{def:packed-representation}).
    For every $(A,B) \in \mathcal{Q}$, the lengths $|A|$ and $|B|$
    are also given and satisfy $|A| + |B| = m$.
  \item[Output:]
    A $\texttt{YES}/\texttt{NO}$ answer indicating whether there exist $(X,Y) \in \mathcal{P}$ and $(A,B) \in \mathcal{Q}$
    such that $A$ is a suffix of $X$ and $B$ is a prefix of $Y$.
  \end{description}
\end{problem}
\end{minipage}\par
\vspace{1ex}

\begin{remark}\label{rm:sn-input-size}
  In algorithms for Dictionary Matching and String Nesting, we use a parameter
  $N\in\Z_{\geq2}$ as an upper bound on the total length of the input strings.
  For Dictionary Matching, this means $\Textlen+km\leq N$.
  For String Nesting, the pairs in $\mathcal P$ contribute $2km$ symbols
  and those in $\mathcal Q$ contribute $km$ symbols, giving the condition
  $3km\leq N$.

  Although this measure counts only string symbols, the entire input can be
  stored in $\bigO(N)$ bits when $w=\Theta(\log N)$.
  Indeed, the parameter choices in both problem definitions ensure that
  the total string length is $\Omega(k\log k)$, which implies
  $k=\bigO(N/\log N)$.
  Each input contains $\bigO(k)$ strings and integer parameters.
  Storing the strings in packed representation and each integer parameter
  in one machine word therefore uses $\bigO(N)$ bits in total.
\end{remark}

\begin{remark}[Choice of the Factor $2$]\label{rm:dm-sn-factor-two}
  In our definition of \probname{Dictionary Matching}, the text has length
  $\Textlen$, every pattern has length exactly $2\lfloor\log\Textlen\rfloor$,
  and the dictionary contains exactly
  $\lfloor\Textlen/\log\Textlen\rfloor$ patterns. For
  \probname{String Nesting}, both collections contain $k$ pairs, and the
  length parameter is fixed to $m=2\lfloor\log k\rfloor$. Thus, both strings
  in each pair $(X,Y)$ have this length, while each query pair $(A,B)$
  satisfies $|A|+|B|=m$. These requirements specify concrete computational
  problems without leaving constants implicit in asymptotic restrictions
  on the input. However, fixing the lengths raises a question:
  could allowing longer patterns or string pairs produce a harder problem?

  To address this question, we show how to transform more general inputs
  into instances satisfying the exact definitions above:
  \begin{itemize}
  \item For \probname{Dictionary Matching}, we show that a binary text
    of length $\Textlen\geq2$ and a nonempty dictionary containing $k$
    distinct patterns of any common length $m\geq1$ can be transformed
    into a constant number of new text--dictionary pairs, each satisfying
    our definition of \probname{Dictionary Matching} and containing
    $\bigO(\Textlen+km)$ binary symbols
    (\cref{pr:dm-length-reduction-general,th:dm-length-reduction}).
    There is no restriction on the number of patterns. A pattern occurs
    in the original text if and only if a pattern occurs in at least one
    of the new pairs.
    In particular, this applies to dictionaries with common pattern length
    $m=\Theta(\log\Textlen)$, regardless of the constants in this bound.
    The reduction produces a constant number of instances whose patterns
    have length exactly $2\lfloor\log n_j\rfloor$, where $n_j$ is the text
    length of the corresponding instance.
  \item For \probname{String Nesting}, we show that inputs with
    $m\leq c\log k$, for any fixed constant $c>0$, can be transformed
    into a constant number of instances satisfying our exact
    \probname{String Nesting} definition, each containing $\bigO(km)$
    binary symbols (\cref{pr:sn-length-reduction,th:sn-length-reduction}).
    A matching pair exists in the original input if and only if one
    exists in at least one of the constructed instances.
  \end{itemize}

  Thus, choosing the factor $2$ already captures the difficulty of
  \probname{Dictionary Matching} even when both the dictionary size and
  the common pattern length are arbitrary. For \probname{String Nesting},
  the same choice captures the difficulty of allowing any fixed constant
  in the logarithmic length bound. In either case, other fixed constants
  may yield equally hard problems, but none yields a harder one under
  these reductions. Furthermore, all our reductions from
  \probname{Dictionary Matching} to other problems in the hierarchy are
  proved for patterns of length $2\lfloor\log\Textlen\rfloor$, and all
  our reductions from \probname{String Nesting} are proved for
  $m=2\lfloor\log k\rfloor$. Thus, our reductions start from the hardest
  versions of \probname{Dictionary Matching} and \probname{String Nesting}
  within the classes described above, and composing them with the length
  reductions gives reductions from every version in those classes to the
  respective target problems.
\end{remark}

\subsection{Preliminaries}\label{sec:equiv-dm-and-sn-preliminaries}

\begin{proposition}\label{pr:dm-to-sn-short}
  Let $\Textlen,\AlphabetSize,m,N$ be positive integers satisfying
  $2\leq\AlphabetSize\leq\Textlen\leq N$ and
  $m\leq\frac12\log_{\AlphabetSize}\Textlen$.
  Consider the word RAM model with word size $w\geq1+\log N$.
  Given $N$, $\AlphabetSize$, $\Textlen$, $m$, the packed representation
  $\PackedRepresentation{w}{\AlphabetSize}{\Text}$
  (\cref{def:packed-representation}) of a text
  $\Text\in\IntegerAlphabet^{\Textlen}$, and an array containing the packed representation
  $\PackedRepresentation{w}{\AlphabetSize}{\Pat}$ of every pattern
  $\Pat$ in a set $\mathcal D\subseteq\IntegerAlphabet^m$, we can determine whether
  some $\Pat\in\mathcal D$ occurs in $\Text$ in
  $\bigO(\Textlen/\log_{\AlphabetSize}\Textlen)$ time using
  $\bigO(\Textlen/\log_{\AlphabetSize}\Textlen)$ words of space.
\end{proposition}
\begin{proof}
  We proceed in two steps.
  \begin{enumerate}
  \item First, we compute the frequencies of length-$m$ substrings
    of $\Text$. To this end, apply
    \cref{th:packed-substring-frequencies} to $\Text$ with $\ell=m$.
    In particular, this gives an array $F_m[0\dd\AlphabetSize^m)$
    satisfying $F_m[\Val{\AlphabetSize}{X}]=|\OccTwo{X}{\Text}|$
    for every $X\in\IntegerAlphabet^m$.
    The theorem applies because
    $1\leq m\leq\frac12\log_{\AlphabetSize}\Textlen$ and
    $w\geq1+\log N\geq1+\log\Textlen$.
  \item We now test whether any dictionary pattern occurs in $\Text$.
    Construct the structure from \cref{th:val-encoding}.
    For every $\Pat\in\mathcal D$, use this structure with input
    $m$ and $\PackedRepresentation{w}{\AlphabetSize}{\Pat}$ to
    compute $v=\Val{\AlphabetSize}{\Pat}$.
    If $F_m[v]>0$, return \texttt{YES}.
    After testing all patterns without finding an occurrence,
    return \texttt{NO}.
  \end{enumerate}
  Correctness follows because $F_m[\Val{\AlphabetSize}{\Pat}]>0$
  holds exactly when $\Pat$ occurs in $\Text$.
  Both applied results use size parameter $\Textlen$ and alphabet
  size $\AlphabetSize$.
  The integer-value structure supports all dictionary patterns because
  $m\leq\lfloor\log_{\AlphabetSize}\Textlen\rfloor$.
  Since $\mathcal D$ is a set of length-$m$ strings, we have
  $|\mathcal D|\leq\AlphabetSize^m\leq\sqrt{\Textlen}$.
  Computing $F_m$ takes $\bigO(\Textlen/\log_{\AlphabetSize}\Textlen)$
  time, constructing the integer-value structure takes
  $\bigO(\sqrt{\Textlen})$ time, and testing all patterns takes
  $\bigO(|\mathcal D|)$ time. The total is
  \[
    \bigO(\Textlen/\log_{\AlphabetSize}\Textlen
      +\sqrt{\Textlen}+|\mathcal D|)
      =\bigO(\Textlen/\log_{\AlphabetSize}\Textlen).
  \]
  Both tools use $\bigO(\Textlen/\log_{\AlphabetSize}\Textlen)$ words.
  Each packed pattern occupies $\bigO(1)$ words, and the packed text
  occupies $\bigO(\Textlen/\log_{\AlphabetSize}\Textlen)$ words.
  Thus, the total space, including the input, satisfies the same bound.
\end{proof}

\begin{lemma}[Lower bounds for Dictionary Matching]\label{lm:dm-lower-bounds}
  Let $N\in\Z_{\geq2}$. Consider the word RAM model with word size
  $w\geq1+\log N$.
  Any algorithm for \probname{Dictionary Matching}
  (\cref{prob:dm}) that, given $N$, solves every valid input of
  at most $N$ symbols over alphabet $\BinaryAlphabet$ requires
  $\Omega(N/\log N)$ time and $\Omega(N)$ bits of space
  in the worst case.
\end{lemma}
\begin{proof}
  It suffices to consider $N\geq6$. Let
  \[
    \Textlen=2^{\lfloor\log(N/3)\rfloor},\qquad
    k=\left\lfloor\frac{\Textlen}{\log\Textlen}\right\rfloor,\qquad
    m=2\lfloor\log\Textlen\rfloor.
  \]
  We have $N/6<\Textlen\leq N/3$ and $m\leq\Textlen$.
  Consider
  \[
    \Text=\zero^{\Textlen},\qquad
    \mathcal D=\{\bin{m}{i}:i\in[1\dd k]\}
  \]
  (see \cref{def:bin}).
  Since $k\leq\Textlen<2^m$, the dictionary contains $k$ distinct
  patterns of length $m$, none equal to $\zero^m$.
  This input satisfies $\Textlen+km\leq3\Textlen\leq N$ and all
  requirements of \cref{prob:dm}.
  Its answer is \texttt{NO}.
  Replacing any one pattern by $\zero^m$ preserves distinctness,
  lengths, and cardinality, but changes the answer to \texttt{YES}.

  Each pattern is supplied in a separate packed representation occupying
  $\bigO(1)$ words,
  since $m=\bigO(\log N)=\bigO(w)$.
  If an algorithm reads no word of one of these representations on the
  \texttt{NO} input, replacing that pattern by $\zero^m$ would change the answer
  without affecting the computation.
  Consequently, every algorithm requires
  $\Omega(k)=\Omega(N/\log N)$ time on this input.
  The representations of the $k$ nonempty patterns occupy at least $k$
  words in total, giving the space lower bound
  $\Omega(kw)=\Omega(N)$ bits (see \cref{sec:model}).
\end{proof}

\begin{lemma}[Lower bounds for String Nesting]\label{lm:sn-lower-bounds}
  Let $N\in\Z_{\geq2}$. Consider the word RAM model with word size
  $w\geq1+\log N$.
  Any algorithm for \probname{String Nesting}
  (\cref{prob:sn}) that, given $N$, solves every valid input of
  at most $N$ symbols over alphabet $\BinaryAlphabet$ requires
  $\Omega(N/\log N)$ time and $\Omega(N)$ bits of space
  in the worst case.
\end{lemma}
\begin{proof}
  It suffices to consider $N\geq256$. Let
  \[
    K=\left\lfloor\frac{N}{6\lceil\log N\rceil}\right\rfloor,
    \qquad s=2\lfloor\log K\rfloor.
  \]
  We have $K\geq4$ and $K=\Theta(N/\log N)$.
  In particular, $s-1\geq\lceil\log K\rceil$.
  Let $C_1,\ldots,C_K$ be distinct strings in
  $\BinaryAlphabet^{s-1}$, and let
  \[
    \mathcal P=\{(\zero C_i,\zero^s):i\in[1\dd K]\},
    \qquad
    \mathcal Q=\{(\one C_i,\emptystring):i\in[1\dd K]\}.
  \]
  We have $s\leq2\log K$ and $3Ks\leq6K\log N\leq N$.
  Thus, these sets satisfy all input requirements of \cref{prob:sn}.
  Their answer is \texttt{NO}.
  Changing the first symbol of any string $\one C_i$ in
  $\mathcal Q$ to $\zero$ preserves distinctness and all length
  conditions, but changes the answer to \texttt{YES}.

  Since $s=\bigO(\log N)=\bigO(w)$, each string $\one C_i$ is supplied in a
  separate packed representation occupying $\bigO(1)$ words.
  On the \texttt{NO} input, if no word of the representation of some
  $\one C_i$ were read, replacing that string by $\zero C_i$ would
  change the answer without affecting the computation.
  Hence, every algorithm must read at least one word of each of these
  $K$ representations and requires
  $\Omega(K)=\Omega(N/\log N)$ time on this input.
  These $K$ nonempty string representations occupy at least $K$ words
  in total, giving the space lower bound
  $\Omega(Kw)=\Omega(N)$ bits (see \cref{sec:model}).
\end{proof}

\subsection{Length Reduction for Dictionary Matching}
\label{sec:dm-length-reduction}

\begin{proposition}\label{pr:dm-length-reduction-general}
  Let $N\in\Z_{\geq2}$, and consider the word RAM model with word size
  $w\geq1+\log N$.
  There exists a deterministic algorithm that, given $N$, integers
  $\Textlen\in\Z_{\geq2}$ and $k,m\in\Z_{\geq1}$ satisfying
  \[
    s=\Textlen+km\leq N,
  \]
  the packed representation of a text $\Text\in\BinaryAlphabet^{\Textlen}$,
  and an array containing the packed representations of all patterns in
  a set $\mathcal D\subseteq\BinaryAlphabet^m$ of size $k$
  (\cref{def:packed-representation}), computes $\bigO(1)$ pairs
  $(\Text_j,\mathcal D_j)$ with the following properties.
  Each $\Text_j$ is a binary text of length $n_j$, and each
  $\mathcal D_j$ is a set of $k_j$ binary patterns of common length $m_j$,
  where
  \[
    2\leq n_j\leq s,\qquad
    k_j=\left\lfloor\frac{n_j}{\log n_j}\right\rfloor,\qquad
    m_j=2\lfloor\log n_j\rfloor.
  \]
  A pattern from $\mathcal D$ occurs in $\Text$ if and only if, for some
  $j$, a pattern from $\mathcal D_j$ occurs in $\Text_j$.
  Each output dictionary is returned in an array listing its patterns.
  All output strings are returned in packed form, together with their lengths.
  The algorithm takes $\bigO(s/\log s)$ time and uses
  $\bigO(s/\log s)$ words of space.
\end{proposition}
\begin{proof}
  We begin by outlining the reduction for long patterns. Short patterns
  can be handled directly using \cref{pr:dm-to-sn-short}.
  For longer patterns, we first compute a set of text positions called
  \emph{anchors}, which allows us to restrict the search for pattern
  occurrences to a small number of possible starting positions.

  Specifically, using \cref{th:sparse-anchors-short-offsets}, we compute
  a set of $\bigO(s/\log s)$ anchors and assign a nonnegative offset of
  $\bigO(\log s)$ to each pattern. These satisfy the following property:
  if a pattern occurs in the text, then at least one occurrence starts
  at its assigned offset before an anchor. Thus, for each pattern, it
  suffices to check the starting positions obtained by subtracting its
  offset from the anchors. The remaining task is to perform these checks
  efficiently.

  The anchors reduce the number of positions at which we need to check
  for occurrences, but the patterns themselves may still be long.
  The main idea is to shorten each pattern by replacing a long middle
  substring with an integer identifying that substring, while retaining
  a short prefix and suffix.

  To explain how we choose these three parts, recall that all assigned
  offsets are bounded by a common parameter $\tau=\bigO(\log s)$.
  For each pattern, we retain a prefix whose length equals its assigned
  offset and a suffix chosen so that the two retained parts have total
  length $\tau$. The middle substring then has length $m-\tau$, which
  is the same for every pattern. Moreover, if a pattern occurs at its
  assigned offset before an anchor $c$, its middle substring begins
  exactly at $c$. Consequently, for each anchor, we only need to
  consider one text substring of length $m-\tau$, regardless of which
  pattern we are testing. We discard anchors at which this substring
  extends beyond the text, since the middle substring of a pattern
  occurrence must lie entirely inside the text.

  We now assign distinct integers to the distinct middle substrings of
  the patterns. Each text substring receives the integer assigned to the
  equal pattern substring, if one exists, and a separate value otherwise.
  Equality between the middle substring of a pattern and a text substring
  can therefore be tested by comparing their assigned integers. These integers require
  only $\bigO(\log s)$ bits, and sparse suffix sorting and
  longest-common-extension queries
  (\cref{th:sparse-suffix-sorting,th:lce}) allow us to compute all
  assignments in $\bigO(s/\log s)$ time.

  We then construct a short binary pattern from each retained prefix,
  assigned integer, and retained suffix. For each anchor, we similarly
  construct a short binary text string from its assigned integer and the
  $\tau$ symbols immediately before and after the corresponding long
  substring. To obtain these contexts near the text boundaries, we pad
  each end of the text with $\tau$ copies of a symbol outside the binary
  alphabet. The encoding ensures that a transformed pattern occurs in
  this string exactly when the original pattern occurs at its assigned
  offset before the anchor. It also prevents occurrences from crossing
  boundaries when these short text strings are concatenated. Finally,
  we group and pad the constructed strings to obtain a constant number
  of texts and dictionaries with exactly the lengths and cardinalities
  required in the claim.

  We now give the construction in detail.
  Whenever we determine the answer directly, we return
  $(\mathtt{00},\{\mathtt{00},\mathtt{01}\})$ for \texttt{YES} and
  $(\mathtt{00},\{\mathtt{01},\mathtt{10}\})$ for \texttt{NO}.
  Both pairs satisfy the output requirements with $n_j=k_j=m_j=2$.
  If $m>\Textlen$, return the fixed \texttt{NO} pair.
  We assume henceforth that $m\leq\Textlen$ and $\log s\geq64$
  (for smaller $s$, direct comparison takes $\bigO(1)$ time).
  Denote the input patterns by $\Pat_1,\ldots,\Pat_k$, in their
  array order. Compute
  \[
    b=2\left\lfloor\frac12\log s\right\rfloor,\qquad
    \tau=\left\lfloor\frac17\log_3 s\right\rfloor.
  \]
  We proceed in four steps.
  \begin{enumerate}
  \item First, we select text positions at which to test the patterns
    and determine where to split each pattern for these tests.
    To this end, construct the data structure from
    \cref{th:alphabet-lifting}.
    Apply this structure to compute the packed representation
    $\PackedRepresentation{w}{3}{\Text}$ of $\Text$ and the packed
    representations $\PackedRepresentation{w}{3}{\Pat_i}$ of $\Pat_i$
    for every $i\in[1\dd k]$.
    We then distinguish two cases.
    \begin{enumerate}
    \item\label{step:dm-length-reduction-short}
      If $m<3\tau-1$, construct the packed representation of
      $\Text\two^{s-\Textlen}$ by initialization and concatenation using
      \cref{pr:packed-representation}.
      Apply \cref{pr:dm-to-sn-short} to this text and $\mathcal D$
      with text length $s$.
      Return the fixed pair corresponding to the computed answer.
    \item\label{step:dm-length-reduction-anchors}
      Otherwise, apply \cref{th:sparse-anchors-short-offsets}
      to $\Text$ and $\Pat_1,\ldots,\Pat_k$ with parameter $\tau$.
      This gives a set $C\subseteq[1\dd\Textlen]$ of size at most
      $31s/\tau$ and an array $\Delta[1\dd k]$ with entries in
      $[0\dd\tau]$ such that, for every $i\in[1\dd k]$,
      \[
        \OccTwo{\Pat_i}{\Text}\neq\emptyset
        \quad\Longleftrightarrow\quad
        \text{there exists }c\in C\text{ with }
        c-\Delta[i]\in\OccTwo{\Pat_i}{\Text}.
      \]
      Remove from $C$ every position
      $c>\Textlen-(m-\tau)+1$.
      If $C=\emptyset$, return the fixed \texttt{NO} pair.
    \end{enumerate}
  \item\label{step:dm-length-reduction-integers}
    Next, we assign integers to length-$(m-\tau)$ substrings of
    the patterns and text.
    For every $i\in[1\dd k]$ and $c\in C$, denote
    \[
      H_i=\Pat_i[\Delta[i]+1\dd\Delta[i]+1+m-\tau),
      \qquad G_c=\Text[c\dd c+m-\tau).
    \]
    All these strings have length $m-\tau$.
    To test whether $\Pat_i$ occurs starting at $c-\Delta[i]$, we
    compare $H_i$ with $G_c$. The remaining comparisons concern the
    first $\Delta[i]$ and last $\tau-\Delta[i]$ symbols of $\Pat_i$,
    whose total length is $\tau$.
    We assign distinct integers from $[0\dd k)$ to the distinct strings
    among $H_1,\ldots,H_k$.
    Each $G_c$ receives the same integer as $H_i$ if $G_c=H_i$ for
    some $i\in[1\dd k]$, and the integer $k$ otherwise.
    We compute these assignments in three substeps.
    \begin{enumerate}
    \item\label{step:dm-length-reduction-positions}
      We first construct a single text containing every $H_i$ and
      $G_c$ as a substring.
      Using \cref{pr:packed-representation}, concatenate the packed
      representations of $\Text,\Pat_1,\ldots,\Pat_k$ to obtain
      \[
        \Text_{\rm all}=\Text\Pat_1\cdots\Pat_k,
        \qquad |\Text_{\rm all}|=s.
      \]
      Denote $t=|C|$ and the elements of $C$ by $c_1,\ldots,c_t$,
      in the order returned by the anchor algorithm.
      Construct an array $A_{\rm pos}[1\dd k+t]$ satisfying
      \[
        \begin{aligned}
          A_{\rm pos}[i]
            &=\Textlen+(i-1)m+1+\Delta[i]
              &&\text{for }i\in[1\dd k],\\
          A_{\rm pos}[k+j]
            &=c_j
              &&\text{for }j\in[1\dd t].
        \end{aligned}
      \]
      These are the starting positions in $\Text_{\rm all}$ of $H_i$ and $G_{c_j}$,
      respectively.
    \item\label{step:dm-length-reduction-groups}
      We now group the indices of equal middle strings.
      Denote $Q=\{A_{\rm pos}[j]:j\in[1\dd k+t]\}$.
      Apply \cref{th:sparse-suffix-sorting} to $\Text_{\rm all}$ and $A_{\rm pos}$,
      obtaining
      an array $A_{\rm lex}[1\dd k+t]$ containing $\LexSorted{Q}{\Text_{\rm all}}$.
      We convert this order of positions into a permutation
      $\pi[1\dd k+t]$ of the corresponding indices as follows.
      Construct arrays $E[1\dd k+t]$ and $F[1\dd k+t]$ with
      $E[j]=(A_{\rm pos}[j],j)$ and $F[j]=(A_{\rm lex}[j],j)$ for every
      $j\in[1\dd k+t]$.
      Sort both arrays in ascending order of their first coordinates.
      For each pair of corresponding entries $(p,i)$ in $E$ and
      $(p,j)$ in $F$, set $\pi[j]=i$.
      Construct the LCE data structure from \cref{th:lce} on $\Text_{\rm all}$.
      Scan $\pi[1],\ldots,\pi[k+t]$, starting a new group before
      $\pi[r]$, for $r\in[2\dd k+t]$, precisely when
      \[
        \LCE{\Text_{\rm all}}{A_{\rm pos}[\pi[r-1]]}{A_{\rm pos}[\pi[r]]}<m-\tau.
      \]
    \item\label{step:dm-length-reduction-assign}
      Finally, we assign the integers in an array $R[1\dd k+t]$.
      Initialize $v=0$. For each group, first check whether it contains
      an index $i\in[1\dd k]$, representing a pattern substring $H_i$.
      Then assign its entries as follows.
      \begin{itemize}
      \item If it does, set $R[j]=v$ for every index $j$ in the group,
        and increase $v$ by one.
      \item Otherwise, set $R[j]=k$ for every index $j$ in the group.
      \end{itemize}
      Denote $\rho(H_i)=R[i]$ for $i\in[1\dd k]$ and
      $\rho(G_{c_j})=R[k+j]$ for $j\in[1\dd t]$.
    \end{enumerate}
  \item We are now ready to construct the transformed binary patterns
    and text pieces. For every $i\in[1\dd k]$, denote
    $\Pat_i=A_iH_iB_i$, where $|A_i|=\Delta[i]$ and
    $|B_i|=\tau-\Delta[i]$. Denote $W=\two^\tau\Text\two^\tau$.
    For every $c\in C$, denote also
    \[
      X_c=W[c\dd c+\tau),\qquad
      Y_c=W[c+m\dd c+m+\tau).
    \]
    Denote by $\mu$ the mapping given by
    \[
      \mu(\zero)=\mathtt{000},\qquad
      \mu(\one)=\mathtt{010},\qquad
      \mu(\two)=\mathtt{100},
    \]
    extended to strings by concatenation.
    Compute $z=b/2-3\tau-5$.
    For $v\in[0\dd k]$, denote by $\gamma(v)$ the string obtained from
    $\bin{b}{v}$ (\cref{def:bin}) by appending $\zero$ after
    each consecutive block of two bits. The required strings are
    \begin{align*}
      \Pat_i'&=\mu(A_i)\cdot\mathtt{01110}\cdot
        \gamma(\rho(H_i))\cdot\zero^z\cdot\mu(B_i)
          &&\text{for }i\in[1\dd k],\\*
      S_c&=\mu(X_c)\cdot\mathtt{01110}\cdot
        \gamma(\rho(G_c))\cdot\zero^z\cdot\mu(Y_c)
          &&\text{for }c\in C.
    \end{align*}
    We construct their packed representations in three substeps.
    \begin{enumerate}
    \item First, we compute the representations under $\mu$.
      Initialize the packed representation of $\two^\tau$ and
      concatenate it with those of $\Text$ and $\two^\tau$ to obtain
      the packed representation of $W$, using \cref{pr:packed-representation}.
      Extract $A_i$ and $B_i$ from each pattern using
      \cref{pr:packed-representation}.
      Extract $X_c$ and $Y_c$ from $W$ at positions
      $c$ and $c+m$, respectively,
      taking $\tau$ symbols in each case.
      Construct the structure from \cref{th:packed-binary-encoding}
      and the binary structure from
      \cref{th:packed-pointwise-concatenation-partition}.
      For each extracted string $U$, use the binary-mapping structure
      to obtain the packed representation of the sequence
      $\bin{2}{U[1]},\ldots,\bin{2}{U[|U|]}$.
      Using \cref{pr:packed-representation}, initialize the packed
      representation of $\zero^{|U|}$, viewed as a sequence of $|U|$
      one-symbol strings.
      Apply the pointwise concatenation structure to these two
      sequences to append a zero to each two-bit string, obtaining
      the packed representation of $\mu(U)$.
    \item Next, we compute the representations of the required strings
      $\gamma(v)$. Construct the binary data structure from
      \cref{th:val-decoding}.
      Initialize the packed representation of $\zero^{b/2}$ using
      \cref{pr:packed-representation}.
      For each assigned integer $v$, use the decoding structure to
      obtain the packed representation of $\bin{b}{v}$, viewed as a
      sequence of $b/2$ strings of length two.
      Apply the pointwise concatenation structure to this sequence and
      the sequence of $b/2$ one-symbol strings represented by
      $\zero^{b/2}$. This appends a zero to each two-bit string,
      producing the packed representation of $\gamma(v)$.
    \item Finally, we construct the packed representations of
      $\Pat_i'$ and $S_c$ as follows.
      Initialize the packed representations of $\mathtt{01110}$ and
      $\zero^z$, and concatenate the five representations in each
      formula using \cref{pr:packed-representation}.
      Every $\Pat_i'$ has length $2b$, and every $S_c$ has length
      $\ell=2b+3\tau$.
    \end{enumerate}
  \item Finally, we assemble the output texts and dictionaries.
    Compute $M=2^b$ and $K=\lfloor M/b\rfloor$.
    \begin{enumerate}
    \item We first construct the texts of length $M$.
      Partition the array containing the representations of
      $S_{c_1},\ldots,S_{c_t}$ into consecutive groups of
      $\lfloor M/\ell\rfloor$ entries, with the last possibly smaller.
      For each group, concatenate its strings and append enough zeros
      to obtain a text of length exactly $M$, using
      \cref{pr:packed-representation}.
    \item Next, we construct dictionaries containing $K$ patterns each.
      Partition the array containing the representations of
      $\Pat_1',\ldots,\Pat_k'$ into consecutive groups of $K$ entries,
      with the last possibly smaller.
      To each dictionary containing $k'$ transformed patterns, add
      \[
        \mathtt{1111}\cdot\zero^{b-4}\cdot\bin{b}{v},
        \qquad v\in[0\dd K-k').
      \]
      Compute their packed representations using the binary-decoding
      structure already constructed, followed by initialization and
      concatenation from \cref{pr:packed-representation}.
    \end{enumerate}
    Return every pair consisting of one constructed text and one
    constructed dictionary.
  \end{enumerate}

  We first justify the early returns and the restriction of $C$.
  In Step~\ref{step:dm-length-reduction-short}, appending copies of
  $\two$ creates no occurrence of a binary pattern, and hence the
  returned pair has the correct answer.
  In Step~\ref{step:dm-length-reduction-anchors}, removing positions
  from $C$ preserves the equivalence returned by
  \cref{th:sparse-anchors-short-offsets}: at an occurrence starting
  at $c-\Delta[i]$, the length-$(m-\tau)$ substring starting at $c$
  lies entirely inside that occurrence. In particular, if $C$ becomes
  empty, then no pattern from $\mathcal D$ occurs in $\Text$.

  We prove the correctness of the remaining construction in four steps.
  \begin{itemize}
  \item First, we show that
    Step~\ref{step:dm-length-reduction-integers} assigns the integers
    described there. The positions $A_{\rm pos}[1],\ldots,
    A_{\rm pos}[k+t]$ constructed in
    Step~\ref{step:dm-length-reduction-positions} are distinct: the
    positions for different patterns lie in disjoint pattern blocks,
    while those for distinct anchors lie in the initial copy of $\Text$.
    Since all middle strings have the same length, their equality
    classes are exactly the consecutive groups obtained in
    Step~\ref{step:dm-length-reduction-groups}.
    At most $k$ groups contain indices of the pattern substrings
    $H_1,\ldots,H_k$, and therefore all integers assigned in
    Step~\ref{step:dm-length-reduction-assign} belong to
    $[0\dd k]$. In particular,
    $\rho(H_i)=\rho(G_c)$ if and only if $H_i=G_c$.
  \item Next, we show that the binary construction preserves the
    comparisons around each anchor, i.e., for every $i\in[1\dd k]$
    and $c\in C$, we have
    \[
      \OccTwo{\Pat_i'}{S_c}\neq\emptyset
      \quad\Longleftrightarrow\quad
      \Pat_i=W[\tau+c-\Delta[i]\dd\tau+c-\Delta[i]+m).
    \]
    The definitions of the pattern pieces and the strings around $c$ give
    \[
      \Pat_i=W[\tau+c-\Delta[i]\dd\tau+c-\Delta[i]+m)
      \quad\Longleftrightarrow\quad
      \begin{cases}
        A_i\text{ is a suffix of }X_c,\\
        H_i=G_c,\\
        B_i\text{ is a prefix of }Y_c.
      \end{cases}
    \]
    Every nonempty string of the form $\mu(X)$ or $\gamma(v)$ ends
    in $\zero$ and contains no three consecutive ones.
    Consequently, $\mathtt{01110}$ occurs exactly once in each
    transformed pattern and text piece, at positions
    $3\Delta[i]+1$ and $3\tau+1$, respectively.
    Concatenating the text pieces or appending zeros creates no
    further occurrence of $\mathtt{01110}$.
    Thus, a match of $\Pat_i'$ in an output text must align this string
    with its occurrence in some $S_c$.
    The pattern has $3\Delta[i]\leq3\tau$ symbols before
    $\mathtt{01110}$ and
    $\frac32b+z+3(\tau-\Delta[i])\leq\frac32b+z+3\tau$
    symbols after it. These lengths do not exceed the corresponding
    lengths in $S_c$, and hence the match lies entirely inside $S_c$.
    At this alignment, injectivity of $\mu$ and $\gamma$, together
    with $\rho(H_i)=\rho(G_c)\Longleftrightarrow H_i=G_c$, makes
    the match equivalent to the three conditions above.
  \item Next, we verify that each output dictionary is a set of
    $K$ distinct patterns. If $\Pat_i'=\Pat_j'$, the positions of
    their unique occurrences of $\mathtt{01110}$ give
    $\Delta[i]=\Delta[j]$.
    Equality of the following length-$\frac32b$ strings gives
    $\rho(H_i)=\rho(H_j)$ and hence $H_i=H_j$, since distinct strings
    among $H_1,\ldots,H_k$ receive different integers.
    Injectivity of $\mu$ gives $A_i=A_j$ and $B_i=B_j$.
    Therefore $\Pat_i=\Pat_j$, which implies $i=j$ because the
    input patterns are pairwise distinct.
    Every occurrence of $\mathtt{111}$ in a transformed pattern or
    constructed text is immediately preceded and followed by $\zero$.
    In particular, none contains $\mathtt{1111}$.
    Each added pattern starts with $\mathtt{1111}$, and the added
    patterns have distinct length-$b$ suffixes $\bin{b}{v}$.
    Thus, the added patterns are distinct from one another and from
    the transformed patterns, and none occurs in a constructed text.
  \item Finally, we prove that some output pair has a matching pattern
    if and only if a pattern from $\mathcal D$ occurs in $\Text$.
    For every $i\in[1\dd k]$ and $c\in C$, we have
    \[
      \begin{gathered}
        1\leq c-\Delta[i]\leq\Textlen-m+1,\\
        \Pat_i=\Text[c-\Delta[i]\dd c-\Delta[i]+m)
      \end{gathered}
      \quad\Longleftrightarrow\quad
      \Pat_i=W[\tau+c-\Delta[i]\dd\tau+c-\Delta[i]+m).
    \]
    Indeed, a binary pattern cannot contain any symbol from either
    padding string in $W$, and the copy of $\Text$ starts at
    position $\tau+1$.
    The occurrence equivalence preserved when restricting $C$ shows that
    $\Pat_i$ occurs in $\Text$ if and only if $\Pat_i'$ occurs in
    some $S_c$. Every $S_c$ appears in a constructed text, every
    $\Pat_i'$ appears in a constructed dictionary, and all pairs
    of these texts and dictionaries are returned.
    As shown above, every occurrence of a transformed pattern $\Pat_i'$
    in an output text lies entirely within one of the concatenated strings
    $S_c$. Since the added patterns do not occur in the output texts,
    this proves the required equivalence.
  \end{itemize}

  We next verify the numerical bounds and the tool assumptions.
  Distinctness of the binary input patterns gives
  $k\leq\min\{s/m,2^m\}=\bigO(s/\log s)$.
  Indeed, if $m\geq\frac12\log s$, then $k\leq2s/\log s$,
  and otherwise $k<\sqrt s$.
  In the short-pattern case, $m<3\tau-1<\frac12\log_3 s$ and
  $w\geq1+\log N\geq1+\log s$, as required by
  \cref{pr:dm-to-sn-short}.
  The application of \cref{th:sparse-anchors-short-offsets} uses the
  original text and patterns, whose total length is $s$, and the value
  of $\tau$ computed above.

  For $m\geq3\tau-1$, we have $m\geq4$.
  Since $m\leq\Textlen$, we have $s\geq(k+1)m$.
  The choice of $b$ gives $s/4<M\leq s$, and hence
  \[
    k+1\leq s/4<M=2^b.
  \]
  Thus, every assigned integer can be represented using $b$ bits.
  The choices of $b$ and $\tau$ give $10\tau\leq b$.
  Together with $b\geq64$, this implies $3\tau+5\leq b/2$
  and therefore $z\geq0$.
  Moreover, $\ell=2b+3\tau\leq3b\leq M$.
  Each output text has length $M$, and each dictionary contains
  $K$ patterns of length $2b$.
  Since $\log M=b$ and $M\leq s$, the output parameters have
  the required values, and each pair contains at most
  $M+2bK\leq3M\leq3s$ symbols.
  The bounds $t\leq31s/\tau=\bigO(s/\log s)$ and
  $\ell=\bigO(\log s)$ give $t\ell=\bigO(s)$.
  Since $\lfloor M/\ell\rfloor\geq M/(2\ell)$, the partition
  of text pieces produces $\bigO(1)$ texts.
  Likewise, $K=\Theta(s/\log s)$ and $k=\bigO(s/\log s)$ give
  $\bigO(1)$ dictionaries and hence $\bigO(1)$ output pairs.

  Since $0\leq\Delta[i]\leq\tau<m$, each $H_i$ lies inside $\Pat_i$.
  For every $c\in C$, the bound $c\leq\Textlen-(m-\tau)+1$
  ensures that $G_c$ lies inside $\Text$ and that
  $c+m+\tau-1\leq\Textlen+2\tau=|W|$.
  Thus, $X_c$ and $Y_c$ lie inside $W$.
  Moreover, $m\geq3\tau-1\geq2\tau$ gives
  $|W|=\Textlen+2\tau\leq\Textlen+m\leq s$.
  We also have $|\Text_{\rm all}|=s$ and $k+t\leq k+\Textlen\leq s$.
  Hence, \cref{th:sparse-suffix-sorting,th:lce} apply to $\Text_{\rm all}$,
  since $w\geq1+\log N\geq1+\log s$.
  Our direct calls to \cref{pr:packed-representation} use length
  bound $s$ and symbol widths $1$ and $2$ for alphabets of sizes
  $2$ and $3$, respectively.
  Every other result applied in this proof uses size parameter $s$.
  The alphabet-lifting structure uses source alphabet size $2$ and
  target alphabet size $3$, the decoding and pointwise-concatenation
  structures use alphabet size $2$, and all other applied results use
  alphabet size $3$.
  The alphabet-lifting structure receives strings of length at most $s$.
  The binary-encoding queries produce strings of length at most
  $2\tau\leq s$.
  The decoding queries use length
  $b\leq\lfloor\log s\rfloor$ and integer arguments in $[0\dd2^b)$.
  The pointwise concatenation queries have component lengths two and one.
  Their output lengths are at most $3\tau\leq s$ for the strings
  $\mu(U)$ and equal to $3b/2\leq s$ for the strings $\gamma(v)$.

  Finally, we bound the running time and space.
  Computing the parameters takes $\bigO(\log s)$ time, and constructing
  the alphabet-lifting, binary-mapping, decoding, and
  pointwise-concatenation structures takes
  $\bigO(\sqrt s)$ time in total.
  Lifting the input and constructing $\Text_{\rm all}$ and $W$ take
  $\bigO(k+s/\log s)=\bigO(s/\log s)$ time.
  The short-pattern computation or anchor construction, including
  the scan removing positions from $C$, takes
  $\bigO(s/\log s)$ time.
  Sparse suffix sorting and LCE preprocessing take
  $\bigO(s/\log s+k+t)
    =\bigO(s/\log s)$ time.
  Sorting $E$ and $F$ using radix sort takes
  $\bigO(k+t+\sqrt s)=\bigO(s/\log s)$ time with digits of
  $\lfloor\frac12\log s\rfloor$ bits.
  Grouping the indices and assigning the integers take
  $\bigO(k+t)$ time, including both scans of each group.
  Each extraction and encoding of a string of length $\bigO(\log s)$
  takes $\bigO(1)$ time. Thus, constructing the binary patterns and
  text pieces takes $\bigO(k+t)$ time.
  Assembling and padding the output texts takes
  $\bigO(t+s/\log s)$ time, while constructing the added patterns
  takes $\bigO(K)$ time per dictionary.
  These bounds, including the constant number of output copies, give
  $\bigO(s/\log s)$ total time.
  The input representations occupy
  $\bigO(k+s/w)=\bigO(s/\log s)$ words, and the running time bounds
  the remaining space, including the output.
\end{proof}

The following proposition strengthens the preceding reduction by also
bounding the total length of the text and dictionary in each output instance.

\begin{proposition}
  \label{pr:dm-size-reduction}
  Let $c\geq1$ be a fixed constant and $N\in\Z_{\geq2}$.
  Consider the word RAM model with word size $w\geq1+\log N$.
  There exists a deterministic algorithm that, given $N$, integers
  $\Textlen\in\Z_{\geq2}$ and $k,m\in\Z_{\geq1}$ satisfying
  $5c\leq s=\Textlen+km\leq N$, the packed representation of a text
  $\Text\in\BinaryAlphabet^{\Textlen}$, and an array containing the
  packed representations of all patterns in a set
  $\mathcal D\subseteq\BinaryAlphabet^m$ of size $k$
  (\cref{def:packed-representation}), computes a nonempty list of
  $\bigO(1)$ valid \probname{Dictionary Matching} inputs
  (\cref{prob:dm}) with the following properties:
  \begin{itemize}
  \item Each returned input has total string length at most $s/c$.
  \item A pattern from $\mathcal D$ occurs in $\Text$ if and only if
    at least one returned input is a \texttt{YES}-instance.
  \end{itemize}
  All output strings are returned in packed form, together with their
  lengths, and each dictionary is returned in an array listing its patterns.
  The algorithm takes $\bigO(s/\log s)$ time and uses
  $\bigO(s/\log s)$ words of space. The constants in these bounds
  and in the number of output instances may depend on $c$.
\end{proposition}
\begin{proof}
  We first apply the length reduction to obtain instances whose patterns
  have logarithmic length. We then divide each text into overlapping
  substrings and each dictionary into smaller groups. Applying the
  length reduction to these smaller pairs restores the exact pattern
  length and dictionary size required by \cref{prob:dm}.

  The pairs
  $(\mathtt{000},\{\mathtt{00}\})$ and
  $(\mathtt{000},\{\mathtt{11}\})$ are fixed \texttt{YES}- and
  \texttt{NO}-instances, respectively. Both satisfy \cref{prob:dm}
  with text length three, dictionary size one, and pattern length two.
  Each contains five symbols and therefore satisfies the bound $s/c$.
  If $s<\lceil18c\rceil$, determine the answer by direct comparison
  and return a list containing the corresponding fixed instance.
  This takes $\bigO(1)$ time because $c$ is fixed.
  Henceforth, assume that $s\geq\lceil18c\rceil$.
  Compute $q=\lceil18c\rceil$ and initialize an empty list of output
  instances. We proceed as follows.
  \begin{enumerate}
  \item Apply \cref{pr:dm-length-reduction-general} to the given
    text and dictionary. Denote its returned
    pairs by $(\Text_j,\mathcal D_j)$ for $j\in[1\dd t]$, where
    $t=\bigO(1)$. For each $j$, denote the text length, dictionary
    size, and pattern length by $n_j$, $k_j$, and $m_j$, respectively.
    These satisfy $n_j\leq s$,
    $k_j=\lfloor n_j/\log n_j\rfloor$, and
    $m_j=2\lfloor\log n_j\rfloor$.

  \item For each $j\in[1\dd t]$, we construct smaller instances
    such that a pattern from $\mathcal D_j$ occurs in $\Text_j$
    if and only if at least one of them is a \texttt{YES}-instance.
    We distinguish two cases.
    \begin{enumerate}
    \item\label{step:dm-size-reduction-direct}
      If $\lfloor n_j/q\rfloor<2m_j$, compare every pattern
      in $\mathcal D_j$ with every length-$m_j$ substring of $\Text_j$.
      Add the fixed \texttt{YES}-instance to the output list if a
      comparison finds equal strings, and the fixed \texttt{NO}-instance
      otherwise.

    \item Otherwise, we divide the text and dictionary as follows.
      \begin{enumerate}
      \item\label{step:dm-size-reduction-texts}
        Compute $h_j=\lceil n_j/q\rceil$ and partition
        $[1\dd n_j]$ into consecutive intervals of length $h_j$,
        except possibly for the last interval. For each such interval
        $[a\dd b]$, extract the packed representation of
        $\Text_j[a\dd\min\{n_j,b+m_j-1\}]$.
      \item\label{step:dm-size-reduction-groups}
        Partition the array listing $\mathcal D_j$ into
        consecutive groups of $\lceil k_j/q\rceil$ patterns,
        except possibly for the last group. Copy the packed patterns
        of each group into a separate array.
      \item For every extracted substring $U$ of length at least
        $m_j$ and every dictionary group $\mathcal E$, apply
        \cref{pr:dm-length-reduction-general} with text $U$ and
        dictionary $\mathcal E$. Supply the text length $|U|$,
        dictionary size $|\mathcal E|$, common pattern length $m_j$,
        and their packed representations. Add all returned instances
        to the output list. Omit the substrings shorter than $m_j$.
      \end{enumerate}
    \end{enumerate}
  \end{enumerate}
  If the output list is empty, add the fixed \texttt{NO}-instance.
  Return the list.

  To prove correctness for a pair $(\Text_j,\mathcal D_j)$ handled
  by the partition, consider an occurrence of a pattern
  $P\in\mathcal D_j$ in $\Text_j$, starting at a position in the
  interval $[a\dd b]$ used in the partition. The occurrence ends
  at or before $b+m_j-1$ and at or before $n_j$. It is therefore
  contained in the extracted substring associated with $[a\dd b]$,
  and this substring has length at least $m_j$.
  Together with the group containing $P$, this substring gives one
  of the pairs passed to \cref{pr:dm-length-reduction-general}.
  Conversely, an occurrence of a pattern from a group
  $\mathcal E\subseteq\mathcal D_j$ in an extracted substring is
  also an occurrence in $\Text_j$.
  The calls to \cref{pr:dm-length-reduction-general} preserve these
  answers, and the fixed instances preserve the answers obtained by
  direct comparison. The correctness of the first call then proves
  the required equivalence for the original input.
  If the output list is empty, this equivalence implies that no
  original pattern occurs, and adding the fixed \texttt{NO}-instance
  preserves the answer.

  We next verify the sizes and word-size assumptions. In the case
  using the partition, we have $2m_j\leq n_j/q$ and
  $k_jm_j\leq2n_j$.
  Thus, each pair $(U,\mathcal E)$ has total string length
  $u=|U|+|\mathcal E|m_j$ satisfying
  \[
    u\leq\left\lceil\frac{n_j}{q}\right\rceil+m_j-1
      +\left\lceil\frac{k_j}{q}\right\rceil m_j
      \leq\frac{n_j+k_jm_j}{q}+2m_j
      \leq\frac{4n_j}{q}
      \leq\frac{s}{3c}.
  \]
  Every instance returned by the corresponding length-reduction call
  has text length at most $u$ and total string length at most $3u$,
  which is at most $s/c$.
  The first call to \cref{pr:dm-length-reduction-general} uses size
  parameter $s$. Each further call has $|U|\geq m_j\geq2$ and uses size
  parameter $u\leq s\leq N$, and hence the original word size suffices.
  The parameter $q$ is at most $s$, as are the group sizes and
  the total lengths $u$ supplied to the length-reduction calls.
  The packed extractions and direct comparisons use
  \cref{pr:packed-representation} with alphabet size two,
  symbol width one, and length bound $s$.

  Finally, the first length reduction takes $\bigO(s/\log s)$ time
  and words. In Step~\ref{step:dm-size-reduction-direct}, the condition
  $\lfloor n_j/q\rfloor<2m_j$ implies
  $n_j<2qm_j\leq4q\log n_j$, which bounds $n_j$ by a constant
  depending only on $c$. Hence, this case takes $\bigO(1)$ time.
  Step~\ref{step:dm-size-reduction-texts} extracts at most $q$
  substrings, and Step~\ref{step:dm-size-reduction-groups} produces at
  most $q$ nonempty dictionary groups. Thus, there are $\bigO(1)$
  pairs $(\Text_j,\mathcal D_j)$ and at most $q^2=\bigO(1)$ further
  calls for each pair.
  Each further call takes $\bigO(u/\log u)=\bigO(s/\log s)$
  time and words. The extracted substrings have total length
  $\bigO(n_j+qm_j)=\bigO(n_j)$, and extracting them takes
  $\bigO(q+n_j/w)$ time. Every packed pattern occupies
  $\bigO(1)$ words, and constructing the dictionary arrays takes
  $\bigO(k_j)$ time. Since $n_j\leq s$ and
  $k_j=\bigO(n_j/\log n_j)=\bigO(s/\log s)$,
  these operations take $\bigO(s/\log s)$ time in total.
  The same bound holds for space, including the original input,
  all intermediate representations, and the output.
\end{proof}

\begin{theorem}\label{th:dm-length-reduction}
  Let $N\in\Z_{\geq2}$. Consider the word RAM model with word size
  $w$ satisfying $1+\log N\leq w=\bigO(\log N)$. Suppose that there
  exists an algorithm for \probname{Dictionary Matching}
  (\cref{prob:dm}) that, given $N$ and any valid input of at most $N$ symbols
  over alphabet $\BinaryAlphabet$, achieves the following complexities:
  \begin{itemize}
  \item running time $T(N)$,
  \item space usage $S(N)$ bits.
  \end{itemize}
  Then, there exists an algorithm that,
  given $N$, integers $\Textlen\in\Z_{\geq2}$ and $k,m\in\Z_{\geq1}$ satisfying
  $\Textlen+km\leq N$, the packed representation of any text
  $\Text\in\BinaryAlphabet^{\Textlen}$, and an array containing the packed
  representations of all patterns in any set
  $\mathcal D\subseteq\BinaryAlphabet^m$ of size $k$
  (\cref{def:packed-representation}), decides whether a pattern from
  $\mathcal D$ occurs in $\Text$, with the following complexities:
  \begin{itemize}
  \item running time $\bigO(T(N))$,
  \item space usage $\bigO(S(N))$ bits.
  \end{itemize}
\end{theorem}
\begin{proof}
  Denote $s=\Textlen+km$. If $s<5$, direct comparison determines
  the answer in $\bigO(1)$ time using $\bigO(1)$ words.
  Assume henceforth that $s\geq5$.
  Apply \cref{pr:dm-size-reduction} with $c=1$
  to the given text and dictionary.
  Each returned pair is a valid Dictionary Matching input with
  total string length at most $s\leq N$.
  Run the assumed algorithm on every returned pair and return
  \texttt{YES} if at least one call returns \texttt{YES}, and
  \texttt{NO} otherwise. Correctness follows from the occurrence
  equivalence in \cref{pr:dm-size-reduction}.
  The size reduction and every call to the assumed algorithm use
  size parameter $N$.

  The construction takes $\bigO(s/\log s)\subseteq\bigO(N/\log N)$
  time and uses $\bigO(N/\log N)$ words, including the input and output.
  Since $w=\bigO(\log N)$, this space occupies $\bigO(N)$ bits.
  Together with the constant number of calls to the assumed algorithm,
  the total running time is $\bigO(N/\log N+T(N))$, and the space
  usage is $\bigO(N+S(N))$ bits.
  By \cref{lm:dm-lower-bounds}, we have
  $T(N)=\Omega(N/\log N)$ and $S(N)=\Omega(N)$.
  Hence, the claimed bounds follow.
\end{proof}

\subsection{Length Reduction for String Nesting}
\label{sec:sn-length-reduction}

\begin{proposition}\label{pr:sn-length-reduction}
  Let $N\in\Z_{\geq2}$.
  Consider the word RAM model with word size $w\geq1+\log N$.
  There exists a deterministic algorithm that, given $N$, integers $k\in\Z_{\geq2}$ and
  $m\in\Z_{\geq1}$ satisfying
  \[
    3km\leq N,\qquad m=\bigO(\log k),
  \]
  and arrays listing two sets $\mathcal P$ and $\mathcal Q$ of binary
  string pairs satisfying
  \begin{align*}
    |\mathcal P|&=k,& |X|=|Y|&=m
      &&\text{for every }(X,Y)\in\mathcal P,\\*
    |\mathcal Q|&=k,& |A|+|B|&=m
      &&\text{for every }(A,B)\in\mathcal Q,
  \end{align*}
  computes $\bigO(1)$ pairs of sets $(\mathcal P_t,\mathcal Q_t)$
  of binary string pairs with the following properties.
  For each $t$, there are integers $K_t\geq2$ and $s_t\geq1$ such that
  \begin{gather*}
    |\mathcal P_t|=|\mathcal Q_t|=K_t,\qquad
    s_t=2\lfloor\log K_t\rfloor,\qquad 3K_ts_t\leq90km,\\*
    |X|=|Y|=s_t\quad\text{for every }(X,Y)\in\mathcal P_t,\qquad
    |A|+|B|=s_t\quad\text{for every }(A,B)\in\mathcal Q_t.
  \end{gather*}
  Furthermore, there exist $(X,Y)\in\mathcal P$ and $(A,B)\in\mathcal Q$ such that
  $A$ is a suffix of $X$ and $B$ is a prefix of $Y$ if and only if
  such pairs exist in $\mathcal P_t$ and $\mathcal Q_t$ for some $t$.
  Each input string is supplied by its length and packed representation
  (\cref{def:packed-representation}).
  The output sets are returned in the same form, in arrays listing their pairs.
  The algorithm takes $\bigO(k)$ time and uses
  $\bigO(k)$ words of space. The constants in these bounds and in the
  number of output instances may depend on the constant implicit in
  $m=\bigO(\log k)$.
\end{proposition}
\begin{proof}
  We begin by outlining how we obtain the required string lengths while
  preserving the suffix and prefix comparisons. The main idea is to
  replace part of each pair by a short binary identifier. To do this,
  we first need to account for the fact that the division between $A$
  and $B$ may differ between pairs, even though $|A|+|B|=m$ always holds.

  We therefore group the pairs in $\mathcal Q$ according to $|A|$,
  using intervals of width $\Theta(\log k)$.
  Since $m=\bigO(\log k)$, this produces only a constant number of
  groups. We process each group together with all pairs
  in $\mathcal P$.

  Within one group, the lengths satisfy $L\leq|A|\leq L+d$, where
  $d=\bigO(\log k)$. Denote $R=m-L-d$.
  Every $A$ in this group has a suffix of length $L$, and every $B$
  has a prefix of length $R$. For each pair $(A,B)$, we concatenate
  these two substrings to obtain a string of length $m-d$.
  The parts left out are a prefix of $A$ and a suffix of $B$, whose
  total length is only $d$.

  For each pair $(X,Y)\in\mathcal P$, we similarly concatenate the
  length-$L$ suffix of $X$ and the length-$R$ prefix of $Y$.
  Equality of the two concatenated strings means precisely that the
  corresponding suffixes of $A$ and $X$ agree and that the corresponding
  prefixes of $B$ and $Y$ agree. The remaining comparisons concern
  the prefix of $A$ and suffix of $B$ that we left out.
  Crucially, $L$ and $R$ are fixed throughout the group, and therefore
  the string constructed from $(X,Y)$ can be used for every query pair
  in that group.

  We use packed string sorting (\cref{pr:packed-string-sorting}) to
  assign equal binary identifiers exactly to equal concatenated strings.
  Each identifier has $\bigO(\log k)$ bits.
  We then construct new pairs that retain the short strings needed for
  the remaining comparisons and place the identifier in the left
  component. The suffix comparison between the new left components
  checks both equality of the identifiers and the remaining prefix
  of $A$. The prefix comparison between the new right components
  checks the remaining suffix of $B$.

  Finally, we choose the group width and padding lengths so that the
  transformed strings have exactly the lengths required for a common
  output cardinality $K=\Theta(k)$. In particular, each data string has
  length $2\lfloor\log K\rfloor$, and the two strings in each query
  pair have this total length.
  We include an encoding of the original index in each transformed data
  pair to keep these pairs distinct. This encoding lies beyond every
  prefix tested by a query pair.
  We then add distinct pairs that cannot participate in a match until
  both sets contain exactly $K$ pairs. Packed string sorting and the
  remaining construction take $\bigO(k)$ time per group and in total.

  We now give the construction in detail.
  Denote $M=3km$. If $k<8$, then the assumption $m=\bigO(\log k)$
  gives $m=\bigO(1)$, and direct comparison determines the answer in
  $\bigO(1)$ time. In this case, return the fixed data set
  $\{(\mathtt{00},\mathtt{00}),(\mathtt{01},\mathtt{00})\}$
  together with $\{(\mathtt{00},\emptystring),(\mathtt{11},\emptystring)\}$
  for \texttt{YES} or
  $\{(\mathtt{10},\emptystring),(\mathtt{11},\emptystring)\}$
  for \texttt{NO}.
  These sets satisfy the output requirements with $K_t=s_t=2$.
  Henceforth, assume that $k\geq8$.
  Denote the input pairs by $(X_j,Y_j)$ for $j\in[1\dd k]$ and
  $(A_i,B_i)$ for $i\in[1\dd k]$, in their array order.
  The algorithm proceeds as follows.
  \begin{enumerate}
  \item First, we group the query pairs according to the lengths of
    their first components.
    To this end, compute
    \[
      f=\lceil\log(2k)\rceil,\qquad
      K=2^f,\qquad h=f-2,\qquad s=2f,
    \]
    and construct the groups
    $I_r=\{i\in[1\dd k]:\lfloor |A_i|/h\rfloor=r\}$ for
    $r\in[0\dd\lfloor m/h\rfloor]$.
    We also construct the data structure from \cref{th:val-decoding},
    shared by all groups.
  \item Next, for each $r\in[0\dd\lfloor m/h\rfloor]$ with
    $I_r\neq\emptyset$, we construct one output pair of sets.
    Let $I=I_r$ and
    \[
      L=rh,\qquad d=\min\{h,m-L\},\qquad
      R=m-L-d.
    \]
    Thus, $L\leq|A_i|\leq L+d$ for every $i\in I$.
    We process this group as follows.
    \begin{enumerate}
    \item First, we construct the packed representations of the pieces
      used for this group.
      For every $i\in I$, let $A_i=Z_iU_i$ and $B_i=V_iW_i$,
      where $|U_i|=L$ and $|V_i|=R$, and let $H_i=U_iV_i$.
      In particular, $|Z_i|+|W_i|=d$ and $|H_i|=m-d$.
      For every $j\in[1\dd k]$, let
      \begin{align*}
        X_j^0&=X_j[m-L-d+1\dd m-L],\\*
        G_j&=X_j[m-L+1\dd m]\cdot Y_j[1\dd R],\\*
        Y_j^0&=Y_j[R+1\dd R+d].
      \end{align*}
      Here, $|X_j^0|=|Y_j^0|=d$ and $|G_j|=m-d$.
      Extracting and concatenating substrings with
      \cref{pr:packed-representation} gives all required representations.
    \item We now replace the middle strings by their binary identifiers.
      The equality $H_i=G_j$ will be represented by equal binary
      identifiers, while the remaining suffix and prefix comparisons
      involve $Z_i$ and $W_i$, respectively.
      Sort the sequence consisting of the strings $H_i$ for $i\in I$
      followed by the strings $G_j$ for $j\in[1\dd k]$, using
      \cref{pr:packed-string-sorting}.
      Denote the returned rank of a string $H$ by $\rho(H)$.
      Since $k+|I|\leq2k\leq K$, these ranks belong to $[0\dd K)$.
      Let $g(H)=\bin{f}{\rho(H)}$ (\cref{def:bin}).
      We compute its packed representation using the shared data structure
      from \cref{th:val-decoding}.
      Using the same structure, compute the packed representation of
      $\bin{f}{j-1}$ for every $j\in[1\dd k]$.
      The transformed pairs are given by
      \begin{align*}
        X_j'&=X_j^0\cdot g(G_j)\cdot\zero^{f-d},&
        Y_j'&=Y_j^0\cdot\bin{f}{j-1}\cdot\zero^{f-d}
          &&\text{for }j\in[1\dd k],\\*
        A_i'&=Z_i\cdot g(H_i)\cdot\zero^{f-d},&
        B_i'&=W_i
          &&\text{for }i\in I.
      \end{align*}
      The string $\bin{f}{j-1}$ distinguishes the transformed data pairs.
      Since $d\leq h=f-2$, we have $f-d\geq2$.
      Initialize the required strings of zeros and concatenate the
      representations using \cref{pr:packed-representation}.
      The strings $X_j'$ and $Y_j'$ have length $s$, while
      $|A_i'|+|B_i'|=s$ for every $i\in I$.
    \item Finally, we form two sets of exactly $K$ pairs for this group.
      Let
      \[
        \mathcal P'=\{(X_j',Y_j'):j\in[1\dd k]\},\qquad
        \mathcal Q'=\{(A_i',B_i'):i\in I\}.
      \]
      These sets have cardinalities $k$ and $|I|$, respectively,
      as proved below.
      The resulting sets are
      \begin{align*}
        \mathcal P_I&=\mathcal P'\cup
          \{(\zero^{s-2}\mathtt{01},\bin{s}{v}):v\in[0\dd K-k)\},\\*
        \mathcal Q_I&=\mathcal Q'\cup
          \{(\bin{s-2}{v}\cdot\mathtt{10},\emptystring):
            v\in[0\dd K-|I|)\}.
      \end{align*}
      For each required $v<K$, obtain the packed representation of
      $\bin{f}{v}$ using \cref{th:val-decoding}.
      The identities
      $\bin{s}{v}=\zero^f\cdot\bin{f}{v}$ and
      $\bin{s-2}{v}=\zero^{f-2}\cdot\bin{f}{v}$ allow us to construct
      the added pairs with \cref{pr:packed-representation}.
      Add $(\mathcal P_I,\mathcal Q_I)$ to the output.
    \end{enumerate}
  \end{enumerate}

  We first verify the construction of $\mathcal P_I$ and $\mathcal Q_I$
  in three steps.
  \begin{itemize}
  \item First, we show that the transformation preserves every matching
    pair: for every group $I$, every $i\in I$, and every $j\in[1\dd k]$,
    \[
      \begin{aligned}
        &A_i\text{ is a suffix of }X_j
          \text{ and }B_i\text{ is a prefix of }Y_j\\
        &\quad\Longleftrightarrow\quad
          A_i'\text{ is a suffix of }X_j'
          \text{ and }B_i'\text{ is a prefix of }Y_j'.
      \end{aligned}
    \]
    By the definitions of the pieces, the condition on the left is
    equivalent to
    \[
      Z_i\text{ is a suffix of }X_j^0,\qquad
      H_i=G_j,\qquad W_i\text{ is a prefix of }Y_j^0.
    \]
    Indeed, the length-$L$ suffix of $A_i$ must equal the length-$L$
    suffix of $X_j$, and the length-$R$ prefix of $B_i$ must equal
    the length-$R$ prefix of $Y_j$.
    Together these equalities are precisely $H_i=G_j$.
    The remaining comparisons concern $Z_i$ and $W_i$.
    For the transformed left components, the final $f-d$ zeros agree,
    and the preceding $f$ bits agree if and only if $H_i=G_j$,
    by the rank assignment.
    The remaining suffix comparison is between $Z_i$ and $X_j^0$.
    Since $|W_i|\leq d$, the comparison of $B_i'=W_i$ with $Y_j'$
    only concerns its prefix $Y_j^0$, and is unaffected by the
    following string $\bin{f}{j-1}$.
    This proves the equivalence, including when some pieces are empty.
  \item Next, we verify that the pairs in each transformed set are distinct.
    For every $j\in[1\dd k]$, the string
    $Y_j'[d+1\dd d+f]=\bin{f}{j-1}$ determines $j$ uniquely.
    Thus, the transformed data pairs are distinct.
    For the query pairs, suppose that
    $(A_i',B_i')=(A_{i'}',B_{i'}')$ for $i,i'\in I$.
    Equality of the left-component lengths gives $|Z_i|=|Z_{i'}|$.
    Comparing the prefixes and the $f$ bits preceding the final $f-d$
    zeros gives $Z_i=Z_{i'}$ and $g(H_i)=g(H_{i'})$.
    Equal ranks identify equal middle strings, and hence $H_i=H_{i'}$.
    Splitting these strings after $L$ symbols gives
    $U_i=U_{i'}$ and $V_i=V_{i'}$.
    Moreover, equality of the right components gives $W_i=W_{i'}$.
    Consequently, $(A_i,B_i)=(A_{i'},B_{i'})$.
    Since the input query pairs are distinct, we obtain $i=i'$.
  \item Finally, we show that the added pairs do not create any match.
    Every left component in $\mathcal P'$ and $\mathcal Q'$ ends
    in $\mathtt{00}$, while the added data and query left components
    end in $\mathtt{01}$ and $\mathtt{10}$, respectively.
    Every query left component has length at least two.
    Thus, any comparison involving an added pair fails the suffix test.
    The added pairs are distinct within each set because their binary
    representations of $v$ are distinct.
    Their left components' final two symbols also distinguish them
    from the pairs in $\mathcal P'$ and $\mathcal Q'$, respectively.
    Hence, $|\mathcal P_I|=|\mathcal Q_I|=K$.
    Together with the first two steps, this proves that a matching
    pair exists in these sets if and only if an input query with
    index in $I$ matches an input data pair.
  \end{itemize}
  The output lengths satisfy $s_t=s=2f=2\lfloor\log K\rfloor$,
  and $K_t=K$.
  Every query belongs to one group, proving the claimed equivalence.

  We next verify the parameter bounds and the assumptions of the tools.
  By definition, $2k\leq K<4k$ and $f\geq4$.
  Thus, $h=f-2=\Theta(\log k)$, and the assumption
  $m=\bigO(\log k)$ gives at most
  $\lfloor m/h\rfloor+1=\bigO(1)$ nonempty groups.
  The data pairs are distinct elements of
  $\BinaryAlphabet^m\times\BinaryAlphabet^m$, giving $k\leq2^{2m}$.
  Hence, $f=\lceil\log(2k)\rceil\leq2m+1\leq3m$.
  Together with $K<4k$, this implies that each output pair of sets satisfies
  \[
    3Ks=6Kf<72km\leq90km.
  \]
  Since $k\geq8$ and $k\leq2^{2m}$, we have $m\geq2$, and hence
  $K<4k\leq M$ and $M\geq48$.
  All packed operations use length bound $M$ and alphabet size $2$.
  The sorting calls and the integer-decoding structure use size
  parameter $M$ and alphabet size $2$.
  The middle strings have length $m-d\leq m\leq M/3$, and the
  transformed strings have length at most $s=2f\leq2\log M\leq M$.
  Every sorting call has at most $2k\leq M$ strings.
  The integer-decoding queries request strings of length
  $f=\log K\leq\lfloor\log M\rfloor$ and integers less than $K=2^f$.
  Since $M\leq N$, we have $w\geq1+\log N\geq1+\log M$,
  as required by every tool call.

  It remains to bound the running time and space usage.
  Since $M=3km=\bigO(k\log k)$, parameter computation and construction
  of the integer-decoding structure take $\bigO(\sqrt M)=\bigO(k)$ time.
  All strings processed within a group have length
  $\bigO(\log k)$, and extracting, concatenating, or encoding each
  of them takes $\bigO(1)$ time.
  The sorting call for each group involves $k+|I|=\Theta(k)$
  middle strings of common length $m-d=\bigO(\log k)$.
  Thus, \cref{pr:packed-string-sorting} gives $\bigO(k)$ time per call.
  Constructing the additional pairs takes $\bigO(K)=\bigO(k)$ time per group.
  Since the number of groups is constant, the total running time is $\bigO(k)$.
  The input occupies $\bigO(k+M/w)=\bigO(k)$ words,
  and the running time bounds the remaining space, including the output.
\end{proof}

The following proposition strengthens the preceding reduction by also
bounding the total length of the strings in each output instance.

\begin{proposition}\label{pr:sn-size-reduction}
  Let $c\geq1$ be a fixed constant and $N\in\Z_{\geq2}$.
  Consider the word RAM model with word size $w\geq1+\log N$.
  There exists a deterministic algorithm that, given $N$, integers
  $k\in\Z_{\geq2}$ and $m\in\Z_{\geq1}$ satisfying
  $12c\leq s=3km\leq N$ and $m=\bigO(\log k)$, and arrays
  listing two sets $\mathcal P$ and $\mathcal Q$ of binary string
  pairs satisfying
  \begin{align*}
    |\mathcal P|&=k,& |X|=|Y|&=m
      &&\text{for every }(X,Y)\in\mathcal P,\\*
    |\mathcal Q|&=k,& |A|+|B|&=m
      &&\text{for every }(A,B)\in\mathcal Q,
  \end{align*}
  computes a nonempty list of $\bigO(1)$ valid
  \probname{String Nesting} inputs (\cref{prob:sn}), each of
  total string length at most $s/c$.
  There exist $(X,Y)\in\mathcal P$ and $(A,B)\in\mathcal Q$
  such that $A$ is a suffix of $X$ and $B$ is a prefix of $Y$
  if and only if at least one output is a \texttt{YES}-instance.
  Each input string is supplied by its length and packed representation
  (\cref{def:packed-representation}). The output sets are returned
  in the same form, in arrays listing their pairs.
  The algorithm takes $\bigO(k)$ time and uses $\bigO(k)$ words
  of space. The constants in these bounds and in the number of output
  instances may depend on $c$ and on the constant implicit in
  $m=\bigO(\log k)$.
\end{proposition}
\begin{proof}
  We form smaller subsets of equal cardinality in each input
  collection and apply \cref{pr:sn-length-reduction} to every
  combination of one set from each collection. The groups are small
  enough that the length reduction's increase in size still leaves
  each output within the required bound.

  Denote $q=\lceil60c\rceil$. We use the sets
  $\{(\mathtt{00},\mathtt{00}),(\mathtt{01},\mathtt{00})\}$ and
  $\{(\mathtt{00},\emptystring),(\mathtt{11},\emptystring)\}$
  as a fixed \texttt{YES}-instance. Replacing the second set with
  $\{(\mathtt{10},\emptystring),(\mathtt{11},\emptystring)\}$
  gives a fixed \texttt{NO}-instance.
  Both satisfy \cref{prob:sn} with cardinality two and length
  parameter two, and contain twelve symbols in total.
  They therefore satisfy the bound $s/c$.
  If $k<q^2$, test every
  combination $(X,Y)\in\mathcal P$ and $(A,B)\in\mathcal Q$
  by comparing $A$ with the length-$|A|$ suffix of $X$ and $B$
  with the length-$|B|$ prefix of $Y$.
  Return a list containing the fixed \texttt{YES}-instance if both
  comparisons succeed for some combination, and a list containing
  the fixed \texttt{NO}-instance otherwise.
  In this case, $k$ is bounded by a constant depending
  on $c$, and $m=\bigO(\log k)=\bigO(1)$. Thus, direct comparison
  and construction of the fixed output take $\bigO(1)$ time.

  Henceforth, assume that $k\geq q^2$. Initialize an empty list of
  output instances. We proceed in two steps.
  \begin{enumerate}
  \item First, we construct smaller sets of equal cardinality as follows.
    Compute $b=\lceil k/q\rceil$ and $t=\lceil k/b\rceil$.
    Partition each input array into $t$ consecutive groups of $b$
    pairs, except possibly for the last group.
    If the last group has fewer than $b$ pairs, replace it with the
    last $b$ pairs of the same input array.
    Copy the packed strings and their lengths from each resulting
    group into a separate array. Denote the resulting sets by
    $\mathcal P_1,\ldots,\mathcal P_t$ and
    $\mathcal Q_1,\ldots,\mathcal Q_t$, respectively.
    Each set contains $b$ distinct original pairs.
  \item Next, we apply the length reduction to these sets.
    For every $i,j\in[1\dd t]$, apply
    \cref{pr:sn-length-reduction} to $\mathcal P_i$ and
    $\mathcal Q_j$.
    Add all returned instances to the output list.
  \end{enumerate}
  If the output list is empty, add the fixed \texttt{NO}-instance.
  Return the list.

  To prove correctness, first observe that every original pair
  belongs to at least one of the constructed sets. In particular,
  replacing a short final group preserves all its pairs, since
  they are among the last $b$ pairs of the input array.
  Therefore, if $(X,Y)\in\mathcal P$ and $(A,B)\in\mathcal Q$
  satisfy the required suffix and prefix conditions, they belong
  to $\mathcal P_i$ and $\mathcal Q_j$ for some $i,j\in[1\dd t]$.
  Conversely, every pair in a constructed set belongs to the
  corresponding original set, and hence any match in
  $(\mathcal P_i,\mathcal Q_j)$ is also an original match.
  By \cref{pr:sn-length-reduction}, a match exists in
  $(\mathcal P_i,\mathcal Q_j)$ if and only if one of its returned
  instances has answer \texttt{YES}.
  This proves the claimed equivalence. In particular, an empty
  output list implies that the original input has answer
  \texttt{NO}, and adding the fixed \texttt{NO}-instance preserves
  the answer.

  We next verify the assumptions of the length reduction and the
  sizes of its outputs. Since $k\geq q^2$, we have
  $b\geq k/q\geq\sqrt{k}\geq2$ and
  $b\leq2k/q\leq k$.
  Consequently, $m=\bigO(\log k)=\bigO(\log b)$, and each call
  uses size parameter $3bm\leq s\leq N$.
  The original word size therefore suffices for every call.
  Each output instance has total string length at most
  \[
    90bm\leq\frac{180km}{q}\leq\frac{3km}{c}=\frac{s}{c}.
  \]
  The initial cases $k<q^2$ are recognized before computing $q$, by a
  fixed sequence of tests $k=2,\ldots,q^2-1$, stopping upon equality.
  Every tested integer is at most $k$ and fits in a word.
  In the remaining case, $q\leq k\leq N$ also fits in a word.
  The direct comparisons and construction of the fixed instances
  use \cref{pr:packed-representation} with alphabet size two,
  symbol width one, and length bound $N$.

  Finally, $t=\lceil k/b\rceil\leq q=\bigO(1)$, and each call
  to \cref{pr:sn-length-reduction} returns $\bigO(1)$ instances
  in $\bigO(b)$ time and words.
  There are $t^2=\bigO(1)$ such calls, giving $\bigO(k)$ total
  time and words. Each input pair occupies $\bigO(1)$ words,
  since $m=\bigO(\log k)$ and $w\geq1+\log N$.
  Copying the groups takes $\bigO(tb)=\bigO(k)$ time and words.
  Thus, the claimed bounds include the original input, all
  intermediate arrays, and the output.
\end{proof}

\begin{theorem}\label{th:sn-length-reduction}
  Let $N\in\Z_{\geq2}$. Consider the word RAM model with word size
  $w$ satisfying $1+\log N\leq w=\bigO(\log N)$. Suppose that there
  exists an algorithm for \probname{String Nesting}
  (\cref{prob:sn}) that, given $N$ and any valid input of at most $N$ symbols
  over alphabet $\BinaryAlphabet$, achieves the following complexities:
  \begin{itemize}
  \item running time $T(N)$,
  \item space usage $S(N)$ bits.
  \end{itemize}
  Then, there exists an algorithm that,
  given $N$, integers $k\in\Z_{\geq2}$ and $m\in\Z_{\geq1}$ satisfying
  $3km\leq N$ and $m=\bigO(\log k)$, and arrays listing two sets
  $\mathcal P$ and $\mathcal Q$ of binary string pairs satisfying
  \begin{align*}
    |\mathcal P|&=k,& |X|=|Y|&=m
      &&\text{for every }(X,Y)\in\mathcal P,\\*
    |\mathcal Q|&=k,& |A|+|B|&=m
      &&\text{for every }(A,B)\in\mathcal Q,
  \end{align*}
  decides whether there exist $(X,Y)\in\mathcal P$ and
  $(A,B)\in\mathcal Q$ such that $A$ is a suffix of $X$ and $B$ is a
  prefix of $Y$.
  Each input string is supplied by its length and packed representation
  (\cref{def:packed-representation}).
  The algorithm achieves the following complexities:
  \begin{itemize}
  \item running time $\bigO(T(N))$,
  \item space usage $\bigO(S(N))$ bits.
  \end{itemize}
  The constants in these bounds may depend on the constant implicit in
  $m=\bigO(\log k)$.
\end{theorem}
\begin{proof}
  Denote $s=3km$. If $s<12$, direct comparison determines the
  answer in $\bigO(1)$ time using $\bigO(1)$ words.
  Assume henceforth that $s\geq12$.
  Apply \cref{pr:sn-size-reduction} with $c=1$
  to the given sets.
  Each returned pair of sets is a valid String Nesting input with
  total string length at most $s\leq N$.
  Run the assumed algorithm on every returned pair and return
  \texttt{YES} if at least one call returns \texttt{YES}, and
  \texttt{NO} otherwise. Correctness follows from
  \cref{pr:sn-size-reduction}.
  The size reduction and every call to the assumed algorithm use
  size parameter $N$.

  Since $\mathcal P$ contains $k$ distinct pairs of binary strings
  of length $m$, we have $k\leq2^{2m}$ and hence
  $m\geq\frac12\log k$.
  Together with $m=\bigO(\log k)$, this gives
  $s=\Theta(k\log k)$ and
  $k=\bigO(s/\log s)\subseteq\bigO(N/\log N)$.
  The construction takes $\bigO(k)$ time and uses $\bigO(k)$ words,
  including the input and output.
  Since $w=\bigO(\log N)$, this space occupies $\bigO(N)$ bits.
  Together with the constant number of calls to the assumed algorithm,
  the total running time is $\bigO(N/\log N+T(N))$, and the space
  usage is $\bigO(N+S(N))$ bits.
  By \cref{lm:sn-lower-bounds}, we have
  $T(N)=\Omega(N/\log N)$ and $S(N)=\Omega(N)$.
  Hence, the claimed bounds follow.
\end{proof}

\subsection[Reducing Dictionary Matching to String Nesting]{%
  \texorpdfstring{\phantomsection\label{sec:dm-to-sn-problem-reduction}}{}%
  Reducing Dictionary Matching to String Nesting}

\begin{theorem}\label{th:dm-to-sn}
  Let $N\in\Z_{\geq2}$. Consider the word RAM model with word size
  $w$ satisfying $1+\log N\leq w=\bigO(\log N)$. Suppose that there
  exists an algorithm for \probname{String Nesting}
  (\cref{prob:sn}) that, given $N$ and any valid input of
  at most $N$ symbols over alphabet $\BinaryAlphabet$, achieves
  the following complexities:
  \begin{itemize}
  \item running time $T_{\rm SN}(N)$,
  \item space usage $S_{\rm SN}(N)$ bits.
  \end{itemize}
  Then, there exists an algorithm
  for \probname{Dictionary Matching}
  (\cref{prob:dm}) that, given $N$ and any valid input of
  at most $N$ symbols over alphabet $\BinaryAlphabet$, achieves
  the following complexities:
  \begin{itemize}
  \item running time $\bigO(T_{\rm SN}(N))$,
  \item space usage $\bigO(S_{\rm SN}(N))$ bits.
  \end{itemize}
\end{theorem}
\begin{proof}
  Consider the given Dictionary Matching input with text length
  $\Textlen_0$, dictionary size $k_0=\lfloor\Textlen_0/\log\Textlen_0\rfloor$,
  and pattern length $m_0=2\lfloor\log\Textlen_0\rfloor$.
  Denote its total string length by $L_0=\Textlen_0+k_0m_0\leq N$.
  If $L_0<5\cdot2^{18}$, direct comparison determines the answer in
  $\bigO(1)$ time and words of space.
  Otherwise, apply \cref{pr:dm-size-reduction} with $c=2^{18}$
  to the given text and dictionary. This produces
  $\bigO(1)$ valid Dictionary Matching inputs, each with total string
  length at most $L_0/2^{18}$.

  We describe a procedure for solving one of the returned inputs.
  Denote its text by $\Text\in\BinaryAlphabet^{\Textlen}$ and its dictionary by
  $\mathcal D\subseteq\BinaryAlphabet^m$ of size $k$, where
  $k=\lfloor\Textlen/\log\Textlen\rfloor$ and
  $m=2\lfloor\log\Textlen\rfloor$.
  These strings have total length
  $L=\Textlen+km\leq L_0/2^{18}\leq N/2^{18}$.
  Assume that $\log\Textlen\geq14$ (for smaller $\Textlen$, direct
  comparison determines the answer in $\bigO(1)$ time).
  Denote $\mathcal D=\{\Pat_1,\ldots,\Pat_k\}$, with patterns indexed in
  their input order.
  We proceed in six steps.
  \begin{enumerate}
  \item\label{step:dm-to-sn-select} First, we select text positions at which to test the patterns
    and determine where to split each pattern for these tests.
    Compute $\tau=\lfloor\frac17\log L\rfloor$, and apply
    \cref{th:sparse-anchors-short-offsets} with parameter $\tau$
    to $\Text$ and the patterns in $\mathcal D$.
    This gives a set $C\subseteq[1\dd\Textlen]$ and an array
    $\Delta[1\dd k]$ with entries in $[0\dd\tau)$ such that,
    for every $i\in[1\dd k]$,
    \[
      \OccTwo{\Pat_i}{\Text}\neq\emptyset
      \quad\Longleftrightarrow\quad
      \text{there exists }c\in C\text{ with }
      c-\Delta[i]\in\OccTwo{\Pat_i}{\Text}.
    \]
    Thus, it suffices to test whether the first $\Delta[i]$ symbols
    of $\Pat_i$ occur immediately before some $c\in C$ and its
    remaining symbols occur starting at $c$.
    If $C=\emptyset$, return \texttt{NO}.
  \item\label{step:dm-to-sn-pairs} Next, we construct the text-context pairs and pattern splits.
    \begin{enumerate}[beginpenalty=10000]
    \item We first construct the packed representations of
      $W=\two^m\Text\two^m$ and all patterns over alphabet $[0\dd3)$.
      To this end, construct the data structure from
      \cref{th:alphabet-lifting}.
      Query it to obtain the packed representations of $\Text$
      and all patterns over $[0\dd3)$.
      Using \cref{pr:packed-representation}, initialize the packed
      representation of $\two^m$ and concatenate the representations
      of $\two^m$, $\Text$, and $\two^m$, in this order, to obtain
      the packed representation of $W$.
    \item We now construct a pair of length-$m$ strings for each selected
      text position: one immediately before that position and one starting
      there, using the padded text $W$.
      For every $c\in C$, denote
      \[
        X_c=W[c\dd c+m),\qquad
        Y_c=W[c+m\dd c+2m).
      \]
      Denote $\mathcal P=\{(X_c,Y_c):c\in C\}$.
      Using \cref{pr:packed-representation}, extract these substrings
      from $W$ to obtain their packed representations.
      To remove repeated context pairs when $|C|\geq2$, construct the
      packed representations of $X_cY_c$, sort them with
      \cref{pr:packed-string-sorting}, and keep one pair for each distinct
      concatenation. This produces an array listing $\mathcal P$ because
      every $X_c$ has length $m$, and hence distinct concatenations
      $X_cY_c$ correspond to distinct pairs $(X_c,Y_c)$.
      Denote $p=|\mathcal P|$.
    \item Finally, we split each input pattern into two strings at the
      offset computed in Step~\ref{step:dm-to-sn-select}.
      For every $i\in[1\dd k]$, denote
      $A_i=\Pat_i[1\dd\Delta[i]]$ and
      $B_i=\Pat_i[\Delta[i]+1\dd m]$.
      The query set is $\mathcal Q=\{(A_i,B_i):i\in[1\dd k]\}$.
      For every $i\in[1\dd k]$, split the packed representation of
      $\Pat_i$ after $\Delta[i]$ symbols using \cref{pr:packed-representation}
      to obtain those of $A_i$ and $B_i$.
      The query pairs are distinct because their concatenations are the
      distinct input patterns.
    \end{enumerate}
  \item We now enlarge both sets to cardinality $r=p+k$.
    To this end, denote
    \begin{align*}
      \mathcal P^+&=\mathcal P\cup
        \{(\zero^{m-1}\two,\two\cdot\bin{m-1}{v}):v\in[0\dd k)\},\\*
      \mathcal Q^+&=\mathcal Q\cup
        \{(\emptystring,\two\two\cdot\bin{m-2}{v}):v\in[0\dd p)\}.
    \end{align*}
    Since $C\subseteq[1\dd\Textlen]$, we have $p\leq|C|\leq\Textlen$.
    The assumptions $m=2\lfloor\log\Textlen\rfloor$ and
    $\log\Textlen\geq14$ give $m-2\geq\lceil\log\Textlen\rceil$.
    Thus, $p,k\leq\Textlen\leq2^{m-2}$, and all binary representations
    in these definitions are well-defined.
    Denote $M=6rm$ and $f=\lceil\log r\rceil$.
    Construct the data structure from \cref{th:val-decoding}.
    For each required integer $v$, use this structure to obtain the
    packed representation of $\bin{f}{v}$.
    Prepend zeros to obtain the representations of $\bin{m-1}{v}$
    and $\bin{m-2}{v}$ as needed, using \cref{pr:packed-representation},
    and convert these strings to alphabet $[0\dd3)$ with the
    alphabet-lifting structure from Step~\ref{step:dm-to-sn-pairs}.
    Initialization and concatenation with \cref{pr:packed-representation}
    then give the representations of the added pairs.
    Append them to the arrays listing $\mathcal P$ and $\mathcal Q$
    to obtain arrays listing $\mathcal P^+$ and $\mathcal Q^+$.
  \item Next, we convert the two enlarged sets to binary.
    Apply \cref{th:sn-alphabet-reduction} to
    $\mathcal P^+$ and $\mathcal Q^+$.
    Denote the returned sets by $\widehat{\mathcal P}$ and
    $\widehat{\mathcal Q}$, and denote $\ell=2m$.
    Both returned sets have size $r$.
    We have $|X|=|Y|=\ell$ for every
    $(X,Y)\in\widehat{\mathcal P}$ and $|A|+|B|=\ell$ for every
    $(A,B)\in\widehat{\mathcal Q}$.
  \item We then apply the String Nesting length reduction.
    Apply \cref{pr:sn-length-reduction} to
    $\widehat{\mathcal P}$ and $\widehat{\mathcal Q}$.
    Denote the resulting $\bigO(1)$ pairs of sets by
    $(\mathcal P_t,\mathcal Q_t)$, their common cardinalities by $K_t$,
    and their string-length parameters by $s_t$.
    Each output pair of sets satisfies $s_t=2\lfloor\log K_t\rfloor$,
    as required by \cref{prob:sn}.
  \item Finally, we determine whether a pattern in $\mathcal D$ occurs
    in $\Text$. For every $t$, run the assumed String Nesting algorithm on
    $\mathcal P_t$ and $\mathcal Q_t$.
    Return \texttt{YES} if at least one call returns \texttt{YES},
    and \texttt{NO} otherwise.
  \end{enumerate}

  Apply this procedure to every returned Dictionary Matching input,
  and return \texttt{YES} if at least one has answer \texttt{YES},
  and \texttt{NO} otherwise.

  We prove correctness of the procedure in three steps.
  \begin{itemize}[beginpenalty=10000]
  \item First, we show that the context pairs preserve pattern occurrences.
    For every $i\in[1\dd k]$ and $c\in C$, the definitions give
    \[
      \Pat_i=W[m+c-\Delta[i]\dd m+c-\Delta[i]+m)
      \quad\Longleftrightarrow\quad
      A_i\text{ is a suffix of }X_c
        \text{ and }B_i\text{ is a prefix of }Y_c.
    \]
    Since $\Pat_i$ is binary, the equality on the left is equivalent to
    \[
      1\leq c-\Delta[i]\leq\Textlen-m+1,\qquad
      \Pat_i=\Text[c-\Delta[i]\dd c-\Delta[i]+m).
    \]
    By \cref{th:sparse-anchors-short-offsets}, $\Pat_i$ occurs in
    $\Text$ if and only if these conditions hold for some $c\in C$.
    This also justifies returning \texttt{NO} when
    $C=\emptyset$.
    Removing repeated context pairs does not change whether any of
    them matches a query in $\mathcal Q$.
  \item Next, we show that $\mathcal P^+$ and $\mathcal Q^+$ contain
    $r$ distinct pairs each and preserve the matching condition.
    Every original right context $Y_c$ begins with
    $W[c+m]=\Text[c]\in\BinaryAlphabet$, whereas the right
    component of each added context pair begins with $\two$.
    Hence, no added context pair belongs to $\mathcal P$.
    The added query pairs contain $\two\two$, whereas every original
    query pair is binary.
    The binary representations of $v$ distinguish the added pairs
    within each collection. Thus, both collections have size $p+k=r$.
    Every original match remains present, and no match involving an
    added pair is possible, for the following reasons.
    \begin{itemize}
    \item An original query $(A,B)$ cannot match an added context pair.
      If $A\neq\emptystring$, its last binary symbol differs from
      the final $\two$ of the left context.
      Otherwise, $B$ is nonempty and its first symbol differs from
      the initial $\two$ of the right context.
    \item An added query requires the right context to begin with
      $\two\two$, whereas every original right context begins with
      a binary symbol.
    \item The right context of an added pair begins with $\two$
      followed by a binary symbol, and therefore cannot match an
      added query either.
    \end{itemize}
  \item Finally, we verify that binary conversion and length reduction
    preserve the matching condition.
    By \cref{th:sn-alphabet-reduction}, there exist
    $(X,Y)\in\mathcal P^+$ and $(A,B)\in\mathcal Q^+$ such that $A$ is
    a suffix of $X$ and $B$ is a prefix of $Y$ if and only if
    such pairs exist in $\widehat{\mathcal P}$ and
    $\widehat{\mathcal Q}$.
    By \cref{pr:sn-length-reduction}, the latter condition holds
    if and only if such pairs exist in some returned pair of sets
    $(\mathcal P_t,\mathcal Q_t)$.
    The last step tests exactly this condition.
  \end{itemize}
  By \cref{pr:dm-size-reduction}, at least one of the returned
  Dictionary Matching inputs has answer \texttt{YES} if and only
  if the original input does.

  We now verify the size bounds and the assumptions of the tools.
  Since $\Textlen\leq L\leq3\Textlen$ and
  $m=2\lfloor\log\Textlen\rfloor$, we have $\tau\geq1$ and
  $3\tau-1<\log\Textlen\leq m$.
  Thus, \cref{th:sparse-anchors-short-offsets} applies, giving
  $p\leq|C|\leq31L/\tau$.
  Since $\tau\geq\frac1{14}\log L$ and
  $k\leq\Textlen/\log\Textlen\leq L/\log L$, we obtain
  \[
    r=p+k\leq31L/\tau+k<2^9L/\log L.
  \]
  Together with $m\leq2\log L$, this gives
  \[
    M=6rm<2^{13}L\leq N/32.
  \]
  Since $k=\Theta(\Textlen/\log\Textlen)$, we have
  $m=\Theta(\log k)$ and $L=\Theta(k\log k)$.
  The anchor bound implies $|C|=\bigO(k)$ and $r=p+k=\Theta(k)$,
  and hence $M=6rm=\Theta(L)$.
  The alphabet reduction receives $3rm=M/2$ symbols, whose binary
  encodings have total length $6rm=M$, which matches its size parameter.
  It preserves the common cardinality $r$, and the returned binary
  strings have total length $3r\ell=M$.
  Moreover, $\ell=2m=\bigO(\log r)$.
  The length reduction therefore applies, and every resulting pair
  of sets satisfies $3K_ts_t\leq90r\ell=30M\leq N$.
  Thus, each pair is a valid input to the assumed String Nesting algorithm.
  For the added pairs, $p\leq|C|\leq\Textlen$ gives $r\leq2\Textlen$ and
  $f\leq\lfloor\log\Textlen\rfloor+2\leq m-2$.
  Furthermore, $2r\leq M$ gives
  $f\leq\lfloor\log M\rfloor$,
  as required by the integer-decoding structure.
  The basic packed operations and the sorting call use length bound
  $L$: we have $|W|=\Textlen+2m\leq L$ because $k\geq2$,
  and both $|C|\leq\Textlen\leq L$ and $2m\leq L$ hold.
  The symbol widths for alphabets of sizes $2$ and $3$ are $1$ and $2$,
  respectively.
  The size reduction and the calls to the assumed String Nesting
  algorithm use size parameter $N$. The anchor computation and the
  lifting structure use size parameter $L$, while the decoding
  structure, the alphabet reduction, and the length reduction use
  size parameter $M$. The anchor computation and the decoding
  structure use alphabet size $2$, the lifting structure uses source
  alphabet size $2$ and target alphabet size $3$, and the sorting call
  and the alphabet reduction use alphabet size $3$.
  Since $L,M\leq N$, the assumption $w\geq1+\log N$ suffices
  for every invoked tool.

  It remains to bound the running time and space usage.
  The initial application of \cref{pr:dm-size-reduction} takes
  $\bigO(L_0/\log L_0)=\bigO(N/\log N)$ time and words of space,
  including the original input and all returned inputs.
  For each returned input, the six-step procedure has the following costs.
  Parameter computation and construction of the lifting and
  integer-decoding structures take
  $\bigO(\sqrt L)=\bigO(k)$ time.
  Lifting the text and dictionary, constructing $W$, and computing
  the anchors also take $\bigO(k)$ time in total.
  The context extractions, pattern splits, and construction of the
  added pairs take $\bigO(1)$ time per pair, or $\bigO(k)$ time in total.
  For $q=|C|\geq2$, constructing the context concatenations, sorting
  them, and removing repeated pairs take
  \[
    \bigO\!\left(q\left(1+\frac{m}{\log q}\right)\right)
      =\bigO(k).
  \]
  For $q\geq3$, this follows from $q=\bigO(k)$,
  $m=\bigO(\log k)$, and the monotonicity of $x/\log x$ for $x\geq3$.
  For $q=2$, the running time is $\bigO(m)=\bigO(\log k)$.
  The alphabet reduction takes
  $\bigO(r+M/\log M)=\bigO(k)$ time,
  and the length reduction takes $\bigO(r)=\bigO(k)$ time.
  Thus, the entire construction takes $\bigO(k)$ time.
  Each returned Dictionary Matching input occupies
  $\bigO(k+L/w)=\bigO(k)$ words,
  and the construction time bounds the remaining space, including
  all intermediate representations.
  Since $k=\bigO(N/\log N)$ and $w=\bigO(\log N)$, the construction
  takes $\bigO(N/\log N)$ time and uses $\bigO(N)$ bits of space.
  There are $\bigO(1)$ returned Dictionary Matching inputs, each
  requiring $\bigO(1)$ calls to the assumed String Nesting algorithm.
  Including these calls and the initial size reduction,
  the total running time is $\bigO(N/\log N+T_{\rm SN}(N))$, and
  the space usage is $\bigO(N+S_{\rm SN}(N))$ bits.
  By \cref{lm:sn-lower-bounds}, we have
  $T_{\rm SN}(N)=\Omega(N/\log N)$ and
  $S_{\rm SN}(N)=\Omega(N)$.
  Thus, the total running time is $\bigO(T_{\rm SN}(N))$, and the
  space usage is $\bigO(S_{\rm SN}(N))$ bits.
\end{proof}

\subsection{Reducing String Nesting to Dictionary Matching}\label{sec:sn-to-dm-problem-reduction}

\begin{theorem}\label{th:sn-to-dm}
  Let $N\in\Z_{\geq2}$. Consider the word RAM model with word size
  $w$ satisfying $1+\log N\leq w=\bigO(\log N)$. Suppose that there
  exists an algorithm for \probname{Dictionary Matching}
  (\cref{prob:dm}) that, given $N$ and any valid input of
  at most $N$ symbols over alphabet $\BinaryAlphabet$, achieves
  the following complexities:
  \begin{itemize}
  \item running time $T_{\rm DM}(N)$,
  \item space usage $S_{\rm DM}(N)$ bits.
  \end{itemize}
  Then, there exists an algorithm
  for \probname{String Nesting}
  (\cref{prob:sn}) that, given $N$ and any valid input of
  at most $N$ symbols over alphabet $\BinaryAlphabet$, achieves
  the following complexities:
  \begin{itemize}
  \item running time $\bigO(T_{\rm DM}(N))$,
  \item space usage $\bigO(S_{\rm DM}(N))$ bits.
  \end{itemize}
\end{theorem}
\begin{proof}
  Consider the given String Nesting input with parameters $k_0$ and
  $m_0=2\lfloor\log k_0\rfloor$. Denote its total string length by
  $L_0=3k_0m_0\leq N$.
  If $L_0<432$, direct comparison determines the answer in
  $\bigO(1)$ time and words of space.
  Otherwise, apply \cref{pr:sn-size-reduction} with $c=36$
  to the given sets. This produces $\bigO(1)$ valid
  String Nesting inputs, each with total string length at most $L_0/36$.

  We describe a procedure for solving one of the returned inputs.
  Denote its sets by $\mathcal P$ and $\mathcal Q$, and its parameters
  by $k$ and $m=2\lfloor\log k\rfloor$.
  Denote also the pairs in $\mathcal P$ and $\mathcal Q$ by
  $(X_i,Y_i)$ and $(A_i,B_i)$, respectively, for $i\in[1\dd k]$,
  in their array order. These strings have total length
  $L=3km\leq L_0/36\leq N/36$.
  We proceed in four steps.
  \begin{enumerate}
  \item First, we construct the packed representations of the ternary
    text and patterns
    \[
      \Text=X_1\two Y_1\cdots X_k\two Y_k,
      \qquad \Pat_i=A_i\two B_i\quad(i\in[1\dd k]).
    \]
    Denote $\mathcal D=\{\Pat_1,\ldots,\Pat_k\}$ and
    $\Textlen=|\Text|=k(2m+1)$. We proceed as follows.
    \begin{enumerate}
    \item We first obtain the packed representations of all strings
      in $\mathcal P$ and $\mathcal Q$ over alphabet $[0\dd3)$.
      To this end, construct the data structure from
      \cref{th:alphabet-lifting}, and query it for each string.
    \item Next, we assemble the representations of $\Text$ and
      $\Pat_1,\ldots,\Pat_k$.
      Using \cref{pr:packed-representation}, initialize the packed
      representation of the one-symbol string $\two$ and concatenate
      the representations in the order specified by the definitions of
      $\Text$ and $\Pat_i$ above.
      Store the pattern representations in an array listing $\mathcal D$.
    \end{enumerate}
  \item Next, we convert $\Text$ and $\mathcal D$ to binary.
    Apply \cref{th:dm-alphabet-reduction} to $\Text$ and $\mathcal D$.
    Denote the returned text and dictionary by $\Text_{\rm bin}$ and
    $\mathcal D_{\rm bin}$.
    The text has length $7\Textlen$, and $\mathcal D_{\rm bin}$
    contains $k$ patterns of length $\ell=7(m+1)$.
  \item We now construct the final Dictionary Matching instances.
    To this end, apply \cref{pr:dm-length-reduction-general}
    to $\Text_{\rm bin}$ and $\mathcal D_{\rm bin}$.
    Denote the $\bigO(1)$ returned pairs by $(\Text_j,\mathcal D_j)$,
    with text lengths $n_j$, dictionary sizes $k_j$, and common pattern
    lengths $m_j$.
  \item Finally, we determine whether $(\mathcal P,\mathcal Q)$ has
    answer \texttt{YES}.
    Run the assumed Dictionary Matching algorithm on each
    $(\Text_j,\mathcal D_j)$, and return \texttt{YES} if at least
    one call returns \texttt{YES}, and \texttt{NO} otherwise.
  \end{enumerate}

  Apply this procedure to every returned String Nesting input, and
  return \texttt{YES} if at least one has answer \texttt{YES}, and
  \texttt{NO} otherwise.

  Since $\two$ occurs in none of the input strings, the equality
  $\Pat_i=\Pat_{i'}$ implies $(A_i,B_i)=(A_{i'},B_{i'})$.
  Hence, $|\mathcal D|=k$.
  If $A_i$ is a suffix of $X_j$ and $B_i$ is a prefix of $Y_j$,
  then $\Pat_i=A_i\two B_i$ is a substring of $X_j\two Y_j$
  and hence occurs in $\Text$.
  Conversely, an occurrence of $\Pat_i=A_i\two B_i$ in $\Text$
  must align its only symbol $\two$ with the separator in some
  block $X_j\two Y_j$. Since $|A_i|,|B_i|\leq m$, the occurrence
  lies entirely inside that block. Thus, $A_i$ is a suffix of
  $X_j$ and $B_i$ is a prefix of $Y_j$.
  Therefore, $(\mathcal P,\mathcal Q)$ has answer \texttt{YES}
  if and only if a pattern in $\mathcal D$ occurs in $\Text$.
  The alphabet reduction preserves pattern occurrences, and
  \cref{pr:dm-length-reduction-general} guarantees that a pattern from
  $\mathcal D_{\rm bin}$ occurs in $\Text_{\rm bin}$ if and only if,
  for some $j$, a pattern from $\mathcal D_j$ occurs in $\Text_j$.
  Hence, the final step returns \texttt{YES} if and only if
  $(\mathcal P,\mathcal Q)$ has answer \texttt{YES}.
  By \cref{pr:sn-size-reduction}, at least one of the returned
  String Nesting inputs has answer \texttt{YES} if and only if the
  original input does.

  We now verify the size bounds and the assumptions of the tools.
  Since $m=2\lfloor\log k\rfloor=\Theta(\log k)$, we have
  $L=\Theta(k\log k)$ and $\Textlen=\Theta(k\log k)$.
  Denote $s=|\Text_{\rm bin}|+k\ell$.
  The total length of the ternary strings is
  $\Textlen+k(m+1)=L+2k$. Consequently,
  \[
    s=7(L+2k)\leq\tfrac{35}{3}L<12L\leq N/3.
  \]
  Thus, both the alphabet reduction and the length reduction satisfy
  their input-size requirements.
  Each returned pair $(\Text_j,\mathcal D_j)$ is a valid Dictionary
  Matching input of size $n_j+k_jm_j\leq3n_j\leq3s\leq N$.
  The basic packed operations use length bound $L$: each concatenation
  has at most $3k\leq L$ strings and produces a string of length at
  most $\Textlen\leq L$.
  The lifting structure uses size parameter $L$, source alphabet size
  $2$, and target alphabet size $3$. The alphabet reduction uses size
  parameter $N$ and alphabet size $3$, and the size reduction, the
  length reduction, and the assumed Dictionary Matching algorithm use
  size parameter $N$.
  All invoked tools use size parameters at most $N$,
  and hence $w\geq1+\log N$ suffices for their word-size requirements.

  The initial application of \cref{pr:sn-size-reduction} takes
  $\bigO(k_0)=\bigO(L_0/\log L_0)=\bigO(N/\log N)$ time and
  words of space, including the original input and all returned inputs.
  For each returned input, the four-step procedure has the following costs.
  The lifting structure and its queries take
  $\bigO(\sqrt L+k+L/\log L)=\bigO(k)$ time.
  The concatenations take $\bigO(k+L/w)=\bigO(k)$ time, and
  the alphabet reduction takes
  $\bigO(k+N/\log N)=\bigO(N/\log N)$ time, since
  $k=\Theta(L/\log L)$ and $L\leq N$.
  Since $s=\Theta(L)=\Theta(k\log k)$, the length reduction takes
  $\bigO(s/\log s)=\bigO(k)$ time and uses $\bigO(k)$ words,
  including the returned instances.
  Since $k=\Theta(L/\log L)$ and $L\leq N$, the construction takes
  $\bigO(N/\log N)$ time.
  There are $\bigO(1)$ returned String Nesting inputs, and the final
  step of each procedure takes $\bigO(T_{\rm DM}(N))$ time, giving total time
  $\bigO(N/\log N+T_{\rm DM}(N))$.
  Each returned String Nesting input occupies
  $\bigO(k+L/w)=\bigO(L/\log L)$ words,
  and the construction time bounds the remaining space, including
  all intermediate representations and the returned instances.
  Thus, the construction and all its input and output representations
  occupy $\bigO(N/\log N)$ words, or $\bigO(N)$ bits since
  $w=\bigO(\log N)$.
  Together with the space used in the final step, this gives
  $\bigO(N+S_{\rm DM}(N))$ bits of space.
  By \cref{lm:dm-lower-bounds}, we have
  $T_{\rm DM}(N)=\Omega(N/\log N)$ and
  $S_{\rm DM}(N)=\Omega(N)$.
  Thus, the total running time is $\bigO(T_{\rm DM}(N))$, and the
  space usage is $\bigO(S_{\rm DM}(N))$ bits.
\end{proof}

\subsection{Summary}\label{sec:dm-to-sn-summary}

\begin{theorem}[Equivalence of Dictionary Matching and String Nesting]
  \label{th:equivalence-dm-sn}
  Let $N\in\Z_{\geq2}$. In the word RAM model with word size $w$
  satisfying $1+\log N\leq w=\bigO(\log N)$, the problems of
  \probname{Dictionary} \probname{Matching} and
  \probname{String} \probname{Nesting}
  (\cref{prob:dm,prob:sn}) are asymptotically equivalent:
  if there exists an algorithm for one of these two problems that,
  given $N$ and any valid input of at most $N$ symbols over alphabet
  $\BinaryAlphabet$,\footnote{\phantomsection\label{fn:dm-sn-input}For
  Dictionary Matching, the input consists of the packed representation of a binary text
  $\Text$ of length $\Textlen\geq2$ and an array containing the packed representations
  of all patterns in a dictionary
  $\mathcal D\subseteq\BinaryAlphabet^m$, where
  $k=|\mathcal D|=\lfloor\Textlen/\log\Textlen\rfloor$,
  $m=2\lfloor\log\Textlen\rfloor$, and $\Textlen+km\leq N$, together with
  the integers $\Textlen$, $k$, and $m$.
  For String Nesting, the input consists of two sets $\mathcal P$
  and $\mathcal Q$ of $k\geq2$ binary string pairs each,
  where $m=2\lfloor\log k\rfloor$ and $3km\leq N$.
  Both strings in every pair of $\mathcal P$ have length $m$,
  and each $(A,B)\in\mathcal Q$ satisfies $|A|+|B|=m$.
  The sets are supplied in arrays listing their pairs, with each string
  in packed form, together with the integers
  $k,m$ and the lengths $|A|,|B|$.
  The packed strings and integer parameters together occupy
  $\bigO(N)$ bits under the stated word-size assumption; see also
  \cref{rm:sn-input-size}.}
  achieves the following complexities:
  \begin{itemize}
  \item running time $T(N)$,
  \item space usage $S(N)$ bits,
  \end{itemize}
  then there exists an algorithm
  for the other problem that, given $N$ and any valid input of at most $N$
  symbols over alphabet $\BinaryAlphabet$, achieves the following
  complexities:
  \begin{itemize}
  \item running time $\bigO(T(N))$,
  \item space usage $\bigO(S(N))$ bits.
  \end{itemize}
\end{theorem}
\begin{proof}
  An algorithm for binary \probname{String Nesting} yields an algorithm
  for binary \probname{Dictionary Matching} by \cref{th:dm-to-sn}.
  The reverse direction follows from \cref{th:sn-to-dm}.
\end{proof}

\section{Reducing Dictionary Matching to String Problems}\label{sec:dm-to-string-problems}

\subsection{Reducing Dictionary Matching to LZ77 Parity}\label{sec:lz}

\subsubsection{Problem Definitions}\label{sec:lz-problem}

\begin{problem}[LZ77 Parity]\label{prob:lz77-parity}
  \begin{description}[style=sameline,itemsep=0ex,font={\normalfont\bf}]
  \item[Input:]
    Let $\Textlen \in \Z_{\geq 1}$. Consider the word RAM model with
    word size $w > \log \Textlen$. The input consists of $\Textlen$ and
    the packed representation $\PackedRepresentation{w}{2}{\Text}$
    (\cref{def:packed-representation}) of a string
    $\Text \in \BinaryAlphabet^{\Textlen}$.
  \item[Output:]
    The value $\LZSize{\Text} \bmod 2$,
    i.e., the parity of the number of phrases in the LZ77 factorization of $\Text$
    (\cref{def:lz77}).
  \end{description}
\end{problem}

\begin{problem}[LZ77 Size]\label{prob:lz77-size}
  \begin{description}[style=sameline,itemsep=0ex,font={\normalfont\bf}]
  \item[Input:]
    Let $\Textlen \in \Z_{\geq 1}$. Consider the word RAM model with
    word size $w > \log \Textlen$. The input consists of $\Textlen$ and
    the packed representation $\PackedRepresentation{w}{2}{\Text}$
    (\cref{def:packed-representation}) of a string
    $\Text \in \BinaryAlphabet^{\Textlen}$.
  \item[Output:]
    The value $\LZSize{\Text}$,
    i.e., the number of phrases in the LZ77 factorization of $\Text$
    (\cref{def:lz77}).
  \end{description}
\end{problem}

\begin{problem}[LZ77 Factorization]\label{prob:lz77-factorization}
  \begin{description}[style=sameline,itemsep=0ex,font={\normalfont\bf}]
  \item[Input:]
    Let $\Textlen \in \Z_{\geq 1}$. Consider the word RAM model with
    word size $w > \log \Textlen$. The input consists of $\Textlen$ and
    the packed representation $\PackedRepresentation{w}{2}{\Text}$
    (\cref{def:packed-representation}) of a string
    $\Text \in \BinaryAlphabet^{\Textlen}$.
  \item[Output:]
    An LZ77 representation of $\Text$ (\cref{def:lz77-representation}).
  \end{description}
\end{problem}

The variants of these problems, in which the
classical LZ77 is replaced with the non-overlapping
variant (\cref{def:nonoverlapping-lz77,def:nonoverlapping-lz77-representation}), are called
\probname{Non-overlapping} \probname{LZ77} \probname{Parity},
\probname{Non-overlapping} \probname{LZ77} \probname{Size}, and
\probname{Non-overlapping} \probname{LZ77} \probname{Factorization}.

\subsubsection{Preliminaries}\label{sec:lz-preliminaries}

\begin{lemma}[Lower bounds for LZ77 Parity]\label{lm:lz77-parity-lower-bounds}
  Let $N\in\Z_{\geq2}$. Consider the word RAM model with word size
  $w$ satisfying $1+\log N\leq w=\bigO(\log N)$.
  For either parsing convention, any algorithm for \probname{LZ77 Parity}
  (\cref{prob:lz77-parity}) that, given $N$, solves every valid input of
  at most $N$ symbols over alphabet $\BinaryAlphabet$ requires
  $\Omega(N/\log N)$ time and $\Omega(N)$ bits of space
  in the worst case.
\end{lemma}
\begin{proof}
  Let $L$ be the largest power of two with $L+1\leq N$, and assume
  $N\geq5$. Consider $S={\tt 0}^L{\tt 1}$. For every integer
  $a\in(L/2\dd L)$, changing $S[a+1]$ to ${\tt 1}$ yields
  $S_a={\tt 0}^a{\tt 1}{\tt 0}^{L-a-1}{\tt 1}$.
  The final suffix ${\tt 0}^{L-a-1}{\tt 1}$ of $S_a$ forms one
  repeat phrase whose earlier occurrence ends at the first
  ${\tt 1}$ and does not overlap this suffix. Thus, for either
  parsing convention, the phrase count of $S_a$ is the phrase
  count of ${\tt 0}^a$ plus two, whereas that of $S$ is the
  phrase count of ${\tt 0}^L$ plus one. These two zero strings
  have the same phrase count: two with overlaps, and
  $1+\log L$ without overlaps. Indeed,
  $\LZNonOvSize{{\tt 0}^r}=1+\lceil\log r\rceil$ for every
  integer $r\geq1$, and $L/2<a<L$ with $L$ a power of two
  implies $\lceil\log a\rceil=\log L$.
  Consequently, every such change
  reverses the parity. On input $S$, the algorithm must therefore
  read every word containing one of these $\Theta(N)$ positions.
  This requires $\Omega(N/w)=\Omega(N/\log N)$ time for either
  parsing convention. Since $|S|=L+1=\Theta(N)$, the packed input
  occupies $\Omega(N)$ bits, which gives the claimed space lower bound
  (see \cref{sec:model}).
\end{proof}

\subsubsection{Combinatorial Results}\label{sec:lz-combinatorial-results}

\begin{lemma}\label{lm:dm-to-lz}
  Let $\Textlen\in\Z_{\geq1}$, $k\in\Z_{\geq0}$, and $m\in\Z_{\geq3}$, and let
  $\Text \in \BinaryAlphabet^{\Textlen}$ be a text and $\mathcal{D} =
  \{\Pat_1, \Pat_2, \dots, \Pat_k\} \subseteq \BinaryAlphabet^{m}$
  be a collection of $k$ pairwise distinct patterns of common length $m$.
  For every $i \in [1 \dd k]$, let $\Pat'_i \in \{{\tt 2}, {\tt 3}\}^{m}$
  be the string satisfying $\Pat'_i[j]=\Pat_i[j]+2$ for every $j\in[1\dd m]$.
  Let also $\Pat_{k+1} = {\tt 0}^{m}$. Denote $S = S_1S_2S_3 \in \{{\tt 0}, {\tt 1}, {\tt 2}, {\tt 3}, {\tt 4}, {\tt 5}, {\tt 6}\}^{*}$,
  where (brackets added for clarity):
  \begin{align*}
    S_1 &= \Text \cdot {\tt 5},\\
    S_2 &= \big(\textstyle\bigodot_{i=1}^{k}
      \Pat_i(1 \dd m] \cdot {\tt 4} \cdot \Pat_i[1 \dd m) \cdot {\tt 4} \cdot
      \Pat_i[m] \cdot
      \Pat'_i[1 \dd m] \cdot {\tt 4} \cdot \Pat'_i[1 \dd m] \cdot \Pat_{i+1}[1] \cdot {\tt 4}\big) \cdot
      {\tt 6},\\
    S_3 &= \big(\textstyle\bigodot_{i=1}^{k}
      \Pat_i[1 \dd m] \cdot \Pat'_i[1 \dd m] \big) \cdot \Pat_{k+1}[1].
  \end{align*}
  If there exists $i \in [1 \dd k]$ such that $\OccTwo{\Pat_i}{\Text} \neq \emptyset$, then
  $\LZSize{S} - \LZSize{S_1S_2} = \LZNonOvSize{S} - \LZNonOvSize{S_1S_2} = 2k$. Otherwise,
  $\LZSize{S} - \LZSize{S_1S_2} = \LZNonOvSize{S} - \LZNonOvSize{S_1S_2} = 2k + 1$.
\end{lemma}
\begin{proof}
  If $k=0$, then $S=\Text\cdot{\tt 560}$ and $S_1S_2=\Text\cdot{\tt 56}$.
  For either variant, the symbol ${\tt 6}$ forms a literal phrase, and the
  final ${\tt 0}$ forms one additional phrase. Assume henceforth that $k\geq1$.
  Denote $\delta=|S_1S_2|$ and $p_i=\delta+2(i-1)m+1$ for every
  $i\in[1\dd k]$. The copy of $\Pat_i\Pat'_i$ in $S_3$ starts at
  position $p_i$ in $S$. We proceed in two steps.
  \begin{enumerate}
  \item\label{lm:dm-to-lz-step-1}
    First, we determine the values of $\LPF{S}$ and $\LPnF{S}$ at
    positions $p_i$, $p_i+1$, $p_i+m-1$, and $p_i+m$ for every $i\in[1\dd k]$.
    Since $\LPnF{S}[j]\leq\LPF{S}[j]$ for every $j\in[1\dd|S|]$,
    a lower bound for $\LPnF{S}[j]$ and a matching upper bound for
    $\LPF{S}[j]$ determine both values.
    An occurrence of a string over $[0\dd4)$ cannot cross either boundary
    between $S_1$, $S_2$, and $S_3$, because $S_1$ and $S_2$ end with
    ${\tt 5}$ and ${\tt 6}$, respectively.
    \begin{enumerate}
    \item\label{lm:dm-to-lz-step-1a}
      We begin by proving $\LPF{S}[p_i+1]=\LPnF{S}[p_i+1]=m-1$.
      The string $\Pat_i[2\dd m]$ occurs entirely in $S_2$, giving
      $\LPnF{S}[p_i+1]\geq m-1$.
      An earlier match of length $m$ would be an earlier occurrence of
      $X=\Pat_i[2\dd m]\cdot\Pat'_i[1]$.
      This string cannot occur in $S_1$, which contains no symbols from
      $\{{\tt 2},{\tt 3}\}$. It cannot occur in $S_2$ either: $X$ has
      $m-1\geq2$ binary symbols immediately before its last symbol,
      whereas in $S_2$ every binary symbol immediately followed by
      ${\tt 2}$ or ${\tt 3}$ is preceded by ${\tt 4}$.
      In $S_3$, an earlier occurrence must start at $p_j+1$ for some
      $j\in[1\dd i)$, since each binary block $\Pat_j$ has length $m$.
      This would imply $\Pat_j[2\dd m]=\Pat_i[2\dd m]$ and
      $\Pat'_j[1]=\Pat'_i[1]$.
      Since $\Pat'_j[1]=\Pat_j[1]+2$ and $\Pat'_i[1]=\Pat_i[1]+2$,
      the latter equality gives $\Pat_j[1]=\Pat_i[1]$,
      contradicting $\Pat_j\neq\Pat_i$.
    \item\label{lm:dm-to-lz-step-1b}
      Next, we prove that
      \[
        \LPF{S}[p_i]=\LPnF{S}[p_i]=
        \begin{cases}
          m & \text{if }\OccTwo{\Pat_i}{\Text}\neq\emptyset,\\
          m-1 & \text{otherwise}.
        \end{cases}
      \]
      By \cref{ob:LPF} and Step~\ref{lm:dm-to-lz-step-1a},
      $\LPF{S}[p_i]\leq\LPF{S}[p_i+1]+1=m$.
      The occurrence of $\Pat_i[1\dd m)$ in $S_2$ gives
      $\LPnF{S}[p_i]\geq m-1$.
      Every substring of $S_2$ over $\BinaryAlphabet$ has length at most
      $m-1$, and the length-$m$ binary substrings of $S_3$ are exactly
      the distinct patterns $\Pat_1,\ldots,\Pat_k$.
      Thus, $\Pat_i$ has an occurrence before position $p_i$ in $S$ if
      and only if it occurs in $\Text$.
      If it does, that occurrence lies entirely in $S_1$ and gives
      $\LPnF{S}[p_i]\geq m$. Otherwise, $\LPF{S}[p_i]<m$.
    \item\label{lm:dm-to-lz-step-1c}
      Finally, we prove $\LPF{S}[q]=\LPnF{S}[q]=m+1$ for both
      $q=p_i+m-1$ and $q=p_i+m$.
      In $S_2$ and $S_3$, the symbols from $\{{\tt 2},{\tt 3}\}$ occur
      in blocks of exactly $m$ symbols, each equal to some $\Pat'_j$.
      Since $\Pat'_1,\ldots,\Pat'_k$ are pairwise distinct, the only
      occurrences of $\Pat'_i$ before position $p_i+m$ are its two copies
      in $S_2$.
      The symbols adjacent to these copies are as follows:
      \[
        \Pat_i[m]\cdot\Pat'_i\cdot{\tt 4},
        \qquad
        {\tt 4}\cdot\Pat'_i\cdot\Pat_{i+1}[1]\cdot{\tt 4}.
      \]
      We consider the two positions separately:
      \begin{itemize}
      \item At position $p_i+m-1$, the string $\Pat_i[m]\Pat'_i$ has a
        previous occurrence entirely in $S_2$, given by the first string
        above. Its following symbol is ${\tt 4}$, whereas the corresponding
        symbol in $S_3$ is $\Pat_{i+1}[1]\in\BinaryAlphabet$.
        The second copy of $\Pat'_i$ is preceded by ${\tt 4}$ and cannot
        give an occurrence of $\Pat_i[m]\Pat'_i$.
        Hence, $\LPnF{S}[p_i+m-1]\geq m+1$ and
        $\LPF{S}[p_i+m-1]\leq m+1$.
      \item At position $p_i+m$, the string $\Pat'_i\Pat_{i+1}[1]$ has
        a previous occurrence entirely in $S_2$, given by the second
        string above. The first copy of $\Pat'_i$ is followed by
        ${\tt 4}$ rather than $\Pat_{i+1}[1]$.
        If $i<k$, the second copy is followed by $\Pat_{i+1}[1]{\tt 4}$,
        whereas $\Pat'_i$ in $S_3$ is followed by $\Pat_{i+1}[1\dd2]$.
        If $i=k$, the remaining suffix has length $m+1$.
        Thus, $\LPnF{S}[p_i+m]\geq m+1$ and
        $\LPF{S}[p_i+m]\leq m+1$.
      \end{itemize}
    \end{enumerate}
  \item\label{lm:dm-to-lz-step-2}
    We now describe the phrases following $S_1S_2$ under either convention.
    The first occurrence of ${\tt 6}$ is at position $\delta$, where it
    forms a literal phrase. No repeat phrase can contain this occurrence,
    and the factorizations of $S$ and $S_1S_2$ agree through position
    $\delta$. It remains to count the phrases in $S_3$.
    \begin{itemize}
    \item Suppose that no pattern occurs in $\Text$.
      For each $i\in[1\dd k]$, the two consecutive phrases are
      $\Pat_i[1\dd m)$ and $\Pat_i[m]\Pat'_i$, starting at positions
      $p_i$ and $p_i+m-1$, respectively. Their lengths $m-1$ and $m+1$
      follow from Steps~\ref{lm:dm-to-lz-step-1b}
      and~\ref{lm:dm-to-lz-step-1c}.
      The concatenation of these two phrases is $\Pat_i\Pat'_i$.
      The final ${\tt 0}$ forms one additional phrase, giving $2k+1$
      phrases in total.
    \item Otherwise, let $t=\min\{i\in[1\dd k]:
      \OccTwo{\Pat_i}{\Text}\neq\emptyset\}$.
      We prove that the following table lists the consecutive phrases
      in $S_3$, with the pairs taken in the order $i=1,\ldots,k$:
      \[
        \begin{array}{c|c|c}
          & \text{First string} & \text{Second string}\\ \hline
          i<t & \Pat_i[1\dd m) & \Pat_i[m]\Pat'_i\\
          i=t & \Pat_t & \Pat'_t\Pat_{t+1}[1]\\
          i>t & \Pat_i[2\dd m] & \Pat'_i\Pat_{i+1}[1]
        \end{array}
      \]
      We verify this in two steps:
      \begin{enumerate}
      \item First, we show that the concatenation of the listed strings
        is $S_3$. For $i<t$, the concatenation of the two strings is
        $\Pat_i\Pat'_i$. For $i\geq t$, the second string includes
        $\Pat_{i+1}[1]$, which is omitted from the first string for index
        $i+1$ when $i<k$. For $i=k$, the second string ends with
        $\Pat_{k+1}[1]={\tt 0}$. Thus, the listed strings cover exactly
        $\Pat_1\Pat'_1\cdots\Pat_k\Pat'_k\cdot{\tt 0}=S_3$.
      \item Next, we check that each listed string has the greedy phrase
        length at its starting position. We consider the two columns
        separately:
        \begin{itemize}
        \item The first strings start at $p_i$ for $i\leq t$ and at
          $p_i+1$ for $i>t$. By the definition of $t$ and
          Step~\ref{lm:dm-to-lz-step-1b}, the greedy phrase length at
          $p_i$ is $m-1$ for $i<t$ and $m$ for $i=t$.
          For $i>t$, Step~\ref{lm:dm-to-lz-step-1a} gives the greedy
          phrase length $m-1$ at $p_i+1$.
        \item The second strings start at $p_i+m-1$ for $i<t$ and at
          $p_i+m$ for $i\geq t$. Step~\ref{lm:dm-to-lz-step-1c}
          gives the greedy phrase length $m+1$ at all these positions.
        \end{itemize}
      \end{enumerate}
      Consequently, the listed strings are the $2k$ phrases following
      $S_1S_2$ under either convention.\qedhere
    \end{itemize}
  \end{enumerate}
\end{proof}

\subsubsection{Algorithms}\label{sec:lz-algorithms}

\begin{theorem}\label{th:lz}
  Let $N\in\Z_{\geq2}$. Consider the word RAM model with word size
  $w$ satisfying $1+\log N\leq w=\bigO(\log N)$. Suppose that there
  exists an algorithm for \probname{LZ77} \probname{Parity}
  (resp.\ its non-overlapping variant, see \cref{prob:lz77-parity})
  that, given $N$ and any valid input of at most $N$ symbols over alphabet
  $\BinaryAlphabet$,\footnote{\phantomsection\label{fn:lz-input}
    The input consists of the packed representation
    $\PackedRepresentation{w}{2}{\Text}$ of a binary text $\Text$
    with $1\leq|\Text|\leq N$, together with its length $|\Text|$.
    The complete input occupies $\bigO(N)$ bits at the stated word size.} achieves
  the following complexities:
  \begin{itemize}
  \item running time $T_{\rm LZ}(N)$ (resp.\ $T_{\rm LZno}(N)$),
  \item space usage $S_{\rm LZ}(N)$ bits (resp.\ $S_{\rm LZno}(N)$ bits).
  \end{itemize}
  Then, there exists an
  algorithm for \probname{Dictionary} \probname{Matching}
  (\cref{prob:dm}) that, given $N$ and any valid input of
  at most $N$ symbols over alphabet $\BinaryAlphabet$, achieves
  the following complexities:
  \begin{itemize}
  \item running time $\bigO(T_{\rm LZ}(N))$
    (resp.\ $\bigO(T_{\rm LZno}(N))$),
  \item space usage $\bigO(S_{\rm LZ}(N))$ bits
    (resp.\ $\bigO(S_{\rm LZno}(N))$ bits).
  \end{itemize}
\end{theorem}
\begin{proof}
  Consider the given Dictionary Matching input with text length
  $\Textlen_0$, dictionary size $k_0=\lfloor\Textlen_0/\log\Textlen_0\rfloor$,
  and pattern length $m_0=2\lfloor\log\Textlen_0\rfloor$.
  Denote its total string length by $L_0=\Textlen_0+k_0m_0\leq N$.
  If $L_0<1280$, direct comparison determines the answer in
  $\bigO(1)$ time and words of space.
  Otherwise, apply \cref{pr:dm-size-reduction} with $c=256$
  to the given text and dictionary. This produces
  $\bigO(1)$ valid Dictionary Matching inputs, each with total string
  length at most $L_0/256$.

  We describe a procedure for solving one of the returned inputs.
  Denote its text by $\Text\in\BinaryAlphabet^{\Textlen}$ and its dictionary by
  $\mathcal D=\{\Pat_1,\ldots,\Pat_k\}\subseteq\BinaryAlphabet^m$,
  where $k=\lfloor\Textlen/\log\Textlen\rfloor$,
  $m=2\lfloor\log\Textlen\rfloor$, and
  $\Textlen+km\leq L_0/256\leq N/256$.
  We assume that $\Textlen\geq8$ (for smaller $\Textlen$, direct
  comparison solves the instance in $\bigO(1)$ time).
  Let $S=S_1S_2S_3$ and $\Pat'_i$ be defined as in
  \cref{lm:dm-to-lz}. We proceed in four steps.
  \begin{enumerate}

  \item We first construct the packed representations of
    $\Text,\Pat_1,\ldots,\Pat_k,\Pat'_1,\ldots,\Pat'_k$
    over alphabet $[0\dd7)$.
    Let $M=\Textlen+km=\Theta(\Textlen)$.
    Build the data structure from \cref{th:alphabet-lifting}.
    Query it for $\Text,\Pat_1,\ldots,\Pat_k$, and then query each
    pattern $\Pat_i$ again with offset $2$ to obtain the packed
    representation of $\Pat'_i$.
    Including preprocessing, this takes
    $\bigO(\sqrt M+k+M/\log M)=\bigO(\Textlen/\log\Textlen)$ time.

  \item We now assemble the packed representation of $S$.
    Denote
    \[
      U=|S|=\Textlen+k(6m+4)+3,
      \qquad \delta=|S_1S_2|=\Textlen+k(4m+4)+2.
    \]
    We proceed as follows:
    \begin{enumerate}
    \item First, we prepare the remaining pieces.
      For every $i\in[1\dd k]$, extract the packed representations
      of $\Pat_i[2\dd m]$, $\Pat_i[1\dd m)$, and the one-symbol
      strings $\Pat_i[m]$ and $\Pat_i[1]$ from the representation
      of $\Pat_i$, using
      \cref{pr:packed-representation}\eqref{pr:packed-representation-substring}.
      Each extraction takes $\bigO(1)$ time because
      $m=\bigO(\log\Textlen)=\bigO(w)$.
      Also construct the packed representations of the one-symbol
      strings ${\tt 0}$, ${\tt 4}$, ${\tt 5}$, and ${\tt 6}$,
      using ${\tt 0}$ for $\Pat_{k+1}[1]$.
    \item Next, concatenate the prepared representations in the order
      defining $S$ in \cref{lm:dm-to-lz}, using
      \cref{pr:packed-representation}\eqref{pr:packed-representation-concat}.
    \end{enumerate}
    The $\bigO(k)$ pieces have total length $U=\Theta(\Textlen)$.
    Hence, this step takes $\bigO(k+U/w)=\bigO(\Textlen/\log\Textlen)$ time.

  \item\label{th:dm-to-lz-binary-step-parities} Next, we compute the
    parities of the phrase counts of $R_1=S$ and
    $R_2=S[1\dd\delta]=S_1S_2$ using binary encodings.
    First, extract the packed representation of $R_2$ from $S$ using
    \cref{pr:packed-representation}\eqref{pr:packed-representation-substring}.
    This takes $\bigO(1+\delta/w)=\bigO(\Textlen/\log\Textlen)$ time.
    For each $t\in\{1,2\}$, we proceed as follows:
    \begin{enumerate}
    \item Apply \cref{th:lz-alphabet-reduction-poly-to-binary}
      to $R_t$.
      In $\bigO(N/\log N)$ time,
      it returns the length and packed representation of a binary
      string $B_t$, together with an integer $h_t\in[1\dd|B_t|]$,
      satisfying
      \[
        \LZSize{R_t}=\LZSize{B_t}-\LZSize{B_t[1\dd h_t]},
        \qquad
        \LZNonOvSize{R_t}=\LZNonOvSize{B_t}
          -\LZNonOvSize{B_t[1\dd h_t]}.
      \]
    \item Construct the packed representation of
      $L_t=B_t[1\dd h_t]$ using
      \cref{pr:packed-representation}\eqref{pr:packed-representation-substring}.
      This takes $\bigO(1+h_t/w)=\bigO(\Textlen/\log\Textlen)$ time.
      Run the assumed binary algorithm on the packed representations
      of $B_t$ and $L_t$ to obtain
      \[
        a_t=\LZSize{B_t}\bmod2,
        \qquad b_t=\LZSize{L_t}\bmod2.
      \]
      Compute $z_t=(a_t+b_t)\bmod2$ in $\bigO(1)$ time.
      The first identity gives
      $z_t=\LZSize{R_t}\bmod2$, since subtraction and addition
      agree modulo $2$.
    \end{enumerate}
    The four calls take $\bigO(T_{\rm LZ}(N))$ time and use
    $\bigO(S_{\rm LZ}(N)+w)$ bits in addition to the constructed
    representations and data structures.

  \item Finally, we decide whether any dictionary pattern occurs in
    $\Text$. Output \texttt{YES} if $z_1=z_2$, and \texttt{NO} otherwise.
    By \cref{lm:dm-to-lz}, the difference between the phrase counts
    of $R_1=S$ and $R_2=S_1S_2$ is $2k$ if some pattern occurs
    and $2k+1$ otherwise.
    Thus, the parities agree exactly when a pattern occurs in $\Text$.
  \end{enumerate}

  Apply this procedure to each returned Dictionary Matching input,
  and return \texttt{YES} if and only if at least one of them is a
  \texttt{YES}-instance. Correctness for the original input follows
  from \cref{pr:dm-size-reduction}.
  The size reduction, the alphabet reduction, and the calls to the
  assumed algorithm use size parameter $N$, and the alphabet reduction
  uses alphabet size $7$. The lifting structure uses size parameter $M$,
  source alphabet size $2$, and target alphabet size $7$.
  Since $\Textlen\geq8$, the strings supplied to the alphabet reduction satisfy
  \[
    16\leq\delta<U\leq8M.
  \]
  For each $t\in\{1,2\}$, we have
  \[
    18|R_t|+148\leq32|R_t|
      \leq32U\leq256M\leq N.
  \]
  Thus, the alphabet reduction applies and returns
  a string of length $|B_t|=18|R_t|+148\leq N$.
  Since $1\leq h_t\leq|B_t|\leq N$, all four calls to the
  assumed binary algorithm are valid.
  The operations over alphabet $[0\dd7)$ use length bound $U$,
  and the binary substring operations use length bound $N$.
  All packed-string tools satisfy their assumptions because
  $7\leq M,U\leq N$ and $w\geq1+\log N\geq1+\log M$.
  The initial size reduction takes
  $\bigO(L_0/\log L_0)=\bigO(N/\log N)$ time and words of space.
  Each of its $\bigO(1)$ returned inputs requires
  $\bigO(N/\log N)$ additional time
  and at most four calls to the assumed algorithm.
  The total time is therefore $\bigO(N/\log N+T_{\rm LZ}(N))$.
  Including the original input, all returned inputs, the data structures,
  and all constructed representations, the reduction uses
  $\bigO(N/\log N)$ words in addition to the storage of the assumed algorithm.
  Since $w=\bigO(\log N)$, the total space is
  $\bigO(N+S_{\rm LZ}(N))$ bits.
  For the non-overlapping variant, use the assumed binary algorithm
  for that variant in Step~\ref{th:dm-to-lz-binary-step-parities}.
  The second encoding identity yields
  $z_t=\LZNonOvSize{R_t}\bmod2$ for $t\in\{1,2\}$.
  The corresponding identity in \cref{lm:dm-to-lz} proves the same
  test $z_1=z_2$ correct, with $T_{\rm LZno}(N)$ and
  $S_{\rm LZno}(N)$ in the same time and space bounds.
  By \cref{lm:lz77-parity-lower-bounds}, the assumed algorithm
  requires $\Omega(N/\log N)$ time and $\Omega(N)$ bits of space
  for either parsing convention. These bounds absorb the additive
  construction costs, yielding the claimed bounds.
\end{proof}

\begin{remark}\leavevmode\phantomsection\label{rm:lz77-parity-size-factorization}
  For a binary input text $\Text$, the reductions from
  \probname{LZ77 Parity} to \probname{LZ77 Size}
  and from \probname{LZ77 Size} to \probname{LZ77 Factorization}
  (\cref{prob:lz77-parity,prob:lz77-size,prob:lz77-factorization})
  preserve the asymptotic time and space bounds:
  \begin{itemize}
  \item Given an algorithm for \probname{LZ77 Size}, compute
    $z=\LZSize{\Text}$ and return $z\bmod2$.
    This requires $\bigO(1)$ additional time and words of space.
  \item Given an algorithm for \probname{LZ77 Factorization},
    construct an LZ77 representation of $\Text$ and count its pairs.
    If the representation contains $z$ pairs, counting them takes
    $\bigO(z)$ time and $\bigO(1)$ additional words of space.
    This time is bounded by the time required to output the
    representation.
  \end{itemize}
  Both arguments also apply to the non-overlapping variants.
  Consequently, the Dictionary Matching reduction from \cref{th:lz}
  also holds for \probname{LZ77 Size} and \probname{LZ77 Factorization}
  under either parsing convention.
\end{remark}

\subsection{Reducing Dictionary Matching to RLBWT Size}\label{sec:bwt}

\subsubsection{Problem Definitions}\label{sec:bwt-problem}

\begin{problem}[RLBWT Size]\label{prob:rlbwt-size}
  \begin{description}[style=sameline,itemsep=0ex,font={\normalfont\bf}]
  \item[Input:]
    Let $\Textlen \in \Z_{\geq 1}$. Consider the word RAM model with
    word size $w > \log \Textlen$. The input consists of $\Textlen$ and
    the packed representation $\PackedRepresentation{w}{2}{\Text}$
    (\cref{def:packed-representation}) of a string
    $\Text \in \BinaryAlphabet^{\Textlen}$.
  \item[Output:]
    The value $\RLBWTSize{\Text}$, i.e., the number of runs in the BWT of $\Text$
    (\cref{def:bwt}).
  \end{description}
\end{problem}

\begin{problem}[BWT Construction]\label{prob:bwt-construction}
  \begin{description}[style=sameline,itemsep=0ex,font={\normalfont\bf}]
  \item[Input:]
    Let $\Textlen \in \Z_{\geq 1}$. Consider the word RAM model with
    word size $w > \log \Textlen$. The input consists of $\Textlen$ and
    the packed representation $\PackedRepresentation{w}{2}{\Text}$
    (\cref{def:packed-representation}) of a string
    $\Text \in \BinaryAlphabet^{\Textlen}$.
  \item[Output:]
    The packed representation
    $\PackedRepresentation{w}{2}{\BWT{\Text}[1 \dd \Textlen]}$
    of $\BWT{\Text}[1 \dd \Textlen]$ (\cref{def:bwt}).
  \end{description}
\end{problem}

\subsubsection{Preliminaries}\label{sec:bwt-preliminaries}

\begin{lemma}[Lower bounds for RLBWT Size]\label{lm:rlbwt-size-lower-bounds}
  Let $N\in\Z_{\geq2}$. Consider the word RAM model with word size
  $w$ satisfying $1+\log N\leq w=\bigO(\log N)$.
  Any algorithm for \probname{RLBWT Size}
  (\cref{prob:rlbwt-size}) that, given $N$, solves every valid input of
  at most $N$ symbols over alphabet $\BinaryAlphabet$ requires
  $\Omega(N/\log N)$ time and $\Omega(N)$ bits of space
  in the worst case.
\end{lemma}
\begin{proof}
  On input $\Text={\tt 0}^N$, every word of its packed representation
  must be inspected: changing a symbol in an uninspected word to
  ${\tt 1}$ changes the answer from one run to at least two.
  Indeed, the BWT contains exactly the same symbols as the text.
  This requires $\Omega(N/w)=\Omega(N/\log N)$ time.
  The packed input occupies $\Omega(N)$ bits, which gives the claimed
  space lower bound (see \cref{sec:model}).
\end{proof}

\subsubsection{Combinatorial Results}\label{sec:bwt-combinatorial-results}

\begin{lemma}\label{lm:dm-to-bwt}
  Let $\Textlen,m\in\Z_{\geq1}$ and $k\in\Z_{\geq0}$, and let
  $\Text \in \BinaryAlphabet^{\Textlen}$ be a text and $\mathcal{D} =
  \{\Pat_1, \Pat_2, \dots, \Pat_k\} \subseteq \BinaryAlphabet^{m}$
  be a collection of $k$ pairwise distinct patterns of common length $m$.
  For any $q\in\Z_{\geq1}$ and string $S = a_1 a_2 \cdots a_q\in\BinaryAlphabet^q$,
  we define $\widetilde{S} := a_1 {\tt 4} a_2 {\tt 4} a_3 {\tt 4} \cdots {\tt 4} a_q$.
  Consider the following strings over alphabet $\{{\tt 0}, {\tt 1}, \ldots, {\tt 9}\}$ (brackets are added for clarity):
  \begin{align*}
    S_1 &= {\tt 4}\widetilde{T}{\tt 4} \cdot
      \textstyle\bigodot_{i=1}^{k} \big({\tt 4}\widetilde{\Pat_i}{\tt 2}\cdot
                                        {\tt 8}\widetilde{\Pat_i}{\tt 5}\cdot
                                        {\tt 4}\widetilde{\Pat_i}{\tt 6}\cdot
                                        {\tt 4}\widetilde{\Pat_i}{\tt 7}\cdot
                                        {\tt 9}\widetilde{\Pat_i}{\tt 3}
                                   \big),\\
    S_2 &= {\tt 4}\widetilde{T}{\tt 4} \cdot
      \textstyle\bigodot_{i=1}^{k} \big({\tt 4}\widetilde{\Pat_i}{\tt 2}\cdot
                                        {\tt 8}\widetilde{\Pat_i}{\tt 6}\cdot
                                        {\tt 4}\widetilde{\Pat_i}{\tt 5}\cdot
                                        {\tt 4}\widetilde{\Pat_i}{\tt 7}\cdot
                                        {\tt 9}\widetilde{\Pat_i}{\tt 3}
                                   \big).
  \end{align*}
  Then, it holds $\RLBWTSize{S_2} - \RLBWTSize{S_1} =
  |\{i \in [1 \dd k] : \OccTwo{\Pat_i}{\Text} = \emptyset\}|$.
\end{lemma}
\begin{proof}
  The claim holds trivially if $k=0$. Let us thus assume that $k \geq 1$.
  Denote $\Sigma = \{{\tt 0}, {\tt 1}, \ldots, {\tt 9}\}$,
  $\Textlen' := |S_1| = |S_2|$, $\Delta := 2\Textlen + 1$ and $\delta := 2m + 1$,
  and note that then $\Textlen' = \Delta + 5k \delta$. Let $\mathcal{L}$ (resp.\ $\mathcal{R}$)
  denote the set of all $i \in [1 \dd \Textlen' - 2m + 1]$ such that $S_1[i \dd i + 2m)$ contains
  the symbol ${\tt 5}$ (resp.\ ${\tt 6}$). It is easy to check that
  \begin{align*}
    \mathcal{L} &= \textstyle\bigcup_{i \in [1 \dd k]} [\Delta + 5(i{-}1)\delta + \delta + 2 \dd \Delta + 5(i{-}1)\delta + 2\delta],\\
    \mathcal{R} &= \textstyle\bigcup_{i \in [1 \dd k]} [\Delta + 5(i{-}1)\delta + 2\delta + 2 \dd \Delta + 5(i{-}1)\delta + 3\delta].
  \end{align*}
  We also define $\mathcal{P} := [1 \dd \Textlen'] \setminus (\mathcal{L} \cup \mathcal{R})$.
  Note that $\mathcal{L} \cap \mathcal{R} = \emptyset$ and $\{i + \delta : i \in \mathcal{L}\} = \mathcal{R}$.
  Let $f : [1 \dd \Textlen'] \rightarrow
  [1 \dd \Textlen']$ be such that for every $x \in [1 \dd \Textlen']$,
  \[
    f(x) =
      \begin{cases}
        x + \delta & \text{if }x \in \mathcal{L}, \\
        x - \delta & \text{if }x \in \mathcal{R}, \\
        x          & \text{if }x \in \mathcal{P}.
      \end{cases}
  \]
  For $i \in \{1, 2\}$, we denote $B_i = \BWT{S_i}$.
  For every $c \in \Sigma$, we define $b_c = |\{i \in [1 \dd \Textlen'] : S_1[i] \prec c\}|$ and
  $e_c = |\{i \in [1 \dd \Textlen'] : S_1[i] \preceq c\}|$.
  Note that, since for every $c \in \Sigma$, the number of occurrences of $c$ in $S_1$ is the same as in $S_2$,
  it holds that for every $i \in \{1, 2\}$ and $c \in \Sigma$, $B_i(b_c \dd e_c]$ is the block in the BWT of $S_i$
  containing all symbols preceding suffixes of $S_i$ starting with symbol $c$
  (where the symbol preceding $S_i[1]$ is $S_i[\Textlen']$).

  The proof consists of six steps:
  \begin{enumerate}
  \item\label{lm:dm-to-bwt-step-1}
    In the first step, we show that for every $i, j \in [1 \dd \Textlen']$, $S_1[i \dd \Textlen'] \prec S_1[j \dd \Textlen']$ holds if and only
    if $S_2[f(i) \dd \Textlen'] \prec S_2[f(j) \dd \Textlen']$. Note that the claim holds trivially if $i = j$. Let us thus assume that
    $i \neq j$. Denote $\ell = \LCE{S_1}{i}{j}$.
    Consider three cases:
    \begin{enumerate}
    \item Assume that $\max(i,j) + \ell = \Textlen' + 1$. We begin by observing that in that case, we have $\max(i,j) \geq \Textlen' - \delta + 1$ (otherwise,
      both $S_1[i \dd i + \ell)$ and $S_1[j \dd j + \ell)$ would have an occurrence of substring ${\tt 9}\widetilde{\Pat}_k{\tt 3}$
      starting at the same offset, which is not possible since it only occurs once in $S_1$). Consequently,
      \[
        \ell
          = \Textlen' + 1 - \max(i,j)
          \leq \Textlen' + 1 - (\Textlen' - \delta + 1) = \delta.
      \]
      By definition of $\ell$, it holds $S_1[\min(i,j) + \ell - 1] =
      S_1[\max(i,j) + \ell - 1] = S_1[\Textlen'] = {\tt 3}$. Combining this with $\ell \leq \delta$
      implies that the distance between $\min(i,j)$ and the nearest ${\tt 3}$ on the right is bounded by $\delta-1$.
      Thus, $\min(i,j) \in \mathcal{P}$. On the other hand, $\max(i,j) \geq \Textlen'-\delta+1$
      also implies $\max(i,j) \in \mathcal{P}$. We thus obtain that $i,j \in \mathcal{P}$, and hence, $f(i) = i$ and $f(j) = j$.
      It remains to observe that by $\max(i,j) > \Textlen' - \delta$, the fragment
      $S_1[\max(i,j) \dd \Textlen']$ lies within the final block ${\tt 9}\widetilde{\Pat}_k{\tt 3}$.
      Hence, the string $S_1[i \dd i + \ell) = S_1[j \dd j + \ell)$ contains neither ${\tt 5}$ nor ${\tt 6}$.
      Since the only difference between $S_1$ and $S_2$ is
      at positions in $\OccTwo{{\tt 5}}{S_1} \cup \OccTwo{{\tt 6}}{S_1}$, we thus have
      $S_2[i \dd i + \ell) = S_1[i \dd i + \ell)$ and $S_2[j \dd j + \ell) = S_1[j \dd j + \ell)$.
      By $\max(i,j) + \ell = \Textlen' + 1$, this implies that $S_1[i \dd \Textlen'] \prec S_1[j \dd \Textlen']$ holds if and only if
      $S_2[i \dd \Textlen'] \prec S_2[j \dd \Textlen']$, which is equivalent to
      $S_2[f(i) \dd \Textlen'] \prec S_2[f(j) \dd \Textlen']$.
    \item\label{lm:dm-to-bwt-case-1b}
      Assume that $\max(i,j)+\ell\leq\Textlen'$ and
      $i,j \in \mathcal{L} \cup \mathcal{R}$.
      For any $x \in \mathcal{L} \cup \mathcal{R}$, let $d_x$ denote the smallest non-negative integer such that
      $S_1[x + d_x] \in \{{\tt 5}, {\tt 6}\}$, and let $\ell_x = d_x + \delta$. Observe two properties of $\ell_x$:
      \begin{itemize}
      \item First, note that if $x \in \mathcal{L}$ (resp.\ $x \in \mathcal{R}$), then there exists an index
        $t \in [1 \dd k]$ such that the string ${\tt 5}{\tt 4}\widetilde{\Pat}_t$ (resp.\ ${\tt 6}{\tt 4}\widetilde{\Pat}_t$) is
        a suffix of $S_1[x \dd x + \ell_x)$. Since for every $t \in [1 \dd k]$, ${\tt 5}{\tt 4}\widetilde{\Pat}_t$ and
        ${\tt 6}{\tt 4}\widetilde{\Pat}_t$ each have only a single occurrence in $S_1$, this implies that for every $x' \in [1 \dd \Textlen']$
        such that $x' \neq x$, it holds $\LCE{S_1}{x}{x'} < \ell_x$.
      \item Second, observe that
        for every $x \in \mathcal{L} \cup \mathcal{R}$, it holds $S_1[x \dd x + \ell_x) = S_2[f(x) \dd f(x) + \ell_x)$. To
        see this, note that if an occurrence of $c \in \{{\tt 5}, {\tt 6}\}$ (in either $S_1$ or $S_2$)
        is preceded by $\widetilde{\Pat}_t$, then it is always followed by ${\tt 4}\widetilde{\Pat}_t$.
      \end{itemize}
      By the above properties, we have $\ell < \min(\ell_i, \ell_j)$.
      Consequently, letting $\ell' = \min(\ell_i, \ell_j)$,
      $S_1[i \dd \Textlen'] \prec S_1[j \dd \Textlen']$ holds if and only if
      $S_1[i \dd i + \ell') \prec S_1[j \dd j + \ell')$. By the second property,
      $S_2[f(i) \dd f(i) + \ell_i) = S_1[i \dd i + \ell_i)$ and $S_2[f(j) \dd f(j) + \ell_j) = S_1[j \dd j + \ell_j)$.
      In particular, $S_2[f(i) \dd f(i) + \ell') \neq S_2[f(j) \dd f(j) + \ell')$. Thus,
      $S_2[f(i) \dd \Textlen'] \prec S_2[f(j) \dd \Textlen']$ is equivalent to
      $S_2[f(i) \dd f(i) + \ell') \prec S_2[f(j) \dd f(j) + \ell')$. This in turn holds if and only if
      $S_1[i \dd i + \ell') \prec S_1[j \dd j + \ell')$, which is equivalent to $S_1[i \dd \Textlen'] \prec S_1[j \dd \Textlen']$.
    \item\label{lm:dm-to-bwt-case-1c}
      Assume that $\max(i,j)+\ell\leq\Textlen'$ and
      that either $i \in \mathcal{P}$ or $j \in \mathcal{P}$.
      We begin by proving two auxiliary properties:
      \begin{itemize}
      \item First, we prove that for every $c \in \{{\tt 5}, {\tt 6}\}$,
        $\OccTwo{c}{S_1[i \dd i+\ell)} = \OccTwo{c}{S_1[j \dd j + \ell)} = \emptyset$. Suppose that this does not hold.
        Since the two strings are equal, there exists a position $q \in [0 \dd \ell)$ satisfying
        $S_1[i + q] = S_1[j + q] \in \{{\tt 5}, {\tt 6}\}$. Observe that it is not possible that $q < 2m$, since
        then, by definition of $\mathcal{L}$ and $\mathcal{R}$, we would either have $i,j \in \mathcal{L}$ or
        $i,j \in \mathcal{R}$, contradicting the assumption $\{i,j\} \cap \mathcal{P} \neq \emptyset$. Thus, $q \geq 2m$.
        By the placement of symbols ${\tt 5}$ and ${\tt 6}$ in $S_1$, this implies that there exists $x \in [1 \dd k]$
        such that $\widetilde{\Pat}_{x}{\tt 5}$ or $\widetilde{\Pat}_{x}{\tt 6}$ occurs in both
        $S_1[i \dd i+\ell)$ and $S_1[j \dd j+\ell)$ at the same position.
        Since, for every $y \in [1 \dd k]$, $\widetilde{\Pat}_{y}{\tt 5}$
        and $\widetilde{\Pat}_{y}{\tt 6}$ each occur only once
        in $S_1$, this is not possible.
      \item Next, we prove that either
        $S_1[i+\ell] \not\in \{{\tt 5}, {\tt 6}\}$
        or $S_1[j+\ell] \not\in \{{\tt 5}, {\tt 6}\}$. Suppose that this does not hold,
        and assume without loss of generality
        that $S_1[i+\ell] = {\tt 5}$ and $S_1[j+\ell] = {\tt 6}$. Consider two cases:
        \begin{enumerate}
        \item Assume that $\ell < 2m$. Then we have
          $i \in \mathcal{L}$ and $j \in \mathcal{R}$, which
          contradicts $\{i,j\} \cap \mathcal{P} \neq \emptyset$.
        \item Assume that $\ell \geq 2m$. Then, for
          some $x \in [1 \dd k]$, ${\tt 8}\widetilde{\Pat}_x$ is a substring occurring at the same position
          in $S_1[i \dd i + \ell)$ and $S_1[j \dd j + \ell)$. This contradicts the uniqueness of ${\tt 8}\widetilde{\Pat}_x$
          in $S_1$.
        \end{enumerate}
      \end{itemize}
      We are now ready to prove the main claim, i.e., that $S_1[i \dd \Textlen'] \prec S_1[j \dd \Textlen']$ holds if and only if
      $S_2[f(i) \dd \Textlen'] \prec S_2[f(j) \dd \Textlen']$. To this end, we will first show that
      $S_2[f(i) \dd f(i) + \ell] \neq S_2[f(j) \dd f(j) + \ell]$, and
      $S_1[i \dd i+\ell] \prec S_1[j \dd j+\ell]$ holds if and only if
      $S_2[f(i) \dd f(i)+\ell] \prec S_2[f(j) \dd f(j)+\ell]$.
      Let $t$ be the total number of occurrences of symbols ${\tt 5}$ and ${\tt 6}$
      in $S_1[i \dd i + \ell]$ and $S_1[j \dd j + \ell]$. By the two auxiliary properties proved above, we have $t \leq 1$. Moreover,
      if such an occurrence exists, it is at position $i+\ell$ or $j+\ell$.
      Consider two cases:
      \begin{itemize}
      \item Let us first assume that $i,j \in \mathcal{P}$.
        Consider two subcases:
        \begin{itemize}
        \item If $t=0$, then by definition of $S_2$,
          $S_2[i \dd i + \ell] = S_1[i \dd i + \ell]$ and $S_2[j \dd j + \ell] = S_1[j \dd j + \ell]$.
          Thus, $S_2[i \dd i + \ell] \neq S_2[j \dd j + \ell]$, and
          $S_1[i \dd i + \ell] \prec S_1[j \dd j + \ell]$ holds if and only if
          $S_2[i \dd i + \ell] \prec S_2[j \dd j + \ell]$.
        \item Let us now assume $t=1$.
          Without loss of generality, assume that $S_1[i + \ell] \in \{{\tt 5}, {\tt 6}\}$ and
          $S_1[j + \ell] \not\in \{{\tt 5}, {\tt 6}\}$. By definition of $S_2$,
          we then have $S_2[i \dd i + \ell) = S_1[i \dd i + \ell)$, $S_2[j \dd j + \ell) = S_1[j \dd j + \ell)$,
          $S_2[i + \ell] \in \{{\tt 5}, {\tt 6}\}$, and $S_2[j + \ell] = S_1[j + \ell] \not\in \{{\tt 5}, {\tt 6}\}$.
          Recall that ${\tt 5}$ and ${\tt 6}$ are consecutive symbols in $\Sigma$.
          Thus, $S_2[i \dd i + \ell] \neq S_2[j \dd j + \ell]$, and
          $S_1[i \dd i+\ell] \prec S_1[j \dd j + \ell]$ holds if and only if $S_2[i \dd i + \ell] \prec S_2[j \dd j + \ell]$.
        \end{itemize}
        In both cases, $S_2[i \dd i + \ell] \neq S_2[j \dd j + \ell]$, and
        $S_1[i \dd i+\ell] \prec S_1[j \dd j+\ell]$ holds if and only if
        $S_2[i \dd i+\ell] \prec S_2[j \dd j+\ell]$.
        By $i,j \in \mathcal{P}$, we have $f(i) = i$ and $f(j) = j$. Thus, the last inequality is equivalent to
        $S_2[f(i) \dd f(i)+\ell] \prec S_2[f(j) \dd f(j)+\ell]$. We also have
        $S_2[f(i) \dd f(i)+\ell] \neq S_2[f(j) \dd f(j)+\ell]$.
      \item Let us now assume that $|\{i,j\} \cap \mathcal{P}| = 1$. Without loss of generality, let us assume
        that $i \in \mathcal{P}$ and $j \in \mathcal{L} \cup \mathcal{R}$. Note that then $f(i) = i$.
        Consider two subcases:
        \begin{itemize}
        \item Let us first assume that $S_1[i+\ell] \not\in \{{\tt 5}, {\tt 6}\}$.
          Since $S_1$ and $S_2$ only differ at positions $\OccTwo{{\tt 5}}{S_1} \cup \OccTwo{{\tt 6}}{S_1}$,
          we then immediately obtain $S_2[f(i) \dd f(i)+\ell] = S_2[i \dd i+\ell] = S_1[i \dd i+\ell]$.
          On the other hand, by the above properties we know that $S_1[j \dd j + \ell)$ does not contain ${\tt 5}$ or ${\tt 6}$.
          By definition of $S_1$ and $S_2$, and the function $f$, it is easy to see that regardless of whether
          $S_1[j+\ell] \in \{{\tt 5}, {\tt 6}\}$, we have $S_2[f(j) \dd f(j) + \ell] = S_1[j \dd j + \ell]$.
          Putting everything together, we thus obtain that $S_1[i \dd i + \ell] = S_2[f(i) \dd f(i) + \ell]$ and
          $S_1[j \dd j + \ell] = S_2[f(j) \dd f(j) + \ell]$. In particular, $S_2[f(i) \dd f(i) + \ell] \neq S_2[f(j) \dd f(j) + \ell]$,
          and $S_1[i \dd i + \ell] \prec S_1[j \dd j + \ell]$ holds if and only if $S_2[f(i) \dd f(i) + \ell] \prec
          S_2[f(j) \dd f(j) + \ell]$.
        \item Let us now assume $S_1[i+\ell] \in \{{\tt 5}, {\tt 6}\}$.
          By the two auxiliary properties proved above, $S_1[j \dd j + \ell]$ does not contain ${\tt 5}$ or ${\tt 6}$.
          By definition of $S_1$, $S_2$, and the function $f$, this implies that $S_2[f(j) \dd f(j) + \ell] = S_1[j \dd j + \ell]$.
          On the other hand, since $f(i) = i$, and in $S_2$ every ${\tt 5}$ (resp.\ ${\tt 6}$) from $S_1$ is replaced with
          ${\tt 6}$ (resp.\ ${\tt 5}$), it follows that $S_2[f(i) \dd f(i) + \ell) = S_1[i \dd i + \ell)$ and
          $S_2[f(i) + \ell] \in \{{\tt 5}, {\tt 6}\}$. This implies that $S_2[f(i) \dd f(i) + \ell] \neq S_2[f(j) \dd f(j) + \ell]$
          and $S_2[f(i) \dd f(i) + \ell] \prec S_2[f(j) \dd f(j) + \ell]$ holds if and only if $S_1[i \dd i + \ell] \prec
          S_1[j \dd j + \ell]$.
        \end{itemize}
      \end{itemize}
      We have thus proved that in all cases it holds $S_2[f(i) \dd f(i) + \ell] \neq S_2[f(j) \dd f(j) + \ell]$, and
      $S_1[i \dd i+\ell] \prec S_1[j \dd j+\ell]$ is equivalent to $S_2[f(i) \dd f(i)+\ell] \prec S_2[f(j) \dd f(j)+\ell]$.
      Since the compared prefixes have equal length and are distinct,
      their lexicographic order determines that of the corresponding
      full suffixes. This proves the claim.
    \end{enumerate}
  \item\label{lm:dm-to-bwt-step-2}
    In this step, we relate the suffix arrays and BWTs of $S_1$ and $S_2$.
    First, we show that for every $j \in [1 \dd \Textlen']$, it holds
    $\ISA{S_1}[j] = \ISA{S_2}[f(j)]$.
    By Step~\ref{lm:dm-to-bwt-step-1} and the bijectivity of $f$,
    the definition of the inverse suffix array gives
    \begin{align*}
      \ISA{S_1}[j]
        &= 1 + |\{i \in [1 \dd \Textlen'] : S_1[i \dd \Textlen'] \prec S_1[j \dd \Textlen']\}|\\
        &= 1 + |\{i \in [1 \dd \Textlen'] : S_2[f(i) \dd \Textlen'] \prec S_2[f(j) \dd \Textlen']\}|\\
        &= 1 + |\{i \in [1 \dd \Textlen'] : S_2[i \dd \Textlen'] \prec S_2[f(j) \dd \Textlen']\}|
        = \ISA{S_2}[f(j)].
    \end{align*}
    For every $i \in [1 \dd \Textlen']$, letting $j=\SA{S_1}[i]$,
    we thus have $i=\ISA{S_2}[f(j)]$, and hence $\SA{S_2}[i]=f(j)$.
    By definition of the BWT, this gives
    \[
      B_1[i]=S_1^{\infty}[j-1],
      \qquad B_2[i]=S_2^{\infty}[f(j)-1].
    \]
  \item\label{lm:dm-to-bwt-step-3}
    In the next step, we prove that it holds
    \[
      |\RL{B_1(b_2 \dd e_9]}| = |\RL{B_2(b_2 \dd e_9]}|,
    \]
    i.e., the substrings of the two BWTs corresponding to suffixes starting
    with a symbol in $\{{\tt 2},\ldots,{\tt 9}\}$ have the same number of runs.
    We proceed in three steps:
    \begin{enumerate}
    \item First, we prove that $B_1(b_2 \dd e_9]$ has the form
      $X_0 {\tt 5}{\tt 6} X_1 {\tt 5}{\tt 6} X_2 \cdots X_{k-1} {\tt 5}{\tt 6} X_{k}$, where
      for every $i \in [0 \dd k]$, $X_i \in \{{\tt 0}, {\tt 1}, {\tt 2}, {\tt 3}, {\tt 4}, {\tt 7}\}^{*}$. To
      show this it suffices to observe that if $j \in [1 \dd \Textlen']$ is such that
      $S_1^{\infty}[j - 1] = {\tt 5}$, then there exists $i \in [1 \dd k]$ such
      that $S_1[j \dd \Textlen']$ has the string ${\tt 4}\widetilde{\Pat}_i {\tt 6}$ as a prefix.
      The suffix $S_1[j + \delta \dd \Textlen']$ has ${\tt 4}\widetilde{\Pat}_i {\tt 7}$ as a prefix.
      Each of these two prefixes occurs only once in $S_1$. Since ${\tt 6}$ and ${\tt 7}$ are
      consecutive alphabet symbols, it follows that $j$ and $j+\delta$ occur consecutively
      in the suffix array of $S_1$. Thus, every occurrence of $S_1^{\infty}[j-1] = {\tt 5}$ in the BWT of $S_1$ is followed by
      $S_1^{\infty}[j+\delta-1] = {\tt 6}$.
      Since this exhausts all occurrences of ${\tt 5}$ and ${\tt 6}$ in $S_1$, we obtain the claim.
    \item Next, we prove that $B_2(b_2 \dd e_9] = X_0 {\tt 6}{\tt 5} X_1 {\tt 6}{\tt 5} X_2 \cdots X_{k-1} {\tt 6}{\tt 5} X_k$,
      where $X_i$ is as in the previous step for all $i \in [0 \dd k]$. Denote
      $Q = \{j \in [1 \dd \Textlen'] : S_1[j] \in \{{\tt 2}, {\tt 3}, \dots, {\tt 9}\}\}$. Moreover, let
      $Q_{\tt 5} = \{j \in Q : S_1^{\infty}[j-1] = {\tt 5}\}$,
      $Q_{\tt 6} = \{j \in Q : S_1^{\infty}[j-1] = {\tt 6}\}$, and $Q_{\rm rest} = Q \setminus
      (Q_{\tt 5} \cup Q_{\tt 6})$. Note that $Q_{\tt 5}$ contains precisely all positions in the block $\SA{S_1}(b_2 \dd e_9]$
      for which the symbol in the BWT is ${\tt 5}$. An analogous property holds for $Q_{\tt 6}$.
      The set $Q_{\rm rest}$ consists of the suffix starting positions
      $\SA{S_1}[i]$ for which position $i$ in $B_1$ lies in one of the
      blocks $X_0,\ldots,X_k$.
      Let $i \in (b_2 \dd e_9]$ and denote $j=\SA{S_1}[i]\in Q$.
      By the definitions of $S_1$, $S_2$, and $f$, and the BWT identities
      from Step~\ref{lm:dm-to-bwt-step-2}, we obtain the following:
      \begin{enumerate}
      \item If $j \in Q_{\rm rest}$, then
        $S_1^{\infty}[j-1]=S_2^{\infty}[f(j)-1]$, and hence $B_2[i]=B_1[i]$.
      \item If $j \in Q_{\tt 5}$, then
        $S_1^{\infty}[j-1]={\tt 5}$ and $S_2^{\infty}[f(j)-1]={\tt 6}$,
        yielding $B_1[i]={\tt 5}$ and $B_2[i]={\tt 6}$.
      \item If $j \in Q_{\tt 6}$, then
        $S_1^{\infty}[j-1]={\tt 6}$ and $S_2^{\infty}[f(j)-1]={\tt 5}$,
        from which we obtain $B_1[i]={\tt 6}$ and $B_2[i]={\tt 5}$.
      \end{enumerate}
      We thus obtain the claim, i.e.,
      $B_2(b_2 \dd e_9] = X_0 {\tt 6}{\tt 5} X_1 {\tt 6}{\tt 5} X_2 \cdots X_{k-1} {\tt 6}{\tt 5} X_k$.
    \item Since neither ${\tt 5}$ nor ${\tt 6}$ occurs in any $X_i$,
      every occurrence of either symbol in $B_1(b_2 \dd e_9]$ and
      $B_2(b_2 \dd e_9]$ forms a run of length one. We thus obtain
      \begin{align*}
      |\RL{B_1(b_2 \dd e_9]}|
        &= |\RL{X_0{\tt 5}{\tt 6}X_1 \cdots X_{k-1}{\tt 5}{\tt 6}X_k}|
        = 2k + \textstyle\sum_{i=0}^{k}|\RL{X_i}|\\
        &= |\RL{X_0{\tt 6}{\tt 5}X_1 \cdots X_{k-1}{\tt 6}{\tt 5}X_k}|
        = |\RL{B_2(b_2 \dd e_9]}|.
      \end{align*}
    \end{enumerate}
  \item\label{lm:dm-to-bwt-step-4}
    In the next step, we prove that
    \[
      |\RL{B_1(b_0 \dd e_1]}| = 3k + 1 +|\{i \in [1 \dd k] : \OccTwo{\Pat_i}{\Text} \neq \emptyset\}|.
    \]
    The proof proceeds in two steps:
    \begin{enumerate}
    \item We prove that for every $i \in [1 \dd k]$, letting
      $b = \RangeBegTwo{\widetilde{\Pat}_i}{S_1}$ and $e = \RangeEndTwo{\widetilde{\Pat}_i}{S_1}$, it holds
      $B_1(b \dd e] = {\tt 4}{\tt 9}{\tt 4}^{n_i}{\tt 8}{\tt 4}{\tt 4}$, where $n_i = |\OccTwo{\Pat_i}{\Text}|$.
      To show this, observe that since $\widetilde{\Pat}_i$ contains only symbols from $\{{\tt 0}, {\tt 1}, {\tt 4}\}$,
      it follows by definition of $S_1$ that
      \[
        \OccTwo{\widetilde{\Pat}_i}{S_1} =
          \{2j : j \in \OccTwo{\Pat_i}{\Text}\} \cup \{\Delta + 5(i-1)\delta + 2 + j\delta : j \in [0 \dd 4]\}.
      \]
      Exactly $n_i$ of these occurrences lie in ${\tt 4}\widetilde{T}{\tt 4}$
      and are both preceded and followed by ${\tt 4}$.
      The five remaining occurrences are followed by ${\tt 2}$, ${\tt 3}$,
      ${\tt 5}$, ${\tt 6}$, and ${\tt 7}$.
      Ordering the suffixes starting at these $n_i+5$ occurrences and
      reading their preceding symbols gives
      $B_1(b\dd e]={\tt 4}{\tt 9}{\tt 4}^{n_i}{\tt 8}{\tt 4}{\tt 4}$.
    \item Denote $Q = \{j \in [1 \dd \Textlen'] : S_1[j] \in \{{\tt 0}, {\tt 1}\}\}$. Let
      $Q_{\rm occ} = \bigcup_{i=1}^{k} \OccTwo{\widetilde{\Pat}_i}{S_1}$ and $Q_{\rm rest} = Q \setminus Q_{\rm occ}$.
      Since $Q_{\rm occ} \subseteq Q$, it follows that $Q$ is a disjoint union of $Q_{\rm occ}$ and $Q_{\rm rest}$.
      Note also that $e_1 - b_0 = |Q|$. Since for every $i,i' \in [1 \dd k]$ such that $i \neq i'$, it holds that
      the set $\{\widetilde{\Pat}_{i}, \widetilde{\Pat}_{i'}\}$ is prefix-free (i.e., neither of the strings is a prefix of the other),
      it follows that suffix array ranges corresponding to all strings in the collection $\{\widetilde{\Pat}_i\}_{i \in [1 \dd k]}$
      are disjoint. On the other hand, observe that for every $j \in Q_{\rm rest}$, it holds $S_1^{\infty}[j-1] = {\tt 4}$.
      Since each block in the BWT of $S_1$ corresponding to suffixes beginning with
      $\widetilde{\Pat}_i$ begins and ends with ${\tt 4}$, the symbols in the BWT
      corresponding to suffixes starting at positions in $Q_{\rm rest}$ do not create
      new runs. Therefore, letting $(a_1, \ldots, a_k)$ be a permutation of $\{1, \ldots, k\}$ such that
      $\widetilde{\Pat}_{a_1} \prec \widetilde{\Pat}_{a_2} \prec \cdots \prec \widetilde{\Pat}_{a_k}$, it follows that
      for some sequence $(r_0, r_1, \ldots, r_k)$ of positive integers, it holds
      \[
        B_1(b_0 \dd e_1] = {\tt 4}^{r_0}{\tt 9}{\tt 4}^{n_{a_1}}{\tt 8}{\tt 4}^{r_1}{\tt 9}{\tt 4}^{n_{a_2}}{\tt 8}{\tt 4}^{r_2} \cdots
            {\tt 4}^{r_{k-1}}{\tt 9}{\tt 4}^{n_{a_k}}{\tt 8}{\tt 4}^{r_{k}}.
      \]
      This immediately implies that
      \[
        |\RL{B_1(b_0 \dd e_1]}|
              = 3k + 1 + |\{i \in [1 \dd k] : n_i > 0\}|
              = 3k + 1 + |\{i \in [1 \dd k] : \OccTwo{\Pat_i}{\Text} \neq \emptyset\}|.
      \]
    \end{enumerate}
  \item\label{lm:dm-to-bwt-step-5}
      Next, we show that
      \[
        |\RL{B_2(b_0 \dd e_1]}| = 4k + 1.
      \]
    We follow the same argument in two steps:
    \begin{enumerate}
    \item First, we observe that by an argument analogous to the one above, for every $i \in [1 \dd k]$, letting
      $b = \RangeBegTwo{\widetilde{\Pat}_i}{S_2}$ and $e = \RangeEndTwo{\widetilde{\Pat}_i}{S_2}$, it holds
      $B_2(b \dd e] = {\tt 4}{\tt 9}{\tt 4}^{n_i}{\tt 4}{\tt 8}{\tt 4} =
      {\tt 4}{\tt 9}{\tt 4}^{n_i+1}{\tt 8}{\tt 4}$, where $n_i = |\OccTwo{\Pat_i}{\Text}|$.
    \item Second, we observe that
      each block in the BWT of $S_2$ corresponding to suffixes beginning with
      $\widetilde{\Pat}_i$ begins and ends with ${\tt 4}$.
      Therefore, letting $(a_1, \ldots, a_k)$ be a permutation of $\{1, \ldots, k\}$ such that
      $\widetilde{\Pat}_{a_1} \prec \widetilde{\Pat}_{a_2} \prec \cdots \prec \widetilde{\Pat}_{a_k}$, it follows that
      for some sequence $(r_0, r_1, \ldots, r_k)$ of positive integers, it holds
      \[
        B_2(b_0 \dd e_1] =
            {\tt 4}^{r_0}{\tt 9}{\tt 4}^{n_{a_1}+1}{\tt 8}{\tt 4}^{r_1}{\tt 9}{\tt 4}^{n_{a_2}+1}{\tt 8}{\tt 4}^{r_2} \cdots
            {\tt 4}^{r_{k-1}}{\tt 9}{\tt 4}^{n_{a_k}+1}{\tt 8}{\tt 4}^{r_{k}}.
      \]
      This immediately implies the claim, i.e.,
      $|\RL{B_2(b_0 \dd e_1]}| = 4k + 1$.
    \end{enumerate}
  \item\label{lm:dm-to-bwt-step-6}
    We are now ready to prove the main claim. By the decompositions in
    Steps~\ref{lm:dm-to-bwt-step-4} and~\ref{lm:dm-to-bwt-step-5}, we have
    $B_1(b_0 \dd e_1], B_2(b_0 \dd e_1] \in \{{\tt 4}, {\tt 8}, {\tt 9}\}^{+}$.
    On the other hand, every occurrence of ${\tt 2}$ is preceded in $S_1$ and $S_2$ by either ${\tt 0}$ or ${\tt 1}$, and hence
    $B_1(b_2 \dd e_2], B_2(b_2 \dd e_2] \in \{{\tt 0}, {\tt 1}\}^{+}$. This implies that position $e_1$ is the end of a run
    both in the BWT of $S_1$ and $S_2$, i.e.,
    \begin{align*}
      |\RL{B_1(b_0 \dd e_9]}| &= |\RL{B_1(b_0 \dd e_1]}| + |\RL{B_1(b_2 \dd e_9]}|,\\
      |\RL{B_2(b_0 \dd e_9]}| &= |\RL{B_2(b_0 \dd e_1]}| + |\RL{B_2(b_2 \dd e_9]}|.
    \end{align*}
    Subtracting these equations and applying
    Steps~\ref{lm:dm-to-bwt-step-3},~\ref{lm:dm-to-bwt-step-4}, and~\ref{lm:dm-to-bwt-step-5},
    we obtain
    \begin{align*}
      \RLBWTSize{S_2}-\RLBWTSize{S_1}
        &= |\RL{B_2(b_0 \dd e_1]}|-|\RL{B_1(b_0 \dd e_1]}|\\
        &= (4k+1)-(3k+1+|\{i \in [1 \dd k] : \OccTwo{\Pat_i}{\Text} \neq \emptyset\}|)\\
        &= |\{i \in [1 \dd k] : \OccTwo{\Pat_i}{\Text} = \emptyset\}|.
      \qedhere
    \end{align*}
  \end{enumerate}
\end{proof}

\begin{remark}\label{rm:dm-to-bwt-binary}
  The next lemma gives a version of \cref{lm:dm-to-bwt} in which
  all strings are binary. We retain the ten-symbol construction
  because its proof is simpler. For arguments shared by the two
  constructions, the proof below describes only the necessary changes.
\end{remark}

\begin{lemma}\label{lm:dm-to-bwt-binary}
  Let $\Textlen,m\in\Z_{\geq1}$ and $k\in\Z_{\geq0}$, and let
  $\Text \in \BinaryAlphabet^{\Textlen}$ be a text and $\mathcal{D} =
  \{\Pat_1, \Pat_2, \dots, \Pat_k\} \subseteq \BinaryAlphabet^{m}$
  be a collection of $k$ pairwise distinct patterns of common length $m$.
  For any $q\in\Z_{\geq1}$ and string $S = a_1 a_2 \cdots a_q\in\BinaryAlphabet^q$,
  we define $\widetilde{S} := a_1 {\tt 10} a_2 {\tt 10} a_3 {\tt 10} \cdots a_q {\tt 10}$.
  Consider the following strings over alphabet $\BinaryAlphabet$ (brackets are added for clarity):
  \begin{align*}
    S_1 &= {\tt 10}\widetilde{T} \cdot
      \textstyle\bigodot_{i=1}^{k} \big({\tt 10}\widetilde{\Pat_i}{\tt 0}^{7}{\tt 1}^{4}\cdot
                                        {\tt 00}\widetilde{\Pat_i}{\tt 0}^{6}{\tt 1}^{5}\cdot
                                        {\tt 00}\widetilde{\Pat_i}{\tt 1}^{3}{\tt 0}^{3}{\tt 1}^{5}\cdot
                                        {\tt 00}\widetilde{\Pat_i}{\tt 1}^{4}{\tt 0}^{3}{\tt 1}^{4}\cdot
                                        {\tt 00}\widetilde{\Pat_i}{\tt 0}^{8}{\tt 1}^{3}
                                   \big),\\
    S_2 &= {\tt 10}\widetilde{T} \cdot
      \textstyle\bigodot_{i=1}^{k} \big({\tt 10}\widetilde{\Pat_i}{\tt 0}^{6}{\tt 1}^{5}\cdot
                                        {\tt 00}\widetilde{\Pat_i}{\tt 0}^{7}{\tt 1}^{4}\cdot
                                        {\tt 00}\widetilde{\Pat_i}{\tt 1}^{3}{\tt 0}^{3}{\tt 1}^{5}\cdot
                                        {\tt 00}\widetilde{\Pat_i}{\tt 1}^{4}{\tt 0}^{3}{\tt 1}^{4}\cdot
                                        {\tt 00}\widetilde{\Pat_i}{\tt 0}^{8}{\tt 1}^{3}
                                   \big).
  \end{align*}
  Then, it holds $\RLBWTSize{S_2} - \RLBWTSize{S_1} =
  2|\{i \in [1 \dd k] : \OccTwo{\Pat_i}{\Text} = \emptyset\}|$.
\end{lemma}
\begin{proof}
  The claim holds trivially if $k=0$. Let us thus assume that $k \geq 1$.
  Denote $\Textlen' := |S_1| = |S_2|$, $\Delta := 3\Textlen + 2$, and $\delta := 3m + 13$,
  and note that then $\Textlen' = \Delta + 5k \delta$.
  Denote
  \begin{align*}
    P_{\mathcal{L}}
      &= \{\Delta + 5(i-1)\delta + 3m + 9 : i \in [1 \dd k]\},\\
    P_{\mathcal{R}}
      &= \{\Delta + 5(i-1)\delta + \delta + 3m + 9 : i \in [1 \dd k]\}.
  \end{align*}
  Observe that $P_{\mathcal{L}} \cup P_{\mathcal{R}}$ is precisely the set of positions on which $S_1$ differs from $S_2$
  (specifically, $P_{\mathcal{L}}$ is the set where an occurrence of ${\tt 0}$ in $S_1$ changes into ${\tt 1}$ in $S_2$, and
  $P_{\mathcal{R}}$ is symmetrical).
  Let $\mathcal{L}$ (resp.\ $\mathcal{R}$) be the set of all $j \in [1 \dd \Textlen']$ satisfying
  $[j \dd j + 3m + 8) \cap P_{\mathcal{L}} \neq \emptyset$ (resp.\ $[j \dd j + 3m + 8) \cap P_{\mathcal{R}} \neq \emptyset$).
  It is easy to check that
  \begin{align*}
    \mathcal{L}
       &= \textstyle\bigcup_{i \in [1 \dd k]}
            [\Delta + 5(i{-}1)\delta + 2 \dd
             \Delta + 5(i{-}1)\delta + 3m + 9],\\
    \mathcal{R}
      &= \textstyle\bigcup_{i \in [1 \dd k]}
            [\Delta + 5(i{-}1)\delta + \delta + 2 \dd
             \Delta + 5(i{-}1)\delta + \delta + 3m + 9].
  \end{align*}
  We also define $\mathcal{P} := [1 \dd \Textlen'] \setminus (\mathcal{L} \cup \mathcal{R})$.
  Note that $\mathcal{L} \cap \mathcal{R} = \emptyset$ and $\{i + \delta : i \in \mathcal{L}\} = \mathcal{R}$.
  Let $f : [1 \dd \Textlen'] \rightarrow
  [1 \dd \Textlen']$ be such that for every $x \in [1 \dd \Textlen']$,
  \[
    f(x) =
      \begin{cases}
        x + \delta & \text{if }x \in \mathcal{L}, \\
        x - \delta & \text{if }x \in \mathcal{R}, \\
        x          & \text{if }x \in \mathcal{P}.
      \end{cases}
  \]
  For $i \in \{1, 2\}$, we denote $B_i = \BWT{S_i}$.
  For every $c \in \BinaryAlphabet$, we define $b_c = |\{i \in [1 \dd \Textlen'] : S_1[i] \prec c\}|$ and
  $e_c = |\{i \in [1 \dd \Textlen'] : S_1[i] \preceq c\}|$.
  Since for every $c \in \BinaryAlphabet$, the number of occurrences of $c$ in $S_1$ is the same as in $S_2$,
  it holds that for every $i \in \{1, 2\}$ and $c \in \BinaryAlphabet$, $B_i(b_c \dd e_c]$
  is the block in the BWT of $S_i$ containing all symbols preceding suffixes of $S_i$
  starting with symbol $c$ (where the symbol preceding $S_i[1]$ is $S_i[\Textlen']$).

  The proof consists of five steps:
  \begin{enumerate}
  \item\label{lm:dm-to-bwt-binary-step-1}
    In the first step, we show that for every $i, j \in [1 \dd \Textlen']$,
    $S_1[i \dd \Textlen'] \prec S_1[j \dd \Textlen']$ holds if and only
    if $S_2[f(i) \dd \Textlen'] \prec S_2[f(j) \dd \Textlen']$.
    Note that the claim holds trivially if $i = j$.
    Let us thus assume that $i \neq j$. Denote $\ell = \LCE{S_1}{i}{j}$.
    Consider three cases:
    \begin{enumerate}
    \item Assume that $\max(i,j) + \ell = \Textlen' + 1$. Assume without loss of generality
      that $i < j$. Note that we then
      have $j + \ell = \Textlen' + 1$ and $S_1[i \dd i + \ell) = S_1[j \dd j + \ell) = S_1[j \dd \Textlen']$.
      Thus, by $i \neq j$, it holds
      $S_1[i \dd \Textlen'] \succ S_1[j \dd \Textlen']$. Our goal is therefore to prove that
      $S_2[f(i) \dd \Textlen'] \succ S_2[f(j) \dd \Textlen']$. We proceed in three
      steps:
      \begin{itemize}
      \item First, observe that $\ell \leq 3m + 10$, since otherwise we would
        have two occurrences in $S_1$ of a unique substring $\widetilde{\Pat}_k{\tt 0}^{8}{\tt 1}^{3}$.
        Consequently, $j \in \mathcal{P}$, $f(j) = j$, and $S_2[j \dd \Textlen'] = S_1[j \dd \Textlen']$.
      \item In the second step, we prove that $S_1[i \dd i + \ell) = S_2[f(i) \dd f(i) + \ell)$.
        Observe that for every $x \in [1 \dd \Textlen' - (3m + 7) + 1]$, it holds
        $S_1[x \dd x + 3m + 7) = S_2[f(x) \dd f(x) + 3m + 7)$. This implies that if $\ell \leq 3m + 7$, the
        claim follows immediately. It thus remains
        to prove that $S_1[i \dd i + \ell) = S_2[f(i) \dd f(i) + \ell)$ for $\ell = 3m + \alpha$, where $\alpha
        \in \{8,9,10\}$. We consider two cases:
        \begin{itemize}
        \item If $\alpha \geq 9$ or $m \geq 2$, then $3m + \alpha \geq 12$, and
          hence ${\tt 0}^{9}{\tt 1}^{3}$ is a suffix of $S_1[i \dd i + \ell)$.
          By $\OccTwo{{\tt 0}^{9}{\tt 1}^{3}}{S_1} = \{\Delta + 5(t-1)\delta + 4\delta + 3m + 2 : t \in [1 \dd k]\}$,
          we thus obtain that for some $t \in [1 \dd k]$, it
          holds $\Delta + 5(t-1)\delta + 4\delta < i < i + \ell \leq \Delta + 5t\delta + 1$. For
          any such $i$, we clearly have $f(i) = i$ and $S_1[i \dd i + \ell) = S_2[i \dd i + \ell) =
          S_2[f(i) \dd f(i) + \ell)$.
        \item It remains to consider the case $\alpha = 8$ and $m = 1$. Then, $\ell = 11$ and
          $S_1[i \dd i + \ell) = {\tt 0}^{8}{\tt 1}^{3}$. Note that $\OccTwo{{\tt 0}^{8}{\tt 1}^{3}}{S_1} =
          \{\Delta + 5(t-1)\delta + 3m + 2 : t \in [1 \dd k]\} \cup \{\Delta + 5(t-1)\delta + 4\delta + 3m + 3 : t \in [1 \dd k]\}$.
          It is easy to check that for any position $x$ in this set, it holds $S_1[x \dd x + 11) = S_2[f(x) \dd f(x) + 11)$.
        \end{itemize}
      \item By the above two steps, it holds
        $S_2[f(i) \dd f(i) + \ell) = S_1[i \dd i + \ell) = S_1[j \dd j + \ell) = S_1[j \dd \Textlen'] =
        S_2[j \dd \Textlen'] = S_2[f(j) \dd \Textlen']$. By $f(i) \neq f(j)$, we thus obtain
        $S_2[f(i) \dd \Textlen'] \succ S_2[f(j) \dd \Textlen']$.
      \end{itemize}
    \item Assume that $\max(i,j)+\ell\leq\Textlen'$ and
      $i,j \in \mathcal{L} \cup \mathcal{R}$.
      For any $x \in \mathcal{L} \cup \mathcal{R}$, let $d_x$ denote the
      smallest non-negative integer such that $x + d_x \in P_{\mathcal{L}} \cup P_{\mathcal{R}}$,
      and let $\ell_x = d_x + 3m + 7$.
      Observe two properties of $\ell_x$:
      \begin{itemize}
      \item First note that if
        $x \in \mathcal{L}$ (resp.\ $x \in \mathcal{R}$)
        then for some $t \in [1 \dd k]$, ${\tt 0}{\tt 1}^{4}{\tt 00}\widetilde{\Pat}_t$ (resp.\ ${\tt 1}^{5}{\tt 00}\widetilde{\Pat}_t$)
        is a suffix of $S_1[x \dd x + \ell_x)$, and moreover, this suffix starts at a position in $\mathcal{L} \cup \mathcal{R}$.
        Since both
        ${\tt 0}{\tt 1}^{4}{\tt 00}\widetilde{\Pat}_t$ and
        ${\tt 1}^{5}{\tt 00}\widetilde{\Pat}_t$ each have only one occurrence in $S_1$ starting in $\mathcal{L} \cup \mathcal{R}$,
        it follows that for every $x, x' \in \mathcal{L} \cup \mathcal{R}$ with $x\neq x'$, it holds $S_1[x \dd x + \ell_x) \neq S_1[x' \dd x' + \ell_{x'})$,
        and moreover, the set $\{S_1[x \dd x + \ell_x), S_1[x' \dd x' + \ell_{x'})\}$ is prefix-free.
        Indeed, if $S_1[x \dd x + \ell_x)$ were a prefix of $S_1[x' \dd x' + \ell_{x'})$, then
        $d_x \leq d_{x'}$, and the above suffix of $S_1[x \dd x + \ell_x)$, which starts at
        $x + d_x$, would also occur at $x' + d_x \in [x' \dd x' + d_{x'}]$. Every position $y$ in this
        interval belongs to $\mathcal{L} \cup \mathcal{R}$, because $d_{x'} < 3m + 8$ and
        $x' + d_{x'} \in [y \dd y + 3m + 8) \cap (P_{\mathcal{L}} \cup P_{\mathcal{R}})$.
        The uniqueness of the occurrence would then give $x = x'$.
        This implies that $\LCE{S_1}{x}{x'} < \min(\ell_x, \ell_{x'})$.
      \item Second, observe that for every $x \in \mathcal{L} \cup \mathcal{R}$, it holds
        $S_1[x \dd x + \ell_{x}) = S_2[f(x) \dd f(x) + \ell_{x})$.
      \end{itemize}
      Using the above properties and the same argument as in the proof of Step~\ref{lm:dm-to-bwt-case-1b} in \cref{lm:dm-to-bwt},
      it follows that $S_1[i \dd \Textlen'] \prec S_1[j \dd \Textlen']$ holds if and only
      if $S_2[f(i) \dd \Textlen'] \prec S_2[f(j) \dd \Textlen']$.
    \item Assume that $\max(i,j)+\ell\leq\Textlen'$ and
      that either $i \in \mathcal{P}$ or $j \in \mathcal{P}$.
      We begin by proving three auxiliary properties:
      \begin{itemize}
      \item First, we prove that for every $x \in [1 \dd \Textlen']$ and $y \in \mathcal{P}$ with $x\neq y$,
        letting $h = \LCE{S_1}{x}{y}$, it holds
        $[y \dd y + h) \cap (P_{\mathcal{L}} \cup P_{\mathcal{R}}) = \emptyset$.
        Suppose that this does not hold. Let $q \in [0 \dd h)$ be such
        that $y + q \in P_{\mathcal{L}} \cup P_{\mathcal{R}}$. Observe that we must have $q \geq 3m + 8$, since otherwise
        by definition of $\mathcal{L}$ and $\mathcal{R}$ we would have $y \in \mathcal{L} \cup \mathcal{R}$.
        Observe now that if $y + q \in P_{\mathcal{L}}$, then there exists $t \in [1 \dd k]$ such that
        ${\tt 10}\widetilde{\Pat}_t{\tt 0}^7$ is a substring of $S_1[y \dd y + h)$. By $h = \LCE{S_1}{x}{y}$ and $x \neq y$,
        this implies, however, that ${\tt 10}\widetilde{\Pat}_t{\tt 0}^{7}$ has at least two occurrences in $S_1$, which does
        not hold. Similarly, if $y + q \in P_{\mathcal{R}}$, then for some $t \in [1 \dd k]$ we obtain two
        occurrences of ${\tt 00}\widetilde{\Pat}_t{\tt 0}^{6}{\tt 1}$ in $S_1$, which again is a contradiction.
      \item Second, we prove that for every $x \in [1 \dd \Textlen']$ and $y \in \mathcal{P}$ with $x\neq y$,
        letting $h = \LCE{S_1}{x}{y}$, it holds $y+h \not\in P_{\mathcal{L}}$.
        Suppose that this does not hold.
        Similarly, as above we must have $h \geq 3m + 8$ (otherwise, we would
        have $y \in \mathcal{L}$). This, however, implies that for some $t \in [1 \dd k]$,
        ${\tt 10}\widetilde{\Pat}_{t}{\tt 0}^{6}$ occurs at the same position in
        $S_1[y \dd y + h)$ and
        $S_1[x \dd x + h)$.
        This is a contradiction, since ${\tt 10}\widetilde{\Pat}_{t}{\tt 0}^{6}$ occurs in $S_1$ only once.
      \item Third, we prove that for every $x \in [1 \dd \Textlen']$ and $y \in \mathcal{P}$ with $x\neq y$,
        letting $h = \LCE{S_1}{x}{y}$, $y + h \in P_{\mathcal{R}}$ implies that $x \in \mathcal{P}$.
        First, we observe that $y \in \mathcal{P}$ and $y+h \in P_{\mathcal{R}}$,
        together with the first auxiliary property $[y \dd y+h) \cap (P_{\mathcal{L}} \cup P_{\mathcal{R}}) = \emptyset$, imply
        that there exists an index $t \in [1 \dd k]$ and an integer $\alpha \in [0 \dd 5)$ such that
        $y = \Delta + 5(t-1)\delta + \delta + 1 - \alpha$ and
        ${\tt 1}^{\alpha}{\tt 00}\widetilde{\Pat}_t {\tt 0}^{6}$ is a prefix of $S_1[y \dd y + h)$.
        Since ${\tt 00}\widetilde{\Pat}_t {\tt 0}^{6}$ occurs in $S_1$ only twice, this
        implies that $x = \Delta + 5(t-1)\delta + 4\delta + 1 - \alpha$.
        In particular, $x \in \mathcal{P}$.
      \end{itemize}
      We are now ready to prove the main claim, i.e., that
      $S_1[i \dd \Textlen'] \prec S_1[j \dd \Textlen']$ holds if and only
      if $S_2[f(i) \dd \Textlen'] \prec S_2[f(j) \dd \Textlen']$.
      We consider two cases:
      \begin{enumerate}
      \item First, let us assume that $i,j \in \mathcal{P}$.
        By the above properties, $[i \dd i + \ell) \cap (P_{\mathcal{L}} \cup P_{\mathcal{R}}) = \emptyset$
        and $[j \dd j + \ell) \cap (P_{\mathcal{L}} \cup P_{\mathcal{R}}) = \emptyset$.
        We also have $\{i+\ell,j+\ell\} \cap P_{\mathcal{L}} = \emptyset$. Observe also that we
        have $|\{i+\ell,j+\ell\} \cap P_{\mathcal{R}}| \leq 1$, since all symbols of $S_1$ at positions in $P_{\mathcal{R}}$ are
        ${\tt 1}$, and by definition of $\ell$ we must have $S_1[i+\ell] \neq S_1[j+\ell]$.
        We consider two subcases:
        \begin{itemize}
        \item First, assume that $\{i+\ell,j+\ell\} \cap P_{\mathcal{R}} = \emptyset$. We thus have
          $[i \dd i + \ell] \cap (P_{\mathcal{L}} \cup P_{\mathcal{R}}) = \emptyset$ and
          $[j \dd j + \ell] \cap (P_{\mathcal{L}} \cup P_{\mathcal{R}}) = \emptyset$.
          Recalling that $S_1$ and $S_2$ differ only on positions in $P_{\mathcal{L}} \cup P_{\mathcal{R}}$,
          we thus have $S_1[i \dd i + \ell] = S_2[i \dd i + \ell]$ and $S_1[j \dd j + \ell] = S_2[j \dd j + \ell]$.
          On the other hand, by $i,j \in \mathcal{P}$, we have $f(i) = i$ and $f(j) = j$.
          Consequently, $S_1[i \dd i + \ell] = S_2[f(i) \dd f(i) + \ell]$ and $S_1[j \dd j + \ell] = S_2[f(j) \dd f(j) + \ell]$.
          Since by definition of $\ell$, $S_1[i \dd \Textlen'] \prec S_1[j \dd \Textlen']$ holds if and only if
          $S_1[i \dd i + \ell] \prec S_1[j \dd j + \ell]$, we thus obtain the claim.
        \item Let us now assume that $\{i+\ell, j+\ell\} \cap P_{\mathcal{R}} \neq \emptyset$.
          Without loss of generality, assume that $i+\ell \in P_{\mathcal{R}}$.
          As noted above, it then follows that
          there exists an index $t \in [1 \dd k]$ and an integer $\alpha \in [0 \dd 5)$ such that
          $i = \Delta + 5(t-1)\delta + \delta + 1 - \alpha$ and
          ${\tt 1}^{\alpha}{\tt 00}\widetilde{\Pat}_t {\tt 0}^{6}$ is a prefix of $S_1[i \dd i + \ell)$.
          Since ${\tt 00}\widetilde{\Pat}_t {\tt 0}^{6}$ occurs in $S_1$ only twice, this
          implies that $j = \Delta + 5(t-1)\delta + 4\delta + 1 - \alpha$.
          This in turn implies that $\ell = \alpha + 3m + 8$,
          $S_1[i \dd i + \ell] = {\tt 1}^{\alpha}{\tt 00}\widetilde{\Pat}_t {\tt 0}^{6}{\tt 1}$, and
          $S_1[j \dd j + \ell] = {\tt 1}^{\alpha}{\tt 00}\widetilde{\Pat}_t {\tt 0}^{7}$.
          Therefore, we obtain $S_1[i \dd \Textlen'] \succ S_1[j \dd \Textlen']$.
          It now remains to verify that $S_2[f(i) \dd \Textlen'] \succ S_2[f(j) \dd \Textlen']$.
          To this end, first recall that $f(i) = i$ and $f(j) = j$.
          Hence, ${\tt 1}^{\alpha}{\tt 00}\widetilde{\Pat}_t {\tt 0}^{7}{\tt 1}$ is a prefix of $S_2[f(i) \dd \Textlen']$.
          On the other hand, ${\tt 1}^{\alpha}{\tt 00}\widetilde{\Pat}_t {\tt 0}^{8}$ is a prefix of
          $S_2[f(j) \dd \Textlen']$. Thus, indeed $S_2[f(i) \dd \Textlen'] \succ S_2[f(j) \dd \Textlen']$.
        \end{itemize}
      \item Let us now assume that $|\{i,j\} \cap \mathcal{P}| = 1$. Without loss of generality,
        assume that $i \in \mathcal{P}$ and $j \in \mathcal{L} \cup \mathcal{R}$. Note that then $f(i) = i$.
        Note also that by the three properties above, we then have $[i \dd i + \ell] \cap
        (P_{\mathcal{L}} \cup P_{\mathcal{R}}) = \emptyset$.
        Since $S_1$ and $S_2$ differ only on positions in $P_{\mathcal{L}} \cup P_{\mathcal{R}}$, it follows
        that $S_2[f(i) \dd f(i) + \ell] = S_1[i \dd i + \ell]$.
        Observe that if $\ell < \ell_j$ (where $\ell_j$ is defined as above),
        then by $S_1[j \dd j + \ell_j) = S_2[f(j) \dd f(j) + \ell_j)$ it follows that
        $S_1[j \dd j + \ell] = S_2[f(j) \dd f(j) + \ell]$. Combining this with $S_1[i \dd i + \ell] = S_2[f(i) \dd f(i) + \ell]$
        implies the claim, since comparing $S_1[i \dd i + \ell]$ and $S_1[j \dd j + \ell]$ is by definition of $\ell$
        equivalent to comparing $S_1[i \dd \Textlen']$ and $S_1[j \dd \Textlen']$.
        Let us thus assume $\ell \geq \ell_j$.
        Consider now two cases:
        \begin{itemize}
        \item First, assume that $j \in \mathcal{L}$.
          Let $d_j$ denote the smallest nonnegative integer
          such that $j + d_j \in P_{\mathcal{L}}$.
          We first show that $d_j < 3$. Otherwise, the substring
          ${\tt 0}^{4}{\tt 1}^{4}{\tt 00}\widetilde{\Pat}_t$ (for some $t \in [1 \dd k]$) would be a substring of
          $S_1[j \dd j + \ell)$. Thus, by $S_1[j \dd j + \ell) = S_1[i \dd i + \ell)$, it would have at least two
          occurrences in $S_1$, which is not possible, since for every $t \in [1 \dd k]$,
          $|\OccTwo{{\tt 0}^{4}{\tt 1}^{4}{\tt 00}\widetilde{\Pat}_t}{S_1}| = 1$. Consequently, there exist
          $t \in [1 \dd k]$ and $\alpha \in [1 \dd 3]$ such that $j = \Delta + 5(t-1)\delta + 3m + 10 - \alpha$ and
          ${\tt 0}^{\alpha}{\tt 1}^{4}{\tt 00}\widetilde{\Pat}_t$ is a prefix
          of $S_1[j \dd j + \ell)$. Note that since ${\tt 0}{\tt 1}^{4}{\tt 00}\widetilde{\Pat}_t$ occurs in
          $S_1$ exactly twice, it follows
          that $i = \Delta + 5(t-1)\delta + 3\delta + 3m + 10 - \alpha$. This in turn implies that
          $\ell = \alpha + 3m + 12$,
          $S_1[j \dd j + \ell] = {\tt 0}^{\alpha}{\tt 1}^{4}{\tt 00}\widetilde{\Pat}_t{\tt 0}^{6}{\tt 1}$, and
          $S_1[i \dd i + \ell] = {\tt 0}^{\alpha}{\tt 1}^{4}{\tt 00}\widetilde{\Pat}_t{\tt 0}^{7}$.
          Therefore, we obtain $S_1[j \dd \Textlen'] \succ S_1[i \dd \Textlen']$.
          It now remains to verify that $S_2[f(j) \dd \Textlen'] \succ S_2[f(i) \dd \Textlen']$.
          First, note that $f(j) = j + \delta = \Delta + 5(t-1)\delta + \delta + 3m + 10 - \alpha$.
          Hence ${\tt 0}^{\alpha}{\tt 1}^{4}{\tt 00}\widetilde{\Pat}_t{\tt 1}$ is a prefix of $S_2[f(j) \dd \Textlen']$.
          On the other hand, recall that $f(i) = i$, and hence
          ${\tt 0}^{\alpha}{\tt 1}^{4}{\tt 00}\widetilde{\Pat}_t{\tt 0}$ is a prefix of $S_2[f(i) \dd \Textlen']$.
          Thus, indeed $S_2[f(j) \dd \Textlen'] \succ S_2[f(i) \dd \Textlen']$.
        \item Let us now assume that $j \in \mathcal{R}$.
          Let $d_j$ denote the smallest nonnegative integer
          such that $j + d_j \in P_{\mathcal{R}}$.
          We first show that $d_j < 4$. Otherwise, the substring
          ${\tt 0}^{4}{\tt 1}^{5}{\tt 00}\widetilde{\Pat}_t$ (for some $t \in [1 \dd k]$) would be a substring of
          $S_1[j \dd j + \ell)$. Thus, by $S_1[j \dd j + \ell) = S_1[i \dd i + \ell)$, it would have at least two
          occurrences in $S_1$, which is not possible, since for every $t \in [1 \dd k]$,
          $|\OccTwo{{\tt 0}^{4}{\tt 1}^{5}{\tt 00}\widetilde{\Pat}_t}{S_1}| = 1$. Consequently, there exist
          $t \in [1 \dd k]$ and $\alpha \in [0 \dd 3]$ such that $j = \Delta + 5(t-1)\delta + \delta + 3m + 9 - \alpha$ and
          ${\tt 0}^{\alpha}{\tt 1}^{5}{\tt 00}\widetilde{\Pat}_t$ is a prefix
          of $S_1[j \dd j + \ell)$. Note that since ${\tt 1}^{5}{\tt 00}\widetilde{\Pat}_t$ occurs in
          $S_1$ exactly twice, it follows
          that $i = \Delta + 5(t-1)\delta + 2\delta + 3m + 9 - \alpha$. This in turn implies that
          $\ell = \alpha + 3m + 10$,
          $S_1[j \dd j + \ell] = {\tt 0}^{\alpha}{\tt 1}^{5}{\tt 00}\widetilde{\Pat}_t{\tt 1}^{3}{\tt 0}$, and
          $S_1[i \dd i + \ell] = {\tt 0}^{\alpha}{\tt 1}^{5}{\tt 00}\widetilde{\Pat}_t{\tt 1}^{4}$.
          Therefore, we obtain $S_1[j \dd \Textlen'] \prec S_1[i \dd \Textlen']$.
          It now remains to verify that $S_2[f(j) \dd \Textlen'] \prec S_2[f(i) \dd \Textlen']$.
          First, note that $f(j) = j - \delta = \Delta + 5(t-1)\delta + 3m + 9 - \alpha$.
          Hence ${\tt 0}^{\alpha}{\tt 1}^{5}{\tt 00}\widetilde{\Pat}_t{\tt 0}$ is a prefix of $S_2[f(j) \dd \Textlen']$.
          On the other hand, recall that $f(i) = i$, and hence
          ${\tt 0}^{\alpha}{\tt 1}^{5}{\tt 00}\widetilde{\Pat}_t{\tt 1}$ is a prefix of $S_2[f(i) \dd \Textlen']$.
          Thus, indeed $S_2[f(j) \dd \Textlen'] \prec S_2[f(i) \dd \Textlen']$.
        \end{itemize}
      \end{enumerate}
    \end{enumerate}
  \item\label{lm:dm-to-bwt-binary-step-2}
    In the second step, we relate the suffix arrays and BWTs of $S_1$ and $S_2$.
    By Step~\ref{lm:dm-to-bwt-binary-step-1} and the bijectivity of $f$,
    for every $j \in [1 \dd \Textlen']$, the definition of the inverse suffix array gives
    \begin{align*}
      \ISA{S_1}[j]
        &= 1 + |\{i \in [1 \dd \Textlen'] : S_1[i \dd \Textlen'] \prec S_1[j \dd \Textlen']\}|\\
        &= 1 + |\{i \in [1 \dd \Textlen'] : S_2[f(i) \dd \Textlen'] \prec S_2[f(j) \dd \Textlen']\}|\\
        &= 1 + |\{i \in [1 \dd \Textlen'] : S_2[i \dd \Textlen'] \prec S_2[f(j) \dd \Textlen']\}|
        = \ISA{S_2}[f(j)].
    \end{align*}
    For every $i \in [1 \dd \Textlen']$, letting $j=\SA{S_1}[i]$,
    we thus have $i=\ISA{S_2}[f(j)]$, and hence $\SA{S_2}[i]=f(j)$.
    By definition of the BWT, this gives
    \[
      B_1[i]=S_1^{\infty}[j-1],
      \qquad B_2[i]=S_2^{\infty}[f(j)-1].
    \]
  \item\label{lm:dm-to-bwt-binary-step-3}
    In the third step, we prove that
    \[
      |\RL{B_1(b_0 \dd e_0]}| - |\RL{B_2(b_0 \dd e_0]}| = 2|\{i \in [1 \dd k] : \OccTwo{\Pat_i}{\Text} \neq \emptyset\}|.
    \]
    For every $i \in [1 \dd k]$, denote $\Pat'_i = {\tt 0}\widetilde{\Pat}_i$.
    We begin by observing that for every $i \in [1 \dd k]$,
    \[
      \OccTwo{\Pat'_i}{S_1} =
        \{3j-1 : j \in \OccTwo{\Pat_i}{\Text}\}
        \cup
        \{\Delta + 5(i-1)\delta + 2 + j\delta : j \in [0 \dd 4]\}.
    \]
    Furthermore, note that for every $i \in [1 \dd k]$, we also have
    $\OccTwo{\Pat'_i}{S_1} = \OccTwo{\Pat'_i}{S_2}$.
    Note also that for every $i \in [1 \dd k]$,
    $\{f(j) : j \in \OccTwo{\Pat'_i}{S_1}\} = \OccTwo{\Pat'_i}{S_2}$.
    By Step~\ref{lm:dm-to-bwt-binary-step-2}, this implies that
    \[
      \{\ISA{S_1}[j] : j \in \OccTwo{\Pat'_i}{S_1}\}
        = \{\ISA{S_2}[f(j)] : j \in \OccTwo{\Pat'_i}{S_1}\}
        = \{\ISA{S_2}[j] : j \in \OccTwo{\Pat'_i}{S_2}\}.
    \]
    Thus, for every $i \in [1 \dd k]$, it holds
    $\RangeBegTwo{\Pat'_i}{S_1} = \RangeBegTwo{\Pat'_i}{S_2}$ and
    $\RangeEndTwo{\Pat'_i}{S_1} = \RangeEndTwo{\Pat'_i}{S_2}$.
    Let $(a_1,\ldots,a_k)$ denote a permutation of $[1\dd k]$ satisfying
    $\Pat'_{a_1}\prec\cdots\prec\Pat'_{a_k}$. For every $i\in[1\dd k]$, let
    $s_i=\RangeBegTwo{\Pat'_{a_i}}{S_1}$ and
    $t_i=\RangeEndTwo{\Pat'_{a_i}}{S_1}$. Since the patterns are distinct and have
    equal lengths, these suffix-array intervals are disjoint and occur in this order.
    Let also $t_0 = b_0$ and $s_{k+1} = e_0$.
    For every $p \in \{1,2\}$ and $i \in [1 \dd k]$, denote
    $Y_{p,i} = B_p(s_i \dd t_i]$. For every $p \in \{1,2\}$ and $i \in [0 \dd k]$, let
    also $X_{p,i} = B_p(t_{i} \dd s_{i+1}]$. Then,
    \begin{align*}
      B_1(b_0 \dd e_0]
        &= X_{1,0} Y_{1,1} X_{1,1} Y_{1,2} \cdots X_{1,k-1} Y_{1,k} X_{1,k},\\
      B_2(b_0 \dd e_0]
        &= X_{2,0} Y_{2,1} X_{2,1} Y_{2,2} \cdots X_{2,k-1} Y_{2,k} X_{2,k}.
    \end{align*}
    By the above discussion, in the above decompositions, for every $p \in \{1,2\}$ and $i \in [1 \dd k]$,
    the substring $Y_{p,i}$ contains symbols preceding suffixes starting in $\OccTwo{\Pat'_{a_i}}{S_p}$,
    and $X_{p,0}, X_{p,1}, \ldots, X_{p,k}$ contain symbols preceding suffixes starting in
    $\OccTwo{{\tt 0}}{S_p} \setminus \mathcal{U}$, where $\mathcal{U} =
    \bigcup_{i=1}^{k}\OccTwo{\Pat'_i}{S_1} = \bigcup_{i=1}^{k}\OccTwo{\Pat'_i}{S_2}$.
    We consider each group of substrings separately:
    \begin{itemize}
    \item Let $i \in [1 \dd k]$. First, we characterize substrings $Y_{1,i}$ and $Y_{2,i}$.
      We begin by inspecting the suffixes starting at positions in $\OccTwo{\Pat'_{a_i}}{S_1}$.
      Observe that every occurrence of $\Pat'_{a_i}$ corresponding to an occurrence of
      $\Pat_{a_i}$ in $\Text$ is followed by
      a substring $\alpha \in \{{\tt 010}, {\tt 10}, {\tt 110}\}$.
      Since regardless of what $\alpha$ is, it holds
      $
      {\tt 0}\widetilde{\Pat}_{a_i}{\tt 0}^{8} \prec
      {\tt 0}\widetilde{\Pat}_{a_i}{\tt 0}^{7}{\tt 1} \prec
      {\tt 0}\widetilde{\Pat}_{a_i}{\tt 0}^{6}{\tt 1} \prec
      {\tt 0}\widetilde{\Pat}_{a_i}\alpha \prec
      {\tt 0}\widetilde{\Pat}_{a_i}{\tt 1}^{3}{\tt 0} \prec
      {\tt 0}\widetilde{\Pat}_{a_i}{\tt 1}^{4}
      $, it follows that the symbols preceding the lexicographically sorted suffixes starting in
      $\OccTwo{\Pat'_{a_i}}{S_1}$ are
      $Y_{1,i} = {\tt 010}{\tt 1}^{n_i}{\tt 00}$, where $n_i = |\OccTwo{\Pat_{a_i}}{\Text}|$.
      By analogously inspecting $S_2$, we obtain $Y_{2,i} = {\tt 00}{\tt 1}^{n_i+1}{\tt 00}$.
    \item We now characterize strings $X_{p,i}$, where $p \in \{1,2\}$ and $i \in [0 \dd k]$.
      Consider any $y \in (b_0 \dd e_0]$ satisfying $\SA{S_1}[y] \not\in \mathcal{U}$.
      Denote $x = \SA{S_1}[y]$ and observe that by definition of $S_1$,
      for any such $x$, it holds $S_1^{\infty}[x-1] = S_2^{\infty}[f(x)-1]$.
      By Step~\ref{lm:dm-to-bwt-binary-step-2}, we thus obtain
      $B_2[y]=S_2^{\infty}[f(x)-1]=S_1^{\infty}[x-1]=B_1[y]$.
      This implies that for every $i \in [0 \dd k]$, it holds $X_{1,i} = X_{2,i}$.
    \end{itemize}
    By the above, letting $X_{i} := X_{1,i} = X_{2,i}$ for every $i \in [0 \dd k]$, it holds
    \begin{align*}
      B_1(b_0 \dd e_0] &= X_{0}   \cdot {\tt 010}{\tt 1}^{n_1}{\tt 00} \cdot
                          X_{1}   \cdot {\tt 010}{\tt 1}^{n_2}{\tt 00} \cdot X_{2} \cdots
                          X_{k-1} \cdot {\tt 010}{\tt 1}^{n_k}{\tt 00} \cdot X_{k},\\
      B_2(b_0 \dd e_0] &= X_{0}   \cdot {\tt 00}{\tt 1}^{n_1+1}{\tt 00} \cdot
                          X_{1}   \cdot {\tt 00}{\tt 1}^{n_2+1}{\tt 00} \cdot X_{2} \cdots
                          X_{k-1} \cdot {\tt 00}{\tt 1}^{n_k+1}{\tt 00} \cdot X_{k}.
    \end{align*}
    Each of the blocks $Y_{1,i}$ and $Y_{2,i}$ starts and ends with ${\tt 0}$.
    Since the strings $X_i$ are identical in both decompositions,
    concatenation merges the same number of runs at block boundaries,
    even when some $X_i$ are empty.
    The block $Y_{2,i}$ has three runs, whereas $Y_{1,i}$ has three runs
    if $n_i=0$ and five otherwise.
    Therefore, $|\RL{B_1(b_0 \dd e_0]}| - |\RL{B_2(b_0 \dd e_0]}| =
    2|\{i \in [1 \dd k] : \OccTwo{\Pat_i}{\Text} \neq \emptyset\}|$, i.e., the claim.

  \item\label{lm:dm-to-bwt-binary-step-4}
    In the fourth step, we prove that
    \[
      |\RL{B_2(b_1 \dd e_1]}| - |\RL{B_1(b_1 \dd e_1]}| = 2k.
    \]
    For every $i \in [1 \dd k]$, denote $\Pat'_i = {\tt 1}^{4}{\tt 0}^{2}\widetilde{\Pat}_i$.
    We begin by observing that for every $i \in [1 \dd k]$,
    \[
      \OccTwo{\Pat'_i}{S_1} = \{\Delta + 5(i-1)\delta + 3m + 10 + j\delta : j \in [0 \dd 3]\}.
    \]
    Furthermore, note that for every $i \in [1 \dd k]$, we also have
    $\OccTwo{\Pat'_i}{S_1} = \OccTwo{\Pat'_i}{S_2}$.
    Note also that $\OccTwo{\Pat'_i}{S_1} \subseteq \mathcal{P}$, which implies that for every
    $i \in [1 \dd k]$ and $j \in \OccTwo{\Pat'_i}{S_1}$,
    it holds $f(j) = j$, and hence (by Step~\ref{lm:dm-to-bwt-binary-step-2})
    $\ISA{S_1}[j] = \ISA{S_2}[f(j)] = \ISA{S_2}[j]$.
    Thus, for every $i\in[1\dd k]$, the suffix-array intervals of
    $\Pat'_i$ in $S_1$ and $S_2$ have the same endpoints.
    Let $(a_1,\ldots,a_k)$ denote a permutation of $[1\dd k]$ satisfying
    $\Pat'_{a_1}\prec\cdots\prec\Pat'_{a_k}$. For every $i\in[1\dd k]$, let
    $s_i=\RangeBegTwo{\Pat'_{a_i}}{S_1}$ and
    $t_i=\RangeEndTwo{\Pat'_{a_i}}{S_1}$. Since the patterns are distinct and have
    equal lengths, these suffix-array intervals are disjoint and occur in this order.
    Let also $t_0 = b_1$ and $s_{k+1} = e_1$.
    For every $p \in \{1,2\}$ and $i \in [1 \dd k]$, denote
    $Y_{p,i} = B_p(s_i \dd t_i]$. For every $p \in \{1,2\}$ and $i \in [0 \dd k]$, let
    also $X_{p,i} = B_p(t_{i} \dd s_{i+1}]$. Then,
    \begin{align*}
      B_1(b_1 \dd e_1]
        &= X_{1,0} Y_{1,1} X_{1,1} Y_{1,2} \cdots X_{1,k-1} Y_{1,k} X_{1,k},\\
      B_2(b_1 \dd e_1]
        &= X_{2,0} Y_{2,1} X_{2,1} Y_{2,2} \cdots X_{2,k-1} Y_{2,k} X_{2,k}.
    \end{align*}
    By the above discussion, in the above decompositions, for every $p \in \{1,2\}$ and $i \in [1 \dd k]$,
    the substring $Y_{p,i}$ contains symbols preceding suffixes starting in $\OccTwo{\Pat'_{a_i}}{S_p}$,
    and $X_{p,0}, X_{p,1}, \ldots, X_{p,k}$ contain symbols preceding suffixes starting in
    $\OccTwo{{\tt 1}}{S_p} \setminus \mathcal{U}$, where $\mathcal{U} =
    \bigcup_{i=1}^{k}\OccTwo{\Pat'_i}{S_1} = \bigcup_{i=1}^{k}\OccTwo{\Pat'_i}{S_2}$.
    We consider each group of substrings separately:
    \begin{itemize}
    \item Let $i \in [1 \dd k]$. First, we characterize substrings $Y_{1,i}$ and $Y_{2,i}$.
      We begin by inspecting the suffixes starting at positions in $\OccTwo{\Pat'_{a_i}}{S_1}$.
      Since
      ${\tt 1}^{4}{\tt 0}^{2}\widetilde{\Pat}_{a_i}{\tt 0}^{8} \prec
      {\tt 1}^{4}{\tt 0}^{2}\widetilde{\Pat}_{a_i}{\tt 0}^{6}{\tt 1} \prec
      {\tt 1}^{4}{\tt 0}^{2}\widetilde{\Pat}_{a_i}{\tt 1}^{3}{\tt 0} \prec
      {\tt 1}^{4}{\tt 0}^{2}\widetilde{\Pat}_{a_i}{\tt 1}^{4}$,
      it follows that the symbols preceding the lexicographically sorted suffixes starting in
      $\OccTwo{\Pat'_{a_i}}{S_1}$ are $Y_{1,i} = {\tt 0011}$.
      By analogously inspecting $S_2$, we obtain $Y_{2,i} = {\tt 0101}$.
    \item We now characterize strings $X_{p,i}$, where $p \in \{1,2\}$ and $i \in [0 \dd k]$.
      Consider any $y \in (b_1 \dd e_1]$ satisfying $\SA{S_1}[y] \not\in \mathcal{U}$.
      Denote $x = \SA{S_1}[y]$ and observe that by definition of $S_1$,
      for any such $x$, it holds $S_1^{\infty}[x-1] = S_2^{\infty}[f(x)-1]$.
      By Step~\ref{lm:dm-to-bwt-binary-step-2}, we thus obtain
      $B_2[y]=S_2^{\infty}[f(x)-1]=S_1^{\infty}[x-1]=B_1[y]$.
      This implies that for every $i \in [0 \dd k]$, it holds $X_{1,i} = X_{2,i}$.
    \end{itemize}
    By the above, letting $X_{i} := X_{1,i} = X_{2,i}$ for every $i \in [0 \dd k]$, it holds
    \begin{align*}
      B_1(b_1 \dd e_1] &= X_{0}   \cdot {\tt 0011} \cdot X_{1} \cdot {\tt 0011} \cdot X_{2} \cdots
                          X_{k-1} \cdot {\tt 0011} \cdot X_{k},\\
      B_2(b_1 \dd e_1] &= X_{0}   \cdot {\tt 0101} \cdot X_{1} \cdot {\tt 0101} \cdot X_{2} \cdots
                          X_{k-1} \cdot {\tt 0101} \cdot X_{k}.
    \end{align*}
    The blocks ${\tt 0011}$ and ${\tt 0101}$ both start with ${\tt 0}$
    and end with ${\tt 1}$. Boundary merges therefore again affect the two
    concatenations equally. Since these blocks have two and four runs,
    respectively, we obtain
    $|\RL{B_2(b_1 \dd e_1]}| - |\RL{B_1(b_1 \dd e_1]}| = 2k$, i.e., the claim.

  \item\label{lm:dm-to-bwt-binary-step-5}
    We are now ready to prove the main claim.
    First, observe that the string ${\tt 0}{\tt 1}^{5}{\tt 0}$ is a substring of both $S_1$ and $S_2$.
    On the other hand, ${\tt 0}{\tt 1}^{6}$ does not occur in $S_1$ or $S_2$. This implies
    that ${\tt 0}{\tt 1}^{5}{\tt 0}$ is a prefix of $S_1[\SA{S_1}[e_0] \dd \Textlen']$ and
    $S_2[\SA{S_2}[e_0] \dd \Textlen']$. Finally, note that every occurrence of ${\tt 0}{\tt 1}^{5}{\tt 0}$ in $S_1$ and $S_2$
    is preceded by a symbol ${\tt 0}$. This implies that $B_1[e_0] = B_2[e_0] = {\tt 0}$.
    Next, observe that, since ${\tt 1}$ is a suffix of both $S_1$ and $S_2$, we must have
    $S_1[\SA{S_1}[b_1+1] \dd \Textlen'] = S_2[\SA{S_2}[b_1+1] \dd \Textlen'] = {\tt 1}$.
    Thus, $B_1[b_1+1] = S_1[\Textlen' - 1] = {\tt 1}$ and $B_2[b_1+1] = S_2[\Textlen' - 1] = {\tt 1}$.
    Hence, $B_1[b_1+1] = B_2[b_1+1] = {\tt 1}$. Putting this together with
    $B_1[e_0] = B_2[e_0] = {\tt 0}$, we obtain that there is a run ending at
    position $e_0$ in both $B_1$ and $B_2$.
    Consequently,
    \begin{align*}
      |\RL{B_1(b_0 \dd e_1]}| &= |\RL{B_1(b_0 \dd e_0]}| + |\RL{B_1(b_1 \dd e_1]}|,\\
      |\RL{B_2(b_0 \dd e_1]}| &= |\RL{B_2(b_0 \dd e_0]}| + |\RL{B_2(b_1 \dd e_1]}|.
    \end{align*}
    Subtracting these equations and applying
    Steps~\ref{lm:dm-to-bwt-binary-step-3} and~\ref{lm:dm-to-bwt-binary-step-4},
    we obtain
    \begin{align*}
      \RLBWTSize{S_2}-\RLBWTSize{S_1}
        &= 2k-2|\{i \in [1 \dd k] : \OccTwo{\Pat_i}{\Text} \neq \emptyset\}|\\
        &= 2|\{i \in [1 \dd k] : \OccTwo{\Pat_i}{\Text} = \emptyset\}|.
      \qedhere
    \end{align*}
  \end{enumerate}
\end{proof}

\subsubsection{Algorithms}\label{sec:bwt-algorithms}

\begin{theorem}\label{th:bwt}
  Let $N\in\Z_{\geq2}$. Consider the word RAM model with word size
  $w$ satisfying $1+\log N\leq w=\bigO(\log N)$. Suppose that there
  exists an algorithm for \probname{RLBWT} \probname{Size}
  (\cref{prob:rlbwt-size}) that, given $N$ and any valid input of at most
  $N$ symbols over alphabet $\BinaryAlphabet$,\footnote{\phantomsection\label{fn:bwt-input}
    The input consists of the packed representation
    $\PackedRepresentation{w}{2}{\Text}$ of a binary text $\Text$
    with $1\leq|\Text|\leq N$, together with its length $|\Text|$.
    The complete input occupies $\bigO(N)$ bits at the stated word size.} achieves the
  following complexities:
  \begin{itemize}
  \item running time $T_{\rm RLBWT}(N)$,
  \item space usage $S_{\rm RLBWT}(N)$ bits.
  \end{itemize}
  Then, there exists an
  algorithm for \probname{Dictionary} \probname{Matching}
  (\cref{prob:dm}) that, given $N$ and any valid input of
  at most $N$ symbols over alphabet $\BinaryAlphabet$, achieves
  the following complexities:
  \begin{itemize}
  \item running time $\bigO(T_{\rm RLBWT}(N))$,
  \item space usage $\bigO(S_{\rm RLBWT}(N))$ bits.
  \end{itemize}
\end{theorem}
\begin{proof}
  Consider the given Dictionary Matching input with text length
  $\Textlen_0$, dictionary size $k_0=\lfloor\Textlen_0/\log\Textlen_0\rfloor$,
  and pattern length $m_0=2\lfloor\log\Textlen_0\rfloor$.
  Denote its total string length by $L_0=\Textlen_0+k_0m_0\leq N$.
  If $L_0<400$, direct comparison determines the answer in
  $\bigO(1)$ time and words of space.
  Otherwise, apply \cref{pr:dm-size-reduction} with $c=80$
  to the given text and dictionary. This produces
  $\bigO(1)$ valid Dictionary Matching inputs, each with total string
  length at most $L_0/80$.

  We describe a procedure for solving one of the returned inputs.
  Denote its text by $\Text\in\BinaryAlphabet^{\Textlen}$ and its dictionary by
  $\mathcal D=\{\Pat_1,\ldots,\Pat_k\}\subseteq\BinaryAlphabet^m$,
  where $k=\lfloor\Textlen/\log\Textlen\rfloor$,
  $m=2\lfloor\log\Textlen\rfloor$, and
  $\Textlen+km\leq L_0/80\leq N/80$.
  We assume that $\Textlen\geq16$ (for smaller $\Textlen$, direct
  comparison solves the instance in $\bigO(1)$ time).
  Let $S_1$ and $S_2$ be the strings from \cref{lm:dm-to-bwt-binary},
  and denote their common length by
  $U=3\Textlen+5k(3m+13)+2=\Theta(\Textlen)$.
  We proceed in four steps.
  \begin{enumerate}

  \item We first construct the packed representations of the encoded
    text $\widetilde{\Text}$ and patterns $\widetilde{\Pat}_i$.
    The encoding is the morphism $\mu$ given by
    $\mu(\zero)=\zero\one\zero$ and $\mu(\one)=\one\one\zero$.
    Using \cref{pr:packed-representation}, construct an array containing
    the packed representations of $\mu(\zero)$ and $\mu(\one)$.
    Build the data structure from \cref{th:packed-morphism}
    with this array.
    Query it for $\Text,\Pat_1,\ldots,\Pat_k$.
    Construction takes $\bigO(\sqrt U)$ time, and each query for
    a string $Y$ takes $\bigO(1+|Y|/\log U)$ time.
    Thus, including preprocessing, this step takes
    \[
      \bigO\bigl(\sqrt U+k+(\Textlen+km)/\log U\bigr)
        =\bigO(\Textlen/\log\Textlen)
    \]
    time.

  \item Next, we assemble the packed representations of
    $S_1$ and $S_2$.
    Construct the packed representations of the constant strings
    in \cref{lm:dm-to-bwt-binary} using \cref{pr:packed-representation}.
    For each of $S_1$ and $S_2$, concatenate the representations
    of the pieces in its defining order using
    \cref{pr:packed-representation}\eqref{pr:packed-representation-concat}.
    Each concatenation has $\bigO(k)$ pieces of total length $U$.
    The resulting time bound is $\bigO(k+U/w)=\bigO(\Textlen/\log\Textlen)$.

  \item We now compute the two BWT run counts.
    Run the assumed algorithm on the packed representations of
    $S_1$ and $S_2$ to obtain
    \[
      r_1=\RLBWTSize{S_1},\qquad r_2=\RLBWTSize{S_2}.
    \]
    The two calls take $\bigO(T_{\rm RLBWT}(N))$ time and use
    $\bigO(S_{\rm RLBWT}(N)+w)$ bits in addition to the data structure
    and constructed representations.

  \item Finally, we decide whether any dictionary pattern occurs in
    $\Text$. Return \texttt{YES} if $r_2-r_1<2k$, and \texttt{NO}
    otherwise. By \cref{lm:dm-to-bwt-binary},
    \[
      r_2-r_1=2|\{i\in[1\dd k]:\OccTwo{\Pat_i}{\Text}=\emptyset\}|.
    \]
    Hence, $r_2-r_1<2k$ holds precisely when at least one pattern
    occurs in $\Text$.
  \end{enumerate}

  Apply this procedure to each returned Dictionary Matching input,
  and return \texttt{YES} if and only if at least one of them is a
  \texttt{YES}-instance. Correctness for the original input follows
  from \cref{pr:dm-size-reduction}.
  The size reduction and the calls to the assumed algorithm use size
  parameter $N$, and the morphism structure uses size parameter $U$ and
  alphabet size $2$.
  Since $m\geq1$ and $\Textlen+km\leq N/80$, the common length of the
  constructed strings satisfies
  \[
    U=3\Textlen+15km+65k+2\leq3\Textlen+80km+2\leq80(\Textlen+km)\leq N.
  \]
  Every string $Y$ supplied to the morphism satisfies $3|Y|\leq U$,
  and the basic packed operations use length bound $U$.
  Since $w\geq1+\log N\geq1+\log U$, both packed-string tools and
  the assumed BWT run-count algorithm apply.
  The initial size reduction takes
  $\bigO(L_0/\log L_0)=\bigO(N/\log N)$ time and words of space.
  Each of its $\bigO(1)$ returned inputs requires
  $\bigO(\Textlen/\log\Textlen)=\bigO(N/\log N)$ additional time
  and at most two calls to the assumed algorithm.
  The total time is therefore $\bigO(N/\log N+T_{\rm RLBWT}(N))$.
  Including the original input, all returned inputs, the data structure,
  and all constructed representations, the reduction uses
  $\bigO(N/\log N)$ words in addition to the storage of the assumed algorithm.
  Since $w=\bigO(\log N)$, the total space is
  $\bigO(N+S_{\rm RLBWT}(N))$ bits.
  By \cref{lm:rlbwt-size-lower-bounds}, the assumed algorithm requires
  $\Omega(N/\log N)$ time and $\Omega(N)$ bits of space.
  These bounds absorb the additive construction costs, proving the claim.
\end{proof}

\begin{remark}\label{rm:rlbwt-size-to-bwt-construction}
  Let $N\in\Z_{\geq2}$, and consider the word RAM model with
  $1+\log N\leq w=\bigO(\log N)$.
  The problem \probname{RLBWT Size} (\cref{prob:rlbwt-size}) reduces
  to \probname{BWT Construction} (\cref{prob:bwt-construction})
  on the same input as follows.
  Given a binary string $\Text$ of length $\Textlen\leq N$ in packed
  form, first construct the packed representation of $\BWT{\Text}$.
  For $\Textlen\geq2$, apply \cref{th:packed-run-counting} with
  alphabet size $2$, supplying $\Textlen$ and this representation, to
  compute $\RLBWTSize{\Text}$.
  For $\Textlen=1$, the run count is $1$.
  The additional time is $\bigO(N/\log N)$, and the additional
  space is $\bigO(N)$ bits.
  Writing the packed BWT of a length-$N$ string requires
  $\Omega(N/w)=\Omega(N/\log N)$ time, and the input occupies
  $\Omega(N)$ bits. Thus, the reduction preserves the asymptotic
  time and space bounds of the BWT construction algorithm.
  Composing it with \cref{th:bwt} gives a reduction from
  \probname{Dictionary Matching} to \probname{BWT Construction}
  with the same asymptotic time and bit-space guarantees.
\end{remark}

\subsection{Reducing Dictionary Matching to Batched ISA Queries}\label{sec:isa}

\subsubsection{Problem Definition}\label{sec:isa-problem}

\begin{problem}[Batched ISA Queries]\label{prob:batched-isa}
  \begin{description}[style=sameline,itemsep=0ex,font={\normalfont\bf}]
  \item[Input:]
    Let $\Textlen \in \Z_{\geq 2}$ and
    $q=\lfloor\Textlen/\log\Textlen\rfloor$. Consider the word RAM model
    with word size $w > \log \Textlen$. The input consists of $\Textlen$,
    $q$, the packed representation $\PackedRepresentation{w}{2}{\Text}$
    (\cref{def:packed-representation}) of a string
    $\Text \in \BinaryAlphabet^{\Textlen}$, and an array containing a sequence
    $(j_1, \ldots, j_q) \in [1 \dd \Textlen]^q$, stored with one entry per word.
  \item[Output:]
    An array containing the sequence $(\ISA{\Text}[j_1], \ldots, \ISA{\Text}[j_q])$
    (\cref{def:isa}), stored with one entry per word.
  \end{description}
\end{problem}

\subsubsection{Combinatorial Results}\label{sec:isa-combinatorial-results}

\begin{lemma}\label{lm:dm-to-isa}
  Let $\Textlen,m\in\Z_{\geq1}$ and $k\in\Z_{\geq0}$, and let
  $\Text \in \BinaryAlphabet^{\Textlen}$ be a text and $\mathcal{D} =
  \{\Pat_1, \Pat_2, \dots, \Pat_k\} \subseteq \BinaryAlphabet^{m}$
  be a collection of $k$ pairwise distinct patterns of common length $m$.
  For any $c \in \BinaryAlphabet$, let $f(c) := c + 1$. For any
  $X \in \BinaryAlphabet^{*}$, we then denote $f(X) := \bigodot_{i=1}^{|X|} f(X[i])$.
  Consider the following string over alphabet $\{{\tt 0}, {\tt 1}, {\tt 2}, {\tt 3}, {\tt 4}\}$
  (brackets added for clarity):
  \[
    S = f(\Text) \cdot {\tt 3} \cdot \textstyle\bigodot_{i=1}^{k} \big(f(\Pat_i) \cdot {\tt 0} \cdot f(\Pat_i) \cdot {\tt 4} \big).
  \]
  Denote $\Delta = \Textlen + 1$ and $\delta = 2(m+1)$. Then,
  for every $i \in [1 \dd k]$, $\OccTwo{\Pat_i}{\Text} \neq \emptyset$
  holds if and only if
  \[
    \ISA{S}[\Delta + (i-1)\delta + 1] + 1 < \ISA{S}[\Delta + (i-1)\delta + m + 2].
  \]
\end{lemma}
\begin{proof}
  Let $i\in[1\dd k]$, and denote $j_1=\Delta+(i-1)\delta+1$
  and $j_2=\Delta+(i-1)\delta+m+2$.
  We first show that
  \[
    \OccTwo{f(\Pat_i)}{S}
      =\OccTwo{\Pat_i}{\Text}\cup\{j_1,j_2\}.
  \]
  Since $f(\Pat_i)$ contains only symbols from $\{{\tt 1},{\tt 2}\}$,
  no occurrence can contain a separating symbol ${\tt 0}$, ${\tt 3}$,
  or ${\tt 4}$.
  Its occurrences in $f(\Text)$ are exactly the positions in
  $\OccTwo{\Pat_i}{\Text}$, because $f$ is injective.
  In $S[\Delta+1\dd|S|]$, the copies of the encoded patterns
  have length $m$ and are separated by ${\tt 0}$ or ${\tt 4}$.
  An occurrence must therefore be one of these complete copies.
  Since the patterns are distinct, only the copies at $j_1$ and $j_2$ match.

  The suffixes of $S$ beginning with $f(\Pat_i)$ occupy consecutive
  positions in $\SA{S}$ (\cref{rm:occ}).
  The suffixes at $j_1$ and $j_2$ begin with
  $f(\Pat_i)\cdot{\tt 0}$ and $f(\Pat_i)\cdot{\tt 4}$, respectively.
  For every $j\in\OccTwo{\Pat_i}{\Text}$, the symbol $S[j+m]$
  belongs to $\{{\tt 1},{\tt 2},{\tt 3}\}$, with ${\tt 3}$ following
  an occurrence ending at position $\Textlen$.
  Thus, the suffixes at $j_1$ and $j_2$ are the first and last in
  this interval, and all suffixes between them start at positions
  in $\OccTwo{\Pat_i}{\Text}$.
  Counting these positions gives
  \[
    \ISA{S}[j_2]-\ISA{S}[j_1]-1=|\OccTwo{\Pat_i}{\Text}|.
  \]
  Consequently, $\OccTwo{\Pat_i}{\Text}\neq\emptyset$ if and only if
  $\ISA{S}[j_1]+1<\ISA{S}[j_2]$, as claimed.
\end{proof}

\subsubsection{Algorithms}\label{sec:isa-algorithms}

\begin{theorem}\label{th:isa}
  Let $N\in\Z_{\geq2}$. Consider the word RAM model with word size
  $w$ satisfying $1+\log N\leq w=\bigO(\log N)$. Suppose that there
  exists an algorithm for \probname{Batched ISA Queries}
  (\cref{prob:batched-isa}) that, given $N$ and any valid input of at most
  $N$ symbols over alphabet $\BinaryAlphabet$,\footnote{\phantomsection\label{fn:isa-input}
    That is, the packed representation $\PackedRepresentation{w}{2}{\Text}$
    of a binary string $\Text$ with $2\leq|\Text|\leq N$, an array
    containing a sequence of query positions
    $(j_1,\ldots,j_q)\in[1\dd|\Text|]^q$, one entry per word,
    and the integers $|\Text|$ and $q$, where
    $q=\lfloor|\Text|/\log|\Text|\rfloor$.
    The complete input occupies $\bigO(N)$ bits at the stated word size.} achieves the
  following complexities:
  \begin{itemize}
  \item running time $T_{\rm ISA}(N)$,
  \item space usage $S_{\rm ISA}(N)$ bits.
  \end{itemize}
  Then, there exists an algorithm
  for \probname{Dictionary Matching}
  (\cref{prob:dm}) that, given $N$ and any valid input of
  at most $N$ symbols over alphabet $\BinaryAlphabet$, achieves
  the following complexities:
  \begin{itemize}
  \item running time $\bigO(T_{\rm ISA}(N))$,
  \item space usage $\bigO(S_{\rm ISA}(N))$ bits.
  \end{itemize}
\end{theorem}
\begin{proof}
  Consider the given Dictionary Matching input with text length
  $\Textlen_0$, dictionary size $k_0=\lfloor\Textlen_0/\log\Textlen_0\rfloor$,
  and pattern length $m_0=2\lfloor\log\Textlen_0\rfloor$.
  Denote its total string length by $L_0=\Textlen_0+k_0m_0\leq N$.
  If $L_0<180$, direct comparison determines the answer in
  $\bigO(1)$ time and words of space.
  Otherwise, apply \cref{pr:dm-size-reduction} with $c=36$
  to the given text and dictionary. This produces
  $\bigO(1)$ valid Dictionary Matching inputs, each with total string
  length at most $L_0/36$.

  We describe a procedure for solving one of the returned inputs.
  Denote its text by $\Text\in\BinaryAlphabet^{\Textlen}$ and its dictionary by
  $\mathcal D=\{\Pat_1,\ldots,\Pat_k\}\subseteq\BinaryAlphabet^m$,
  where $k=\lfloor\Textlen/\log\Textlen\rfloor$,
  $m=2\lfloor\log\Textlen\rfloor$, and
  $\Textlen+km\leq L_0/36\leq N/36$.
  We assume that $\Textlen\geq8$ (for smaller $\Textlen$, direct
  comparison solves the instance in $\bigO(1)$ time).
  We proceed in five steps.
  \begin{enumerate}

  \item We first construct the packed representations of
    $f(\Text),f(\Pat_1),\ldots,f(\Pat_k)$ over alphabet $[0\dd5)$,
    where $f$ adds $1$ to each symbol as in \cref{lm:dm-to-isa}.
    Let $M=\Textlen+km=\Theta(\Textlen)$.
    Build the data structure from \cref{th:alphabet-lifting},
    and query it for each input string with offset $1$.
    Including preprocessing, this takes
    $\bigO(\sqrt M+k+M/\log M)=\bigO(\Textlen/\log\Textlen)$ time.

  \item Next, we construct the packed representation of
    \[
      S=f(\Text)\cdot{\tt 3}\cdot
        \bigodot_{i=1}^k(f(\Pat_i)\cdot{\tt 0}\cdot f(\Pat_i)\cdot{\tt 4}).
    \]
    Denote $U=|S|=\Textlen+2k(m+1)+1=\Theta(\Textlen)$.
    Construct the packed representations of the one-symbol separators
    ${\tt 0}$, ${\tt 3}$, and ${\tt 4}$, and then concatenate the pieces
    in the order specified by the definition of $S$ above using
    \cref{pr:packed-representation}\eqref{pr:packed-representation-concat}.
    With $\bigO(k)$ pieces, this step takes
    $\bigO(k+U/w)=\bigO(\Textlen/\log\Textlen)$ time.

  \item\label{step:dm-to-isa-encoding} We now construct a binary encoding of $S$ together with
    the offset needed to recover its suffix ranks.
    Apply \cref{th:isa-alphabet-reduction-linear-to-binary} to $S$.
    In $\bigO(N/\log N)$ time,
    it returns an integer $\Delta$, the length $U'=9U$, and
    the packed representation $\PackedRepresentation{w}{2}{S'}$
    of a binary string $S'$ satisfying, for every $j\in[1\dd U]$,
    \[
      \ISA{S}[j]=\ISA{S'}[9(j-1)+1]-\Delta.
    \]

  \item We answer the ISA queries at the positions in $S'$
    corresponding to the two copies of each $f(\Pat_i)$ in $S$.
    Let $q'=\lfloor U'/\log U'\rfloor$ and construct an array
    containing the sequence $(j'_1,\ldots,j'_{q'})$, where, for
    every $i\in[1\dd k]$,
    \[
      \begin{aligned}
        j'_{2i-1}&=9\bigl(\Textlen+1+2(i-1)(m+1)\bigr)+1,\\
        j'_{2i}&=j'_{2i-1}+9(m+1).
      \end{aligned}
    \]
    Fill the remaining entries with $j'_t=1$ for $t\in(2k\dd q']$.
    Constructing this array takes
    $\bigO(q')=\bigO(\Textlen/\log\Textlen)$ time.
    Run the assumed algorithm on the packed representation of $S'$
    and this array to obtain the answers $x_t=\ISA{S'}[j'_t]$
    for $t\in[1\dd q']$.
    The call takes $T_{\rm ISA}(N)$ time and $S_{\rm ISA}(N)$ bits of space.

  \item Finally, subtract $\Delta$ from the first $2k$ returned ranks.
    By the identity from Step~\ref{step:dm-to-isa-encoding}, these are the ISA values
    at the starts of the two copies of each $f(\Pat_i)$ in $S$.
    Return \texttt{YES} if $x_{2i-1}+1<x_{2i}$ for some $i\in[1\dd k]$,
    and \texttt{NO} otherwise. Correctness follows from \cref{lm:dm-to-isa}.
    The updates and scan take $\bigO(k)$ time.
  \end{enumerate}

  Apply this procedure to each returned Dictionary Matching input,
  and return \texttt{YES} if and only if at least one of them is a
  \texttt{YES}-instance. Correctness for the original input follows
  from \cref{pr:dm-size-reduction}.
  The size reduction, the alphabet reduction, and the calls to the
  assumed algorithm use size parameter $N$, and the alphabet reduction
  uses alphabet size $5$. The lifting structure uses size parameter $M$,
  source alphabet size $2$, and target alphabet size $5$.
  The concatenation uses length bound $U$.
  Since $k\leq km$ and $1\leq\Textlen$, the constructed strings satisfy
  \[
    5\leq U=\Textlen+2k(m+1)+1\leq2\Textlen+4km\leq4M,
    \qquad U'=9U\leq36M\leq N.
  \]
  Since $x/\log x$ is increasing for $x\geq3$, $U'\geq9\Textlen$,
  and $\log(9\Textlen)\leq3\log\Textlen$, we have
  \[
    \frac{U'}{\log U'}\geq
    \frac{9\Textlen}{\log(9\Textlen)}\geq
    \frac{3\Textlen}{\log\Textlen}\geq2k.
  \]
  Thus, the $2k$ useful query positions fit in the array of length $q'$.
  All query positions belong to $[1\dd U']$, which ensures that the
  assumed binary algorithm applies.
  The packed-string tools and alphabet reduction satisfy their assumptions
  because $5\leq M,U\leq N$ and $w\geq1+\log N$.
  The initial size reduction takes
  $\bigO(L_0/\log L_0)=\bigO(N/\log N)$ time and words of space.
  Each of its $\bigO(1)$ returned inputs requires
  $\bigO(N/\log N)$ additional time
  and at most one call to the assumed algorithm.
  The total time is therefore $\bigO(N/\log N+T_{\rm ISA}(N))$.
  Including the original input, all returned inputs, the tool workspaces,
  the constructed strings, and both arrays, the reduction uses
  $\bigO(N/\log N)$ words in addition to the storage of the assumed algorithm.
  Since $w=\bigO(\log N)$, the total space is
  $\bigO(N+S_{\rm ISA}(N))$ bits.

  The assumed algorithm supports binary texts of length $N$
  with $\lfloor N/\log N\rfloor=\Theta(N/\log N)$ queries.
  It explicitly returns one word per answer. Writing the output
  requires $\Omega(N/\log N)$ time, and storing it requires
  $\Omega((N/\log N)w)=\Omega(N)$ bits. Thus,
  $T_{\rm ISA}(N)=\Omega(N/\log N)$ and $S_{\rm ISA}(N)=\Omega(N)$.
  These bounds absorb the additive terms in the time and space bounds above.
\end{proof}

\subsection{Reducing Dictionary Matching to Batched LPF Queries}\label{sec:lpf}

\subsubsection{Problem Definition}\label{sec:lpf-problem}

\begin{problem}[Batched LPF Queries]\label{prob:batched-lpf}
  \begin{description}[style=sameline,itemsep=0ex,font={\normalfont\bf}]
  \item[Input:]
    Let $\Textlen \in \Z_{\geq 2}$ and
    $q=\lfloor\Textlen/\log\Textlen\rfloor$. Consider the word RAM model
    with word size $w > \log \Textlen$. The input consists of $\Textlen$,
    $q$, the packed representation $\PackedRepresentation{w}{2}{\Text}$
    (\cref{def:packed-representation}) of a string
    $\Text \in \BinaryAlphabet^{\Textlen}$, and an array containing a sequence
    $(j_1, \ldots, j_q) \in [1 \dd \Textlen]^q$, stored with one entry per word.
  \item[Output:]
    An array containing the sequence $(\LPF{\Text}[j_1], \ldots, \LPF{\Text}[j_q])$
    (\cref{def:lpf}), stored with one entry per word.
  \end{description}
\end{problem}

By \probname{Batched LPnF Queries} we define the variant of the above problem in which
the $\LPF{\Text}$ array is replaced with its non-overlapping variant $\LPnF{\Text}$
(\cref{def:lpnf}).

\subsubsection{Combinatorial Results}\label{sec:lpf-combinatorial-results}

\begin{observation}\label{ob:dm-to-lpf}
  Let $\Textlen,k\in\Z_{\geq0}$ and $m\in\Z_{\geq1}$, and let
  $\Text \in \BinaryAlphabet^{\Textlen}$ be a text and $\mathcal{D} =
  \{\Pat_1, \Pat_2, \dots, \Pat_k\} \subseteq \BinaryAlphabet^{m}$
  be a collection of $k$ pairwise distinct patterns of common length $m$.
  Denote
  \[
    \Text' = \Text \cdot {\tt 2} \cdot \big(\textstyle\bigodot_{i=1}^{k} \big(\Pat_i \cdot {\tt 3} \big)\big).
  \]
  Denote $\Delta = \Textlen + 1$ and $\delta = m + 1$.
  Then, for every $i \in [1 \dd k]$, the following three conditions are equivalent:
  \begin{enumerate}
  \item $\OccTwo{\Pat_i}{\Text} \neq \emptyset$,
  \item $\LPF{\Text'}[\Delta + (i-1)\delta + 1] = m$,
  \item $\LPnF{\Text'}[\Delta + (i-1)\delta + 1] = m$.
  \end{enumerate}
\end{observation}
\begin{proof}
  Denote $j=\Delta+(i-1)\delta+1$, so that $\Text'[j\dd j+m)=\Pat_i$
  and $\Text'[j+m]={\tt 3}$. An occurrence of $\Pat_i$ in $\Text'$
  starting before position $j$ consists of symbols ${\tt 0}$ and
  ${\tt 1}$ only. It therefore lies within $\Text$ or within a block
  $\Pat_t$ with $t<i$. In the latter case, it equals $\Pat_t$, which
  contradicts $\Pat_t\neq\Pat_i$. Consequently, $\Pat_i$ occurs in
  $\Text'$ before position $j$ if and only if it occurs in $\Text$, and
  every such occurrence ends before position $\Delta<j$. Moreover,
  $\Pat_i\cdot{\tt 3}$ does not occur before position $j$, because such
  an occurrence would start at a block $\Pat_t$ with $t<i$ and give
  $\Pat_t=\Pat_i$. By \cref{def:lpf,def:lpnf}, we obtain
  $\LPnF{\Text'}[j]\leq\LPF{\Text'}[j]\leq m$, and both values equal
  $m$ if and only if $\Pat_i$ occurs in $\Text$.
\end{proof}

\subsubsection{Algorithms}\label{sec:lpf-algorithms}

\begin{theorem}\label{th:lpf}
  Let $N\in\Z_{\geq2}$. Consider the word RAM model with word size
  $w$ satisfying $1+\log N\leq w=\bigO(\log N)$. Suppose that there
  exists an algorithm for \probname{Batched} \probname{LPF} \probname{Queries}
  (\cref{prob:batched-lpf}) that, given $N$ and any valid input of at most
  $N$ symbols over alphabet $\BinaryAlphabet$,\footnote{\phantomsection\label{fn:lpf-input}
    That is, the packed representation $\PackedRepresentation{w}{2}{\Text}$
    of a binary string $\Text$ with $2\leq|\Text|\leq N$, an array
    containing a sequence of query positions
    $(j_1,\ldots,j_q)\in[1\dd|\Text|]^q$, one entry per word,
    and the integers $|\Text|$ and $q$, where
    $q=\lfloor|\Text|/\log|\Text|\rfloor$.
    The complete input occupies $\bigO(N)$ bits at the stated word size.} achieves the
  following complexities:
  \begin{itemize}
  \item running time $T_{\rm LPF}(N)$,
  \item space usage $S_{\rm LPF}(N)$ bits.
  \end{itemize}
  Then, there exists an
  algorithm for \probname{Dictionary} \probname{Matching}
  (\cref{prob:dm}) that, given $N$ and any valid input of
  at most $N$ symbols over alphabet $\BinaryAlphabet$, achieves
  the following complexities:
  \begin{itemize}
  \item running time $\bigO(T_{\rm LPF}(N))$,
  \item space usage $\bigO(S_{\rm LPF}(N))$ bits.
  \end{itemize}
  The same statement
  holds for \probname{Batched} \probname{LPnF} \probname{Queries}, with $T_{\rm LPnF}$
  and $S_{\rm LPnF}$ in place of $T_{\rm LPF}$ and $S_{\rm LPF}$
  throughout.
\end{theorem}
\begin{proof}
  Consider the given Dictionary Matching input with text length
  $\Textlen_0$, dictionary size $k_0=\lfloor\Textlen_0/\log\Textlen_0\rfloor$,
  and pattern length $m_0=2\lfloor\log\Textlen_0\rfloor$.
  Denote its total string length by $L_0=\Textlen_0+k_0m_0\leq N$.
  If $L_0<80$, direct comparison determines the answer in
  $\bigO(1)$ time and words of space.
  Otherwise, apply \cref{pr:dm-size-reduction} with $c=16$
  to the given text and dictionary. This produces
  $\bigO(1)$ valid Dictionary Matching inputs, each with total string
  length at most $L_0/16$.

  We describe a procedure for solving one of the returned inputs.
  Denote its text by $\Text\in\BinaryAlphabet^{\Textlen}$ and its dictionary by
  $\mathcal D=\{\Pat_1,\ldots,\Pat_k\}\subseteq\BinaryAlphabet^m$,
  where $k=\lfloor\Textlen/\log\Textlen\rfloor$,
  $m=2\lfloor\log\Textlen\rfloor$, and
  $\Textlen+km\leq L_0/16\leq N/16$.
  We assume that $\Textlen\geq4$ (for smaller $\Textlen$, direct
  comparison solves the instance in $\bigO(1)$ time).
  We proceed in five steps.
  \begin{enumerate}

  \item We first construct the packed representations of
    $\Text,\Pat_1,\ldots,\Pat_k$ over alphabet $[0\dd4)$.
    To this end, let $M=\Textlen+km=\Theta(\Textlen)$.
    Build the data structure from \cref{th:alphabet-lifting},
    and query it for each input string.
    Including preprocessing, this takes
    $\bigO(\sqrt M+k+M/\log M)=\bigO(\Textlen/\log\Textlen)$ time.

  \item Next, we construct the packed representation of
    \[
      S=\Text\cdot{\tt 2}\cdot
        \bigodot_{i=1}^k(\Pat_i\cdot{\tt 3}).
    \]
    Denote $U=|S|=\Textlen+k(m+1)+1=\Theta(\Textlen)$.
    Construct the packed representations of the one-symbol separators
    ${\tt 2}$ and ${\tt 3}$, and then concatenate the pieces in the
    order specified by the definition of $S$ above using
    \cref{pr:packed-representation}\eqref{pr:packed-representation-concat}.
    The $\bigO(k)$ pieces can be concatenated in
    $\bigO(k+U/w)=\bigO(\Textlen/\log\Textlen)$ time.

  \item\label{step:dm-to-lpf-encoding} We then construct a binary encoding of $S$.
    Apply \cref{th:lpf-alphabet-reduction-linear-to-binary} to $S$.
    In $\bigO(N/\log N)$ time,
    it returns the length $U'=8U$ and the packed representation
    $\PackedRepresentation{w}{2}{S'}$ of a binary string $S'$
    satisfying, for every $j\in[1\dd U]$,
    \[
      \LPF{S}[j]
        =\left\lfloor\frac{\LPF{S'}[8(j-1)+1]}{8}\right\rfloor,
      \qquad
      \LPnF{S}[j]
        =\left\lfloor\frac{\LPnF{S'}[8(j-1)+1]}{8}\right\rfloor.
    \]

  \item\label{pr:dm-to-lpf-step-queries} We now answer the LPF
    queries at the positions in $S'$ corresponding to the starts of
    the patterns in $S$.
    Let $q'=\lfloor U'/\log U'\rfloor$ and construct an array
    containing the sequence $(j'_1,\ldots,j'_{q'})$, where
    \[
      j'_i=8\bigl(\Textlen+1+(i-1)(m+1)\bigr)+1
        \qquad\text{for }i\in[1\dd k].
    \]
    Fill the remaining entries with $j'_i=1$ for $i\in(k\dd q']$.
    Constructing this array takes
    $\bigO(q')=\bigO(\Textlen/\log\Textlen)$ time.
    Run the assumed algorithm on the packed representation of $S'$
    and this array to obtain the answers $x_i=\LPF{S'}[j'_i]$
    for $i\in[1\dd q']$.
    The call takes $T_{\rm LPF}(N)$ time and $S_{\rm LPF}(N)$ bits of space.

  \item Finally, replace each of the first $k$ returned values $x_i$
    with $\lfloor x_i/8\rfloor$.
    By the first identity from Step~\ref{step:dm-to-lpf-encoding}, these are the LPF
    values at the starts of the patterns in $S$.
    Return \texttt{YES} if $x_i=m$ for some $i\in[1\dd k]$,
    and \texttt{NO} otherwise. Correctness follows from \cref{ob:dm-to-lpf}.
    The updates and scan take $\bigO(k)$ time.
  \end{enumerate}

  Apply this procedure to each returned Dictionary Matching input,
  and return \texttt{YES} if and only if at least one of them is a
  \texttt{YES}-instance. Correctness for the original input follows
  from \cref{pr:dm-size-reduction}.
  The size reduction, the alphabet reduction, and the calls to the
  assumed algorithm use size parameter $N$, and the alphabet reduction
  uses alphabet size $4$. The lifting structure uses size parameter $M$,
  source alphabet size $2$, and target alphabet size $4$.
  The concatenation uses length bound $U$.
  The constructed strings satisfy
  \[
    4\leq U\leq2M,
    \qquad U'=8U\leq16M\leq N.
  \]
  Since $x/\log x$ is increasing for $x\geq3$ and
  $U'\geq\Textlen\geq4$, we have $k\leq q'$.
  All query positions belong to $[1\dd U']$, which ensures that the
  assumed binary algorithm applies.
  The packed-string tools and alphabet reduction satisfy their assumptions
  because $4\leq M,U\leq N$ and $w\geq1+\log N$.
  The initial size reduction takes
  $\bigO(L_0/\log L_0)=\bigO(N/\log N)$ time and words of space.
  Each of its $\bigO(1)$ returned inputs requires
  $\bigO(N/\log N)$ additional time
  and at most one call to the assumed algorithm.
  The total time is therefore $\bigO(N/\log N+T_{\rm LPF}(N))$.
  Including the original input, all returned inputs, the tool workspaces,
  the constructed strings, and both arrays, the reduction uses
  $\bigO(N/\log N)$ words in addition to the storage of the assumed algorithm.
  Since $w=\bigO(\log N)$, the total space is
  $\bigO(N+S_{\rm LPF}(N))$ bits.
  For \probname{Batched LPnF Queries}, use the assumed binary LPnF
  algorithm in Step~\ref{pr:dm-to-lpf-step-queries}.
  The second identity from Step~\ref{step:dm-to-lpf-encoding} shows that the same division
  recovers the LPnF values at the starts of the patterns in $S$.
  The test $x_i=m$ is correct by the third condition of
  \cref{ob:dm-to-lpf}, and the time and space bounds follow in the same way.

  The assumed algorithm supports binary texts of length $N$
  with $\lfloor N/\log N\rfloor=\Theta(N/\log N)$ queries.
  It explicitly returns one word per answer. Writing the output
  requires $\Omega(N/\log N)$ time, and storing it requires
  $\Omega((N/\log N)w)=\Omega(N)$ bits. Thus,
  $T_{\rm LPF}(N)=\Omega(N/\log N)$ and $S_{\rm LPF}(N)=\Omega(N)$.
  The same argument applies to \probname{Batched LPnF Queries},
  since its output has the same representation. These lower bounds
  absorb the additive terms in the time and space bounds above.
\end{proof}

\subsection{Reducing Dictionary Matching to Longest Common Factor}\label{sec:lcf}

\subsubsection{Problem Definition}\label{sec:lcf-problem}

\begin{problem}[Longest Common Factor (LCF)]\label{prob:lcf}
  \begin{description}[style=sameline,itemsep=0ex,font={\normalfont\bf}]
  \item[Input:]
    Let $\Textlen_1, \Textlen_2 \in \Z_{\geq 0}$. Consider the word RAM
    model with word size $w > \log(1 + \Textlen_1 + \Textlen_2)$.
    The input consists of $\Textlen_1$, $\Textlen_2$, and the packed
    representations $\PackedRepresentation{w}{2}{S_1}$ and
    $\PackedRepresentation{w}{2}{S_2}$ (\cref{def:packed-representation})
    of strings $S_1 \in \BinaryAlphabet^{\Textlen_1}$ and
    $S_2 \in \BinaryAlphabet^{\Textlen_2}$.
  \item[Output:]
    The value $\LCF{S_1}{S_2}$ (\cref{sec:basic-definitions}).
  \end{description}
\end{problem}

\subsubsection{Preliminaries}\label{sec:lcf-preliminaries}

\begin{lemma}[Lower bounds for Longest Common Factor]\label{lm:lcf-lower-bounds}
  Let $N\in\Z_{\geq2}$. Consider the word RAM model with word size
  $w$ satisfying $1+\log N\leq w=\bigO(\log N)$.
  Any algorithm for \probname{Longest Common Factor}
  (\cref{prob:lcf}) that, given $N$, solves every valid input of
  at most $N$ symbols over alphabet $\BinaryAlphabet$ requires
  $\Omega(N/\log N)$ time and $\Omega(N)$ bits of space
  in the worst case.
\end{lemma}
\begin{proof}
  On input $(S_1,S_2)=({\tt 0}^{N-1},{\tt 1})$, every word
  containing a symbol of $S_1$ must be inspected: changing an
  uninspected symbol to ${\tt 1}$ changes the LCF length from
  zero to one. This requires $\Omega(N/w)=\Omega(N/\log N)$ time.
  The packed input occupies $\Omega(N)$ bits, which gives the claimed
  space lower bound (see \cref{sec:model}).
\end{proof}

\subsubsection{Combinatorial Results}\label{sec:lcf-combinatorial-results}

\begin{observation}\label{ob:dm-to-lcf}
  Let $\Textlen,m\in\Z_{\geq1}$ and $k\in\Z_{\geq0}$, and let
  $\Text \in \BinaryAlphabet^{\Textlen}$ be a text and $\mathcal{D} =
  \{\Pat_1, \Pat_2, \dots, \allowbreak \Pat_k\} \subseteq \BinaryAlphabet^{m}$
  be a collection of $k$ patterns of common length $m$.
  Denote $S_1 = \Text$ and $S_2 = \bigodot_{i=1}^{k} \big(\Pat_i \cdot {\tt 2} \big)$ (brackets added for clarity).
  Then, there exists $i \in [1 \dd k]$ such that $\OccTwo{\Pat_i}{\Text} \neq \emptyset$
  if and only if $\LCF{S_1}{S_2} = m$.
\end{observation}

\subsubsection{Algorithms}\label{sec:lcf-algorithms}

\begin{theorem}\label{th:lcf}
  Let $N\in\Z_{\geq2}$. Consider the word RAM model with word size
  $w$ satisfying $1+\log N\leq w=\bigO(\log N)$. Suppose that there
  exists an algorithm for \probname{Longest} \probname{Common} \probname{Factor}
  (\cref{prob:lcf}) that, given $N$ and any valid input of at most
  $N$ symbols over alphabet $\BinaryAlphabet$,\footnote{\phantomsection\label{fn:lcf-input}
    The input consists of the packed representations
    $\PackedRepresentation{w}{2}{S_1}$ and
    $\PackedRepresentation{w}{2}{S_2}$ of two binary strings $S_1,S_2$
    satisfying $|S_1|+|S_2|\leq N$, together with their lengths
    $|S_1|$ and $|S_2|$. The strings may be empty.
    The complete input occupies $\bigO(N)$ bits at the stated word size.} achieves the
  following complexities:
  \begin{itemize}
  \item running time $T_{\rm LCF}(N)$,
  \item space usage $S_{\rm LCF}(N)$ bits.
  \end{itemize}
  Then, there exists an
  algorithm for \probname{Dictionary} \probname{Matching}
  (\cref{prob:dm}) that, given $N$ and any valid input of
  at most $N$ symbols over alphabet $\BinaryAlphabet$, achieves
  the following complexities:
  \begin{itemize}
  \item running time $\bigO(T_{\rm LCF}(N))$,
  \item space usage $\bigO(S_{\rm LCF}(N))$ bits.
  \end{itemize}
\end{theorem}
\begin{proof}
  Consider the given Dictionary Matching input with text length
  $\Textlen_0$, dictionary size $k_0=\lfloor\Textlen_0/\log\Textlen_0\rfloor$,
  and pattern length $m_0=2\lfloor\log\Textlen_0\rfloor$.
  Denote its total string length by $L_0=\Textlen_0+k_0m_0\leq N$.
  If $L_0<160$, direct comparison determines the answer in
  $\bigO(1)$ time and words of space.
  Otherwise, apply \cref{pr:dm-size-reduction} with $c=32$
  to the given text and dictionary. This produces
  $\bigO(1)$ valid Dictionary Matching inputs, each with total string
  length at most $L_0/32$.

  We describe a procedure for solving one of the returned inputs.
  Denote its text by $\Text\in\BinaryAlphabet^{\Textlen}$ and its dictionary by
  $\mathcal D=\{\Pat_1,\ldots,\Pat_k\}\subseteq\BinaryAlphabet^m$,
  where $k=\lfloor\Textlen/\log\Textlen\rfloor$,
  $m=2\lfloor\log\Textlen\rfloor$, and
  $\Textlen+km\leq L_0/32\leq N/32$.
  We assume that $\Textlen\geq4$ (for smaller $\Textlen$, direct
  comparison solves the instance in $\bigO(1)$ time).
  We proceed in four steps.
  \begin{enumerate}

  \item\label{step:dm-to-lcf-input} We first construct the packed representations of
    $\Text,\Pat_1,\ldots,\Pat_k$ over alphabet $[0\dd3)$.
    To this end, let $M=\Textlen+km=\Theta(\Textlen)$.
    Build the data structure from \cref{th:alphabet-lifting},
    and query it for each input string.
    Including preprocessing, this takes
    $\bigO(\sqrt M+k+M/\log M)=\bigO(\Textlen/\log\Textlen)$ time.

  \item Next, we construct the packed representations of
    \[
      S_1=\Text,\qquad
      S_2=\bigodot_{i=1}^k(\Pat_i\cdot{\tt 2}).
    \]
    The representation of $S_1$ is already available from Step~\ref{step:dm-to-lcf-input}.
    Denote $U=|S_1|+|S_2|=\Textlen+k(m+1)=\Theta(\Textlen)$.
    To obtain the representation of $S_2$, construct the packed
    representation of the one-symbol separator ${\tt 2}$, and then
    concatenate the pieces in the order specified by the definition
    of $S_2$ above using
    \cref{pr:packed-representation}\eqref{pr:packed-representation-concat}.
    Concatenating these $\bigO(k)$ pieces takes
    $\bigO(k+|S_2|/w)=\bigO(\Textlen/\log\Textlen)$ time.

  \item We now construct the packed representations of binary
    encodings of $S_1$ and $S_2$.
    Apply \cref{th:lcf-alphabet-reduction-linear-to-binary} to $S_1$
    and $S_2$.
    In $\bigO(N/\log N)$ time,
    it returns the lengths $|S'_i|=8|S_i|+4$ and the packed
    representations $\PackedRepresentation{w}{2}{S'_i}$ of
    binary strings $S'_i$ for $i\in\{1,2\}$, satisfying
    \[
      \LCF{S_1}{S_2}
        =\left\lfloor\frac{\LCF{S'_1}{S'_2}-4}{8}\right\rfloor.
    \]

  \item Finally, we determine whether any dictionary pattern occurs
    in $\Text$. Run the assumed binary algorithm on the packed
    representations of $S'_1$ and $S'_2$ to obtain
    $\lambda'=\LCF{S'_1}{S'_2}$.
    This call takes $T_{\rm LCF}(N)$ time and uses
    $S_{\rm LCF}(N)$ bits of space.
    Since $\LCF{S_1}{S_2}\geq0$, the preceding identity implies
    $\lambda'\geq4$.
    Compute $\lambda=\lfloor(\lambda'-4)/8\rfloor$ in
    $\bigO(1)$ time, and output \texttt{YES} if $\lambda=m$
    and \texttt{NO} otherwise.
    The identity gives $\lambda=\LCF{S_1}{S_2}$, and
    \cref{ob:dm-to-lcf} proves the decision correct.
  \end{enumerate}

  Apply this procedure to every returned Dictionary Matching input,
  and return \texttt{YES} if at least one has answer \texttt{YES},
  and \texttt{NO} otherwise.
  Correctness for the original input follows from
  \cref{pr:dm-size-reduction}.
  The size reduction, the alphabet reduction, and the calls to the
  assumed algorithm use size parameter $N$, and the alphabet reduction
  uses alphabet size $3$. The lifting structure uses size parameter $M$,
  source alphabet size $2$, and target alphabet size $3$.
  The concatenation uses length bound $U$.

  Both strings supplied to the alphabet reduction have length at least $3$,
  and
  \[
    U=M+k\leq2M.
  \]
  Their binary encodings have combined length
  \[
    |S'_1|+|S'_2|=8U+8\leq16U\leq32M\leq N.
  \]
  Thus, the alphabet reduction applies, and
  $w\geq1+\log N>\log(1+|S'_1|+|S'_2|)$ ensures that the
  assumed binary algorithm applies as well.
  The packed-string tools also satisfy their assumptions because
  $3\leq M,U\leq N$ and $w\geq1+\log N\geq1+\log M$.
  The initial application of \cref{pr:dm-size-reduction} takes
  $\bigO(L_0/\log L_0)=\bigO(N/\log N)$ time and words of space,
  including the original input and all returned inputs.
  For each returned input, the construction takes
  $\bigO(N/\log N)$ time.
  There are $\bigO(1)$ returned Dictionary Matching inputs, each
  requiring at most one call to the assumed Longest Common Factor algorithm.
  Thus, the total time is $\bigO(N/\log N+T_{\rm LCF}(N))$.
  Including the original input, the data structures, and all
  constructed representations, the reduction uses $\bigO(N/\log N)$
  words in addition to the storage of the assumed algorithm.
  Since $w=\bigO(\log N)$, the total space is
  $\bigO(N+S_{\rm LCF}(N))$ bits.
  By \cref{lm:lcf-lower-bounds}, the assumed algorithm requires
  $\Omega(N/\log N)$ time and $\Omega(N)$ bits of space.
  These bounds absorb the additive construction costs, proving the claim.
\end{proof}

\section{Reducing String Nesting to Range Prefix Search}\label{sec:sn-to-rpm}

\subsection{Problem Definition}\label{sec:range-prefix-matching-problem-def}

\begin{problem}[Range Prefix Search]\label{prob:rps}
  \begin{description}[style=sameline,itemsep=0ex,font={\normalfont\bf}]
  \item[Input:]
    Let $m \in \Z_{\geq 2}$ and $\ell=2\lfloor\log m\rfloor$.
    Consider the word RAM model with word size $w>\log m$.
    The input consists of $m$, $\ell$,
    an array $S[1 \dd m]$ of strings in $\BinaryAlphabet^{\ell}$, given by
    the packed representation
    $\PackedRepresentation{w}{2}{S[j]}$ (\cref{def:packed-representation})
    of $S[j]$ for every $j \in [1 \dd m]$,
    and a sequence
    $(b_1,e_1,Q_1), \ldots, \allowbreak (b_m,e_m,Q_m)$ of $m$ triples.
    For every $i \in [1 \dd m]$, we have $b_i,e_i\in[0\dd m]$,
    $b_i\leq e_i$, and $Q_i \in \BinaryAlphabet^{\leq \ell}$, given by its length $|Q_i|$ and
    its packed representation
    $\PackedRepresentation{w}{2}{Q_i}$.
  \item[Output:]
    A $\texttt{YES}/\texttt{NO}$ answer indicating whether there exists $i \in [1 \dd m]$ such that
    $Q_i$ is a prefix of at least one string in $S(b_i \dd e_i]$.
  \end{description}
\end{problem}

\begin{remark}[Choice of the Factor $2$]\label{rm:rps-length-factor}
  The choice of length $2\lfloor\log m\rfloor$ in \cref{prob:rps}
  captures the difficulty of Range Prefix Search throughout the
  logarithmic-length regime. By \cref{pr:rps-length-reduction}, instances
  with any common string length $\ell=\bigO(\log m)$ reduce to a constant
  number of instances satisfying this exact length requirement, in
  $\bigO(m)$ time and words of space. Thus, other fixed constants in
  the logarithmic length bound may yield equally hard problems, but none
  yields a harder one under these reductions.
\end{remark}

\begin{remark}\label{rm:rps-input-size}
  In algorithms for Range Prefix Search, we use a parameter
  $N\in\Z_{\geq2}$ as an upper bound on the total length of the
  input strings, i.e.,
  $m\ell+\sum_{i=1}^{m}|Q_i|\leq N$.
  Since $\ell=2\lfloor\log m\rfloor=\Theta(\log m)$, such an input
  contains $\bigO(N/\log N)$ strings.
  This implies that, in the word RAM model with word size
  $w=\Theta(\log N)$, all string and sequence lengths, as well as
  the query endpoints, can be stored with one integer per machine
  word while the entire input still occupies $\bigO(N)$ bits.
\end{remark}

\subsection{Preliminaries}\label{sec:sn-to-rps-preliminaries}

\begin{lemma}[Lower bounds for Range Prefix Search]\label{lm:rps-lower-bounds}
  Let $N\in\Z_{\geq2}$. Consider the word RAM model with word size
  $w\geq1+\log N$.
  Any algorithm for \probname{Range} \probname{Prefix} \probname{Search}
  (\cref{prob:rps}) that, given $N$, solves every valid input of
  at most $N$ symbols over alphabet $\BinaryAlphabet$ requires
  $\Omega(N/\log N)$ time and $\Omega(N)$ bits of space
  in the worst case.
\end{lemma}
\begin{proof}
  For all sufficiently large $N$, let
  $m=\lfloor N/(4\lceil\log N\rceil)\rfloor$ and
  $\ell=2\lfloor\log m\rfloor$.
  Take $S[j]=\texttt{0}^{\ell}$ for every $j\in[1\dd m]$ and
  $(b_i,e_i,Q_i)=(0,m,\texttt{0}^{\ell-1}\texttt{1})$
  for every $i\in[1\dd m]$.
  This is a valid input with $2m\ell\leq4m\log N\leq N$ binary symbols and
  answer \texttt{NO}.
  Changing the last symbol of any $Q_i$ to zero changes the
  answer to \texttt{YES}.
  Each query string $Q_i$ is supplied in a separate packed representation
  occupying $\bigO(1)$ words. If an algorithm reads no word of one of
  these representations on the \texttt{NO} input, replacing that query
  string by $\zero^\ell$ would change the answer without affecting the
  computation. Thus, at least one word of each of the
  $m=\Theta(N/\log N)$ representations must be read.
  These nonempty string representations occupy at least $m$ words in
  total, giving the space lower bound
  $\Omega(mw)=\Omega(N)$ bits (see \cref{sec:model}).
\end{proof}

\subsection{Length Reduction for Range Prefix Search}
\label{sec:rps-length-reduction}

\begin{proposition}\label{pr:rps-length-reduction}
  Let $N\in\Z_{\geq2}$, and consider the word RAM model with word size
  $w\geq1+\log N$.
  There exists a deterministic algorithm that, given $N$, integers
  $m\in\Z_{\geq2}$ and $\ell\in\Z_{\geq1}$ with
  $\ell=\bigO(\log m)$, an array $S[1\dd m]$ of strings in
  $\BinaryAlphabet^\ell$, and a sequence of $m$ triples
  $(b_i,e_i,Q_i)$, $i\in[1\dd m]$, satisfying
  $0\leq b_i\leq e_i\leq m$, $Q_i\in\BinaryAlphabet^{\leq\ell}$, and
  \[
    m\ell+\sum_{i=1}^{m}|Q_i|\leq N,
  \]
  computes a nonempty list of $\bigO(1)$ valid
  \probname{Range Prefix Search} inputs (\cref{prob:rps}) with the
  following properties:
  \begin{itemize}
  \item Each returned input contains $m$ strings of length
    $2\lfloor\log m\rfloor$ and $m$ queries.
  \item There exist $i\in[1\dd m]$ and $j\in(b_i\dd e_i]$ such that
    $Q_i$ is a prefix of $S[j]$ if and only if at least one returned
    input is a {\tt YES}-instance.
  \end{itemize}
  Each input string is supplied by its length and packed representation
  (\cref{def:packed-representation}). All output strings are returned
  in the same form, together with the query endpoints and the parameters
  of each instance.
  The algorithm takes $\bigO(m)$ time and uses $\bigO(m)$ words of space.
  The constants in these bounds and in the number of output instances
  may depend on the constant implicit in $\ell=\bigO(\log m)$.
\end{proposition}
\begin{proof}
  We group the queries according to their lengths. Within each group,
  we replace prefixes of a common length by binary identifiers, assigning
  equal identifiers exactly to equal prefixes. The queries keep their
  remaining symbols, and each string in $S$ keeps enough of its following
  symbols to test these remaining prefix comparisons. We choose the group
  width and identifier length so that the resulting strings have exactly
  the required length.

  Compute $\ell'=2\lfloor\log m\rfloor$.
  If $m<8$, then $\ell=\bigO(1)$, and we determine the answer by
  comparing each $Q_i$ with the strings in $S(b_i\dd e_i]$.
  Construct one instance with $m$ copies of $\zero^{\ell'}$ and
  $m$ queries $(0,0,\emptystring)$. If the answer is \texttt{YES},
  replace the first query by $(0,1,\emptystring)$.
  This gives the required output in $\bigO(1)$ time, using
  \cref{pr:packed-representation} with alphabet size two and length
  bound $N$.
  Assume henceforth that $m\geq8$.

  If $\ell\leq3$, there are at most fifteen binary strings of length
  at most $\ell$. For each such string $P$, construct its packed
  representation using initialization and symbol updates from
  \cref{pr:packed-representation}, and compute an array $C_P[0\dd m]$
  as follows. Set $C_P[0]=0$. For $j=1,\ldots,m$, set
  $C_P[j]=C_P[j-1]+1$ if $P$ is a prefix of $S[j]$, and
  $C_P[j]=C_P[j-1]$ otherwise. Thus, $C_P[j]$ counts the strings
  in $S[1\dd j]$ having prefix $P$.
  Test whether $C_{Q_i}[e_i]>C_{Q_i}[b_i]$ for any $i\in[1\dd m]$;
  this holds exactly when the original answer is \texttt{YES}.
  Construct one output instance with $m$ copies of $\zero^{\ell'}$
  and $m$ queries $(0,0,\emptystring)$, replacing the first query
  by $(0,1,\emptystring)$ if the answer is \texttt{YES}.
  Return the list containing this instance.
  Each prefix test takes $\bigO(1)$ time using substring extraction
  and comparison from \cref{pr:packed-representation}, and each
  output string takes $\bigO(1)$ time to initialize.
  These operations use alphabet size two and length bound $N$,
  since $\ell\leq3$, $\ell'\leq m\leq N$, and
  $\ell'=\bigO(\log m)=\bigO(w)$.
  This case takes $\bigO(m)$ time and uses $\bigO(m)$ words of space.
  Assume henceforth that $\ell\geq4$.

  Denote $M=m\ell$. All packed-string tools below use alphabet size
  two and size parameter $M$; the symbol width is $\lceil\log2\rceil=1$.
  Construct the data structure from \cref{th:val-decoding} with these
  parameters. We use it to compute the packed binary representations
  of the identifiers. Initialize an empty list of output instances.
  The remaining construction proceeds as follows.
  \begin{enumerate}
  \item First, we group the query indices by query length.
    Compute $f=\lceil\log(2m)\rceil$ and $h=\ell'-f$, and form the groups
    \[
      I_r=\{i\in[1\dd m]:\lfloor |Q_i|/h\rfloor=r\}
      \qquad\text{for }r\in[0\dd\lfloor\ell/h\rfloor].
    \]
    To construct arrays listing these groups, apply stable counting
    sort to the sequence $1,\ldots,m$, using
    $\lfloor |Q_i|/h\rfloor$ as the key of index $i$.
    Thus, the indices within each group are listed in increasing order.

  \item Next, for every $r$ with $I_r\neq\emptyset$, we construct one
    output instance. Denote $I=I_r$ and $p=rh$.
    Every query in this group satisfies $p\leq|Q_i|<p+h$.
    We process the group in two steps.
    \begin{enumerate}
    \item First, we assign binary identifiers to the length-$p$ prefixes
      of the strings in $S$ and of the queries with indices in $I$.
      Using \cref{pr:packed-representation}, extract the packed
      representations of $S[j][1\dd p]$ for every $j\in[1\dd m]$
      and of $Q_i[1\dd p]$ for every $i\in I$.
      Form the sequence listing the former prefixes in increasing order
      of $j$, followed by the latter in increasing order of $i$.
      Apply \cref{pr:packed-string-sorting} to this sequence, and
      denote the returned rank of a string $X$ by $\rho(X)$.
      There are $m+|I|\leq2m\leq2^f$ strings, and therefore
      $0\leq\rho(X)<2^f$.
      Let $g(X)=\bin{f}{\rho(X)}$ (\cref{def:bin}).
      For every prefix in the sequence, compute the packed representation
      of its identifier using the data structure from \cref{th:val-decoding}.

    \item We now construct the strings and queries of the output instance.
      Denote $d=\min\{h,\ell-p\}$. For each string in $S$, we keep the
      next $d$ symbols after its length-$p$ prefix and append zeros
      if fewer than $h$ symbols remain. The output array $S'[1\dd m]$
      and transformed query strings are given by
      \begin{align*}
        S'[j]&=g(S[j][1\dd p])\cdot S[j][p+1\dd p+d]
          \cdot\zero^{h-d}
          &&\text{for }j\in[1\dd m],\\*
        Q_i'&=g(Q_i[1\dd p])\cdot Q_i[p+1\dd|Q_i|]
          &&\text{for }i\in I.
      \end{align*}
      Extract the indicated substrings, construct the required zeros,
      and concatenate them with the identifiers, using
      \cref{pr:packed-representation}.
      The query list consists of $(b_i,e_i,Q_i')$ for $i\in I$,
      in increasing order of $i$, followed by $m-|I|$ copies of
      $(0,0,\emptystring)$. Append this instance to the output list,
      supplying the packed representations of $S'[1],\ldots,S'[m]$,
      all query strings and their lengths, and the parameters $m$ and $\ell'$.
    \end{enumerate}
  \end{enumerate}
  Return the list of output instances.
  We first verify that the returned instances are valid and preserve
  the answer. Consider a group with prefix length $p$.
  Each string in $S'$ has length $f+h=\ell'$.
  For every $i\in I$, the number $|Q_i|-p$ of symbols after its
  length-$p$ prefix is smaller than $h$ and at most $\ell-p$.
  Hence, $|Q_i|-p\leq d$ and $|Q_i'|=f+|Q_i|-p<\ell'$.
  The output contains exactly $m$ strings and $m$ queries, and all
  endpoints remain in $[0\dd m]$.

  Equal identifiers correspond exactly to equal length-$p$ prefixes.
  After comparing the identifiers, the prefix test compares the
  remaining $|Q_i|-p\leq d$ symbols of $Q_i$ with the symbols of
  $S[j]$ at the same positions. It never reaches the appended zeros.
  Consequently, for every $i\in I$ and $j\in[1\dd m]$, it holds
  \[
    Q_i'\text{ is a prefix of }S'[j]
      \quad\Longleftrightarrow\quad
    Q_i\text{ is a prefix of }S[j].
  \]
  Since the endpoints $b_i$ and $e_i$ are unchanged, the transformed
  query has the same answer as the original query. The added queries
  have empty intervals and cannot yield a match.
  Every original query belongs to one group, proving the claimed
  equivalence. Since $m\geq2$, at least one group is nonempty.

  We next check the parameter bounds and the assumptions of the tools.
  Writing $t=\lfloor\log m\rfloor\geq3$, we have $f\leq t+2$ and
  $h=2t-f\geq t-2\geq1$. Moreover, $h=\Theta(\log m)$.
  Thus, there are at most $\lfloor\ell/h\rfloor+1=\bigO(1)$ groups.
  All prefixes have length at most $\ell\leq M$, while
  $\ell'=2\lfloor\log m\rfloor\leq m\leq M$.
  Every sorting call contains at most $2m\leq M$ strings.
  Since $2^f<4m\leq M$, the integer-decoding queries have length
  $f\leq\lfloor\log M\rfloor$ and values less than $2^f$.
  Finally, $M=m\ell\leq N$ implies $1+\log M\leq1+\log N\leq w$.
  These bounds justify all tool invocations on the original word RAM.

  Computing the parameters and grouping the queries takes
  $\bigO(m)$ time. The integer-decoding structure takes
  $\bigO(\sqrt M)=\bigO(m)$ time to construct, since
  $M=\bigO(m\log m)$.
  Within a group, all strings have length $\bigO(\log m)=\bigO(w)$,
  and each extraction, concatenation, initialization, or decoding
  takes $\bigO(1)$ time.
  The sorting call involves $m+|I|=\Theta(m)$ strings of common
  length $p\leq\ell=\bigO(\log m)$ and takes $\bigO(m)$ time by
  \cref{pr:packed-string-sorting}.
  Constructing the output strings and queries also takes $\bigO(m)$
  time per group. There are $\bigO(1)$ groups, giving $\bigO(m)$
  total time. The input occupies $\bigO(m(1+\ell/w))=\bigO(m)$ words,
  and the running time bounds the remaining space, including the output.
\end{proof}

We next use the length reduction to obtain instances with smaller total
string length.

\begin{proposition}\label{pr:rps-size-reduction}
  Let $c\geq1$ be a fixed constant and $N\in\Z_{\geq2}$.
  Consider the word RAM model with word size $w\geq1+\log N$.
  There exists a deterministic algorithm that, given $N$, integers
  $m\in\Z_{\geq2}$ and $\ell\in\Z_{\geq1}$ satisfying
  $\ell=\Theta(\log m)$, an array $S[1\dd m]$ of strings in
  $\BinaryAlphabet^\ell$, and $m$ triples $(b_i,e_i,Q_i)$,
  $i\in[1\dd m]$, satisfying $0\leq b_i\leq e_i\leq m$,
  $Q_i\in\BinaryAlphabet^{\leq\ell}$, and
  \[
    4c\leq s=m\ell+\sum_{i=1}^{m}|Q_i|\leq N,
  \]
  computes a nonempty list of $\bigO(1)$ valid
  \probname{Range Prefix Search} inputs (\cref{prob:rps}) with the
  following properties:
  \begin{itemize}
  \item Each output instance has total string length at most $s/c$.
  \item There exist $i\in[1\dd m]$ and $j\in(b_i\dd e_i]$ such that
    $Q_i$ is a prefix of $S[j]$ if and only if at least one output
    is a {\tt YES}-instance.
  \end{itemize}
  Each input string is supplied by its length and packed representation
  (\cref{def:packed-representation}). The output strings are returned
  in the same form, together with the query endpoints and the parameters
  of each instance.
  The algorithm takes $\bigO(m)$ time and uses $\bigO(m)$ words of
  space. The constants in these bounds and in the number of output
  instances may depend on $c$ and on the constants implicit in
  $\ell=\Theta(\log m)$.
\end{proposition}
\begin{proof}
  We partition the array of strings and the sequence of queries into
  smaller blocks. For each pair of blocks, we restrict the queries
  to the strings in that block and apply \cref{pr:rps-length-reduction}.
  This restores the required relation between the string length and
  the number of strings.

  Since $\ell=\Theta(\log m)$, there is a fixed constant
  $\gamma\geq1$ such that $\log m\leq\gamma\ell$.
  Denote $q=\lceil8c\gamma\rceil$.
  If $m<2q$, test each query against every string in its interval.
  In this case, $m$ and $\ell$ are bounded by constants depending
  only on $c$ and the constants in $\ell=\Theta(\log m)$.
  Thus, these comparisons take $\bigO(1)$ time.
  Construct one instance with two copies of
  $\mathtt{00}$ and two queries $(0,0,\emptystring)$.
  If the original answer is \texttt{YES}, replace the first query
  by $(0,1,\emptystring)$. Return the list containing this instance.
  It is a valid input with four symbols in total, which is at most $s/c$.

  Assume henceforth that $m\geq2q$. Compute $t=\lceil m/q\rceil$.
  Partition $S$ into consecutive blocks of $t$ strings and the query
  sequence into consecutive blocks of $t$ queries; in each partition,
  the last block may be smaller. Initialize an empty list of output
  instances. For every pair consisting of one string block and one
  query block, proceed as follows. Denote the string block by
  $S[a+1\dd a+u]$, where $1\leq u\leq t$.
  \begin{enumerate}
  \item First, we construct an array $S'[1\dd t]$ containing the
    strings from the block, followed by additional strings if $u<t$.
    Copy the packed representations of $S[a+1],\ldots,S[a+u]$ into
    the first $u$ positions of $S'$, and fill its remaining positions
    with copies of $\zero^\ell$, using \cref{pr:packed-representation}.

  \item Next, we restrict the queries to this block of strings.
    For each query index $i$ in the query block, compute endpoints
    $b_i'$ and $e_i'$ by replacing each endpoint $x\in\{b_i,e_i\}$ with
    \[
      \begin{cases}
        0 & \text{if }x\leq a,\\
        x-a & \text{if }a<x<a+u,\\
        u & \text{if }x\geq a+u.
      \end{cases}
    \]
    Form the sequence of queries $(b_i',e_i',Q_i)$ in the original
    order, copying the packed query strings and their lengths.
    Append copies of $(0,0,\emptystring)$ until this sequence
    contains exactly $t$ queries.

  \item Finally, apply \cref{pr:rps-length-reduction} to $S'$ and
    the constructed query sequence, with cardinality $t$ and common
    string length $\ell$.
    Add all returned instances to the output list.
  \end{enumerate}
  Return the list of output instances.

  To prove correctness, consider an original query $i$ and a block
  $S[a+1\dd a+u]$. Its new endpoints satisfy $0\leq b_i'\leq e_i'\leq u$,
  and for every $j\in[1\dd u]$, it holds
  $j\in(b_i'\dd e_i']$ if and only if $a+j\in(b_i\dd e_i]$.
  Since $S'[j]=S[a+j]$, the restricted query has answer \texttt{YES}
  exactly when the original query has a match in this block.
  No query interval includes any of the appended strings, and the
  added queries have empty intervals. Every original query and every
  original string belong to exactly one block of the corresponding
  partition. Consequently, the original input has answer \texttt{YES}
  if and only if one of the inputs supplied to
  \cref{pr:rps-length-reduction} has answer \texttt{YES}.
  That proposition preserves this equivalence for the returned inputs.
  There is at least one pair of blocks, and every call returns a
  nonempty list, proving that our output list is nonempty.

  We next verify the length-reduction assumptions and the output sizes.
  Since $m\geq2q$, we have $2\leq t\leq2m/q\leq m$.
  Moreover, $m\leq qt$ and $q$ is a constant, giving
  $\log m\leq\log q+\log t=\bigO(\log t)$.
  Hence, $\ell=\bigO(\log t)$.
  Each input supplied to the length reduction has $t$ strings and
  $t$ queries, all of length at most $\ell$, and its total string
  length is at most $2t\ell\leq4m\ell/q\leq m\ell\leq N$.
  This justifies using the original size bound $N$ and word size $w$
  in every call.
  Each returned instance has $t$ strings of length
  $2\lfloor\log t\rfloor$ and $t$ queries of at most that length.
  Its total string length is therefore at most
  \[
    4t\log t\leq\frac{8m\log m}{q}
      \leq\frac{m\ell}{c}\leq\frac{s}{c},
  \]
  where the second inequality follows from
  $q\geq8c\gamma$ and $\log m\leq\gamma\ell$.

  There are at most $\lceil m/t\rceil\leq q$ blocks in each
  partition, and hence at most $q^2=\bigO(1)$ pairs of blocks.
  All copying, comparisons, and construction of repeated symbols
  use \cref{pr:packed-representation} with alphabet size two,
  symbol width $\lceil\log2\rceil=1$, and length bound $N$.
  Its assumptions hold since $\ell\leq N$ and $w\geq1+\log N$.
  The initial cases $m<2q$ can be recognized by a fixed sequence
  of tests $m=2,\ldots,2q-1$, stopping upon equality.
  Every tested integer is at most $m$ and fits in a word; in the
  remaining case, $q\leq m/2\leq N$ also fits in a word.
  Since $\ell=\bigO(\log m)=\bigO(w)$, forming the arrays and
  query sequences takes $\bigO(t)$ time per pair of blocks.
  Each length-reduction call takes $\bigO(t)$ time and returns
  $\bigO(1)$ instances. The total time and number of output instances
  are therefore $\bigO(m)$ and $\bigO(1)$, respectively.
  The input occupies $\bigO(m)$ words, and the running time bounds
  the remaining space, including the output.
\end{proof}

\begin{theorem}
  \leavevmode\phantomsection\label{th:rps-length-reduction}
  Let $N\in\Z_{\geq2}$. Consider the word RAM model with word size
  $w$ satisfying $1+\log N\leq w=\bigO(\log N)$. Suppose that there
  exists an algorithm for \probname{Range Prefix Search}
  (\cref{prob:rps}) that, given $N$ and any valid input of at most $N$ symbols
  over alphabet $\BinaryAlphabet$, achieves the following complexities:
  \begin{itemize}
  \item running time $T(N)$,
  \item space usage $S(N)$ bits.
  \end{itemize}
  Then, there exists an algorithm that, given $N$, integers
  $m\in\Z_{\geq2}$ and $\ell\in\Z_{\geq1}$ satisfying
  $\ell=\Theta(\log m)$, an array $S[1\dd m]$ of strings in
  $\BinaryAlphabet^\ell$, and $m$ triples $(b_i,e_i,Q_i)$,
  $i\in[1\dd m]$, satisfying $0\leq b_i\leq e_i\leq m$,
  $Q_i\in\BinaryAlphabet^{\leq\ell}$, and
  $m\ell+\sum_{i=1}^{m}|Q_i|\leq N$, decides whether there exist
  $i\in[1\dd m]$ and $j\in(b_i\dd e_i]$ such that $Q_i$ is a
  prefix of $S[j]$.
  Each input string is supplied by its length and packed representation
  (\cref{def:packed-representation}).
  The algorithm achieves the following complexities:
  \begin{itemize}
  \item running time $\bigO(T(N))$,
  \item space usage $\bigO(S(N))$ bits.
  \end{itemize}
  The constants in these bounds may depend on the constants implicit
  in $\ell=\Theta(\log m)$.
\end{theorem}
\begin{proof}
  Denote $s=m\ell+\sum_{i=1}^{m}|Q_i|$.
  If $s<4$, direct comparison determines the answer in
  $\bigO(1)$ time using $\bigO(1)$ words.
  Assume henceforth that $s\geq4$.
  Apply \cref{pr:rps-size-reduction} with $c=1$
  to the given array and queries.
  Each returned instance is a valid Range Prefix Search input with
  total string length at most $s\leq N$.
  Run the assumed algorithm on each returned instance and return
  \texttt{YES} if at least one call returns \texttt{YES}, and
  \texttt{NO} otherwise. Correctness follows from
  \cref{pr:rps-size-reduction}.
  The size reduction and every call to the assumed algorithm use
  size parameter $N$.

  Since $m\ell\leq s\leq2m\ell$ and $\ell=\Theta(\log m)$,
  we have $s=\Theta(m\log m)$ and
  $m=\bigO(s/\log s)\subseteq\bigO(N/\log N)$.
  The construction takes $\bigO(m)$ time and uses $\bigO(m)$ words,
  including the input and output.
  Since $w=\bigO(\log N)$, this space occupies $\bigO(N)$ bits.
  Together with the constant number of calls to the assumed algorithm,
  the total running time is $\bigO(N/\log N+T(N))$, and the space
  usage is $\bigO(N+S(N))$ bits.
  By \cref{lm:rps-lower-bounds}, we have
  $T(N)=\Omega(N/\log N)$ and $S(N)=\Omega(N)$.
  Hence, the claimed bounds follow.
\end{proof}

\subsection{Combinatorial Results}\label{sec:sn-to-rps-combinatorial-results}

\begin{lemma}\label{lm:sn-to-rps}
  Let $m,q,\ell\in\Z_{\geq1}$, and let
  $(X_1,Y_1),\ldots,(X_m,Y_m)$ be a sequence of pairs of strings
  in $\BinaryAlphabet^\ell$ such that $Y_1\preceq\cdots\preceq Y_m$.
  Let $(A_1,B_1),\ldots,(A_q,B_q)$ be a sequence of pairs of strings
  in $\BinaryAlphabet^{\leq\ell}$ with $|A_i|+|B_i|=\ell$
  for every $i\in[1\dd q]$.
  Let $S[1\dd m]$ be the array with $S[j]=\revstr{X_j}$ for
  every $j\in[1\dd m]$. For every $i\in[1\dd q]$, denote
  $Q_i=\revstr{A_i}$, $L_i=B_i\cdot\zero^{\ell-|B_i|}$,
  $U_i=B_i\cdot\one^{\ell-|B_i|}$, and
  \[
    b_i=|\{j\in[1\dd m]:Y_j\prec L_i\}|,
    \qquad
    e_i=|\{j\in[1\dd m]:Y_j\preceq U_i\}|.
  \]
  Then, for every $i\in[1\dd q]$ and $j\in[1\dd m]$, the following
  conditions are equivalent:
  \begin{enumerate}
  \item $A_i$ is a suffix of $X_j$ and $B_i$ is a prefix of $Y_j$,
  \item $j\in(b_i\dd e_i]$ and $Q_i$ is a prefix of $S[j]$.
  \end{enumerate}
\end{lemma}
\begin{proof}
  Since $|Y_j|=\ell$, $B_i$ is a prefix of $Y_j$ if and only if
  $L_i\preceq Y_j\preceq U_i$.
  The order $Y_1\preceq\cdots\preceq Y_m$ and the definitions of
  $b_i$ and $e_i$ make this equivalent to $j\in(b_i\dd e_i]$.
  Moreover, $A_i$ is a suffix of $X_j$ if and only if
  $Q_i=\revstr{A_i}$ is a prefix of $S[j]=\revstr{X_j}$.
  Combining these equivalences proves the claim.
\end{proof}

\subsection{Algorithms}\label{sec:sn-to-rps-algorithms}

\begin{theorem}\label{th:rpm}
  Let $N\in\Z_{\geq2}$.
  Consider the word RAM model with word size
  $1+\log N\leq w=\bigO(\log N)$.
  If there exists an algorithm for
  \probname{Range} \probname{Prefix} \probname{Search}
  that, given $N$ and any valid input of at most $N$ symbols over alphabet
  $\BinaryAlphabet$ (\cref{prob:rps}),
  achieves the following complexities:
  \begin{itemize}
  \item running time $T_{\rm RPS}(N)$,
  \item space usage $S_{\rm RPS}(N)$ bits,
  \end{itemize}
  then there exists an
  algorithm for \probname{String} \probname{Nesting}
  that, given $N$ and any valid input of at most $N$ symbols over alphabet
  $\BinaryAlphabet$ (\cref{prob:sn}),
  achieves the following complexities:
  \begin{itemize}
  \item running time $\bigO(T_{\rm RPS}(N))$,
  \item space usage $\bigO(S_{\rm RPS}(N))$ bits.
  \end{itemize}
\end{theorem}
\begin{proof}
  Consider a valid String Nesting input consisting
  of sets $\mathcal P$ and $\mathcal Q$, where
  $q=|\mathcal P|=|\mathcal Q|$ and $\ell=2\lfloor\log q\rfloor$.
  The total length of the input strings is $3q\ell\leq N$.
  Let $m=q$, and let $(A_1,B_1),\ldots,(A_q,B_q)$ denote
  the sequence listing the pairs of $\mathcal Q$ in their input order.
  All packed-string tools below use alphabet size $2$ and
  size parameter $M=m+q=2q$.
  The algorithm proceeds as follows.
  \begin{enumerate}

  \item First, we sort the pairs $(X,Y)\in\mathcal P$ by
    their second components $Y$.
    We apply \cref{pr:packed-string-sorting} to the sequence of
    second components and reorder the pairs using the returned
    permutation. Let
    $(X_1,Y_1),\ldots,(X_m,Y_m)$ denote the resulting sequence.
    Then $Y_1\preceq\cdots\preceq Y_m$.
    Since $\ell=\Theta(\log m)$, this takes
    $\bigO(m)=\bigO(q)$ time.

  \item\label{step:sn-to-rps-lower-boundaries}
    Next, we compute
    $b_i=|\{j\in[1\dd m]:Y_j\prec L_i\}|$ for every $i\in[1\dd q]$,
    where $L_i=B_i\cdot\texttt{0}^{\ell-|B_i|}$.
    To this end, we first construct the packed representations of $L_i$
    by appending $\texttt{0}^{\ell-|B_i|}$ to each $B_i$, using
    \cref{pr:packed-representation}. We then sort the sequence
    $L_1,\ldots,L_q$ using \cref{pr:packed-string-sorting}.
    The construction and sorting take $\bigO(q)$ time.
    Let $\lambda[1\dd q]$ be the returned permutation. Then
    \[
      L_{\lambda[1]}\preceq\cdots\preceq L_{\lambda[q]}.
    \]
    Initialize $j=0$.
    For $r=1,\ldots,q$, increment $j$ while $j<m$ and
    $Y_{j+1}\prec L_{\lambda[r]}$, and then set $b_{\lambda[r]}=j$.
    Each comparison takes $\bigO(1+\ell/w)=\bigO(1)$ time by
    \cref{pr:packed-representation}\eqref{pr:packed-representation-compare}
    since $\ell=\Theta(\log q)=\bigO(w)$.
    Since $j$ never decreases, the scan takes
    $\bigO(m+q)=\bigO(q)$ time.

  \item We now compute
    $e_i=|\{j\in[1\dd m]:Y_j\preceq U_i\}|$ for every $i\in[1\dd q]$,
    where $U_i=B_i\cdot\texttt{1}^{\ell-|B_i|}$.
    To construct these strings, for every $i\in[1\dd q]$, we initialize
    the packed representation of $\texttt{1}^{\ell-|B_i|}$ and append it
    to that of $B_i$, using \cref{pr:packed-representation}.
    Since $\ell=\bigO(w)$, the construction takes $\bigO(q)$ total time.
    We sort the sequence $U_1,\ldots,U_q$ and let
    $\mu[1\dd q]$ be the returned permutation. It satisfies
    \[
      U_{\mu[1]}\preceq\cdots\preceq U_{\mu[q]}.
    \]
    Repeat the scan from Step~\ref{step:sn-to-rps-lower-boundaries}
    with $j$ reset to zero, replacing
    $Y_{j+1}\prec L_{\lambda[r]}$ by $Y_{j+1}\preceq U_{\mu[r]}$
    and storing the result as $e_{\mu[r]}=j$.
    Strings equal to $U_i$ are now included in the count $e_i$.
    As in Step~\ref{step:sn-to-rps-lower-boundaries}, the construction,
    sorting, and scan together take $\bigO(q)$ time.

  \item We next construct the packed representations of the strings
    \[
      S[j]=\revstr{X_j}\quad(j\in[1\dd m]),
      \qquad
      Q_i=\revstr{A_i}\quad(i\in[1\dd q]).
    \]
    We build the data structure from \cref{th:string-reversal}
    in $\bigO(\sqrt M)=\bigO(\sqrt q)$ time,
    and then query it for $X_1,\ldots,X_m$ and $A_1,\ldots,A_q$.
    Since each string has length at most
    $\ell=\Theta(\log q)=\Theta(\log M)$, each query takes
    $\bigO(1+\ell/\log M)=\bigO(1)$ time.
    Thus, the entire step takes $\bigO(q)$ time.

  \item Finally, we run the assumed Range Prefix Search algorithm
    on $S[1\dd m]$ and the queries $(b_i,e_i,Q_i)_{i\in[1\dd q]}$,
    and return its answer.
    By \cref{lm:sn-to-rps}, for every $i\in[1\dd q]$ and
    $j\in[1\dd m]$, the string $Q_i$ is a prefix of $S[j]$ with
    $j\in(b_i\dd e_i]$ if and only if $A_i$ is a suffix of $X_j$
    and $B_i$ is a prefix of $Y_j$. Thus, the returned answer
    solves the String Nesting instance.
  \end{enumerate}
  The constructed instance is valid because
  $m=q$, $\ell=2\lfloor\log m\rfloor$, $0\leq b_i\leq e_i\leq m$,
  and $|Q_i|\leq\ell$ for every $i\in[1\dd q]$;
  here, $b_i\leq e_i$ holds because $L_i\preceq U_i$.
  Its number of binary symbols satisfies
  \[
    m\ell+\sum_{i=1}^q|Q_i|
      =q\ell+\sum_{i=1}^q|A_i|\leq3q\ell\leq N.
  \]
  The assumed algorithm therefore applies with size parameter $N$.
  All strings used by the packed-string tools have length at most
  $\ell\leq q\leq M$. Their word-size condition holds because
  $M\leq N$ and $w\geq1+\log N\geq1+\log M$.
  Apart from the call to the assumed algorithm, the reduction takes
  $\bigO(q)$ time and uses $\bigO(q)$ words, including its input.
  Since $3q\ell\leq N$ and $\ell=\Theta(\log q)$, we have
  $q=\bigO(N/\log N)$.
  Together with $w=\bigO(\log N)$, this gives total time
  $\bigO(N/\log N+T_{\rm RPS}(N))$ and total space
  $\bigO(N+S_{\rm RPS}(N))$ bits.

  By \cref{lm:rps-lower-bounds}, we have
  $T_{\rm RPS}(N)=\Omega(N/\log N)$ and $S_{\rm RPS}(N)=\Omega(N)$.
  These bounds absorb the additive terms, proving the claim.
\end{proof}

\section{Reducing Range Prefix Search to Counting Colored Inversions}\label{sec:rpm-to-cci}

\subsection{Problem Definition}\label{sec:cci-problem}

\begin{problem}[Counting Colored Inversions]\label{prob:cci}
  \begin{description}[style=sameline,itemsep=0ex,font={\normalfont\bf}]
  \item[Input:]
    Let $n \in \Z_{\geq 1}$. Consider the word RAM model with word size
    $w > \log n$. The input consists of $n$ and arrays $C[1 \dd n]$ and
    $A[1 \dd n]$, stored with one entry per machine word, such that
    $C[i] \in \{0,1\}$ and $A[i] \in [0 \dd n)$ for every $i \in [1 \dd n]$.
  \item[Output:]
    The number of colored inversions in $A$, i.e., the value
    $|\{(i,j) \in [1 \dd n]^2 : i<j,\,C[i] \neq C[j],\text{ and }A[i] > A[j]\}|$.
  \end{description}
\end{problem}

\begin{remark}[Representing inversion counts]\label{rm:inversion-count-representation}
  An inversion count for an array of length $n$ is smaller than $n^2$.
  Since both inversion-counting problems assume $w>\log n$
  (\cref{prob:cci,prob:ci}), every answer fits in two machine words.
  Throughout, we represent such a count $x$ by a pair
  $(x_{\rm hi},x_{\rm lo})$ satisfying
  $x=2^w x_{\rm hi}+x_{\rm lo}$, where
  $x_{\rm hi},x_{\rm lo}\in[0\dd2^w)$.
  We compare two counts by comparing their high words first and,
  when these agree, their low words.
  Addition and nonnegative subtraction take $\bigO(1)$ time using
  word arithmetic modulo $2^w$ with a carry or borrow, provided
  the result fits in two words.
  To compute $\lfloor x/2\rfloor$, shift both words right by one
  position and set the most significant bit of the low word to the
  least significant bit of the original high word.
  This also takes $\bigO(1)$ time.
\end{remark}

\subsection{Preliminaries}\label{sec:rps-to-cci-preliminaries}

\begin{lemma}[Lower bounds for Counting Colored Inversions]\label{lm:cci-lower-bounds}
  Let $N\in\Z_{\geq2}$. Consider the word RAM model with word size
  $w$ satisfying $1+\log N\leq w=\bigO(\log N)$.
  Any algorithm for \probname{Counting Colored Inversions}
  (\cref{prob:cci}) that, given $N$, solves every valid input of
  at most $N$ bits requires
  $\Omega(N/\log N)$ time and $\Omega(N)$ bits of space
  in the worst case.
\end{lemma}
\begin{proof}
  For sufficiently large $N$, let
  $r=\lfloor N/(4w)\rfloor\geq2$.
  Consider a Counting Colored Inversions input consisting of $r$,
  an all-zero value array $B[1\dd r]$, and a color array
  $D[1\dd r]$ with $D[i]=0$ for $i\in[1\dd r)$ and $D[r]=1$.
  This is a valid input of $(2r+1)w\leq N$ bits, including the
  length parameter, and its answer is zero.
  Changing any entry $B[i]$, with $i\in[1\dd r)$, to one creates
  the colored inversion $(i,r)$. Thus, an algorithm must read
  the first $r-1$ entries of $B$ on the original input.
  Since each entry occupies a separate word and
  $r=\Theta(N/w)=\Theta(N/\log N)$, this proves the time lower bound.
  The input itself occupies $\Theta(rw)=\Theta(N)$ bits, which gives
  the claimed space lower bound (see \cref{sec:model}).
\end{proof}

\begin{lemma}\label{lm:cci-size-extension}
  Let $c\geq1$ be a fixed constant and $N\in\Z_{\geq2}$.
  Consider the word RAM model with word size $w$ satisfying
  $1+\log N\leq w=\bigO(\log N)$. Suppose that there exists an algorithm
  for \probname{Counting} \probname{Colored} \probname{Inversions}
  (\cref{prob:cci}) that,
  given $N$ and any valid input of at most $N$ bits, achieves the following
  complexities:
  \begin{itemize}
  \item running time $T(N)$,
  \item space usage $S(N)$ bits.
  \end{itemize}
  Then, using the same word size, there exists an algorithm for
  \probname{Counting Colored Inversions} that, given $N$ and any valid input
  of at most $cN$ bits, achieves the following complexities:
  \begin{itemize}
  \item running time $\bigO(T(N))$,
  \item space usage $\bigO(S(N))$ bits.
  \end{itemize}
  The constants in these bounds may depend on $c$.
\end{lemma}
\begin{proof}
  We divide the input into consecutive blocks small enough that any two
  blocks together fit within $N$ bits. The colored inversion count of
  their concatenation, minus the two counts within the individual blocks,
  gives the number of colored inversions between them.

  Consider an input consisting of $n$ and arrays $C[1\dd n]$ and
  $A[1\dd n]$, with $(2n+1)w\leq cN$.
  For sufficiently large $N$, we have $w\geq c$ and
  $k=\lfloor N/w\rfloor\geq5$. We assume these inequalities below;
  for the remaining bounded values of $N$, checking all pairs of
  positions and accumulating their contributions in two words
  (\cref{rm:inversion-count-representation}) takes $\bigO(1)$ time and words.
  The algorithm proceeds as follows.
  \begin{enumerate}
  \item First, we construct the blocks.
    Compute $h=\lfloor(k-1)/4\rfloor$, and divide $A$ and $C$
    into corresponding consecutive blocks of $h$ positions, except
    possibly for the last block. Copy these blocks into arrays
    $A_i[1\dd n_i]$ and $C_i[1\dd n_i]$ for $i\in[1\dd t]$,
    where $t=\lceil n/h\rceil$ and $n_i$ is the length of block $i$.

  \item Next, we compute the colored inversion counts of the blocks
    and their pairwise concatenations. Denote the required counts by
    $x_i=\ColoredInversionCount{C_i}{A_i}$ for $i\in[1\dd t]$ and
    $x_{i,j}=\ColoredInversionCount{C_iC_j}{A_iA_j}$ for
    $1\leq i<j\leq t$.
    For each count, first copy the corresponding colors and values
    into arrays $D[1\dd m]$ and $B[1\dd m]$, preserving their order.
    Before calling the assumed algorithm, we replace the values in $B$
    by integers in $[0\dd m)$ as follows.
    Initialize an array $R[0\dd n)$ to zero and set $R[B[r]]=1$
    for every $r\in[1\dd m]$.
    Scan $R$ in ascending order of its indices, replacing its marked
    values by consecutive integers starting at zero.
    For every $r\in[1\dd m]$, replace $B[r]$ by $R[B[r]]$.
    Equal original values receive the same integer, and distinct values
    receive integers in their original order. Thus, this replacement
    preserves the colored inversion count, and all values now belong
    to $[0\dd m)$.
    Run the assumed algorithm on $m$, $D$, and $B$, and store its
    answer as $x_i$ or $x_{i,j}$, respectively.

  \item Finally, return
    \[
      \sum_{i\in[1\dd t]}x_i
      +\sum_{1\leq i<j\leq t}(x_{i,j}-x_i-x_j).
    \]
    For each pair of blocks, compute the parenthesized difference
    before adding it to the sum.
  \end{enumerate}
  Every colored inversion either has both positions in one block or
  has its positions in two distinct blocks $i<j$.
  In the latter case, it is counted by $x_{i,j}-x_i-x_j$, since
  subtracting $x_i$ and $x_j$ removes exactly the inversions within
  those two blocks. Hence, the returned sum counts every original
  colored inversion exactly once.

  Each call uses size parameter $N$: it has $m\leq2h$ and occupies,
  including its length parameter, $(2m+1)w\leq(4h+1)w\leq kw\leq N$ bits.
  The same word size is used throughout. Since $w\geq c$, the original
  size bound gives $2n+1\leq N$. All array lengths and values
  therefore fit in one word. Every count fits in two words by
  \cref{rm:inversion-count-representation}.
  Moreover, $x_i+x_j\leq x_{i,j}$ for every $i<j$.
  Both subtractions in $x_{i,j}-x_i-x_j$ are nonnegative, and the
  accumulated sum never exceeds the final count.
  Thus, each addition and subtraction takes $\bigO(1)$ time.

  Finally, $h=\Theta(k)=\Theta(N/w)$ and
  $n\leq cN/(2w)$ imply that $t=\bigO(1)$.
  There are $\bigO(1)$ calls, and preparing each input takes
  $\bigO(n)$ time and words, including the array $R$.
  Executing the calls sequentially gives total time
  $\bigO(N/\log N+T(N))$ and space
  $\bigO(N+S(N))$ bits, including the original input and all
  constructed arrays and counts.
  By \cref{lm:cci-lower-bounds}, we have $T(N)=\Omega(N/\log N)$
  and $S(N)=\Omega(N)$, which yields the claimed bounds.
\end{proof}

\subsection{Combinatorial Results}\label{sec:rps-to-cci-combinatorial-results}

\begin{definition}[Deletion]\label{def:delete}
  Let $S$ be a finite sequence of elements from a set $\mathcal X$,
  and let $Q \subseteq [1 \dd |S|]$.
  By $\DeleteSubseq{S}{Q}$ we denote the sequence obtained by deleting
  the elements of $S$ at positions in $Q$, preserving the order of
  the remaining elements.
\end{definition}

\begin{definition}[Insertion]\label{def:insert}
  Let $S$ be a finite sequence over a set $\mathcal X$,
  and let $Q = \{(p_1,c_1), \ldots, (p_k,c_k)\}
  \subseteq \Z_{\geq1} \times \mathcal X$, where
  $k\in\Z_{\geq0}$ and $1 \leq p_1 < p_2 < \cdots < p_k \leq |S|+k$
  whenever $k>0$.
  We define $\InsertSubseq{S}{Q}$ as the sequence $S'$ of length $|S|+k$
  satisfying the following conditions:
  \begin{enumerate}
  \item For every $i \in [1 \dd k]$, it holds $S'[p_i] = c_i$,
  \item $\DeleteSubseq{S'}{\{p_1, p_2, \ldots, p_k\}} = S$.
  \end{enumerate}
  For arrays, deletion and insertion denote the arrays containing the
  resulting sequences of entries.
\end{definition}

\begin{example}\label{ex:deletion-insertion}
  Let $S=(3,1,4,1,5)$. Deleting the elements at positions $2$ and $4$
  gives $S'=\DeleteSubseq{S}{\{2,4\}}=(3,4,5)$.
  Conversely, $\InsertSubseq{S'}{\{(2,1),(4,1)\}}=S$.
  The insertion positions $2$ and $4$ refer to the resulting sequence $S$.
\end{example}

\begin{lemma}\label{lm:ci-reduction}
  Let $m\in\Z_{\geq0}$ and $\ell\in\Z_{\geq1}$.
  Consider an array $A[1 \dd m]$ of $m$ strings of length $\ell$ over alphabet $\Sigma$.
  Let also $C[1 \dd m] = [0, 0, \ldots, 0]$. Let $Q \in \Sigma^{<\ell}$ and
  $b,e \in [0 \dd m]$ be such that $b \leq e$.
  Let $\dol$ and $\hash$ be symbols such that for every
  $a \in \Sigma$, it holds $\dol \prec a \prec \hash$.
  Further, let $Q_{\rm low} = Q \cdot \dol$ and $Q_{\rm high} = Q \cdot \hash$.
  We define the following arrays (for the colored inversion counts below, we use
  \cref{def:colored-inv-count} with $\mathcal{X}=(\Sigma\cup\{\dol,\hash\})^*$
  equipped with the lexicographic order):
  \begin{itemize}
  \item $C' = \InsertSubseq{C}{\mathcal{I}}$, where $\mathcal{I} = \{(b+1,1), (e+2,1)\}$,
  \item $A^{\rm sub} = \InsertSubseq{A}{\mathcal{I}^{\rm sub}}$,
    where $\mathcal{I}^{\rm sub} = \{(b+1,Q_{\rm low}), (e+2,Q_{\rm high})\}$,
  \item $A^{\rm add} = \InsertSubseq{A}{\mathcal{I}^{\rm add}}$,
    where $\mathcal{I}^{\rm add} = \{(b+1, Q_{\rm high}), (e+2, Q_{\rm low})\}$.
  \end{itemize}
  Then, letting $n_{\rm occ} = |\{i \in (b \dd e] : Q\text{ is a prefix of }A[i]\}|$, it holds
  \begin{align*}
    \ColoredInversionCount{C'}{A^{\rm add}} -
    \ColoredInversionCount{C'}{A^{\rm sub}} = 2 \cdot n_{\rm occ}.
  \end{align*}
\end{lemma}
\begin{proof}
  In both $A^{\rm add}$ and $A^{\rm sub}$, the original entries have
  color $0$, and the two inserted entries have color $1$.
  Every colored inversion therefore involves exactly one original
  entry and one inserted entry. We compare the numbers of these
  inversions for each original index $j\in[1\dd m]$:
  \begin{itemize}
  \item If $j\in[1\dd b]$, the original entry remains at position $j$,
    with $A^{\rm add}[j]=A^{\rm sub}[j]=A[j]$.
    Both inserted entries follow it. In each array, the number of
    colored inversions involving this entry equals the number of
    strings among $Q_{\rm low}$ and $Q_{\rm high}$ that are
    lexicographically smaller than $A[j]$. Exchanging the positions
    of the two inserted strings leaves this number unchanged.
  \item For $j\in(b\dd e]$, the original entry is at position $j+1$,
    with $A^{\rm add}[j+1]=A^{\rm sub}[j+1]=A[j]$.
    Since $b+1<j+1<e+2$, one inserted entry precedes it and the
    other follows it. There are two cases:
    \begin{itemize}
    \item If $Q$ is a prefix of $A[j]$, then
      $Q_{\rm low}\prec A[j]\prec Q_{\rm high}$.
      In $A^{\rm add}$, the preceding inserted entry is
      $Q_{\rm high}$ and the following one is $Q_{\rm low}$,
      giving two colored inversions involving the entry at position
      $j+1$. In $A^{\rm sub}$, their values are interchanged,
      giving no colored inversion involving that entry.
    \item If $Q$ is not a prefix of $A[j]$, their first differing
      symbols imply that either $A[j]\prec Q_{\rm low}$ or
      $Q_{\rm high}\prec A[j]$.
      In either case, exactly one of the two inserted entries forms
      a colored inversion with the entry at position $j+1$ in each
      array.
    \end{itemize}
  \item If $j\in(e\dd m]$, the original entry is at position $j+2$,
    with $A^{\rm add}[j+2]=A^{\rm sub}[j+2]=A[j]$.
    Both inserted entries precede it. In each array, the number of
    colored inversions involving this entry equals the number of
    strings among $Q_{\rm low}$ and $Q_{\rm high}$ that are
    lexicographically greater than $A[j]$. This number is again
    unchanged by exchanging their positions.
  \end{itemize}
  Summing the differences over all $j\in[1\dd m]$ gives
  \[
    \ColoredInversionCount{C'}{A^{\rm add}}
      -\ColoredInversionCount{C'}{A^{\rm sub}}
      =2\cdot n_{\rm occ}.
    \qedhere
  \]
\end{proof}

\begin{example}\label{ex:ci-reduction}
  Let $Q=\texttt{10}$, $b=2$, and $e=6$.
  For the array $A[1\dd8]$ in the two left columns below and the alphabet order
  $\dol\prec\zero\prec\one\prec\hash$, the construction in
  \cref{lm:ci-reduction} gives the arrays on the right:
  \[
    \begin{array}{r@{\quad}l|r@{\quad}c@{\quad}l@{\quad}l}
      j & A[j] & i & C'[i] & A^{\rm add}[i] & A^{\rm sub}[i]\\\hline
      1 & \texttt{1011} & 1 & 0 & \texttt{1011} & \texttt{1011}\\
      2 & \texttt{0110} & 2 & 0 & \texttt{0110} & \texttt{0110}\\
        &                & 3 & 1 & \texttt{10}\hash & \texttt{10}\dol\\
      3 & \texttt{1001} & 4 & 0 & \texttt{1001} & \texttt{1001}\\
      4 & \texttt{0011} & 5 & 0 & \texttt{0011} & \texttt{0011}\\
      5 & \texttt{1100} & 6 & 0 & \texttt{1100} & \texttt{1100}\\
      6 & \texttt{1010} & 7 & 0 & \texttt{1010} & \texttt{1010}\\
        &                & 8 & 1 & \texttt{10}\dol & \texttt{10}\hash\\
      7 & \texttt{1000} & 9 & 0 & \texttt{1000} & \texttt{1000}\\
      8 & \texttt{1110} & 10 & 0 & \texttt{1110} & \texttt{1110}
    \end{array}
  \]
  The entries $A[3]=\texttt{1001}$ and $A[6]=\texttt{1010}$ are the only
  entries in $A[3\dd6]$ with prefix $Q$, giving $n_{\rm occ}=2$.
  Each forms two colored inversions with the inserted entries in $A^{\rm add}$
  and none in $A^{\rm sub}$.
  The entries $A[1]$ and $A[7]$ have prefix $Q$ but lie outside $(b\dd e]$,
  while $A[4]$ and $A[5]$ lie inside this interval but do not have prefix $Q$.
  Each forms one colored inversion in each array.
  The remaining entries $A[2]$ and $A[8]$ form none.
  Thus, $\ColoredInversionCount{C'}{A^{\rm add}}=8$ and
  $\ColoredInversionCount{C'}{A^{\rm sub}}=4$, giving a difference
  of $4=2n_{\rm occ}$.
\end{example}

\begin{lemma}\label{lm:inv}
  Let $m,q\in\Z_{\geq0}$ and $\ell\in\Z_{\geq1}$.
  Consider an array $A[1 \dd m]$ of $m$ strings of length $\ell$ over alphabet $\Sigma$.
  Let $(b_i,e_i,Q_i)_{i \in [1 \dd q]}$ be a sequence such that, for every $i \in [1 \dd q]$, it holds
  $b_i,e_i\in[0\dd m]$, $b_i\leq e_i$, and $Q_i \in \Sigma^{<\ell}$.
  Let $\dol$ and $\hash$ be symbols, and assume that $\dol \prec a \prec \hash$ holds for
  every $a \in \Sigma$.
  Let $(p_r,U_r,V_r)_{r\in[0 \dd 2q)}$ be the result of sorting
  the elements of the sequence $(b_1+1, Q_1 \hash, Q_1 \dol), (e_1 + 1, Q_1 \dol, Q_1 \hash), \ldots, (b_q+1, Q_{q} \hash, Q_{q} \dol),
  (e_q+1, Q_{q} \dol, Q_{q} \hash)$ by the first coordinate.
  Let also $Z[1 \dd m] = [0, 0, \ldots, 0]$. Then, it holds (for the
  colored inversion counts, we use \cref{def:colored-inv-count} with
  $\mathcal{X}=(\Sigma\cup\{\dol,\hash\})^*$ equipped with the
  lexicographic order)
  \begin{align*}
    \ColoredInversionCount{C}{A^{\rm add}} - \ColoredInversionCount{C}{A^{\rm sub}} = 2 \cdot \sum_{i=1}^{q} n_i,
  \end{align*}
  where:
  \begin{itemize}
  \item $C = \InsertSubseq{Z}{\{(p_r+r,1)\}_{r \in[0 \dd 2q)}}$,
  \item $A^{\rm add} = \InsertSubseq{A}{\{(p_r+r,U_r)\}_{r \in[0 \dd 2q)}}$,
  \item $A^{\rm sub} = \InsertSubseq{A}{\{(p_r+r,V_r)\}_{r \in[0 \dd 2q)}}$, and
  \item $n_i = |\{t \in (b_i \dd e_i] : Q_i\text{ is a prefix of }A[t]\}|$.
  \end{itemize}
\end{lemma}
\begin{proof}
  If $q=0$, both colored inversion counts and the sum in the claim
  are zero. Henceforth, assume that $q\geq1$.
  For every $p\in[1\dd m+1]$ and
  $X\in(\Sigma\cup\{\dol,\hash\})^*$, let
  \[
    f(p,X)
      =|\{j\in[1\dd p):A[j]\succ X\}|
       +|\{j\in[p\dd m]:A[j]\prec X\}|.
  \]
  This is the number of inversions between $X$ and the entries of $A$
  when $X$ is inserted after the first $p-1$ entries of $A$.

  We first determine which original entries precede each insertion
  in $A^{\rm add}$ and $A^{\rm sub}$.
  Since $(p_r)_{r\in[0\dd2q)}$ is nondecreasing, the positions
  $p_0,p_1+1,\ldots,p_{2q-1}+2q-1$ are strictly increasing.
  Hence, exactly $r$ inserted entries precede position $p_r+r$,
  leaving exactly $(p_r+r-1)-r=p_r-1$ original entries before it.
  By \cref{def:insert}, these are $A[1],\ldots,A[p_r-1]$,
  and the remaining original entries follow it in their original order.
  The inserted entry at position $p_r+r$ therefore forms
  $f(p_r,U_r)$ colored inversions with original entries in
  $A^{\rm add}$, and $f(p_r,V_r)$ in $A^{\rm sub}$.
  No pair of inserted entries is a colored inversion, since both
  entries have color $1$. Likewise, no pair of original entries
  is a colored inversion, since both entries have color $0$.
  Thus, every colored inversion is counted exactly once in
  \[
    \begin{aligned}
      \ColoredInversionCount{C}{A^{\rm add}}
        &=\sum_{r=0}^{2q-1}f(p_r,U_r),\\
      \ColoredInversionCount{C}{A^{\rm sub}}
        &=\sum_{r=0}^{2q-1}f(p_r,V_r).
    \end{aligned}
  \]

  For every $i\in[1\dd q]$, denote
  \[
    \begin{aligned}
      d_i^{\rm add}
        &=f(b_i+1,Q_i\hash)+f(e_i+1,Q_i\dol),\\
      d_i^{\rm sub}
        &=f(b_i+1,Q_i\dol)+f(e_i+1,Q_i\hash).
    \end{aligned}
  \]
  Sorting preserves the triples in the original sequence, including
  their multiplicities. Regrouping the two sums by the two triples
  specified for each $i\in[1\dd q]$ therefore gives
  \[
    \ColoredInversionCount{C}{A^{\rm add}}
      -\ColoredInversionCount{C}{A^{\rm sub}}
      =\sum_{i=1}^{q}\bigl(d_i^{\rm add}-d_i^{\rm sub}\bigr).
  \]

  It remains to determine $d_i^{\rm add}-d_i^{\rm sub}$.
  Apply \cref{lm:ci-reduction} with $Q=Q_i$, $b=b_i$, and $e=e_i$.
  In the array $A^{\rm add}$ defined in that lemma, the two inserted
  entries have values $Q_i\hash$ and $Q_i\dol$ at positions
  $b_i+1$ and $e_i+2$. They have $b_i$ and $e_i$ original entries
  before them, respectively, since one of the entries preceding
  the second insertion is the first insertion.
  Their contributions to the colored inversion count are therefore
  $f(b_i+1,Q_i\hash)$ and $f(e_i+1,Q_i\dol)$, whose sum is
  $d_i^{\rm add}$.
  In the lemma's array $A^{\rm sub}$, the inserted values are
  interchanged, giving the sum
  $f(b_i+1,Q_i\dol)+f(e_i+1,Q_i\hash)=d_i^{\rm sub}$.
  These sums include every colored inversion: all original entries
  have color $0$, and both inserted entries have color $1$.
  Consequently, the conclusion of \cref{lm:ci-reduction} gives
  $d_i^{\rm add}-d_i^{\rm sub}=2n_i$.
  This also covers $b_i=e_i$, when the two defining sums are
  identical and $n_i=0$.
  Substituting $d_i^{\rm add}-d_i^{\rm sub}=2n_i$ into the sum above
  proves the claim.
\end{proof}

\subsection{Algorithms}\label{sec:rps-to-cci-algorithms}

\begin{theorem}\label{th:rpm-to-cci}
  Let $N\in\Z_{\geq2}$. Consider the word RAM model with word size
  $w$ satisfying $1+\log N\leq w=\bigO(\log N)$.
  Suppose that there exists an algorithm for
  \probname{Counting} \probname{Colored} \probname{Inversions}
  (\cref{prob:cci}) that, given $N$ and any valid input of at most
  $N$ bits, achieves the following complexities:
  \begin{itemize}
  \item running time $T_{\rm CCI}(N)$,
  \item space usage $S_{\rm CCI}(N)$ bits.
  \end{itemize}
  Then, there exists an algorithm for
  \probname{Range} \probname{Prefix} \probname{Search}
  (\cref{prob:rps}) that, given $N$ and any
  valid input of at most $N$ symbols over alphabet
  $\BinaryAlphabet$, achieves the following complexities:
  \begin{itemize}
  \item running time $\bigO(T_{\rm CCI}(N))$,
  \item space usage $\bigO(S_{\rm CCI}(N))$ bits.
  \end{itemize}
\end{theorem}
\begin{proof}
  Let $d\geq2$ be a fixed integer such that $w\leq d\log N$,
  and let $c=26d$.
  We assume that $N$ is sufficiently large, handling bounded values
  by direct comparison in $\bigO(1)$ time and words of space.
  Denote by $m_0$ the number of strings in the given Range Prefix Search
  input and by $L_0\leq N$ the total length of its strings and query strings.
  If $L_0<4c$, direct comparison again determines the answer in
  $\bigO(1)$ time and words of space.
  Otherwise, apply \cref{pr:rps-size-reduction} with factor $c$
  to the given input. This produces $\bigO(1)$
  valid Range Prefix Search inputs, each with total string length
  at most $L_0/c$.

  We describe a procedure for solving one of the returned inputs.
  Denote its array of strings by $S[1\dd m]$, its queries by
  $(b_i,e_i,Q_i)_{i\in[1\dd q]}$, and its common string length
  by $\ell=2\lfloor\log m\rfloor$, where $q=m$.
  The total length of its strings is
  $L=m\ell+\sum_{i=1}^{q}|Q_i|\leq L_0/c\leq N/c$.
  Since $\ell=\Theta(\log m)$ and $q=m$, we have
  $L=\Theta(m\log m)$. Let $m'=m+2q=3m$ and $M=m'$.
  All packed-string tools below use size parameter $M$.
  The algorithm proceeds in four steps.
  \begin{enumerate}

  \item First, we construct the packed representations of the strings
    over alphabet $[0\dd5)$ used in the reduction.
    For each input string $R$, let $\widehat R$ be obtained by
    replacing each symbol $a$ with $a+2$.
    To compute these representations, we build the data structure
    from \cref{th:alphabet-lifting} in
    $\bigO(\sqrt M)=\bigO(\sqrt m)$ time.
    We query it with offset $2$ for each input string $R$.
    Since $|R|\leq\ell=\Theta(\log M)$, each query takes
    $\bigO(1+|R|/\log M)=\bigO(1)$ time.
    Let
    \[
      \widetilde S[j]=\widehat{S[j]}\cdot\texttt{1}
        \quad(j\in[1\dd m]),
      \qquad
      \widetilde Q_i=\widehat{Q_i}\quad(i\in[1\dd q]).
    \]
    Using \cref{pr:packed-representation}, we append
    $\texttt{1}$ to each $\widehat{S[j]}$ and construct the packed
    representations of $\widetilde Q_i\cdot\texttt{0}$ and
    $\widetilde Q_i\cdot\texttt{4}$ for every $i\in[1\dd q]$.
    The entire step takes $\bigO(m+q)=\bigO(m)$ time.
    A query $Q_i$ is a prefix of $S[j]$ if and only if
    $\widetilde Q_i$ is a prefix of $\widetilde S[j]$, and
    $|\widetilde Q_i|\leq\ell<|\widetilde S[j]|=\ell+1$.

  \item Next, we construct the color array $C$ and auxiliary string arrays
    $W^{\rm add}$ and $W^{\rm sub}$.
    To this end, construct the sequence consisting, for every $i\in[1\dd q]$,
    of the two triples
    $(b_i+1,\widetilde Q_i\cdot\texttt{4},
        \widetilde Q_i\cdot\texttt{0})$ and
    $(e_i+1,\widetilde Q_i\cdot\texttt{0},
        \widetilde Q_i\cdot\texttt{4})$.
    Sort the triples by their first coordinate using counting
    sort over $[1\dd m+1]$, in $\bigO(m+q)$ time,
    and denote the result by $(p_r,U_r,V_r)_{r\in[0\dd2q)}$.
    Construct arrays $C$, $W^{\rm add}$, and $W^{\rm sub}$
    of length $m'$ as follows.
    For every $r\in[0\dd2q)$, place $1$ at position $p_r+r$ of $C$,
    $U_r$ at this position of $W^{\rm add}$, and $V_r$ at this
    position of $W^{\rm sub}$.
    Fill the remaining positions of $C$ with zero and the
    remaining positions of both other arrays with
    $\widetilde S[1],\ldots,\widetilde S[m]$, in this order.
    As the positions $p_r+r$ are strictly increasing and belong
    to $[1\dd m']$, this takes $\bigO(m+q)=\bigO(m)$ time.
    By \cref{lm:inv}, applied with alphabet $\{\texttt{1},\texttt{2},\texttt{3}\}$
    and separating symbols $\texttt{0}$ and $\texttt{4}$, the resulting
    arrays satisfy
    \[
      \ColoredInversionCount{C}{W^{\rm add}}-
      \ColoredInversionCount{C}{W^{\rm sub}}
        =2\sum_{i=1}^{q}n_i,
    \]
    where $n_i=|\{j\in(b_i\dd e_i]:Q_i\text{ is a prefix of }S[j]\}|$.

  \item We now compute the integer arrays $A^{\rm add}$ and
    $A^{\rm sub}$ by replacing the strings with their lexicographic ranks.
    To this end, for each $s\in\{{\rm add},{\rm sub}\}$, construct
    the packed representations of the strings
    \[
      Z_j=W^s[j]\cdot\texttt{0}^{\ell+1-|W^s[j]|}
        \qquad(j\in[1\dd m'])
    \]
    using \cref{pr:packed-representation}.
    Every string in $W^s$ belongs to
    $\{\texttt{2},\texttt{3}\}^*\{\texttt{0},\texttt{1},\texttt{4}\}$.
    No two distinct strings in this language are prefixes of one another.
    Thus, appending zeros preserves all their lexicographic
    comparisons and equalities.
    Apply \cref{pr:packed-string-sorting} to
    $Z_1,\ldots,Z_{m'}$, and let $A^s$ be the returned rank array.
    Constructing and sorting these representations takes
    $\bigO(m')=\bigO(m)$ time because their common string length is
    $\ell+1=\bigO(\log_5 m')$.
    Each entry $A^s[j]$ is the number of distinct strings in $W^s$
    lexicographically smaller than $W^s[j]$.
    Thus, the entries belong to $[0\dd m')$ and preserve all
    strict comparisons and equalities among the strings in $W^s$.
    Both replacements preserve the respective colored inversion
    counts, giving
    \[
      \ColoredInversionCount{C}{A^{\rm add}}-
      \ColoredInversionCount{C}{A^{\rm sub}}
        =2\sum_{i=1}^{q}n_i.
    \]

  \item Finally, we determine whether any query matches.
    Make a copy $C'$ of $C$ to preserve the color array for the
    second call. Run the assumed algorithm on $(m',C',A^{\rm add})$
    and $(m',C,A^{\rm sub})$, and let $z_{\rm add}$ and $z_{\rm sub}$
    denote the returned counts, respectively.
    Return \texttt{YES} if $z_{\rm add}>z_{\rm sub}$, and
    \texttt{NO} otherwise. The identity above proves correctness.
  \end{enumerate}

  Apply this procedure to every returned Range Prefix Search input,
  and return \texttt{YES} if at least one has answer \texttt{YES},
  and \texttt{NO} otherwise. Correctness for the original input follows
  from \cref{pr:rps-size-reduction}.
  The lifting structure uses source alphabet size $2$ and target
  alphabet size $5$, and the sorting calls use alphabet size $5$.
  The size reduction and the calls to the assumed algorithm use size
  parameter $N$.

  Since $m\geq2$ and $\ell\leq m$, we have $5\leq M$ and
  $\ell+1\leq M$. Moreover, $L\geq m\ell\geq2m$ and $c\geq2$ give
  $M=3m\leq3L/2\leq3N/(2c)<N$.
  Hence, $w\geq1+\log N\geq1+\log M$.
  Thus, all packed-string tools satisfy their size and word-size assumptions.

  We now verify that both Counting Colored Inversions instances
  have at most $N$ bits.
  Each instance occupies $(2m'+1)w$ bits, including its length
  and both arrays. Since $\log m\leq\ell\leq m$ and
  $L\geq m\ell\geq4$, monotonicity of $x/\log x$ for $x\geq4$ gives
  $L/\log L\geq m\ell/\log(m\ell)\geq m/2$.
  Hence, $2m'+1=6m+1\leq13L/\log L$.
  For sufficiently large $N$, we have $N/c\geq4$ and
  $\log N\leq2\log(N/c)$. Since $4\leq L\leq N/c$ and
  $x/\log x$ is increasing for $x\geq4$, each instance occupies at most
  \[
    (2m'+1)w\leq\frac{13Lw}{\log L}
      \leq\frac{13(N/c)w}{\log(N/c)}
      \leq\frac{26dN}{c}=N
  \]
  bits. Both calls are therefore valid.
  Each returned count fits in two words and can be compared with
  the other count in $\bigO(1)$ time
  (\cref{rm:inversion-count-representation}).

  The initial application of \cref{pr:rps-size-reduction} takes
  $\bigO(m_0)=\bigO(L_0/\log L_0)=\bigO(N/\log N)$ time and words
  of space, including the original input and all returned inputs.
  For each returned input, the four-step procedure takes
  $\bigO(m)=\bigO(L/\log L)=\bigO(N/\log N)$ time and words of space
  apart from the two calls to the assumed algorithm.
  This includes the copy of $C$.
  There are $\bigO(1)$ returned inputs, and we process them one at a time.
  Thus, the total time is $\bigO(N/\log N+T_{\rm CCI}(N))$.
  Since $w=\bigO(\log N)$, the total space is
  $\bigO(N+S_{\rm CCI}(N))$ bits.
  By \cref{lm:cci-lower-bounds}, we have
  $T_{\rm CCI}(N)=\Omega(N/\log N)$ and
  $S_{\rm CCI}(N)=\Omega(N)$. Thus, the total time is
  $\bigO(T_{\rm CCI}(N))$ and the total space is
  $\bigO(S_{\rm CCI}(N))$ bits, as claimed.
\end{proof}

\section{Equivalence of Counting Inversions and Counting Colored Inversions}\label{sec:equiv-ci-and-cci}

\subsection{Problem Definition}\label{sec:ci-problem}

\begin{problem}[Counting Inversions]\label{prob:ci}
  \begin{description}[style=sameline,itemsep=0ex,font={\normalfont\bf}]
  \item[Input:]
    Let $n \in \Z_{\geq 1}$. Consider the word RAM model with word size
    $w > \log n$. The input consists of $n$ and an array $A[1 \dd n]$, stored
    with one entry per machine word, such that $A[i] \in [0 \dd n)$
    for every $i \in [1 \dd n]$.
  \item[Output:]
    The number of inversions in $A$, i.e., the value
    $|\{(i,j) \in [1 \dd n]^2 : i<j\text{ and }A[i] > A[j]\}|$.
  \end{description}
\end{problem}

\subsection{Preliminaries}\label{sec:equiv-ci-and-cci-preliminaries}

\begin{lemma}[Lower bounds for Counting Inversions]\label{lm:ci-lower-bounds}
  Let $N\in\Z_{\geq2}$. Consider the word RAM model with word size
  $w$ satisfying $1+\log N\leq w=\bigO(\log N)$.
  Any algorithm for \probname{Counting Inversions}
  (\cref{prob:ci}) that, given $N$, solves every valid input of
  at most $N$ bits requires
  $\Omega(N/\log N)$ time and $\Omega(N)$ bits of space
  in the worst case.
\end{lemma}
\begin{proof}
  For sufficiently large $N$, let
  $r=\lfloor N/(4w)\rfloor\geq2$.
  Consider an all-zero array $B[1\dd r]$. Together with its length,
  this is a valid Counting Inversions input of $(r+1)w\leq N$ bits.
  Its inversion count is zero. Changing $B[i]$ to one for any
  $i\in[1\dd r)$ creates the inversion $(i,r)$.
  Hence, an algorithm must inspect all of the first $r-1$ words
  of $B$ on the original input.
  Since $r=\Theta(N/w)=\Theta(N/\log N)$, this proves the time lower bound.
  The input occupies $\Theta(rw)=\Theta(N)$ bits, which gives
  the claimed space lower bound (see \cref{sec:model}).
\end{proof}

\subsection{Reducing Counting Colored Inversions to Counting Inversions}\label{sec:cci-to-ci}

\subsubsection{Combinatorial Results}\label{sec:cci-to-ci-combinatorial-results}

\begin{lemma}\label{lm:colored-inv-to-inv}
  Let $m\in\Z_{\geq0}$, and let $\mathcal{X}$ be a set equipped with a total order $\prec$.
  Consider an array $A[1 \dd m]$ of elements from $\mathcal{X}$
  and let $C[1 \dd m]$ be such that $C[i] \in \{0,1\}$ for all $i \in [1 \dd m]$.
  For any $b \in \{0,1\}$, denote $\mathcal{I}_b = \{i \in [1 \dd m] : C[i] = b\}$
  and $A_b = \DeleteSubseq{A}{\mathcal{I}_{1-b}}$ (\cref{def:delete}).
  Then, it holds (see \cref{def:inv-count,def:colored-inv-count}):
  \begin{align*}
    \ColoredInversionCount{C}{A}
      &= \InversionCount{A} - (\InversionCount{A_0} + \InversionCount{A_1}).
  \end{align*}
\end{lemma}
\begin{proof}
  The inversions in $A$ whose entries have different colors are counted by
  $\ColoredInversionCount{C}{A}$. For each $b\in\{0,1\}$, the array $A_b$
  contains exactly the entries of $A$ with color $b$, in their original order.
  Hence, its inversions correspond bijectively to inversions in $A$ whose
  entries both have color $b$. These three classes partition the inversions
  in $A$. It follows that $\InversionCount{A}=\ColoredInversionCount{C}{A}
  +\InversionCount{A_0}+\InversionCount{A_1}$, which implies the claim.
\end{proof}

\subsubsection{Algorithms}\label{sec:cci-to-ci-algorithms}

\begin{theorem}\label{th:cci-to-ci}
  Let $N\in\Z_{\geq2}$. Consider the word RAM model with word size
  $w$ satisfying $1+\log N\leq w=\bigO(\log N)$. Suppose that there
  exists an algorithm for \probname{Counting Inversions}
  (\cref{prob:ci}) that, given $N$ and any valid input of at most
  $N$ bits, achieves the following complexities:
  \begin{itemize}
  \item running time $T_{\rm CI}(N)$,
  \item space usage $S_{\rm CI}(N)$ bits.
  \end{itemize}
  Then, there exists an algorithm
  for \probname{Counting Colored Inversions}
  (\cref{prob:cci}) that, given $N$ and any valid input of at most
  $N$ bits, achieves the following complexities:
  \begin{itemize}
  \item running time $\bigO(T_{\rm CI}(N))$,
  \item space usage $\bigO(S_{\rm CI}(N))$ bits.
  \end{itemize}
\end{theorem}
\begin{proof}
  Consider a valid Counting Colored Inversions input of at most $N$ bits,
  consisting of $n$ and arrays $C[1\dd n]$ and $A[1\dd n]$.
  The algorithm proceeds as follows:
  \begin{enumerate}

  \item\label{step:cci-to-ci-arrays} First, we construct arrays $B_0[1\dd n]$ and $B_1[1\dd n]$
    by prepending zeros to the two color subsequences.
    To this end, we count the entries of each color and denote
    their number by $n_c$ for $c\in\{0,1\}$.
    For each $c\in\{0,1\}$, construct $B_c$ by first writing
    $n-n_c$ zeros and then copying the values $A[i]$ with $C[i]=c$,
    scanning $A$ and $C$ in increasing index order.
    Prepending these zeros does not change the inversion count
    of the copied subsequence, because all entries of $A$ are
    nonnegative. This also applies when $n_c=0$.
    Constructing both arrays takes $\bigO(n)$ time and uses
    $\bigO(n)$ words of space.

  \item We now compute the three inversion counts using the assumed
    algorithm on $A$, $B_0$, and $B_1$.
    Denote the returned counts by $x$, $x_0$, and $x_1$, respectively.
    We return
    \[
      x-(x_0+x_1).
    \]
    By \cref{lm:colored-inv-to-inv} and the padding property
    from Step~\ref{step:cci-to-ci-arrays}, this is $\ColoredInversionCount{C}{A}$.
    The three calls take $\bigO(T_{\rm CI}(N))$ time and use
    $\bigO(S_{\rm CI}(N))$ bits of space.
  \end{enumerate}
  The original input occupies $M=(2n+1)w\leq N$ bits,
  including the length parameter $n$.
  Since all values lie in $[0\dd n)$, padding the color subsequences
  to length $n$ makes them valid Counting Inversions inputs.
  The three calls use size parameter $N$: each input, including its
  length parameter, occupies $(n+1)w\leq M\leq N$ bits.
  The same word size $w$ is used throughout. Since $n\leq N$,
  all array lengths and values fit in one word.
  Each count fits in two words (\cref{rm:inversion-count-representation}).
  Moreover, $x_0+x_1\leq x$ by \cref{lm:colored-inv-to-inv}.
  Thus, both the sum and the nonnegative difference fit in two words,
  and the addition and subtraction take $\bigO(1)$ time.
  As $nw\leq M\leq N$, we have
  $n=\bigO(N/w)=\bigO(N/\log N)$.
  The total time is therefore $\bigO(N/\log N+T_{\rm CI}(N))$,
  and the total space is $\bigO(N+S_{\rm CI}(N))$ bits,
  including the original input, constructed arrays, and returned counts.
  By \cref{lm:ci-lower-bounds}, we have
  $T_{\rm CI}(N)=\Omega(N/\log N)$ and
  $S_{\rm CI}(N)=\Omega(N)$. Thus, the total time is
  $\bigO(T_{\rm CI}(N))$ and the total space is
  $\bigO(S_{\rm CI}(N))$ bits, as claimed.
\end{proof}

\subsection{Reducing Counting Inversions to Counting Colored Inversions}\label{sec:ci-to-cci}

\subsubsection{Combinatorial Results}\label{sec:ci-to-cci-combinatorial-results}

\begin{lemma}\label{lm:inv-to-colored-inv}
  Let $m\in\Z_{\geq0}$, and let $\mathcal{X}$ be a set equipped with a total order $\prec$.
  Consider an array $A[1 \dd m]$ of elements from $\mathcal{X}$.
  Let $C[1 \dd 2m]$ and $A'[1 \dd 2m]$ be such that, for every
  $i \in [1 \dd m]$, it holds
  $C[2i-1] = 0$, $C[2i] = 1$, and $A'[2i-1] = A'[2i] = A[i]$.
  Then, it holds:
  \begin{align*}
    \InversionCount{A} &= \frac{1}{2} \cdot \ColoredInversionCount{C}{A'}.
  \end{align*}
\end{lemma}
\begin{proof}
  Since $A'[2i-1]=A'[2i]$ for every $i\in[1\dd m]$, no inversion
  in $A'$ uses both copies of the same entry of $A$.
  For any $i,j\in[1\dd m]$ with $i<j$, both copies of $A[i]$ precede
  both copies of $A[j]$ in $A'$, since $2i<2j-1$.
  Of the four pairs formed by these copies, exactly $(2i-1,2j)$ and
  $(2i,2j-1)$ have different colors. Each of these two pairs is an
  inversion if and only if $A[i]\succ A[j]$.
  Every colored inversion in $(C,A')$ arises from exactly one such pair
  $(i,j)$. Consequently, $\ColoredInversionCount{C}{A'}=2\cdot\InversionCount{A}$,
  proving the claim.
\end{proof}

\subsubsection{Algorithms}\label{sec:ci-to-cci-algorithms}

\begin{theorem}\label{th:ci-to-cci}
  Let $N\in\Z_{\geq2}$. Consider the word RAM model with word size
  $w$ satisfying $1+\log N\leq w=\bigO(\log N)$. Suppose that there
  exists an algorithm for \probname{Counting Colored Inversions}
  (\cref{prob:cci}) that, given $N$ and any valid input of at most
  $N$ bits, achieves the following complexities:
  \begin{itemize}
  \item running time $T_{\rm CCI}(N)$,
  \item space usage $S_{\rm CCI}(N)$ bits.
  \end{itemize}
  Then, there exists an algorithm
  for \probname{Counting Inversions}
  (\cref{prob:ci}) that, given $N$ and any valid input of at most
  $N$ bits, achieves the following complexities:
  \begin{itemize}
  \item running time $\bigO(T_{\rm CCI}(N))$,
  \item space usage $\bigO(S_{\rm CCI}(N))$ bits.
  \end{itemize}
\end{theorem}
\begin{proof}
  By \cref{lm:cci-size-extension} with $c=4$, the assumed algorithm
  yields an algorithm for Counting Colored Inversions on inputs of
  at most $4N$ bits, using the same word size, in
  $\bigO(T_{\rm CCI}(N))$ time and $\bigO(S_{\rm CCI}(N))$ bits
  of space. Consider a valid
  Counting Inversions input of at most $N$ bits, consisting
  of $n$ and an array $A[1\dd n]$.
  The algorithm proceeds as follows:
  \begin{enumerate}

  \item First, we construct the Counting Colored Inversions instance
    by duplicating each entry of $A$ and assigning different colors
    to its two copies.
    To this end, construct arrays $C[1\dd2n]$ and $A'[1\dd2n]$
    by setting, for every $i\in[1\dd n]$,
    \[
      A'[2i-1]=A'[2i]=A[i],\qquad C[2i-1]=0,\quad C[2i]=1.
    \]
    This takes $\bigO(n)$ time and uses $\bigO(n)$ words of space.

  \item Next, we compute the colored inversion count and recover
    the required answer.
    Run the algorithm obtained from \cref{lm:cci-size-extension}
    on $2n$, $C$, and $A'$, and denote the
    returned count by $x$. Return $x/2$.
    By \cref{lm:inv-to-colored-inv}, we have
    $x=2\cdot\InversionCount{A}$. Hence, $x$ is even and the
    returned value is $\InversionCount{A}$.
    The call takes $\bigO(T_{\rm CCI}(N))$ time and uses
    $\bigO(S_{\rm CCI}(N))$ bits of space.
  \end{enumerate}
  The original input occupies $M=(n+1)w\leq N$ bits,
  including the length parameter $n$.
  The constructed input is valid because its values belong to
  $[0\dd n)\subseteq[0\dd2n)$. Including its length parameter,
  it occupies $(4n+1)w\leq4M\leq4N$ bits, and hence the algorithm from
  \cref{lm:cci-size-extension} applies to it with size parameter $N$.
  The same word size $w$ is used throughout. Since $2n\leq N$,
  all array lengths and values fit in one word.
  The returned count fits in two words, and computing $x/2$ takes
  $\bigO(1)$ time (\cref{rm:inversion-count-representation}).
  As $nw\leq M\leq N$, we have
  $n=\bigO(N/w)=\bigO(N/\log N)$.
  The total time is therefore $\bigO(N/\log N+T_{\rm CCI}(N))$,
  and the total space is $\bigO(N+S_{\rm CCI}(N))$ bits,
  including the original input, constructed arrays, and returned count.
  By \cref{lm:cci-lower-bounds}, we have
  $T_{\rm CCI}(N)=\Omega(N/\log N)$ and
  $S_{\rm CCI}(N)=\Omega(N)$. Thus, the total time is
  $\bigO(T_{\rm CCI}(N))$ and the total space is
  $\bigO(S_{\rm CCI}(N))$ bits, as claimed.
\end{proof}

\subsection{Summary}\label{sec:ci-and-cci-summary}

\begin{theorem}\label{th:inv-problems-equivalence}
  Let $N\in\Z_{\geq2}$. In the word RAM model with word size $w$
  satisfying $1+\log N\leq w=\bigO(\log N)$, the problems of
  \probname{Counting} \probname{Inversions} (\cref{prob:ci}) and
  \probname{Counting} \probname{Colored} \probname{Inversions}
  (\cref{prob:cci}) are asymptotically equivalent:
  if there exists an algorithm for one of these two problems that,
  given $N$ and any valid input of at most $N$ bits,\footnote{\phantomsection\label{fn:ci-cci-input}
  Both problems take as input an integer $n\geq1$ and a value array
  $A[1\dd n]$ with entries in $[0\dd n)$.
  The colored variant additionally takes a color array $C[1\dd n]$
  with entries in $\{0,1\}$.
  Each input integer occupies one machine word.
  The size condition is $(n+1)w\leq N$ for Counting Inversions
  and $(2n+1)w\leq N$ for Counting Colored Inversions.}
  achieves the following
  complexities:
  \begin{itemize}
  \item running time $T(N)$,
  \item space usage $S(N)$ bits,
  \end{itemize}
  then there exists an algorithm
  for the other problem that, given $N$ and any valid input of at most $N$
  bits, achieves the following complexities:
  \begin{itemize}
  \item running time $\bigO(T(N))$,
  \item space usage $\bigO(S(N))$ bits.
  \end{itemize}
\end{theorem}
\begin{proof}
  An algorithm for \probname{Counting Inversions} yields an algorithm
  for \probname{Counting Colored Inversions} by \cref{th:cci-to-ci}.
  The reverse direction follows from \cref{th:ci-to-cci}.
\end{proof}

\paragraph*{AI Disclaimer.}
All algorithms and data structures were developed by the authors.
AI (OpenAI GPT-6) was used solely to help with proofreading,
refactoring, and exposition.

\bibliographystyle{alphaurl}
\bibliography{paper}

\newcommand{\etalchar}[1]{$^{#1}$}
\begin{thebibliography}{DKK{\etalchar{+}}04}

\bibitem[ABM08]{bwtbook}
Donald Adjeroh, Tim Bell, and Amar Mukherjee.
\newblock {\em The {B}urrows-{W}heeler Transform: Data Compression, Suffix
  Arrays, and Pattern Matching}.
\newblock Springer, New York, NY, USA, 2008.
\newblock \href {https://doi.org/10.1007/978-0-387-78909-5}
  {\path{doi:10.1007/978-0-387-78909-5}}.

\bibitem[AC75]{AhoC75}
Alfred~V. Aho and Margaret~J. Corasick.
\newblock Efficient string matching: An aid to bibliographic search.
\newblock {\em Communications of the {ACM}}, 18(6):333--340, 1975.
\newblock \href {https://doi.org/10.1145/360825.360855}
  {\path{doi:10.1145/360825.360855}}.

\bibitem[AZM{\etalchar{+}}02]{adjeroh2002dna}
Donald Adjeroh, Yong Zhang, Amar Mukherjee, Matt Powell, and Tim Bell.
\newblock {DNA} sequence compression using the {B}urrows-{W}heeler transform.
\newblock In {\em 1st {IEEE} Computer Society Bioinformatics Conference ({CSB}
  2002)}, pages 303--313. {IEEE} Computer Society, 2002.
\newblock \href {https://doi.org/10.1109/csb.2002.1039352}
  {\path{doi:10.1109/csb.2002.1039352}}.

\bibitem[BCR24]{BannaiCR24}
Hideo Bannai, Panagiotis Charalampopoulos, and Jakub Radoszewski.
\newblock Maintaining the size of {LZ77} on semi-dynamic strings.
\newblock In Shunsuke Inenaga and Simon~J. Puglisi, editors, {\em 35th Annual
  Symposium on Combinatorial Pattern Matching ({CPM} 2024)}, volume 296 of {\em
  Leibniz International Proceedings in Informatics ({LIPIcs})}, pages
  3:1--3:20. Schloss Dagstuhl -- Leibniz-Zentrum f{\"{u}}r Informatik, 2024.
\newblock \href {https://doi.org/10.4230/LIPIcs.CPM.2024.3}
  {\path{doi:10.4230/LIPIcs.CPM.2024.3}}.

\bibitem[BGK26]{BonehGK26}
Itai Boneh, Shay Golan, and Matan Kraus.
\newblock The complexity of dynamic {LZ77} is {$\widetilde{\Theta}(n^{2/3})$}.
\newblock In Kasper~Green Larsen and Barna Saha, editors, {\em 37th Annual
  {ACM-SIAM} Symposium on Discrete Algorithms ({SODA} 2026)}, pages 1859--1872.
  {SIAM}, 2026.
\newblock \href {https://doi.org/10.1137/1.9781611978971.67}
  {\path{doi:10.1137/1.9781611978971.67}}.

\bibitem[BGKS15]{WaveletSuffixTree}
Maxim Babenko, Pawe{\l} Gawrychowski, Tomasz Kociumaka, and Tatiana
  Starikovskaya.
\newblock Wavelet trees meet suffix trees.
\newblock In Piotr Indyk, editor, {\em 26th Annual {ACM-SIAM} Symposium on
  Discrete Algorithms ({SODA} 2015)}, pages 572--591. {SIAM}, 2015.
\newblock \href {http://arxiv.org/abs/1408.6182v4} {\path{arXiv:1408.6182v4}},
  \href {https://doi.org/10.1137/1.9781611973730.39}
  {\path{doi:10.1137/1.9781611973730.39}}.

\bibitem[BP16]{BelazzouguiP16}
Djamal Belazzougui and Simon~J. Puglisi.
\newblock Range predecessor and {L}empel-{Z}iv parsing.
\newblock In Robert Krauthgamer, editor, {\em 27th Annual {ACM-SIAM} Symposium
  on Discrete Algorithms ({SODA} 2016)}, pages 2053--2071. {SIAM}, 2016.
\newblock \href {http://arxiv.org/abs/1507.07080} {\path{arXiv:1507.07080}},
  \href {https://doi.org/10.1137/1.9781611974331.CH143}
  {\path{doi:10.1137/1.9781611974331.CH143}}.

\bibitem[BW94]{bwt}
Michael Burrows and David~J. Wheeler.
\newblock A block-sorting lossless data compression algorithm.
\newblock Technical Report 124, Digital Equipment Corporation, Palo Alto,
  California, 1994.
\newblock URL:
  \url{https://www.hpl.hp.com/techreports/Compaq-DEC/SRC-RR-124.pdf}.

\bibitem[CBJR12]{cox2012large}
Anthony~J. Cox, Markus~J. Bauer, Tobias Jakobi, and Giovanna Rosone.
\newblock Large-scale compression of genomic sequence databases with the
  {B}urrows--{W}heeler transform.
\newblock {\em Bioinformatics}, 28(11):\mbox{1415--1419}, 2012.
\newblock \href {https://doi.org/10.1093/bioinformatics/bts173}
  {\path{doi:10.1093/bioinformatics/bts173}}.

\bibitem[Cha88]{Chazelle88}
Bernard Chazelle.
\newblock A functional approach to data structures and its use in
  multidimensional searching.
\newblock {\em {SIAM} Journal on Computing}, 17(3):427--462, 1988.
\newblock \href {https://doi.org/10.1137/0217026} {\path{doi:10.1137/0217026}}.

\bibitem[CHL07]{AlgorithmsOnStrings}
Maxime Crochemore, Christophe Hancart, and Thierry Lecroq.
\newblock {\em Algorithms on Strings}.
\newblock Cambridge University Press, Cambridge, UK, 2007.
\newblock \href {https://doi.org/10.1017/cbo9780511546853}
  {\path{doi:10.1017/cbo9780511546853}}.

\bibitem[CI08]{CrochemoreI08}
Maxime Crochemore and Lucian Ilie.
\newblock Computing longest previous factor in linear time and applications.
\newblock {\em Information Processing Letters}, 106(2):75--80, 2008.
\newblock \href {https://doi.org/10.1016/J.IPL.2007.10.006}
  {\path{doi:10.1016/J.IPL.2007.10.006}}.

\bibitem[CKPR21]{Charalampopoulos21}
Panagiotis Charalampopoulos, Tomasz Kociumaka, Solon~P. Pissis, and Jakub
  Radoszewski.
\newblock Faster algorithms for longest common substring.
\newblock In Petra Mutzel, Rasmus Pagh, and Grzegorz Herman, editors, {\em 29th
  Annual European Symposium on Algorithms ({ESA} 2021)}, volume 204 of {\em
  Leibniz International Proceedings in Informatics ({LIPIcs})}, pages
  30:1--30:17. Schloss Dagstuhl -- Leibniz-Zentrum f{\"{u}}r Informatik, 2021.
\newblock \href {https://doi.org/10.4230/LIPIcs.ESA.2021.30}
  {\path{doi:10.4230/LIPIcs.ESA.2021.30}}.

\bibitem[Cla96]{Clark96}
David~R. Clark.
\newblock {\em Compact {Pat} Trees}.
\newblock PhD thesis, University of Waterloo, 1996.

\bibitem[CLZ02]{CrochemoreLZ02}
Maxime Crochemore, Gad~M. Landau, and Michal Ziv{-}Ukelson.
\newblock A sub-quadratic sequence alignment algorithm for unrestricted cost
  matrices.
\newblock In David Eppstein, editor, {\em 13th Annual {ACM-SIAM} Symposium on
  Discrete Algorithms ({SODA} 2002)}, pages 679--688. {SIAM}, 2002.
\newblock URL: \url{http://dl.acm.org/citation.cfm?id=545381.545472}.

\bibitem[CP10]{ChanP10}
Timothy~M. Chan and Mihai P{\u{a}}tra{\c{s}}cu.
\newblock Counting inversions, offline orthogonal range counting, and related
  problems.
\newblock In Moses Charikar, editor, {\em 21st Annual {ACM-SIAM} Symposium on
  Discrete Algorithms ({SODA} 2010)}, pages 161--173. {SIAM}, 2010.
\newblock \href {https://doi.org/10.1137/1.9781611973075.15}
  {\path{doi:10.1137/1.9781611973075.15}}.

\bibitem[CPR22]{CPR22}
Panagiotis Charalampopoulos, Solon~P. Pissis, and Jakub Radoszewski.
\newblock Longest palindromic substring in sublinear time.
\newblock In Hideo Bannai and Jan Holub, editors, {\em 33rd Annual Symposium on
  Combinatorial Pattern Matching ({CPM} 2022)}, volume 223 of {\em Leibniz
  International Proceedings in Informatics ({LIPIcs})}, pages 20:1--20:9.
  Schloss Dagstuhl -- Leibniz-Zentrum f{\"{u}}r Informatik, 2022.
\newblock \href {https://doi.org/10.4230/LIPICS.CPM.2022.20}
  {\path{doi:10.4230/LIPICS.CPM.2022.20}}.

\bibitem[CPS07]{ChenPS07}
Gang Chen, Simon~J. Puglisi, and William~F. Smyth.
\newblock Fast and practical algorithms for computing all the runs in a string.
\newblock In Bin Ma and Kaizhong Zhang, editors, {\em 18th Annual Symposium on
  Combinatorial Pattern Matching ({CPM} 2007)}, volume 4580 of {\em Lecture
  Notes in Computer Science}, pages 307--315. Springer, 2007.
\newblock \href {https://doi.org/10.1007/978-3-540-73437-6_31}
  {\path{doi:10.1007/978-3-540-73437-6_31}}.

\bibitem[DKK{\etalchar{+}}04]{DuvalKKLL04}
Jean{-}Pierre Duval, Roman~M. Kolpakov, Gregory Kucherov, Thierry Lecroq, and
  Arnaud Lefebvre.
\newblock Linear-time computation of local periods.
\newblock {\em Theoretical Computer Science}, 326(1-3):229--240, 2004.
\newblock \href {https://doi.org/10.1016/J.TCS.2004.06.024}
  {\path{doi:10.1016/J.TCS.2004.06.024}}.

\bibitem[DN23]{DiazDominguezN23}
Diego D{\'{\i}}az{-}Dom{\'{\i}}nguez and Gonzalo Navarro.
\newblock Efficient construction of the {BWT} for repetitive text using string
  compression.
\newblock {\em Information and Computation}, 294:105088, 2023.
\newblock \href {https://doi.org/10.1016/J.IC.2023.105088}
  {\path{doi:10.1016/J.IC.2023.105088}}.

\bibitem[EGG23]{EllertGG23}
Jonas Ellert, Pawe{\l} Gawrychowski, and Garance Gourdel.
\newblock Optimal square detection over general alphabets.
\newblock In Nikhil Bansal and Viswanath Nagarajan, editors, {\em 34th Annual
  {ACM-SIAM} Symposium on Discrete Algorithms ({SODA} 2023)}, pages 5220--5242.
  {SIAM}, 2023.
\newblock \href {https://doi.org/10.1137/1.9781611977554.CH189}
  {\path{doi:10.1137/1.9781611977554.CH189}}.

\bibitem[Ell23]{Ellert23}
Jonas Ellert.
\newblock Sublinear time {L}empel-{Z}iv {(LZ77)} factorization.
\newblock In Franco~Maria Nardini, Nadia Pisanti, and Rossano Venturini,
  editors, {\em 30th International Symposium on String Processing and
  Information Retrieval ({SPIRE} 2023)}, volume 14240 of {\em Lecture Notes in
  Computer Science}, pages 171--187. Springer, 2023.
\newblock \href {https://doi.org/10.1007/978-3-031-43980-3_14}
  {\path{doi:10.1007/978-3-031-43980-3_14}}.

\bibitem[ELMT19]{EgidiLMT19}
Lavinia Egidi, Felipe~A. Louza, Giovanni Manzini, and Guilherme~P. Telles.
\newblock External memory {BWT} and {LCP} computation for sequence collections
  with applications.
\newblock {\em Algorithms for Molecular Biology}, 14(1):6:1--6:15, 2019.
\newblock \href {https://doi.org/10.1186/S13015-019-0140-0}
  {\path{doi:10.1186/S13015-019-0140-0}}.

\bibitem[FGGK15]{FischerGGK15}
Johannes Fischer, Travis Gagie, Pawe{\l} Gawrychowski, and Tomasz Kociumaka.
\newblock Approximating {LZ77} via small-space multiple-pattern matching.
\newblock In Nikhil Bansal and Irene Finocchi, editors, {\em 23rd Annual
  European Symposium on Algorithms ({ESA} 2015)}, volume 9294 of {\em Lecture
  Notes in Computer Science}, pages 533--544. Springer, 2015.
\newblock \href {https://doi.org/10.1007/978-3-662-48350-3_45}
  {\path{doi:10.1007/978-3-662-48350-3_45}}.

\bibitem[FGM12]{FerraginaGM12}
Paolo Ferragina, Travis Gagie, and Giovanni Manzini.
\newblock Lightweight data indexing and compression in external memory.
\newblock {\em Algorithmica}, 63(3):707--730, 2012.
\newblock \href {https://doi.org/10.1007/S00453-011-9535-0}
  {\path{doi:10.1007/S00453-011-9535-0}}.

\bibitem[FH11]{FischerH11}
Johannes Fischer and Volker Heun.
\newblock Space-efficient preprocessing schemes for range minimum queries on
  static arrays.
\newblock {\em {SIAM} Journal on Computing}, 40(2):465--492, 2011.
\newblock \href {https://doi.org/10.1137/090779759}
  {\path{doi:10.1137/090779759}}.

\bibitem[FM05]{FerraginaM05}
Paolo Ferragina and Giovanni Manzini.
\newblock Indexing compressed text.
\newblock {\em Journal of the {ACM}}, 52(4):552--581, 2005.
\newblock \href {https://doi.org/10.1145/1082036.1082039}
  {\path{doi:10.1145/1082036.1082039}}.

\bibitem[FNN20]{SepulvedaNN20}
Jos{\'{e}} Fuentes{-}Sep{\'{u}}lveda, Gonzalo Navarro, and Yakov Nekrich.
\newblock Parallel computation of the {B}urrows {W}heeler transform in compact
  space.
\newblock {\em Theoretical Computer Science}, 812:123--136, 2020.
\newblock \href {https://doi.org/10.1016/J.TCS.2019.09.030}
  {\path{doi:10.1016/J.TCS.2019.09.030}}.

\bibitem[FW65]{periodicitylemma}
Nathan~J. Fine and Herbert~S. Wilf.
\newblock Uniqueness theorems for periodic functions.
\newblock {\em Proceedings of the American Mathematical Society},
  16(1):109--114, 1965.
\newblock \href {https://doi.org/10.2307/2034009} {\path{doi:10.2307/2034009}}.

\bibitem[GHN20]{Gao0N20}
Younan Gao, Meng He, and Yakov Nekrich.
\newblock Fast preprocessing for optimal orthogonal range reporting and range
  successor with applications to text indexing.
\newblock In Fabrizio Grandoni, Grzegorz Herman, and Peter Sanders, editors,
  {\em 28th Annual European Symposium on Algorithms ({ESA} 2020)}, volume 173
  of {\em Leibniz International Proceedings in Informatics ({LIPIcs})}, pages
  54:1--54:18. Schloss Dagstuhl -- Leibniz-Zentrum f{\"{u}}r Informatik, 2020.
\newblock \href {https://doi.org/10.4230/LIPICS.ESA.2020.54}
  {\path{doi:10.4230/LIPICS.ESA.2020.54}}.

\bibitem[GJKT24]{quantumlz}
Daniel Gibney, Ce~Jin, Tomasz Kociumaka, and Sharma~V. Thankachan.
\newblock Near-optimal quantum algorithms for bounded edit distance and
  {L}empel-{Z}iv factorization.
\newblock In David~P. Woodruff, editor, {\em 35th Annual {ACM-SIAM} Symposium
  on Discrete Algorithms ({SODA} 2024)}, pages 3302--3332. {SIAM}, 2024.
\newblock \href {https://doi.org/10.1137/1.9781611977912.118}
  {\path{doi:10.1137/1.9781611977912.118}}.

\bibitem[GNP20]{Gagie2020}
Travis Gagie, Gonzalo Navarro, and Nicola Prezza.
\newblock Fully functional suffix trees and optimal text searching in
  {BWT}-runs bounded space.
\newblock {\em Journal of the {ACM}}, 67(1):2:1--2:54, 2020.
\newblock \href {https://doi.org/10.1145/3375890} {\path{doi:10.1145/3375890}}.

\bibitem[GS04]{GusfieldS04}
Dan Gusfield and Jens Stoye.
\newblock Linear time algorithms for finding and representing all the tandem
  repeats in a string.
\newblock {\em Journal of Computer and System Sciences}, 69(4):525--546, 2004.
\newblock \href {https://doi.org/10.1016/j.jcss.2004.03.004}
  {\path{doi:10.1016/j.jcss.2004.03.004}}.

\bibitem[Hag98]{Hagerup98}
Torben Hagerup.
\newblock Sorting and searching on the word {RAM}.
\newblock In Michel Morvan, Christoph Meinel, and Daniel Krob, editors, {\em
  15th Annual Symposium on Theoretical Aspects of Computer Science ({STACS}
  1998)}, volume 1373 of {\em Lecture Notes in Computer Science}, pages
  366--398. Springer, 1998.
\newblock \href {https://doi.org/10.1007/BFb0028575}
  {\path{doi:10.1007/BFb0028575}}.

\bibitem[HLN22]{HanLN22}
Ling~Bo Han, Bin Lao, and Ge~Nong.
\newblock Succinct parallel {L}empel-{Z}iv factorization on a multicore
  computer.
\newblock {\em The Journal of Supercomputing}, 78(5):7278--7303, 2022.
\newblock \href {https://doi.org/10.1007/S11227-021-04165-W}
  {\path{doi:10.1007/S11227-021-04165-W}}.

\bibitem[Jac89]{Jac89}
Guy Jacobson.
\newblock Space-efficient static trees and graphs.
\newblock In {\em 30th Annual {IEEE} Symposium on Foundations of Computer
  Science ({FOCS} 1989)}, pages 549--554. {IEEE} Computer Society, 1989.
\newblock \href {https://doi.org/10.1109/SFCS.1989.63533}
  {\path{doi:10.1109/SFCS.1989.63533}}.

\bibitem[Kem19]{Kempa19}
Dominik Kempa.
\newblock Optimal construction of compressed indexes for highly repetitive
  texts.
\newblock In Timothy~M. Chan, editor, {\em 30th Annual {ACM-SIAM} Symposium on
  Discrete Algorithms ({SODA} 2019)}, pages 1344--1357. {SIAM}, 2019.
\newblock \href {https://doi.org/10.1137/1.9781611975482.82}
  {\path{doi:10.1137/1.9781611975482.82}}.

\bibitem[KK99]{KolpakovK99}
Roman~M. Kolpakov and Gregory Kucherov.
\newblock Finding maximal repetitions in a word in linear time.
\newblock In {\em 40th Annual {IEEE} Symposium on Foundations of Computer
  Science ({FOCS} 1999)}, pages 596--604. {IEEE} Computer Society, 1999.
\newblock \href {https://doi.org/10.1109/SFFCS.1999.814634}
  {\path{doi:10.1109/SFFCS.1999.814634}}.

\bibitem[KK00]{KolpakovK00}
Roman~M. Kolpakov and Gregory Kucherov.
\newblock Finding repeats with fixed gap.
\newblock In Pablo de~la Fuente, editor, {\em 7th International Symposium on
  String Processing and Information Retrieval ({SPIRE} 2000)}, pages 162--168.
  {IEEE} Computer Society, 2000.
\newblock \href {https://doi.org/10.1109/SPIRE.2000.878192}
  {\path{doi:10.1109/SPIRE.2000.878192}}.

\bibitem[KK03]{KolpakovK03}
Roman~M. Kolpakov and Gregory Kucherov.
\newblock Finding approximate repetitions under {H}amming distance.
\newblock {\em Theoretical Computer Science}, 303(1):135--156, 2003.
\newblock \href {https://doi.org/10.1016/S0304-3975(02)00448-6}
  {\path{doi:10.1016/S0304-3975(02)00448-6}}.

\bibitem[KK17]{KempaK17b}
Dominik Kempa and Dmitry Kosolobov.
\newblock {LZ}-{E}nd parsing in compressed space.
\newblock In Ali Bilgin, Michael~W. Marcellin, Joan Serra{-}Sagrist{\`{a}}, and
  James~A. Storer, editors, {\em Data Compression Conference ({DCC} 2017)},
  pages 350--359. {IEEE}, 2017.
\newblock \href {https://doi.org/10.1109/DCC.2017.73}
  {\path{doi:10.1109/DCC.2017.73}}.

\bibitem[KK19]{sss}
Dominik Kempa and Tomasz Kociumaka.
\newblock String synchronizing sets: Sublinear-time {BWT} construction and
  optimal {LCE} data structure.
\newblock In Moses Charikar and Edith Cohen, editors, {\em 51st Annual {ACM}
  {SIGACT} Symposium on Theory of Computing ({STOC} 2019)}, pages 756--767.
  {ACM}, 2019.
\newblock \href {http://arxiv.org/abs/1904.04228} {\path{arXiv:1904.04228}},
  \href {https://doi.org/10.1145/3313276.3316368}
  {\path{doi:10.1145/3313276.3316368}}.

\bibitem[KK20]{resolution}
Dominik Kempa and Tomasz Kociumaka.
\newblock Resolution of the {B}urrows-{W}heeler transform conjecture.
\newblock In Sandy Irani, editor, {\em 61st Annual {IEEE} Symposium on
  Foundations of Computer Science ({FOCS} 2020)}, pages 1002--1013. {IEEE},
  2020.
\newblock \href {https://doi.org/10.1109/FOCS46700.2020.00097}
  {\path{doi:10.1109/FOCS46700.2020.00097}}.

\bibitem[KK22]{dynsa}
Dominik Kempa and Tomasz Kociumaka.
\newblock Dynamic suffix array with polylogarithmic queries and updates.
\newblock In Stefano Leonardi and Anupam Gupta, editors, {\em 54th Annual {ACM}
  {SIGACT} Symposium on Theory of Computing ({STOC} 2022)}, pages 1657--1670.
  {ACM}, 2022.
\newblock \href {http://arxiv.org/abs/2201.01285} {\path{arXiv:2201.01285}},
  \href {https://doi.org/10.1145/3519935.3520061}
  {\path{doi:10.1145/3519935.3520061}}.

\bibitem[KK23a]{breaking}
Dominik Kempa and Tomasz Kociumaka.
\newblock Breaking the ${O(n)}$-barrier in the construction of compressed
  suffix arrays and suffix trees.
\newblock In Nikhil Bansal and Viswanath Nagarajan, editors, {\em 34th Annual
  {ACM-SIAM} Symposium on Discrete Algorithms ({SODA} 2023)}, pages 5122--5202.
  {SIAM}, 2023.
\newblock \href {http://arxiv.org/abs/2106.12725} {\path{arXiv:2106.12725}},
  \href {https://doi.org/10.1137/1.9781611977554.ch187}
  {\path{doi:10.1137/1.9781611977554.ch187}}.

\bibitem[KK23b]{collapsing}
Dominik Kempa and Tomasz Kociumaka.
\newblock Collapsing the hierarchy of compressed data structures: Suffix arrays
  in optimal compressed space.
\newblock In {\em 64th Annual {IEEE} Symposium on Foundations of Computer
  Science ({FOCS} 2023)}, pages 1877--1886. {IEEE}, 2023.
\newblock \href {http://arxiv.org/abs/2308.03635} {\path{arXiv:2308.03635}},
  \href {https://doi.org/10.1109/FOCS57990.2023.00114}
  {\path{doi:10.1109/FOCS57990.2023.00114}}.

\bibitem[KK24]{sublinearlz}
Dominik Kempa and Tomasz Kociumaka.
\newblock {L}empel-{Z}iv {(LZ77)} factorization in sublinear time.
\newblock In {\em 65th Annual {IEEE} Symposium on Foundations of Computer
  Science ({FOCS} 2024)}, pages 2045--2055. {IEEE}, 2024.
\newblock \href {http://arxiv.org/abs/2409.12146} {\path{arXiv:2409.12146}},
  \href {https://doi.org/10.1109/FOCS61266.2024.00122}
  {\path{doi:10.1109/FOCS61266.2024.00122}}.

\bibitem[KKP14]{KarkkainenKP14}
Juha K{\"{a}}rkk{\"{a}}inen, Dominik Kempa, and Simon~J. Puglisi.
\newblock {L}empel-{Z}iv parsing in external memory.
\newblock In Ali Bilgin, Michael~W. Marcellin, Joan Serra{-}Sagrist{\`{a}}, and
  James~A. Storer, editors, {\em Data Compression Conference ({DCC} 2014)},
  pages 153--162. {IEEE}, 2014.
\newblock \href {https://doi.org/10.1109/DCC.2014.78}
  {\path{doi:10.1109/DCC.2014.78}}.

\bibitem[KKR{\etalchar{+}}20]{KociumakaKRRW20}
Tomasz Kociumaka, Marcin Kubica, Jakub Radoszewski, Wojciech Rytter, and Tomasz
  Wale{\'{n}}.
\newblock A linear-time algorithm for seeds computation.
\newblock {\em {ACM} Transactions on Algorithms}, 16(2):27:1--27:23, 2020.
\newblock \href {https://doi.org/10.1145/3386369} {\path{doi:10.1145/3386369}}.

\bibitem[Kos15a]{Kosolobov15}
Dmitry Kosolobov.
\newblock Faster lightweight {L}empel-{Z}iv parsing.
\newblock In Giuseppe~F. Italiano, Giovanni Pighizzini, and Donald Sannella,
  editors, {\em 40th International Symposium on Mathematical Foundations of
  Computer Science ({MFCS} 2015)}, volume 9235 of {\em Lecture Notes in
  Computer Science}, pages 432--444. Springer, 2015.
\newblock \href {https://doi.org/10.1007/978-3-662-48054-0_36}
  {\path{doi:10.1007/978-3-662-48054-0_36}}.

\bibitem[Kos15b]{Kosolobov15b}
Dmitry Kosolobov.
\newblock {L}empel-{Z}iv factorization may be harder than computing all runs.
\newblock In Ernst~W. Mayr and Nicolas Ollinger, editors, {\em 32nd
  International Symposium on Theoretical Aspects of Computer Science ({STACS}
  2015)}, volume~30 of {\em Leibniz International Proceedings in Informatics
  ({LIPIcs})}, pages 582--593. Schloss Dagstuhl -- Leibniz-Zentrum f{\"{u}}r
  Informatik, 2015.
\newblock \href {https://doi.org/10.4230/LIPICS.STACS.2015.582}
  {\path{doi:10.4230/LIPICS.STACS.2015.582}}.

\bibitem[KVNP20]{KosolobovVNP20}
Dmitry Kosolobov, Daniel Valenzuela, Gonzalo Navarro, and Simon~J. Puglisi.
\newblock {L}empel-{Z}iv-like parsing in small space.
\newblock {\em Algorithmica}, 82(11):3195--3215, 2020.
\newblock \href {https://doi.org/10.1007/s00453-020-00722-6}
  {\path{doi:10.1007/s00453-020-00722-6}}.

\bibitem[KW05]{KleinW05}
Shmuel~Tomi Klein and Yair Wiseman.
\newblock Parallel {L}empel {Z}iv coding.
\newblock {\em Discrete Applied Mathematics}, 146(2):180--191, 2005.
\newblock \href {https://doi.org/10.1016/J.DAM.2004.04.013}
  {\path{doi:10.1016/J.DAM.2004.04.013}}.

\bibitem[Lar13]{larsen2013models}
Kasper~Green Larsen.
\newblock {\em Models and techniques for proving data structure lower bounds}.
\newblock PhD thesis, Aarhus University, 2013.
\newblock URL: \url{https://cs.au.dk/~larsen/papers/dissertation.pdf}.

\bibitem[LD09]{bwa}
Heng Li and Richard Durbin.
\newblock Fast and accurate short read alignment with {B}urrows-{W}heeler
  transform.
\newblock {\em Bioinformatics}, 25(14):1754--1760, 2009.
\newblock \href {https://doi.org/10.1093/bioinformatics/btp324}
  {\path{doi:10.1093/bioinformatics/btp324}}.

\bibitem[LS12]{bowtie2}
Ben Langmead and Steven~L. Salzberg.
\newblock Fast gapped-read alignment with {B}owtie 2.
\newblock {\em Nature Methods}, 9(4):357--359, 2012.
\newblock \href {https://doi.org/10.1038/nmeth.1923}
  {\path{doi:10.1038/nmeth.1923}}.

\bibitem[LZ76]{LZ76}
Abraham Lempel and Jacob Ziv.
\newblock On the complexity of finite sequences.
\newblock {\em {IEEE} Transactions on Information Theory}, 22(1):75--81, 1976.
\newblock \href {https://doi.org/10.1109/TIT.1976.1055501}
  {\path{doi:10.1109/TIT.1976.1055501}}.

\bibitem[Mah]{ltcb}
Matt Mahoney.
\newblock {L}arge {T}ext {C}ompression {B}enchmark.
\newblock \url{https://mattmahoney.net/dc/text.html}.
\newblock Accessed: 2026-10-26.

\bibitem[Mai89]{main1989detecting}
Michael~G. Main.
\newblock Detecting leftmost maximal periodicities.
\newblock {\em Discrete Applied Mathematics}, 25(1-2):145--153, 1989.
\newblock \href {https://doi.org/10.1016/0166-218X(89)90051-6}
  {\path{doi:10.1016/0166-218X(89)90051-6}}.

\bibitem[Man01]{Manzini01}
Giovanni Manzini.
\newblock An analysis of the {B}urrows-{W}heeler transform.
\newblock {\em Journal of the {ACM}}, 48(3):407--430, 2001.
\newblock \href {https://doi.org/10.1145/382780.382782}
  {\path{doi:10.1145/382780.382782}}.

\bibitem[MI05]{moffat2005word}
Alistair Moffat and R.~Yugo~Kartono Isal.
\newblock Word-based text compression using the {B}urrows--{W}heeler transform.
\newblock {\em Information Processing {\&} Management}, 41(5):1175--1192, 2005.
\newblock \href {https://doi.org/10.1016/j.ipm.2004.08.009}
  {\path{doi:10.1016/j.ipm.2004.08.009}}.

\bibitem[MNN20]{MunroNN20}
J.~Ian Munro, Gonzalo Navarro, and Yakov Nekrich.
\newblock Text indexing and searching in sublinear time.
\newblock In Inge~Li G{\o}rtz and Oren Weimann, editors, {\em 31st Annual
  Symposium on Combinatorial Pattern Matching ({CPM} 2020)}, volume 161 of {\em
  Leibniz International Proceedings in Informatics ({LIPIcs})}, pages
  24:1--24:15. Schloss Dagstuhl -- Leibniz-Zentrum f{\"{u}}r Informatik, 2020.
\newblock \href {https://doi.org/10.4230/LIPICS.CPM.2020.24}
  {\path{doi:10.4230/LIPICS.CPM.2020.24}}.

\bibitem[MNV16]{MunroNV16}
J.~Ian Munro, Yakov Nekrich, and Jeffrey~Scott Vitter.
\newblock Fast construction of wavelet trees.
\newblock {\em Theoretical Computer Science}, 638:91--97, 2016.
\newblock \href {https://doi.org/10.1016/j.tcs.2015.11.011}
  {\path{doi:10.1016/j.tcs.2015.11.011}}.

\bibitem[OS11]{OzsoyS11}
Adnan Ozsoy and D.~Martin Swany.
\newblock {CULZSS:} {LZSS} lossless data compression on {CUDA}.
\newblock In {\em {IEEE} International Conference on Cluster Computing
  ({CLUSTER} 2011)}, pages 403--411. {IEEE} Computer Society, 2011.
\newblock \href {https://doi.org/10.1109/CLUSTER.2011.52}
  {\path{doi:10.1109/CLUSTER.2011.52}}.

\bibitem[OSC14]{OzsoySC14}
Adnan Ozsoy, D.~Martin Swany, and Arun Chauhan.
\newblock Optimizing {LZSS} compression on {GPGPU}s.
\newblock {\em Future Generation Computer Systems}, 30:170--178, 2014.
\newblock \href {https://doi.org/10.1016/J.FUTURE.2013.06.022}
  {\path{doi:10.1016/J.FUTURE.2013.06.022}}.

\bibitem[OST{\etalchar{+}}18]{OhnoSTIS18}
Tatsuya Ohno, Kensuke Sakai, Yoshimasa Takabatake, Tomohiro I., and Hiroshi
  Sakamoto.
\newblock A faster implementation of online {RLBWT} and its application to
  {LZ77} parsing.
\newblock {\em Journal of Discrete Algorithms}, 52-53:18--28, 2018.
\newblock \href {https://doi.org/10.1016/J.JDA.2018.11.002}
  {\path{doi:10.1016/J.JDA.2018.11.002}}.

\bibitem[P{\u{a}}t08]{Patrascu08}
Mihai P{\u{a}}tra{\c{s}}cu.
\newblock {\em Lower bound techniques for data structures}.
\newblock PhD thesis, Massachusetts Institute of Technology, 2008.
\newblock URL: \url{https://hdl.handle.net/1721.1/45866}.

\bibitem[PP15]{PolicritiP15}
Alberto Policriti and Nicola Prezza.
\newblock Fast online {L}empel-{Z}iv factorization in compressed space.
\newblock In Costas~S. Iliopoulos, Simon~J. Puglisi, and Emine Yilmaz, editors,
  {\em 22nd International Symposium on String Processing and Information
  Retrieval ({SPIRE} 2015)}, volume 9309 of {\em Lecture Notes in Computer
  Science}, pages 13--20. Springer, 2015.
\newblock \href {https://doi.org/10.1007/978-3-319-23826-5_2}
  {\path{doi:10.1007/978-3-319-23826-5_2}}.

\bibitem[RR14]{rawat2014evaluation}
C.~D. Rawat and Smita Rao.
\newblock Evaluation of {B}urrows {W}heeler transform based image compression
  algorithm for multimedia applications.
\newblock In {\em International Conference on Advances in Communication and
  Computing Technologies ({ICACACT} 2014)}, pages 1--2. {IEEE}, 2014.
\newblock \href {https://doi.org/10.1109/eic.2015.7230717}
  {\path{doi:10.1109/eic.2015.7230717}}.

\bibitem[Ryt03]{Rytter03}
Wojciech Rytter.
\newblock Application of {L}empel--{Z}iv factorization to the approximation of
  grammar-based compression.
\newblock {\em Theoretical Computer Science}, 302(1-3):211--222, 2003.
\newblock \href {https://doi.org/10.1016/S0304-3975(02)00777-6}
  {\path{doi:10.1016/S0304-3975(02)00777-6}}.

\bibitem[Sad02]{cst}
Kunihiko Sadakane.
\newblock Succinct representations of {LCP} information and improvements in the
  compressed suffix arrays.
\newblock In David Eppstein, editor, {\em 13th Annual {ACM-SIAM} Symposium on
  Discrete Algorithms ({SODA} 2002)}, pages 225--232. {SIAM}, 2002.
\newblock URL: \url{http://dl.acm.org/citation.cfm?id=545381.545410}.

\bibitem[Sta12]{Starikovskaya12}
Tatiana Starikovskaya.
\newblock Computing {L}empel-{Z}iv factorization online.
\newblock In Branislav Rovan, Vladimiro Sassone, and Peter Widmayer, editors,
  {\em 37th International Symposium on Mathematical Foundations of Computer
  Science ({MFCS} 2012)}, volume 7464 of {\em Lecture Notes in Computer
  Science}, pages 789--799. Springer, 2012.
\newblock \href {https://doi.org/10.1007/978-3-642-32589-2_68}
  {\path{doi:10.1007/978-3-642-32589-2_68}}.

\bibitem[SZ13]{ShunZ13}
Julian Shun and Fuyao Zhao.
\newblock Practical parallel {L}empel-{Z}iv factorization.
\newblock In Ali Bilgin, Michael~W. Marcellin, Joan Serra{-}Sagrist{\`{a}}, and
  James~A. Storer, editors, {\em Data Compression Conference ({DCC} 2013)},
  pages 123--132. {IEEE}, 2013.
\newblock \href {https://doi.org/10.1109/DCC.2013.20}
  {\path{doi:10.1109/DCC.2013.20}}.

\bibitem[Vo09]{vo2009image}
Si~Van Vo.
\newblock Image compression using {B}urrows-{W}heeler transform.
\newblock Master's thesis, Helsinki University of Technology, 2009.
\newblock URL: \url{https://urn.fi/URN:NBN:fi:aalto-201203091380}.

\bibitem[ZH14]{ZuH14}
Yuan Zu and Bei Hua.
\newblock {GLZSS:} {LZSS} lossless data compression can be faster.
\newblock In John Cavazos, Xiang Gong, and David~R. Kaeli, editors, {\em 7th
  Workshop on General Purpose Processing Using {GPU}s ({GPGPU} 2014)}, pages
  46--53. {ACM}, 2014.
\newblock \href {https://doi.org/10.1145/2588768.2576785}
  {\path{doi:10.1145/2588768.2576785}}.

\bibitem[ZL77]{LZ77}
Jacob Ziv and Abraham Lempel.
\newblock A universal algorithm for sequential data compression.
\newblock {\em {IEEE} Transactions on Information Theory}, 23(3):337--343,
  1977.
\newblock \href {https://doi.org/10.1109/TIT.1977.1055714}
  {\path{doi:10.1109/TIT.1977.1055714}}.

\end{thebibliography}

\end{document}